\documentclass{book} 

\usepackage[top=2.5cm,bottom=3cm,inner=3.5cm,outer=2.5cm]{geometry}
\usepackage{changepage}   

\usepackage{float} 

\usepackage[ngerman,english]{babel}
\usepackage[utf8]{inputenc}

\usepackage{fourier} 

\usepackage{tocbibind} 

\usepackage{fancyhdr}
\fancypagestyle{plain}{%
	\fancyhf{}
	\fancyfoot[OR]{\thepage}
	\fancyfoot[EL]{\thepage}
}

\usepackage{titlesec}
\titleformat{\chapter}[display]
{\filleft\Large}
{\chaptername\ \thechapter}
{5ex}
{\bfseries\MakeUppercase}
\titlespacing*{\chapter}
{0pt}
{-15pt}
{60pt}

\usepackage{setspace} 
\usepackage{siunitx} 

\usepackage{url}
\usepackage{hyperref}

\usepackage{tocloft}
\cftsetpnumwidth{2em}

\usepackage{amsmath}
\usepackage{amssymb}
\usepackage{bm}		
\usepackage{bbold} 
\usepackage{empheq}  	
\allowdisplaybreaks[1] 

\usepackage{braket} 
\usepackage{mathtools}

\DeclarePairedDelimiter\braketsm{\langle}{\rangle}
\DeclarePairedDelimiterX\braketm[2]{\langle}{\rangle}{#1 \delimsize\vert #2} 

\usepackage{amsthm}

\newtheorem{algorithm}{Algorithm}

\usepackage{graphicx}
\usepackage{xcolor}
\usepackage[percent]{overpic}
\usepackage{caption}
\usepackage{enumitem} 
\usepackage{booktabs}

\usepackage[numbers]{natbib} %
\usepackage{bibentry}
\usepackage{csquotes}

\usepackage{1extra_commands}

\nobibliography*

\begin{document}

	\pagenumbering{Alph}
	\pagestyle{empty}
	\newgeometry{margin=2cm,top=3cm,bottom=2cm}
	
%
\begin{titlepage}
    \begin{center}
    \vspace*{\baselineskip}
%
    \rule{\textwidth}{2pt}\vspace*{-\baselineskip} 
    \\[2.6\baselineskip]
    %
    %
    { \fontsize{30pt}{12pt}\selectfont
    Many Electrons and the Photon Field \\[0.4\baselineskip] 
     \Large
     The many-body structure of\\[0.2\baselineskip] 
      nonrelativistic quantum electrodynamics
    }\\[1.0\baselineskip]
    %
    %
    %

    \rule{\textwidth}{2pt}\\[\baselineskip]
 	
 	\vspace*{2cm}
    
   {\large
   Florian Buchholz}\\[0.09cm]
   Max Planck Institute for the Structure and Dynamics of Matter\\[0.09cm]
   Basic Research Community for Physics \\[1.5cm]
   Dissertation approved by the Faculty II -- Mathematics and Natural Sciences\\
   of {Technische Universit\"at Berlin}\\
   in fulfillment of the requirements for the degree of \\
   {\scshape Doctor of Science}\\
   Dr.rer.nat.\\[0.5cm]
   Digital Edition
    \end{center}
    \vspace*{0.9\baselineskip}
%

\vspace*{10.5pt}

{\noindent \scshape Examination board:}

\begin{flushleft} 
{Chairwomen:} Prof. Dr. Ulrike Woggon  \\[0.2cm]%
{Referee:} Prof. Dr. Andreas Knorr\\[0.2cm]%
{Referee:} Dr. Michael Ruggenthaler\\[0.2cm]%
{Referee:} Prof. Dr. Angel Rubio\\[0.2cm]%
{Referee:} Prof. Dr. Dieter Bauer 
\end{flushleft}
\vspace*{1.1cm}
\begin{center}
Berlin 2021
\end{center}
%
%
    \vfill
 
  \end{titlepage}
	\restoregeometry

	\cleardoublepage
	\pagenumbering{roman}
	\pagestyle{fancy}

	\chapter*{Acknowledgements}
\vspace{-1.8cm}
The here presented work is the result of a long process that naturally involved many different people. They all contributed importantly to it, though in more or less explicit ways.

First of all, I want to express my gratitude to Chiara, all my friends and my (German and Italian) family. Thank you for being there!

Next, I want to name Michael Ruggenthaler, who did not only supervise me over many years, contributing crucially to the here presented research, but also has become a dear friend. I cannot imagine how my time as a PhD candidate would have been without you! In the same breath, I want to thank Iris Theophilou, who saved me so many times from despair over non-converging codes and other difficult moments. Also without you, my PhD would not have been the same.

Then, I thank Angel Rubio for his supervision and for making possible my unforgettable and formative time at the Max-Planck Institute for the Structure and Dynamics of Matter. There are few people who have the gift to make others feel so excited about physics, as Angel does.

Heiko Appel supervised my first steps in the scientific environment and remained also afterwards always close. His excitement for unusual questions and ideas greatly influenced my projects and work. Nobody inspired my idea of what science really is more than Markus Penz. He regularly shows with great confidence how to think out of the box and to question things beyond traditional boundaries. Then I want to name Micael Oliveira the great tamer of the Octopus. He taught me not only to implement, but to develop. I thank Henning Glawe for being patient with me and for using his supernatural powers to make the penguin regain his feet, whenever I made it fall. Warm thanks go to my dear friends Björn Bembnista and Wilhelm Bender for the many stimulating nights of discussion about physics and for sharing their great knowledge about coding and minimization. I also want to thank Vasilis Rokaj, Davis Welakuh, Christian Schäfer, Guillem Albareda, Arun Debnath, Nicolas Tancogne-Dejean, Florian Eich, Michael Sentef, Enrico Ronca, Massimo Altarelli and Martin Lüders for many interesting discussions. 

Warm thanks goes to Uliana Mordovina, Teresa Reinhard, Christian Schäfer, Mary-Leena Tchenkoue Djouom, Norah Hoffmann and Alexandra Göbel for sharing the precious moments with me, where we needed to forget science. I thank Fabio Covito, Simone Latini, Matteo Vandelli, and Enrico Ronca for sharing coffee and the feeling of sunny places. And I thank Ute Ramseger, Graciela Sanguino, Kathja Schroeder and Frauke Kleinwort for their big efforts to support us PhD students at the Institute.

Sarah Loos contributed not only to my scientific world view in many inspiring discussions, but also directly to this thesis by her indispensable feedback, understanding what I wanted to say before I did. I also want to thank David Licht and Chiara for valuable feedback on my thesis, and Uliana for help with the layout. Furthermore, I want to mention Michael Duszat and the ``deep readers,'' who made me aware on how much information only one sentences may contain and who were the very first audience for this thesis. I also want to thank all the many authors that have contributed to making the various \textsc{stackexchange} pages such a valuable platform with answers to so many difficult questions.

Finally, I want to mention the other members of the \emph{Basic Research Community For Physics}. It is so inspiring to know so many people who share the believe in a cooperative, respectful and open-minded scientific research inside and outside traditional institutions.\\
\includegraphics[width=3cm]{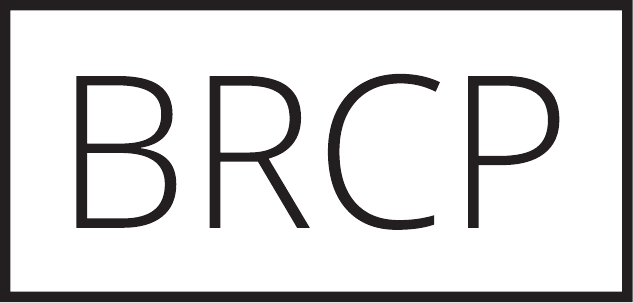} \hfill \includegraphics[width=3cm]{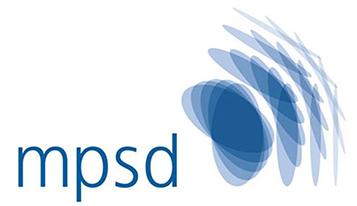}
	\cleardoublepage

  \selectlanguage{ngerman}
  \clearpage	
	 \chapter*{Zusammenfassung}
Neueste experimentelle Fortschritte im Bereich von ``Cavity''-Quantenelektrodynamik  ermöglichen die Erforschung der starken Wechselwirkung zwischen quantisiertem Licht und komplexen Materiesystemen. Aufgrund der kohärenten Kopplung zwischen Photonen und Materiefreiheitsgraden, entstehen Polaritonen, hybride Licht-Materie Quasiteilchen, die dazu beitragen können, Materieeigenschaften und komplexe Prozesse wie chemische Reaktionen entscheidend zu beeinflussen. Dieses Regime der starken Kopplung eröffnet Möglichkeiten zur Kontrolle von Materialien und Chemie in einer beispiellosen weise. Allerdings sind die genauen Mechanismen hinter vielen solcher Phänomene nicht vollständig verstanden. Ein wichtiger Grund dafür ist, dass das physikalische Problem oft mit äußerst vereinfachten Methoden beschrieben wird, wobei die Materie zu wenigen effektiven "Levels" reduziert wird. Akkuratere first-principles Methoden, die Photonen gleichwertig zu Elektronen behandeln entstehen nur langsam, da die Erforschung solcher Methoden sowohl durch die erhöhte Komplexität der kombinierten Elektron-Photon Wellenfunktionen, als auch dem Fakt, dass zwei verschiedene Teilchenspezies miteinbezogen werden müssen, aufgehalten wird.

In dieser Doktorarbeit schlagen wir vor diese Problem zu umgehen, indem das gekoppelte Elektron-Photon Problem exakt in einem anderen zweckgebauten Hilbertraum neuformuliert wird. Dadurch, dass wir ein  System, bestehend aus $N$ Elektronen und $M$ Moden, mit einer $N$-Polaritonen Wellenfunktion  repräsentieren, können wir explizit zeigen wie ein electronic-structure in eine polaritonic-structure Methode umgewandelt werden kann, die für schwache bis hin zu starken Kopplungstärken akkurat ist. Wir rationalisieren diesen Paradigmenwechsel innerhalb einer umfassenden Revision der Licht-Materie Wechselwirkung und indem wir die Verbindung zwischen verschiedenen electronic-structure Methoden und quantenoptischen Modellen hervorheben. Diese ausführliche Diskussion hebt hervor, dass die Polariton-Konstruktion nicht nur ein mathematischer Trick ist, sondern auf einem einfachen und physikalischem Argument basiert: wenn die Anregungen eines Systems einen hybriden Charakter haben, dann ist es nur natürlich, die zugehörige Theorie bezüglich dieser neuen Entitäten zu formulieren.

Schließlich diskutieren wir ausführlich, wie Standard-Algorithmen von electronic-structure Methoden angepasst werden müssen, um der neuen Fermi-Bose Statistik gerecht zu werden. Um die zugehörigen nichtlinearen Ungleichungs-Nebenbedingungen zu garantieren, sind sorgfältige Entwicklung, Implementierung und Validierung der numerischen Algorithmen nötig. Diese zusätzliche numerische Komplexität ist der Preis, den wir zahlen, um das gekoppelte Elektron-Photon Problem zugänglich zu first-principles Methoden zu machen. 
  \selectlanguage{english}
  \cleardoublepage

	\section*{Abstract}
Recent experimental progress in the field of cavity quantum electrodynamics allows to study the regime of strong interaction between quantized light and complex matter systems. Due to the coherent coupling between photons and matter-degrees of freedom, polaritons -- hybrid light-matter quasiparticles -- emerge, which can signi\-ficantly influence matter properties and complex process such as chemical reactions. This strong-coupling regime opens up possibilities to control materials and chemistry in an unprecedented way. However, the precise mechanisms behind many of these phenomena are not yet entirely understood. One important reason is that often the physical problem is described with highly simplified models, where the matter system is reduced to a few effective levels. More accurate first-principles approaches that consider photons on the same footing as electrons only slowly emerge. Their development is hampered by the increase of complexity of the combined electron-photon wave functions and the fact that we have to deal with two different species of particles.


In this thesis we propose a way to overcome these problems by reformulating the coupled electron-photon problem in an exact way in a different, purpose-build Hilbert space, where no longer electrons and photons are the basic physical entities but the polaritons. Representing an $N$-electron-$M$-mode system by an $N$-polariton wave function with hybrid Fermi-Bose statistics, we show explicitly how to turn electronic-structure methods into polaritonic-structure methods that are accurate from the weak to the strong-coupling regime.
We elucidate this paradigmatic shift by a comprehensive review of light-matter coupling, as well as by highlighting the connection between different electronic-structure methods and quantum-optical models. This extensive discussion accentuates that the polariton description is not only a mathematical trick, but it is grounded in a simple and intuitive physical argument: when the excitations of a system are hybrid entities
a formulation of the theory in terms of these new entities is natural.

Finally, we discuss in great detail how to adopt standard algorithms of 
electronic-structure methods to adhere to the new hybrid Fermi-Bose statistics. Guaranteeing the corresponding nonlinear inequality constraints in practice requires a careful development, implementation and validation of numerical algorithms. This extra numerical complexity is the price we pay for making the coupled matter-photon problem feasible for first-principle methods.  
	\cleardoublepage
	
	\chapter*{List of publications}
\addcontentsline{toc}{chapter}{List of publications}

A part of the results of my research as a PhD candidate have been published prior to this thesis. The following publications are part of this thesis:
\begin{itemize}
	\item[\citep{Buchholz2019}] Buchholz, F., Theophilou, I., Nielsen, S. E., Ruggenthaler, M., and Rubio, A.\\
		\emph{Reduced Density-Matrix Approach to Strong Matter-Photon Interaction}\\
		\textbf{ACS Photonics, American Chemical Society, 2019, 6, 2694}\\ \href{https://doi.org/10.1021/acsphotonics.9b00648}{DOI:10.1021/acsphotonics.9b00648}
	\item[\citep{Buchholz2020}] Buchholz, F., Theophilou, I., Giesbertz, K. J. H., Ruggenthaler, M., and Rubio, A.\\
	\emph{Light-matter hybrid-orbital-based first-principles methods: the influence of the polariton statistics}\\
	\textbf{J. Chem. Theory Comput., American Chemical Society, 2020, 16, 24}\\ 
	\href{https://doi.org/10.1021/acs.jctc.0c00469}{DOI:10.1021/acs.jctc.0c00469}
	\item[\citep{Tancogne-Dejean2020}] 
	 Tancogne-Dejean, N., Oliveira, M. J. et al.\\
	\emph{Octopus, a computational framework for exploring light-driven phenomena and quantum dynamics in extended and finite systems}\\
	\textbf{J. Chem. Phys., American Institute of Physics Inc., 2020, 152, 124119}\\
	Ch. 4: \emph{Dressed reduced density matrix functional theory for ultra-strongly coupled light-matter systems}\\
	Ch. 14: \emph{Conjugate gradient implementation in RDMFT}\\
	\href{https://doi.org/10.1063/1.5142502}{DOI:10.1063/1.5142502}
\end{itemize} 
The following publication is not part of this thesis:
\begin{itemize}
	\item[\citep{Theophilou2018}] Theophilou, I., Buchholz, F., Eich, F. G., Ruggenthaler, M., and Rubio, A.\\
	\emph{Kinetic-Energy Density-Functional Theory on a Lattice}\\
	\textbf{J. Chem. Theory Comput., American Chemical Society, 2018, 14, 4072}\\
	\href{https://doi.org/10.1021/acs.jctc.8b00292}{DOI:10.1021/acs.jctc.8b00292}
\end{itemize}
	\newpage	

	\chapter*{Remarks on notation and terminology}
For a better readability, we try whenever possible to refrain from abbreviating words. However, the following few abbreviations are used several times in this text (note that we will sometimes not always use the abbreviated form):
\begin{table}[H]
	\begin{tabular}{lll}
	NR-QED & \hphantom{xxx} & non-relativistic quantum electrodynamics\\
	QED & & quantum electrodynamics\\
	HF && Hartree-Fock \\
	DFT && density functional theory\\
	QEDFT && quantum-electrodynamical density functional theory\\
	KS && Kohn-Sham \\
	RDM && reduced density matrix\\
	RDMFT && reduced density matrix functional theory\\
	MF		&& mean field\\
	pXC 	&& photon-exchange-correlation (only part~\ref{sec:conclusion})
\end{tabular}
\end{table}

We want to comment on the terms electronic-structure theory, many-body theory,  and first-principles, which are used almost interchangeably. However, there is a hierarchy between them, that is the idea of a description of a system from \emph{first-principles} is to make use of as little knowledge as possible that is specific for the scenario. For instance, to describe the equilibrium properties of a Helium atom, the standard first-principles approach would consider a doubly-positively charged nucleus and two electrons. To describe such a 2-electron-1-nucleus system, we need methods that efficiently approximate the interaction between all the particles. We call this in general a many-body problem and the research area connected to that many-body theory. Thus, a first-principles description of microscopic systems is done with many-body methods. Importantly, we often can separate the electronic from the nuclear dynamics (Born-Oppenheimer approximation, see Sec.~\ref{sec:intro:nrqed_longwavelength}) and for a plethora of phenomena, it is sufficient to describe only the former part, i.e., the electronic structure accurately. This research area is generally called electronic-structure theory and it comprises most of the known many-body methods (see also Sec.~\ref{sec:est}).

We sometimes will quote from publications that are written in German. In this case, my translations are provided below (in italic letters).
	\newpage
	

	\tableofcontents
	\cleardoublepage
	
	\pagenumbering{arabic}

	\chapter*{Introduction}
\addcontentsline{toc}{chapter}{Introduction}
\markboth{\MakeUppercase{Introduction}} {\MakeUppercase{Introduction}}
In the natural sciences, the phenomenon of electricity and its relation with charged particles is a research topic at least since the 16th century,\footnote{See for example the very well-documented article on \textsc{Wikipedia}: \url{https://en.wikipedia.org/wiki/Electricity\#History}, \textit{accessed 12.06.2020}.} and its importance has since then continuously increased. In the 19th century, researchers started to understand the connection between charge and electric forces not only more quantitatively but also the relation of both to magnetism and even light. Nowadays, all these phenomena are understood as different aspects of the electromagnetic field, whose dynamics and interaction with charge is well described by one set of four coupled equations, named after \emph{James Clerk Maxwell}.\footnote{Naturally, there were many people involved in the discovery of these equations, but it was Maxwell who added a last term to make the system of equations consistent~\citep[part 6.3]{Jackson2007}. For more information on the history of Maxwell's equations and a well-readable introduction to the topic, we recommend the \textsc{Wikipedia} article and the references therein: \url{https://en.wikipedia.org/wiki/Maxwell's_equations}, \textit{accessed 12.06.2020}.} Rapid advances in experimental techniques at the turn of the twentieth century revealed that charge is a characteristic of \emph{all} materials -- not only of certain materials, as we used to think up until then. To the best of our today's knowledge, all matter consists of atoms, and all atoms consist of negatively charged electrons and positively charged nuclei.\footnote{According to the standard model of particle physics~\citep{Cottingham2007}, nuclei can be even further divided into smaller constituents, but this does not influence our statement.} 
To understand matter on these atomic scales, physicists had to give up the classical concept of point particles governed by Newton's laws, which is accurate only in the macroscopic world, and replace it by the set of tools and laws of quantum mechanics. For a consistent description, not only matter, but also the electromagnetic field had to be quantized and finally in 1938, Wolfgang Pauli and Markus Fierz formulated the theory that accounts for the full quantum nature of electrodynamics and charged particles on atomic scales~\citep{Pauli1938}. Today, we call this theory \emph{non-relativistic quantum electrodynamics} (NR-QED)\footnote{The term ``non-relativistic'' refers here to the matter description that does not include high-energy phenomena like particle creation or annihilation processes. See Ref.~\citep{Greiner2009}.} (or Pauli-Fierz theory after its developers). It is claimed that NR-QED describes  ``any physical phenomenon in between [gravity on the Newtonian level and nuclear- and high-energy physics], including life on Earth''~\citep[p. 157]{Spohn2004}. In other words, physicists have reduced all Life on Earth on the interaction between charged particles and the electromagnetic field.

However, as the full theory of NR-QED is too difficult for strict mathematical deductions, the limit cases have become much more important in the following years. For example, the quantized electromagnetic field, possibly controlled by some classical external charges or charge currents, is very-well understood today.\footnote{Note that this limit case is very important for the theoretical understanding of field quantization~\citep{Greiner2013}, but in practice, it is basically not relevant. The reason is that the quantum nature is barely visible without the interaction with matter. See for example the discussion in part 3 of the book by \citet{Keller2012}.}
A very active field of research is \emph{quantum optics}, where physicists investigate the interaction between simple models of matter and photons, i.e., the quanta of the electromagnetic field. This allows to study fundamental atomic processes, such as light emission and absorption, to characterize the quantum nature of light or to develop important devices, e.g., lasers or single-photon emitters. The crucial approximation behind quantum optical models is the simplification of the matter-degrees of freedom, which allows to study the photon field in detail.

If we instead perform the \emph{static limit}, that is we turn off the interaction to the quantized part of the electromagnetic field and neglect the dynamics of the heavy nuclei, we arrive at today's standard microscopic model of matter: it considers the atomic nuclei as classical charges that create electric potentials and bind the electrons. The electronic degrees of freedom are then governed by the laws of quantum mechanics, including their electric repulsion.\footnote{This model is in impressively many cases sufficient to understand the structure of atoms, molecules, but also condensed matter, their spectra and many more properties. However, there are phenomena such as molecular vibrations or the heat capacity of solids that require to take the dynamics and often also the quantum nature of the nuclei into account and there are generalizations of the model to account for that. The inner structure of the nuclei, e.g., the dynamics of protons, neutrons or even quarks instead influences very rarely the properties of matter in some direct way. It is usually sufficient to describe the atomic nucleus as effective point charge with a mass and a spin.}
To really follow this programme, i.e., to describe matter from first principles in practice, many further approximations are necessary. The static limit of the Pauli-Fierz theory, i.e., the Schrödinger theory of say $N$ interacting electrons is still too difficult for exact solutions if $N$ is large.\footnote{To the best of our knowledge, the largest system that has been described until today exactly, i.e., in the basis set limit consisted of $N=54$ electrons in a very special scenario~\citep{Shepherd2012}.} The reason is that the theory describes the state of the system by the \emph{many-body wave function}, which depends on $N$ coordinates. This means that the configuration space, i.e., the space of all these functions grows \emph{exponentially} with $N$, which makes it for larger $N$ \emph{entirely inaccessible} (we will discuss this in detail in Ch.~\ref{sec:est}). This is the so-called (quantum) \emph{many-body problem}, which is so severe that the Nobel laureate Walter Kohn questioned (for large systems) the legitimacy of the many-electron wave function (and thus of the whole theory) as a scientific concept~\citep{Kohn1999}. However, in the research field of \emph{electronic-structure theory} (or more general \emph{many-body theory}) many accurate approximation strategies and even alternative formulations have been developed to describe matter on the microscopic level, i.e., from \emph{first principles}.\footnote{Note that in the literature the term ab-initio is often used equivalently to first principles.} Nowadays, thanks to these methods and high performance computers, we can predict the structure of many molecules and crystalline solids, calculate many of their properties like excitation spectra, and even understand complex processes like chemical reactions.

Interestingly, quantum opticians and many-body theorists (and the same is true for others) have conducted their research without much overlap despite their common origin. Both communities study different aspects of the quantized theory of charged particles and the electromagnetic field. Only recently, this has changed and the full Pauli-Fierz theory started to raise renewed interest. One reason is that new experimental techniques allow nowadays to probe systems and parameter regimes, where many-body effects and the quantum nature of the electromagnetic field play a role~\citep{Ruggenthaler2018}. In this \emph{strong-coupling regime}, hybrid light-matter particles (so-called polaritons) emerge that are capable to modify the properties of the coupled system significantly in comparison to the separate subsystems. Applications include the possibility of building polariton lasers~\citep{Kena-Cohen2010}, the modification of chemical landscapes~\citep{Hutchison2012}, the control of long-range energy transfer between different matter systems~\citep{Coles2014a} or the emergence of entirely new states of matter~\citep{Ruggenthaler2018,Kiffner2019,Ashida2020}. In such scenarios, fundamental approximations of the traditional methods and models break down  and consequently ``many theoretical works [have been proposed] which diverge significantly in their predictions compared to experiments''~\citep{Hirai2020}.
This suggests to take the more general perspective of Pauli-Fierz theory and reevaluate in a less-biased way the assumptions and approximations of the standard methods. We believe that this is the natural task of a generalized first-principles approach that treats matter and photons on the same footing. 
 
To describe all the degrees of freedom of Pauli-Fierz theory from first principles, it is necessary to find ways to efficiently describe the interaction between electrons or more generally charged particles and photons (in a similar way as researchers in electronic-structure theory once have found ways to deal with the Coulomb interaction between electrons). It is clear that adding the quantized electromagnetic field to the already difficult many-electron problem is a very challenging task.
In fact, the many-body problem in Pauli-Fierz theory is considerably more severe than for matter systems (see Sec.~\ref{sec:qed_est:general}), which opens many research questions already at a very basic level. To get accustomed to the challenges that the new type of interaction poses, we therefore focus in this work on cavity QED, a limit case of NR-QED. We will study many-electron systems that are coupled merely via their dipole to one (or a few) modes of the electro-magnetic field.\footnote{Thus we ignore, e.g., the spatial dependence of the electron-photon interaction.} 
The dipole and few-mode approximation are very common in the field of cavity QED, where researchers study matter systems inside optical cavities, i.e., resonators that ``trap'' photons with selected frequencies~\citep{Dutra2005}.\footnote{In their simplest form, one can imagine a cavity as two (high quality) mirrors that are positioned with a certain distance opposed to each other. The distance selects a light-mode with a certain frequency and the corresponding photons are reflected back and forth very often before they can dissipate. Every time, they cross the volume of the cavity, the photons can interact with the matter system, which can be effectively described by an increased light-matter coupling strength.} 
Most of the aforementioned strong-coupling phenomena have been observed in such cavity settings.\footnote{Note that there are cavity experiments, which require a theoretical description beyond the dipole and few-mode approximation. However, in most cases these approximations are very accurate~\citep{Dutra2005}.}
A very important motivation for our work are some open debates in the only recently established field of \emph{polaritonic chemistry}~\citep{Ebbesen2016}. Here researchers study the (possibly considerable) influence of cavity photons on molecular systems and complex chemical process such as reactions. A reasonable first step to understand this influence, is to study the physical setting of cavity QED from first principles.\footnote{A ``full'' first-principles perspective would explicitly describe the cavity as a part of the matter system. This would however require to describe the electromagnetic field fully spatially resolved. See Ref.~\citep{Jestaedt2019}.} 
Importantly, this limit case of the full Pauli-Fierz theory exhibits already many fundamental issues and challenges of the coupled electron-photon problem and thus, defines a very good starting point for the development of new methods. We will hereby focus on equilibrium scenarios, which play an important role for the actual debates in polaritonic chemistry but are considerably less well studied than the time-dependent case~\citep{DeLiberato2017}.

One of the most important challenges for the description from first principles is the simultaneous inclusion of (at least) two particle species, e.g., electrons and photons. Most existing methods are geared to the accurate description of only one particle species, such as electrons and their interaction. For instance, most many-body methods explicitly take the particle statics, e.g., the Fermi-statistics of electrons into account. This is usually a very important part of an accurate description (see Ch.~\ref{sec:est}). Any kind of generalization of such methods to treat more species faces thus the problem to describe more degrees of freedom, which have different statistics (and other properties) and a different type of interaction. A prominent example is the accurate description of non-adiabatic effects between electrons and nuclei, which is an unresolved problem for many relevant scenarios~\cite{Baer2006}. We face a similar challenge in cavity QED, when we want to describe the physics of polaritons where matter and light degrees of freedom are strongly mixed. For instance, the emergence of polaritons induces correlation between the matter-degrees of freedom~\citep{Buchholz2020, Schaefer2019}, which can make the accurate description considerably more difficult (see Ch.~\ref{sec:qed_est}).

Motivated by the challenges of a multi-species description and the prominent role of polaritons in strong-coupling physics, we propose to reformulate the coupled electron-photon problem in a new purpose-built Hilbert space. The basic entities here are not anymore electrons and photons, but polaritons. This allows us to represent a system that consists of say $N$ electrons and $M$ photon modes in an exact manner by an $N$-polariton wave function that adheres to hybrid Fermi-Bose statistics. Importantly, the new Hamiltonian resembles structurally the Hamiltonian of an $N$-electron system, i.e., both operators consist of a one-body part (kinetic and potential energy) and a two-body part (interaction energy). This allows for a straightforward application of established electronic-structure methods to describe electrons and photons on the same footing and makes this mathematical reformulation of cavity QED practical.

The derivation of this dressed-orbital construction, on the one hand, can be explained purely with mathematical similarities. On the other hand, we can rationalize the approach by combining the basic principles shared by electronic-structure methods with the central insight from quantum-optical models: the fundamental entities in strongly-coupled light-matter systems are polaritons and a first-principles description should therefore be based on these physical degrees of freedom, instead of the individual electrons and photons. We demonstrate that this paradigmatic shift allows to capture the physics from the weak to the strong-coupling regime efficiently and accurately. The reason is that already simple approximations in polariton space correspond to nontrivial (multi-reference) approximations in the standard Hilbert space. Besides others, this allows to show that the (local) details of the electronic-structure strongly influence the effect of the light matter coupling. This is an example of a phenomenon that cannot be captured by usual model approaches. 

The price we pay for making the coupled matter-photon problem accessible to first-principles methods is that standard algorithms need to be extended to account for the new hybrid statistics of the polaritons. While straightforward in principle, the nonlinear nature of the resulting inequality constraints need a careful and test-intensive implementation to provide robust and accurate results. Yet our results show that polaritonic-structure calculations for real molecules are feasible and provide new insights for polaritonic chemistry and material science. With the ever refined control of chemical reactions and material properties by quantum cavities, the presented approach has the potential to become an important tool to design the next generation of cavity-controlled matter.

\vspace{0.5cm}
Before we conclude the introduction with a summary of the structure and contents of this thesis, we want to make a comment on the challenge and opportunity that comes along with the interdisciplinarity in the research on coupled matter-photon systems. 
For instance, from a quantum chemist's point of view, a molecule is a highly complex many-body system whose accurate description usually requires numerically very expensive methods. On the contrary, quantum opticians describe molecules almost exclusively as effective two- (or few-)level systems, which are not further specified. When scientists look at ``the same thing'' from such different perspectives, there is a great potential for misinterpretations and even conflict, which in turn may negatively affect chances of doing good research. At the same time, this challenge bears big opportunities to raise questions from unusual point of views, and rethink established concepts. For example, the debate on the ``correct'' model of a molecule suggests directly the interesting question, for which scenarios the ubiquitous two-level approximation is inadequate.  
Indeed, polaritonic-structure methods provide valuable information regarding this question (Sec.~\ref{sec:dressed:results}). 
At the same time, one should ask, when we can no longer ignore the quantum nature of light, as it is common place in quantum chemistry. Here quantum-optical models highlight under which conditions this assumption breaks down (and, e.g., polaritons emerge). To account for both perspectives, we will explain many standard concepts and tools of the respective research fields thoroughly and accompanied with easy examples. Necessarily, certain introductory parts of this thesis will therefore seem trivial for quantum opticians and others for electronic-structure theorists. We hope that this allows for a more comprehensive perspective on this interdisciplinary and exciting field.

\subsubsection*{Outline}
This thesis consists of three parts that reflect the different aspects of our research. The topic of part~\ref{sec:ch1} is the general theoretical analysis of coupled light-matter systems.
We start (Ch.~\ref{sec:intro}) with the introduction of the physical setting of cavity QED. We discuss the phenomenology of strongly coupled electron-photon systems, the standard way to understand their principal features, and the limitations of this perspective that is illustrated by concrete examples. We then discuss these standard approaches in more detail (quantum optics perspective) and define a framework that allows for a more general description from first principles. 
In the second chapter (Ch.~\ref{sec:est}), we turn to the static limit of NR-QED and introduce first-principles electronic-structure theory. We outline the challenges of the accurate description of many-electron systems and specifically discuss three specific approaches to deal with these. In the last chapter of part~\ref{sec:ch1} (Ch.~\ref{sec:qed_est}), we generalize these approaches to coupled light-matter systems and discuss them in detail. The analysis reveals why in particular in equilibrium scenarios, many powerful concepts of electronic-structure theory are less useful in the coupled setting. 

Motivated from the results of the analysis of the first part, we propose in part~\ref{sec:dressed} a new strategy to describe coupled light-matter problems by introducing a purpose-built Hilbert space (dressed-orbital construction, Ch.~\ref{sec:dressed:construction}). This approach allows to restructure the coupled electron-photon many-body space such that we can describe a system by a ``many-polariton'' wave function. We thus define polaritons as its own particle-species that has electronic (fermionic) and photonic (bosonic) degrees of freedom and consequently adheres to a Fermi-Bose hybrid statistics. In the polariton description, the coupled light-matter Hamiltonian resembles the electronic-structure Hamiltonian, which allows to generalize electronic-structure methods to the coupled problem in a very straightforward way (Ch.~\ref{sec:dressed:est}). We show this explicitly with the example of Hartree-Fock (HF) theory and reduced density matrix functional theory (RDMFT) and present first results for model systems in Ch.~\ref{sec:dressed:results}. Despite their reduced dimensionality, these example systems exhibit already a rich spectrum of nontrivial behavior that is accurately described by the newly proposed methods. This highlights the potential of the polariton description.

After the discussion of the theory and the results, we concentrate in part~\ref{sec:numerics} on the details of the numerical part of the research. The gain of making applicable first-principles methods to strongly-coupled light-matter systems is accompanied by the need for new algorithms and an increased numerical complexity. To do so, we present specific algorithms to solve the electronic HF and RDMFT equations in real space, including a newly developed conjugate-gradients algorithm (Ch.~\ref{sec:numerics:rdmft}). We then explain how to modify these algorithms to describe coupled-light matter systems by means of the dressed construction (Ch.~\ref{sec:numerics:dressed}). We present in great detail the validation of our implementation in the electronic-structure code \textsc{Octopus}~\citep{Tancogne-Dejean2020} and show how the results presented in part~\ref{sec:dressed} have been converged. This implementation was geared toward the two-polariton case. In Ch.~\ref{sec:numerics:hybrid}, we finish the numerical part by presenting an algorithm for the general case.

We finalize the work by presenting in part~\ref{sec:conclusion} the conclusions and perspective.
	\cleardoublepage

\part{Light, Matter and Strong Coupling} 
\label{sec:ch1}

\begin{aquote}{A.J. Coleman, 2007 \citep[Ch. 1]{Mazziotti2007}}
	Few distinctions in quantum mechanics are as important as that between fermions and bosons. [...] I do not have the authority to assert that God agrees with me as to the importance of this distinction, but I am sure that most happy humans will since, as noted by Eddington, if there were no fermions there would be no electrons, so no molecules, so no DNA, no humans!
\end{aquote}

\chapter{Strong light-matter coupling: experiments, theory and more theory} %
\chaptermark{Strong Coupling}
\label{sec:intro}
 
This chapter aims to motivate the need for first-principles approaches to describe strong-coupling phenomena and to define a theoretical framework that is suited to investigate such approaches. This framework is given by the non-relativistic Pauli-Fierz theory, i.e., an interacting quantum-field theory that allows to define the equilibrium properties of coupled many-electron-photon systems. The theory includes as a limit case the physical setting of cavity QED, which entails the dipole-approximation and the restriction to a few effective modes. This level of theory is still enough to capture the possibly strong modifications of atoms, molecules and solids due to the coupling to the modes of an optical cavity.
\section{What is strong light-matter coupling?}
\label{sec:intro:experiment_strong_coupling}

In electrodynamics, we can differentiate between several effects and interactions, whose strength depends on the physical setting.  Electric charges for instance attract or repel each other by the Coulomb force, which is the dominant interaction on atomic scales. It is thus impossible to understand the properties of condensed matter or molecular systems without the Coulomb interaction. However, when we want to study electrically neutral entities (like the atoms of a gas), Coulomb forces typically play a negligible role. Then, there are magnetic forces that are connected to electric currents or spins. Such interactions are crucial to understand phenomena such as ferromagnetism or the quantum-Hall effect, but do not play an important role in, e.g., spin-saturated (closed-shell) systems or in the thermodynamic equilibrium. Besides the role of the electromagnetic field as mediator of interaction, there is another important degree of freedom, which we call electromagnetic radiation or simply light.\footnote{Strictly speaking, light denotes electromagnetic radiation in a certain frequency (or wavelength) interval that can be perceived by the human eye. However, it has become customary to extend this definition and denote, e.g., the spectra with smaller and larger the wavelengths than the visible range as infra-red and ultra-violet light, respectively.} 
Since light can move freely, it is treated in the theoretical description as a separate entity that can interact with matter (charge) via absorption and emission processes. Usually, this interaction is so small in comparison to, e.g., Coulomb or magnetic forces that we can treat absorption and emission processes perturbatively. For example, this means that the emission of a photon by a matter system usually does not influence its properties, i.e., there is a negligible back reaction. This is expressed in the fact that the coupling constant between the free electromagnetic field and charged particles is small \emph{independently} of the system of units.\footnote{It is called the \emph{fine-structure} constant, which is dimensionless and has approximately the value $\alpha\approx1/137$. This statement is textbook knowledge (see, e.g., \citep{Greiner2009}) and to the best of our knowledge unquestioned for the time-scales that we are interested in here. Only for very large say geological or even cosmological time-scales, there are speculations about possible time-variations of $\alpha$. See for example \citep{Webb1999}.}

However, the combined efforts of researchers from many research communities have revealed that although difficult, it is indeed possible to overcome this fundamental limitation and reach strong interaction between light and matter. One possible way for that is due to the very large field strengths of modern ultra-short laser pulses. Striking nonlinear effects have been demonstrated, such as high harmonic generation~\citep{Eden2004}, strong-field ionization~\citep{Ivanov2005} or light-induced superconductivity~\citep{Mitrano2016}. 
A different path to effectively enhance the light-matter interaction is by employing longer (but not as strong) laser pulses with comparatively sharp frequencies. This leads to a periodic driving of the matter system and the related phenomena are subsumed under the term Floquet engineering~\citep{Oka2019}. Examples are the observation of Floquet-Bloch states on the surface of a topological insulator~\citep{Mahmood2016} or the light-induced anomalous Hall effect~\citep{McIver2020}. Importantly, both these pathways to reach strong interaction can be essentially understood in a semiclassical picture, where the quantized matter is driven by a classical external field. The reason is that the quantum fluctuations of the laser photons are negligible in comparison to the large field-strengths.

A complementary approach to reach strong interaction is to control the \emph{coupling} between light and the matter system. This makes (strong) modifications of matter properties possible for very small field strengths or even only the vacuum~\citep{Torma2015,Ebbesen2016}. Thus, strong-interaction phenomena can be studied, but without, e.g., the heating due to strong lasers and also with additional nontrivial quantum effects.
With the first breakthrough experiments~\citep{Lidzey1998,Fujita1998} only about two decades ago, it is a relatively young research topic, but because of its high potential for applications, the investigation of this \emph{strong-coupling} regime literally ``has exploded [...] in the past few years''~\citep{Barnes2018}. Nowadays, strong coupling has been demonstrated in various systems with not only distinct basic entities, i.e., the employed matter system and the degrees of freedom of field and matter that are coupled to each other, but even different mechanisms that allow to reach strong coupling. However, all these systems share two key features, which we can loosely define in the following way: for a matter system to reach strong coupling with certain modes of the electromagnetic field,
\begin{enumerate}
	\item these modes have to be \emph{confined} to very small volumes, and
	\item the matter system has to be chosen such that it \emph{responds especially strong} to these confined modes.
\end{enumerate}
It is difficult to make this definition more concrete, because there are on the one hand so many ways to reach strong coupling. On the other hand, many different communities participate in the research and there is not an ultimate consent on the correct definition of strong coupling.\footnote{As we will see in the following, the topic of strong coupling is located between many established research fields. This sometimes leads to situations, where scientists have to explain and defend basic concepts of their field to (in this regard) non-specialists. This can be a formidable task, as many long and heated discussions at conferences and in peer-review processes have testified.}

Missing such a consent, we start in the next paragraph with the (relatively) unquestioned part: the experimentally determined facts and their basic interpretation. According to this interpretation, all the phenomena of the strong-coupling regime are related by the emergence of hybrid light-matter quasi-particle states, called \emph{polaritons}. These determine the properties of the combined light-matter system, which can be significantly different than the properties of the separate subsystems. This mechanism can be understood impressively well by a minimal model. In fact, most of the strong-coupling effects can be classified according to the different parameters of this model.
These parameters were and still are a very important guideline for experimentalists to design new setups that reach the strong-coupling regime. 

However, the explanatory power of this and similar cavity-QED models is limited, when the matter systems are sufficiently complex.  Take for instance the field of \emph{polaritonic chemistry}, where researchers modify the properties of molecular systems by coupling them to photons, e.g., they control chemical reactions by letting them take place inside a cavity. Nowadays conducted routinely, these experiments usually take place at room temperature, involving complex molecules in some solvent, and often with lossy cavities. Even outside the cavity, the accurate description of such settings require sophisticated first-principles methods. This indicates why the current understanding that is principally based on reduced quantum-optical models is still unsatisfactory~\citep{Hirai2020}. 
Therefore we need first-principles methods, such as quantum-electrodynamical density functional theory (QEDFT)~\citep{Ruggenthaler2014} that take into account not only the matter but also the photon-degrees of freedom on the same footing~\citep{Ruggenthaler2018}. 
\subsubsection*{The experimental breakthrough of strong coupling}
\begin{figure}
	\centering
	\includegraphics[width=0.6\columnwidth]{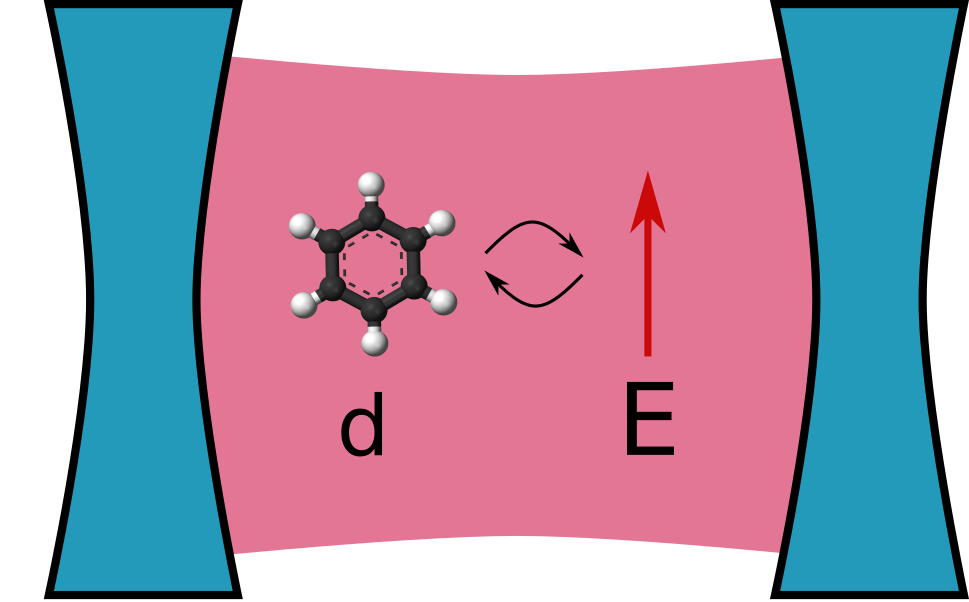}
	\caption{Sketch of a typical strong-coupling experiment: the dipole $d$ of a matter system (here illustrated by a benzene molecule) couples strongly to the electric field $E$ of the confined modes (red) inside a cavity (here illustrated by two concave mirrors in blue).}
	\label{fig:cavity_sketch}
\end{figure}

\begin{figure}
	\begin{tabular}{cc}
		\large\textbf{weak coupling, $g<\gamma$} & \large\textbf{strong coupling, $\gamma>g$}\\
		\begin{overpic}[width=0.49\columnwidth]{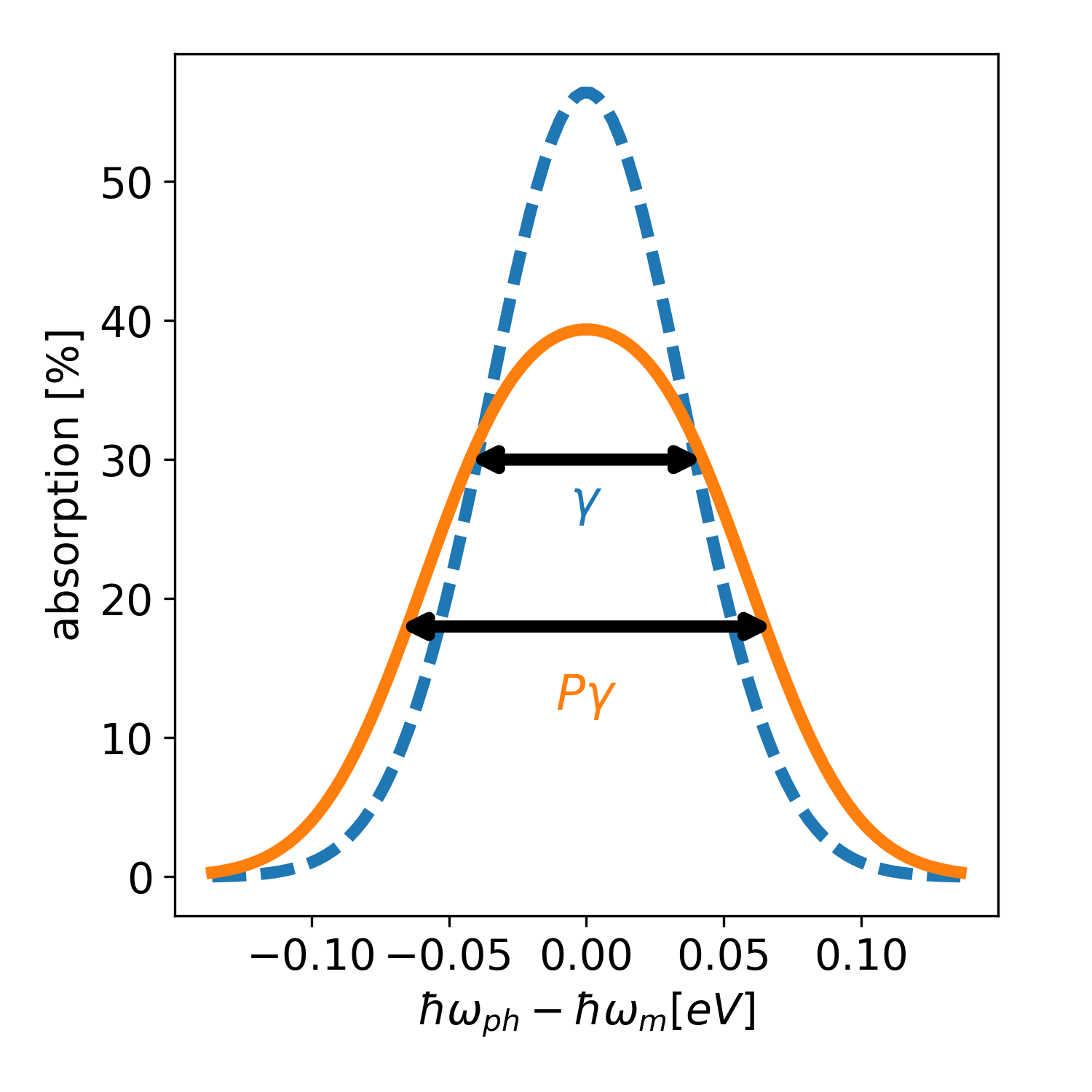}
			\put (22,83) {\textcolor{black}{(a)}}
		\end{overpic}
	&
		\begin{overpic}[width=0.49\columnwidth]{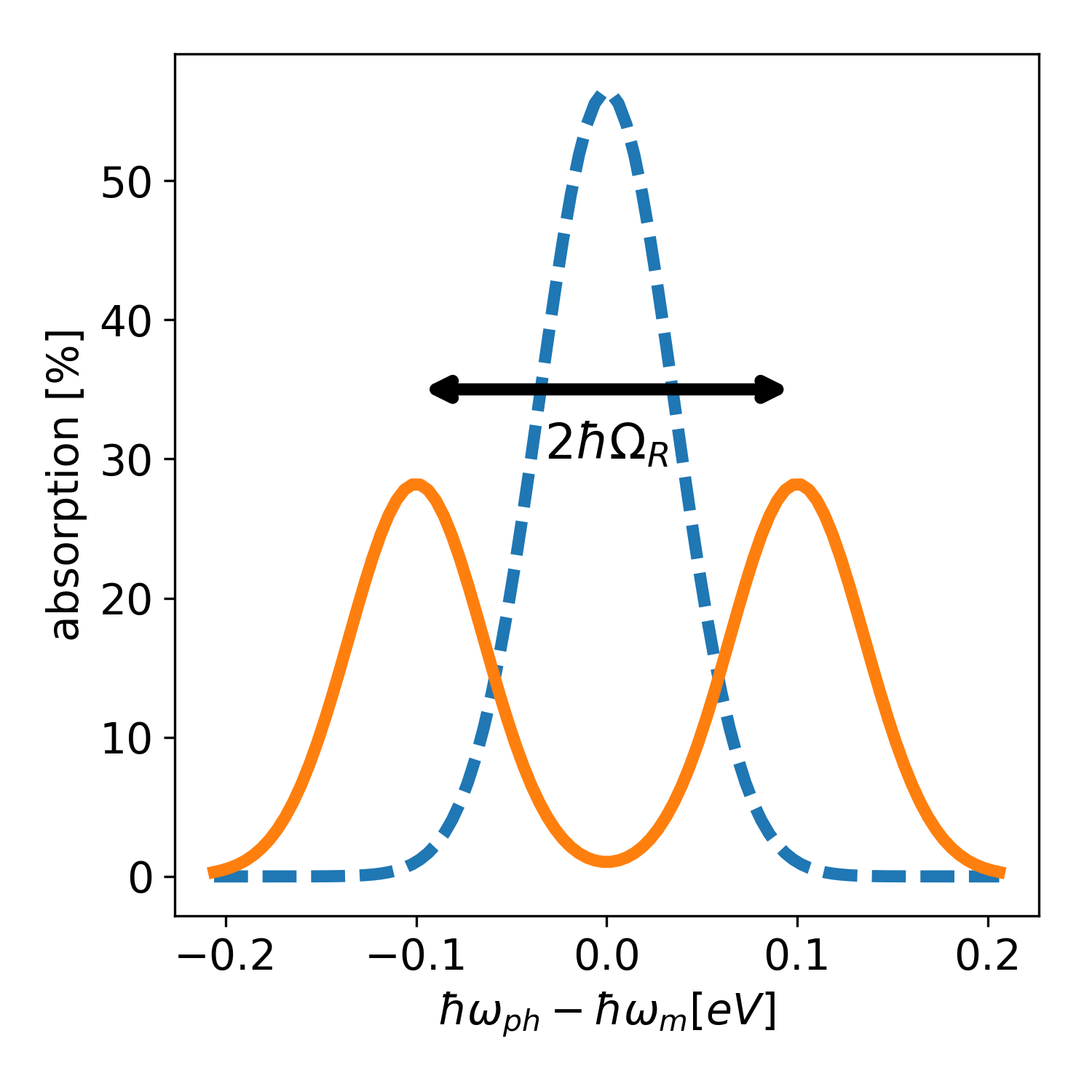}
			\put (26,83) {\textcolor{black}{(b)}}
		\end{overpic}
		\\
		\begin{overpic}[width=0.49\columnwidth]{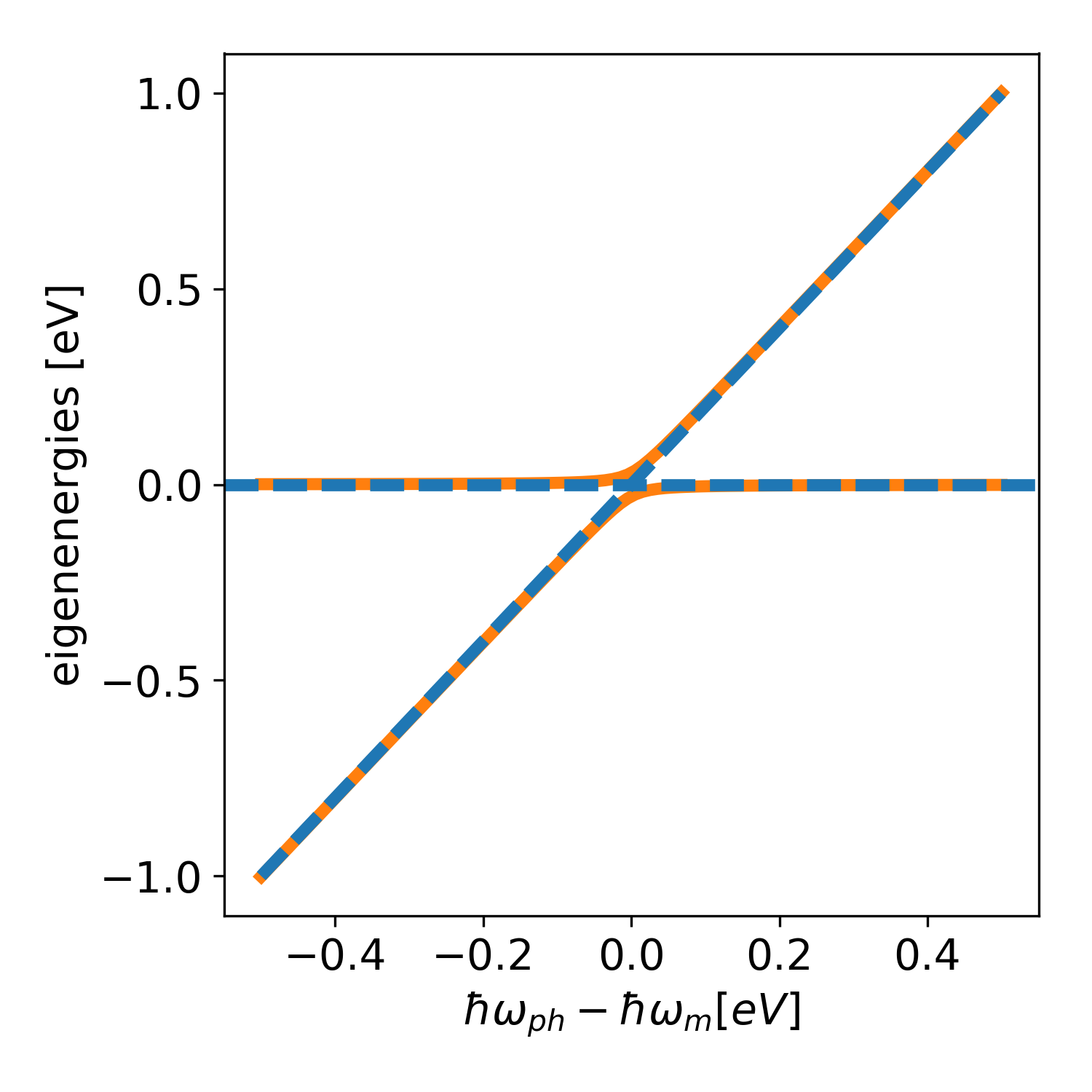}
			\put (22,83) {\textcolor{black}{(c)}}
		\end{overpic}
	&
		\begin{overpic}[width=0.49\columnwidth]{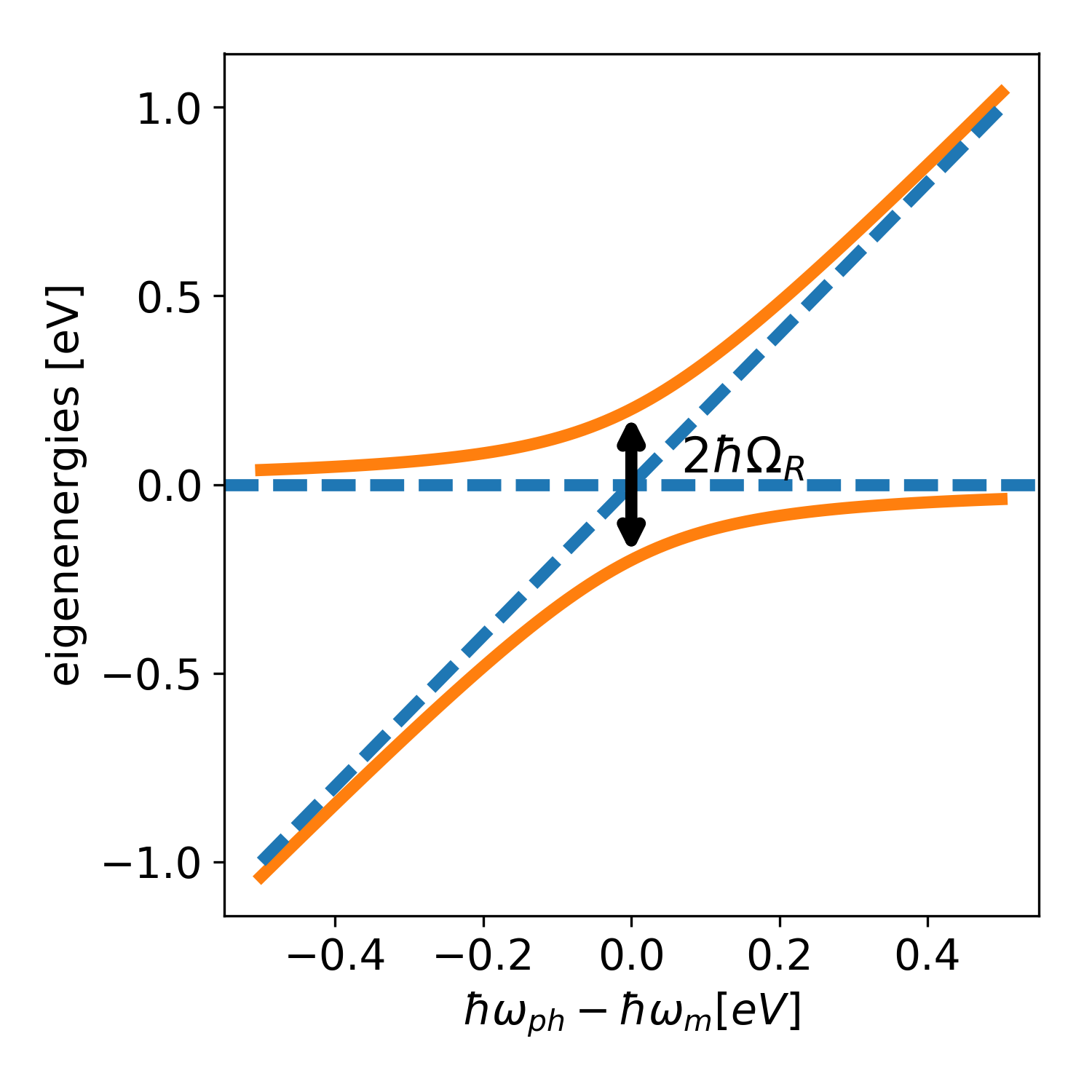}
			\put (26,83) {\textcolor{black}{(d)}}
		\end{overpic}	
	\end{tabular}
	
	\caption{We depict the idealized data of a strong-coupling experiment according to the Jaynes-Cummings model, which provides a good description of the dynamics of coupled light-matter systems close to resonance, i.e., $\omega_m\approx\omega_{ph}$, where $\omega_{ph}$ and $\omega_m$ are the frequencies of the (high-Q) cavity mode and a matter transition, respectively. This allows to differentiate two coupling regimes depending on the ratio between coupling constant $g$ and the spontaneous emission rate $\gamma$. \\ 
	If $g<\gamma$, the light-matter coupling manifests in an effective line-broadening,  i.e., an increase of the spontaneous emission rate  of the combined system $\gamma_P$ (the combined system corresponds in all plots to the orange, solid line) with respect to the matter system outside of the cavity $\gamma$ (in all plots blue, dashed line). The ratio  $P=\gamma_P/\gamma$ is called the Purcell factor and it is related to the coupling strength $P\propto g^2$~\citep{DeLiberato2014}. This can be explained by the energy eigenspectrum of the Jaynes-Cummings Hamiltonian (part c): The light-matter coupling leads to an anti-crossing between the electronic and photonic energy eigenvalues. This is so small that it cannot be resolved spectroscopically, but becomes only visible as the broadening.\\
	If instead $g > \gamma$, two separated peaks can be distinguished (part b), which characterizes the strong-coupling regime. The coupling between light and matter is here so strong that the Rabi splitting $2\hbar\Omega_R\propto \sqrt{\gamma^2+ g^2}$ and thereby the coupling constant $g$ is measurable.
	In the spectrum (part d), we can observe how two clearly separated lines emerge inside the cavity, which describe the dispersion relation of the two polaritons.
}
	\label{fig:intro:strong_coupling_experiment_results}
\end{figure}

In one of the most cited reviews of the field, the experimentalist Thomas Ebbesen names two publications of 1998 as the breakthrough experiments that ``generated increased interest among physicists'' \citep{Ebbesen2016}.\footnote{Note that strong coupling has been already achieved before 1998. See below.} In the first one, \citet{Lidzey1998} achieved strong coupling by fabricating a so-called microcavity, which is a very small resonator that is capable to trap the photons of a mode with frequency $\omega_{ph}$\footnote{To be precise, the cavity does not trap exactly photons of one but a narrow band of modes around a center with frequency $\omega_{ph}$. Considering only $\omega_{ph}$ instead of the full band is for such kind of cavities usually justified.} inside a very small volume of space (key feature one). The breakthrough however was not achieved by their advances in cavity fabrication, but due to the matter system (a special type of organic semiconductor) that they coupled to the cavity modes (key feature two). To get an idea of the setting, we show a simplified sketch in Fig.~\ref{fig:cavity_sketch}.
To prove that their system was able to reach strong coupling, they pumped the cavity mode by an external laser pulse for a series of $\omega_{ph}$.\footnote{In this type of cavity, the trapped mode frequency $\omega_{ph}$ is controlled by the incidence angle of the laser pulse. Thus to perform a measurement series for $\omega_{ph}$, they merely had to vary this incidence angle.} 
In Fig.~\ref{fig:intro:strong_coupling_experiment_results}, we have depicted a sketch of the ``typical'' outcome of this type of experiment.
When they tuned $\omega_{ph}$ close to the frequency $\omega_m$ of a certain excitation of the matter system, they observed a splitting of the absorption peak, i.e., two peaks symmetrically distributed left and right from $\omega_m\approx \omega_{ph}$ (orange solid line in Fig.~\ref{fig:intro:strong_coupling_experiment_results} (a)). When they measured the same absorption spectrum of the matter system, but outside the cavity, they instead observed only one peak at $\omega_m$ (blue dashed line in Fig.~\ref{fig:intro:strong_coupling_experiment_results} (a)). Collecting the position of all those peaks as a function of the mode frequency $\omega_{ph}$ in one graph, they observed two lines that approach each other until they reach a minimal distance at resonance $\omega_{ph}=\omega_m$ and then move again apart as schematically depicted in Fig.~\ref{fig:intro:strong_coupling_experiment_results} (b). Without light-matter coupling, both curves would cross each other (blue, dashed lines in the plot) and thus the observed \emph{anti-crossing} or ``\emph{Rabi splitting}'' is considered as the principal indicator for (strong) coupling.\footnote{Theoretically, the anti-crossing happens for every light-matter system that has a non-vanishing coupling. However, the splitting may be too small to be spectroscopically resolvable, i.e., it is smaller than the linewidth of the two peaks. In this case, the system is said to be in the \emph{weak coupling} regime.} 
The minimal distance $d_{min}=2\hbar \Omega_R$ between the two lines is proportional to the \emph{Rabi frequency} $\Omega_R$, which measures the strength of the light-matter coupling ($\hbar$ denotes as usual the Planck constant). For their experiment, \citeauthor{Lidzey1998} found a maximal value of $\hbar\Omega_R=\SI{160}{\milli\electronvolt}$, which was about 10 times larger then \emph{any} Rabi splitting reported before and which explains the work's impact.

In the other paper that Ebbessen cited, \citet{Fujita1998} measured a Rabi splitting of $\hbar\Omega_R=\SI{100}{\milli\electronvolt}$ in a similar experiment where they put a complex quantum-well structure with organic and inorganic semiconductors into a different type of cavity. 

\subsubsection*{How can we understand the phenomenon of strong coupling?}
Although Rabi splitting has been achieved experimentally already before 1998,\footnote{See for instance Ref.~\citep{Haroche1989,Pockrand1982,Skolnick1998,Raimond2001}. In these experiments lower effective coupling strengths were reached and they were usually conducted under very restrictive conditions such as cryostatic temperatures.} it were especially the new materials, employed in Refs.~\citep{Lidzey1998,Fujita1998}, that allowed for the large values of $\hbar\Omega_R$. The typical (excitonic) transitions in organic semi-conductors respond very strongly to the driving by electromagnetic radiation, which is expressed by a large value of their (dimensionless) oscillator strength $f$. The oscillator strength is thus one (and in fact the most common) measure to quantify the second key feature of strong coupling. A very good measure of the first key feature is the so-called mode volume $V$, which loosely speaking denotes the ``space in which the mode is confined.'' For say a cubic cavity with side-length $a$, we can simply calculate $V=a^3$.\footnote{In other setups, this simple formula is note valid anymore, but has to be replaced by a more general mode volume that can be assigned to, e.g., nanoplasmonic cavities.} Nowadays, we call basically every experimental setup that accomplishes such a confinement a \emph{cavity}. And therefore, strong coupling is often considered as a subfield of cavity QED, which generally deals with coupled matter-cavity systems~\citep{Dutra2005}. We can derive an explicit expression for the Rabi frequency from the oscillator strength and the mode volume by a minimal model. The model considers one matter transition between say an energetically lower $\ket{m_1}$ and higher state $\ket{m_2}$ with energy difference $\hbar\omega_m$ and an electromagnetic mode with frequency $\omega_{ph}$.\footnote{We want to remark that the matter states are often referred to as ground and excited state in the literature. However, this nomenclature is quite misleading, since in most experiments, e.g., for the excitonic transitions of our two examples, both states are actually excited states.} Importantly, we do not need to further specify the nature of this transition and $\ket{m_1},\ket{m_2}$ may represent for example electronic, vibrational or some collective states. The only necessary (external) parameter is the \emph{transition dipole moment} $\mathbf{d}_{12}$ between the two states, whose absolute value is proportional to the square-root of the oscillator strength $|\mathbf{d}_{12}|\propto \sqrt{f}$. A straightforward extension of the model takes $N$ identical matter systems into account by simply exchanging $f\rightarrow Nf$. The matter-dipole couples to the electric field $\mathbf{E}$ of the mode, which has a magnitude $|\mathbf{E}|\propto 1/\sqrt{V}$ proportional to the inverse of the square-root of $V$. Close to resonance and neglecting losses for the moment, the Rabi frequency for this so-called \emph{Jaynes-Cummings model}~\citep{Jaynes1963} (or Tavis-Cummings model for $N>1$ ) is given by
\begin{align}
\label{eq:RabiFrequency_JCM}
	\hbar\Omega_R=2\,\mathbf{d}_{12}\cdot\mathbf{E} \propto \sqrt{N\frac{f}{V}},
\end{align}
Despite its simplicity, this model or one of its  generalizations that we collectively subsume under the term \emph{cavity-QED models}\footnote{\label{fn:Rabi_Dicke_model} Another important example is the Rabi-(Dicke-)model that includes so-called off-resonant terms for one (many identical) matter transition(s). Other generalizations may also include more than two (but not much more) electronic energy levels. See Sec.~\ref{sec:intro:cavity_qed_models} for further details on cavity models.} describe the principal physics of the strong-coupling regime well. This accordance is very important, because it indicates that the dominant phenomenon of the regime is the hybridization between two energy transitions.\footnote{Note that if we assume that both transitions stem from a harmonic oscillator, we cannot differentiate the classical from the quantum description (Hopfield model) anymore in a spectroscopic experiment. Hence, there are still debates on the ``quantumness'' of many of the strong-coupling phenomena~\citep{Scheel2009}.}

The Jaynes-Cummings model, which we will employ as a prototype for cavity-QED models, provides us not only with a handy formula for the Rabi frequency but also with an interpretation of the new transitions, that have a hybrid electron-photon character.\footnote{See Sec.~\ref{sec:intro:cavity_qed_models:justification} for a discussion on the limitations of this and other cavity-QED models.} As it is common in quantum physics, we interpret such effective degrees of freedom as quasiparticles, which in this case are called \emph{polaritons}. In the Jaynes-Cummings model, polaritons ``emerge'' as the eigenstates of the model Hamiltonian\footnote{The Jaynes-Cummings model is one of the very few QED models that can be diagonalized analytically.} and they have the simple form
\begin{align}
\label{eq:Polariton_JCM}
P_{+/-}^n=\alpha (\ket{m_1}\otimes\ket{n+1}) \pm \beta (\ket{m_2}\otimes\ket{n}),
\end{align}
where $\ket{n}$ denotes the $n$-photon state of the mode and the coefficients $\alpha,\beta$ depend on the details of the model. 
The power of the Jaynes-Cummings model lies in its extreme simplification of the in general arbitrarily complex phenomenology of light-matter interaction. It provides us with simple concepts and thus vocabulary to describe and discuss about strong-coupling physics. Until nowadays, it is the most important tool for the interpretation of strong-coupling experiments, which is remarkable in the light of the variety of the field.
An exhaustive overview over this variety is clearly beyond the scope of this thesis and the interested reader is referred to, e.g., the reviews of \citet{Litinskaya2006}, \citet{Torma2015}, \citet{Ebbesen2016}, \citet{Kockum2018}, \citet{Ruggenthaler2018} and the references therein. We content ourselves instead with the presentation of some selected examples to give the reader a flavour of the richness of polaritonic physics and that illustrate the challenges for the theoretical description. 

\subsubsection*{The variety of strong-coupling phenomena: some selected experimental examples}
We start with some general considerations. The relation \eqref{eq:RabiFrequency_JCM} presents the three major ``knobs'' that have been turned in the past twenty years to reach the strong light-matter coupling regime with many different systems inside cavities: 
\begin{enumerate}
	\item the mode volume $V$,
	\item the oscillator strength $f$, and connected to $f$, 
	\item the number of oscillators $N$ that couple collectively to the mode.
\end{enumerate}
From a broad perspective, the most important of these knobs to control the light-matter coupling is the mode volume $V$. As we have mentioned in the introduction, the light-matter coupling strength is fundamentally determined by the fine-structure constant $\alpha$, which is small independently of the system of units used. This is the reason why strong-coupling is very difficult to reach in practice and the phenomena that we discuss here can only occur because of the modern cavities that strongly reduce $V$.
Thus, \emph{all} strong-coupling experiments use some form of a cavity or more precisely, they confine some spectral band of the electro-magnetic field in small volumes, in which they position some matter system.
However, even with state-of-the-art cavities, we cannot reach the strong-coupling regime with \emph{all} materials, but we need to turn also the second and third knob. Regarding the oscillator strength $f$, especially organic materials showed to have large oscillators and most of the following examples employ matter systems of this class. Additionally, most (but not all) of the experiments couple many (a macroscopic number of) oscillators to the photon mode and thus ``make heavily use'' of the third knob.

We start with the two 1998 experiments. They used the aforementioned microcavities to reduce $V$ as much as possible but the crucial step to achieve such large Rabi frequencies was due to the employed materials. They utilized organic semi-conductors, since ``large oscillator strengths are a characteristic feature of these materials''~\citep{Lidzey1998}. In terms of the knobs, this refers not only to a large value of $f$ but also of $N$, since, e.g., all the excitons of the conduction band couple to the mode.\footnote{Though the exact coupling strength depends on the wave vector of the charge carriers.} In the following years, many of the most striking experiments have been conducted with this material class, including polariton lasing~\citep{Kena-Cohen2010} and condensation~\citep{Plumhof2014}, both of which have been predicted before~\citep{Imamoglu1996}. \citet{Orgiu2015} reported that they increased the conductivity of an organic semi-conductor by an order of magnitude by coupling it to only the vacuum field of a cavity and \citet{Coles2014a} modified the energy transfer pathways in a light-harvesting complex with the help of strong coupling. 

\subsubsection*{Strong coupling with molecular systems: polaritonic chemistry}
But not only the excitons in organic semi-conductors are suitable to reach the strong coupling regime. In \citeyear{Schwartz2011}, when \citeauthor{Schwartz2011} used the photochemical properties of an organic molecule to ``switch'' its coupling to a cavity mode, the field of \emph{polaritonic chemistry} emerged. In contrast to semi-conductors, where the most important degrees of freedom often approximately resemble free electrons, molecules exhibit an enormous spectrum of qualitatively different degrees of freedom. \citet{Schwartz2011} used for example an electronic transition for their experiment, but \citet{Thomas2016} coupled the vibronic transition of a molecule to a cavity to demonstrate one of the most striking opportunities that polaritonic chemistry offers: they \emph{changed} the reaction rate of a chemical reaction ``just'' by letting the reaction take place inside a cavity. 
 
Molecular strong coupling has not only been achieved for molecular liquids, where many molecules of the same type are injected in the cavity (like in the latter two examples), but even on the single- and few-molecule level. Since a few molecules have a considerably smaller total oscillator strength than molecular liquids ($N$ is small), either cryostatic conditions are necessary to observe the Rabi splitting \citep{Wang2019} or the mode volume $V$ needs to be substantially decreased. That such single-molecule strong coupling at room temperature is indeed possible has been shown for the first time by \citet{Chikkaraddy2016}. They manufactured a so-called plasmonic nano-cavity, which confined the photon field to effective volumes of $V\leq 40$nm. In fact, such kind of ``cavities'' have nothing in common with the classical picture of two opposite mirrors that we also employed in our sketch of Fig~\ref{fig:cavity_sketch}. Instead, they are based on a striking property of the electromagnetic field close to conductors: singular geometries like edges, small spheres or tips strongly enhance the field density \citep{Pendry2006}. The ``cavity'' of the experiment of \citet{Chikkaraddy2016} was in fact a gold nano-particle. 

To conclude our small review, we want to mention three further strong-coupling setups that do not employ organic materials: superconducting circuits \cite{Niemczyk2010,Yoshihara2017}, 2d materials~\citep{Liu2014,Forn-Diaz2019} and Laundau-polariton systems \cite{Scalari2012,Bayer2017}. These two system classes are according to the review by \citet{Kockum2018} the current ``record holders'' of strong-coupling in the sense that they measured the largest Rabi splitting energies in relation to corresponding transition frequencies.

\subsubsection*{The challenges of an interdisciplinary field}
These (in comparison to the total number of publications) few examples illustrate how diverse the field is. We saw at least the research fields of nanophotonics, plasmonics, quantum optics, material science (including 2d materials), solid-state physics and (quantum) chemistry appearing. All of them play an important role. And all of them have their own focus and consequently also their own point of view on the topic. In the nanophotonics and plasmonics communities, people try to understand the behaviour of the electromagnetic field on the nano-scale. They are interested in understanding and optimizing geometries and as theoretical tool they usually just solve the classical Maxwell's equations, where the matter often only enters in the form of complicated boundary conditions. The fields of materials science, solid state physics and quantum chemistry instead focus on matter properties on atomic length scales. Crucially, this requires an encompassing quantum-mechanical description and thus in principle solving the many-body Schrödinger equation. Since this is impossible in practice, the main challenge here is to find approximations or alternative descriptions that are numerically feasible but still ``sufficiently'' accurate. The question ``what is sufficient?'' is hereby one of the important research questions in the field. Electromagnetic fields normally enter in this description merely as ``external potentials'' or ``perturbations'' without shape and spatial extension. Probably the most important contribution to strong-coupling physics comes from the quantum optics community. Quantum opticians study light-matter interaction on the smallest scales and they developed the first theories to describe the strong-coupling phenomenology, including the aforementioned cavity-QED models. Still, these descriptions usually emphasize the accurate description of the electromagnetic field, including the complex phenomenology of its quantum statistics. Matter though treated quantum mechanically, is almost exclusively reduced to two (or a few) levels. 
Until nowadays, these cavity-QED models are the basis for most of our understanding of strong-coupling physics and chemistry.

\subsubsection*{The theoretical challenge: complex systems exhibit strong-coupling effects that require complex theory}
It is obvious that there are many scenarios where such simplified descriptions cannot account for the complexity of the involved processes. Especially in the realm of polaritonic chemistry, where complex molecular systems are coupled to the photon field, the explanatory power of cavity-QED models is limited. There are several open questions, such as
\begin{itemize}
	\item What is the influence of the strong electron-photon interaction, if the electronic-structure \emph{changes} as it happens in a chemical reaction~\citep{Lethuillier-Karl2019,Lather2019}? 
	\item Can the vacuum fluctuations of a cavity mode really modify equilibrium properties of matter systems as claimed by experimentalists~\citep{Hutchison2012,Thomas2016}? 
	\item Is the mechanism that leads to collective strong coupling really as simple as predicted by the Dicke model~\citep{Keeling2007}? And can local properties be substantially modified, due to this pathway to reach strong coupling~\citep{Martinez-Martinez2018,Galego2019}?
\end{itemize}
To answer such questions, we need genuine first-principles methods such as QEDFT~\citep{Ruggenthaler2014}  that are capable to describe the inner structure of matter systems \emph{and} its interaction with the field~\citep{Ruggenthaler2018}. However, developing such methods is a delicate task that involves finding entirely new approximation strategies (for the electron-photon interaction) and deriving and solving highly nonlinear equations. 
This requires not only an accurate mathematical treatment, but also a careful development of numerical methods. In this thesis, we address and analyze these theoretical and numerical challenges from a very general perspective. Based on this, we present the dressed-orbital construction that allows to circumvent many of the identified difficulties by an exact reformulation of (equilibrium) cavity QED in terms of polaritonic particles. It is noteworthy that in contrast to Eq.~\eqref{eq:Polariton_JCM}, this definition of a polariton in terms of dressed orbitals is general and does not require restrictive assumptions such as the few-level approximation.
%

We present two specific examples of dressed-orbital-based methods including the details of the according numerical algorithms and implementations.
With the help of these methods, we provide explicit evidence for phenomena that cannot be described with standard model approaches. 
For instance, we demonstrate in Sec.~\ref{sec:dressed:results:1chemical_reaction} how a simple chemical reaction in one spatial dimension is influenced differently by the cavity, depending on the reaction coordinate. Missing spatial resolution, such an effect cannot be captured by methods that rely on the few-level approximation to account for the electron-photon interaction. However, there are indications that such local effects might play in important role for the understanding of many strong-coupling phenomena~\citep{Schaefer2019} and our results of Sec.~\ref{sec:dressed:results:3confinement} provide additional evidence for this. We find there that both, electronic localization and correlation, strongly influences the light-matter interaction and points toward a different than the Dicke-type mechanism behind collective strong coupling. 


\subsubsection*{Unresolved questions in polaritonic chemistry: the limitations of cavity-QED models}
Let us illustrate the limitations of model descriptions with a concrete example that concerns the just mentioned ``collective contribution'' to the coupling strength appearing in relation \eqref{eq:RabiFrequency_JCM} by the simple factor $\sqrt{N}$. The assumption that leads to this dependence is that $N$ two-level molecules couple to the same cavity and thus have an $\sqrt{N}$ times larger effect on the Rabi splitting, but at the same time every molecule itself only ``feels'' the small coupling for $N=1$. Accordingly, the coupling does not have strong local effects and cannot significantly change, e.g., the electronic structure. But if this is true, how are chemical reactions modified by strong matter-photon coupling as experimentalists claimed~\citep{Hutchison2012} and asked~\citep{George2015}? 

\citet{Feist2015} and \citet{Cwik2016} tried to answer this question with their models and showed that some observables are collective and others are not, however partly contradicting each other. One prominent controversy arose around the question, whether the \emph{ground-state potential energy surface} (ground-state PES), which is a crucial quantity in chemical reactions (see Sec.~\ref{sec:est}) is modified by the collective or merely the single-molecule coupling. \citet{Feist2015} and \citet{Herrera2016} showed for different models that modifications of the ground-state PES proportional to the collective coupling are possible. \citet{Martinez-Martinez2018} instead showed that such modifications for their model are only proportional to the single-molecule coupling strength. They explicitly state that their results contradict Ref.~\citep{Feist2015} and Ref.~\citep{Herrera2016}, but is in line with Ref.~\citep{Cwik2016}. One year later, \citet{Galego2019} enforced with improved methods their argument that ground-state PES modifications due to collective effects are indeed possible. They argue that Ref.~\citep{Martinez-Martinez2018} did not take ground-state dipole moments into account, which would be crucial for a correct description of chemical reactions under strong coupling.

This debate reflects the inherent challenges of understanding and describing complex systems: there are several effects playing a role at the same time and some of them might be cooperative, others in competition. And especially, this might change with certain system parameters. Thus, it is not enough to identify these effects, but one also needs to accurately quantify their importance. As we have seen, model descriptions like the Jaynes-Cummings model are very efficient and powerful in describing one or a few features of a system. However, with increasing complexity, i.e., with an increasing number of such effects this strength becomes a weakness. We somehow have to decide, which feature we include in the description and which not and depending on that, the model may provide different answers. In some cases this might be quite clear, but the aforementioned example shows that this is not always the case.

Another very old example is the not-yet resolved debate on the (non-)existence of a so-called \emph{superradiant phase} that \citeauthor{Dicke1954} predicted already in \citeyear{Dicke1954}. The there proposed model (which is today known as the Dicke-model) describes $N$ two-level systems coupled to a photon mode and predicts a transition in the superradiant phase \citep{Hepp1973}, where all the dipole moments of the atoms align and the photon mode occupation reaches values much bigger than $N$. There have been several publications trying to answer if a real system can undergo such a phase-transition. Certain ``no-go-theorems'' have been derived, e.g., Refs.~\citep{Rzaewski1975,Viehmann2011}, and contested several times, e.g., Refs.~\citep{Keeling2007,DeBernardis2018}. Recently, the superradiant transition could be demonstrated in artificial realizations of the Dicke model~\citep{Baumann2010,Zhiqiang2017}, but the question whether the transition can occur in more realistic situations that Dicke originally had in mind is still not resolved. For a good summary of the topic, the reader is referred to the recent review by \citet{Kirton2019}.

It is such kind of problems that have motivated our research on a first-principles description of light-matter systems. How can we describe the details of strong-coupling physics in a less-biased way, i.e., \emph{without deciding a priori} which features we include in the description? Cavity-QED models have proven their explanatory power, but what are their limits? And if we reach these limits, how can we improve the models in a systematic way?

\subsubsection*{The range of validity of cavity-QED models}
So one might ask, how it is possible that so many different and highly complex materials can be modelled by cavity-QED models in the moment we put them into a cavity that itself might be a complicated plasmonic nano-structure. It took decades of research to develop the machinery that allows to accurately describe the properties of these materials and cavities. How can an additional interaction between two already complex systems simplify things? In fact, exactly this is what (in many cases) happens. Tuning the cavity in resonance with one matter excitation, can be seen as a sort of selection process. If the other matter transitions are energetically well separated from the selected transition and only this energy range is probed in an experiment, one can observe the Rabi splitting exactly as it is described by the Jaynes-Cummings model. \citet{Wang2019} showed this explicitly in a recent experiment, where they ``turn[ed] a molecule into a coherent two-level quantum system.'' But even if some other matter degrees of freedom play a role, it is often enough to extend the model by, e.g., some more matter or photon levels to match theory and experiment.

Nevertheless, it is clear that there must be a limit to such a procedure. With the increasing complexity of the effects that have to be described, more and more features have to be added to the models to fit the experimental data. This will not only become computationally difficult at a certain point, but more importantly, such a path heads toward a situation, where so many parameters have to be introduced that interpretations and predictions might become difficult. For instance, when \citet{George2016} wanted to interpret their experimental data, they first tried to employ a Jaynes-Cummings-like model, which they could not fit to their results. They say in the publication that it was necessary to add several extra matter and photon states and a more thorough description of the electron-photon interaction to the model to properly interpret their data. The reason for the break-down of the simple model in this case is two-fold: first, they fabricated a system with very strong electron-photon coupling (for smaller couplings in the same experiment, the simple model worked) and second, the energy structure of the molecule they employed was such that \emph{several} matter excitations strongly coupled to the mode (what they called ``multimode splitting effect''). Consequently, they did not only have to describe the two polaritonic states ($P_{+/-}^n$ for a fixed $n$), but a ``genuine ladder of vibrational polaritonic states'' ($P_{+/-}^n$ for a series of $n=1,2,...$) and especially, the cross-talk between these states. 

In this example, the machinery of cavity QED could still describe the experimental data quite satisfactorily, but how much further can we push it? \citet{George2016} stress that they had only one fitting parameter, the Rabi frequency, but will this be enough if even more matter transitions need to be taken into account? And if we are interested in understanding the origins of such large Rabi frequencies, how can cavity-QED models help us, if we need the Rabi frequency itself as a fitting parameter?

\subsubsection*{Perspective: connecting models and first-principles approaches}
It is clear that answering such questions is anything but easy and in many cases first-principles approaches might not be (directly) applicable simply because of numerical limitations. 
This kind of problem is well known in other fields of quantum physics such as strongly-correlated electron systems.
To this category belong materials like high-temperature superconductors~\citep{Fujita2001}, Mott insulators~\citep{Phillips2006} or many important catalysts~\citep{VanSanten2015}, all of which promise huge possibilities for applications. 
Modelling such effects is among the hardest problems of material science, because efficient many-body methods like (approximate) density functional theory are typically too inaccurate.\footnote{This is considered as more or less basic knowledge in quantum chemistry and solid state physics, which is the motivation behind many new theory developments. However, for a recent specification of this statement, the reader is referred to Ref.~\citep{Mardirossian2017}.} Thus, a crucial role for the understanding of strongly-correlated electrons has been played by effective models, most importantly the Hubbard model~\citep{Hubbard1963}. The Hubbard model exhibits a wide range of correlated electron behavior including all the above mentioned phenomena and thus allows to study the basic mechanisms behind these phenomena. Such kind of studies have revealed the complexity of the effects but also their extreme dependence on tiny variations of system parameters, many of which cannot directly be determined by experiments and thus require first-principles calculations.\footnote{To provide an example for this statement, we refer the reader to the very comprehensive study of the two-dimensional Hubbard model with different methods by \citet{LeBlanc2015}.} 
Triggered by this insight, the field of strongly-correlated electrons provides nowadays a plethora of examples, where models and first-principles descriptions have been successfully combined.\footnote{For example in Ref.~\citep{Fanfarillo2012} the author explains very well how to connect models to first-principles calculations. Another even more explicit example is the $LDA+U$ method~\citep{Anisimov1997} that connects the Hubbard model with the local-density approximation of density functional theory. It was built exactly with the purpose to unify the advantages of both, the model and the first-principles world.} 

Judging from the complex phenomenology that experiments have revealed, one can expect that the field of strong electron-photon coupling and especially the sub-field of polaritonic chemistry exhibit a similarly complex phenomenology as strongly-correlated electrons. A general perspective on the problem will allow us to put the connection between electronic strong-correlation and strong coupling even in quite concrete terms (see Sec.~\ref{sec:qed_est:general:coupled_wavefunction}). The aforementioned examples, where different cavity-QED models contradict each other additionally indicate the need for new less-biased methods. We believe that a combination of model and first-principles approaches that was so successful in other areas of physics like strongly-correlated electrons, could also be fruitful in the field of strong electron-photon coupling.

	\newpage
\section{The essence of polaritonic physics: cavity-QED models}
\label{sec:intro:cavity_qed_models}
Before we come to first-principles methods, we want to make a brief detour to the standard way to describe the phenomena of strongly coupled electron-photon systems. As we have mentioned in the last section, the researchers that historically first investigated such phenomena stem from the community of \emph{quantum optics}. They introduced the cavity-QED models such as the Jaynes-Cummings model, which allowed to identify the basic mechanism behind the emergence of polaritons and satisfactorily describe many of the experiments in the field. 

In this section, we briefly present the derivation of this set of models, which has two purposes. First, presenting this standard description helps to acquaint the reader already with the concepts and tools that are necessary to describe light-matter interaction on the quantum-level. In the next section, we present a big part of this derivation a second time, but including all the technicalities that are necessary to derive a proper framework for a first-principles perspective. We hope that the preliminary discussion in this section facilitates reading and understanding of the general derivation. The second reason for such a detailed presentation of cavity-QED models is to make their limitations more concrete. Thus, we put a special emphasis on the approximations that enter the models. After the presentation of the general derivation and some important special cases, we briefly analyze their range of validity. Importantly, the models accurately describe experimental data, even in cases where certain approximations are not strictly justified. This important fact reveals the universality of the models and puts this in the context of their obvious limitations in the realm of quantum chemistry. We then briefly explain, why first-principles methods are a valuable tool to overcome these limitations and finish the subsection with a short summary.

\subsection{The origin of cavity-QED models}
To describe the physics of cavity experiments, we essentially need to model the matter systems, the  photon modes, and the interaction between both. 
The standard starting point for the discussion of \emph{quantum atom-field interaction} is the Hamiltonian~\citep[part 6.1]{Scully1999}
\begin{align}
\label{eq:intro:QO:Hamiltonian_QED_dipole}
	\hat{H} = \hat{H}_m + \hat{H}_{ph} - e \hat{\br}\cdot\hat{\bE},
\end{align}
where $\hat{H}_m$ is the matter Hamiltonian and $\hat{H}_{ph}$ is the electromagnetic-field Hamiltonian. The last term describes the interaction between both subsystems that is given by the inner product of the electron dipole $- e \hat{\br}$ ($e$ denotes the elementary charge and $\hat{\br}$ is the position operator) and the electric field operator $\hat{\bE}$.\footnote{To account for the quantum nature of the electromagnetic field, the electric field vector is promoted to an operator. See also Sec.~\ref{sec:intro:nr_qed}.} 
However, one needs to be aware that Hamiltonian~\eqref{eq:intro:QO:Hamiltonian_QED_dipole} involves already many assumptions (such as the dipole approximation and the neglect of the dipole-self energy), which are in many textbooks discussed in the \emph{semiclassical} theory~\citep[part 5]{Scully1999}. This means that $\hat{H}_{ph}$ is neglected and $\hat{\bE}\rightarrow \bE$ and all other descriptors of the electromagnetic field are treated as external classical vector fields.  

If we consider this semiclassical theory for a single-electron atom, i.e., one electron with mass $m$ that is confined by the electrostatic potential $V$ of the nucleus with mass $m_n\rightarrow \infty$ (Born-Oppenheimer approximation, see Sec.~\ref{sec:intro:nrqed_longwavelength}), the corresponding Hamiltonian reads
\begin{align}
\label{eq:intro:QO:Hamiltonian_general}
	\hat{H}_{sc}=-\frac{\hbar^2}{2m} [\nabla - \I\frac{e}{\hbar}\bA(\br,t)]^2 + e \phi(\br,t) + V(\br),
\end{align}
where $\phi$ and $\bA$ are the scalar and vector potentials of the electromagnetic field, respectively. The form of the matter-field interaction in $\hat{H}_{sc}$ can be derived by basic principles (see \citep[part 5.1.1]{Scully1999}) and has a very large range of validity. 

Then one chooses the Coulomb gauge\footnote{The theory of electrodynamics exhibits (on the classical and on the quantum level) a so-called gauge-symmetry. This means that the theoretical description has a certain redundancy, which usually is removed in a concrete application. This is done by \emph{choosing} one of many possible gauges. In electrodynamics, there are many established standard gauges, e.g., the Coulomb gauge, which can crucially simplify the description of certain problems. We discuss this in more detail in Sec.~\ref{sec:intro:nr_qed}. See also Def.~\ref{def:Maxwell-Lorentz_Equations}.} $\nabla\cdot\bA=0$ and additionally sets $\phi=0$,\footnote{Both together is called radiation gauge in this context.} which is strictly speaking only possible far away from any charge. However, since only one electron is considered, this term contributes only by the constant self-energy of the electron and thus can be neglected as we will discuss in Sec.\ref{sec:intro:nrqed_longwavelength}. For a many-particle system, $\phi$ leads actually to the Coulomb interaction between the (charged) particles. After applying the Coulomb gauge, Hamiltonian~\eqref{eq:intro:QO:Hamiltonian_general} reads
\begin{align}
\label{eq:intro:QO:Hamiltonian_Radiation}
	\hat{H}_{sc}^C=&-\frac{\hbar^2}{2m} \nabla^2 + V(\br) + \I\frac{e}{\hbar}\bA_{\perp}(\br,t) + \frac{e^2}{2m}\bA_{\perp}^2 (\br,t) \nonumber\\
	=&\hat{H}_m + \I\frac{e}{\hbar}\nabla\cdot\bA_{\perp}(\br,t) + \frac{e^2}{2m}\bA_{\perp}^2 (\br,t),
\end{align}
where $\bA_{\perp}$ is the transversal part of the vector potential.\footnote{According to Helmholtz's theorem, every vector field $X=X_{\parallel}+X_{\perp}$ can uniquely be decomposed in its longitudinal $X_{\parallel}$ and transversal $X_{\perp}$ component, with $\nabla\times X_{\parallel}=0$ and $\nabla\cdot X_{\perp}=0$. In the Coulomb gauge, where $\nabla\cdot\bA=0$, thus the longitudinal contribution of $\bA$ is explicitly removed. This also induces $[\nabla,\bA_{\perp}]=0$ in a quantum-mechanical sense, when $-\I\hbar\nabla$ is the particle momentum operator.} 
We can generalize $\hat{H}_{sc}^C$ straightforwardly to $N_e$ electrons, if we drop the assumption $\phi=0$. This generalization is considered as the \emph{basic} Hamiltonian of electronic structure theory (see Ch.~\ref{sec:est}). 

\subsubsection*{The long-wavelength limit and the diamagnetic term}
The next step is to introduce the \emph{dipole approximation} by assuming
\begin{align}
\label{eq:intro:QO:dipole_approximation}
	\bA(\br,t)\approxeq \bA(\br_0,t),
\end{align}
where $\br_0$ is the center of charge (which is equal to the center of mass) of the matter system. This approximation is also called the \emph{long-wavelength limit} and it is well justified, if the spatial extension of the atom is much smaller than the wavelength of the considered modes of the electromagnetic field. This is this case in most cavity-matter systems (see Sec.~\ref{sec:intro:nrqed_longwavelength}). 
The derivation is finished by applying the gauge transformation $U_d=\exp(\I\frac{e}{\hbar}\bA(\br_0,t)\cdot\br)$ to the Hamiltonian~\eqref{eq:intro:QO:Hamiltonian_Radiation}, which after some standard rearrangements has the form
\begin{align}
\label{eq:intro:QO:length_transformation}
	\hat{H}_{sc}^C\underset{U_d}{\overset{\bA(\br,t)\approxeq \bA(\br_0,t)}{\xrightarrow{\hspace*{2cm}}}}\hat{H}_d =& \frac{\hbar^2}{2m} \nabla^2 + e U(\br,t) + V(\br) - e \br\cdot\bE \nonumber\\
	=& \hat{H}_m - e \hat{\br}\cdot\bE.
\end{align}
To arrive from here at the (fully-quantized) starting Hamiltonian~\eqref{eq:intro:QO:Hamiltonian_QED_dipole}, we add $H_{ph}$ and follow the prescription of canonical quantization. We reserve the details of this procedure for Sec.~\ref{sec:intro:nr_qed} and just assume that $(\bE,\bA)\rightarrow(\hat{\bE},\hat{\bA})$ are now operators.

It is important to realize that within the ``rearrangements,'' another approximation has been employed. The diamagnetic contribution $\frac{e^2}{2m}\bA_{\perp}^2 (\br,t)$ that was still present in Eq.~\eqref{eq:intro:QO:Hamiltonian_Radiation} is indeed removed by the transformation $U_d$, but at the same time a new term that is proportional to $\br^2$ is introduced (see Sec.~\ref{sec:intro:nrqed_longwavelength}). This term is called the dipole-self energy and it is obviously neglected in Eq.~\ref{eq:intro:QO:length_transformation}. 
That this standard approximation in quantum-optical models~\citep[p. 150]{Scully1999} is reasonable, is not well visible with the definitions of the scalar and vector potentials that are considered here. In the chosen unit system, the vector potential $\bA\propto 1/c$ is inverse proportional to the speed of light $c$. Thus, the $\bA^2$-term has a prefactor of $1/c^2$ that makes its contribution to the Hamiltonian usually small in comparison to all the other terms and justifies its neglect. 
For instance, if we want to investigate the interaction of an atom with the electromagnetic vacuum to understand  lifetimes or energy level shifts due to the photon field like the Lamb-shift, $\braket{\bA}$ is small and we can perform a calculation in terms of lowest-order perturbation theory, where the $\bA^2$-term naturally would drop out. However, when some of the electromagnetic field modes are strongly enhanced by a cavity, this approximation might break down. We discuss the $\bA^2$-term in Sec~\ref{sec:intro:nrqed_longwavelength} again.

\subsubsection*{The making of the cavity-QED Hamiltonian: the diagonalization of the matter Hamiltonian and the single-mode approximation}
To finish the derivation of the atom-field model of quantum optics, we still need to perform one crucial step, which is the diagonalization of the matter Hamiltonian, i.e.,
\begin{align}
\label{eq:QO_Eigencomp_Matter}
&\hat{H}_m\ket{m_i}=\epsilon_{i}\ket{m_i}.
\end{align}
If this decomposition is known, we can rewrite
\begin{align}
\label{eq:QO:Hm_matrixform}
&\hat{H}_m = \sum_i \epsilon_{i}\ket{m_i}\bra{m_i} \equiv \sum_i \epsilon_{i}\sigma_{ii}
\end{align}
in its diagonal matrix form. For the single-electron system that is considered here, this is not problematic. Although in most cases, there is no analytic solution, we can always find a suitable approximate basis for the problem and diagonalize it numerically (see Sec.~\ref{sec:est:general:example_case}). For more than one or a few particles, the situation changes drastically, because of the already mentioned \emph{many-body problem}. This manifests here in the cost for such a numerical diagonalization, that grows \emph{exponentially} with the particle number (independently of the details of the problem, see Ch.~\ref{sec:est}).  
We discuss some implications of this with respect to cavity-QED models in Sec.~\ref{sec:intro:cavity_qed_models:justification}. 

Additionally, we diagonalize the free photon Hamiltonian which is not as problematic as for the electronic problem. There is no (direct) confining potential and no (direct) interaction in $\hat{H}_{ph}$ and thus its diagonal form can even be derived \emph{analytically} by going to $\bk$-space.\footnote{The connection between the real space, where quantities are parametrized by position vectors $\br$ and $\bk$-space is given by the Fourier transformation. Specifically for a function $f(\br)$, we have $f(\bk)=1/\sqrt{2\pi}^3\int\td\br f(\br)\exp({-\I \br\cdot\bk})$.} We find
\begin{align}
\label{eq:QO:Hph_matrixform}
\hat{H}_{ph}=\sum_{\bk,s}\hbar\omega_{k}(\hat{a}_{\bk,s}^{\dagger}\hat{a}_{\bk,s}^{\dphan}+\frac{1}{2}),
\end{align}
where $\omega_{k}=c |\bk|= c k$ is the mode frequency and $\hat{a}_{\bk,s}^{(\dagger)}$ are the annihilation (creation) operators  for a photon in the mode with wave-vector $\bk$ and the polarizaton-index $s=1,2$.\footnote{For example, $s=1(2)$ could denote left (right) circular polarized light with respect to $\bk$.} The operators $\hat{a}_{\bk,s}^{(\dagger)}$ obey the commutation relations $[\hat{a}_{\bk,s}^{\dphan},\hat{a}_{\bk,s}^{\dagger}]=1$ and $[\hat{a}_{\bk,s}^{(\dagger)},\hat{a}_{\bk,s}^{(\dagger)}]=0$, which defines the quantum mechanical properties of the photon field.

These definitions are sufficient to write the whole Hamiltonian \eqref{eq:intro:QO:Hamiltonian_QED_dipole} in matrix form. For that, we expand the dipole moment
\begin{align}
\label{eq:QO:Hint_matrixform}
	e\hat{\br} = \sum_{i,j}e \ket{m_i}\braket{m_i|\hat{\br}|m_j}\bra{m_j} \equiv \sum_{i,j}e \mathbf{d}_{ij} \sigma_{ij}
\end{align}
and the electric field
\begin{align}
	\hat{\bE} = \sum_{\bk,s} \mathbf{e}_{\bk,s} E_{\omega_{k}} (\hat{a}_{\bk,s}^{\dagger} + \hat{a}_{\bk,s}^{\dphan}),
\end{align}
where $\mathbf{e}_{\bk,s}$ is the polarization vector and $E_{\omega_{k}}=(\hbar\omega_{k}/2\epsilon_0 V)^{1/2}$ is a prefactor that includes the vacuum permittivity $\epsilon_0$ and importantly, the mode volume $V$ that plays a crucial role in polaritonic physics as we have discussed in the previous section.
Using \eqref{eq:QO:Hm_matrixform}, \eqref{eq:QO:Hph_matrixform} and \eqref{eq:QO:Hint_matrixform}, we arrive at the matrix form of \eqref{eq:intro:QO:Hamiltonian_QED_dipole}
\begin{align}
\label{eq:QO_Hamiltonian_QED_matrix}
	\hat{H} = \sum_{\bk}\hbar\omega_{k}\hat{a}_{\bk}^{\dagger}\hat{a}_{\bk}^{\dphan} + \sum_i \epsilon_{i}\sigma_{ii} + \hbar\sum_{i,j} \sum_{\bk} g_{\bk,ij} \sigma_{ij}(\hat{a}_{\bk}^{\dagger} + \hat{a}_{\bk}^{\dphan}),
\end{align}
with the coupling-matrix
\begin{align}
	g_{\bk,ij}=- \tfrac{eE_{\omega_{k}}}{\hbar} \textstyle\sum_s \mathbf{d}_{ij}\cdot \mathbf{e}_{\bk,s},
\end{align}
where we summed over the 2 polarization directions, assuming polarized light, such that only one of the two polarizations contributes. Additionally, we neglected the zero-point energy of the electromagnetic field. Although, it regards only a one-electron system on the matter side, Hamiltonian \eqref{eq:QO_Hamiltonian_QED_matrix} has a broad range of applicability and in general, its diagonalization is nontrivial.

For the cavity setting that we are interested in, we assume that there is one dominant mode with frequency $\omega_{ph}$. The cavity strongly confines the mode to the volume $V$, which leads to a large prefactor $E_{\omega_{k}}\propto \sqrt{1/V}$. Consequently, it holds for the coupling elements $g_{ij} \equiv g_{ij,ph}$ of this mode that $g_{ij}\gg g_{\bk,ij}$ for all other $\bk$ and we can neglect these coupling elements.\footnote{For dynamical problems, we usually have take to the influence of the \emph{bath} represented by the other modes into account. In many cases (for example to include spontaneous emission processes), this can be done approximately by the introduction of a damping factor. The reader is referred to, e.g., the corresponding chapters in Ref.~\citep{Scully1999}.} The resulting Hamiltonian is the origin of the cavity-QED models and it reads
\begin{empheq}{align}
\label{eq:intro:QO:Hamiltonian_cavity_QED}
\hat{H}_{cQED} =& \hbar\omega_{ph}\hat{a}^{\dagger}\hat{a} + \sum_{i=1} \epsilon_{i}\sigma_{ii} + \hbar\sum_{i,j} g_{ij} \sigma_{ij}(\hat{a}^{\dagger} + \hat{a}).
\end{empheq}
Note that we dropped the index of the operators $\hat{a}^{(\dagger)}$ that refer to the cavity mode. 

\subsubsection*{The crucial step: the few-level approximation}
We want to stress again that the assumption of a single-electron atom was not necessary to arrive at $\hat{H}_{cQED}$. In principle, we can assume a Hamiltonian of the same form to study a complicated many-electron system, because the possibility of the underlying eigendecomposition simply results from the linear structure of quantum mechanics. However, the corresponding eigenvalue problems are so high-dimensional that in order to solve them in practice, we have to \emph{significantly} restrict the underlying configuration spaces, i.e., the bases. The choice of the basis is thus a crucial step in any quantum mechanical calculation (see Sec.~\ref{sec:est:general:example_case}).

The most important approximation in cavity-QED models is to restrict this choice from the beginning by assuming that a very small matter basis is sufficient to describe the phenomena that we are interested in. For most methods in quantum optics, it is crucial that this \emph{few-level approximation} reduces the dimension of $\hat{H}_{cQED}$ such that it can be diagonalized \emph{exactly}. The most important special case of this approximation is the two-level system that we discuss in the next paragraph~\citep{Frasca2003}. However, if we wanted to employ more electronic levels, the standard approach consists of a two-step procedure. In the first step, we use some first-principles methods to identify the most important electronic states, their energies and the corresponding coupling elements. These enter in the second step as input parameters in $\hat{H}_{cQED}$, which is diagonalized. In the realm of polaritonic chemistry, several hybrid approaches of this type have been proposed~\citep{Saurabh2016,Luk2017,Vendrell2018,Zhang2019, Groenhof2019}. However, to make the diagonalization of $\hat{H}_{cQED}$ numerically feasible, most of these methods rely on the projection on the single-excitation space, i.e., the rotating-wave approximation that we discuss below. It is very difficult to extend such methods in a simple way.\footnote{In Ref.~\citep{Bennett2016}, the authors propose a method that goes beyond the rotating wave approximation. They discuss the difficulties and the (strong) limitations of such a generalization.}

\subsubsection*{The standard case: the two-level atom}
The most common models reduce the matter description to a minimum and consider a \emph{two-level atom} (or molecule), i.e., they neglect all but two effective matter states $\ket{m_{1/2}}$. Assuming  a real transition dipole $d_{12}=d_{21}$ between these states, we arrive at one of the most important cavity-QED model, the Rabi-model.\footnote{In fact, the original publication by \citet{Rabi1936} from 1936 considered  a nuclear spin and not an atom.} 
The corresponding Hamiltonian reads
\begin{align}
\label{eq:QO_Hamiltonian_Rabi}
\hat{H}_R =& \hbar\omega_{ph}\hat{a}^{\dagger}\hat{a} + \sum_{i=1}^2 \epsilon_{i}\sigma_{ii} + \hbar\sum_{i,j=1}^2 g_{12} \sigma_{ij}(\hat{a}^{\dagger} + \hat{a}) \nonumber\\
\equiv& \hbar\omega_{ph}\hat{a}^{\dagger}\hat{a} +  \hbar\omega_{12}\sigma_{z} + \hbar \Omega_R (\sigma_{+}+\sigma_-)(\hat{a}^{\dagger} + \hat{a}),
\end{align}
where in the second line, we have renamed the coupling constant $g_{12}=g_{21}=\Omega_R$ as usual in this context. $\Omega_R$ is the famous Rabi frequency that we have introduced in the previous section. Since it is common practice, we additionally have introduced the Pauli-matrices $\sigma_z=\sigma_{22}-\sigma_{11}, \sigma_+=\sigma_{12}, \sigma_-=\sigma_{21}$. To rewrite the electronic Hamiltonian in the second line, we have utilized the equality $\sigma_{22}+\sigma_{11}=1$, have introduced the transition frequency $\omega_{12}=(\epsilon_2-\epsilon_1)/\hbar$ and have removed the constant energy contribution of $(\epsilon_2+\epsilon_1)/2$.
Despite its seeming simplicity, the Rabi model has only a semi-analytic solution, which is only known since \citeyear{Braak2011}~\citep{Braak2011}. This reflects how intricate the coupled electron-photon problem really is. 

Finally, there is the aforementioned rotating-wave approximation to $H_R$ that allows for a complete analytical diagonalization. Neglecting the so-called counter-rotating terms $\sigma_+\hat{a}^{\dagger}$ and $\sigma_-\hat{a}$, we arrive at the Jaynes-Cummings Hamiltonian \citep{Jaynes1963}
\begin{align}
\label{eq:QO_Hamiltonian_JC}
	\hat{H}_{JC} = \hbar\omega_{ph}\hat{a}^{\dagger}\hat{a} +  \hbar\omega_{12}\sigma_{z} + \hbar \Omega_R (\sigma_{+}\hat{a}+ \sigma_-\hat{a}^{\dagger}),
\end{align}
that we have introduced in the previous section. It is one of the very few light-matter problems that are analytically solvable, which explains its key-role for the understanding of electron-photon interaction, especially for the cavity-system.

\subsubsection*{Collective coupling: the Dicke construction}
All cavity-QED models can be generalized in a simple manner to describe $N$ identical molecules (or more general matter systems) if we assume that the wave functions of different atoms do not overlap. In this case, one can represent the $N$-molecule-one-mode problem again as single molecule, that is coupled to one mode with an effective coupling strength $g_{ij}\rightarrow \sqrt{N}g_{ij}$.\footnote{A very detailed derivation of this can be found in the review of ~\citet{Kirton2019} (see especially Sec. 3.2).} This assumption is for example well justified in a molecular gas, which was the scenario that \citet{Dicke1954} had in mind, when he introduced his construction for $N$ two-level systems in 1954. In the realm of strong-coupling physics, this collective superposition of the matter systems is regarded as one of the basic mechanisms that lead to strong coupling.\footnote{This is typically regarded as a common fact. See for instance the review of \citet{Kockum2018} or \citet{Keeling2007}.} The \emph{collective} coupling strength is $\sqrt{N}$ times larger than the single-molecule coupling, which leads to a huge increase for a macroscopic number of such systems, i.e., $N\approx 10^{23}$.

\subsection{The limitations of cavity-QED models}
\label{sec:intro:cavity_qed_models:justification}
We conclude with a brief discussion on the justification and range of validity of the cavity-QED models. First of all, we have seen that to arrive at the $\hat{H}_{cQED}$, we have started at a very general level and followed a well-defined hierarchy of approximations. We have discussed already that, as written in Eq.~\eqref{eq:intro:QO:Hamiltonian_cavity_QED}, the Hamiltonian has a wide range of validity and this fact is often stressed in the literature.\footnote{However, it is important to realize that one cannot just increase the number of basis states in Eq.~\eqref{eq:intro:QO:Hamiltonian_cavity_QED}. Due to having no $\br^2$-term, this model has no eigenstates in the limit of large  basis-sets~\citep{Schaefer2020}.} Cavity-QED models are thus often seen as a kind of first-principles method.\footnote{See for example the section on ``models'' in Ref.~\citep{Kockum2018}.} One conclusion from this point of view is that the different perspective that, e.g., first-principles methods would provide is superfluous. The impressive success of cavity-QED models to understand the phenomena of strong-coupling physics support this argument. However, this point of view basically disregards the quantum many-body problem and the complexity that matter systems present.

To illustrate this, let us briefly reflect on the most common cavity-QED models, which describe molecules as two-level systems. Strictly speaking, the two-level approximation is only valid for a single spin, but its validity can be convincingly generalized to any kind of transition in a matter system that is energetically \emph{well-separated} from all other transitions of the system (at least for not too strong coupling-strengths). Nevertheless, the two-level approximation has been extensively used way beyond this range of validity. \citet{Frasca2003} summarizes this fact in his review of the two-level approximation as: ``It is safe to say that the foundations of quantum optics are built on the concept of a few level atom.'' Hence, the justification for the approximation is obviously not its well-defined range of validity, but the very good agreement of the corresponding models with many experimental results. In their review on strong coupling, \citet{Kockum2018} even present the cavity-QED models in a generalized version, explicitly noting that only special parameter choices can be ``derived from first principles.'' Again their justification is coming from experiment. This shows that the strength of cavity-QED models is not their (only sometimes possible and often hardly justifiable) connection to the fundamental level of theory, but their obvious \emph{universality}.

Take for instance the absorption spectrum of a complex systems like a molecule in a cavity in resonance with a certain energy transition of the molecule, which shows a Rabi splitting (see Fig.~\ref{fig:intro:strong_coupling_experiment_results} (a)). If we can describe this accurately by the Jaynes-Cummings model, then we have learned that the principal physics of this process can be understood in terms of the hybridization between one electronic transition and a harmonic oscillator.\footnote{The model additionally allows to differentiate the ``character'' of the electronic transition. For example, if the dominant matter oscillator is a two-level system, a so-called \emph{quantum-blockade} \citep{Imamoglu1997}, which is experimentally measurable, occurs.} This is extremely valuable, because it reveals the underlying mechanism of this phenomenon.


\subsubsection*{Why the first-principles perspective is useful}
We have seen in the previous subsection, how important this principal understanding of polariton emergence was for the progress in strong-coupling physics. However, we have also discussed the (current) limitations of the models, which are difficult to define but non-negligible. The controversy about the collective nature of strong-coupling effects in the realm of polaritonic chemistry exemplified some of these limitations: to describe the complex setting of a chemical reaction that takes place inside a cavity, simple approaches like the Jaynes-Cummings model have to be extended, because there are several matter-degrees of freedom that play a role. However, such extensions are not straightforward, but require a detailed knowledge of the matter system, which often is not provided by experimental data. Sometimes this situation can be remedied by physical argumentation, but the discussed controversies show that this is not always the case. 

The history of the field of quantum chemistry knows a multitude of such kind of controversies and in the majority of the resolved cases, the answers were nontrivial.\footnote{For a general introduction to the theoretical description of chemical reactions and its challenges, the reader is referred to, e.g., the textbook by ~\citet{Moore1981}.} Properties of molecules might depend on tiny variations of the geometry or the electronic configuration, which in turn might be the result of an intricate interplay of different effects that counteract each other. Many open questions regarding chemical reactions could only be resolved by an exhaustive use of first-principles methods, many of which nowadays have become standard tools. Clearly, all practical methods to describe quantum many-particle systems have to employ approximations because of the many-body problem. But the first-principles approach can provide (if applicable) a less-biased perspective than effective models, because the explicit description of the particles allows for approximations an a more general level. In the case of coupled electron-photon systems, this means that we treat electrons and their Coulomb interaction on the \emph{same footing} as photons and the electron-photon interaction. Thus, we are able to study the \emph{interplay} of both forms of interaction. We will see in Ch.~\ref{sec:dressed:results} how this perspective allows to identify new effects that are not easy to describe with methods that are based on cavity-QED models. For instance, we will present some results suggesting a further mechanism that influences the coupling strength between electrons and photons. This puts the Dicke-type mechanism of collective strong coupling in a different perspective.

\subsubsection*{The problem of standard model approximations: can we neglect the dipole-self energy?}
Finally, we want to remark on one other approximation that we have done for the derivation of the models. This is the neglect of the dipole-self energy, i.e., the $\br^2$-term. We will see in the next sections that in order to find the ground state of a coupled electron-photon system with a first-principles method, this term is of utmost importance and its neglect would lead to useless results. The reason is that the corresponding Hamiltonian is unbounded without this term~\citep{Rokaj2018}. In a practical calculation, this means that the ground state is not well-defined but depends on the basis and its energy can in principle be shifted to arbitrarily low values. Interestingly, this is a considerably less severe issues for cavity-QED models. The reason is that within the few-level approximation, every system becomes finite and if we fit the electronic energy levels to some experimental data, there is basically no issue. However, neglecting the diamagnetic term can also lead to issues for few-level systems. The Rabi model for example loses its gauge invariance for large coupling strengths~\citep{DeBernardis2018,DiStefano2019}. And connected to that, there are several indications that the transition to the superradiant phase, predicted by the Dicke model cannot occur in equilibrium if the dipole-self energy is properly taken into account~\citep{Andolina2019}.

In contrast, one of the main goals of first-principles methods is to determine these electronic energy levels, which crucially requires the diamagnetic term. This is an example of how intricate the interplay of approximations can be. It shows on the one hand the need for a very careful analysis of the mathematical and physical framework in which a new method is developed (and which is the topic of the next chapters). On the other hand, this example illustrates the challenges of an interdisciplinary field. The neglect of the diamagnetic term is a standard approximation in cavity QED and in usual publications, is not even discussed anymore. On the contrary, in Ref.~\citep{Rokaj2018}, the authors have shown possible severe consequences of this approximation such as the non-existence of a ground state. Which consequences these findings have with respect to the validity of cavity-QED models, is still not entirely resolved.\footnote{See for example the discussion on p.3 in Ref.~\citep{Galego2019}, where the authors defend the neglect of the dipole self-energy in their model. They refer explicitly to Ref.~\citep{Rokaj2018}.} An exhaustive discussion about the diamagnetic term can be found in Ref.~\citep{Schaefer2020}


\subsubsection*{Summary}
In summary, the set of cavity-QED models constitutes a very powerful tool box that has been successfully used to accurately describe a wide range of phenomena in coupled light-matter systems. With only very few fitting parameters, cavity-QED models describe many experimental results \emph{quantitatively}. Their simplicity is thereby an important strength, because it allows for the definition of simple and clear concepts to interpret and not only to fit the data, e.g., a polariton-model of the Jaynes-Cummings model. However, there are limits to the applicability of cavity-QED models and determining and overcoming these limits is crucial for progress in the research field of cavity QED. In the next section, we thus want to approach the problem from ``the other side,'' and introduce the framework for a first-principles description of coupled matter-cavity systems.

\clearpage
	\section{The (quantum) theory of light-matter interaction}
\label{sec:intro:light_matter}
The goal of this section is to find a good starting point for our first-principles description of coupled light-matter systems. 
For that \citet{Schaefer2020} have defined the following three ``basic constraints [...] a theory of light-matter interactions [should] adhere to'':\footnote{Note that many cavity QED models and also typical approaches based on perturbation theory do \emph{not} adhere to all of these constraints. This is in many cases related to the neglect of the diamagnetic contribution~\citep{Schaefer2020}.}
\begin{enumerate}
	\item All physical observables should be independent of the gauge choice and of the choice of coordinate system (for instance, it would be unphysical that the properties of atoms and molecules would depend on the choice of the origin of the laboratory reference frame). 
	\item The theory should support stable ground states (else we could not define equilibrium properties and identify specific atoms and molecules).
	\item The coupled light-matter ground state should have a zero transversal electric field (else the system would radiate and cascade into lower-energy states).
\end{enumerate}
Having these constraints in mind, we briefly discuss in the following subsections the principles and mathematical issues that enter the (quantum) theory of light-matter interaction on different levels of accuracy. We start at the most fundamental level, which is the (full-relativistic)\footnote{This means that matter as well as photon degrees of freedom are described within the framework of the theory of \emph{special relativity}. See for example Ref.~\citep{Greiner1990}.} theory of quantum electrodynamics  (QED).\footnote{Where with most general, we mean to include all effects that are somehow related to the electromagnetic interaction. Other fundamental forms of interaction like gravity, weak or strong interaction are excluded in this picture.} 
We keep the discussion of QED very short, summarizing the basic concepts of QED, its fundamental character and especially its mathematical challenges. We will see that formulating a theory in the realm of QED that adheres to the three basic constraints is very difficult. 
However, the wide range of energy-scales that QED theoretically covers is not necessary for an accurate description of the typical phenomena of the field of polaritonic chemistry. We therefore reserve the principal part of the section for the \mbox{(semi-)}nonrelativistic limit of QED, where we assume that the active particles of the considered matter systems have considerably lower energies than their rest-energy. This approximation is very well justified for most processes in condensed matter and chemical systems. 
In the last subsection, we derive the \emph{long-wavelength limit} of the Pauli-Fierz Hamiltonian and introduce the \emph{Born-Oppenheimer approximation}. The resulting cavity-QED Hamiltonian and the corresponding mathematical framework serves as the basis for our discussion of first-principle methods.  

\subsection{The most fundamental theory: QED}
\label{sec:intro:full_qed}
\begin{aquote}{\citeauthor{Dirac1929}, \citeyear{Dirac1929} \citep{Dirac1929}}
	The general theory of quantum mechanics is now almost complete ... The underlying physical laws necessary for the mathematical theory of a large part of physics and the whole of chemistry are thus completely known, and the difficulty is only that the exact application of these laws leads to equations much too complicated to be soluble. It therefore becomes desirable that approximate practical methods of applying quantum mechanics should be developed, which can lead to an explanation of the main features of complex atomic systems without too much computation.
\end{aquote}
The interaction between light and matter as we understand it today from a physics perspective is such a vast topic that it concerns to a smaller or greater extend \emph{all} natural sciences. The usual explanation behind this is simply that all known matter consists of charged particles, which constantly radiate and interact via some form of electro-magnetic fields, i.e., light. The natural conclusion from this (reductionist) point of view has been summarized concisely in the above quotation by Dirac. Transferred to our problem, this means that in order to understand strong-coupling physics, we just need to find the most precise, i.e., quantum description of light-matter interaction. The rest is ``just'' mathematics. This ``most precise'' theory of light-matter interaction is  QED, which usually is regarded as the unification of the theory of special relativity, quantum mechanics, quantum field theory and classical electrodynamics~\citep{Greiner1990,Greiner2009}. 

That is why we call QED a \emph{relativistic quantum field theory}, which means that it is consistent with the geometry of special relativity, accounts for quantum effects of the microscopical world and is formulated only in terms of fields (instead of, e.g., particles that here emerge as effective degrees of freedom from the field). There are many reasons for this unification of theories with the most important being \emph{consistency}. Since Maxwell's equations of the electromagnetic field (Eq.~\ref{eq:intro:MWEquations}) are relativistically invariant,
it seems to be a logical step to also generalize the quantum description of matter to the full-relativistic level, i.e., to consider the \emph{Dirac equation}. Combining both theories requires then to leave the realm of standard quantum mechanics based on wave functions, and generalize Dirac's theory to a \emph{quantum-field theory}~\citep{Greiner2013}. For further details on these considerations and other very interesting discussions, the reader is referred to one of the many textbooks on this topic, e.g., Refs.~\citep{Greiner1990,Greiner2009,Greiner2013}.

Guided by the idea of consistency, researchers have pursued the unification of quantum mechanics and electrodynamics at least since the 1930s. In this process, they encountered so many severe mathematical and connected conceptual problems that first successful calculations have not been undertaken before the late 1940s.\footnote{We recommend the very-well documented article on the history of QED on \textsc{Wikipedia}: \url{https://en.wikipedia.org/wiki/Quantum_electrodynamics\#History}.}
The last important steps have been made by Tomonaga, Schwinger and Feynman, which were rewarded with the Noble prize ``for their fundamental work in quantum electrodynamics''.\footnote{\url{https://www.nobelprize.org/prizes/physics/1965/summary/}} 
Most strikingly, on the basis of QED one can explain why the ``electron's magnetic moment proved to be somewhat larger than expected,''\footnote{See facts about J. Schwinger on \url{https://www.nobelprize.org/prizes/physics/1965/summary/}} -- with twelve decimal places precision.  This and also many other impressive predictions\footnote{There is set of standard effects that can be described and understood by QED and that are nowadays textbook knowledge. See for example \citep[part 6]{Kaku1993}. For a recent application of QED, see \citep{Aoyama2012}.} established undoubtedly today's reputation of QED of being the ``best-tested'' theory or the ``jewel'' of physics."\footnote{These kinds of statements can be found in basically every book about QED or other quantum field theories. For a good summary of this point of view and the according predictions, we recommend, e.g., Ref.~\citep{Henson2017}.}
This success story clearly illustrates the power of the idea of consistency between different physical theories. And indeed, encouraged by the success of QED, the story of unification went on constantly leading to  the \emph{standard model of particle physics}. This theoretical framework unifies three of the four known fundamental interactions (electromagnetic, weak and strong). Until today, many researchers work on including the missing gravitational interaction, which is described by the theory of general relativity. 

However, there is also an opposing point of view on the consistency of QED, which puts the theory in a different light. Most importantly, there is still no practical way known how to remove the occurring divergences in a non-perturbative approach~\citep{Berestetskii1982}. Additionally, the mere concept of a many-particle wave function leads to several inconsistencies in a relativistic situation, since each particle should have its own time-coordinate~\citep{Lienert2017}. 
Many years after the formulation by Tomonaga, Feynman and Schwinger (1948/1949), Iwo and Zofia Bia{\l}ynicki-Birula\footnote{Note that Iwo and Zofia Bia{\l}ynicki-Birula are strong supporters of ``this beautiful theory[, i.e., QED]''~\citep{Biaynicki-Birula1996}. They contributed importantly to formulate QED more rigorously.} summarize the state of the research on QED in the following way:
\begin{aquote}{\citeauthor{Bialynicki1975}, \citeyear{Bialynicki1975} \citep{Bialynicki1975}}
	In quantum electrodynamics, we have a theory that is not complete. Owing to the enormous mathematical difficulties, it is not yet known whether the set of postulates adopted is not inconsistent and whether it determines the theory uniquely. All we know is an approximate scheme for carrying out calculations based on perturbation theory.
\end{aquote}
They conclude that ``Quantum electrodynamics thus constitutes a programme rather than a closed theory.'' The problems, Iwo and Zofia Bia{\l}ynicki-Birula are referring to are especially severe in the relativistic domain, where fermion-pair-creation or other high-energy processes have to be taken into account. Although there has been progress in the mathematical formulation of QED~\citep{Tachibana2017}, it has shown its practical value until nowadays predominantly as basis for perturbative corrections of scattering processes.\footnote{Note that basically all of the aforementioned impressive predictions belong into this group.}
However, scattering processes happen typically in very special experimental situations or for very high energies. Their theoretical description is based on some well-defined and simple reference states, which describe the system before and after the scattering event has taken place~\citep{Greiner2009}. But not all types of systems can be represented within the strict boundaries of this assumption. Most importantly, it is very unpractical to describe ground states of (strongly) coupled light-matter systems as a scattering process.\footnote{In principle, one could perform a scattering description of bound states as well. The problem is then more practical. Assume we would have a QED theory that captures bound-state resonances (live long, but not infinitely long). The spectrum goes (from -$\infty$ if no positron interpretation is done) to $\infty$ and we have no idea where the special "ground-state resonance" is lying. So we would need to probe all in/out states and minimization is useless. If we instead have a Hamiltonian that is bounded from below, the variational principle tells us ``where to look'' for the ground state, i.e., we just need to minimize the energy.}
One (but not the only) important reason that hampers applying QED to other than scattering processes is that the standard technique of renormalization can only be performed order by order in a perturbation series~\citep{Greiner2009}. 

One of the few attempts to go beyond scattering problems by introducing a ``space-time resolved approach'' to QED has been undertaken by \citet{Wagner2011}. They also investigated within this approach the possibility of a purely computational renormalization scheme~\citep{Wagner2013}, yet only for very simplified models. To employ the usual methods of electronic-structure theory, we would need a well-defined Hamiltonian formulation of QED. Such formulations indeed do exist\footnote{See for example Ref.~\citep[part 5]{Barut1984} for a physicist's definition and Ref.~\citep{Takaesu2009} for a mathematically rigorous discussion and definition of the QED Hamiltonian.}, but the problems that Iwo and Zofia Bia{\l}ynicki-Birula mentioned remain. There is for example no definite answer on how to deal with the many of the occurring divergences, besides removing them by artificial cutoffs (which makes calculations depend on these cutoffs)~\citep{Takaesu2009}.

\subsection{The solid ground: the Pauli-Fierz Hamiltonian}
\label{sec:intro:nr_qed}
The energy-scales that are involved in usual chemical and materials-science processes are usually small in comparison to the involved rest masses, such that relativistic effects of the matter-degrees of freedom can be neglected. However, the description of the electromagnetic field needs to be relativistic on these energy scales.\footnote{Note that there are also non-relativistic formulations of the electrodynamics, which however have a limited practical applicability. See, e.g., Ref.~\citep{LeBellac1973}.} This suggests to regard the (semi-)nonrelativistic ``limit'' of QED (NR-QED) that considers nonrelativistic charged particles and the electromagnetic field. This scenario is very well studied and thus especially its mathematical properties are comparatively well understood. We summarize in the following how NR-QED can be constructed and discuss the most important conceptual and mathematical issues. We follow hereby the excellent book by \citet{Spohn2004}. 

\begin{fdef}{0.9\columnwidth}{Maxwell-Lorentz Equations}
	\label{def:Maxwell-Lorentz_Equations}
	For a charge density of $N_e$ electrons and $N_n$ nuclei with masses $m_i$, charges $q_i$ and positions $\br_i, i=1,...,N_e+N_n$
	\begin{align}
	\rho(\br,t)=\sum_{i=1}^{N_e+N_n} q_i \delta(\br-\br_i),
	\end{align}
	we associate a current $\bj(\br,t)$. Both are linked by the continuity equation
	\begin{align}
	\partial_t \rho(\br,t) + \nabla\cdot\bj(\br,t)=0.
	\end{align}
	For simplicity, we constrain the description explicitly to length scales, were the contribution of the aforementioned form-factor is negligible~\citep{Spohn2004} and thus consider point-particles in this definition. The evolution of the $N_e+N_n$ particles and the electric field $\bE$ and the magnetic field $\bB$ is governed by the 4 Maxwell's equations and Newton-Lorentz's equations for the particles.	The first two Maxwell's equations describe the evolution of $\bE$ and $\bB$ the under the influence of the current by
	\begin{subequations}
		\label{eq:intro:MWEquations}
		\begin{align}
		\partial_t \bB (\br,t) =& - \nabla \times \bE(\br,t), \\
		\partial_t \bE (\br,t) =& c^2 \nabla \times \bB(\br,t) + \mu_0c^2 \bj(\br,t),
		\intertext{where  $c$ is the speed of light and $\mu_0$ the vacuum permeability. The other two Maxwell's equations are constraints to the evolution that read}
		\label{eq:MW_Gauss}
		\nabla\cdot\bE =&\rho(\br,t),\\
		\nabla\cdot\bB =& 0.
		\end{align}
	\end{subequations}
	Newton-Lorentz's equation for the evolution of the i-th particle under the influence of $\bE$ and $\bB$ reads
	\begin{align}
	m_i \tfrac{\td^2}{\td t^2}\br_i(t) = e\bE(\br_i,t) + c^{-1} \tfrac{\td}{\td t}\br_i \times \bB(\br_i,t).
	\end{align} 
\end{fdef}

\subsubsection*{The fundamental inconsistency of the classical theory of coupled light-matter systems}
To derive the Hamiltonian or equivalently the equations of motion of NR-QED, we can apply the \emph{correspondence principle}~\citep{VanVleck1928} to its classical analogue. Thus, there is no need of, e.g., performing a limiting procedure from a more fundamental theory and we will follow this route here. The classical correspondence to NR-QED is \emph{Lorentz-Maxwell} theory that defines a set of coupled  equations  of motion (Def.~\ref{def:Maxwell-Lorentz_Equations}) which govern the dynamics of $N$ charged particles that are coupled to their own electro-magnetic radiation field. Already on the classical level, this coupling is problematic. To see this, let us shortly consider the well-defined equations of motion of a point charge in an \emph{external} electromagnetic field (Newton-Lorentz equations). Neglecting magnetic fields for the moment, we can write the Newton-Lorentz equation for a nonrelativistic point charge of the mass $m$ and charge $e$ as
\begin{align}
\label{eq:NewtonLorentz}
m\frac{\td^2}{\td t^2}\br(t) = e \bE(\br,t),
\end{align} 
where $\bE(\br,t)$ is the electric field. Since we assumed that $\bE(\br,t)$ is external, i.e., prescribed, this equation together with a suitable initial condition (e.g., $\br=\dot{\br}=0$) constitutes a well-defined initial value problem. It is however important to know $\bE(\br,t)$ exactly at the position $\br$ of the particle for all times $t\geq t_0$ bigger or equal to the initial time $t_0$. In Lorentz-Maxwell theory, we then want to couple Eq.~\eqref{eq:NewtonLorentz} with Maxwell's equations (see Def.~\ref{def:Maxwell-Lorentz_Equations}). This means that $\bE(\br,t)$ is not anymore some prescribed function, but a dynamical quantity that has to be \emph{self-consistently} determined by the coupled equations (this is of course also true for all other variables). This allows for example for back-reactions by the field on the particles and vice versa. 
On the other hand, there is the well-defined theory of Maxwell's equations that react to a prescribed charge density $\rho(\br,t)$. For instance, the electric field of a static charge distribution is obtained simply by solving Gauss's law, cf. Eq.~\eqref{eq:MW_Gauss} (the other equations do not play a role in this simple case)
\begin{align}
\label{eq:CoulombLaw}
\bE(\br,t)=\int\td\br' \rho(\br')\br'/|\br-\br'|^3.
\end{align}
In the case that $\rho(\br)=\delta(\br-\br')$ is just a point charge, we $\bE$ is not anymore well-defined at every point, but \emph{diverges} exactly at the position of the charge. We have to conclude that connecting Eq.~\eqref{eq:CoulombLaw} and Eq.~\eqref{eq:NewtonLorentz} leads to an inconsistency. Note that this issue is resolved in Schrödinger quantum mechanics for matter systems with the Coulomb interaction, but without taking into account the transversal degrees of freedom of the electromagnetic interaction. The debate on this has a long and interesting history and for further details, the reader is referred to, e.g., the essay by the famous mathematical physicist John \citet{Baez2016}. 

\subsubsection*{The resolution: regularization and renormalization}
Nevertheless, we need to ``cure'' this inconsistency for our theory and the usual way to do so is by a so-called \emph{regularization} procedure for short distances. This means in practice that we consider a small sphere instead of a point charge, which directly removes the divergence in Eq.~\eqref{eq:CoulombLaw}. Still, we need to decide about the exact shape and size of this sphere, which is formally done by introducing a \emph{form-factor} in the equations.\footnote{There are different ways to do this and certain connected issues. See \citep[part 2/part 4]{Spohn2004} for details}
By the introduction of the cutoff, we remove small length scales from the description that ``come from a more refined theory''~\citep[p. 15]{Spohn2004}. 
The regularization removes the divergence of the self-energy, i.e., electrostatic energy of the particles corresponding to the Coulomb force~\eqref{eq:CoulombLaw}. But it still contributes to the energy-momentum relation of the coupled theory, which consequently differs from the one of a free particle. Since the latter is an experimental fact, we need to ``fix'' the coupled energy-momentum relation by performing a \emph{renormalization}. We see here that although mostly discussed in the area of quantum field theory, the technique of renormalization is already required on the classical level. 
In the case of Lorentz-Maxwell theory, this means that we choose the particle's mass as a parameter that we utilize to fit the theory to the experiment~\citep[part 6 and part 7]{Spohn2004}.

We see that the description of coupled light-matter systems is a very challenging topic, even on the ``simplest,'' i.e., the classical, level and there are many other subtleties and interesting issues.

\subsubsection*{The quantization of the classical theory: gauge dependence}
Starting from the classical Lorentz-Maxwell theory, we now introduce NR-QED by the correspondence principle. The standard way to do this is to employ the canonical quantization procedure, which requires the \emph{Hamiltonian} formulation of the classical theory. Regarding the matter part, this is straightforward, but for the Hamiltonian formulation of Maxwell's theory, care has to be taken. To describe the electromagnetic field in terms of canonical variables, one usually introduces the vector and scalar potential $\bA$ and $\phi$. These are connected to the fields $\bE,\bB$ by
\begin{subequations}
	\begin{align}
	\label{eq:EB_connection_APhi}
		\bE(\br,t)&=-\tfrac{1}{c}\partial_t \bA(\br,t) -c\nabla \phi(\br,t)\\
		\label{Def_vector_scalar_potential_B}
		\bB(\br,t)&=\tfrac{1}{c}\nabla\times\bA(\br,t),
	\end{align}
\end{subequations}
which is not a unique correspondence but leaves a so-called gauge freedom. This means that the quantization procedure \emph{depends} on the gauge~\citep{Greiner2013}. 
In the nonrelativistic description, it is very convenient to employ the so-called \emph{Coulomb gauge}
\begin{align}
\label{eq:Coulomb_gauge}
\nabla\cdot\bA(\br,t) = 0.
\end{align}
The Coulomb gauge removes the longitudinal part of the vector potential, which by means of Gauss's law (Eq. \eqref{eq:MW_Gauss}) also fixes the scalar potential $\phi(\br, t)$.\footnote{To see this, compare Eq.~\eqref{eq:MW_Gauss} with the definition of $\bE$ in terms of $bA,\phi$, cf. Eq.~\eqref{eq:EB_connection_APhi}: $\rho=\nabla\cdot\bE=\nabla\cdot(-1/c \partial_t\bA-c\nabla \phi)=-c\nabla^2 \phi$.} The remaining two transversal degrees of freedom of $\bA=\bA_{\perp}$ describe the two known polarizations of the electromagnetic field.
This means that in the Coulomb gauge, there is no ``superfluous'' field component and thus it can be seen as kind of maximal gauge. That is why basically all derivations of NR-QED are done in the Coulomb gauge. However, there are also other gauges that are maximal in the same sense, e.g., the Poincare gauge and employing these might be advantageous in certain scenarios~\citep{Keller2012}. 

\subsubsection*{The Hamiltonian of NR-QED: how to construct a well-defined mathematical theory of light-matter interaction}
However, to discuss the quantization of the Maxwell-Lorentz theory we stick to the standard case. We employ Coulomb gauge and follow the prescription of the canonical quantization, which leads the so-called \emph{Pauli-Fierz Hamiltonian}. Since this is a standard procedure that can be found in several textbooks, we only summarize the most important aspects. We again follow closely the book of Herbert Spohn~\citep{Spohn2004}. The Pauli-Fierz Hamiltonian describes the quantum dynamics of $N$ charged particles that are coupled to the (vacuum) electromagnetic field. Since we are interested in molecular systems, we further divide $N=N_e+N_n$ in $N_e$ electrons and $N_n$ nuclei. 
The quantum formulation does not remedy the inconsistencies of the classical theory, but adds further problems. The most important property of a Hamiltonian of a quantum theory is hermiticity (which in mathematical literature is usually called \emph{self-adjointness}). Importantly, it is possible to prove the self-adjointness of the Pauli-Fierz Hamiltonian under the following two conditions~\citep[part 13.3]{Spohn2004}. 
\begin{enumerate}
	\item We need to remove the self-energy of the particles analogously to the classical case.
	\item We need to introduce a further ``suitable'' ultraviolet cutoff. This comes \emph{on top} of the classical cutoff that we already included by quantizing extended instead of point charges. Still, the new cutoff just further renormalizes the mass and in most situations, we can include it as in the classical case by using the physical instead of the bare mass of the charge carriers. However, although the proof for self-adjointness provides a definition of the cutoff, it remains arbitrary up to a certain degree.
\end{enumerate}

A Hamiltonian that obeys these conditions is the Pauli-Fierz Hamiltonian of Def.~\ref{def:Pauli-Fierz_Hamiltonian}
\begin{fdef}{0.92\columnwidth}{Pauli-Fierz Hamiltonian}
		\label{def:Pauli-Fierz_Hamiltonian}
			A system consisting of $N_e$ electrons and $N_n$ nuclei with masses $m_i$ and charges $q_i, i=1,...,N_n+N_e$ that are minimally coupled to the electromagnetic field in the Coulomb gauge are described in the nonrelativistic energy regime by the \emph{Pauli-Fierz Hamiltonian}
			\begin{align}
			\label{eq:Hamiltonian_Pauli-Fierz}
			\begin{split}
				\hat{H}_{PF}=&\sum_i^{N_e+N_n} \tfrac{1}{2m_i} \left(-\I\hbar\nabla_i - \frac{q_i}{c} \hat{\bA}_{\perp}(\br_i)\right)^2 + \frac{1}{8\pi\epsilon_0}\sum_{i\neq j}^{N_e+N_n} \frac{q_iq_j}{|\br_i-\br_j|} \\
				&+ \frac{1}{2}\sum_{\bk,s}\left((-\I\partial_{q_{\bk,s}})^2 + \omega_{k}^2 q_{\bk,s}^2\right)
			\end{split}
			\end{align}
			Here $\I$ is the imaginary unit, $\hbar$ the Planck constant and $\epsilon_0$ the vacuum permittivity. We denoted the field operator of the (transversal) vector potential with $\hat{\bA}_{\perp}$ that reads
			\begin{align}
				\hat{\bA}_{\perp}(\br)=\sqrt{\tfrac{\hbar c^2}{\epsilon_0}}\sum_{\bk,s} \frac{1}{\omega_k} \left(\hat{a}_{\bk,s}^{\dphan}\bS_{\bk,s}(\br) + \hat{a}_{\bk,s}^{\dagger}\bS^*_{\bk,s}(\br) \right),
			\end{align}
			where 
			\begin{align}
			\label{eq:intro:mode_function}
				\bS_{\bk,s}(\br) = \tfrac{1}{\sqrt{V}}\bn_{\bk,s} e^{\I \bk\cdot\br}
			\end{align}
			are the mode functions for a plane wave with wave-vector $\bk$ and frequency $\omega_k = c k$ ($k\equiv |\bk|$). We denote with $V=a^3$ the volume of the quantization box with side-length $a$.  
			Furthermore, $\bn_{\bk,s}$ denotes the two polarization vectors ($s=1,2$) for every mode. To arrive at the form \eqref{eq:Hamiltonian_Pauli-Fierz}, we have defined the displacement coordinates
			\begin{align}
				q_{\bk,s}=&\sqrt{\tfrac{\hbar}{2\omega_k}}(\hat{a}_{\bk,s}^{\dagger}+\hat{a}_{\bk,s}^{\dphan})\\
				-\I\partial_{q_{\bk,s}}=&\I\sqrt{\tfrac{\hbar\omega_k}{2}}(\hat{a}_{\bk,s}^{\dagger}-\hat{a}_{\bk,s}^{\dphan}).
			\end{align}
\end{fdef}


\subsubsection*{A remark on the validity of Pauli-Fierz theory}
We want to conclude this subsection with a small remark on the interpretation Pauli-Fierz theory. This paragraph is not necessary to follow the course of this text and can be skipped by the fast reader.

Pauli-Fierz theory as defined in Def.~\ref{def:Pauli-Fierz_Hamiltonian} is mathematically a well-defined quantum theory, i.e., the Hamiltonian of definition~\ref{def:Pauli-Fierz_Hamiltonian} is self-adjoint and it has eigenstates that can be described by a many-particle wave function. However, to remove the occurring divergences of the theory, we have to introduce a cutoff that is in principle arbitrary. Thus, for high-energy processes the accuracy of the theory is somewhat unclear.
Nevertheless, the range of validity, on which these mathematical issues are not problematic, is huge:
\begin{aquote}{\citet[p. 157]{Spohn2004}}
	The claimed range of validity of the Pauli-Fierz Hamiltonian is flabbergasting... As the bold claim goes, any physical phenomenon in between [gravity on the Newtonian level and nuclear- and high-energy physics], including life on Earth, is accurately described through the Pauli-Fierz Hamiltonian.
\end{aquote}
The conclusion that Pauli-Fierz theory provides a useful description of, e.g., molecules as complex as DNA or whole organisms is indeed ``flabbergasting.''
Thus, it seems that Dirac's vision has become true (see the quotation in the beginning of Sec.~\ref{sec:intro:full_qed}): we have found a mathematical theory that describes ``a large part of physics and the whole of chemistry.''
We just need to solve this theory by ``approximate practical methods of applying quantum mechanics [...], which can lead to an explanation of the main features of complex atomic systems without too much computation.''  

Interestingly, although Pauli-Fierz theory is known since the 1930s, no practical method has yet been developed that is capable to solve the theory for any relativistic scenarios. Spohn remarks on this reductionist's perspective:
\begin{aquote}{\citet[p. 157]{Spohn2004}}
	Of course, our trust is not based on strict mathematical deductions from the Pauli-Fierz Hamiltonian. This is too difficult a program. Our confidence comes from well-studied limit cases. [...]
	In the static limit we imagine turning off the interaction to the quantized part of the Maxwell field. This clearly results in Schrödinger particles interacting through a purely Coulombic potential, for which many predictions are accessible to experimental verification. 
	But beware, even there apparently simple questions remain to be better understood. For example, the size of atoms as we see them in nature remains mysterious if only the Coulomb interaction and the Pauli exclusion principle are allowed.
\end{aquote}
Another example for such a limit is the standard form of cavity QED that we will derive in the next section from Pauli-Fierz theory by performing the long-wavelength approximation. We will see in the course of this thesis, that only finding the (approximate) ground state of cavity QED is an utmost difficult task. Without the great knowledge of the many-electron problem that we have nowadays, this would probably be an impossible task. This explicitly shows the limitation of the very common (reductionist's) point of view that defines a hierarchy between more general theories and their limit cases. Although Pauli-Fierz theory is more fundamental in the sense that it is more general than, e.g., many-electron Schrödinger theory, it is the predictions of the lower member that are the most important justification for the higher lying member.

This illustrates how long the way from writing down a Hamiltonian until describing nature really is. And most importantly, it is not ``just mathematics'' that is required to go this way as the reductionist's hypothesis suggest. This has been pointed out already in 1972 by \citet{Anderson1972} with several examples from molecular and solid-state physics. The impressive success of the cavity-QED models is another examples that prove this far too simple idea wrong:
the power of these models lies especially in their validity in settings, where the explicit mathematical derivation from first-principles, i.e., the Pauli-Fierz Hamiltonian is not possible (see Sec.~\ref{sec:intro:cavity_qed_models}).
Spohn summarizes this in the following way:
\begin{aquote}{\citet[p. 158]{Spohn2004}}
	[If t]he Hamiltonian is a self-adjoint linear operator, [... o]f course this does not mean that we have solved any physical problem. It just assures us of a definite mathematical framework within which consequences can be explored.
\end{aquote}
This is the reason for the long discussion of the mathematical foundations of Pauli-Fierz theory in this subsection. To develop new methods that accurately describe electron-photon interaction from first principles, we need a (mathematically and physically) firm ground. With the corresponding methods, we will however not ``solve'' Pauli-Fierz theory in the sense that we diagonalize $\hat{H}_{PF}$. There are no analytical solutions known and even the simplest case of one particle, coupled to the Maxwell-field is basically numerically inaccessible. But we need such a ``definite mathematical framework'' to derive approximate theories. 

	\newpage
\subsection{NR-QED in the Long-Wavelength Limit}
\label{sec:intro:nrqed_longwavelength}
In this subsection, we derive the cavity-QED Hamiltonian from Pauli-Fierz theory. The physical setting, defined by this limit is the basis for our analysis in the following chapters. Since we aim to understand strong-coupling phenomena that require the strong mode confinement of cavities, we constrain the electromagnetic field description to only a \emph{few modes}. These modes are assumed to have long wavelengths in comparison to the spatial extension of the considered matter systems, such that the \emph{dipole approximation} is valid.\footnote{Note that the dipole approximation is accurate also in more general settings in the context of cavity QED, e.g., nanoplasmonic environments.} Regarding the matter-description, we only consider the (adiabatic) electronic structure in the electrostatic potential of (quasi)static or clamped nuclei. This means that we constrain our description to settings, where the \emph{Born-Oppenheimer} approximation can be applied and disregard the dynamics of the nuclei and their quantum nature. This setting is however sufficient to describe a large range of chemical processes~\citep{Worth2004}.\footnote{See also Ch.~\ref{sec:est}.}
 
Despite these many simplifications, we are still confronted with an extraordinarily difficult problem. Also in this limit, there are currently only two special cases, for which analytical solutions are known.\footnote{The first considers one electron in a harmonic potential~~\citep[part13.7]{Spohn2004} and the second $N$ electrons without interaction in a box with periodic boundary conditions~\citep{Rokaj2020}.} 

\subsubsection*{The starting point: the macroscopic description of Pauli-Fierz theory}
The starting point of the derivation is the non-relativistic theory of quantum electrodynamics, which is described by the Pauli-Fierz Hamiltonian $\hat{H}_{PF}$ (Def. \ref{def:Pauli-Fierz_Hamiltonian}). 
Since we are interested in processes with subatomic length and energy scales, we use from now on a suitable unit system, so-called atomic units (see Def.~\ref{def:atomic_units})
\begin{fdef}{\columnwidth}{Atomic units}
\label{def:atomic_units}
		The system of \emph{atomic units} (denoted by \emph{[a.u.]}) is defined by the following base units:
\begin{center}
	\fbox{
		\begin{tabular}{lcl}
			mass & $m_e$ & the electron mass\\
			charge &$e$ & the electronic charge\\
			action &$\hbar$ & the reduced Planck constant\\
			permittivity &$4\pi\epsilon_0$ & the inverse Coulomb constant
		\end{tabular}
	}
\end{center}
		This means that we measure mass in multiples of the electron mass etc. This is formally done by setting these four fundamental constants to one.
		Derived from these base units, we have the following derived units (we list only the most important ones)
		\begin{center}
			\fbox{
				\begin{tabular}{llll}
					length & $\si{\bohr}=4\pi\epsilon_0\hbar^2/(m_e e^2)$ & $1 \si{\bohr}=0.529 \si{\angstrom}$ & ``bohr''\\
					energy & $\si{\hartree}=\hbar^2/(m_ea_0^2)$ & $1 \si{\hartree}=27.211 \si{\electronvolt}$ & ``Hartree''
				\end{tabular}
			}
		\end{center}
	Furthermore, we can derive the numerical value of the speed of light $c=1/\alpha\approx 137$ \emph{[a.u.]}, where we denoted with $\alpha$ the fine-structure constant $\alpha=e^2/(4\pi\epsilon_0\hbar c)$. 
\end{fdef}

We start the derivation with the \emph{macroscopic description} of Maxwell's equations, which separates the charge current
\begin{align*}
	\bj(\br,t)=\bj^b(\br,t) + \bj^f(\br,t)
\end{align*}
into its bound part $\bj^b$ and the free current $\bj^f$. The former is generated by the magnetization $\bM$ and polarization $\bP$ of the matter system as
\begin{align*}
	\bj^b(\br,t) = \nabla \times \bM(\br,t) + \partial_t \bP(\br,t).
\end{align*}
We can then define the displacement field $\mathbf{D}=\epsilon_0\bE + \bP$ and the magnetization field $\mathbf{H}=\frac{1}{\mu_0}\bB - \bM$, which are generated by the free current as
\begin{align*}
	\bj^f(\br,t) = - \nabla \times \mathbf{H}(\br,t) + \partial_t \mathbf{D}(\br,t).
\end{align*}
This construction is in a sense artificial, because the actual forces that are exerted by the field are only due to the total electric and magnetic fields $\bE,\bB$. But it is very suited for the following theoretical considerations, because it separates out the part of the electromagnetic field that is exclusively produced by the (bound) charges of our system.

The connection to the macroscopic form of \eqref{eq:Hamiltonian_Pauli-Fierz} is given by the unitary Power-Zienau-Woolley (PZW) transformation
\begin{align*}
	\hat{U}_{PZW} = \exp\left(-\I \alpha \int\td\br \hat{\bP}_{\perp}(\br)\cdot\hat{\bA}_{\perp}(\br)\right),
\end{align*}
where we only need to transform the transversal fields, because of the Coulomb gauge.
With the choice of our matter system, we can explicitly define the polarization
\begin{align}
\label{eq:Polarization}
	\bP(\br)=-\sum_{i=1}^{N_e}\br_i \int_{0}^{1}\delta^3(\br-s\br_i)\td s + \sum_{i=1}^{N_n} Z_i \bR_i \int_{0}^{1}\delta^3(\bR-s\bR_i)\td s,
\end{align}
where we separated electronic $\br_i$ and nuclear coordinates $\bR_i$ and denoted the charge of the $i$-th nucleus by $Z_i$. Since we included all physical charges in the definition of $\bP$, we trivially have $\bj^f=0$.\footnote{This follows, because we treat a closed system here. In a more general setting, the free current could, e.g., represent an external current that pumps the cavity mode.} 

\subsubsection*{The dipole approximation}
We can now introduce the dipole approximation by disregarding the integration over s in the definition of $\bP$ (Eq. \eqref{eq:Polarization}) or equivalently by setting
\begin{align}
\tag{cf. \ref{eq:intro:QO:dipole_approximation}}
	\bA(\br) \approxeq \bA(0),
\end{align}
where we assumed that the coordinate frame origin is located at the center of charge. Specifically, this leads to the transformation
\begin{align}
\label{eq:Trafo_dipole_length_gauge}
	\hat{U}_{PZW,dip} = \exp\left(-\I \sqrt{4\pi} \sum_{\bk,s} S_{\bk,s}(0) \bn_{\bk,s}\cdot\bR \right)q_{\bk,s},
\end{align}
where $\bn_{\bk,s}$ is the polarization vector and $S_{\bk,s}(0)\propto 1/\sqrt{V}$ the (absolute value of the) mode function of the photon mode with wave vector $\bk$ and polarization $s$ (see Def.~\ref{def:Pauli-Fierz_Hamiltonian}). 
Additionally, we have defined the total dipole of the matter system
\begin{align}
	\bR=-\sum_{i=1}^{N_e}\br_i  + \sum_{i=1}^{N_n} Z_i \bR_i.
\end{align}
Further, we constrain the description to a finite number of $M$ photon modes, where $M$ cannot be chosen arbitrarily, since the dipole approximations breaks down for very large $M$~\citep{Rokaj2020}. In the following, we will usually assume that $M$ is very small (corresponding to the enhanced modes of an optical cavity) and thus, this issue will not be problematic.  Note that Eq. \eqref{eq:Trafo_dipole_length_gauge} is also called \emph{length-gauge} transformation and it is the generalized form of the transformation $U_d$ (defined above Eq. \eqref{eq:intro:QO:length_transformation}) that was employed in the derivation of the cavity-QED models. Furthermore, we subsume $\alpha={\bk,s}$ and apply a canonical transformation that exchanges the role of the photon coordinates and momenta
\begin{subequations}
	\begin{align*}
		-\I\partial_{q_{\alpha}} &\rightarrow -\omega_{\alpha} p_{\alpha}\\
		q_{\alpha} &\rightarrow -\I\omega_{\alpha}^{-1}\partial_{p_{\alpha}}.
	\end{align*}
\end{subequations}
The non-relativistic QED Hamiltonian in the long-wavelength limit reads then
\begin{align}
\label{eq:Hamiltonian_NRQED_dipole}
	\hat{H}_{LW}=\hat{H}_{n}+\hat{H}_{e}+\hat{H}_{ne}+\hat{H}_{p} +\hat{H}_{ep}+\hat{H}_{np},
\end{align}
where we defined the nuclear Hamiltonian 
\begin{align*}
	\hat{H}_{n} =&\sum_i^{N_n} \frac{1}{2 M_i} \nabla_{\bR_i}^2 + \frac{1}{2} \sum_{i\neq j}^{N_n}  \frac{Z_iZ_j}{|\bR_i-\bR_j|},
\intertext{with the nuclear masses $M_i$. The electronic Hamiltonian reads}
	\hat{H}_{e} =&\sum_i^{N_e} \frac{1}{2} \nabla_{\br_i}^2 + \frac{1}{2} \sum_{i\neq j}^{N_e}  \frac{1}{|\br_i-\br_j|}
\intertext{and the interaction Hamiltonian between both matter degrees of freedom is given by}
	\hat{H}_{ne} =& \frac{1}{2} \sum_{i\neq j}^{N_n}  \frac{Z_j}{|\br_i-\bR_j|}.
\intertext{The Hamiltonians for the contributions of all the $M$ photon modes can be collected together by}
	\hat{H}_{p} +\hat{H}_{ep}+\hat{H}_{np}=&\frac{1}{2}\sum_{\alpha}^{M}\left[-\partial_{p_{\alpha}}^2 + \omega_{\alpha}^2 \left(p_{\alpha} - \frac{\blambda_{\alpha}}{\omega_{\alpha}}\cdot \bR \right)^2\right],
\end{align*}
where the light matter interaction becomes explicit by the appearance of the total dipole $\bR$. We also introduced the \emph{fundamental light-matter coupling constant} for mode $\alpha$
\begin{empheq}[box=\fbox]{align}
\label{eq:intro:lambda_definition}
	\blambda_{\alpha}=\sqrt{4\pi} S_{\alpha}(0) \bepsilon_{\alpha}.
\end{empheq}

\subsubsection*{The Born-Oppenheimer approximation}
The Hamiltonian~\eqref{eq:Hamiltonian_NRQED_dipole} describes the electrons and nuclei of a matter systems, which are basically \emph{all} degrees of freedom molecules and solids.\footnote{According to the standard microscopic model of matter, that neglects the inner structure of the nuclei.} It is thus a suited basis for a first-principles description of, e.g., the phenomena of polaritonic chemistry.
Nevertheless, the accurate description of these degrees of freedom is very challenging and in many cases, the electronic and nuclear dynamics can be separated (Born-Oppenheimer approximation). The electrons are then described in the electrostatic potential of the nuclei with fixed coordinates $\bR_i$ (clamped nuclei). The electronic energy as a function of the $\bR_i$ (potential energy surface) defines then an effective potential for the nuclear dynamics. In this work, we focus on the electronic part of the matter-description (electronic structure, see also Ch.~\ref{sec:est}). However, since electrons and nuclei couple in exactly the same way (with their dipole operator) to the photon modes, a large part of our discussion can be generalized straightforwardly to the nuclear theory. For further details about additional implications of the approximation for the coupled light-matter systems and possible improvements, the reader is referred to Ref.~\citep{Schaefer2018}.  

To derive the electron-photon theory formally, we let $M_i\rightarrow\infty$ for $i=1,...,N_n$. Since $m_e\ll M_i$, this approximation has a wide range of validity.\footnote{However, there are also many situations, when the Born-Oppenheimer approximation fails, but these cases are beyond the scope of this work. The interested reader is referred to, e.g., the review by \citet{Worth2004}.} Thus, the kinetic energy of the nuclei $\sum_i\tfrac{1}{2M_i}\nabla^2_{\bR_i}\rightarrow 0$ and we can treat them as fixed classic charges with coordinates $\bR_i$. 
Under the Born-Oppenheimer approximation (abbreviated by BOA in the equations), the nuclear contribution
\begin{align*}
	\hat{H}_{n} \overset{BOA}{\longrightarrow}& \frac{1}{2} \sum_{i\neq j}^{N_n}  \frac{Z_iZ_j}{|\bR_i-\bR_j|}
\intertext{is merely a constant that we can neglect in many scenarios. We will consider the term only for comparisons of different nuclear configurations, e.g., to study simple chemical reactions.
The electron-nuclear interaction has then a prescribed local potential $\hat{H}_{ne}\overset{BOA}{\longrightarrow} v(\br)=\sum_{i=1}^{N_n} \frac{1}{|\bR_i-\br|}$ that we include in the electronic Hamiltonian}
	\hat{H}_{e} \overset{BOA}{\longrightarrow}& \hat{H}_{e} + v(\br).
\end{align*}
We also neglect the constant contribution of the nuclear charge to the total dipole $\bR\overset{BOA}{\longrightarrow}-\sum_{i=1}^{N}\br_i$, where we have removed the index of $N_e\rightarrow N$ for convenience.
We note that the Hamiltonian~\eqref{eq:Hamiltonian_NRQED_dipole} is symmetric with respect to the simultaneous inversion of $\br \rightarrow -\br$ and $p_{\alpha}\rightarrow-p_{\alpha}$. Additionally, there is the ambiguity of the sign of the electric charge~\citep{Kirilyuk1997}. That is why one can find different definitions of the bilinear coupling term in the literature. Importantly both choices, $p_{\alpha} \pm \frac{\blambda_{\alpha}}{\omega_{\alpha}}\cdot \bR$, lead to the \emph{same} expectation values of the observables. We make use of this freedom and redefine 
\begin{align*}
\bR=+\sum_{i=1}^{N}\br_i,
\end{align*}
which is the more common choice in the literature.

Collecting all the terms, we arrive at the coupled electron-photon Hamiltonian in the long-wavelength limit or the \emph{cavity-QED Hamiltonian} that is explicitly given in Def.~\ref{def:cavity-QED_Hamiltonian}. It will be the basis for (most of) the discussions of the rest of this thesis. 


\begin{fdef}{0.92\columnwidth}{Cavity-QED Hamiltonian}
	\label{def:cavity-QED_Hamiltonian}
A system consisting of $N$ nonrelativistic electrons that are coupled via their dipole to $M$ modes of the electromagnetic field in the length gauge are described by the \emph{cavity-QED Hamiltonian}
\begin{empheq}[box=\fbox]{align}
\label{eq:Hamiltonian_BO}
\hat{H}=\sum_i^{N} \left[\frac{1}{2} \nabla_{\br_i}^2 + v(\br_i)\right] + \frac{1}{2}\sum_{i\neq j}^{N} \frac{1}{|\br_i-\br_j|}+ \frac{1}{2}\sum_{\alpha}^{M}\left[-\partial_{p_{\alpha}}^2 + \omega_{\alpha}^2 \left(p_{\alpha} + \frac{\blambda_{\alpha}}{\omega_{\alpha}}\cdot \bR \right)^2\right]
\end{empheq}
Here $\bR=+\sum_{i=1}^{N}\br_i$ is the total dipole of the matter systems and for each mode $\alpha$, $\omega_{\alpha}$ denotes the frequency and
\begin{align}
	\blambda_{\alpha}=\sqrt{4\pi} S_{\alpha} \bepsilon_{\alpha}
\end{align}
is the light-matter coupling constant, that includes the polarization vector $\epsilon_{\alpha}$ and the (effective) mode function $S_{\alpha}$. In free space, the modes are plane waves and thus, $S_{\alpha}$ is given by the simple expression $S_{\alpha}=\tfrac{1}{\sqrt{V}}$ (cf. Eq.~\eqref{eq:intro:mode_function}). However, inside a cavity the mode functions may have a nontrivial form~\citep{Power1982}.

For later convenience, we further differentiate the contributions
\begin{align*}
	\hat{H}=\hat{T}+\hat{V}+\hat{W}+\hat{H}_{ph}+ \hat{H}_{int}+\hat{H}_{self},
\end{align*}
where 
\begin{align*}
	&\hat{T}=\sum_i^{N} \frac{1}{2} \nabla_{\br_i}^2, \quad
	\hat{V}=\sum_i^{N} v(\br_i), \quad
	\hat{W}= \frac{1}{2}\sum_{i\neq j}^{N} \frac{1}{|\br_i-\br_j|},\\
	&\hat{H}_{ph}= \frac{1}{2}\sum_{\alpha}^{M}\left[ -\partial_{p_{\alpha}}^2 + \omega_{\alpha}^2 p_{\alpha}^2\right], \quad
	\hat{H}_{int}=\sum_i^{N} \sum_{\alpha}^{M} \omega_{\alpha} p_{\alpha} \blambda_{\alpha}\cdot \br_i, \quad
	\hat{H}_{self}=\frac{1}{2} \sum_{i,j=1}^{N} \left( \blambda_{\alpha}\cdot \br_i \right) \left( \blambda_{\alpha}\cdot \br_j \right)
\end{align*}
\end{fdef}

\subsubsection*{A remark on the gauge choice}
We conclude this subsection with a remark on the specific gauge choice in $\hat{H}$. In the length (PZW) gauge, we describe the photon degrees of freedom with the (transversal) displacement field
\begin{align*}
	\hat{\mathbf{D}}_{\perp} =&\sum_{\alpha=1}^{M}\frac{\omega_{\alpha}}{4\pi}\blambda_{\alpha}p_{\alpha}
\intertext{as canonical variable with corresponding momentum}
	\hat{\mathbf{B}}=& \sum_{\alpha=1}^{M}(-\I\partial_{p_{\alpha}}) \blambda_{\alpha}\times\epsilon_{\alpha}.
\intertext{We recover the electric field as}
	\hat{\bE} =&4\pi \left(\hat{\mathbf{D}}_{\perp} - \hat{\bP}_{\perp}\right)
\intertext{with the (transversal) polarization}
	\hat{\bP}_{\perp}=&\sum_{\alpha} \frac{1}{4\pi} \blambda_{\alpha}(\blambda_{\alpha}\cdot\bR).
\end{align*}
The original creation and annihilation operators for the electromagnetic field modes have now the form\footnote{Note that $-\I\partial_{p_{\alpha}}^+=-\I\partial_{p_{\alpha}}$ is the canonical momentum operator that is hermitian.}
\begin{subequations}
		\begin{align*}
			\hat{a}_{\alpha}=&\frac{1}{\sqrt{2\omega_{\alpha}}}\left(-\I\partial_{p_{\alpha}} -\I\omega_{\alpha}p_{\alpha} + \I\blambda_{\alpha}\cdot\bR  \right)\\ 
			\hat{a}_{\alpha}^+=&\frac{1}{\sqrt{2\omega_{\alpha}}}\left(-\I\partial_{p_{\alpha}} +\I\omega_{\alpha}p_{\alpha} - \I\blambda_{\alpha}\cdot\bR  \right)
		\end{align*}
\end{subequations}
This has consequences for the interpretation of photon-observables. Most importantly, the expectation value
\begin{align}
\label{eq:PhotonEnergy_displacement}
	\tilde{E}_{ph}=&\braketsm*{\hat{H}_{ph}}=\braketsm*{\tfrac{1}{2}\sum_{\alpha}^{M}\left[ -\partial_{p_{\alpha}}^2 + \omega_{\alpha}^2 p_{\alpha}^2\right]}
\intertext{does not correspond to the energy of the electromagnetic field that is defined as $\tfrac{1}{8\pi}\int\td\br (\bE^2+\bB^2)$ in terms of the electric field, but is instead connected to the displacement-field $\mathbf{D}$. 
To calculate the photon energy (in its standard definition), we have to consider instead}
	E_{ph}=& \braketsm*{\hat{H}_{ph} + \hat{H}_{int} + \hat{H}_{self}}
\intertext{Correspondingly, the photon number of mode $\alpha$ is calculated as $N_{ph}=\frac{1}{\omega_{\alpha}}E_{ph}-\frac{1}{2}= \sum_{\alpha}\braket{\hat{N}_{ph,\alpha}}$, where we defined the photon number operator for mode $\alpha$ }
\label{eq:PhotonNumber_length}
\hat{N}_{ph,\alpha} =& \frac{1}{\omega_{\alpha}} \left(- \frac{1}{2} \frac{\partial^2}{\partial p_{\alpha}^2}+ \frac{\omega_{\alpha}^2}{2}\left(p_{\alpha} - \frac{\blambda_{\alpha}}{\omega_{\alpha}} \cdot \bR \right)^2\right) -\frac{1}{2},
\intertext{For the sake of completeness, let us also define to ``photon-number'' operator with respect to Eq.~\eqref{eq:PhotonEnergy_displacement}, i.e.,}
	\tilde{N}_{ph,\alpha}=& \frac{1}{2\omega_{\alpha}} \left[ -\partial_{p_{\alpha}}^2 + \omega_{k}^2 p_{\alpha}^2\right] -\frac{1}{2},
\end{align}
that we call the \emph{mode occupation} to differentiate it from $N_{ph,\alpha}$.

This illustrates that in the length gauge, the electronic and photonic degrees of freedom are (partly) mixed. Consequently, a part of the electron-photon interaction is described only by matter quantities as, e.g., the occurrence of the dipole-self-interaction $\hat{H}_{self}$ shows. This is a convenient starting point to develop new first-principles methods for light-matter interaction, since the description of matter is much better understood. For instance, one could approximate the light-matter interaction by only taking $\hat{H}_{self}$ into account and would end up with a matter-only theory, which provides a good (qualitative) description of certain scenarios~\citep{Flick2017cavity,Schaefer2018}. Without the PWZ transformation, cf. Eq.~\eqref{eq:Trafo_dipole_length_gauge}, such a mixing does not occur and we describe the electromagnetic field by the vector potential and the electric field exactly as in the Coulomb gauge of NR-QED. Within the long-wavelength description, this is called the \emph{velocity gauge}. Instead of $\hat{H}_{self}$, we would find in this gauge the \emph{diamagnetic term}, mentioned in Sec.~\ref{sec:intro:cavity_qed_models}, that is proportional to $\hat{A}^2$.

\clearpage
	
	\chapter[Electronic-Structure Theory]{Matter from first principles: electronic-structure theory}
\label{sec:est}
Before we investigate how the coupled light-matter problem can be approached from a first-principles perspective (which is the topic of Sec.~\ref{sec:qed_est}), we discuss in this chapter the ``simpler'' problem of matter-only systems. How to accurately describe the microscopic details of matter is a topic so well studied that many techniques have already become basic textbook knowledge. Nevertheless, we discuss this topic in the following in great detail and with a particular focus on the conceptual level. This serves on the one hand to provide a complete picture also for the unacquainted reader. On the other hand, we aim at highlighting for each method the essential ingredients that makes it numerically efficient, yet accurate. We then show that this efficiency is lost, once we consider the standard form of the light-matter problem based on individual electrons and photons (Sec.~\ref{sec:qed_est}). An analysis of this drawback provides already the physical rationale why a polariton picture would be preferable (part~\ref{sec:dressed}).


\vspace{0.5cm}
The ``keystone''~\citep{Worth2004} of the first-principles description of matter is the Born-Oppenheimer approximation that we also have employed in the derivation of the cavity-QED Hamiltonian in the previous section. Within the Born-Oppenheimer approximation, the dynamics of the electrons and nuclei are decoupled, which allows to separate the description of matter into methods that focus on nuclear effects (such as vibrations or rotations) and so-called \emph{electronic-structure theory}. The latter comprises all approaches to describe the electronic degrees of freedom of matter systems including the majority of all known first-principles methods.\footnote{Note that there are also first-principles methods that, e.g., focus explicitly on the description of the coupled electron-nuclei dynamics~\citep{Albareda2019} or only the nuclei~\citep{Beck2000mctdh}.} 
A very prominent role in electronic-structure theory is played by the ground state, that determines fundamental properties of matter systems such as the equilibrium geometry or the effective screening of the Coulomb interaction inside a solid. It also plays a key-part in understanding chemical processes like reactions that often can be described in the quasistatic picture of the Born-Oppenheimer approximation. Many first-principles methods exclusively aim to accurately describe electronic ground states and we put the focus on such methods in the following.


The accurate description of the ground state of a many-electron system is an extraordinary challenging problem that is a research topic since the beginnings of quantum mechanics. Consequently, many approaches have been developed for this task and it depends strongly on the specific scenario, which one is the best choice. However, there are certain generic features that play a role in all approaches.

The most important example for that is the complexity that arises from the \emph{many-body problem} and that basically prohibits the exact description of most many-electron systems. The only explicitly known way for such a description is to solve the Schrödinger equation for the many-body wave function. For a system consisting of $N$ electrons, the wave function depends on $4N$ coordinates,\footnote{Three spatial and one spin coordinate for each electron.} and thus,  the configuration space, i.e., the space of all these wave functions grows exponentially with $N$. For large $N$, this makes the concept of the many-body wave function in practice useless. The complexity of the many-body problem occurs in some way or the other in all electronic-structure approaches and it is the reason, why practical methods always rely on approximations.

Another example of a generic feature of all electronic-structure methods is the \emph{particle-exchange symmetry}. Electrons are fermions and a fermionic wave functions is antisymmetric with respect to the exchange of the coordinates of any two electrons. As a consequence, fermions adhere to the Pauli principle, i.e., two fermions cannot have the same quantum number. Accounting for the Pauli principle is fundamental for an accurate description of the electronic structure.

The third example is the \emph{variational principle} that defines the ground state as the minimizer of the total energy. This suggests to determine the ground state by an explicit minimization of the energy, which is the starting point of every electronic-structure method. To carry out this minimization, it is necessary to \emph{characterize the state} of the system, which in turn defines the energy functional. In fact, there are several ways to do this characterization and each of these defines a class of electronic-structure methods. 
%
In the following, we present three different electronic-structure approaches that adhere to the above principles:
\begin{enumerate}
		\item Hartree-Fock (HF) theory (Sec.~\ref{sec:est:hf})
		\item Density functional theory (DFT, Sec.~\ref{sec:est:dft})
		\item Reduced density matrices (RDMs) that are the basis of variational RDM theory (Sec.~\ref{sec:est:rdms:general}) or one-body RDM functional theory (RDMFT, Sec.~\ref{sec:est:rdms:rdmft})
\end{enumerate}
Let us now briefly explain these approaches, highlighting the just defined generic principles.

HF theory is one of the simplest examples of the class of wave-function methods, that directly approximate the many-body wave function. The HF wave function is one Slater determinant (single-reference ansatz), which is the simplest many-body wave function that is antisymmetric and already provides a good qualitative description of many systems. In general, wave function methods guarantee the particle-exchange symmetry by employing Slater determinants. One can improve on the accuracy of HF by considering wave functions built from more than one Slater determinant (multi-reference ansatz). 
Important examples for this approach are \emph{configuration interaction}, \emph{coupled cluster theory}\footnote{Note however that in the most common practical coupled cluster methods, the wave function is not entirely reconstructable~\cite{Bartlett2007}.} or \emph{multiconfigurational self-consistent field theory}~\citep[part 5]{Helgaker2000}. More recently, tensor-network methods such as \emph{density matrix renormalization group}~\citep{Chan2011} theory have additionally enriched the spectrum of this approach. Such \emph{wave-function methods} yield until today the most precise results for many systems. However, their high precision usually comes with equivalently high numerical costs. Systems with larger particle numbers of say $N\gtrapprox 100$ can rarely be described accurately with such methods. This is the manifestation of the many-body problem in wave function methods.

In DFT instead, we drop the concept of the many-body wave function and describe the state of a system with the (one-body) electron density $\rho(\br)$. The many-body problem manifests here more indirectly than for wave-function methods. The reformulation of many-body quantum mechanics in terms of $\rho$ leads to an unknown quantity, which is called the \emph{exchange-correlation functional}. Constructing approximations for this functional is a very difficult task and severely limits the applicability of functional methods to certain system classes. The most common starting point for the development of functionals is the Kohn-Sham (KS) construction that describes the state of the system with one Slater determinant exactly as in HF theory. An important reason for this choice is that the Slater determinant accounts for the fermionic antisymmetry. Usual approximate DFTs have numerical costs that scale much better with the particle number than wave-function methods and thus can in principle achieve accurate results for very large systems.\footnote{State of the art DFT methods can describe systems with more than $1000$ particles. See for instance Ref.~\citep{Kanungo2017}, where the authors present results for more then $8000$ electrons.}

Employing RDMs as a basic variable provides yet another perspective on many-electron systems. For instance, we can describe the expectation value of the energy of an $N$-electron system exactly in terms of the 2-body RDM (2RDM). The configuration space of such a description does not grow with the particle number and thus a variational minimization in terms of the 2RDM is in principle feasible. However, the many-body problem also arises in this description in the form of conditions that determine the exact configuration space, i.e., the set of all 2RDMs. The number of these so-called $N$-representability conditions grows exponentially with the particle number $N$. Thus, in practical variational minimizations of the energy with respect to the 2RDM (variational 2RDM theory) only a subset of conditions is considered. 

A special role is here taken by the 1-body RDM (1RDM) $\gamma$ that has comparatively simple $N$-representability conditions. These reflect the exchange symmetry of the particle species and thus provide an alternative to employing Slater-determinants to enforce the antisymmetry of electrons. Although $\gamma$ is not sufficient to describe the energy of a many-electron system in a linear way, it carries all information on the system by virtue of a generalized Hohenberg-Kohn theorem (Gilbert's theorem). This establishes RDMFT as an alternative to DFT that employs $\gamma$ instead of $\rho$ as the basic variable. Usual approximate RDMFTs are numerically more expensive than DFT, but conversely, they can account for multi-reference effects in an easier way. 

We want to remark on a fourth and last common feature of all electronic-structure methods, which is their computational complexity. Independently of the details of a certain method, we always exchange a linear problem on a huge configuration space, i.e., solving the many-body Schrödinger equation, with a nonlinear problem on a comparatively small configuration space. Importantly, the equations of electronic-structure (and in general many-body) methods involve usually nonlinear operators and some form of self-consistency, which is a challenging combination. 
To do this in practice, it is not sufficient to employ, e.g., a solver for partial differential equations from a standard library. But we usually need specific algorithms that have been designed to solve the equations of a specific theory. Thus, electronic-structure methods are fundamentally connected to computational mathematics. That we are nowadays able to apply a method like HF to problems that involve many hundred electrons is a consequence, more of algorithmic improvements than of the increase of computer power. 

	\newpage
\section{The wavefunction, symmetry and the many-body problem}
\label{sec:est:general}
We start our survey on (equilibrium) electronic-structure theory with the proper definition of the problem that we want to solve. We aim to accurately describe the ground state of a system, consisting of $N$ electrons that repel each other via the Coulomb interaction exposed to some local potential $v(\br)$. Usually $v$ is thought as the electrostatic potential of some classic charges at fixed positions that represent the nuclei in the Born-Oppenheimer approximation (see Sec.~\ref{sec:intro:nrqed_longwavelength}), but we do not have to specify this for the general theory. This setting is described by the (electronic-structure) Hamiltonian
\begin{align}
\label{eq:est:general:hamiltonian}
\hat{H}=\sum_i^{N} \left[\frac{1}{2} \nabla_{\br_i}^2 + v(\br_i)\right] + \frac{1}{2}\sum_{i\neq j}^{N} \frac{1}{|\br_i-\br_j|},
\end{align}
that can be derived in several ways, e.g., by the canonical quantization of the corresponding classical theory or as the static limit of the Pauli-Fierz Hamiltonian (Def.~\ref{def:Pauli-Fierz_Hamiltonian}), i.e., by turning off the interaction to the quantized part of the Maxwell field.\footnote{Equivalently, we can derive Hamiltonian~\eqref{eq:est:general:hamiltonian} from the cavity-QED Hamiltonian, cf. Eq.~\eqref{eq:Hamiltonian_BO}.}
Eq.~\eqref{eq:est:general:hamiltonian} defines a linear operator\footnote{We generally indicate operator by the '$\hat{\phantom{a}}$' symbol. However, because we work in the real-space picture, we will often drop the symbol for operators that obviously only depend on the position, such as $\hat{\br}\equiv \br$ or $\hat{v}(\hat{\br})\equiv v(\br)$. The reason is that they act multiplicative and thus are represented by a function.} that has a well-defined spectrum and is bounded from below.\footnote{Both follows from the proof for self-adjointness of $\hat{H}$ that is valid for a very large class of local potentials $v(\br)$ (Kato-Rellich theorem). See for example Ref.~[part 6]\citep{DeOliveira2009}.}
Hence, we can define the ground state $\Psi_0$ that by a variational minimization of the energy expectation value $E=\braket{\Psi|\hat{H}\Psi}$ over all ``physical'' wave functions $\Psi$, i.e.,
\begin{align}
\label{eq:est:general:var_principle_est}
E_0=\braket{\Psi_0|\hat{H}\Psi_0}=\inf_{\Psi}\braket{\Psi|\hat{H}\Psi},
\end{align}
where we will discuss the term ``physical'' in the following paragraphs.\footnote{Note that for certain potentials $v$, it can be proven that the ground state exists and thus we can exchange the $\inf$ by a $\min$~\citep[part 6.2]{DeOliveira2009}. This includes the important case where $v(\br)=\sum_i^{N_n} 1/|\br-\bR_i|$ describes the electrostatic potential of $N_n$ nuclei at the positions $\bR_i$.}
This \emph{variational principle} is of central importance in electronic-structure theory and it will play a fundamental role for all methods that we present in the following. Sloppily we can say that the goal of (equilibrium) electronic-structure theory is to (approximately) solve the minimization problem~\eqref{eq:est:general:var_principle_est} with Hamiltonian~\eqref{eq:est:general:hamiltonian}.


\subsection{The ground state of a two-electron problem}
\label{sec:est:general:example_case}
The general definition of the electronic-structure problem by Eqs.~\eqref{eq:est:general:hamiltonian} and \eqref{eq:est:general:var_principle_est} is quite abstract and for the unacquainted reader, it might be hard to imagine how difficult solving this problem really is. To make things more tangible, we will put Hamiltonian~\eqref{eq:est:general:hamiltonian} aside for a moment and consider the  problem of two electrons in one spatial dimension and without spin. 

\subsubsection*{The one-particle problem}
We start with one particle in a ``box,'' that we define simply by constraining the domain of all quantities to an interval $L\subset\mathbb{R}$ and zero-boundary conditions. The energy of this system is described by the Hamiltonian
\begin{align}
\label{eq:est:general:Box_Hamiltonian_one_particle}
\hat{H}=\hat{H}(x)= -\tfrac{1}{2} \partial_x^2,
\end{align}
that is just the kinetic energy operator $\hat{t}(x)=-\tfrac{1}{2} \partial_x^2$ on the domain.
According to the prescription of standard many-body quantum mechanics, we can describe the state of the system by a wave function
\begin{align}
\Psi(x)\in \mathfrak{h}^1[L],
\end{align}
where we denote with $\mathfrak{h}^1[L]$ the Hilbert space of square-integrable functions.\footnote{A function  $f:[a,b]\rightarrow \mathbb{C}$ is square-integrable if its ($L^2$-)norm is finite, i.e., $\int_{a}^{b}|f(x)|^2\td x<\infty$. This is a necessary condition for the interpretation of $|\psi(x)|^2$ as probability amplitude.}

To determine the ground state $\Psi_0$ of this system, we employ the variational principle
\begin{align} 
\tag{cf. \ref{eq:est:general:var_principle_est}}
E_0=\braket{\Psi_0|\hat{H}\Psi_0}=\inf_{\Psi\in\mathcal{C}}\braket{\Psi|\hat{H}\Psi}.
\end{align}
This means (assuming no degeneracy) that $\Psi_0$ is the one (normalized) wave function\footnote{Note that if $\Psi_0$ exists and is not degenerate, it must be an eigenstate of $\hat{H}$, because $\hat{H}$ is self-adjoint on $\mathfrak{h}^1[L]$ and thus has according to the \emph{spectral theorem}~\citep[part 1]{DeOliveira2009} an eigenrepresentation that is bounded from below, i.e., $\hat{H}=\sum_{i=0}E_i \ket{\tilde{\Psi}_i}\bra{\tilde{\Psi}_i}$ with $E_0 < E_1\leq E_2 ...$. If we expand $\Psi_0=\sum_i c_i \tilde{\Psi}_i$, we see that the lowest energy is given by  $\Psi_0\equiv \tilde{\Psi}_0$. If there is degeneracy, i.e., $E_0=E_1=...+E_m$ for some finite $m$, then $\Psi_0=\sum_{i=1}^m c_i \tilde{\Psi}_i$. However, it might be that $\Psi_0\notin \mathfrak{h}^1[L]$ and thus, the system cannot reach its infimum. For instance, this happens in the free-space case, where $L\rightarrow\infty$.} 
with the lowest energy expectation value chosen from the set $\mathcal{C}\subset \mathfrak{h}^1[L]$ of all \emph{allowed} wave functions, which we call the \emph{configuration space}.

We can solve this minimization problem simply by diagonalizing the kinetic energy operator, i.e.,
\begin{align}
\label{eq:est:general:ewp_one_particle_box}
-\tfrac{1}{2}\nabla_{x}^2\psi_{i}(x)=k_i^2/2 \psi_i(x),
\end{align}
where the $\psi_{i}(x)=\sqrt{2/L}\sin(x k_i)$ are the ``box states'' with the momenta $k_i=i\pi/L$.
Thus, the configuration space is $\mathcal{C}=\text{span}(\psi_1,\psi_2,...)$ and the ground state $\Psi_0=\psi_{1}(x)$ is just the box-state with the lowest allowed momentum $k_1=\pi/L$. 

\subsubsection*{The two-particle problem on a first glance}
Let us now consider two particles in a box. Their energy is described by the Hamiltonian
\begin{align}
\label{eq:est:ParticleBox_Hamiltonian}
\hat{H}=\hat{H}(\br_1,\br_2)=\sum_{i=1}^{2} -\tfrac{1}{2} \partial_x^2,
\end{align}
that is just the sum of the kinetic energy operators for each particle (with index $i$) on the domain. To describe the state of the system, we need a wave function that depends on two coordinates, i.e., 
\begin{align}
	\Psi(x_1,x_2)\in \mathfrak{h}^2.
\end{align}
The linear structure of the theory suggests now that we define the two-body Hilbert space
\begin{align}
	\mathfrak{h}^2[L]= \mathfrak{h}^1[L] \otimes\mathfrak{h}^1[L]
\end{align}
simply as the product of the one-particle space $\mathfrak{h}^1[L]$. 
This is equivalent to employing the wave function ansatz
\begin{align}
\label{eq:est:general:wf_ansatz_2particles_no_symmetry}
	\Psi(x_1,x_2)=\sum_{i,j=1}^{\infty} c_{ij} \psi_i(x_1)\psi_j(x_2),
\end{align}
where we introduced the two-index expansion-coefficient $c_{ij}\in \mathbb{C}$. To guarantee the normalization $|\Psi|^2=1$ of the wave function, the coefficients need to satisfy the sum rule $\sum_i|c_i|^2=1$.
Thus, if we perform the minimization over $\mathcal{C}=\mathfrak{h}^2[L]$, we find the ground state
\begin{align*}
\Psi_0(x_1,x_2)=\psi_1(x_1)\psi_1(x_2),
\end{align*}
with eigenvalue $E_0=2\pi/L$. Obviously, this state does not describe a valid configuration of electrons, because it does \emph{not adhere to the Pauli-principle}.

\subsubsection*{What differentiates many from one: exchange symmetry}
The reason is that we have not performed the minimization over the correct configuration space. Characterizing this space is a crucial step to solve the many-electron minimization problem.
In fact, the physically correct configuration space is \emph{considerably smaller} than $\mathfrak{h}^2$. The true ground state of the ``two electrons in a box'' problem is the lowest eigenstate of $\hat{H}$ that is also \emph{antisymmetric} under the exchange of the coordinates, i.e.,
\begin{align}
\label{eq:est:general:antisymmetry_2particles}
	\Psi(x_1,x_2)=-\Psi(x_2,x_1).
\end{align}
We call this the \emph{exchange symmetry}, which plays a fundamental role all theoretical descriptions of quantum many-body systems. This is especially true in the realm of first-principles methods, where we do not only utilize the variational principle as a formal definition, but actually perform the minimizations of the form~\eqref{eq:est:general:var_principle_est} with large-scale numerical algorithms. To do so, we need to \emph{parametrize} $\mathcal{C}$ in some way and for that, the antisymmetry of electrons is one of the most important tools as we will see in the following.

\subsubsection*{How to parametrize the antisymmetric space: Slater determinants}
Let us therefore regard again our wave function ansatz~\eqref{eq:est:general:wf_ansatz_2particles_no_symmetry} with the expansion coefficients $c_{ij}$. Since the indices correspond explicitly to the particle coordinates, we can simply transfer the condition~\eqref{eq:est:general:antisymmetry_2particles} to the coefficients by demanding
\begin{align*}
c_{ij}=-c_{ji}.
\end{align*}
We can subsume this condition into the expansion and get the more common form\footnote{We redefine here the coefficients $c_{ij}\rightarrow c_{ij} \sqrt{2}$ slightly such that the normalization conditions /  sum rules do not change.}
\begin{align}
\label{eq:est:general:wf_ansatz_2particles_antisymmetric}
\begin{split}
\Psi(x_1,x_2)=&\tfrac{1}{\sqrt{2}}\sum_{i,j}c_{ij} \left(\psi_i(x_1)\psi_j(x_2)- \psi_j(x_1)\psi_i(x_2)\right)\\
=&1/\sqrt{2}\sum_{i,j}c_{ij} |\psi_i(x_1)\psi_j(x_2)|_-
\end{split}
\end{align} 
of an expansion in terms of the (two-body) \emph{Slater determinants} $\Psi_{ij}(x_1,x_2)=1/\sqrt{2}\left(\psi_i(x_1)\psi_j(x_2)- \psi_j(x_1)\psi_i(x_2)\right)=1/\sqrt{2}|\psi_i\psi_j|_-$.  The set of all these two-body Slater determinants spans the antisymmetric two-particle Hilbert space 
\begin{align}
\label{eq:est:general:2particle_hilbert_space_antisymmetric}
\begin{split}
	\mathfrak{h}^2_A =&\mathfrak{h}^1 \wedge \mathfrak{h}^1\\
	\equiv&\mathcal{A}\mathfrak{h}^2,
\end{split}
\end{align}
where the operator $\mathcal{A}$ turns the tensor product $\otimes$ into an antisymmetric tensor product $\wedge$ or ``antisymmetrizes'' a given tensor~\citep{Coleman1963}.

Thus, we can parametrize the correct configuration space by employing the ansatz~\eqref{eq:est:general:wf_ansatz_2particles_antisymmetric}. This means that we calculate the matrix elements $H_{ij}=\braket{\psi_i|\hat{H}\phi_j}$ and reformulate the original minimization problem~\eqref{eq:est:general:var_principle_est} in terms of the coefficients:  
\begin{align}
\label{est:energy_functional_wf_2particles}
E[c_{ij}] = \braket{\Psi[c_{ij}]|\hat{H}\Psi[c_{ij}]}.
\end{align}
To calculate the 2-body matrix elements $H_{ij}$, let us see how $\hat{H}$ acts on one Slater determinant, built by the box states. We have
\begin{align*}
\hat{H}\left(\psi_i(x_1)\psi_j(x_2) - \psi_j(x_1)\psi_i(x_2)\right)=&\left[\sum_{i=1}^{2} -\tfrac{1}{2} \partial_{x_i}^2\right]\left(\sumphan\psi_i(x_1)\psi_j(x_2) - \psi_j(x_1)\psi_i(x_2)\right)\\
=&\left[-\tfrac{1}{2} \partial_{x_1}^2\psi_i(x_1)\right]\psi_j(x_2) - \left[-\tfrac{1}{2} \partial_{x_1}^2\psi_j(x_1)\right]\psi_i(x_2) \\
&+ \left[-\tfrac{1}{2} \partial_{x_2}^2\psi_j(x_2)\right]\psi_i(x_1) - \left[-\tfrac{1}{2} \partial_{x_2}^2\psi_i(x_2)\right]\psi_j(x_1)\\
=& 2(k_i^2/2+k_j^2/2) \left(\sumphan\psi_i(x_1)\psi_j(x_2)  - \psi_j(x_1)\psi_i(x_2)\right).
\end{align*}
We see that Slater determinants, constructed from the eigenstates of $\hat{t}(x)$, are the eigenfunctions of the many-body Hamiltonian $\hat{H}(x_1,x_2)=\hat{t}(x_1)+\hat{t}(x_2)$. Thus, we know that the ground state is given by $c_{ij}=1$ for $i=0, j=1$ (or vice versa) and $c_{kl}=0$ for all other combinations of indices: the ground state of $\hat{H}$ is a single Slater determinant, constructed from the lowest two eigenstates of $\hat{t}(x)$. 

\subsubsection*{Including a local potential}
Let us now generalize the description and consider a local potential $\hat{v}(x)$. Since the potential now should describe the geometry of the problem, we can formally assume $L\rightarrow\infty$. 
The many-body Hamiltonian reads then
\begin{align}
\label{eq:est:Hamiltonian_onebody_2particles}
\hat{H}=\sum_{i=1}^{2} \hat{t}(x_i) + \hat{v}(x_i).
\end{align}
We can straightforwardly generalize our solution strategy from before, if instead of Eq.~\eqref{eq:est:general:ewp_one_particle_box}, we consider the eigenvalue problem
\begin{align}
\label{eq:est:EWP_one-body}
[\hat{t}(x)+\hat{v}(x)]\psi_i(x)=e_i\psi_i(x).
\end{align}
The physical ground state is simply given as the Slater determinant $\Psi_0(x_1,x_2)=1/\sqrt{2}|\psi_0 \psi_1|_-$ with ground state energy $E_0=e_0+e_1$. 

\subsubsection*{Solving a minimization problem in practice: the basis set}
For almost all choices of $v$, Eq.~\eqref{eq:est:EWP_one-body} cannot be solved analytically and requires a numerical solver. This means that we need to parametrize the single-particle states (orbitals) $\psi_i$ in a \emph{known basis set} with say $B$ elements, i.e., $\psi_i=\sum_{j=1}^{B}b_{i}^j \xi_j$. An obvious choice would be a (sufficiently large) discretized interval in real space, such that $\xi_j(x)=\theta_{j}(x)$, where $j=1,\dotsc,B$ denote the $B$ grid points and $\theta_{j}(x)$ defines the discretization.\footnote{A possible choice would be $\theta_{x_j}=d_{x}^{-1/2}\Theta(d_{x}/2-|x_i-x|)$, where  $d_x$ is the spacing and $\Theta$ denotes the Heaviside step-function.}  
The differential operators would then be approximated by, e.g., \emph{finite-differences} (see Ch.~\ref{sec:numerics:rdmft}). This will introduce an error, which however can be controlled, i.e., with increasing $B$, the approximate solutions will converge to the exact results. We call such a procedure convergence with respect to the basis set.

Of course, we can also choose a different basis and in fact, discretizing the real-space is often not the best choice. For example, we might need very large $B$ to converge with respect to the basis set. 
Another option is to employ the eigenfunctions of the kinetic energy operator, which for $L\rightarrow\infty$ are \emph{plane waves}~\citep[part 2]{Altland2010}. Employing this basis is equivalent to describing the whole problem in Fourier space. The obvious advantage of such a choice is the analytically known form of the basis for all $x\in \mathbb{R}$ and thus, the description is not constrained by a box of finite volume. However, the accurate description of small distances will be inefficient in this basis. In fact, plane waves are a standard tool to describe the bulk of many condensed-matter systems, which are modelled as infinitely extended periodic systems~\citep{Altland2010}. These two examples demonstrate how strongly the basis choice depends on our \emph{knowledge} of the problem. 

Plane waves and discretized real-space bases are in this sense the most generic bases. To describe for instance finite-systems, usually neither of the two is used, but so-called \emph{atomic orbitals}. These are especially designed bases with the aim to reduce the necessary number of orbitals to a minimum. The basic idea is to make use of the analytically known eigenstates of the Hydrogen atom and the most straightforward approach is thus to use the lowest say $B_a$ of these eigenfunctions centered at the positions of all the $N_n$ nuclei of the system. The basis set consists then of $B=B_aN_n$ of such atomic orbitals, which is in most cases considerably smaller then any basis set obtained from a parametrization of the discretized real space. Another very common approach is to employ Gaussian functions instead of the Hydrogen orbitals because of their convenient mathematical properties. These are the two most employed strategies to obtain quantum chemical basis-sets.\footnote{The interested reader is referred to, e.g., the corresponding chapters in the textbook of \citet{Helgaker2000}.} 

However, we want to stress that also real-space bases are used in practice\footnote{Also in this work, see Ch.~\ref{sec:dressed:results}.} because they allows for a description of systems with less foreknowledge and a considerably easier visualization~\citep{Beck2000}. 

\subsubsection*{The interacting problem}
To finalize our survey, there is one last missing piece, which is the inclusion of the interaction. The repulsion between two electrons in 3d is described by the Coulomb operator that reads (employing the coordinates $\br_1,\br_2$)
\begin{align}
\hat{w}(\br_1,\br_2)=\frac{1}{|\br_1-\br_2|}.
\end{align}
This is a two-body operator, because it depends on two coordinates, but in contrast to, e.g., Eq.~\eqref{eq:est:Hamiltonian_onebody_2particles}, it \emph{cannot} be expressed as a sum of two one-body operators. We call this the \emph{order} of an operator: $\hat{w}$ has the order two, but for instance $\sum_{i=1}^{2} [\hat{t}(x_i) + \hat{v}(x_i)]$ has the order one. It is a fundamental property of interaction operators that their order is larger than one. Importantly, this spoils our solution strategy from above. 

To demonstrate this, let us consider a generic interaction operator in 1d $w(x_1,x_2)$\footnote{Note that the one-dimensional version of the Coulomb operator $w(x_1,x_2)=\frac{1}{|x_1-x_2|}$ has very different properties than its 3d version. Thus, in 1d-model calculations the soft-Coulomb potential $w(x_1,x_2)=\frac{1}{\sqrt{(x_1-x_2)^2 + \epsilon^2}}$ with $\epsilon>0$ is usually considered (see also Sec.~\ref{sec:est:comparison}).} that we add to Hamiltonian~\eqref{eq:est:Hamiltonian_onebody_2particles}, i.e.,
\begin{align}
\hat{H}=\sum_{i=1}^{2} \hat{t}(x_i) + \hat{v}(x_i) + \hat{w}(x_1,x_2).
\end{align}
We start with the orbital set that we need to construct the Slater determinants.
A reasonable choice could be the eigenstates of $\hat{t}+\hat{v}$ that we know to calculate numerically. However, Slater determinants of these orbitals cannot be eigenstates of the interacting problem and thus, we have to consider the general ansatz
\begin{align}
\tag{cf. \ref{eq:est:general:wf_ansatz_2particles_antisymmetric}}
\Psi(x_1,x_2)=& \tfrac{1}{\sqrt{2}}\sum_{i,j}c_{ij} \left(\psi_i(x_1)\psi_j(x_2) - \psi_j(x_1)\psi_i(x_2)\right)\\
=& \sum_{i,j}c_{ij}\Psi_{ij}. \nonumber
\end{align}
If we apply the full Hamiltonian to one of these Slater determinants $\Psi_{ij}$,
\begin{align*}
\hat{H}\Psi_{ij}=&\left[\sum_{i=1}^{2} \hat{t}(\br_i) + \hat{v}(\br_i) + \hat{w}(\br_1,\br_2)\right]\Psi_{ij}\\
=&\underbrace{\left[\sum_{i=1}^{2} \hat{t}(\br_i) + \hat{v}(\br_i)\right]\Psi_{ij}}_{2(e_1+e_2) \Psi_{ij}}  + \hat{w}(\br_1,\br_2) \tfrac{1}{\sqrt{2}}\left(\psi_i(x_1)\psi_j(x_2) - \psi_j(x_1)\psi_i(x_2)\right),
\end{align*}
we see that the interaction term \emph{couples} the two orbitals of $\Psi_{ij}$ for every $i,j$. This means that (without further information on the systems like symmetries etc.) we cannot simplify the general ansatz~\eqref{eq:est:general:wf_ansatz_2particles_antisymmetric} for an interacting problem. Even if we chose a highly optimized orbital basis, we have to perform the minimization
\begin{align}
\tag{cf. \ref{est:energy_functional_wf_2particles}}
E[c_{ij}] = \braket{\Psi[c_{ij}]|\hat{H}\Psi[c_{ij}]}.
\end{align}
with respect to in principle \emph{all} entries of the coefficient matrix $c_{ij}$.


\subsection{Solving a generic many-electron problem and the exponential wall}
\label{sec:est:general:general_case}
Let us now transfer the concepts and tools from the two-electron example to the general case of $N$ electrons in three spatial dimensions, including spin. Thus, we consider spin-spatial coordinates $\bx=(\br,\sigma)$, where $\br\in \mathbb{R}^3$ describes a position and $\sigma\in\{\uparrow,\downarrow\}$ is the spin-quantum number. 

\subsubsection*{The antisymmetric $N$-electron space}
We start with the definition of the antisymmetric $N$-particle space, i.e., the generalization of Eq.~\eqref{eq:est:general:2particle_hilbert_space_antisymmetric}. Therefore, we consider again the single-particle Hilbert space $\mathfrak{h}^1$ that is spanned by some spin-spatial orbital basis $\{\bar{\psi}_i(\bx)\}$ (where we assume basis states to be orthonormalized, i.e., $\int\td\br \bar{\psi}_i^*(\br)\bar{\psi}_j(\br)=\delta_{ij}$), such that every single-particle state can be expressed as
\begin{align}
\Psi(\bx)=\sum_{i=1}^{\infty} c_i \bar{\psi}_i(\bx),
\end{align}
where the $c_i\in \mathbb{C}$ are the expansion coefficients (with corresponding sum rule $\sum_{i,j}|c_{ij}|^2=1$ for normalization). From this starting point, we can now successively build Hilbert spaces $\mathfrak{h}^2, \mathfrak{h}^3$ for more particles as antisymmetric product spaces, i.e.,
\begin{align}
\label{eq:est:general:antisymmetric_hilbert_space_N}
\mathfrak{h}_A^N=\bigwedge\limits_{l=1}^{N} \mathfrak{h}^1.
\end{align}
The basis states of $\mathfrak{h}^N$ are the $N$-body Slater determinants
\begin{align}
\label{eq:est:general:def_Slaterdeterminant}
\Psi_I(\bx_1,\dotsc,\bx_N)=&\frac{1}{\sqrt{N!}}\sum_{\pi_j\in P_N} (-1)^{j}{\bar{\psi}}_{\pi_j(I_1)}(\bx_1) \dotsb{\bar{\psi}}_{\pi_j(I_N)}(\bx_N) \nonumber\\
\equiv&\frac{1}{\sqrt{N!}}|\bar{\psi}_{I_1},\dotsc, \bar{\psi}_{I_N}|_-,
\end{align}
where $P_N$ denotes the permutation group on $N$ elements and the index $j$ is chosen such that it is even (odd) for an even (odd) permutation $\pi_j \in S_N$. Thus, every wave function $\Psi(\bx_1,\dotsc,\bx_N)\in \mathfrak{h}_A^N$ can be parametrized as 
\begin{align}
\label{eq:est:wavefunction_Nparticles_slater_expansion}
\begin{split}
\Psi(\bx_1,\dotsc,\bx_N) =& \sum_I c^I \Psi_I(\bx_1,\dotsc,\bx_N)\\
	=& \Psi[c^I],
\end{split}
\end{align}
where $c^I=c^{I}_{I_1,\dotsc,I_N}\in \mathbb{C}$ is the coefficient of the $I$-th Slater determinant including the orbitals with indices $I_1,\dots,I_N$. The according sum rule is $\sum_I |c^I|=1$.
Note that the introduction of Slater determinants is one of the few known ways to guarantee the antisymmetry of a wave function and it is thus a fundamental tool of most electronic-structure theories.

\subsubsection*{The closed-shell setting}
In the following, we will usually consider the special case of \emph{closed-shell} systems, which is very common in electronic-structure theory~\citep{Helgaker2000} and covers a broad range of in practice relevant systems.\footnote{For instance, this excludes systems where certain magnetic effects or spin-orbit coupling play a role~\citep{Altland2010}. The generalization to odd $N$ and spin-dependent Hamiltonians is in principle straightforward. However, approximate methods for the spin-restricted case are generally better understood.} 
For that, we assume that $N$ is even and define
\begin{align}
\label{eq:est:spin_orbitals_closed_shell}
\begin{split}
\bar{\psi}_{2i}(\br,\sigma)=&\psi_i(\br)\alpha(\sigma)\\
\bar{\psi}_{2i+1}(\br,\sigma)=&\psi_i(\br)\beta(\sigma),
\end{split}
\end{align}
where $\{\alpha,\beta\}\equiv \{\uparrow,\downarrow\}$ denote the two spin functions.\footnote{We choose this nomenclature, because it is very common in the quantum chemistry literature.}

This means that each spatial orbital is occupied twice and the system is spin-saturated. We can thus remove the spin-coordinate from our description (by a trivial integration).

\subsubsection*{The ground state of a many-electron system}
With the ansatz~\eqref{eq:est:wavefunction_Nparticles_slater_expansion}, we can reformulate the original minimization problem~\eqref{eq:est:general:var_principle_est} in the following way:  First, we choose a specific \emph{basis} $\{\psi_j(x)\}$, expand $\Psi=\Psi[c^I]$ in Slater determinants of this basis according to Eq.~\eqref{eq:est:wavefunction_Nparticles_slater_expansion} and calculate the general energy expression as a functional of the expansion coefficients $c^I$, i.e.,
\begin{align}
\label{est:energy_functional_wf_general}
	E[c^I] = \braket{\Psi[c^{I}]|\hat{H}\Psi[c^{I}]}.
\end{align}
To find the ground state, we then minimize $E[c^I]$ with respect to the $c^I$. The solution $c^I_0$ defines $\Psi_0=\Psi[c^I_0]$.


\subsubsection*{The non-interacting case}
Importantly, the just described strategy strongly simplifies for every systems that is described by operators of order one, i.e., operators  that can be decomposed as
\begin{align}
\label{eq:est:general:DefOneBodyOperator}
\hat{O}(\br_1,...,\br_N) = \sum_{i=1}^{N} \hat{o}(\br_i).
\end{align} 
An important example for this are non-interacting systems with the Hamiltonian
\begin{align}
\hat{H}=\sum_i^{N} \left[\frac{1}{2} \nabla_{\br_i}^2 + v(\br_i)\right].
\end{align}
To find the ground state of $\hat{H}$, we simply have to calculate the $N/2$ (assuming that $N$ is even) lowest eigenstates of $\frac{1}{2} \nabla_{\br}^2 + v(\br)$ and construct a Slater determinant out of those, using Eq.~\eqref{eq:est:spin_orbitals_closed_shell}. This construction of a many-body state by successively ``filling up'' one-body states is often called \emph{aufbau principle} (see Sec.~\ref{sec:est:excursion_symmetry}). 

Thus, the non-interacting $N$-electron problem (and any other problem of order one) \emph{reduces} effectively to $N$ one-electron or orbital problems. With state-of-the-art algorithms and higher-performance clusters, such (and nonlinear versions of such) eigenvalue problems for orbitals  can be very efficiently solved even for spatially very extended systems (that in turn require large orbital bases). We call a wave function consisting only of one Slater determinant a \emph{single-reference} wave function.

\subsubsection*{Interaction and the exponential wall}
As in the two-particle case, any form of interaction spoils this simple solution strategy. For instance, the electronic-structure Hamiltonian (Eq.~\eqref{eq:est:general:hamiltonian}) includes the Coulomb interaction operator $\frac{1}{2}\sum_{i\neq j}^{N} \frac{1}{|\br_i-\br_j|}$ that has order two. This operator in principle couples all Slater determinants of the general (or \emph{multi-reference}) ansatz
\begin{align}
\tag{cf. \ref{eq:est:wavefunction_Nparticles_slater_expansion}}
\Psi(\bx_1,\dotsc,\bx_N) = \sum_I c^I \Psi_N^I(\bx_1,\dotsc,\bx_N),
\end{align}
with the $N$-dimensional coefficient tensor
\begin{align}
\label{eq:est:general:coefficient_tensor}
	c^I=c^{I}_{j_1,\dotsc,j_N}.
\end{align}
Thus, we cannot simplify the expansion as we have done for the non-interacting case and need to consider the full coefficient tensor.
If we consider $M$ basis states, this tensor has a dimension of
\begin{align}
\label{eq:est:exponential_scaling_wf}
D=M^N,
\end{align}
that grows \emph{exponentially} with the particle number. If we wanted to compute the wave function for say a Benzene molecule ($C_6H_6$) that has $N=6^*6+6=42$ electrons with a small basis set of say $M=10$ orbitals, the coefficient tensor would have $D=10^{42}$ entries, which could hardly be determined by any imaginable computation system. For a system with $N=80$ electrons, we approach the approximated number of particles of the universe.\footnote{Source: \url{https://www.universetoday.com/36302/atoms-in-the-universe/amp/}, accessed 15.07.2020.} Clearly, we can exclude some of this entries by smart basis choices and symmetry considerations. State-of-the-art methods have been employed to calculate the (real-space) many-body wave function (for certain systems) of about 50 electrons~\citep{Shepherd2012}.\footnote{In certain (usually effectively one-dimensional) model descriptions, the many-body wave function has been calculated even for larger $N$. See for example \citet{Motta2017}.} However, the \emph{scaling} of Eq.~\eqref{eq:est:exponential_scaling_wf} remains and represents an ``exponential wall'' as Nobel laureate Walter Kohn called it~\citep{Kohn1999}.

The problem of the exponential wall is so severe that the concept of the many-body wave function for larger systems has basically no value beyond its use for certain theoretical considerations. Walter Kohn went so far as to suggest that 
\begin{aquote}{\citeauthor{Kohn1999},\citeyear{Kohn1999} \citep{Kohn1999}}
	``in general the many-electron wavefunction $\Psi(\br_1,...,\br_N)$ for a system of $N$ electrons is not a legitimate concept, when $N>N_0$, where $N_0\approx 1000$.''
\end{aquote}
Regarding the above-mentioned current state-of-the-art, Kohn's guess of $N_0$ seems even understated. In practice, this means that most of our knowledge on electronic-structure relies on alternative descriptions that carry less information than the full many-body wave function.

\subsection{Excursion: exchange symmetry in quantum mechanics}
\label{sec:est:excursion_symmetry}
Since exchange symmetry will play a central role in the course of this thesis, we want to make a little detour recapitulating the concept of exchange symmetry. This more historically excursion supplements the chapter and is not necessary to follow the course of this text. The fast reader can directly continue with Sec.~\ref{sec:est:hf}.

\subsubsection*{The origin of quantum indistinguishability}
To the best of our knowledge, all electrons have identical physical properties, i.e., the same mass, charge and spin and thus, a valid definition of a many-body Hamiltonian like Eq.~\eqref{eq:est:ParticleBox_Hamiltonian} or Eq.~\eqref{eq:est:general:hamiltonian} must be \emph{symmetric} with respect to the exchange of particles coordinates. From the symmetry of the Hamiltonian we can deduce the exchange symmetry or ``special circumstances'' of the wave function as Wolfgang Pauli called it in one of his famous articles on quantum mechanics:
\begin{aquote}{\citeauthor{Pauli1933}, \citeyear{Pauli1933} \citep{Pauli1933}}
	Wenn wir es mit mehreren gleichartigen Teilchen zu tun haben, treten besondere Verhältnisse ein, die daher rühren, daß der Hamiltonoperator stets invariant ist  bei irgendwelchen Vertauschungen der Teilchen. \\
	\emph{If we deal with many similar [indistinguishable] particles, special circumstances occur that arise from the Hamilton operator being invariant under any permutations of particles.}
\end{aquote}
It is clear that also a classical system consisting of many identical  particles would be described by a symmetric Hamiltonian. However, such classical particles (imagine billiard balls) are distinguishable, even if they have identical physical properties. The reason is that we can in principle locate their position at any instance of time and collect them in a trajectory. This allows to define a label for every trajectory, which in turn distinguishes the balls. The situation changes, if we employ, e.g., a  \emph{statistical} description. This means that instead of describing explicitly the trajectory of each billiard ball, we employ a probability distribution that describes the probability to find \emph{a} billiard ball at a certain position at a given time instance. In such a description, we obviously lose the distinguishability. 

The most common formulations\footnote{For a discussion on possible alternative approaches, see for example Ref.~\citep{Holland1995}.} of quantum mechanics are also probabilistic and thus they are (at least in practice) similar to the latter case.\footnote{However, ontologically there are huge differences between classical mechanics and quantum mechanics, were the details depend on the specific interpretation of quantum mechanics. We refer the interested reader to the well-documented \textsc{Wikipedia} article on the topic and the reference therein: \url{https://en.wikipedia.org/wiki/Interpretations_of_quantum_mechanics}.}A typical argument why quantum particles must be indistinguishable extends on the trajectory example above: the knowledge of a trajectory in time and space contradicts Heisenberg's uncertainty principle~\citep{Gross1991}.\footnote{From this argument follows that quantum particles that are very far away from each other (such that they can be described by separate wave functions that do not overlap) are indeed distinguishable. Dicke used this argument, when he introduced his model of $N$ two-level system coupled to one photon mode~\citep{Dicke1954}.}
A less ``hand-wavy'' argument by means of the counting of states in a fictive experiment was given by Dirac already in the very beginnings of the development of quantum mechanics~\citep{Dirac1926}. For a very detailed discussion of this argument, the reader is referred to the book by \citet[part 1]{Stefanucci2013}.

\subsubsection*{The consequence of indistinguishability: particle statistics}
To the best of our knowledge, the indistinguishability of quantum particles is unquestioned until nowadays and usually regarded as a fact~\citep{Gross1991,Altland2010,Stefanucci2013}. This means in practice that although we formally need to introduce (labelled) coordinates to describe $N$ electrons, we have to make sure that the predictions of the theory cannot be used to differentiate the electrons. Since only \emph{expectation values} of observable operators correspond to possible measurements in experiments, we must guarantee indistinguishability on this level. For instance, to obtain the expectation value of the energy of a two-particle system, we need to calculate the integral  
\begin{align}
E=\int\td\bx_1\td\bx_2 \Psi^*(\bx_1,\bx_2) \hat{H}(\bx_1,\bx_2)  \Psi(\bx_1,\bx_2).
\end{align}
Since the wave function occurs quadratically in such expressions, we have more freedom to do such a symmetrization on the level of $\Psi$ than on the operator level. \citet{Dirac1926} concluded already in 1926 that as a consequence, many-body wave functions need to be either symmetric or antisymmetric\footnote{Note that there are effective Hamiltonians that have even more complicated symmetries, e.g., the effective Hamiltonian of certain topologically nontrivial systems. See Ref.~\citep[part 9]{Altland2010} for a good overview about such scenarios.}  under a pair-wise permutation $\mathcal{P}_{\bx,\by}$ of two coordinates $\bx,\by$:
\begin{subequations}
	\begin{alignat}{2}
	\mathcal{P}_{\bx_1,\bx_2} \Psi(\bx_1,\bx_2) &= \Psi(\bx_2,\bx_1) \overset{!}{=} \Psi(\bx_1,\bx_2) \quad &&\longleftrightarrow\quad\text{symmetry}\\
	\mathcal{P}_{\bx_1,\bx_2} \Psi(\bx_1,\bx_2) &= \Psi(\bx_2,\bx_1) \overset{!}{=} -\Psi(\bx_1,\bx_2) \quad &&\longleftrightarrow\quad\text{antisymmetry}.
	\end{alignat}
\end{subequations}
These symmetry properties \emph{define} the two fundamental particle classes, that we consider until today:\footnote{Note that particles with different symmetry properties can be defined (See also the previous footnote). However, such definitions require some kind of effective theory and thus the according particles are usually not considered as fundamental. This holds also for the polaritons that we will introduce in part~\ref{sec:dressed}. They are electron-photon hybrid particles.} we call particles that are described by a symmetric or antisymmetric many-particle wave function \emph{bosons} or \emph{fermions}, respectively. In Ref.~\citep{Dirac1926}, Dirac further connected  the fundamental exchange symmetry of the particle to its statistical properties: symmetric particles follow Bose-Einstein statistics~\citep{Bose1924,Einstein1925}
and antisymmetric particles follow Fermi-Dirac~\citep{Fermi1927} statistics. The different particle statistics are undoubtedly one of the most important elements of quantum theory and alone, i.e., \emph{without} taking into account many other elements of the theory, they have an impressive explanatory power. Einstein showed for example already in 1925~\citep{Einstein1925}, i.e., \emph{before} Heisenberg and Schrödinger formulated the first well-defined theories of quantum mechanics~\citep{Heisenberg1925,Schrodinger1926}, that Bose-Einstein gases exhibit a phase state, the \emph{condensate}, which is characterised by extreme coherence between all the gas particles, letting them act as one large object.\footnote{Already Einstein predicted this peculiar phase state in the cited paper but it took until 1995 for its first experimental realization \citep{Anderson1995}. Since then, many more Bose-Einstein condensates have been realized experimentally and one can find  an considerable amount of literature about the topic. A good introduction to the topic can be found in the textbook by \citet[part 6.3]{Altland2010}.}

\subsubsection*{Pauli's principle}
Regarding the explanatory potential of fermion statistics, many phenomena could be named, especially in the realm of electronic-structure theory, where electrons are described as fermions. Most importantly, one can show that fermions obey \emph{Pauli's exclusion principle}, which is a crucial element of the modern model of atoms.
Pauli defined his principle in the following way:
\begin{aquote}{\citeauthor{Pauli1925}, \citeyear{Pauli1925} \citep{Pauli1925}}
	Es kann niemals zwei oder mehrere äquivalente Elektronen
	im Atom geben, für welche [...] die Werte aller Quantenzahlen [...] übereinstimmen. Ist ein Elektron im Atom vorhanden, für das diese Quantenzahlen bestimmte Werte haben, so ist dieser Zustand ``besetzt''.\\
	\textit{There are never two or more equivalent electrons in the atom which have the same quantum number. If there is an electron in the atom with certain quantum numbers, this state is ``occupied.''}
\end{aquote}
A remarkable deduction from Pauli's principle is for example the explanation of the \emph{stability of matter}. \citet{Lieb2005} showed under very weak assumptions employing the atom model of electrons and static nuclei that Pauli's principle is necessary and sufficient to guarantee that \emph{any} matter system built by such constituents is stable, i.e., it has a well-defined ground state. 

\subsubsection*{A historical example of the importance of Pauli's principle}
To illustrate the importance of Pauli's principle, let us make a very brief detour to the 1930ies that is based on the first chapters of the book by \citet{Gavroglu2012}. At that time, chemistry and physics were two (almost completely) separate research fields. But with the formulation of quantum mechanics and the corresponding model of the atom, physicists paved the way for establishing \emph{quantum chemistry}. In this since then ever growing research field, researchers try to understand physical models and experiments of chemical systems by connecting quantum theory with the theoretical knowledge of chemistry. However, the development of quantum chemistry was initially very slow and it took a lot of time and especially modern computers until the real success story of the field started. Physicists like Dirac started to promote the point of view that chemistry was merely an application of physical laws, but they had to realize very fast that this application is much more difficult than anybody could have ever imagined. Walter Kohn summarized this in his Nobel lecture in the following way:
\begin{aquote}{\citeauthor{Kohn1999},\citeyear{Kohn1999} \citep{Kohn1999}}
	``There is an oral tradition that [...] Dirac declared that chemistry had come to an end - its content was entirely contained in that powerful [i.e., the Schrödinger] equation. Too bad, he is said to have added, that in almost all cases, this equation was far too complex to allow solution.''
\end{aquote}
On the other hand, Chemistry always relied (and is still relying) on heuristic rules and tables, like Mendeleev's perodic table \citep{Mendelejew1869}, Hund's rule \citep{Hund1925} or Mulliken's ``correlation diagram'' \citep{Mulliken1932}. And most chemists doubted that the mathematically very challenging quantum theory could be ever useful for understanding such kind of rules. The only exception is Pauli's principle, which is somewhat special in comparison to other concepts like the wave function or the Hamilton operator. In fact, Pauli's principle is a rule: Only by utilizing the principle and a simple version of the physicist's atom model, chemists could generalize and combine several of their own rules and concepts. Thus, from all the complex apparatus of the newly developed quantum theory, it was the conceptually most simple part, Pauli's principle, that played a key role in establishing the fruitful connection between physics and chemistry. It is needless to say that this connection played a very important role in the further development of quantum mechanics itself. Nowadays, we know that solving Schrödinger's equation for many-particle systems like larger molecules is de facto impossible and a first-principles description of such systems requires methods that have been developed by mathematicians, physicists and chemists, often in close collaboration. We will see in the following that Pauli's principle plays an important role in all these methods.

\subsubsection*{Bohr's aufbau principle}
Another consequence of fermion statistics that plays an important role in electronic-structure theory is \emph{Bohr's aufbau principle} (building-up principle), which Pauli defined in the following way:
\begin{aquote}{\citeauthor{Pauli1925}, \citeyear{Pauli1925} \citep{Pauli1925}}
	Dieses von Bohr aufgestellte Prinzip besagt, daß bei Anlagerang eines weiteren Elektrons an ein Atom die Quantenzahlen der schon gebundenen Elektronen dieselben Werte behalten, die ihnen im zugehörigen stationären Zustand des freien Atomrestes zukommen.\\
	\textit{This by Bohr established principle states that when a further electron is absorbed, the quantum numbers of an atom's bound electron keep the values, that they would have in the corresponding stationary state of the free atom rest.}
\end{aquote}
The strength of the aufbau principle is especially visible in effective single-particle methods such as HF theory (see Sec.~\ref{sec:est:hf}) to describe many-body systems. In such methods, we solve (nonlinear) eigenvalue equations for single-particle states (orbitals). The aufbau principle then tells us how to ``occupy'' these orbitals with electrons to obtain the many-body state: if our systems consists of $N$ electrons, we will occupy the $N$ energetically lowest orbitals. In this sense, the aufbau principle combines the exclusion principle (we only populate every state by one electron) with the variational principle (we try to obtain the lowest energy). However, other aufbau principles inspired by Bohr's principle have been developed, which in one or the other way fill up some levels.\footnote{See for instance Refs.~\citep{Giesbertz2010aufbau,Piris2009,Rao2001}.} This shows the strength of the very concept of an aufbau principle in electronic-structure theory.
	\newpage
\section{Hartree-Fock theory}
\label{sec:est:hf}
In the last section, we have discussed the problems connected to the $N$-electron wave function that for a general interacting system is a superposition of arbitrary Slater determinants, i.e.,
\begin{align}
\tag{cf. \ref{eq:est:wavefunction_Nparticles_slater_expansion}}
\Psi(\bx_1,\dotsc,\bx_N) = \sum_I c^I \Psi_N^I(\bx_1,\dotsc,\bx_N).
\end{align}
There is however the special setting of non-interacting electrons, for which the $N$-body problem basically reduces to a one-body problem. The corresponding many-body state is given by one Slater determinant. In HF theory, we try to find the ``best fit'' between the non-interacting and interacting problem by simply truncating the  expansion~\eqref{eq:est:wavefunction_Nparticles_slater_expansion} after the first element, i.e., we consider a single-reference wave function ansatz
\begin{align}
	\Psi(\bx_1,\dotsc,\bx_N)\approx \Psi^{HF}(\bx_1,\dotsc,\bx_N)=\tfrac{1}{\sqrt{N!}}|\bar{\psi}_1\dotsc, \bar{\psi}_N|_-.
\end{align}
to describe the interacting problem. 


\subsubsection*{The HF minimization problem}
For our derivations, we assume \emph{spin-restriction}, i.e., we consider systems with even particle number and define
\begin{align}
\tag{cf.	\ref{eq:est:spin_orbitals_closed_shell}}
\begin{split}
	\bar{\psi}_{2i}(\br,\sigma)=&\psi_i(\br)\alpha(\sigma)\\
	\bar{\psi}_{2i+1}(\br,\sigma)=&\psi_i(\br)\beta(\sigma),
\end{split}
\end{align}
where $\{\alpha,\beta\}$ denote the two spin functions (see the discussion around Eq.~\eqref{eq:est:spin_orbitals_closed_shell}). This \emph{restricted HF} is the most common version of HF theory. 
We saw already that wave functions like $\Psi^{HF}$ are the solutions of non-interacting problems, i.e., $\Psi^{HF}$ is the ground state of any Hamiltonian of the form
\begin{align}
	\hat{H}_{ni}=\sum_{i=1}^{N} \hat{t}(\br_i)+\hat{v}(\br_i),
\end{align}
if we identify $\psi_1,..,\psi_{N/2}$ according to the aufbau principle with the lowest $N/2$ eigenstates of $\hat{t}+\hat{v}$. One can thus say that in single-reference methods, we approximate the interacting with a non-interacting wave function.\footnote{This statement is independent of the assumed spin-restriction of our example.} However, there are many possibilities for such an approximation and we have to define a qualifier that determines one out of these. In wave-function methods such as HF theory, this qualifier is always the energy (we will employ a different qualifier in Sec.~\ref{sec:est:dft}). We thus can define the HF ground state by the variational principle, i.e.,
\begin{align}
\label{eq:est:hf:variational_principle}
	E_{0}^{HF} = \braket{\Psi^{HF}_0|\hat{H}\Psi^{HF}_0}=\inf_{\Psi^{HF}} \braket{\Psi^{HF}|\hat{H}\Psi^{HF}},
\end{align}
with the $N$-electron electronic-structure Hamiltonian that we defined in the previous section
\begin{align}
\tag{cf. \ref{eq:est:general:hamiltonian}}
	\hat{H}=\sum_{i=1}^{N} \hat{t}(\br_i) + \hat{v}(\br_i) + \tfrac{1}{2}\sum_{i,j=1}^{N} \hat{w}(\br_i,\br_j).
\end{align}

\subsubsection*{The features of HF theory: the upper energetic bound and the definition of correlation}
It is clear that the ansatz $\Psi\approx\Psi^{HF}$ constrains the full configuration space that we have to consider to find the exact ground state of $\hat{H}$. Since due the variational principle, $\Psi$ must be the wave function with the lowest possible energy expectation value, we can deduce that the HF energy
\begin{align}
	E_{0}^{HF}\geq E_0
\end{align}
is an upper bound of the exact many-body energy $E_0$. We call methods with this property \emph{variational}. The difference between exact and HF energy, the so-called correlation energy
\begin{align}
	E_c=E_0-E_{0}^{HF}\leq 0,
\end{align}
thus must be zero or negative. Consequently, HF defines a kind of ``baseline'' for all electronic-structure methods that from this perspective ``only'' aim to accurately describe $E_c$. Because the HF wave function corresponds to a non-interacting system that as we have seen in the last section can be reduced to an effective one-body problem, the correlation energy is often called \emph{many-body energy}.

\subsubsection*{The HF energy: the ``quantum'' contribution and exchange symmetry}
Let us now calculate this energy expression explicitly. We have
\begin{align*}
	E_{HF}=&\braket{\Psi^{HF}|\hat{H}\Psi^{HF}}\\
	=&\sum_{i=1}^{N} \braket{\Psi^{HF}|\left[\hat{t}(\br_i) + \hat{v}(\br_i)\right]\Psi^{HF}} + \tfrac{1}{2}\sum_{i,j=1}^{N} \braket{\Psi^{HF}|\hat{w}(\br_i,\br_j)\Psi^{HF}},
\end{align*}
where the first, i.e., the one-body part (after some algebra, see Ref.~\citep[part 5.4]{Helgaker2000}) reads
\begin{align*}
	E^{(1)}[\{\psi_i\}]=& 2\sum_{i=1}^{N/2} \int\td\br \psi_i^*(\br) \left[\hat{t}(\br) + \hat{v}(\br)\right] \psi_i(\br) \nonumber\\
	=& 2\sum_{i=1}^{N/2}\braket{\psi_i|\hat{t}+\hat{v}|\psi_i},
\end{align*}
where we have defined the one-body integrals $\braket{\psi_i|\hat{o}^{(1)}|\psi_i}=\int\td\br \psi_i^*(\br) \left[\hat{o}(\br) + \hat{v}(\br)\right] \psi_i(\br)$  for a given one-body $\hat{o}^{(1)}$. Note that the factor 2 is due to the spin-restriction.
The second term involves the two-body operator $\hat{w}(\br,\br')$ and reads~\citep[part 5.4]{Helgaker2000}
\begin{align*}
	E^{(2)}[\{\psi_i\}] =& \sum_{i,j=1}^{N/2} \left[2\int\td\br\td\br' \psi_i^*(\br)\psi_j^*(\br')\hat{w}(\br,\br')\psi_i(\br)\psi_j(\br') \right.\nonumber\\
	&\left.-\int\td\br\td\br' \psi_i^*(\br)\psi_j^*(\br')\hat{w}(\br,\br')\psi_j(\br)\psi_i(\br')  \right] \nonumber\\
	\equiv& \sum_{i,j=1}^{N/2} \left[ 2\braket{\psi_i\psi_j|\hat{w}|\psi_i\psi_j} -\braket{\psi_i\psi_j|\hat{w}|\psi_j\psi_i}\right].
\end{align*}
Here, we have introduced the notation $\braket{\psi_i\psi_j|\hat{o}^{(2)}|\psi_k\psi_l}=\int\td\br\td\br' \psi_i^*(\br)\psi_j^*(\br')\hat{o}^{(2)}(\br,\br')\psi_i(\br)\psi_j(\br')$ to abbreviate the integrals with respect to a two-body operator $\hat{o}^{(2)}$. We call the first term of $E^{(2)}$ the \emph{Hartree} or mean-field energy contribution,
\begin{align}
\label{eq:est:hf:energy_Hartree}
	E_H=2\sum_{i,j=1}^{N/2} \braket{\psi_i\psi_j|\hat{w}|\psi_i\psi_j},
\end{align}
that can be evaluated in two steps by first calculating the Hartree potential
\begin{align}
\label{eq:est:hf:potential_hartree}
	\hat{v}_H(\br)=\sum_{j=1}^{N/2} \int\td\br' \psi_j^*(\br')\hat{w}(\br,\br')\psi_j(\br')
\end{align}
and then the energy expression $E_H=2\sum_{i=1}^{N/2} \braket{\psi_i|\hat{v}_H|\psi_i}$. The name mean-field stems from the usual interpretation of $\braket{\hat{v}_H}$ as the mean Coulomb potential of all the electrons. Importantly, $\braket{\hat{v}_H}$ is equivalent to the Coulomb potential of a classical charge distribution. The second term of $E^{(2)}$ instead is called the \emph{exchange} (or Fock) contribution,
\begin{align}
\label{eq:est:hf:energy_exchange}
	E_X=-\sum_{i,j=1}^{N/2} \braket{\psi_i\psi_j|\hat{w}|\psi_j\psi_i}
\end{align}
that arises because of the negative terms of the Slater determinant and thus has no classical counter-part. The name stems from the fact the two body integrals occurring in Eq.~\eqref{eq:est:hf:energy_exchange} can be obtained from the integrals in Eq.~\eqref{eq:est:hf:energy_Hartree} by exchanging the indices on one of the two sides. The exchange contribution represents thus a quantum-mechanical correction to the classical or mean-field energy. Although this correction is usually much smaller than the other energy terms in $E^{HF}$, it makes the HF description  considerably more accurate than the older Hartree theory that only takes the mean-field into account~\citep{Helgaker2000}. 

Collecting all the terms, we get the following expression for the HF energy
\begin{empheq}[box=\fbox]{align}
\label{eq:est:hf:energy}
	E^{HF}[\{\psi_i\}]=2\sum_{i=1}^{N/2}\braket{\psi_i|\hat{t}+\hat{v}|\psi_i} +\sum_{i,j=1}^{N/2} \left[ 2\braket{\psi_i\psi_j|\hat{w}|\psi_i\psi_j} -\braket{\psi_i\psi_j|\hat{w}|\psi_j\psi_i}\right]
\end{empheq}
We see that the HF energy is a functional of the $N/2$ orbitals $\{\psi_1,...,\psi_{N/2}\}$. 

\subsubsection*{The HF minimization problem and the Fock operator}
To minimize $E^{HF}$ with respect to the orbitals, we need to guarantee their orthonormality, i.e.,
\begin{align}
\label{eq:est:hf:orthonormality_condition}
	\braket{\psi_i|\psi_j}=\delta_{ij}\quad \forall\,i,j.
\end{align}
The standard way to do so, is to constrain the minimization of the functional by introducing a Lagrangian multiplier, $\tilde{\epsilon}_{ij}$, for every condition. Instead of the original minimization problem of Eq.~\eqref{eq:est:hf:variational_principle}, we minimize the Lagrangian
\begin{align}
	L^{HF}[\{\psi_i\}, (\lambda_{ij})] = E^{HF}[\{\psi_i\}] - 2\sum_{i,j=1}^{N/2} \tilde{\epsilon}_{ij} (\braket{\psi_i|\psi_j}-\delta_{ij})
\end{align}
with respect to the orbitals \emph{and} the Lagrange multipliers.\footnote{Note that we introduced the factor 2 in front of the Langrange-multiplier term for later convenience.} As it is customary, we treat the orbitals $\psi_i$ and their complex conjugates $\psi^*_i$ as independent and consider $L^{HF}[\{\psi_i,\psi_i^*\}, (\lambda_{ij})]$.\footnote{It is important to realize that our goal is to perform a free minimization over all possible functions $f(\br)$ that depend on one variable. In this space, there are in principle functions with $f(\br)\neq f^*(\br)$. However, our formalism is constructed such that the solution $f_0$ has the correct property $f_0=f_0^*$.} A necessary condition for a minimum of $L^{HF}$ is \emph{stationarity} with respect to all variables 
\begin{align}
	0=\delta L^{HF}[\{\psi_i\}, (\lambda_{ij})] = \sum_{i=1}^{N/2} \left[\int\td\br \frac{\delta L^{HF}}{\delta \psi_i(\br)} \delta\psi_i (\br)+ \int\td\br \frac{\delta L^{HF}}{\delta \psi_i^*(\br)} \delta\psi_i^* (\br)\right] + \sum_{i,j=1}^{N/2} \frac{\partial L^{HF}}{\partial \lambda_{ij}} \td \lambda_{ij}.
\end{align}
This leads to three independent sets of conditions, where the latter just gives back the orthonormality conditions of Eq.~\eqref{eq:est:hf:orthonormality_condition} and the former two are equivalent to each other. For notational convenience, we utilize the second set of conditions (with respect to $\psi_i^*$) and obtain
\begin{align}
\label{eq:est:hf:stationarity_orbitals}
	\sum_{j=1}^{N/2} \tilde{\epsilon}_{ij} \psi_j(\br)= \left[\hat{t}+\hat{v}\right] \psi_i(\br) + \sum_{j=1}^{N/2} \left[ 2\int\td\br \psi_j^*(\br')\hat{w}(\br,\br')\psi_j(\br') \psi_i(\br) -\int\td\br \psi_j^*(\br')\hat{w}(\br,\br')\psi_i(\br') \psi_j(\br)\right].
\end{align}
To simplify this expression, we can define the Coulomb-operator $\hat{J}_j$\footnote{Note that the Coulomb-operator  is connected to the Hartree potential by $\hat{v}_H=\sum_j \hat{J}_j$.} that acts as
\begin{align}
	\hat{J}_j\psi_i(\br) &= \int\td\bz' \psi_j^*(\br')w(\br,\br') \psi_j(\br') \psi_i(\br)
\intertext{and the Exchange-operator $\hat{K}_j$ that acts as~\cite{Szabo2012}}
	\hat{K}_j\psi_i(\br) &= \int\td\bz' \psi_j^*(\br')w(\br,\br') \psi_i(\br') \psi_j(\br). \label{eq:est:hf:exchange_operator}
\end{align}
The right-hand side of Eq.~\eqref{eq:est:hf:stationarity_orbitals} can then be expressed as
\begin{empheq}[box=\fbox]{align}
\label{eq:est:hf:Fock_operator}
	\hat{H}^{1}\psi_i(\br)=\left[\hat{t}+\hat{v}\right] \psi_i(\br) + \sum_{j=1}^{N/2} \left[ 2\hat{J}_j-\hat{K}_j\right] \psi_i(\br),
\end{empheq}
where we introduced the \emph{Fock-operator} $\hat{H}^{1}$, that acts on one orbital and thus is can be seen as an effective \emph{one-body Hamiltonian}. 

\subsubsection*{Solving the HF equations: a nonlinear orbital eigenvalue problem}
Importantly, $\hat{H}^{1}$ is hermitian and thus, we can apply a unitary transformation to Eq.~\eqref{eq:est:hf:stationarity_orbitals} such that $\tilde{\epsilon}_{ij}=\epsilon_i\delta_{ij}$ becomes diagonal. In this canonical formulation, Eq.~\eqref{eq:est:hf:stationarity_orbitals} obtains the form of an \emph{eigenvalue equation} 
\begin{empheq}[box=\fbox]{align}
\label{eq:est:hf:HF_equations}
	\hat{H}^{1}\psi_i(\br)=\epsilon_i\psi_i(\br).
\end{empheq}
for HF orbitals $\psi_i$ and the orbital energies
\begin{align}
	\epsilon_i =& \braket{\psi_i|\hat{H}^1\psi_i} \nonumber\\
	=&\braket{\psi_i|\left[\hat{t}+\hat{v}\right] \psi_i} + \braket{\psi_i|\sum_{j=1}^{N/2} \left[ 2\hat{J}_j-\hat{K}_j \psi_j(\br)\right] \psi_i} \nonumber \\
	=& \braket{\psi_i|\left[\hat{t}+\hat{v}\right]\psi_i} +\sum_{j=1}^{N/2} \left[ 2\braket{\psi_i\psi_j|\hat{w}|\psi_i\psi_j} -\braket{\psi_i\psi_j|\hat{w}|\psi_j\psi_i}\right].
\end{align}
The set of these equations for all $N/2$ orbitals are called the \emph{Hartree-Fock equations}, which have to be solved (numerically) to obtain the Hartee-Fock ground state of the Hamiltonian~\eqref{eq:est:general:hamiltonian}.

Importantly, Eq.~\eqref{eq:est:hf:HF_equations} is not a usual, i.e., linear eigenvalue equation. The Fock-operator $\hat{H}^{1}=\hat{H}^{1}[\{\psi_i\}]$ \emph{depends} on its own eigenstates and thus the eigenvalue problem of Eq.~\eqref{eq:est:hf:HF_equations} is in fact \emph{nonlinear}. We call $\hat{H}^{1}$ a nonlinear one-body operator, which has very different properties from linear operators. We can utilize $\hat{H}^{1}$ to find the HF ground state or to determine approximate ionization potentials and electron affinities (Koopman's theorem~\citep[Sec. 10.5]{Helgaker2000}). However, the higher lying ``eigenstates'' of $\hat{H}^{1}$, the so-called unoccupied orbitals, have only limited physical meaning. In fact, the validity Koopman's theorem is a very special feature of the Fock-matrix and similar nonlinear one-body operators (see Sec.~\ref{sec:est:rdms}) in other electronic-structure theories do not have the same level of explanatory power. In comparison, the eigenfunctions and eigenvalues of the many-body Hamiltonian have a clear physical interpretation: they simply denote the possible states with corresponding energy levels that the physical system can adopt. 

Another practical consequence of the nonlinearity of $\hat{H}^{1}$ is that solving the HF equations is numerically a very challenging task that is not comparable with an ordinary eigenvalue equation. For instance, there are simple many-body systems, for which the Schrödinger equation is analytically solvable.\footnote{For example the \emph{Harmonium} model that consists of two interacting electrons in a Harmonic potential.} But there is \emph{no known analytic solution} of the HF equations. The only known way to solve Eq.~\eqref{eq:est:hf:HF_equations} is a so-called \emph{self-consistent field} (SCF) procedure. For that, we linearize $\hat{H}^{1}[\{\psi_i\}]\approx \hat{H}^{1}[\{\psi_i^0\}]$ locally around some starting guess of orbitals $\{\psi_i^0\}$, solve Eq.~\eqref{eq:est:hf:HF_equations} for this $\hat{H}^{1}[\{\psi_i^0\}]$ to obtain new orbitals $\{\psi_i^1\}$ and update $\hat{H}^{1}[\{\psi_i^1\}]$. We proceed with this iterative procedure until the orbitals $\{\psi_i^{n+1}\}\approx\{\psi_i^n\}$ do not change significantly, i.e., until self consistence of Eq.~\eqref{eq:est:hf:HF_equations}.\footnote{In part~\ref{sec:numerics}, we will present some algorithms that are capable to perform HF minimizations.}

Although we know from experience that this procedure converges in most cases, we can neither guarantee convergence, nor can we be sure that the converged result is the global minimum of $E_{HF}$ that we are searching for. The reason is again the nonlinearity of the problem that makes a mathematical analysis very difficult. There are certain hints that $E_{HF}$ is not convex, which would suggest that solutions of the HF equations could correspond to local minima, but this has not yet been proven.\footnote{See for example the discussion in Ref.~\citep{Frank2007} and the references therein.}

\subsubsection*{The cornerstone of (molecular) electronic-structure theory}
Let us finish our brief discussion of HF theory with a quotation of one of the most important textbooks on molecular electronic-structure theory:
\begin{aquote}{\citeauthor{Helgaker2000}, \citeyear{Helgaker2000} \citep[p. 433]{Helgaker2000}}
	``The Hartree-Fock wave function is the cornerstone of ab initio electronic-structure theory. [... It] yields total electronic energies that are in error by less than 1\% and a wide range of important molecular properties such as dipole moments, electric polarizabilities, electronic excitation energies, magnetizabilities, force constants and nuclear magnetic shieldings are usually reproduced to within 5-10 \% accuracy. Molecular geometries are particularly well reproduced and are mostly within a few picometres of the true equilibrium structure.
	
	The Hartree-Fock wave function is often used in qualitative studies of molecular systems, particularly larger systems. Indeed, the Hartree-Fock wave function is still the only wave function that can be applied routinely to large systems, and systems containing several hundred atoms have been studied at this level of approximation. For accurate, quantitative studies of molecular systems, the Hartree-Fock wave function is by itself not useful but it constitutes the starting point for more accurate treatments.''
	
\end{aquote}
This summarizes the performance of HF theory and its importance in the big field of electronic-structure methods.\footnote{Note that HF plays a less pronounced role for extend systems, such as solids. One reason is simply that performing HF-calculations in $k$-space is in most cases numerically very expensive. Here, the local-density approximation of KS-DFT has a comparable role to HF for molecular systems (see next section).} Being one of the conceptually and computationally simplest electronic-structure methods, we will repeatedly consider HF theory in the following chapters, when we discuss the coupled electron-photon problem.

However, as \citeauthor{Helgaker2000} have pointed out, in many cases, HF is not sufficient for an accurate description of many-body systems and we need more accurate methods . In the next two sections, we will thus discuss some approaches to go beyond HF theory and try to include the aforementioned correlation energy in the description.

	\section{Density functional theory}
\label{sec:est:dft}
In the last section, we have discussed how we can describe an interacting many-body system by a single Slater determinant, i.e., an effective non-interacting wave function. This is an approximate description that neglects correlation, i.e., all effects that intrinsically require a multi-reference description.
However, if we give up the concept of the many-body wave function (and with this the standard description of quantum states), we can in principle describe any many-particle system by such a single-reference wave function. This is called the \emph{Kohn-Sham} (KS) construction of DFT and it represents for a huge class of systems the best trade-off between efficiency and accuracy in comparison to all alternative descriptions. It is not overstated to say that KS-DFT ``is indispensable for modern quantum-chemical modeling of materials and molecules''~\citep{Medvedev2017}. 

Because of the single-reference ansatz, KS-DFT is in practice very similar to HF. A KS-DFT calculation means to solve the KS equations, which constitute a nonlinear eigenvalue equation. In many cases, this is computationally even less demanding than solving the HF equations. Conceptually however, DFT is very distinguished from any type of wave-function method. In fact, one can see DFT as an alternative formulation of quantum mechanics.\footnote{For instance, \citeauthor{Tokatly2005a} argues that DFT can be interpreted as a \emph{hydrodynamic formulation of quantum mechanics}~\citep{Tokatly2005a,Tokatly2005b}.}


\subsubsection*{The density as basic variable}
As the name suggests, DFT is based on the (single-particle) electron density, that is defined as
\begin{align}
\label{eq:est:dft:one_density_definition}
\rho(\br)=N\sum_{\sigma_1,...,\sigma_N}\int\td\br_2\cdots\td\br_N \Psi_0^*(\bx,\bx_2,...,\bx_N)\Psi_0(\bx,\bx_2,...,\bx_N)
\end{align}
for an $N$-electron system with ground state wave function $\Psi_0(\bx_1,...,\bx_N)$ that is the lowest eigenstate of the electronic-structure Hamiltonian, cf. \eqref{eq:est:general:hamiltonian},
\begin{align}
\label{eq:est:dft:hamiltonian}
\hat{H}=&\sum_{i=1}^{N} \left[\hat{t}(\br_i) + v(\br_i)\right] + \tfrac{1}{2}\sum_{i,j=1}^{N} w(\br_i,\br_j)\\
=& \hat{T} + \hat{V} + \hat{W}. \nonumber
\end{align}
We renamed the three occurring terms, $\hat{T}=\sum_{i=1}^{N} \hat{t}(\br_i)$, etc. for later convenience.
Already in 1964, Walter Kohn and Pierre Hohenberg have proven that $\rho(\br)$ \emph{entirely} determines the ground state of $\hat{H}$~\citep{Hohenberg1964}. On a first glance, it might sound counter-intuitive that all the relevant information of the $4N$-dimensional wave function $\Psi_0$ is contained in a merely 3-dimensional quantity. However, the wave function is simply an auxiliary quantity that allows us to formulate the standard theory of quantum mechanics, based on linear operators. The information that determines the physical setting in the electronic-structure Hamiltonian is actually the local potential $v(\br)$ that is also merely three-dimensional.

For instance, if we want to describe a diatomic molecule with nuclei at the positions $\mathbf{R}_1,\mathbf{R}_2$ and charges $Z_1,Z_2$, we would model this by considering the local potential $v(\br)=\frac{Z_1}{|\br-\mathbf{R}_1|}+\frac{Z_2}{|\br-\mathbf{R}_2|}$. We can arbitrarily enlarge this system with further nuclei by adding Coulomb terms to the local potential. But for all these systems, the kinetic and the interaction operators are the same. These operators are thus said to be \emph{universal}. Any many-electron system in the non-relativistic setting is thus completely defined by the 3-dimensional local potential and the number of electrons $N$ and judged from this perspective, the concept of a theory based on the electronic density might sound more reasonable. 

In the realm of DFT, $v(\br)$ is called the \emph{external} variable that determines the corresponding \emph{internal} variable $\rho(\br)$. This variable pair $(\rho,v)$ is defined by the structure of the Hamiltonian~\eqref{eq:est:dft:hamiltonian}. To see this, let us calculate the expectation value of the potential term
\begin{align*}
	\braket{\hat{V}}=&\braket{\Psi|\sum_{i=1}^{N}v(\br_i)\Psi}\\
	=&\sum_{\sigma_1,...,\sigma_N}\int\td\br_1\br_2\cdots\td\br_N \Psi^*(\bx_1,\bx_2,...,\bx_N)\left[\sum_{i=1}^{N}v(\br_i)\right]\Psi(\bx_1,\bx_2,...,\bx_N)\\
	=&\int\td\br v(\br) N\sum_{\sigma_1,...,\sigma_N}\int\br_2\cdots\td\br_N \Psi^*(\bx,\bx_2,...,\bx_N)\Psi(\bx,\bx_2,...,\bx_N)\\
	=&\int\td\br v(\br) \rho(\br),
\end{align*}
where we used the exchange symmetry of $\Psi$ from the second to the third line. This term  \emph{defines} the pair of external and internal variable and it is at the center of the Hohenberg-Kohn proof~\citep{Penz2019} of the existence of the one-to-one mapping
\begin{align}
\label{eq:est:dft:external_internal_correspondence}
	v(\br) \overset{1:1}{\longleftrightarrow} \rho(\br).
\end{align}
The simplicity of this argument illustrates the fact that there are many possible functional theories, depending on the structure of the Hamiltonian. A famous example is \emph{current-density functional theory}~\citep{Vignale1987} that can describe systems with external magnetic field or \emph{reduced density matrix functional theory} that we will introduce in Sec.~\ref{sec:est:rdms:rdmft}. All these functional theories are based on the same type of proof,\footnote{For very general proof of this statement, see \citet{Penz2019}.} which we do not recapitulate at this point. We will instead present (one version of) the proof in the realm of coupled electron-photon systems in Sec.~\ref{sec:qed_est:rdms:qed_rdmft}. 

\subsubsection*{The universal density functional}
The Hohenberg-Kohn theorem establishes DFT with the density $\rho$ as the fundamental variable to describe many-electron systems. We thus can replace the wave function in the variational principle, cf. Eq.~\eqref{eq:est:general:var_principle_est}, by the density and write
\begin{align}
\label{eq:est:dft:var_principle}
	E_0=\inf_{\rho} E[\rho],
\end{align}
where $E[\rho]$ describes the energy of the system with Hamiltonian~\eqref{eq:est:dft:hamiltonian} as a functional of the density $\rho$. We can further specify this expression, because also the ground state wave function $\Psi=\Psi[\rho]$ is uniquely connected to $\rho$ and thus we have
\begin{align}
\label{eq:est:dft:energy_functional}
\begin{split}
E[\rho]=&\braket{\Psi[\rho]|\hat{H}\Psi[\rho]}\\
			=&\braket{\Psi[\rho]|\hat{T}\Psi[\rho]} + \braket{\Psi[\rho]|\hat{V}\Psi[\rho]} + \braket{\Psi[\rho]|\hat{W}\Psi[\rho]}\\
			=& T[\rho]+ \int\td\br v(\br)\rho(\br)  + W[\rho].
\end{split}
\end{align}
We used here that the potential energy $\braket{\hat{V}}=\int\td\br v(\br)\rho(\br)$ can be explicitly written in terms of $\rho$.
The remaining part of $E[\rho]$ is called the \emph{universal functional}\footnote{For details on mathematical issues of the universal functional as defined by Hohenberg and Kohn and an alternative more general definition, the reader is referred to the publications by \citet{Levy1979} and \citet{Lieb1983}.}
\begin{align}
\label{eq:est:dft:universal_functional}
	F[\rho]= \hat{T}[\rho] + \hat{W}[\rho].
\end{align}
This reformulation has however a \emph{fundamental problem}:
Although the Hohenberg-Kohn theorem proves the existence of $F$, it does \emph{not provide an explicit expression} for it, because the proof is not constructive (see Sec.~\ref{sec:qed_est:rdms:qed_rdmft}). As a matter of fact, most of the research in the field of DFT aims on understanding how this functional looks like. This is a very difficult task and only many years of research lead to the remarkable accuracy of state-of-the-art DFT. 

\subsubsection*{The Kohn-Sham construction}
The first crucial step toward a useful density-functional is the KS construction~\citep{Sham1966} that makes use of the increased flexibility of the density description. The basic idea is to employ as in HF, a non-interacting wave function, the KS wave function $\Psi^s$ (we denote all quantities in the auxiliary system with an s).  We have seen in HF theory, that this is a very good starting point to describe the interacting problem and at the same time, a single Slater determinant can be calculated relatively cheaply by effective orbital equations. However, in contrast to HF theory, the fundamental variable in DFT is the density and not the wave function. Thus, the KS wave function is a purely auxiliary quantity that in principle is \emph{not related} to the many-body wave function of the system. The connection between the non-interacting \emph{Kohn-Sham system} described by the Hamiltonian 
\begin{align}
	\hat{H}^s=\sum_{i=1}^{N}\hat{t}(\br_i)+v^s(\br_i)
\end{align}
and the physical system described by Hamiltonian~\eqref{eq:est:dft:hamiltonian} is given by their densities, i.e., we demand
\begin{align}
\label{eq:est:dft:KS_density_equality}
	\rho^s(\br)=\rho(\br).
\end{align}
Since the connection between density and potential is unique, i.e., there is one and only one interacting system with the density $\rho$, there is also one and only one non-interacting system with the density $\rho^s=\rho$. Thus, the Hohenberg-Kohn theorem also proves that the KS potential is unique.

\subsubsection*{Connecting the physical and the KS system}
The original construction due to Kohn and Sham~\citep{Sham1966} to connect the KS and the physical system is based on the energy. For that we define the Kohn-Sham energy (using Eq.~\eqref{eq:est:dft:KS_density_equality})
\begin{align}
\label{eq:est:dft:KS_energy}
	E^s[\rho]=T^s[\rho]+\int\td\br v^s(\br) \rho(\br)
\end{align}
and then re-express the physical (many-body) energy as
\begin{align}
\label{eq:est:dft:energy_functional_ehxc}
\begin{split}
	E[\rho]=& \int\td\br v(\br)\rho(\br) + F[\rho]\\
	=& T^s[\rho] + \int\td\br v(\br)\rho(\br) + F[\rho] - T^s[\rho]\\
	\equiv& T^s[\rho] + \int\td\br v(\br)\rho(\br) + E_{Hxc}[\rho].
\end{split}
\end{align}
This defines the Hartree-exchange-correlation energy 
\begin{align}
	E_{Hxc}[\rho]=& T[\rho]- T^s[\rho] + W[\rho] \nonumber\\
	=& E_{H}[\rho] + E_{xc}[\rho],
\end{align}
that usually is further divided in the mean-field or Hartree part $E_{H}[\rho]=\frac{1}{2}\int\td\br\br' \rho(\br)w(\br,\br')\rho(\br')$, that is an explicit functional of $\rho$, and the remaining unknown part $E_{xc}$, that is called \emph{exchange-correlation energy} (in analogy to the HF energy definitions, see Sec.~\ref{sec:est:hf}). If we assume that this functional is differentiable with respect to $\rho$-variations,\footnote{In fact, it can be shown that this functional is not differentiable. There are ways to circumvent this problem (see, e.g., \citep{Penz2019}), but from a mathematical point of view this issue is not yet resolved. In practice, this seems to play only a secondary role. However, one should not underrate such mathematical issues. Many advances in DFT were in fact triggered by mathematical research such as the Levy-Lieb constrained search formulation~\citep{Levy1979,Levy1982,Lieb1983}. For a good introduction in this topic from a physicist's perspective, the reader is referred to the respective section in the textbook by \citet{Dreizler1990}.}
we can define the ground state by the stationarity condition
\begin{align*}
	0&=\frac{\delta E[\rho]}{\delta \rho}\\
	&= \frac{\delta T^s[\rho]}{\delta \rho} + v(\br) + \frac{\delta E_{H}[\rho]}{\delta \rho} + \frac{\delta E_{xc}[\rho]}{\delta \rho}.
\end{align*}
The first term
\begin{align*}
	\frac{\delta T^s[\rho]}{\delta \rho}= -v^s[\rho]
\end{align*}
can be calculated by Eq.~\eqref{eq:est:dft:KS_energy} (see \citep[part 4.1]{Dreizler1990} for details), the third term is just the Hartree-potential identical to the definition of HF theory,
\begin{align}
\tag{cf. \ref{eq:est:hf:potential_hartree}}
	v_H[\rho]= \int\td\br' \rho(\br')w(\br,\br'),
\end{align}
and the last term defines the unknown \emph{exchange-correlation potential}
\begin{align}
	v_{xc}[\rho]= \frac{\delta E_{Hxc}[\rho]}{\delta \rho}.
\end{align}
It follows that the potential of the auxiliary system is
\begin{align}
	v^s[\rho](\br)= v(\br) + 	v_H[\rho](\br) + v_{xc}[\rho](\br).
\end{align}
Under our assumption that $E[\rho]$ is differentiable, this relation uniquely defines the KS system. We want to remark at this point that there are alternative constructions to define the KS system avoid certain mathematical subtleties. For instance via the force-balance equations~\citep{Tchenkoue2019} or by considering more general definition spaces~\citep{Penz2019}.

\subsubsection*{The KS equations}
Since $\hat{H}^s$ does not contain interaction terms, its ground state is (assuming no degeneracy\footnote{This assumption makes the derivations in the following simpler, but it is not necessary for the KS construction. For details, see Ref.~\citep{Dreizler1990}.}) described by one Slater determinant 
\begin{align}
\Psi^s(\bx_1,...,\bx_N)=|\psi_1 \cdots \psi_N |_-,
\end{align}
which can be determined by a set of orbital equations
\begin{align}
\left[ \hat{t}(\br) + v^s(\br)\right]\psi_i(\bx) =\epsilon_i\psi_i(\bx).
\end{align}
These are called the \emph{KS equations}. We see that in contrast to HF theory, the KS equations do not include a nonlocal term like the Exchange operator (cf. Eq.\eqref{eq:est:hf:exchange_operator}) and are thus formally simpler. However, approximations to the unknown exchange correlation potential usually have strong nonlinear dependencies on the density, or derivatives of the density. A very sophisticated class of approximate functionals, so-called \emph{hybrids}, even lift the assumption of a local $v^s$ completely and include the HF exchange energy.\footnote{See for example the review by \citet{Perdew2003} for an overview on the main classes of known functionals.}
Nevertheless, the KS equations define a nonlinear eigenvalue problem for the effective one-body Hamiltonian $\hat{H}^s$. This is very similar to HF theory and accordingly the solution strategies are basically the same.\footnote{Although the locality of most exchange-correlation potentials makes a large difference in practical calculations. Especially, KS-DFT can be applied to condensed matter systems considerably more easily than HF.}
For example KS-DFT calculations also require an SCF procedure (see Sec.~\ref{sec:est:hf}) that is called the \emph{KS scheme}.

\subsubsection*{The quest for the universal functional: approximations to the exchange-correlation potential}
Although the KS scheme simplifies the functional construction, we have not gained much so far with respect to HF theory, because we do not know the form of the exchange-correlation functional. The formalism itself does not give us any hint, how $v_{xc}$ could look like and thus, there is literally nothing else than ``intuition'' or'' experience'' to construct functionals. The historically first functional has been proposed by Kohn and Sham in Ref.~\citep{Sham1966}, where they introduced the KS scheme. This \emph{local-density approximation} (LDA) is based on one of the few many-electron systems that can be solved analytically, i.e., the homogeneous electron gas (HEG). 
Importantly, the electron-density of the HEG $\rho^{HEG}$ is constant and thus, we can explicitly parametrize the total energy $E^{HEG}(\rho^{HEG})$ as a function of the density.\footnote{Note that $E^{HEG}(\rho^{HEG})$ is really a function and not a functional, because $\rho^{HEG}$ is just a number.} Utilizing definition~\eqref{eq:est:dft:energy_functional_ehxc}, we can then (analytically) derive the exchange-correlation energy expression~\citep{Parr1980}
\begin{align}
\label{eq:est:dft:LDA_xc_energy}
	E_{xc}^{HEG}(\rho)\equiv E_X^{LDA}[\rho]= -\frac{3}{4}\left(\frac{3}{\pi}\right)^{1/3} \int\td\br\rho^{4/3}(\br),
\end{align}
where we formally reintroduced the spatial dependence of $\rho(\br)=\rho^{HEG}=const.$ Additionally, we removed the index ``C,'' because a non-interacting system as the HEG per definition does not have correlation. This procedure can be generalized to the interacting homogeneous electron gas, yielding a functional $E_{xc}^{LDA}$ that includes correlation effects.\footnote{This model cannot be solved analytically anymore, but it still can be reduced to the numerical calculation of one integral in the many-body coordinate space. This can be done very efficiently by Monte-Carlo integration. For an exhaustive discussion on the homogeneous electron gas, the reader is referred to Ref.~\citep{Giuliani2005}.} 
The local-density approximation is then simply to apply $E_{xc}^{LDA}$ to systems that is not homogeneous, i.e., to consider $\rho=\rho(\br)$ as we have  already anticipated in Eq.~\eqref{eq:est:dft:LDA_xc_energy}. Despite its simplicity, the LDA is impressively accurate for a large range of systems, especially in the realm of condensed matter, where the LDA has a similar significance as HF for molecular systems. But also for certain classes of chemical systems, the LDA is very accurate and often also better than HF~\citep{Perdew2003}. However, the LDA is by far not sufficient to reach ``chemical accuracy,'' i.e., the accuracy required in electronic-structure calculations to make realistic predictions.\footnote{See for example the Nobel-lecture of \citet{Pople1999}.} This has only been achieved with the historically second class of functionals, which are the \emph{generalized gradient approximations} (GGAs). The GGA functionals  are conceptually a straightforward generalization of the LDA, because they consider besides $\rho$ also its gradient, i.e.,
\begin{align}
E_{xc}^{GGA}=E_{xc}^{GGA}[\rho,\nabla\rho].
\end{align}
However, from the definition of the LDA in 1966 until 1985, when \citet{Perdew1985} constructed the first GGA functional that provided significantly better results, passed \emph{almost 20 years}. This illustrates how difficult functional construction in DFT really is. However, once an accurate functional is constructed, its potential is enormous because the the KS equations that have to be solved remain basically the same. Thus it is probably no wonder that the publication of Perdew marks a turning point in the history of DFT. Since 1985, the number of proposed exchange-correlation functionals and publications has grown exponentially~\citep{Jones2015} and started the ``incredible success story''~\citep{Burke2012} of DFT. Modern functionals include, e.g., even higher derivatives of the density  (for example the \emph{meta-GGAs}) and the Fock or exchange contribution of HF (\emph{hybrid functionals})~\citep{Perdew2003,Dreizler1990}. Such theoretical ideas define the principal form of the functionals, but the precise contribution of the different theoretical levels usually cannot be predicted. Many successful KS-DFT functionals have thus been constructed by fitting to experimental data.\footnote{The interested reader is referred to the recent report by \citet{Medvedev2017} that highlights the history  of functional construction and reflects critically on the strong focus on fitting that has become common practice nowadays.} 

Despite these huge efforts and certain systematic improvements, it is difficult to compare and judge the performance of the many existing functionals. Theoretically, there is one universal functional and one could expect that the known functionals somehow ``converge'' into one direction, i.e., they approximate increasingly better the exact functional. However, the opposite seems to be true:  there is basically no functional that is accurate for all material classes, but instead, there are many specific problems that require specific functionals.\footnote{See for example the extensive assessment of density functionals in quantum chemistry by \citet{Mardirossian2017}.} Consequently, applying KS-DFT in practice is everything but easy and requires experience and a careful literature research. Discussing functional construction in more depth is beyond the scope of this text and thus, we refer the reader for further details to the already cited literature. To name a few further examples, the review of \citet{Jones2015} provides a very good introduction to the topic of density functionals including its very interesting history and the review by \citet{Burke2012} provides a perspective on the recent state of DFT.
 
As a last example for an exchange correlation functional, we want to mention the \emph{exact exchange} (EXX)  approximation which will play an important role in the generalization of DFT to coupled electron-photon systems (see Sec.~\ref{sec:qed_est:qedft}). The EXX approximation is exceptional in comparison to most other known KS-DFT functionals, because it does not rely on empirical (or in the special case of the LDA analytical) knowledge. The idea is to approximate the interacting system with a single-reference wave function exactly as in HF. However, the EXX functional is \emph{local}, which requires to solve an auxiliary equation inside the SCF routine. This is a non-unique problem and thus there are several possibilities to define this auxiliary equation. The most common way is the \emph{optimized effective potential} (OEP) approach together with the Krieger-Li-Iafrate (KLI) approximation~\citep{Dreizler1990}, but there are also other possible constructions~\citep{Ruggenthaler2009exx}. Importantly, the EXX functional is usually more accurate than HF but at the same time as generic. This makes it an important tool for developing first-principles methods in settings that are not covered by the electronic-structure Hamiltonian.
 
We conclude our brief survey on KS-DFT functionals with a small example that shows the significance of density functionals in modern material science and chemistry. Despite all the problems and difficulties with KS-DFT, accurate functionals are known nowadays and they are basically the only tool that exists to calculate the properties of many realistic matter systems. This is well-reflected in the extraordinary number of citations of the corresponding publications. The most famous example is the paper~\citep{Perdew1996} in which Perdew, Burke and Enzerhof propose their PBE functional, which has become within merely two decades the \emph{most cited paper of all physics}. According to Google scholar, there are more than 110000 publications that refer to this publication.

\subsubsection*{A remark on the KS scheme and alternatives}
With the KS scheme, we have reduced the problem of calculating the $4N$-dimensional wave function on finding the $N$ lowest eigenstates of the nonlinear one-particle operator $\hat{H}^s$. This is on obvious computational advantage over the exponential wall of the many-body problem. Still, it is much more expensive than our starting point, i.e., a direct minimization of $E[\rho]$ (Eq.~~\eqref{eq:est:dft:var_principle}). In fact, people have tried to construct theories that are directly based on $\rho$ (nowadays known under the term \emph{orbital-free} DFT) even before DFT was developed. A famous example is \emph{Thomas-Fermi} theory,\footnote{See for example the review by \citet{Jones2015} and the references therein.} developed already in 1927, which provides however a very poor description of quantum systems. For example, matter is not stable in Thomas-Fermi theory~\citep{Lieb1981tf}. There have been advances in the field, but still the accuracy of known explicit density functionals is very low.\footnote{See Ref.~\citep{Karasiev2015} for a recent review on the topic.} 
The crucial point of the KS construction is that the contribution of the unknown $v_{xc}[\rho]$ to the total energy is \emph{considerably smaller} than the contribution of the universal functional $F[\rho]$ (cf. Eq.~\eqref{eq:est:dft:energy_functional}). The reason is that the Slater-determinant includes already a very large part of the quantum mechanical problem into our description (as we know from HF theory). Thus, DFT is almost exclusively utilized in the KS picture and often DFT and KS-DFT are used interchangeably.

\subsubsection*{When DFT fails: the strong-correlation regime}
In this last paragraph, we want to discuss the (actual) limitations of KS-DFT (and other single-reference methods). Although density functionals can in principle exactly describe every physical setting that is included in the electronic-structure Hamiltonian (Eq.~\eqref{eq:est:general:hamiltonian}), there are scenarios, where all known KS-DFT functionals are severely inaccurate. The research on functional construction has shown that in such settings, the form of the exact functional is very intricate. For instance, the exact functional of the non-interacting HEG is given by a formula as simple as Eq.~\eqref{eq:est:dft:LDA_xc_energy}. Systems that do not deviate too much from the HEG like typical conductors are usually very well described by KS-DFT. However, for other seemingly simple systems, such as two separate Hydrogen atoms, it turns our that the  exact functional has a very complicated form.

The principal reason for this difference is the single Slater determinant ansatz in the KS system that is a very good starting point for the HEG, but a very bad starting point for two separate atoms. The (spatial part of the) KS wave function for the ground state of two electrons reads simply
\begin{align}
\label{eq:esy:dft:wf_sc_KS}
	\Psi^s(\br_1,\br_2)=\psi(\br_1)\psi(\br_2),
\end{align}
where $\psi$ is the doubly-occupied spatial orbital of a spin-singlet.\footnote{Ground-states are normally singlets, i.e., spin-saturated.}
The exact ground state of two separate Hydrogen atoms after a proper symmetrization reads instead
\begin{align}
\label{eq:esy:dft:wf_sc_exct}
	\Psi(\br_1,\br_2)=& \tfrac{1}{\sqrt{2}}\left(\psi_a(\br_1)\psi_b(\br_2) +\psi_a(\br_2)\psi_b(\br_1)\right),
\end{align}
where $\psi_{a(b)}=\exp(-(r-R_{a(b)}))/\sqrt{\pi}$ is the ground state of the Hydrogen located at $R_a (R_b)$, where $|R_a-R_b|>> 1$ such that the orbitals do not overlap. Also $\Psi$ describes a spin-singlet,\footnote{Note that the antisymmetry is contained in the spin-function: $\tfrac{1}{\sqrt{2}} (\alpha(\sigma_1)\beta(\sigma_2)-\alpha(\sigma_2)\beta(\sigma_1))$.} but it requires two \emph{different} spatial orbitals, which are degenerate, i.e., they have the same energy eigenvalue.
From the wave function perspective, this special situation requires thus a multi-reference ansatz, i.e., a wave function constructed from more than one Slater determinant.
This is clearly an extreme case but the problem is quite generic: whenever the (valence) electrons of a system are sufficiently localized, we can assign orbitals to them that have only small overlaps and we recover a similar scenario as described by Eq.~\eqref{eq:esy:dft:wf_sc_exct}. This happens in chemistry, whenever orbital energies are (nearly) degenerate, e.g., when bonds are stretched~\citep{Chan2011}. Another example are materials with partially filled electron shells such as transition metals or their oxides~\citep{Georges1996}, which exhibit very special properties~\citep{Vollhardt2012} such as high-temperature superconductivity~\citep{Fujita2001}, Mott metal-insulator~\citep{Phillips2006} transitions or colossal magnetoresistance~\citep{Imada1998}.  We call such physical systems \emph{strongly correlated}\footnote{Note that in the realm of electronic-structure theory, one often calls the contribution of such near-degeneracies static correlation(see for instance Ref.~\citep{Scuseria2011}). Additionally, there is the dynamical contribution to the correlation energy that corrects the HF energy of a tightly bound electron pair such as in Helium~\citep{Mok1996}. This type of correlation can be usually very well described with single-reference methods such as KS-DFT.} and they constitute the hardest challenge for electronic-structure methods.\footnote{A description of the electronic-structure in full real-space is usually not possible for strongly-correlated systems. Instead, they are often described with effective models like the Hubbard model~\citep{Hubbard1963} (see also the discussion in Sec.~\ref{sec:intro:experiment_strong_coupling}) or a lattice of Hydrogen molecules with stretched bonds~\citep{Chan2011}. However, even such simplified models are extremely challenging for electronic-structure methods.} 


We have so far only discussed the wave-function perspective, where strong correlation manifests in the multi-reference character of the ansatz. In KS-DFT however, we describe every system with an single-reference ansatz of the form~\eqref{eq:esy:dft:wf_sc_KS}. The multi-reference character needs thus to be captured by the exchange correlation potential. In this simple case, the potential would need to create a kind of barrier between the two atoms, the so-called \emph{intra-system steepening}~\citep{Dimitrov2016}. This leads to one effective orbital of the form
\begin{align}
	\psi(\br)\approx \psi_a(\br) + \psi_b(\br),
\end{align}
which then is occupied twice. It is clear that the exact form of this potential is very sensitive to specific parameters such as the precise distance between the atoms and the orbitals that are involved. Consequently, it is very difficult to construct general exchange correlation potentials for strongly-correlated electrons.

However, the failure for the strongly correlated system that we just described is related to the KS construction and not to DFT per se.
Complementarily to the non-interacting KS system that is obtained by setting the interaction $W=0$ to zero in the Hamiltonian~\eqref{eq:est:general:hamiltonian}, there is the \emph{strong-correlation limit} with zero kinetic energy, $T=0$. For Coulomb systems, this limit can even be solved analytically and one can define a DFT based on this auxiliary system~\citep{Seidl1999,Gori-Giorgi2009}. This approach has several promising analytically derivable features, but the functional construction has not been very fruitful so far. As the long history of KS-DFT shows, functional construction is a very delicate task and one basically has to start from scratch in every new setting. In part~\ref{sec:dressed}, we will also make use of the flexibility of DFT and construct an auxiliary system for coupled electron-photon systems that is explicitly correlated. This unusual auxiliary system will, in contrast to strongly-correlated DFT, even facilitate functional construction.

Discussing strong correlation and the corresponding methods in more detail is clearly beyond the scope of this work and we instead refer the reader to the literature, e.g., Refs.~\citep{Altland2010,Chan2011,Phillips2006,LeBlanc2015,Georges1996}. However, we will discuss in the next section another approach to describe many-body problems that in fact has been used to construct methods that showed excellence accuracy in certain strongly-correlated systems. This approach will play an important role for the auxiliary construction in part~\ref{sec:dressed}.

	\section{Reduced density matrices in electronic-structure theory}
\label{sec:est:rdms}

To complete our discussion of electronic-structure methods, we will introduce in this section the concept of \emph{reduced density matrices} (RDMs). We can utilize these to describe many-body systems in an exact way without the explicit use of the wave function. 
For instance, we can describe the expectation value of the electronic-structure Hamiltonian exactly in terms of the 2-body RDM (2RDM). The configuration space of such a description does not grow with the particle number and thus a variational minimization in terms of the 2RDM is in principle feasible. However, the many-body problem also arises in this description in the form of conditions that determine the exact configuration space, i.e., the set of all 2RDMs. The number of these so-called $N$-representability conditions grows exponentially with the particle number $N$ and thus, one only can consider a subset of conditions in practice.

A special role is here taken by the 1-body RDM (1RDM) $\gamma$ that has comparatively simple $N$-representability conditions. These reflect the exchange symmetry of the particle species and thus provide an alternative to employing Slater-determinants to ensure, e.g., the Pauli-principle. 
Although $\gamma$ is not sufficient to describe the energy of a many-electron system in a linear way, it carries all information on the system by virtue of a generalized Hohenberg-Kohn theorem (Gilbert's theorem). This establishes RDMFT as an alternative to DFT, employing $\gamma$ instead of $\rho$ as the basic variable. 

Importantly, methods based on RDMs are usually more efficient than wave-function methods for describing strongly correlated systems beyond models~\citep{Mazziotti2007,Pernal2016}. 


\subsection{What are reduced density matrices?}
\label{sec:est:rdms:general}
To illustrate the role of RDMs, we consider a system consisting of two electrons that can move freely in some volume $V\subset \mathbb{R}^3$. The corresponding Hamiltonian is simply the kinetic energy operator $\hat{T}$. 
Let us now calculate the energy
\begin{align}
\label{eq:est:rdms:energy_box_2particles}
	E =& \braket{\Psi | \hat{T}\Psi} \nonumber\\
	=& \int\td\bx_1\td\bx_2  \Psi^*(\bx_1,\bx_2) \left[\sum_{i=1}^{2} -\tfrac{1}{2} \nabla^2_{\br_i}\right] \Psi(\bx_1,\bx_2) \nonumber\\
	=& \int\td\bx_1\td\bx_2  \Psi^*(\bx_1,\bx_2) \left[ -\tfrac{1}{2} \nabla^2_{\br_1}\right] \Psi(\bx_1,\bx_2) + \int\td\bx_1\td\bx_2  \Psi^*(\bx_1,\bx_2) \left[ -\tfrac{1}{2} \nabla^2_{\br_2}\right] \Psi(\bx_1,\bx_2).
\end{align}
We see that the energy expression separates in one term that is \emph{only} dependent on $\bx_1$ and another term that is \emph{only} dependent on $\bx_2$. However, we know that the indices are just place-holders and we can arbitrarily exchange them if we respect the exchange symmetry. If we exchange $\bx_1$ and $\bx_2$ in the second term, we get
\begin{align}
\label{eq:est:rdms:ExpValInvarianceParticleExchange}
	\int\td\bx_1\td\bx_2  \Psi^*(\bx_1,\bx_2) \left[ -\tfrac{1}{2} \nabla^2_{\br_2}\right] \Psi(\bx_1,\bx_2) =& - \int\td\bx_1\td\bx_2  \Psi^*(\bx_1,\bx_2) \left[ -\tfrac{1}{2} \nabla^2_{\br_2}\right] \Psi(\bx_2,\bx_1) \nonumber\\
	=& \int\td\bx_1\td\bx_2  \Psi^*(\bx_2,\bx_1) \left[ -\tfrac{1}{2} \nabla^2_{\br_2}\right] \Psi(\bx_2,\bx_1) \nonumber\\
	=& \int\td\bx_1\td\bx_2  \Psi^*(\bx_1,\bx_2) \left[ -\tfrac{1}{2} \nabla^2_{\br_1}\right] \Psi(\bx_1,\bx_2),
\end{align}
where in the last line we renamed the variables for our convenience (we also changed the index of the $\nabla$-operator). We see that exchanging the indices of the (antisymmetric) wave function \emph{within} an expectation value does not change its sign. Inserting this equality \eqref{eq:est:rdms:ExpValInvarianceParticleExchange} back in the energy expression \eqref{eq:est:rdms:energy_box_2particles}, we get the simplified expression,
\begin{align*}
	E=& \int\td\bx_1\td\bx_2  \Psi^*(\bx_1,\bx_2) \left[ -\tfrac{1}{2} \nabla^2_{\br_1}\right] \Psi(\bx_1,\bx_2) + \int\td\bx_1\td\bx_2  \Psi^*(\bx_1,\bx_2) \left[ -\tfrac{1}{2} \nabla^2_{\br_2}\right] \Psi(\bx_1,\bx_2) \nonumber\\
	=& \int\td\bx_1\td\bx_2  \Psi^*(\bx_1,\bx_2) \left[ -\tfrac{1}{2} \nabla^2_{\br_1}\right] \Psi(\bx_1,\bx_2) + \int\td\bx_1\td\bx_2  \Psi^*(\bx_1,\bx_2) \left[ -\tfrac{1}{2} \nabla^2_{\br_1}\right] \Psi(\bx_1,\bx_2)\\
	=& 2 \int\td\bx_1\td\bx_2  \Psi^*(\bx_1,\bx_2) \left[ -\tfrac{1}{2} \nabla^2_{\br_1}\right] \Psi(\bx_1,\bx_2).
\end{align*}
It is obvious that we can generalize this example for \emph{any} many-body operator of order one $\hat{O}= \sum_{i=1}^{N} \hat{o}(\br_i)$, cf. Eq.~\eqref{eq:est:general:DefOneBodyOperator}.
The expectation value of $\hat{O}$ reads for two particles
\begin{align}
\label{eq:est:rdms:ExpValOneBodyOperator}
	O =& \braket{\Psi |\hat{O}\Psi} \nonumber\\
	=& 2 \int\td\bx_1\td\bx_2  \Psi^*(\bx_1,\bx_2) \left[ \hat{o} (\br_1)\right] \Psi(\bx_1,\bx_2).
\end{align}
Importantly, in order to get from the first to the second line in this equation, we do not need to have \emph{any} knowledge about the system besides the exchange-symmetry of $\Psi$, the number of particles $N$ and the order of $\hat{O}$.  This is the \emph{key insight} to understand the role of RDMs: Although we have to introduce coordinates for all the $N$ particles of a system to define the quantum-mechanical wave function and corresponding operators, the \emph{expectation value} of any of these operators depends only on the \emph{order} of the operator. For operators of type \eqref{eq:est:general:DefOneBodyOperator}, we only need one spatial coordinate and this motivates the definition of the (spin-summed) \emph{one-body reduced density matrix} (1RDM)
\begin{align}
\label{eq:est:rdms:Def1RDM2particles}
	\Gamma^{(1)}(\br;\br') = 2 \sum_{\sigma_1,\sigma_2}\int\td\br_2  \Psi^*(\br',\sigma,\bx_2) \Psi(\br, \sigma,\bx_2).
\end{align}
If we know $\Gamma^{(1)}$, we can calculate any expectation value
\begin{align}
O = \int\td\br \hat{o}(\br) \Gamma^{(1)}(\br;\br')|_{\br'=\br},
\end{align}
where the subscript $|_{\br'=\br}$ means that we first apply $\hat{o}$ and then set $\br'=\br$. The 1RDM is exactly the part of the expectation value \eqref{eq:est:rdms:ExpValOneBodyOperator} that is \emph{independent} of the actual operator and in this sense, it carries all ``one-body information'' of a system.

\subsubsection*{The 1RDM with respect to local and nonlocal operators}
We want to remark briefly on the one-body nature of $\Gamma^{(1)}=\Gamma^{(1)}(\br;\br')$ that in fact depends on two coordinates. The definition \eqref{eq:est:rdms:Def1RDM2particles} explicitly differentiates between the coordinates of the wave function and its conjugate in the expectation value. This is crucial to calculate expectation values of so-called \emph{nonlocal} one-body operators like the kinetic energy 
\begin{align*}
 E = \int\td\br \hat{t}(\br) \Gamma^{(1)}(\br;\br')|_{\br'=\br}
 = 2 \sum_{\sigma_1}\int\td\br\td\bx_2  \Psi^*(\br'\sigma_1,\bx_2) \left[ -\tfrac{1}{2} \nabla^2_{\br}\right] \Psi(\br\sigma_1,\bx_2).
\end{align*}
The differential operator only acts on $\Psi$, but not on $\Psi^*$ and thus, we need to introduce a second coordinate $\br'$ that is not affected by $\hat{t}(\br)$. If we instead want to calculate the expectation value of the local potential $\hat{V}=\sum_{i=1}^{2} v(\br_i)$, we find
\begin{align*}
	V =& \braket{\Psi|\hat{V}|\Psi} \\
	=&  2 \int\td\bx_1\td\bx_2  \Psi^*(\bx_1,\bx_2) \left[ v(\br)\right] \Psi(\bx_1,\bx_2)\\
	=& \int\td\br \hat{o}(\br) \Gamma^{(1)}(\br;\br).
\end{align*}
This is nothing else than the one-body density,
\begin{align}
\Gamma^{(1)}(\br;\br) =\rho^{(1)}(\br) (\equiv\rho(\br)),
\end{align}
that we defined in the Sec.~\ref{sec:est:dft}. We see that the expectation values of local operators can be calculated with the one-body density, but for nonlocal operators, we need the full 1RDM.\footnotemark{} This seemingly subtle difference plays an important role in the many-body description. For instance, the exact kinetic energy is not a functional of the density and thus part of the universal functional $F[\rho]$ of DFT  (Eq.~\eqref{eq:est:dft:universal_functional}). In fact, this kinetic part of $F$ typically makes up the largest contribution and thus, approximations have to model especially this part. This insight is the basis of \emph{1RDM functional theory} (RDMFT), where the 1RDM is employed as basic variable instead of the density (see Sec.~\ref{sec:est:rdms:rdmft}). The unknown part of the corresponding universal function has thus a considerably smaller contribution to the total energy than in DFT.
\footnotetext{Note that the calculation of the kinetic energy expectation value does not require the full 1RDM. This can be illustrated in a discretized picture, where we approximate the Laplace operator with finite-differences of some order. The matrix form of the 1RDM becomes then explicit by, e.g., defining $G_{ij}=\Gamma^{(1)}(\br_i;\br_j)$. The density, i.e., the local part of $G$ is the diagonal $\rho=G_{ii}$ and to apply the Laplace operator, we would need the first $n$ off-diagonals, where $n$ is the order of the finite-differences approximation. Importantly, we do not need all off-diagonals and this is why the Laplace operator is called a \emph{semilocal} operator. This fact inspired the formulation of \emph{kinetic energy functional theory} in which the author was involved~\cite{Theophilou2018}.}

\subsubsection*{The hierarchy of RDMs}
Having motivated the definition of the 1RDM, let us now generalize the concept to systems with N particles and (non-)local operators of any order $p$.
Let 
\begin{align*}
\Psi(\bx_1,...,\bx_N)
\end{align*}
be the wave function of a system of $N$ electrons and
\begin{align}
	\Gamma^N(\bx_1,...,\bx_N;\bx_1',...,\bx_N')= \Psi^*(\bx_1',...,\bx_N')\Psi(\bx_1,...,\bx_N)
\end{align}
the corresponding ($N$-body) density matrix.
We define a general non-local operator of the order $p$
\begin{align}
\hat{O}^{(p)}_{nl} = \frac{1}{p!} \sum_{i_1,...,i_p} o{(p)}_{nl}(\br_{i_1},...,\br_{i_p};\br_{i_1}',...,\br_{i_p}'),
\end{align}
and a general local operator of the order $p$
\begin{align}
\hat{O}^{(p)}_{l} = \frac{1}{p!} \sum_{i_1,...,i_p} o{(p)}_{l}(\br_{i_1},...,\br_{i_p}).
\end{align} 
Accordingly, we define the \emph{$p$-body reduced density matrix} ($p$RDM)
\begin{empheq}[box=\fbox]{align}
\label{eq:est:rdms:pRDM_definition}
\begin{split}
\Gamma^{(p)}(\br_1,..,\br_p;\br_1',..,\br_p') =&
\frac{N!}{(N-p)!} \sum_{\sigma_1,...,\sigma_N}\int\td\br_{(N-p+1)}\cdots\td\br_{N}\\ &\Gamma^N(\br_1\sigma_1,...,\br_p\sigma_p,\bx_{p+1},...,\bx_N;\br_1'\sigma_1,...,\br_p'\sigma_p,\bx_{p+1},...,\bx_N).
\end{split}
\end{empheq}
Additionally, we denote the 1RDM in the following by
\begin{align}
\gamma(\br;\br')\equiv\Gamma^{(1)}(\br;\br')
\end{align}
because of its special importance. We call the diagonal of the $p$RDM,
\begin{align}
\rho^{(p)}(\br_1,..,\br_p) =& \Gamma^{(p)}(\br_1,..,\br_p;\br_1,..,\br_p),
\end{align}
the \emph{$p$-density}. Note that $\rho^{(1)}=\rho$, defined in Eq.~\eqref{eq:est:dft:one_density_definition}.
%
With these definition, the expectation value of any operator $\hat{O}$\footnote{Note that local operators are included in the definition of nonlocal operators. We explicitly differentiate between both to stress the difference between RDMs and densities.} with respect to $\Psi$ can be expressed as
\begin{align}
\label{eq:est:rdms:expectation_value_pure_states}
	O=&\braket{\Psi|\hat{O}\Psi} \nonumber\\
	 =&\int\td\br_1\cdots\td\br_N \hat{O}(\br_{i_1},...,\br_{i_p};\br_{i_1}',...,\br_{i_p}') \Gamma^N(\br_1,..,\br_N;\br_1',..,\br_N')|_{\br_1'=\br_1,\dotsc,\br_N'=\br_N}\\
	 \equiv& \Trace [\Gamma^N \hat{O}], \nonumber
\end{align}
where in the last line we formally rewrote the contraction as a \emph{trace} operation for later convenience.
This expression can be reduced for operators of the order $p$ with the aid of the RDMs. In the nonlocal case, we have for an operator $\hat{O}^{(p)}_{nl}$
\begin{align}
\label{eq:est:rdms:ExpValPBodyOperatorNonlocal}
O^{(p)}_{nl} =& \braket{\Psi | \hat{O}^{(p)}_{nl} \Psi} \nonumber\\
=& \int\td\br_1\cdots\td\br_p \hat{O}^{(p)}_{nl}(\br_{1},...,\br_{p};\br_{1}',...,\br_{p}') \Gamma^{(p)}(\br_1,..,\br_p;\br_1',..,\br_p')|_{\br_1'=\br_1,\dotsc,\br_N'=\br_N}\\
\equiv&\Trace[\hat{O}^{(p)}_{nl}\Gamma^{(p)}]. \nonumber
\end{align}
The expectation value of a local operator $\hat{O}^{(p)}_{l}$ is calculated as
\begin{align}
\label{eq:est:rdms:ExpValPBodyOperatorLocal}
O^{(p)}_{l} =& \braket{\Psi | \hat{O}^{(p)}_{l} \Psi} \nonumber\\
=& \int\td\br_1\cdots\td\br_p \hat{O}^{(p)}_{l}(\br_{1},...,\br_{p}) \rho^{(p)}(\br_1,..,\br_p)\\
\equiv&\Trace[\hat{O}^{(p)}_{l}\rho^{(p)}]. \nonumber
\end{align}
We want to stress that the above definitions are straightforward generalizations of our two-particle example and there is nothing new to understand here. We will need these expressions in the following for some theoretical considerations. In practice, we will mostly be confronted with operators of order 1 and only one local operator of order 2, which is the Coulomb interaction. See Tab. \ref{tab:est:ExamplesPBodyOperatorsRDMs} for an overview about these for us important cases.

From the normalization of $\Psi$, i.e., $\braket{\Psi|\Psi}=1$ we can derive also the normalization of the different $p$RDMs, which corresponds to the following sum rule
\begin{align}
\label{eq:est:rdms:sum_rule}
	\int\td\br_1\cdots\br_p \Gamma^{(p)}(\br_1,..,\br_p;\br_1,..,\br_p) = \frac{N!}{(N-p)!}.
\end{align}
Additionally, all the $p$RDM and the $(p+1)$RDM of one system are related by\footnote{This follows directly from the definition and the sum rule~\eqref{eq:est:rdms:sum_rule}.}
\begin{align}
\label{eq:est:rdms:rdm_connection_formula}
		\Gamma^{(p)}(\br_1',..,\br_p';\br_1,..,\br_p) = \frac{p+1}{N-p} \int\td\br_{p+1}\Gamma^{(p+1)}(\br_1',..,\br_p',\br_{p+1};\br_1,..,\br_p,\br_{p+1}).
\end{align}

We conclude this overview with a comment on the prefactors in Eq.~\ref{eq:est:rdms:pRDM_definition}, which essentially stem from the possible permutations of the coordinates. We chose them such that the expressions for the expectation value, \eqref{eq:est:rdms:ExpValPBodyOperatorNonlocal} and \eqref{eq:est:rdms:ExpValPBodyOperatorLocal} are prefactor free. Other definitions are possible\footnote{See for example the definitions in Ref.~\citep[part 2]{Bonitz2016}.} and employed in the literature.

\begin{table}
	\centering
	\begin{tabular}{lllll}
		\toprule
		operator class & example & & RDM & Expectation value\\
		\midrule
		1-body, local & local potential &$\hat{V}=\sum_{i=1}^{N} v(\br_i)$ & 1-density & $\braket{\hat{V}}=\int\td\br v(\br)\rho(\br)$	\\
		1-body, non-local & kinetic energy &$\hat{T}=\sum_{i=1}^{N} -\tfrac{1}{2} \nabla^2_{\br_i}$ & 1RDM & $\braket{\hat{T}}=-\tfrac{1}{2}\int\td\br  \nabla^2_{\br}\gamma(\br;\br')|_{\br'=\br}$ \\
		2-body, local & Coulomb interaction &$\hat{W}=\tfrac{1}{2}\sum_{i,j}\frac{1}{|\br-\br'|}$ & 2-density & $\braket{\hat{W}}=\tfrac{1}{2}\int\td\br\td\br' \frac{1}{|\br-\br'|}\rho^{(2)}(\br,\br')$
	\end{tabular}
\caption{Examples of operator classes with important examples and corresponding RDM. For instance, to calculate the expectation value of a local operator of order 1 like such as the local potential $v$, we need the 1-density.}
\label{tab:est:ExamplesPBodyOperatorsRDMs}
\end{table}

\subsubsection*{Coulson's Challenge: RDMs as basic variables}
Let us recall at this point the basic task of electronic-structure methods that is to describe the electronic state that minimizes the $N$-electron energy expectation value
\begin{align*}
	E_0=\inf E
\end{align*} 
We have introduced this energy functional in Sec.~\eqref{sec:est:general} in terms of the many-body wave function, i.e., 
\begin{align*}
	E[\Psi]=\braket{\Psi|\hat{H}\Psi}= \braket{\Psi|\hat{T}\Psi}+\braket{\Psi|\hat{V}\Psi}+\braket{\Psi|\hat{W}\Psi}.
\end{align*}
Then, we discussed in Sec.~\ref{sec:est:dft} that by means of the Hohenberg-Kohn theorem, we can identify $\Psi=\Psi[\rho]$ and thus reformulate the energy as a functional of $\rho$:
\begin{align*}
	E=E[\rho]
\end{align*}
However, from Eq.~\ref{eq:est:rdms:pRDM_definition} and Tab.~\ref{tab:est:ExamplesPBodyOperatorsRDMs} we can derive another form of the energy functional, expressed only with respect to RDMs, i.e.,
\begin{align}
\label{eq:est:rdms:energy_functional_rdms_exact}
	E=E[\rho,\gamma,\Gamma^{(2)}]=  T[\gamma]+V[\rho]+W[\rho^{(2)}].
\end{align}
Due to relation~\eqref{eq:est:rdms:rdm_connection_formula}, we can even rewrite the energy
\begin{align*}
	E=E[\Gamma^{(2)}]
\end{align*}
solely in terms of the 2RDM.\footnote{Note that the 2-density $\rho^{(2)}$ is not sufficient for such a re-expression, because the 1RDM cannot be connected to $\rho^{(2)}$ by Eq.~\eqref{eq:est:rdms:rdm_connection_formula}.}
This reformulation is exact and has no unknown part. In fact, this functional is even linear. It is thus very tempting to reformulate the variational principle (Eq.~\eqref{eq:est:general:var_principle_est}) as
\begin{align}
\label{eq:est:rdms:var_principle_v2RDM}
E_0=\inf_{\Gamma^{(2)}\in \mathcal{C}_{\Gamma}^2} E[\Gamma^{(2)}]
\end{align}
and determine $E_0$ by a functional variation over the space of 2RDMs $\mathcal{C}_{\Gamma}^2$. Since $\Gamma^{(2)}$ is a merely four-dimensional quantity (\emph{independently} on the number of electrons that are described) this functional minimization should be possible even for very large systems. Especially, $\Gamma^{(2)}$ is not limited by a single-reference construction and thus, it describes the properties of strongly-correlated systems exactly~\citep{Coleman2000}. This has been first pointed out in 1955 by \citet{Loewdin1955} and in the following years, \emph{variational 2RDM theory} became a highly popular research area~\citep{Coleman2000}. However, the variational calculations based on Eq.~\eqref{eq:est:rdms:var_principle_v2RDM} usually resulted in energies that were considerably smaller than the exact references. The reason was that it was not known how to characterize the configuration space $\mathcal{C}_{\Gamma}^2$. This led to violations fo the Pauli principle by bosonic contributions the energy. C.A. Coulson summarized the problem some years later in the following way:
\begin{aquote}{\citeauthor{Coulson1960}, \citeyear{Coulson1960} \citep{Coulson1960}}
	"It has frequently been pointed out that a conventional many-electron wave function tells us more than we need to know. [...] There is an instinctive feeling that matters such as electron correlation should show up in the two-particle density matrix [...] but we still do not know the conditions that must be satisfied by the density matrix. Until these conditions have been elucidated, it is going to be very difficult to make much progress along these lines."
\end{aquote}
This is what A. J. Coleman, one of the most important protagonists of the research area, later termed ``Coulson's challenge''~\citep{Coleman2000} and it has not been completely resolved until today. In 1995, the National Research Council of the USA named the conditions on the 2RDM, nowadays called \emph{$N$-representability conditions},\footnote{Coleman introduced this name, when he proved the first known set of conditions for $\gamma$~\citep{Coleman1963}. $N$ stands for the number of particles in the system.} as one of the ten most prominent research challenges in quantum chemistry~\citep{NRC1995}.

\subsubsection*{The $N$-representability problem}
Although we have discussed several minimization problems in this chapter, the problem of $N$-representability has not yet been occurred, at least not specifically. Strictly speaking, we could have called the task to parametrize of the space of antisymmetric wave functions, an $N$-representability problem. However, we were able to solve this problem by going to the special basis of Slater determinants (see Sec.~\ref{sec:est:general}).
The next parametrization that we have considered was for the space of single-particle densities to perform DFT minimizations. Clearly, not every function $f(\br)$ that depends on one variable is a one-body density according to Eq.~\ref{eq:est:dft:one_density_definition}.  Without being entirely conscious about this question, practitioners of the field used DFT for many years. However, no issues occurred. The reason is that the $N$-representability conditions of the density are very simple as \citet{Gilbert1975} could prove in 1975: every non-negative function that is finite (and thus can be normalized to $N$) is an $N$-representable density. Astonishingly, there is no trace of quantum effects like the exchange-symmetry or the Pauli principle in these conditions. The electron-density is basically equivalent to a classical charge density. This reflects one important advantage of the KS construction: it allows to include the antisymmetry of the electrons explicitly. In, e.g., orbital-free formulations of DFT this is much more difficult.\footnote{Note that besides the $N$-representability, there is the $v$-representability problem that concerns the more specific question, which densities (or RDMs) can be ``produced'' by physically meaningful potentials $v$. Importantly, it would be sufficient to determine all $v$-representable quantities, since these are automatically $N$-representable. The $v$-representability problem has stimulated many conceptual advances of DFT~\citep[part 2.3]{Dreizler1990} and RDMFT~\citep{Giesbertz2019,Benavides-Riveros2020}, but a direct application similar to the $N$-representability conditions is very difficult. See for example Ref.~\citep{Ayers2005,Verdozzi2008}.}

In this sense, the one-particle density is an exceptional RDM: the properties of all other RDMs are crucially influenced by the exchange symmetry of the system's particles. In fact, the antisymmetry plays the key-role in Coulson's challenge. To see this, let us recapitulate the definition of the $p$RDM
\begin{align}
\tag{cf. \ref{eq:est:rdms:pRDM_definition}}
\begin{split}
\Gamma^{(p)}(\br_1',..,\br_p';\br_1,..,\br_p) =&{\textstyle\binom{N}{N-p}} \sum_{\sigma_1,...,\sigma_N}\int\td\br_{(N-p+1)}\cdots\td\br_{N} \\ &\Psi^*(\br_1'\sigma_1,...,\br_p'\sigma_p,\bx_{p+1},...,\bx_N) \Psi(\br_1\sigma_1,...,\br_p\sigma_p,\bx_{p+1},...,\bx_N)\\
 =&\Gamma^{(p)}[\Psi],
\end{split}
\end{align}
that is a functional of $\Psi$ in the sense that for every $\Psi$, Eq.~\eqref{eq:est:rdms:pRDM_definition} defines the map
\begin{align}
	\Psi \rightarrow \Gamma^{(p)}.
\end{align}
The question of $N$-representability, or more precisely \emph{pure-state $N$-representability} concerns the other direction of this map,
\begin{align}
	\Gamma^{(p)}\rightarrow \Psi.
\end{align}
Given a function $g$ that depends on $2p$ coordinates, the pure-state conditions are necessary and sufficient for the existence of an $N$-body wave function $\Psi$ with $g=\Gamma^{(p)}[\Psi]$ according to Eq.~\eqref{eq:est:rdms:pRDM_definition}.
If there was no exchange symmetry, the integration in Eq.~\eqref{eq:est:rdms:pRDM_definition} would be trivial and the only $N$-representability condition would be the positivity (or positive semi-definiteness)\footnote{A $p$RDM  is positive-semidefinite, if for any wave function $\psi(\bx_1,...,\bx_p)$ it holds that $ \braket{\psi |\Gamma^{(p)} \psi}\geq 0$.} of $\Gamma^{(p)}$ that directly follows from the quadratic structure of the definition. However, the coordinates of a  (single-species) wave function are symmetric or antisymmetric under permutations, which is a crucial feature of the quantum mechanical description as we have discussed in detail in Sec.~\ref{sec:est:general}. 

The pure-state $N$-representability problem\footnote{$N$-representability plays also an important role in quantum information theory, where it is usually called the \emph{quantum marginal problem}.} has only been (formally) solved in its full generality for the 1RDM
\begin{align*}
\gamma(\br,\br') =& N \sum_{\sigma_1,...,\sigma_N}\int\td\br_{2}\cdots\td\br_{N} \Psi^*(\br_1'\sigma_1,\bx_{2},...,\bx_N) \Psi(\br_1\sigma_1,\bx_{2},...,\bx_N).
\end{align*}
\citet{Klyachko2006} published in 2006 a prescription to construct a set of conditions, known as the \emph{generalized Pauli constraints}~\citep{Altunbulak2008}, that guarantee the one-to-one correspondence between $\gamma$ and the (antisymmetric) many-body wave function. However, the number of these conditions grows exponentially with the particle number $N$ and the number of basis states $B$ and hence utilizing them in practice is basically impossible.\footnote{See for example the work by \citet{Theophilou2015}, where the application of the generalized Pauli constraints has been explored numerically.} There is ongoing research on approximation strategies that might result in numerically feasible methods yet only for very simplified model problems~\citep{Schilling2013,Schilling2019}.

\subsubsection*{Ensembles}
The whole problem can be significantly simplified, if we generalize our description from pure states to \emph{statistical ensembles} which occur, e.g., in the theory of \emph{open quantum systems}~\citep{Blum2012}. Such mixed states cannot be described by one wave function $\Psi$, but require the introduction of the \emph{(von-Neumann) density matrix}~\citep{Blum2012} 
\begin{align}
\label{eq:est:rdms:vonNeumann_density_matrix}
\Gamma_E^N = \sum_{i} w_i \Gamma_i^N,
\end{align}
where $\Gamma_i^N$ is the $N$-body density matrix corresponding to a pure state $\Psi_i$, $0\leq w_i\leq 1$ and $\sum_{i}w_i=1$. This defines the $\Gamma^N_E$ as a \emph{convex combinations} of the pure states $\Gamma^N_i$. $\Gamma$ describes the statistically mixed state with probability $w_i$ to find the system in the state $\Psi_i$. A specific example is the canonical ensemble of statistical physics with temperature $T$. The weight functions are then given as $w_i=\exp(- E_i/(k_B T))/Z$, where $Z=\sum_i \exp(- E_i/(k_B T))$ is the partition function $E_i$ is the energy expectation value of system $i$ and $k_B$ is the Boltzmann constant. We can calculate the expectation value of an $N$-body operator $\hat{O}$ in the ensemble by generalizing Eq.~\eqref{eq:est:rdms:expectation_value_pure_states}, i.e.,
\begin{align}
\label{eq:est:rdms:expectation_value_ensembles}
\begin{split}
	O=&\sum_i w_i \braket{\Psi_i|\hat{O}\Psi_i}\\
	=& \sum_i w_i \Trace[\Gamma^N_i\hat{O}]\\
	 \equiv&\Trace [\Gamma^N_E\hat{O}].
\end{split}
\end{align}
If we replace $\Gamma^N$ with $\Gamma^N_E$ in Eq.~\ref{eq:est:rdms:pRDM_definition}, we can straightforwardly generalize the concept of RDMs to ensembles. 

The advantage of the ensemble picture lies in the mathematical structure of Eq.~\eqref{eq:est:rdms:vonNeumann_density_matrix}. The set $\mathcal{C}_{\Gamma_E}^N\equiv\mathcal{E}^N$ of all ensemble $N$-body density matrices $\Gamma^N_E$ is convex and the pure state density matrices $\Gamma^N$ are simply its extreme elements. This follows directly from Eq.~\eqref{eq:est:rdms:vonNeumann_density_matrix} that defines any $\Gamma^N_E$ as a convex combination of the $\Gamma^N$.

Importantly, this property carries over to the sets $\mathcal{E}^p$ of the (ensemble) $p$RDMs, because all RDMs are linearly connected. This makes the parametrization of $\mathcal{E}^p$ considerably easier than the corresponding set of pure state $p$RDMs.\footnote{For a good introduction on the topic of convex sets in the realm of RDMs, the reader is referred to Ref.~\citep{Coleman1977}. Mathematical details are well-describe in Ref.~\citep{Giesbertz2010}.} 

\subsubsection*{Ensemble $N$-representability of the 1RDM}
Let us demonstrate this with the example of the 1RDM $N$-representability conditions that have been derived by \citet{Coleman1963} already in 1963. Given the 1RDM of a fermionic (bosonic)  $\gamma(\br;\br')=\sum_i n_i \phi_i'(\br')\phi_i(\br)$ in its diagonal representation, i.e.,  $\int\td\br'\gamma(\br,\br')\phi_i(\br')=n_i\phi_i(\br)$. Then the (necessary and sufficient) ensemble N-representability conditions are
\begin{subequations}
	\label{eq:est:rdms:Nrep_conditions}
\begin{align}
	&0\leq n_i (\leq 1) \label{eq:est:rdms:Nrep_conditions_a}\\
	&\sum_i n_i = N, \label{eq:est:rdms:Nrep_conditions_b}
\end{align}
\end{subequations}
where the upper bound holds only for fermions. We call $n_i$ and $\phi_i$ natural occupation numbers and natural orbitals respectively. This means, if we have a basis set with $B$ elements, we have exactly $B+1$ conditions on the natural occupation numbers, independently of the particle number $N$. This is enormous simplification in comparison to the generalized Pauli constraints for pure states makes the conditions~\eqref{eq:est:rdms:Nrep_conditions} applicable in practice. Additionally, the conditions have a direct physical interpretation, because we can connect them to the Pauli principle~\citep{Altunbulak2008}: bosonic natural orbitals can be occupied arbitrarily often, but fermionic natural occupation numbers are bounded by a maximal value of one. Here, the special case of $n_i=1$ for all $i\leq N$ and $n_j=0$ else corresponds to a single-reference wave function. Thus, non-integer occupations are a \emph{signature of the multi-reference character} (or the correlation) of a system.

There are several ways to prove the above statement~\citep{Coleman1963,Mazziotti2007,Giesbertz2019} and we want to outline one of them for the fermionic case. This very instructive proof has been published by \citet{Giesbertz2019} for the special case of finite basis sets. The crucial step in the proof is the explicit knowledge of the connection between $\gamma$ and $\Psi$ in the case, when $\Psi$ is a single Slater determinant. We assume a basis set of $B$ orbitals, from which we can choose $\binom{B}{N}$ different combinations to construct an $N$-body Slater determinant. We denote each of these sets with the collective index $I_k=(1_k,\dots,N_k)$, where $k=1,...,\binom{B}{N}$ and define $\Psi_{I_k}=1/\sqrt{N!}|\psi_{1_k}\cdots \psi_{N_k}|_-$. A simple calculation yields the corresponding 1RDM
\begin{align*}
	\gamma_{I_k}(\br;\br')=\sum_{i\in I_k} \psi_{i_k}^*(\br)\psi_{i_k}(\br).
\end{align*}
Thus, the natural orbitals $\phi_{i_k}=\psi_{i_k}$ are identical to the orbitals of $\Psi_{I_k}$ and all natural occupation numbers are one. This establishes for each Slater determinant $\Psi_{I_k}$ a bijective map (or a one-to-one correspondence) between the pure state $N$RDM $\Gamma_{I_k}^N=\Psi_{I_k}^*\Psi_{I_k}$ and its 1RDM $\gamma_{I_k}$
\begin{align}
\label{eq:est:rdms:Nrep_1RDM_Slater}
\gamma_{I_k} \overset{1:1}{\leftrightarrow} \Gamma_{I_k}^N.
\end{align}
Next, we note that the $\gamma_{I_k}$ must be the extreme elements of the space of ensemble 1RDMs $\mathcal{E}^1$, because we can construct any $\gamma\in \mathcal{E}^1$ by
\begin{align*}
	\gamma=\sum_k n_k \gamma_{I_k},
\end{align*}
if $0\leq n_k\leq 1$ and $\sum_k n_k=1$, i.e., by a convex combination of the $\gamma_{I_k}$. On the other hand, we can construct every ensemble $N$RDM
\begin{align*}
	\Gamma^N_E=\sum_k n_k \Gamma_{I_k}^N
\end{align*}
as a convex combination of the $N$-body density matrices $\Gamma_{I_k}^N$, since Slater determinants form a basis of the antisymmetric $N$-body space. Crucially, we can employ the \emph{same} prefactors $n_k$, because of Eq.~\eqref{eq:est:rdms:Nrep_1RDM_Slater} which completes the proof.

\subsubsection*{$N$-representability beyond the 1RDM}
Unfortunately, such a simple connection between RDM space and $\mathcal{E}^N$ is only possible for the 1RDM. To see this, let us recapitulate the essential ingredients of the proof: we need on the one hand the convex structure of the spaces of ensemble RDMs $\mathcal{E}^N,\mathcal{E}^1$ (and the linear connection between these spaces). On the other hand, the proof relies crucially on the properties of Slater determinants, which allow to parametrize the antisymmetric $N$-body space by single-particle wave functions, i.e., orbitals.  Slater determinants thus connect the $N$-body with the one-body space, which transfers to the RDM picture by connecting both the extreme elements of $\mathcal{E}^N$ \emph{and} $\mathcal{E}^1$. 

Generalizing this construction to, e.g., the 2RDM thus does not work. The eigenfunctions of the 2RDM, so-called \emph{geminals}, depend on two coordinates and there is no practical way known to construct an antisymmetric many-body wave function (or ensemble $N$-body density matrix) from geminals.\footnote{At least not without further approximations like the strong orthogonality assumption~\cite{Surjan2012}.} This is one way to understand the nowdays well-known fact that even in the ensemble case, the $N$-representability conditions are only simple for the 1RDM. In 1967, shortly after Coleman published his proof of the 1RDM conditions, \citet{Kummer1967} formally defined these conditions, but it took almost \emph{further 50 years}, until this formal solution could be translated into a practical prescription: \citet{Mazziotti2012} published the solution of the (ensemble) $N$-representability problem only in 2012. However, for all $p$RDMs with $p>1$, the number of conditions grows exponentially\footnote{In fact, the number of conditions grows even factorial, i.e., over-exponentially~\citep{Mazziotti2012details}.} and thus, they are not applicable in practice. The exponential wall of the many-body problem manifests thus in the number of $N$-representability conditions, instead of the dimensionality of the configuration space. Nevertheless, these many years of research have been fruitful and methods that only take a subset of all conditions into account have been successfully developed and implemented into quantum-chemistry codes~\citep{Mazziotti2007,Lackner2015,Fosso-Tande2016}. Such methods have proven to be capable to accurately describe comparatively large strongly-correlated systems.\footnote{For example, \citet{Fosso-Tande2016} applied their variational 2RDM method to a system of 50 strongly-correlated electrons in 50 orbitals, which is compatible with state-of-the-art methods.}

\subsection{One-body reduced density matrix functional theory}
\label{sec:est:rdms:rdmft}
Although $\gamma$ is not sufficient to describe an interacting many-body system in a straightforward linear way such as the 2RDM (Eq.~\eqref{eq:est:rdms:var_principle_v2RDM}), there exists an exact but \emph{nonlinear} energy functional
\begin{align}
	E=E[\gamma].
\end{align}
This was first realized by \citet{Gilbert1975}, who generalized the Hohenberg-Kohn theorem to this case in 1975 and established 1RDM functional theory (RDMFT). Exactly as in DFT, the exact functional $E[\gamma]$ is not known, but the contribution of the unknown part to $E[\gamma]$ is smaller than in DFT. This is obvious from the definition of the exact energy in terms of RDMs, i.e.,
\begin{align}
\tag{cf. \ref{eq:est:rdms:energy_functional_rdms_exact}}
	E=T[\gamma]+V[\rho]+W[\Gamma^{(2)}].
\end{align}
Since $\rho(\br)=\gamma(\br;\br)$ is included in $\gamma$, only the interaction contribution $W$ (and not $T+W$) must be approximated in RDMFT. Additionally, we can generalize the potential term
\begin{align}
	v(\br)\rightarrow v(\br,\br')
\end{align}
to nonlocal potentials such that
\begin{align}
V=\int\td\br v(\br,\br')\gamma(\br,\br')|_{\br'=\br}.
\end{align}
Nonlocal potential thus constitute the ``natural'' external partner to the internal variable $\gamma$ (see Eq.~\eqref{eq:est:dft:external_internal_correspondence} and the surrounding text). Although they do not have a direct physical interpretation, nonlocal potentials occur in some effective descriptions, e.g., pseudopotentials~\citep{Schwerdtfeger2011} and are thus a useful extension. The obvious price to pay for these advantages is the necessity to deal with the full 1RDM as basic variable instead of the simple density. Although not obvious from a first glance, the 1RDM severely complicates both, theory and numerics. 

This complication starts with the generalization of the Hohenberg-Kohn theorem. To make use of the simple $N$-representability conditions~\eqref{eq:est:rdms:Nrep_conditions}, Gilbert's theorem considers ensembles instead of pure states. This means that the many-body system is described by the ensemble $N$-body density matrix $\Gamma_E^N$ and the expectation value of an operator $\hat{O}$ is calculated by $O=\Trace[\Gamma_E^N\hat{O}]$, cf. Eq.\eqref{eq:est:rdms:expectation_value_ensembles}. Conceptually, this is not problematic, because the pure ground state of a system is included in the ensemble representation and thus by means of the variational principle
\begin{align}
\label{eq:est:rdms:var_principle_RDMFT}
	E_0=\inf_{\gamma\in \mathcal{G}_E^1} E[\gamma],
\end{align}
The ground state will be also the solution of a variation over ensembles.\footnote{Note that this is strictly only true for the exact functional and there are indications that the performance of approximate functionals can in some cases be increased by employing the pure state conditions~\citep{Theophilou2015}.}
Further, considering nonlocal potentials $v(\br;\br')$, there is \emph{no full one-to-one correspondence} to $\gamma(\br;\br')$: there are many potentials that lead to the same ground state 1RDM.\footnote{Note that the one-to-one correspondence is restored, if we consider the equilibrium states of grand-canonical ensembles instead of pure ground states~\citep{Giesbertz2019}.} This changes the mathematical details of the construction, but it does not prohibit the definition of the functional $E[\gamma]$, which only requires 
\begin{align}
	\Gamma^N\overset{1:1}{\longleftrightarrow}\gamma.
\end{align}
This is proven by Gilbert's theorem\footnote{See Sec.~\ref{sec:qed_est:rdms:qed_rdmft} for the explicit proof for the more general (but in terms of the proof analogous) case of coupled electron-photon systems.} and allows to define $\Gamma^N=\Gamma^N[\gamma]$ and thus $W=W[\Gamma^N]=W[\gamma]$. 

\subsubsection*{The exchange-correlation functional of RDMFT}
Obviously, the definition of $E[\gamma]$ is not sufficient for a practical electronic-structure method, but we have to find accurate approximate expressions for the unknown part $W[\gamma]$.
Although the 1RDM covers a larger part of the energy than the density in an exact way, it is the quality of the approximation of $W[\gamma]$ that matters. In analogy to DFT and HF, we can define as a first step
\begin{align}
	W[\gamma]=E_H[\gamma]+E_{xc}[\gamma],
\end{align}
where $E_H$ is the classical or Hartree contribution (Eq.~\eqref{eq:est:hf:energy_Hartree}) and $E_{xc}$ denotes the unknown exchange-correlation functional (that is different from $E_{xc}[\rho]$ in DFT, cf. Eq.~\eqref{eq:est:dft:energy_functional_ehxc}). One interesting difference to DFT is that the 1RDM is general enough to include exchange terms like in HF. Thus, HF theory is a special case of RDMFT with the functional
\begin{align}
\label{eq;est:rdms:RDMFT_functionals_HF}
	E_{xc}=E_{HF}= \int\td\br\td\br'\gamma(\br;\br')w(\br,\br')\gamma(\br',\br).
\end{align}
The challenge of RDMFT is thus to find functionals that go beyond $E_{HF}$. Unfortunately, $E_{HF}$ is the only known RDMFT functional that can be expressed explicitly in terms of $\gamma$, i.e., orbital-free. 
All practical RDMFT functionals are instead \emph{implicit} functionals, i.e., they can only be formulated in terms of the natural orbitals and natural occupation numbers,
\begin{align}
E_{xc}[\gamma]=E_{xc}[\phi_i,n_i],
\end{align}
where $\gamma=\sum_i n_i \phi_i^*\phi_i$.  In contrast to DFT, it is very difficult to connect this description in a useful way to a KS system, i.e., a non-interacting auxiliary system. This has very severe consequences: although we solve orbital equations in practical RDMFT algorithms, there is no clear underlying single-particle picture. For example, we cannot define an effective one-body Hamiltonian\footnote{Although we can formally define such a one-body Hamiltonian, it has been proven by \citet{Pernal2005} that the spectrum of this Hamiltonian is infinitely degenerate, if the functional $E_{xc}=E_{xc}[\phi_i,n_i]$ is implicit. This makes the one-body Hamiltonian in practice useless and one important consequence is that standard RDMFT algorithms are usually numerically considerably less efficient than DFT algorithms for the same number of orbitals.} which usually provides good approximations for ionization potentials\footnote{In HF, this is justified by \emph{Koopman's theorem}, which can be generalized (in a slightly modified form) to KS-DFT~\citep{Chong2002}.} and that can be diagonalized efficiently.\footnote{There have been efforts to define a local RDMFT~\citep{Lathiotakis2014} for which such a one-body Hamiltonian can be constructed. A first implementation showed promising results while being computationally considerably more efficient than standard (nonlocal) RDMFT.}

\subsubsection*{Functional construction in RDMFT}
Despite the additional challenges of RDMFT in comparison to DFT, accurate exchange-correlation functionals have been successfully constructed and implemented in electronic-structure codes. A comprehensive discussion of these is beyond the scope of this work (the reader is referred to Ref.~\citep{Pernal2016}), but we want to briefly highlight some properties of typical functionals that will play a role in the subsequent chapters.

The most important feature of the (approximate) RDMFT description is clearly its inherent multi-reference character, which is reflected in the (possible) non-integer occupation numbers. This is a considerable advantage of RDMFT over DFT, when it comes to the description of strong correlation. For instance, with modern RDMFT functionals certain dissociation processes have been described very accurately~\cite{Gritsenko2005} (and many more qualitatively) and the Mott-insulating phase of certain strongly-correlated solids could be predicted~\cite{Sharma2013}. 

Let us therefore briefly discuss the origin of the non-integer occupation numbers $0<n_i<1$ in RDMFT solutions. We know already one functional that leads to 1RDMs with only integer occupation numbers, i.e., the HF functional (Eq.\eqref{eq;est:rdms:RDMFT_functionals_HF}). In fact, it has been shown by \citet{Lieb1981} that this holds for \emph{any} $E_{xc}$ that only depends linear on the occupation numbers. To go beyond single-reference, we therefore need to employ $E_{xc}$ the depend \emph{nonlinearly} on the occupation numbers.\footnote{One way to explain, why it is not possible to define a one-body Hamiltonian in RDMFT is exactly this nonlinearity (see Ref.~\citep{Pernal2005} and Sec.~\ref{sec:numerics:rdmft:cg}).} 

Another very important feature of typical RDMFT functionals is that they are \emph{more generic} than, e.g., density functionals. Most DFT functionals are based on the paradigmatic homogeneous electron gas, which defines a very special physical setting. Also in RDMFT, there is such a paradigmatic study case, which is however much less specific: any two-electron wave function can be exactly parametrized in terms of the 1RDM. In this problem, only the particle number is specified, but the form of the Hamiltonian can be arbitrary. This is a considerable advantage, when we want to apply RDMFT to problems that are not as well studied as the electronic-structure Hamiltonian (Eq.~\eqref{eq:est:general:hamiltonian}). For example, to apply the LDA to a one-dimensional problem as we will do in Sec.~\ref{sec:est:comparison}, we cannot employ Eq.~\eqref{eq:est:dft:LDA_xc_energy}, which has been derived for the Coulomb interaction in 3d, but need a different functional form~\citep{Helbig2011}. Usual RDMFT functionals need instead some two-body matrix elements as input parameter, which can be calculated for an arbitrary type of interaction. For instance, we can apply the same functional in 1d or in 3d or in settings, where the Coulomb interaction is modified~\citep{Power1982}.

The parametrization of the two-electron problem in terms of natural orbitals and natural occupation numbers has been discovered already in 1956 by \citet{Lowdin1956}. 
The corresponding ``Löwdin-Shull'' (LS) exchange-correlation functional reads
\begin{align}
E_{xc}=E_{LS}=\min_{f_i,f_j}-\tfrac{1}{2}\sum_{i,j} f_if_j\sqrt{n_in_j}\braket{\phi_i\phi_j|\hat{w}|\phi_j\phi_i},
\end{align}
where $f_i=\pm 1$ are phase factors that have to be determined for every problem separately. $E_{LS}$ is (up to the phase factors) exact for two-electron systems and has been studied extensively in the last two decades to obtain a general understanding of the universal RDMFT functional~\citep{Pernal2016}. For larger electron numbers, the functional is not exact anymore and needs to be adopted. Especially the in practice quite difficult minimization over the phase factors is either completely removed or usually replaced by simple rules. 

In this work, we will explicitly consider the simplest version of the ``LS-type'' of functional, which is obtained by setting $f_i=1$ for all $i$. This functional has been constructed by \citet{Mueller1984} already in 1984 in a different context and without even a reference to the Löwdin-Shull construction. Later in 2002, when the interest in RDMFT had increased \citet{Buijse2002} rederived the Müller functional from more physical considerations\footnote{The Müller functional is thus often called BB functional.} and it provides a reasonable qualitative description of a large range of electronic-structure problems, including strongly-correlated systems. The Müller functional reads
\begin{align}
\label{eq:est:rdms:Mueller_functional}
E_{xc}=E_{M}=-\tfrac{1}{2}\sum_{i,j} \sqrt{n_in_j}\braket{\phi_i\phi_j|\hat{w}|\phi_j\phi_i}.
\end{align}
Nowadays, $E_M$ has become a reference in RDMFT, comparable to the LDA in DFT. A practical feature of $E_M$ that very few other RDMFT (and DFT) functionals share, is its \emph{convexity}, which guarantees a unique global minimum~\citep{Frank2007}. Thus $E_M$ is especially well suited to test new numerical implementations.

Another interesting property of the Müller and many other RDMFT functionals concerns the corresponding ground state energy, which have been observed to be a \emph{lower limit} to the exact reference~\cite{Goedecker1998,Gritsenko2005}.\footnote{The usual explanation for this is that the exchange-correlation functional needs to enforce (in some indirect way as a functional of the 1RDM) the $N$-representability conditions of the 2RDM. Approximate functionals often do not accomplish this and thus the variation is performed over a too large configuration space, which leads to lower energies than the exact reference.} Thus, energetic improvements have to be positive, which is in a sense the contrary to variational methods such as HF, where improvements are rigorously negative. This ``negative variational'' behavior is an important guiding principle for the RDMFT functional construction~\citep{Pernal2016}.

\subsubsection*{The RDMFT minimization}
In this final paragraph of the subsection, we will discuss the last missing piece to apply RDMFT in practice, which is the minimization algorithm.
We will employ a Lagrangian approach as in Sec.~\ref{sec:est:hf} (cf. Eq.~\eqref{eq:est:hf:energy} and the following paragraph) for HF,\footnote{We want to remind the reader that the HF algorithm can be almost directly transferred to KS-DFT problems, because both theories employ a single-reference picture. See Sec.~\ref{sec:est:dft}.} which for RDMFT is however more involved. We do not calculate the energy with respect to the $N$ orbitals of a Slater determinant, but with respect to $B$ natural orbitals $\{\phi_1,...,\phi_B\}$ and natural occupation numbers $\{n_1,\dotsc,n_B\}$, where $B\geq N$ depends on the system and is thus a convergence parameter. The generic energy functional of RDMFT reads
\begin{align}
\label{eq:est:rdms:rdmft:energy_functional}
E[\{\phi_i\},\{n_i\}]=& \sum_{i=1}^B n_i \braket{\phi_i|[\hat{t}+\hat{v}] \phi_i (\br)}
+ \tfrac{1}{2}\sum_{i,j=1}^B n_in_j \braket{\phi_i\phi_j|\hat{w}|\phi_i\phi_j}+E_{xc}[\{n_i\},\{\phi_i\}],
\end{align}
where $E_{xc}$ has to be replaced by a specific exchange-correlation functional such as $E_M$, cf. Eq.~\eqref{eq:est:rdms:Mueller_functional}. The goal is to minimize $E[\{\phi_i\},\{n_i\}]$ under the constraint that the natural orbitals are orthonormalized, i.e.,
\begin{align}
c_{ij}[\{\phi_j\}]=\int\td\br\phi_i^*(\br)\phi_j(\br) -\delta_{ij}=0,
\end{align}
which is similar to the constraint~\eqref{eq:est:hf:orthonormality_condition} for the HF orbitals. Additionally, we need to enforce the $N$-represent\-ability conditions \eqref{eq:est:rdms:Nrep_conditions} that guarantee that the $\{\phi_i\},\{n_i\}$ are connected to a fermionic 1RDM. The first condition on the individual eigenvalues, cf. Eq.~\eqref{eq:est:rdms:Nrep_conditions_a}, can be incorporated explicitly by, e.g., the substitution
\begin{align}
	n_i=2\sin^2(2\pi\theta_i),
\end{align}
allowing for a free minimization of the $\theta_i$. However, the second condition (Eq.~\eqref{eq:est:rdms:Nrep_conditions_b}),
\begin{align}
	\mathcal{S}[\{n_i\}]=\sum_i^B n_i - N=0,
\end{align}
needs to be enforced during the minimization. To accomplish this, we introduce the Lagrange multipliers $\epsilon_{ij}$ and $\mu$ and consider the Lagrangian
\begin{align}
L[\{\phi_i\},\{\theta_i\};\{\epsilon_{ij}\},\mu]&=E[\{\phi_i\},\{n_i\}]-\mu \mathcal{S}[\{\theta_i\}]-\sum_{i,j}\epsilon_{ij} c_{ij}[\{\phi_i\},\{\phi_j\}].
\end{align}
To find the RDMFT ground state defined by the variational principle, cf. Eq.~\eqref{eq:est:rdms:var_principle_RDMFT}, we optimize \newline $L[\{\phi_i\},\{\theta_i\};\{\epsilon_{ij}\},\mu]$ with respect to all variables. A necessary condition for an optimum of $L$ is stationarity
\begin{align}
\label{eq:est:rdms:rdmft_stationarity}
\delta L=0,
\end{align}
which is also sufficient for a minimum, if a convex functional such as $E_M$ is employed. If we consider $\phi_i$ and $\phi_i^*$ as independent, we find 
\begin{align*}
0=&\delta L \nonumber\\
=& \sum_{i=1}^{B}\frac{\delta L}{\delta \phi_i^*} \delta \phi_i^* + \sum_{i=1}^{B}\frac{\delta L}{\delta \phi_i} \delta \phi_i + \sum_{i=1}^{B}\frac{\partial L}{\partial n_i} \td n_i \nonumber\\
=&8\pi\sum_{i=1}^{B}\sin(\theta_i)\left[\frac{\partial E}{\partial n_i}-\mu\right] \td\theta_i +\sum_{i=1}^{B}\int\td^3r\, \delta\phi_i^*(\br)\left[\frac{\delta E}{\delta \phi_i^*(\br)}-\sum_{k=1}^{B}\epsilon_{ki}\phi_k(\br)\right] \nonumber\\ &+\sum_{i=1}^{B}\int\td^3r\, \left[\frac{\delta E}{\delta \phi_i(\br)}-\sum_{k=1}^{B}\epsilon_{ik}\phi_k^*(\br)\right] \delta\phi_i(\br),
\end{align*}
which leads to the three sets of coupled equations
\begin{subequations}
	\label{eq:est:rdms:rdmft:equations}
	\begin{align}
	0=&\frac{\partial E}{\partial n_i}-\mu \label{eq:est:rdms:ELE_n}\\
	0=&\frac{\delta E}{\delta \phi_i^*(\br)}-\sum_{k=1}^{B}\epsilon_{ki}\phi_k(\br) \label{eq:est:rdms:ELE_phi_star}\\
	0=&\frac{\delta E}{\delta \phi_i(\br)}-\sum_{k=1}^{B}\epsilon_{ik}\phi_k^*(\br) \label{eq:est:rdms:ELE_phi}.
	\end{align}
\end{subequations}
In practice, we will thus need a self-consistent procedure, in which we solve alternately the equations for the $n_i$ and for the $\phi_i$, keeping the other variables constant, respectively. Since the $n_i$ are just numbers, their optimization can usually be done with a routine from a standard library.\footnote{Nevertheless, the equations are nonlinear and thus also here caution is in order. In Sec.~\ref{sec:numerics:dressed:non_assessment}, we present an explicit example that demonstrates the challenges of such nonlinearities.} For the orbital optimization, we have a nonlinear operator equation similar to the HF equations and thus, an SCF procedure is required. Similarly to HF, where the stationarity condition defines the Fock operator (Eq.~\eqref{eq:est:hf:Fock_operator}),  Eq.~\eqref{eq:est:rdms:ELE_phi_star} and Eq.~\eqref{eq:est:rdms:ELE_phi} define a nonlinear one-body operator $\hat{H}^{(1)}$. For Eq.~\eqref{eq:est:rdms:ELE_phi_star}, we have
\begin{align}
	\frac{\delta E}{\delta \phi_i^*(\br)}=\hat{H}^{(1)}\phi_i=n_i[\hat{t}+\hat{v}] \phi_i + n_i \hat{v}_H \phi_i + \frac{\delta E_{xc}}{\delta \phi_i^*}.
\end{align}
Importantly, this operator is \emph{not hermitian} and thus cannot be interpreted as an effective one-body Hamiltonian~\citep{Pernal2005}.\footnote{We will discuss this with the concrete example of the Müller in Sec.~\ref{sec:numerics:rdmft:piris}.} An important practical consequence of this fact is that we cannot transform Eq.~\eqref{eq:est:rdms:ELE_phi_star} (or equivalently Eq.~\eqref{eq:est:rdms:ELE_phi}) into a nonlinear eigenvalue equation as we have done in Sec.~\ref{sec:est:hf} to derive the HF equations, cf. Eq.~\eqref{eq:est:hf:HF_equations}. This means that we have to determine all $B^2$ components of the Lagrange multiplier matrix $(\epsilon_{ij})$. We will discuss how to accomplish this in practice in Sec.~\ref{sec:numerics:rdmft}.

It should be stressed that usual RDMFT minimization algorithms are not only more expensive than DFT calculations, but also considerably less stable, which is probably the most important bottleneck of state-of-the-art RDMFT. The reasons for the unsatisfactory convergence of the proposed algorithms are not entirely resolved and thus numerics is an especially important part of the actual research in the field~\cite[part 4]{Pernal2016}. 

Despite all the challenges of the 1RDM as a basic variable, one should not forget that RDMFT is much younger than DFT and investigated by a considerably smaller community. Thus, many promising research directions have only been identified but not yet fully investigated~\citep{Pernal2016} and further significant improvements of the theory are quite probable. Especially, when it comes to entirely new types of problems like the accurate description of the electron-photon interaction, RDMFT could be an interesting starting point, because the contribution of the unknown part of the universal functional is smaller than it is the case in DFT. We will come back to this idea briefly in Sec.~\ref{sec:qed_est:rdms:qed_rdmft} and more specifically in Sec.~\ref{sec:dressed:est:RDMFT}.

	\section{Comparison of the methods}
\label{sec:est:comparison}
\begin{figure}[ht]
	\centering
	\includegraphics[width=0.49\columnwidth] {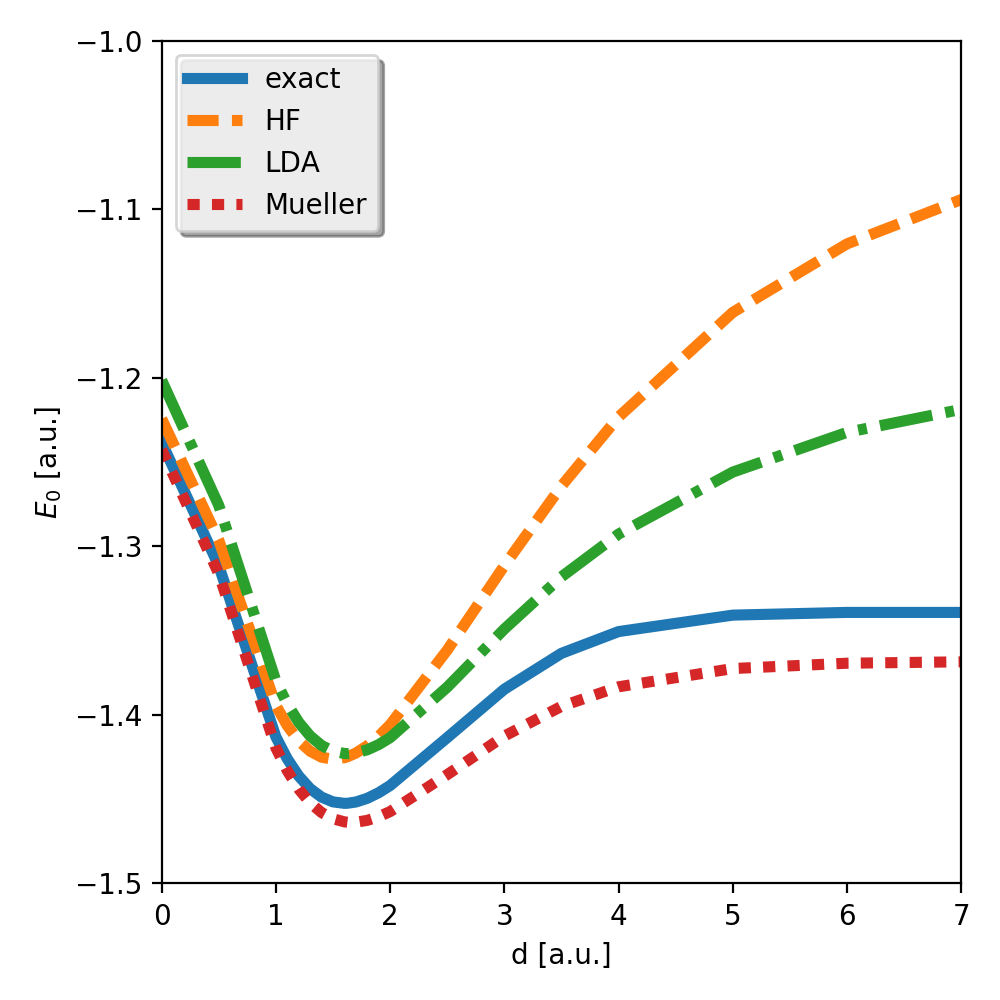}
	\caption{Ground-state potential energy surface $E_0(d)$ of the 1d $H_2$ model (Eq.~\eqref{eq:est:comp:Hamiltonian_H2}), calculated exactly (exact) and with three different electronic-structure methods: HF, KS-DFT with the LDA functional (LDA), and RDMFT with the Müller functional (Mueller). The deviations between the different methods are discussed in detail in the text.}
	\label{fig:est:H2_dissociation}
\end{figure}
To get a feeling for the different approximation schemes that we have defined in the previous sections, we want to conclude our survey on electronic-structure theory with a simple example. We consider a 1d model of a hydrogen molecule ($H_2$) that consists of two hydrogen atoms with distance or \emph{bond length} $2d$. We can describe this scenario by the local potential
\begin{align}
\label{eq:est:comparison:potential_H2}
v_{d}(x) =-\frac{1}{\sqrt{(x-d)^2+1}} -\frac{1}{\sqrt{(x+d)^2+1}},
\end{align} 
where $-1/\sqrt{(x-d)^2+\epsilon^2}$ with $\epsilon=1$\footnote{For nuclei with a bigger charge, the soft Coulomb parameter $\epsilon$ needs to be adjusted to guarantee certain properties. We will make use of this in Sec.~\ref{sec:dressed:results}.} is the ``soft Coulomb potential'' of one elementary charge. The soft Coulomb potential is usually employ in 1d studies, because it resembles many essential features of the standard 3d Coulomb potential~\citep{Helbig2011}. Accordingly, we also approximate the interaction $w$ by a soft Coulomb expression 
\begin{align}
\label{eq:est:comparison:interaction_soft_coulomb}
w(x,x') = 1/\sqrt{|x-x'|^2+1}.
\end{align} 
The many-body Hamiltonian of this problem reads
\begin{align}
\label{eq:est:comp:Hamiltonian_H2}
	\hat{H}=&\hat{T}+\hat{V}+\hat{W} \nonumber\\
	=&\sum_{i=1}^{2}\left[-\tfrac{1}{2}\tfrac{\partial^2}{\partial x_i^2} + v_d(x_i)\right] + w(x_1,x_2)
\end{align}
We now can apply any of our discussed methods to approximate the ground state of $\hat{H}$ for different values of $d$. The resulting energy function $E_0(d)$ is called the \emph{ground-state potential energy surface} (ground-state PES), which plays a central role in electronic-structure theory. One of its most important applications is the structure prediction of matter systems. In our example, this reduces to the \emph{(equilibrium) bond-length} of the $H_2$ molecule, which is simply $2 d_{min}$, where $d_{min}$ is the minimum of $E_0(d)$.\footnote{This can be understood by a simple gedanken experiment: without external driving, any initial configuration of nuclei will relax to the configuration with minimal energy.}  
But this is only one application of the ground state PES and there are many more. For instance, the shape of $E_0(d)$ around the minimum provides information about the energetic costs of \emph{bond stretching} and the large $d$ limit describes the dissociated molecule. Thus, with the knowledge of the full function $E_0(d)$ we can understand complicated processes such as chemical reactions. This shows that there are research questions, that only require the knowledge of a small part of the ground-state PES and consequently methods that are accurate in this part. But there are other problems, such as the description of the complete process of a dissociation, which necessitate methods that accurately describe a large part of the ground-state PES.

Let us see now how the methods that we have discussed in this chapter perform in describing the ground-state PES $E_0(d)$ of the 1d $H_2$ model. Since the Hamiltonian~\eqref{eq:est:comp:Hamiltonian_H2} is very low-dimensional, we can calculate the exact many-body wave function with a simple eigensolver even for large $d$. This makes the model an ideal and often employed system to test the accuracy of new electronic-structure methods~\citep{Helbig2011,Mordovina2019a}. We will consider a generalized from of this model in part~\ref{sec:dressed} to test our new methods for coupled electron-photon problems. 

In this section, we compare three levels of theory, that is HF, and one exemplary functional of KS-DFT and RDMFT, respectively. For the former, we employ a 1d version of the paradigmatic LDA~\citep{Helbig2011} and for the latter, we choose the Müller functional. All calculations are performed in real space in a box with length $L=30$ a.u. and discretized with a spacing of $d_x=0.1$ a.u. In RDMFT, we have obtained converged results with $M=20$ natural orbitals.\footnote{See App.~\ref{sec:numerics:dressed:convergence:RDMFT} for the details on the convergence of RDMFT calculations.} We have plotted $E_0(d)$ for the three cases together with the exact reference in Fig.~\ref{fig:est:H2_dissociation}. 

A first glance on the figure reveals already how challenging an accurate description of the electronic-structure is: there is \emph{no method} that performs well over the whole range of bond lengths. Clearly, there are more sophisticated functionals for DFT and RDMFT that perform much better than our chosen examples, but $H_2$ is also just a very simple problem. The accurate description of the ground-state PES over the whole range of bond lengths is for many systems simply not possible. Methods that are accurate enough are usually numerically too expensive and the most efficient methods can often only describe a small part of the full ground-state PES. 
Close to the equilibrium position, single-reference methods are very accurate with HF providing a reasonable qualitative description\footnote{One of the most precise electronic-structure methods for this regime is coupled cluster theory~\citep{Bartlett2007}, which systematically improves upon the HF ansatz and usually is significantly more accurate then typical DFT functionals. However, coupled cluster is also considerably more expensive than DFT.} and KS-DFT leading the way in terms of efficiency. However, methods based on only one Slater determinant become usually inaccurate for large bond-lengths. We have discussed this in the last paragraph of Sec.~\ref{sec:est:dft} considering the dissociation limit $d\rightarrow\infty$, where we have to describe the two degenerate orbitals of the separate atoms. To describe this limit, we need two Slater determinants, which corresponds to a multi-reference wave function and makes the system strongly-correlated. 

In HF and KS-DFT, we try to describe this inherently multi-reference scenario with only one Slater determinant. We can observe the consequence of this approximation in Fig.~\ref{fig:est:H2_dissociation}: with increasing $d$, the $E^{HF/LDA}(d)$ increases constantly, introducing an artificial attraction. Thus, according to the single-reference description, the molecule would never dissociate but always feel a force that pushes the nuclei back to the equilibrium. The exact solution instead saturates at the \emph{dissociation plateau} with $E_{diss}\approx -1.34$ a.u., where the atoms do not interact anymore. However, since the LDA includes already some correlation, i.e, it corrects the approximation of considering only one Slater determinant, its ``tail'' is closer to the exact result than the one of HF. Around the equilibrium position, HF is however energetically superior to the LDA. Since the energy is an especially important quantity to understand chemical reactions (e.g., to calculate reaction barriers), LDA is not really useful for quantum chemistry.  This illustrates why DFT has only been accepted by the quantum chemistry community after the first generalized gradient approximation has been developed, which significantly increased the accuracy in such scenarios~\citep{Perdew2003}. State-of-art functionals are capable to reproduce the area around $d_{min}$ almost exactly for a large class of molecules~\citep{Mardirossian2017}. In contrast to its bad performance in terms of energetics, the LDA reproduces the correct equilibrium bond length of $d_{eq}=1.63$ a.u. almost exactly, whereas HF underestimates it with a value of $d_{eq}^{HF}\approx 1.5$ a.u. by about $6 \%$.

The most sophisticated method that we have employed in our comparison is RDMFT with the Müller functional and accordingly, $E_0^{\textnormal{Müller}}(d)$ is closest to the exact $E_0(d)$. Importantly, we see that $E_0^{\textnormal{Müller}}(d)$ does not increase arbitrarily with $d$, but saturates at a constant value of about $E_{diss}^{\textnormal{Müller}}=-1.37$. This is still significantly lower than the exact reference, but more sophisticated functionals describe this regime very accurately~\citep{Pernal2019}. This is possible due to the inherent multi-reference character of the formalism and it is the most important strength of RDMFT in general. The strongest limitation of state-of-the-art functionals is the accurate description of the region around the equilibrium~\citep{Pernal2016}. The Müller functional overestimates for our example the equilibrium bond length with $d_{min}^{\textnormal{Müller}}\approx 1.70$ a.u., which is only slightly better than HF. Although simple molecules like $H_2$ are described very well by improved functionals~\citep{Gritsenko2005,Piris2017}, more complicated electronic structures are often better approximated by state-of-the-art DFT functionals~\citep{Pernal2016}. As a last remark, we note that $E_0^{\textnormal{Müller}}(d)$ is always lower than the exact reference, which is a typical feature of many RDMFT functionals (see  Sec.~\ref{sec:est:rdms}).


	\clearpage
\chapter{Light and matter from first principles}
\label{sec:qed_est}

In the previous chapter, we have discussed the electronic many-body problem and presented a selection of  strategies from the big repertoire of state-of-the-art electronic-structure theory to deal with it. In the following, we generalize these strategies to the coupled electron-photon space. We will see that this is straightforward in the sense that all tools and concepts from electronic-structure theory have a clear counterpart in the coupled theory. Also all the features of the matter-only problem reappear, i.e., we will have to deal with the complexity of the many-electron-photon problem. For that, the particle-exchange symmetry plays an important role and we perform variational minimizations to find the ground state, which is in complete analogy to equilibrium electronic-structure theory. This requires to characterize the system's state and also here we can utilize the same concepts: the wave function, the electron density plus the corresponding photon quantity that is the displacement field and (generalized) RDMs.

However, the exploration of the coupled electron-photon space using these tools is considerably more difficult. One important reason is the enormous size of the corresponding Hilbert space that is a direct product of the electronic and photonic Hilbert (sub)spaces. Additionally there is no further symmetry restriction between these subspaces, e.g., there is no exchange symmetry between electronic and photonic coordinates. Thus, the exponential wall grows considerably ``faster'' here than in the separate theories, which in particular limits wave-function methods and the access to exact reference solutions. At the same time, simple approximation schemes, such as the generalization of the Hartree-Fock ansatz (the ``mean field'') do not account for quantum effects of the interaction between electrons and photons, i.e., there is no quantum-mechanical exchange. This is especially severe in equilibrium scenarios, because the mean-field (or classical) contribution of the electron-photon interaction is in many important cases trivially zero. Using the vocabulary of electronic-structure theory, there is only correlation between static electrons and photons.

To accomplish the task to generalize electronic-structure methods to the coupled setting, we are therefore confronted with two complications: the coupled electron-photon space 
is substantially larger than the electronic space alone. At the same time, the mean field description that is one of the most powerful tools of electronic-structure theory, is considerably less useful for coupled systems.\footnote{Note that in time-dependent scenarios, the mean-field approximation of the electron-photon interaction is capable to describe important effects, such as the formation of polaritons~\citep{Jestaedt2019,Flick2019}. Understanding such effects also in equilibrium scenarios is one important motivation for this work.} Thus, we have to approximate a larger configuration space with more expensive tools. A further substantial difference between the electronic and the coupled electron-photon problem is the new type of interaction operator in the latter case: whereas the Coulomb interaction acts as a 2-body operator that pairwise correlates all the particles, the electron-photon interaction acts as a so-called $3/2$-body operator (Sec.~\ref{sec:qed_est:rdms:general}). This shows up in the fact that the photon number of a system is a priori \emph{not determined} - in contrast to the electron number, which is defined by the physical problem. This has important consequences for the characterization of the photonic state, which manifest differently in the wave-function, density-functional or RDM description. 

The purpose of this chapter is thus twofold: on the one hand, we introduce the challenges of the first-principles description of the coupled electron-photon problem. We present already established approaches to deal with that and discuss their limitations. But we also comment on possible improvements and future directions. On the other hand, this analysis provides the background for part~\ref{sec:dressed}. We will connect there many of the here discussed issues and features of the electron-photon problem and introduce a new way to construct the many-body wave function. We will make use of the increased flexibility that the coupled electron-photon Hilbert space offers and employ electron-photon quasi-particles, i.e., polaritons to construct a many-polariton space. This allows to generalize the concepts of electronic-structure theory in yet another way to the coupled electron-photon problem and overcomes several fundamental issues of the straightforward generalizations that we present in this chapter.

	\newpage
\section{The many-electron-photon space}
\label{sec:qed_est:general}
In analogy to Sec.~\ref{sec:est:general}, let us start the discussion with the proper definition of the problem that we aim to solve. We consider the cavity-QED setting (see  Sec.~\ref{sec:intro:nrqed_longwavelength}) that is described by the Hamiltonian (Def.~\ref{def:cavity-QED_Hamiltonian})
\begin{align}
\label{eq:qed_est:general:Hamiltonian}
\hat{H}=\sum_i^{N} \left[\frac{1}{2} \nabla_{\br_i}^2 + v(\br_i)\right] + \frac{1}{2}\sum_{i\neq j}^{N} \frac{1}{|\br_i-\br_j|}+ \frac{1}{2}\sum_{\alpha}^{M}\left[-\partial_{p_{\alpha}}^2 + \omega_{\alpha}^2 \left(p_{\alpha} + \frac{\blambda_{\alpha}}{\omega_{\alpha}}\cdot \bR \right)^2\right],
\end{align}
For the  matter part of the system, this corresponds to the electronic structure Hamiltonian (Eq.~\eqref{eq:est:general:hamiltonian}) and for the photon part, we consider $M$ modes (with known mode functions) that are coupled to the total dipole of the electrons.
For each mode $\alpha$, $\omega_{\alpha}$ denotes the frequency  and $\blambda_{\alpha}=\sqrt{4\pi} S_{\alpha}(0) \bepsilon_{\alpha}$ the coupling vector. Here, $\bepsilon_{\alpha}$ is the polarization direction of the mode and $S_{\alpha}(\br_0)\propto1/\sqrt{V}$  is the mode function at the center of charge $\br_0$ (see Def.~\ref{def:cavity-QED_Hamiltonian}). To control the electron-photon coupling strength, we regard the absolute value of $\lambda_{\alpha}=|\blambda_{\alpha}|=\sqrt{4\pi} S_{\alpha}(0)$ as a tunable parameter.\footnote{This is justified, because via $S_{\alpha}(0)\propto1/\sqrt{V}$ enters the effective mode volume $V$ into the Hamiltonian. This is one of the crucial parameters to reach strong coupling as we have discussed in Sec.~\ref{sec:intro:experiment_strong_coupling}.}
Although $\hat{H}$ can be seen as an abstract operator (and we will sometimes make use of this picture), we have anticipated in Eq.~\eqref{eq:qed_est:general:Hamiltonian} already the coordinate choice that we will employ mostly in the following: as in Sec.~\ref{sec:est}, we will describe the matter system in real space with coordinates $\br$. For the photon modes, we will instead consider the canonical Harmonic-oscillator coordinates $p_{\alpha},-\I\partial_{p_{\alpha}}$ that are in the chosen gauge proportional to the displacement field $p_{\alpha}\propto D_{\alpha}$ and the magnetic field $-\I\partial_{p_{\alpha}}\propto M_{\alpha}$, respectively (see Sec.~\ref{sec:intro:nrqed_longwavelength}). Thus, any eigenstate of $\hat{H}$ can be described by a wave function 
\begin{align}
\label{eq:qed_est:general:wavefunction_qed}
\Psi (\bx_1,...,\bx_N,p_1,...,p_{M}),
\end{align}
that depends on $M$ photon $p_{\alpha}$ and $N$ spin-spatial electron coordinates $\bx=(\br,\sigma)$, where $\sigma\in\{\uparrow,\downarrow\}$ denotes the spin-degree of freedom. Additionally, $\Psi$ is normalized, i.e.,
\begin{align}
	1=&\int\td\br_1\dots\td\br_N\td p_1\td p_M \Psi^*(\bx_1,...,\bx_N,p_1,...,p_{M})\Psi(\bx_1,...,\bx_N,p_1,...,p_{M})\nonumber\\
	\equiv&\braket{\Psi|\Psi},
\end{align}
where we generalized the 'braket' notation to the coupled case in the last line.
With these coordinates, $\Psi$ is antisymmetric under the exchange of electronic coordinates, i.e.,
\begin{align}
\label{eq:qed_est:general:symmetry_wf_full}
\Psi (\bx_1,..,\bx_{i},..,\bx_{j},..,\bx_N,p_1,..,p_{M}) = - \Psi (\bx_1,..,\bx_{j},..,\bx_{i},..,\bx_N,p_1,..,p_{M}) \quad\forall\,i,j=1,...,N.
\end{align}
Importantly, there is no symmetry with respect to the exchange of the displacement field coordinates $p_{\alpha}\leftrightarrow p_{\beta}$, since these are clearly \emph{distinguishable}.\footnote{The indices $\alpha,\beta$ correspond to modes with different frequencies $\omega_{\alpha},\omega_{\beta}$} We discuss this special coordinate choice in the next paragraph.
To complete the problem definition, we collect all $\Psi$ that adhere to Eq.~\eqref{eq:qed_est:general:symmetry_wf_full} in the set $\mathcal{C}$. This constitutes a domain on which $\hat{H}$ is bound from below (and self-adjoint)\footnote{This follows directly from the self-adjointness of the Pauli-Fierz Hamiltonian, see Sec. \ref{sec:intro:nrqed_longwavelength}.} 
and thus, we can define the ground state $\Psi_0$ with energy $E_0=\braket{\Psi_0|\hat{H}\Psi_0}$ via the variational principle
\begin{align}
\label{eq:qed_est:general:var_principle} 	
E_0=\inf_{\Psi\in\mathcal{C}} \braket{\Psi|\hat{H}\Psi}.
\end{align}
Determining $\Psi_0$ will be the topic of the following sections.

\subsection{The multitude of many-body descriptions of photons}
\label{sec:qed_est:general:coupled_many_body}
In Sec.~\ref{sec:est:general}, we have already discussed the antisymmetric many-electron space in great detail. Its parametrization in terms of Slater determinants turned out to be a very important tool for basically all the electronic structure methods that we have discussed afterwards. Thus, we now want to follow the same route for the many-photon space, which however leaves us more options. In Eq.~\eqref{eq:qed_est:general:wavefunction_qed}, we have explicitly parametrized the many-photon space by the displacement coordinates $p_{\alpha}$. The index $\alpha$ corresponds here to the mode and not a specific photon. The advantage of these coordinates is their independence of the photon number. Since the Hamiltonian~\eqref{eq:qed_est:general:Hamiltonian} does not conserve the photon number, these coordinates allow to describe its eigenstates with  only one wave function (and not with a superposition of many wave functions that have a different number of coordinates). This is practical for the first-principles perspective in general and especially in the cavity-QED setting, where the number of modes is not too big. 

However, the most common parametrization of the many-photon space considers directly the photons, which correspond to the quantized excitations of the electromagnetic field modes. It is an empirically known fact that photons are indistinguishable (such as electrons) but they do not adhere to the Pauli-principle, i.e., many photons can occupy the same state. Thus the many-photon wave function must be \emph{symmetric} under particle exchange. If we exchange the antisymmetrization with a symmetrization, we can simply follow the procedure of Sec.~\ref{sec:est:general} to construct the many-electron space.\footnote{This is how the photon many-body space is introduced in most textbooks, e.g., \citep{Greiner2013,Altland2010, Stefanucci2013}.}
To do so, we choose some one-photon space $\mathfrak{h}^1$ and define the $n$-photon space 
\begin{align}
	\mathfrak{H}_S^{n}=\mathcal{S} \bigotimes_{l=1}^{n} \mathfrak{h}^1
\end{align}
as the symmetric product (denoted by $\mathcal{S}$) of orbital spaces. Any $\phi^n \in \mathfrak{H}_S^{n}$ thus describes the state of $n$ photons (or in general bosons), exactly as any $\psi\in \mathfrak{H}_A^{N}$ describes an $N$-electron (fermion) state (cf. Eq.~\eqref{eq:est:general:antisymmetric_hilbert_space_N}). However, when we want to describe the eigenstates of Eq.~\eqref{eq:qed_est:general:Hamiltonian}, we do not know $n$ a priori and thus, we have to allow in principle for all possible values. The appropriate Hilbert space for this is the \emph{Fock space}
\begin{align}
\mathfrak{F}=\bigoplus_{n=0}^{\infty} \mathfrak{H}_S^{n}.
\end{align}
A general state $\Phi\in \mathfrak{F}$ is a superposition of $\phi^n$ with arbitrary $n$.
 
In contrast to the electronic problem, the one-photon or orbital basis of $\mathfrak{h}^1$, i.e., the starting point of the many-body construction is very easy to choose. The reason is that for photons, we usually do not have to consider external currents or variations of the refraction index~\citep{Li2017} that would shape the photon landscape in a similar way as the local potential shapes the electronic orbitals. Once we have solved the classical Maxwell's equations to obtain the mode functions, the quantum mechanical part of the photon problem reduces to a sum of Harmonic oscillators with the Hamiltonian
\begin{align*}
	\hat{H}_{ph}=&\sum_{\alpha}^{M}\left[-\partial_{p_{\alpha}}^2 +p_{\alpha}^2\right]\\
	=&\sum_{\alpha}^{M}\omega_{\alpha}\left(\hat{a}_{\alpha}^{\dagger}\hat{a}_{\alpha}^{\dphan} + \tfrac{1}{2}\right),
\end{align*}
where we have diagonalized $\hat{H}_{ph}$ in the last line by introducing the annihilation $\hat{a}^{\dphan}_{\alpha} =\sqrt{\tfrac{\omega_{\alpha}}{2}}(p_{\alpha}-\tfrac{1}{\omega_{\alpha}}\tfrac{\partial}{\partial p_{\alpha}})$ and creation operators $\hat{a}^{\dagger}_{\alpha} =\sqrt{\tfrac{\omega_{\alpha}}{2}}(p_{\alpha}+\tfrac{1}{\omega_{\alpha}}\tfrac{\partial}{\partial p_{\alpha}})$. The eigenfunctions of $\hat{H}_{ph}$ are known \emph{analytically} (see also Sec.~\ref{sec:dressed:construction:simple:auxiliary_wf}), which makes them the standard basis choice in quantum optics. Additionally, there is usually no photon-photon interaction term such as the Coulomb interaction and thus the description of the \emph{entire} photonic part of the system is basically as simple as the electrons-in-a-box problem (Sec.~\ref{sec:est:general:example_case}).

This analytical structure can be used to describe photon states in a very elegant manner. Any eigenstate $\ket{n_{\alpha}}$ of the individual $\hat{a}_{\alpha}^{\dagger} \hat{a}_{\alpha}$ is given by multiple applications of creation operators to the vacuum state $\ket{0}$, i.e., $\ket{n_{\alpha}} = (\hat{a}^{\dagger}_{\alpha})^{n} \ket{0}$.\footnote{Note that following Refs. \citep{Giesbertz2019,Stefanucci2013}, we chose here explicitly a non-normalized basis $\{\ket{\varphi^n_{\alpha}}\}$ of the $n$-photon sector, with $\braket{\varphi^n_{\alpha}|\varphi^n_{\alpha}}=n!$ for later convenience. The missing normalization factor is shifted to the resolution of identity in this basis, i.e., $\mathbb{1}=\frac{1}{n}\sum_{\alpha_1,...,\alpha_{n}=1}^{M}\ket{\alpha_1,...,\alpha_{n}}\bra{\alpha_1,...,\alpha_{n}}$, where $\ket{\alpha_1,...,\alpha_{n}}=\hat{a}^{\dagger}_{\alpha_1}\cdots \hat{a}^{\dagger}_{\alpha_{n}}\ket{0}$ as defined later in the text.} 
A general $n$ photon state $\ket{\varphi^{n}}$ reads then
\begin{align}
\ket{\varphi^{n}} = \ket{n_1,...,n_M} = (\hat{a}^{\dagger}_{1})^{n_1}...(\hat{a}^{\dagger}_{M})^{n_M}\ket{0},
\end{align}
where $n=n_1+...+n_M$. This representation is already geared to the symmetry of bosons that can occupy one state configuration with several particles and which manifests here in the fact that the mode occupations $n_i$ can have values different from zero or one. By that, we went completely around an explicit coordinate representation and remain in the abstract state space. We see that the problem of the basis set choice, which was an important part of the many-electron description is basically not present for the photon subsystem. This is an important difference between the first-principles description of electrons and photons.

However, this means that any  nontrivial behavior of the photon system stems from the electron-photon interaction\footnote{Note that the form of the electron-photon interaction depends on the gauge. For instance, in the velocity gauge, one would have to consider also the diamagnetic term that, e.g., renormalizes the photon frequency~\citep{Schaefer2020,Rokaj2020}.}
\begin{align*}
	\hat{H}_{int}=-\sum_i^{N} \sum_{\alpha}^{M} \omega_{\alpha} p_{\alpha} \blambda_{\alpha}\cdot \br_i
\end{align*}
that couples the electronic and the photonic Hilbert spaces.\footnote{Note that although we have restricted the discussion to the cavity-QED setting, most of the general features, such as the construction of the coupled Hilbert space can be generalized straightforwardly to full minimal coupling.} In other words, \emph{there is no photon-only many-body theory}, at least if we remain within the Pauli-Fierz picture and do not consider theories with an effective photon-photon interaction.\footnote{See for example Ref.~\citep{Imamoglu1997}.}
The special form of $\hat{H}_{int}$ is also the reason why eigenstates of $\hat{H}$ are superpositions of $\ket{\varphi^{n}}$ with different $n$: the photon-part of the operator $p_{\alpha}=1/\sqrt{2\omega_{\alpha}}(\hat{a}^{\dagger}_{\alpha}+\hat{a}^{\dphan}_{\alpha})$ does not conserve the particle number.

The states $\ket{n_1,...,n_M}$ are connected to the displacement representation that we have employed in Eq.~\eqref{eq:qed_est:general:wavefunction_qed} by $\varphi^n_{\alpha}(p_{\alpha}) = \braket{p_{\alpha}|n_{\alpha}} = \braket{p_{\alpha}|(\hat{a}^{\dagger}_{\alpha})^{n}| 0}$. We can then express an $M$-mode state as
\begin{align}
\label{eq:photon_wf_representation_connection}
\varphi_{n_1,...,n_M}(p_1,...,p_{M}) = \braket{p_1...p_M|n_1,...,n_M} = \bra{p_1...p_M} (\hat{a}^{\dagger}_{1})^{n_1}...(\hat{a}^{\dagger}_{M})^{n_M}\ket{0}.
\end{align}
It is now crucial to realize that although this is a valid representation of the $n=n_1+...+n_M$-photon state, it is not the coordinate representation of the photons. Thus, there is \emph{no exchange symmetry} between the different $p_{\alpha}$. 

Although less common, we can employ also a coordinate representation for photons in analogy to the standard representation of the electronic wave function in terms of electron coordinates. 
For that, we associate every such multi-mode eigenstate $\ket{n_1,...,n_M}$ with a specific photon-number sector, i.e., the zero-photon sector is merely one-dimensional and corresponds to $\ket{0_1,...,0_M} \equiv \ket{0}$, the single-photon sector is $M$-dimensional and corresponds to the span of $\hat{a}^{\dagger}_{\alpha} \ket{0} \equiv \ket{\alpha}$ for all $\alpha$ and so on. For the multi-photon sectors we see due to the commutation relations of the ladder operators the bosonic exchange symmetry appearing, e.g., $ \hat{a}^{\dagger}_{\alpha_1} \hat{a}^{\dagger}_{\alpha_2}\ket{0} = \hat{a}^{\dagger}_{\alpha_2} \hat{a}^{\dagger}_{\alpha_1}\ket{0} \equiv \ket{\alpha_1, \alpha_2}$ for $\alpha_1, \alpha_2 \in \{1,...,M\}$. 
A general photon state can therefore be represented by a sum over all photon-number sectors as 
\begin{align}
\label{eq:qed_est:general:n_photon_wave_function}
\ket{\Phi} = \sum_{n=0}^{\infty} \left(\tfrac{1}{\sqrt{n!}} \sum_{\alpha_1,...,\alpha_{n} = 1}^{M} \tilde{\Phi}(\alpha_1,...,\alpha_{n}) \ket{\alpha_1,...,\alpha_{n}}\right),
\end{align}
where $\tilde{\Phi}(\alpha_1,...,\alpha_{n})=\frac{1}{\sqrt{{n}!}} \braket{\alpha_1,...,\alpha_{n}|\Phi}$.
It is no accident that the bosonic symmetry becomes explicit in this representation since the different modes $\alpha$ determine how the photon wave functions looks like in real space (for further details on this topic, see App. \ref{app:symmetry_photon}).


\subsubsection*{The coupled electron-photon space}
Having illustrated the principal differences between the photonic and the electronic part of the problem, we now want to discuss the coupled many-body Hilbert space.
Since every many-body space depends on the underlying one-body space and thus also on the coordinate choice, we have several options here and we will choose the displacement-coordinates. We choose a photonic one-body (or orbital) basis $\{\chi^{n}_{\alpha}(p_{\alpha})\}$ that corresponds to the Hilbert space $\mathfrak{h}_{\alpha}$ of mode $\alpha$ and construct the photonic space simply as
\begin{align}
	\mathfrak{H}_{ph}=\bigotimes_{\alpha=1}^{M} \mathfrak{h}_{\alpha}.
\end{align}
Importantly, the description in terms of $p_{\alpha}$ allows for an easier definition of the wave function, which is the main reason, why we employ it. The coupled $N$-electron-photon space is hence given by
\begin{align}
\begin{split}
	\mathcal{C}=&\mathfrak{H}_{e}\otimes\mathfrak{H}_{ph}\\
	=&\mathcal{A}\left(\otimes_{i=1}^{N} \mathfrak{h}^1\right) \otimes\bigotimes_{\alpha=1}^{M} \mathfrak{h}_{\alpha},
\end{split}
\end{align}
where $ \mathfrak{h}^1$ is the electronic one-body space and $\mathcal{A}$ is the antisymmetrizer (see Sec.~\ref{sec:est:general}). $\mathcal{C}$ is a proper configuration space of the minimization problem~\eqref{eq:qed_est:general:var_principle}, that we have to explore to find the ground state wave function of the QED Hamiltonian~\eqref{eq:qed_est:general:Hamiltonian}.

\subsection{The many-electron-photon wave function}
\label{sec:qed_est:general:coupled_wavefunction}
Let us now see how we can parametrize wave functions of $\mathcal{C}$ to perform the variational minimization. For simplicity, we consider a system consisting of 2 electrons that are coupled to one cavity mode with frequency $\omega$ and coupling vector $\blambda$. This setting is already sufficient to illustrate the main challenges of the coupled problem. The many-body wave function is represented in real space with two electronic and one photonic variables $\Psi(\bx_1,\bx_2,p)$, where $\bx=(\br,\sigma)$ is a spin-spatial coordinate. We choose an electronic basis set $\{\psi_k(\bx)\}$ and a photonic basis $\{\chi^n(p)\}$ and expand
\begin{align}
\label{eq:qed_est:general:2e1m_many-body-wf_ansatz}
\begin{split}
\Psi(\bx_1,\bx_2,p)=&\frac{1}{\sqrt{2}}\sum_{kl,n}A_{kl}^n\left(\vphantom{\sum}\psi_k(\bx_1)\psi_l(\bx_2) - \psi_k(\bx_2)\psi_l(\bx_1)\right)\chi^n(p)\\
=& \Psi[A_{kl}^n].
\end{split}
\end{align}
The normalization of the wave function manifests then in the sum rule $\sum_{k,l,n}|A_{kl}^n|^2$ for the coefficient tensor elements $A_{kl}^n\in \mathbb{C}$.
We see that additionally to the electronic Slater determinant, that adds one index for each electron to the coefficient tensor, we get one photonic index for every mode.\footnote{We remind the reader that including more modes in this representation would not require any further symmetrization, because the $p$ coordinates are distinguishable. For example, the 2-electron-2-mode wave function reads $\Psi_{2e2m}(\bx_1,\bx_2,p_1,p_2)=\frac{1}{2}\sum_{kl,n_1,n_2}A_{kl}^{n_1n_2}\left(\vphantom{\sum}\psi_k(\bx_1)\psi_l(\bx_2) - \psi_k(\bx_2)\psi_l(\bx_1)\right)\chi^{n_1}(p_1)\chi^{n_2}(p_2)$.}

The Hamiltonian \eqref{eq:qed_est:general:Hamiltonian} for this scenario reads explicitly
\begin{align}
\label{eq:Hamiltonian_2e1m_example_physical}
\hat{H}&=\underbrace{\sum_{k}\left[ -\frac{1}{2} \nabla_k^2+v(\br_k) + \tfrac{1}{2}(\blambda\cdot \br_k)^2 \right]}_{\hat{H}^e=\sum_{k}h^e(\br)= \sum_{k}[t+v+h^{self,1}](\br_k)} \underbrace{-\sum_{k=1}^{2} \omega p \blambda\!\cdot\!\br_k}_{\hat{H}^{ep}=\sum_{k=1}^{2}h^{ep}(\br_k,p)} 
+\underbrace{\tfrac{1}{2}\left(-\tfrac{\td^2}{\td p}+ \omega^2p^2\right)}_{\hat{H}^{ph}}
+\underbrace{\frac{1}{|\br_1-\br_2|} + \left(\blambda\!\cdot\!\br_1\right)\left(\blambda\!\cdot\!\br_2\right).}_{\hat{H}^{ee}=h^{ee}(\br_1,\br_2)=[w+h^{self,2}](\br_1,\br_2)}
\end{align}
For the chosen gauge (see Sec.~\ref{sec:intro:nrqed_longwavelength}), the interaction between electrons and photons appears in the bilinear interaction term $\hat{H}^{ep}$ and in the purely electronic dipole self-interaction $\hat{H}^{self}\equiv \frac{1}{2}\sum_{k,l=1}^{2}\left(\blambda\!\cdot\!\br_k\right)\left(\blambda\!\cdot\!\br_l\right)$ that has a one-body (last term of $\hat{H}^e$) and a two-body contribution (last term of $\hat{H}^{ee}$).\footnote{We remind the reader that this term is gauge-depended. In the velocity gauge for example, the Hamiltonian would exhibit the diamagnetic term instead of the dipole self interaction. However, the bilinear interaction term instead is always present (though the physical meaning of the photon observables change). See Sec.~\ref{sec:intro:nrqed_longwavelength}.}
The energy expectation value of this Hamiltonian computed with the ansatz \eqref{eq:qed_est:general:2e1m_many-body-wf_ansatz} leads to an expression of the form:
\begin{align}
\label{eq:energy_total}
\begin{split}
E=&E[A_{kl}^n]\\
=&\braket{\Psi[A_{kl}^n]|\hat{H} \Psi[A_{kl}^n]}\\
=& E^e[A_{kl}^n] + E^{ph}[A_{kl}^n] + E^{ep}[A_{kl}^n] + E^{ee}[A_{kl}^n],
\end{split}
\intertext{Here, we have defined the electronic and photonic one-body energies $E^e$ and $E^p$}
\label{eq:energy_1body_electronic}
E^e[A_{kl}^n]=&\frac{1}{2}\sum_{k',k}\sum_{l,r}\left(\vphantom{\sum} A_{k'l}^{r*}A_{kl}^r  - A_{k'l}^{r*}A_{lk}^r  \right) \braket{\Psi_{k'}|h^e \Psi_{k} }\\
\label{eq:energy_1body_photonic}
E^{ph}[A_{kl}^n]=&\sum_{r',r}\sum_{kl}A_{kl}^{r'*}A_{kl}^r \braket{\chi^{r'}|h^p\chi^r},
\intertext{the electron-photon interaction energy $E^{ep}$}
\label{eq:energy_electron_photon_interaction}
E^{ep}[A_{kl}^n]=&-\frac{\omega}{2}\sum_{r',r}\sum_{k',k}\sum_{l}\left(\vphantom{\sum} A_{k'l}^{r'*}A_{kl}^r  - A_{k'l}^{r'*}A_{lk}^r  \right) \braket{\Psi_{k'}|\blambda\!\cdot\!\br| \Psi_{k} } \braket{\chi^{r'}|p\chi^r}, 
\intertext{and the electron-electron interaction energy $E^{ee}$}
\label{eq:energy_electron_electron_interaction}
E^{ee}[A_{kl}^n]=& \frac{1}{2}\sum_{r}\sum_{k',k}\sum_{l',l}\left(\vphantom{\sum} A_{k'l'}^{r*}A_{kl}^r  - A_{k'l'}^{r*}A_{lk}^r  \right) \braket{\Psi_{k'}\Psi_{l'}|h^{ee}|\Psi_{k}  \Psi_{l}}.
\end{align}
We see that the (exact) minimization of $E[A_{kl}^n]$ is comparable to the minimization of a 3-electron problem and merely requires the implementation of the extra energy terms. In fact, on the exact wave-function level, electronic and photon degrees of freedom are very similar. However, such a description is numerically infeasible already for most electronic problems. If also the photon modes have to be included, such descriptions will be even more limited. 
This is illustrated by the fact that the largest matter-photon problem that has been solved exactly so far consists of three particles (electron or nuclei) and one photon mode~\citep{Sidler2020}.

\subsubsection*{The mean field approximation: ``Hartree-Fock for QED''}
The substantial differences of the coupled electron-photon space become explicitly visible, when we try to approximate the wave function. Let us as a first example consider the equivalent to the single-Slater-determinant ansatz of HF theory, which is called the \emph{mean field} (MF) approximation. This means that we have to consider the simplest wave function that adheres to the symmetry~\eqref{eq:qed_est:general:symmetry_wf_full}, i.e., one basis element of the coupled many-body space $\mathcal{C}$. This is simply the product of one Slater determinant (the HF ansatz) and a photon orbital
\begin{align}
\label{eq:2e1m_wf_mf}
\begin{split}
\Phi^{MF}(\bx_1,\bx_2,p)=&\Psi^{HF}(\bx_1,\bx_2) \chi(p)\\
=&\frac{1}{\sqrt{2}}\left(\vphantom{\sum}\psi_1(\bx_1)\psi_2(\bx_2) - \psi_2(\bx_2)\psi_1(\bx_1)\right)\chi(p),
\end{split}
\end{align}
with the corresponding energy expression
\begin{align}
\label{eq:qed_est:general:energy_meanfield}
E^{MF}=&\braket{\Phi^{MF}|\hat{H}\Phi^{MF}}\\
=&\sum_{k=1}^{2} \braket{\psi_{k}|h^e \psi_{k} }+ \braket{\chi|h^p\chi} - \omega \sum_{k=1}^{2} \braket{\psi_{k}|\blambda\!\cdot\!\br| \psi_{k} } \braket{\chi|p\chi}+\left(\vphantom{\sum} \braket{\psi_{1}\psi_{2}|h^{ee}|\psi_{1}  \psi_{2}}  - \braket{\psi_{1}\psi_{2}|h^{ee}|\psi_{2}  \psi_{1}} \right). \nonumber
\end{align}
We see that the MF ansatz~\eqref{eq:2e1m_wf_mf} \emph{entirely decouples} electron and photon degrees of freedom, which has severe consequences for the quality of the approximation. This is well-illustrated by discussing the electronic and photonic subsystems separately. 
The electronic part of the energy expression is given by
\begin{align}
	E^{MF,e}=\sum_{k=1}^{2} \braket{\psi_{k}|(h^e + v_{ph}) \psi_{k} } +\left(\vphantom{\sum} \braket{\psi_{1}\psi_{2}|h^{ee}|\psi_{1}  \psi_{2}}  - \braket{\psi_{1}\psi_{2}|h^{ee}|\psi_{2}  \psi_{1}} \right),
\end{align}
where we have defined $v_{ph}(\br)=- \omega \braket{\chi|p\chi} \blambda\!\cdot\!\br$. This is structurally the same expression as in HF, but with a modified local potential and two-body interaction kernel. The photon-part of the energy reads
\begin{align}
	E^{MF,ph}= \braket{\chi|h^p\chi} - \omega\blambda\!\cdot \bR \braket{\chi|p\chi},
\end{align}
where $\bR=\sum_{k=1}^{2}\braket{\psi_{k}|\br \psi_{k} }$ is the total dipole of the electronic system. We see that on the photon side, we have just a harmonic oscillator that is shifted by the electronic total dipole moment.\footnote{Note that the dipole moment becomes the full electronic current in the general minimal coupling setting.}
The eigenstates of shifted oscillators are \emph{coherent states} which are very closely connected to classical fields~\citep{Scully1999}.
%
%
Many systems such as atoms but also many molecules do not have a static dipole, i.e.,  $\bR=0$. In this case,
\begin{align*}
	E^{MF,ph}= \braket{\chi|h^p\chi},
\end{align*}
which yields the photon vacuum-state $\chi^0$ as lowest energy state. The photon contribution to the total energy is the trivial vacuum energy in this case. This important result can be generalized to arbitrary many modes and particles~\citep{Ruggenthaler2014}. We see that the MF approximation entirely \emph{neglects the quantum nature} of the electron-photon interaction.

With the ansatz of the form \eqref{eq:2e1m_wf_mf}, we have derived coupled ``Maxwell-HF'' theory.
Within this approach, we cannot describe  quantum properties of the electro-magnetic field beyond coherent states, which are essentially classical~\citep[Ch. 2.1]{Scully1999}. This is very different for the electronic mean-field theory, i.e., HF. We have seen in Sec.~\ref{sec:est:hf} that the antisymmetry of the HF ansatz alone adds the Fock term to the equations, which is not derivable by any classical theory, and that significantly improves HF over the semiclassical Hartree theory. On the contrary, the description of any ``quantum effect'' due to the electron-photon interaction requires a \emph{multi-reference} (correlated) wave function. 

This does not mean that coupled Maxwell-HF theory or more generally, coupled Maxwell-matter theories are not a reasonable extension of electronic structure theory. A prominent example is the coupled \emph{Maxwell-Kohn-Sham approach}, that calculates the Slater determinant of $\Phi^{MF}$ with the KS-DFT machinery~\citep{Lorin2007,Yabana2012,Jestaedt2019}. Possible applications include the theoretical description of high-harmonic generation \citep{Floss2018}, attosecond physics \cite{Sommer2016} or molecular systems weakly coupled to the modes of a cavity \cite{Flick2019}. The semiclassical approach is especially powerful to investigate \emph{time-dependent} phenomena, where the solution of Maxwell's equations and their self-consistent implementation with the KS equations is highly nontrivial. 

\subsubsection*{Inclusion of correlation: the generalized mean-field ansatz}
To investigate the role of the quantum nature of the photon-field in cavity systems, we have to go beyond the MF approximation. The straightforward way to do so is a more general wave function ansatz, which we briefly want to discuss in the following.

We have concluded already that the configuration space of the ansatz~\eqref{eq:qed_est:general:2e1m_many-body-wf_ansatz} is not practically useful beyond very simple systems. If we nevertheless want to employ a wave function, we therefore need to truncate the configuration space in a reasonable way. For that, we can utilize our knowledge about the electronic system that is quite accurately described by, e.g., one Slater determinant (HF or KS-DFT). To make use of this information, we could try to extend the mean-field wave function in a systematic way with the idea to remain as close as possible to the single-reference ansatz in the electronic subsystem. For instance, we could use the mean-field ansatz $\Phi^{MF}$ as a kind of basis and build the full wave function from superpositions of different such $\Phi^{MF}\rightarrow\Phi^r$ that we denote by an index r. One can see this as the ``many-body generalization'' of the cavity-QED models of Sec. \ref{sec:intro:cavity_qed_models} (from Eq. \eqref{eq:QO_Hamiltonian_Rabi} onwards): the $\Phi^r$ represent the electronic ``levels''  that are coupled to the states of a photon-mode. We call this the generalized MF ansatz and for our 2-electron-1-mode example, the corresponding wave function reads
\begin{align}
\label{eq:qed_est:general:2e1m_gMF_ansatz}
\Phi^{gMF}(\bx_1,\bx_2,p)=&\sum_{r=1}^B A^r\Phi^r(\bx_1,\bx_2,p) \nonumber\\
=&\frac{1}{\sqrt{2}}\sum_{r=1}^BA^r\left(\vphantom{\sum}\psi_1^r(\bx_1)\psi_2^r(\bx_2) - \psi_2^r(\bx_2)\psi_1^r(\bx_1)\right)\chi^r(p),
\end{align}
where the expansion coefficients $A^r\in\mathbb{C}$ satisfy the normalization condition $\sum_{r=1}^B|A^r|^2=1$. In this expansion, we have sorted the problem according to the photon states that define different sectors labelled by $r$. Note that only the full Slater determinants $\braket{\Phi^r|\Phi^{r'}}=\delta_{r,r'}$ between different photon sectors are orthogonal, but not the electronic orbitals alone, i.e., $\braket{\psi_i^r|\psi_j^{r'}}\neq \delta_{r,r'}$. Only within one sector $r$, we have $\braket{\psi_i^r|\psi_j^{r}}=\delta_{ij}$.\footnote{If we assumed instead that $\braket{\psi_i^r|\psi_j^{r'}}=\delta_{ij} \delta_{r,r'}$, the electronic and photon subsystems would decouple as in the MF description.}

Clearly, this expansion is equivalent to the full many-body ansatz \eqref{eq:qed_est:general:2e1m_many-body-wf_ansatz} for a complete basis set ($B\rightarrow \infty$). The central idea behind Eq.~\eqref{eq:qed_est:general:2e1m_gMF_ansatz} is that the number $B$ of included MF states $\Phi^r$ is small.  The corresponding generalized MF energy expression reads
\begin{align*}
E^{gMF}=&\braket{\Phi^{gMF}|\hat{H}\Phi^{gMF}}\nonumber\\
=&E^e_{gMF} + E^{ph}_{gMF} + E^{ep}_{gMF} + E^{ee}_{gMF} \nonumber\\
=&\sum_{r=1}^B\sum_{k=1}^{2} |A^{r}|^2 \braket{\psi_{k}^r|h^e \psi_{k}^r }+ \frac{1}{2}\sum_{r',r=1}^B \sum_{k',k=1}^{2}A^{r'*}A^{r} \braket{\chi^{r'}|h^p\chi^r}\braket{\psi_{k'}^{r'}| \psi_{k}^r } \nonumber\\
&-\frac{\omega}{2}\sum_{r',r=1}^B \sum_{k\neq l=1}^{2} A^{r'*}A^r \left(\braket{\psi_{k}^{r'}|\blambda\!\cdot\!\br| \psi_{k}^r }\braket{\psi_{l}^{r'}| \psi_{l}^r } - \braket{\psi_{k}^{r'}|\blambda\!\cdot\!\br| \psi_{l}^r } \braket{\psi_{l}^{r'}| \psi_{k}^r} \right) \braket{\chi^{r'}|p\chi^r} \nonumber \\
&+\sum_{r=1}^B |A^r|^2 \left(\vphantom{\sum}  \braket{\psi_{1}^{r}\psi_{2}^r|w|\psi_{1}^r  \psi_{2}^r} -\braket{\psi_{1}^{r}\psi_{2}^r|w|\psi_{2}^r  \psi_{1}^r}  \right).
\end{align*}
Let us briefly analyze the occurring terms. The purely electronic contribution is just the sum of the standard MF terms of the involved Slater-determinants,
\begin{align*}
	E_{gMF}^e+E_{gMF}^{ee}=\sum_{r=1}^B |A^{r}|^2 E^e_r + \sum_{r=1}^B |A^r|^2 E^{ee}_r,
\end{align*}
where we have defined $E^e_r=\sum_{k=1}^{2} \braket{\psi_{k}^r|h^e \psi_{k}^r }$ and $E^{ee}_r=\left(\vphantom{\sum}  \braket{\psi_{1}^{r}\psi_{2}^r|h^{ee}|\psi_{1}^r  \psi_{2}^r} -\braket{\psi_{1}^{r}\psi_{2}^r|h^{ee}|\psi_{2}^r  \psi_{1}^r}  \right)$. This corresponds to $B$ HF problems, which is manageable even for large $B$.

We continue with the two terms that involve the photon states. Interestingly, we see that the photon energy term 
\begin{align*}
	E_{gMF}^{ph}=\frac{1}{2}\sum_{r',r=1}^B \sum_{k',k=1}^{2}A^{r'*}A^{r} \braket{\chi^{r'}|h^p\chi^r}\braket{\psi_{k'}^{r'}| \psi_{k}^r }
\end{align*}
that in the full many-body description (Eq. \eqref{eq:energy_1body_photonic}) was a one-body term (involving only the $\braket{\chi^{r'}|h^p\chi^r}$) now also requires the calculation of the overlap integrals $\braket{\psi_{k'}^{r'}| \psi_{k}^r }$ between \emph{all} the electronic basis elements of all the photon sectors. Thus, the $B$ HF equations are seemingly coupled due to this term, which spoils the advantage of the construction. However, we have not yet made use of the freedom to choose the basis functions $\chi^r$. For instance, if we choose the eigenstates of $\hat{H}^{ph}$, i.e., $\hat{H}^{ph}\braket{\chi^r}=\omega(r+\tfrac{1}{2})\braket{\chi^r}$, we have $\braket{\chi^{r'}|\hat{H}^{ph}\chi^r}\propto \delta_{r',r}$ and thus the overlap integrals between the different r-sectors $\braket{\psi_{k'}^{r'}| \psi_{k}^r }=\braket{\psi_{k'}^{r}| \psi_{k}^r }=\delta_{k',k}$ vanish. The energy expression reduces then to
\begin{align}
	E_{gMF}^{ph}\overset{\hat{H}^{ph}\ket{\chi^r}=\omega(r+\tfrac{1}{2})\ket{\chi^r}}{\longrightarrow} \sum_{r=1}^B |A^{r}|^2 \omega(r+\tfrac{1}{2}).
\end{align}
This is exactly the representation that we have employed to derive the cavity-QED models and it makes also the photon-energy term in this generalized MF description manageable. 

The only missing term is the electron-photon interaction that reads
\begin{align}
\label{eq:qed_est:general:gMF_interaction}
E_{gMF}^{ep}=-\frac{\omega}{2}\sum_{r',r=1}^B \sum_{k\neq l=1}^{2} A^{r'*}A^r \left(\braket{\psi_{k}^{r'}|\blambda\!\cdot\!\br| \psi_{k}^r }\braket{\psi_{l}^{r'}| \psi_{l}^r } - \braket{\psi_{k}^{r'}|\blambda\!\cdot\!\br| \psi_{l}^r } \braket{\psi_{l}^{r'}| \psi_{k}^r} \right) \braket{\chi^{r'}|p\chi^r}.
\end{align}
This most-complicated term of the energy has three different contributions: the transition dipole matrix elements $\braket{\psi_{k}^{r'}|\br| \psi_{k}^r }$, again the electronic overlap integrals $\braket{\psi_{l}^{r'}| \psi_{k}^r}$ and the photon-displacement matrix elements $\braket{\chi^{r'}|p\chi^r}$. Since we have chosen a photonic basis, which we know analytically, we can further simplify this expression. We identify $p= 1/\sqrt{2\omega} (\hat{a}+\hat{a}^{\dagger})$ and utilize $\hat{a}^{\dagger}\ket{\chi}^r=\sqrt{r+1}\ket{\chi}^{r+1}, \hat{a}\ket{\chi}^r=\sqrt{r}\ket{\chi}^{r-1}$ to re-express
\begin{align}
	\braket{\chi^{r'}|p\chi^r}\overset{\hat{H}^{ph}\ket{\chi^r}=\omega(r+\tfrac{1}{2})\ket{\chi^r}}{\longrightarrow}\frac{2}{\omega}(\sqrt{r+1}\delta_{r',r+1}+ \sqrt{r}\delta_{r',r-1}),
\end{align}
which reduces the double-sum over $r,r'$ to one sum that couples only neighboring photon sectors. However, we still have to compute the overlap integrals $\braket{\psi_{l}^{r}| \psi_{k}^{r\pm 1}}$ for all combinations of $(k,r)$ and $(l,r+1)$. 

Instead of the eigenstates of $\hat{H}^{ph}$, we could also choose the eigenstates of $\hat{p}$, i.e., coherent states as a photon basis. This would remove all the coupling elements in the $E_{gMF}^{ep}$-expression, $\braket{\chi^{r'}|p\chi^r}\rightarrow \delta_{r,r'}$, in the same way as the eigenstates of $\hat{H}^{ph}$ remove the overlaps in the photon energy before. However, since $[\hat{H}^{ph},p]\neq 0$, there is no basis that diagonalizes \emph{both} operators at the same time and thus, we cannot avoid the calculation of the overlap integrals: we are fundamentally confronted with the problem of determining $B$ coupled Slater determinants. 

This might seem simpler than solving the standard ansatz~\eqref{eq:qed_est:general:2e1m_many-body-wf_ansatz}, but instead the configuration space grows here even faster with the particle number.
To see this, we consider the $N$-electron-1-mode wave function
\begin{align}
\label{eq:Ne-1m_wf_ansatz}
	\Psi(\bx_1,...,\bx_N,p)=&\sum_r A^r \Phi^r(\bx_1,...,\bx_N,p) \nonumber\\
	=& \sum_r A^r \Psi^r(\bx_1,...,\bx_N) \chi^r(p),
\end{align}
with the electronic Slater determinants  
$\Psi^r(\bx_1,...,\bx_N)= \frac{1}{\sqrt{N!}}\sum_{\pi_j\in P_N} (-1)^{j}\psi^r_{\pi_j(1)}(\bx_1)\cdots \psi^r_{\pi_j(N)}(\bx_N)$, where $P_N$ denotes the permutation group on $N$ elements and the index $j$ is chosen such that it is even (odd) for an even (odd) permutation $\pi_i \in S_N$. 
With this ansatz, we calculate the electron-photon interaction term
\begin{align}
E^{ep}=&-\omega\frac{1}{N!}\sum_{r',r}A^{r'*}A^r \braket{\chi^{r'}|p\chi^r} \nonumber\\
& \times\sum_{\pi_i,\pi_j\in P_N} (-1)^{i+j} \left(\braket{\psi_{\pi_i(1)}^{r'}|\blambda\!\cdot\!\br| \psi_{\pi_j(1)}^r }\braket{\psi_{\pi_i(2)}^{r'}| \psi_{\pi_j(2)}^r } \cdots \braket{\psi_{\pi_i(N)}^{r'}| \psi_{\pi_j(N)}^r } \right).
\end{align}
The permutation group $P_N$ has $N!$ many elements over which the double sum in the second line runs. This means that even if we only take two different photon states into account, we have to calculate in principle $N!^2$ many integrals. This number is enormous since the factorial function\footnote{According to Sterling's approximation for factorials~\citep[part 4]{Freitag2006}.} $x!\approx \sqrt{x}x^xe^{-x}$
grows even faster than the exponential. In a practical calculation many of these integrals would be zero, since the coupled Slater determinants usually do not differ in every basis function. Nevertheless, the principal problem remains, which illustrates the \emph{intrinsic} difficulty of the (many-body) description of polaritons. 

The only practical way to utilize the efficient description of electronic structure methods in the coupled setting is to employ the rotating-wave approximation that we have introduced in Sec.~\ref{sec:intro:cavity_qed_models}. This means that we project the wave function on the single excitation space~\citep{Herrera2020review}, which would remove the $r$-index in Eq.~\eqref{eq:qed_est:general:gMF_interaction}. This is a reasonable approach for certain physical settings, especially for time-dependent systems, if the photon mode is close to resonance with a matter excitation. However, for the description of ground states, this approximation breaks down~\cite{DeLiberato2017} and generalizations are very difficult~\citep{Bennett2016}.
\section{Quantum-electrodynamical density functional theory}
\label{sec:qed_est:qedft}
The analysis of the previous section showed that a wave function description of coupled electron-photon problems beyond the mean-field level is quite challenging for larger systems and many modes. However, the mean-field wave function, i.e., the generalization of the HF ansatz cannot capture the quantum effects due to the photon interaction, and is thus not sufficient to describe, e.g., strong-coupling phenomena. For that, we need first-principles methods that are capable to capture electron-photon correlation. 
The most efficient and at the same time very accurate way to describe correlated electronic systems was the KS construction of DFT that we have introduced in Sec.~\ref{sec:est:dft}. Instead of the wave function, DFT employs the electron (one-body) density $\rho(\br)$ as fundamental variable, which can be very accurately approximated by considering the KS system, i.e., a non-interacting auxiliary system. We have mentioned already in Sec.~\ref{sec:est:dft} that the Hohenberg-Kohn theorem that justifies DFT can be generalized to many other scenarios~\citep{Penz2019} including Pauli-Fierz theory and its limits~\citep{Ruggenthaler2014,Ruggenthaler2015}. This allows to define QEDFT, which we want to discuss in the following. 

\citeauthor{Ruggenthaler2014} formally defined time-dependent QEDFT\footnote{For the time-dependent setting is not captured by the Hohenberg-Kohn theorem, but needs to be generalized. This has been shown first by \citet{Runge1984} for the special case of analytical external potentials and was generalized to a very broad range of potentials by \citet{VanLeeuwen1999}.}~\citep{Ruggenthaler2014} and ground-state QEDFT~\citep{Ruggenthaler2015} for the full hierarchy from the full relativistic regime (neglecting the mathematical problems that we mentioned in Sec.~\ref{sec:intro:full_qed}) to model systems in the long-wavelength limit. These works connect the two precursory publications by \citet{Ruggenthaler2011} in the relativistic regime and \citet{Tokatly2013}, who first considered the cavity-QED setting. A possible theoretical framework for the equilibrium phenomena of polaritonic chemistry is thus ground-state QEDFT~\citep{Ruggenthaler2015}. 

\subsubsection*{QEDFT and the KS construction}
Let us start with the basic theory of QEDFT in the cavity-QED setting described by Hamiltonian~\eqref{eq:qed_est:general:Hamiltonian}. We formally include the time-derivative of an additional external mode-resolved current $\dot{j}_{\alpha}$\footnote{Note that we have to consider the time-derivative because of the chosen length-gauge, where $p_{\alpha}$ is proportional to the displacement field. In the velocity gauge for example, we would use the vector potential as principal variable, which couples directly to external currents. Thus, strictly speaking, we consider a quasi-static situation here. In practice, this is just a technical detail and thus, we will not further comment on the time-derivative.} that couples to the photon field. The Hamiltonian then reads
\begin{align}
\label{eq:Hamiltonian_BO_extcurrent}
\hat{H}=\sum_i^{N} \left[\frac{1}{2} \nabla_{\br_i}^2 + v(\br_i)\right] + \frac{1}{2}\sum_{i\neq j}^{N} \frac{1}{|\br_i-\br_j|}+ \frac{1}{2}\sum_{\alpha}^{M}\left[-\partial_{p_{\alpha}}^2 + \omega_{\alpha}^2 \left(p_{\alpha} + \frac{\blambda_{\alpha}}{\omega_{\alpha}}\cdot \bR \right)^2 +\frac{\dot{j}_{\alpha}}{\omega_{\alpha}}p_{\alpha}\right].
\end{align}
The ground state of $\hat{H}$ is described by a wave function of the form
\begin{align}
\Psi (\bx_1,...,\bx_N,p_1,...,p_{M}).
\end{align}
Note that the external current $\dot{j}_{\alpha}$ plays the same role for the photons as $v(\br)$ for the electrons.\footnote{Note that in the time-independent case, $\dot{j}_{\alpha}$ is equivalent to an external vector-potential \citep{Ruggenthaler2015}. We thus do not have to consider external vector-potentials in addition to $\dot{j}_{\alpha}$, which is in contrast to the (most general) time-dependent case.} Correspondingly, the pair of internal and external variables that are in one-to-one correspondence are~\citep{Tokatly2013}
\begin{align}
(\rho(\br),\{p_{\alpha}\}) \overset{1:1}{\longleftrightarrow} (v(\br),\{\dot{j}_{\alpha}\}),
\end{align}
where
\begin{subequations}
	\begin{align}
	\rho(\br)=&\braket{\Psi|\hat{\rho}(\br)\Psi}=\int\td\br^{N-1}\br\td^{M}p \Psi^*(\br,\br_2,...,\br_N,p_1,...,p_{M})\Psi(\br,\br_2,..., \br_N,p_1,...,p_{M}),\\
\intertext{is the one-body density as in standard DFT and}
	p_{\alpha}=&\braket{\Psi|p_{\alpha}\Psi}=\int\td\br^{N}\br\td^{M}p \Psi^*(\br_1,\br_2,...,\br_N,p_1,...,p_{M}) p_{\alpha}\Psi(\br_1,\br_2,..., \br_N,p_1,...,p_{M}).
	\end{align}
\end{subequations}
is the expectation value of the displacement-field of mode $\alpha$.

To facilitate the functional construction, the obvious next step is to consider an auxiliary system that we can describe efficiently. The straightforward generalization of the KS system in DFT, is a non-interacting or mean-field system in the coupled space, i.e., a system described by the mean-field wave function
\begin{align}
	\Phi^s(\br_1,...,\br_N,p_1,...,p_M)=|\psi_1(\br_1)\cdots \psi_N(\br_n)|_- \chi_1(p_1)\cdots \chi_M(p_M)
\end{align} 
that we discussed in the previous section for $N=2$ and $M=1$. 
We call this the KS construction for QEDFT (KS-QEDFT) and for the electronic orbitals $\psi_i$, the according KS equations read
\begin{align}
\label{eq:qed_est:dft:KSequations}
 \epsilon_i\psi_i(\br)=\left[-\tfrac{1}{2}\nabla^2 + v_s(\br)+ \sum_{\alpha}^{M}\left( \omega_{\alpha} \braket{p_{\alpha}}+ \blambda_{\alpha}\cdot \braket{\bR} \right)\blambda_{\alpha}\cdot \hat{\bR} \right]\psi_i(\br).
\end{align}
Note the we explicitly included the expectation value of the displacement $\braket{p_{\alpha}}=\braket{\Phi^s|p_{\alpha}\Phi^s}$ of the photon field modes and the expectation value of the total dipole $\braket{\bR}$ of the matter system. We explicitly added the $\braket{\cdot|\cdot}$ symbol to differentiate these mean values from the operator $\hat{R}$. The whole last term in Eq.~\eqref{eq:qed_est:dft:KSequations} thus represents the mean-field contribution of the electron-photon interaction term (see the according term in Eq.~\eqref{eq:qed_est:general:energy_meanfield}), i.e., the straightforward generalization of the electronic Hartree part potential (see Eq.~\eqref{eq:est:hf:potential_hartree}). The KS potential
\begin{align}
	v_s(\br)=v(\br)+v^{e}_{Hxc}[\rho,\{p_{\alpha}\}](\br)+ \sum_{\alpha}^{M}v^{ph}_{\alpha,xc}[\rho,\{P_{\alpha}\}](\br)
\end{align}
of KS-QEDFT consists of three main parts. The first two terms are equivalent to KS-DFT, where $v(\br)$ is the external potential and $v^{e}_{Hxc}$ is the usual Hartree-exchange-correlation potential  (cf. Eq.~\eqref{eq:est:dft:energy_functional_ehxc}) that describes the Coulomb interaction between the electrons. The third term is new and constitutes the (mode-resolved) photonic exchange-correlation potential $v^{ph}_{\alpha,xc}$ that accounts for the electron-photon correlation.  

The equations for the photonic subsystem are much simpler because the coherent states of the mean-field ansatz can be equivalently described by their mean-value which are the (classical) displacement coordinates $p_{\alpha}$. The corresponding eigenvalue problem is given by the time-independent Maxwell's equations. These reduce to the trivial equality
\begin{align}
\label{eq:qed_est:qedft:MWEquations}
\omega_{\alpha}^2 p_{\alpha} = \omega_{\alpha}\blambda\cdot\braket{\bR}  +  \frac{\dot{j}_{\alpha}}{\omega_{\alpha}}.
\end{align}

\subsubsection*{The problem of a non-interacting KS system}
This shows again the difficulty of the time-independent electron-photon problem: If we consider the case, where $\dot{j}_{\alpha}=0$ and $\bR=0$,\footnote{This is the case for, e.g., all \emph{centrosymmetric} potentials $v(\br)=v(-\br)$, where $\braket{\bR}=\sum_{i=1}^{N}\braket{\br_i-\br_0}\equiv 0$, if we chose the center of mass as the reference $\br_0$ for the dipole-operator.} the photon part of the KS system does not contribute to the solution. Additionally, we lose all the mean-field part of the Kohn-Sham equations that in this scenario read
\begin{align}
	\epsilon_i\phi_i(\br)=\left[-\tfrac{1}{2}\nabla^2 + v_s(\br)\right]\phi_i(\br).
\end{align}
It is important to realize how general this situation is, since it includes \emph{all atoms} and also all molecules that are aligned in a inversion-symmetric way with respect to the polarization direction such that $\lambda\cdot\braket{\bR}\equiv0$. For example, all the concrete problems that we will discuss in the rest of this thesis will be systems of this category.

On the one hand, this means a simplification of the problem. We only need to solve a set of KS equations that are exactly as difficult to solve as the purely electronic system. On the other hand, this means that all quantum effects due the photon field are carried by the new exchange-correlation potential $v^{ph}_{\alpha,xc}$, which we do not know. Additionally, we do not have direct access to non-classical observables of the photon field as, e.g., the photon-number. 
The reason behind this is obviously our choice of a non-interacting auxiliary system. Since there is no symmetrization between the electronic and photonic Hilbert space, we cannot gain ``quantumness'' only by considering a properly symmetrized ansatz. In other words, there is no exchange contribution of the electron-photon interaction. 

\subsubsection*{Dynamics vs Statics in QEDFT}
As we have mentioned in the beginning, QEDFT was first formulated for the time-dependent case, which means that we have to solve the time-dependent version of the coupled KS equations~\eqref{eq:qed_est:dft:KSequations} and~\eqref{eq:qed_est:qedft:MWEquations}  self-consis\-tently. So far QEDFT was mainly applied in this setting, which we therefore briefly want to discuss. Interestingly, the accurate description of many time-dependent problems is actually easier with QEDFT than of ground states. The principal reason is that in the time-dependent case, already the mean field, i.e., the classical electromagnetic field does contribute  nontrivially to the problem. The simplest approximation of such a time-dependent QEDFT just disregards $v^{ph}_{\alpha,xc}$ completely, which is equivalent to the Maxwell-Kohn-Sham approach that we have mentioned in Sec.~\ref{sec:qed_est:general} and which leads to a very good description of the weak (and sometimes intermediate) coupling regime. For instance, the linear response of many systems is captured well~\citep{Flick2019}. It has also been shown how one can in principle include the nuclear dynamics  into the description and present accurate results on the level of Ehrenfest dynamics (nuclear mean field)~\citep{Flick2018abinitio,Flick2018nuclear}. A good overview about the applicability and the actual state of (time-dependent) QEDFT can be found in the review by \citet{Flick2018review}. 

As a final comment for the time-dependent case, we would like to mention the publication of \citet{Wang2020}, in which the authors connect QEDFT to cavity-QED models. They show that indeed for small coupling strengths the cavity-QED models describe the principal physics very accurately. This is a first step in the direction of one of the important applications of first-principles theory, that is to define the \emph{range of applicability} of model descriptions.

\subsubsection*{Photon-exchange-correlation}
Despite the interesting results that have been obtained on the mean-field level and the many conceptual insights that we could gain already only by establishing the general theory of QEDFT, we have to go beyond that for our ground state problem. Thus, we are faced with the most difficult part of every type of DFT, i.e., the approximation of the unknown exchange-correlation functional $v^{ph}_{\alpha,xc}$. Only one functional has been proposed so far by \citet{Pellegrini2015}, which is a generalization of the \emph{optimized effective potential} (OEP) approach to standard DFT (see Sec.~\ref{sec:est:dft}). This approach allows to construct a functional on the basis of perturbation theory by using a connection to the \emph{Green's function} formalism~\citep{Stefanucci2013}. Based on this connection, \citeauthor{Pellegrini2015} developed the \emph{photon OEP}, which considers a lowest order perturbative correction of the effective potential, i.e., it takes one-photon processes into account. This is the straightforward generalization of the \emph{exact-exchange} approximation in KS-DFT that we have mentioned in Sec.~\ref{sec:est:dft}.  As in the electronic case, the photon OEP goes beyond standard perturbation theory, which is used to calculate a correction to an already determined ground state. Instead, the photon OEP is a nonlinear functional of $\rho$ that has to be included self-consistently when we solve the KS equations. 

Unfortunately, using the photon OEP in practice is numerically very expensive and difficult to converge~\citep{Flick2018abinitio}. This is a well-known problem of the OEP approach (regarding photonic or electronic corrections) and the reason why the OEP is rarely used explicitly. Instead, the Krieger-Li-Iafrate approximation~\citep{Krieger1992} is very common that captures the main features of the OEP quite accurately for relatively low computational costs and good convergence properties~\citep{Kim1999}. Unfortunately, the generalization of the KLI-approximation for the photon OEP is severely less accurate than in the electronic case and thus not very useful in practice~\citep{Flick2018abinitio}. Another severe drawback of the OEP functional with regard to strong-coupling physics is its limitation to one-photon processes. \citet{Flick2018abinitio} analyzed this for an exactly solvable system and came to the conclusion that the photon OEP starts to fail, when two-photon process become important in the electron-photon interaction. Since this is a hallmark of the strong-coupling regime, we need functionals that go beyond the photon OEP to describe many of the phenomena of polaritonic chemistry.

Comparing to the history of DFT, it seems very probable that once an accurate photon-exchange-correlation functional has been found, KS-QEDFT will become the standard tool to describe coupled electron-photon problems. However, there are also alternative routes as we will show in part~\ref{sec:dressed} that provide a valuable additional perspective. This might be helpful not only for the functional construction (see also the Outlook in part~\ref{sec:conclusion}) but also for possible scenarios, where such a future functional is inaccurate. For instance, the emergence of polaritons in the strong-coupling regime shows that the electronic and photonic subsystems are strongly mixed, which corresponds to a strongly correlated character of the wave function.
The experience from KS-DFT shows that the non-interacting auxiliary system is not a good starting point to describe such systems. It is thus possible that we will face a similar challenge, when we want to describe strongly-coupled systems with KS-QEDFT. 

	\section{Reduced Density Matrices in QED}
\label{sec:qed_est:rdms}
In this section, we generalize the concept of reduced density matrices (RDMs) to the coupled electron-photon space. Although in principle straightforward, the coupled light-matter problem poses many new challenges to an RDM description, because of the special form of the electron-photon interaction that is not particle-conserving. Nevertheless, it is possible to define a variational RDM theory and RDMFT (QED-RDMFT) also in the coupled setting.

\subsection{Reduced density matrices for the electron-photon space}
\label{sec:qed_est:rdms:general}
We start with the generalization of the concept of RDMs to the coupled electron-photon problem. For that, we analyze an $N$-electron-$M$-mode system in terms of the RDM description, following the same procedure as in Sec.~\ref{sec:est:rdms:general} for the $N$-electron problem. We follow here the section \emph{Reduced density matrices for coupled light-matter systems} of Ref.~\citep{Buchholz2019}. To ease reading, we shortly recapitulate the most important definitions of Sec.~\ref{sec:qed_est:general}.

We consider the cavity-QED Hamiltonian
\begin{align}
\tag{cf. \ref{eq:qed_est:general:Hamiltonian}}
\hat{H}=\sum_i^{N} \left[\frac{1}{2} \nabla_{\br_i}^2 + v(\br_i)\right] + \frac{1}{2}\sum_{i\neq j}^{N_e} \frac{q_iq_j}{|\br_i-\br_j|}+ \frac{1}{2}\sum_{\alpha}^{M}\left[-\partial_{p_{\alpha}}^2 + \omega_{\alpha}^2 \left(p_{\alpha} + \frac{\blambda_{\alpha}}{\omega_{\alpha}}\cdot \bR \right)^2\right],
\end{align}
which we split for later convenience in the contributions
\begin{align}
\hat{H}=\hat{T}+\hat{V}+\hat{W}+\hat{H}_{ph}+\hat{H}_{int} +\hat{H}_{self,1}+\hat{H}_{self,2},
\end{align}
where 
\begin{align*}
&\hat{T}=\sum_i^{N_e} \frac{1}{2} \nabla_{\br_i}^2, \quad
\hat{V}=\sum_i^{N_e} v(\br_i), \quad
\hat{W}= \frac{1}{2}\sum_{i\neq j}^{N_e} \frac{q_iq_j}{|\br_i-\br_j|},\\
&\hat{H}_{ph}= \frac{1}{2}\sum_{\alpha}^{M_{ph}}\left[ -\partial_{p_{\alpha}}^2 + \omega_{k}^2 p_{\alpha}^2\right], \quad
\hat{H}_{int}=-\sum_i^{N_e} \sum_{\alpha}^{M_{ph}} \omega_{\alpha} p_{\alpha} \blambda_{\alpha}\cdot \br_i,\\
& \hat{H}_{self,1}=\frac{1}{2} \sum_{i=1}^{N_e} \left( \blambda_{\alpha}\cdot \br_j \right)^2, \quad
\hat{H}_{self,2}=\frac{1}{2} \sum_{i\neq j=1}^{N_e} \left( \blambda_{\alpha}\cdot \br_i \right) \left( \blambda_{\alpha}\cdot \br_j \right).
\end{align*}
The Hamiltonian \eqref{eq:qed_est:general:Hamiltonian}  has eigenstates of the form
\begin{align}
\tag{cf. \ref{eq:qed_est:general:wavefunction_qed}}
\Psi (\bx_1,...,\bx_N,p_1,...,p_{M}),
\end{align}
where $\bx=(\br,\sigma)$ are spin-spatial coordinates. The wave function $\Psi$ is antisymmetric with respect to the exchange of any two $\bx_j\leftrightarrow\bx_k$ and depends on $M$ photon-mode displacement coordinates $p_{\alpha}$ (see Sec.~\ref{sec:qed_est:general:coupled_many_body}). 

Though RDMs and their properties are quite general objects that can be defined with respect to any wave function or density matrix, their form and their role is most obvious, when we calculate expectation values. The prime example is the energy expression that we use to define the ground state of a system. According to the variational principle, the ground state of the Hamiltonian  \eqref{eq:qed_est:general:Hamiltonian} is the (possibly degenerate) state that has the lowest energy expectation value
\begin{align}
\label{eq:var_principle_psi}
E_0[\Psi]=\inf_{\Psi} \braket{\Psi|\hat{H}\Psi}.
\end{align}
We have discussed already that this minimization is not useful in practice, since it is has to be performed over the configuration space spanned by all possible many-body wave functions, which builds the exponential wall.
However, the full wave function is usually not necessary to compute the energy expectation value but typically only RDMs are sufficient. This reduces the configuration space enormously because the coordinate dependence of RDMs is \emph{independent of the particle number} (see Sec.~\ref{sec:est:rdms}).

\subsubsection*{The RDM perspective and coupled light-matter systems}
\label{eq:qed_est:rdms:qRDM_electron}

In analogy to Def.~\eqref{eq:est:rdms:pRDM_definition}, let us now define the $q$RDM for photons or, more generally, \emph{bosons} by integrating over all but $q$ particle-coordinates. For that, we need an according wave function representation as discussed in Sec.~\ref{sec:qed_est:general:coupled_many_body} ($\tilde{\Phi}$ in Eq.~\eqref{eq:qed_est:general:n_photon_wave_function}). Thus, we consider an $N_b$ boson state in the representation $\psi_b(\alpha_1,...,\alpha_{N_{b}})$ and define the 
corresponding bosonic $q$RDM
\begin{align}
\label{eq:qRDM_boson}
\Gamma_b^{(q)}(\alpha_1,...,\alpha_q;&\alpha_1',...\alpha_q') = \tfrac{N_b!}{(N_b-q)!} 
\sum_{\alpha_{q+1},...,\alpha_{N_b}=1}^{M} \nonumber\\ 
&\psi_b^*(\alpha_1',...,\alpha_q',\alpha_{q+1},...,\alpha_{N_b})  \psi_b(\alpha_1,...,\alpha_q,\alpha_{q+1},...,\alpha_{N_b}).
\end{align}
Equivalently to the electronic case, we denote the 1RDM by $\gamma_{b}(\alpha,\beta) = \Gamma_{b}^{(1)}(\alpha;\beta)$. But this definition is not sufficient to describe expectation values of the Hamiltonian \eqref{eq:Hamiltonian_BO}. Because of the form of the electron-photon interaction, the number of photons is undetermined and we need to work with Fock-space wave functions $\ket{\Phi}$. For the 1RDM, we generalize the definition \eqref{eq:qRDM_boson} as
\begin{align}
\label{eq:qed_est:rdms:1rdm_photonic}
\gamma_{b}(\alpha,\beta)  = \braket{\Phi|\hat{a}^{\dagger}_{\beta} \hat{a}_{\alpha} \Phi} = \sum_{N_b=0}^{\infty} N_b \left( \sum_{\alpha_2,...,\alpha_{N_b}=1}^{M} \psi_b^{*}(\beta, \alpha_2,...,\alpha_{N_b})  \psi_b(\alpha, \alpha_2,...,\alpha_{N_b})\right).
\end{align}
We see here why the coordinate-representation is normally not used for photon states. Since their particle-number is usually not fixed, the description becomes highly cumbersome. The abstract definition of $\ket{\Phi}$ together with the annihilation (creation) operators $\hat{a}_{\alpha}^{(\dagger)}$ are much simpler. Thus, we consider in the following the general bosonic Fock-space $q$RDM directly via
\begin{align}
\label{eq:qed_est:rdms:qRDM_boson_fock}
	\Gamma^{(q)}_{b}(\alpha_1,...,\alpha_q; \alpha_1',...,\alpha_{q}') = \braket{\Phi| \hat{a}^{\dagger}_{\alpha_1'}\cdots\hat{a}^{\dagger}_{\alpha_q'} \hat{a}^{\dphan}_{\alpha_q}\cdots\hat{a}^{\dphan}_{\alpha_1} \Phi }.
\end{align}
The fermionic and bosonic RDMs can be extended to the coupled fermion-boson case straightforwardly by just integrating/summing out the other degrees of freedom. That is, if we have a general electron-boson state of the form of Eq.~\eqref{eq:qed_est:general:wavefunction_qed} we can accordingly define $\Gamma_{e}^{(q)} \equiv \frac{N!}{(N-q)!} \sum_{\sigma_1,...,\sigma_{N}} \int \td^{3(N-q)}r \; \td^{M} p \; \Psi^* \Psi $ as well as $\Gamma^{(q)}_{b} \equiv \braket{\Psi| \hat{a}^{\dagger}_{\alpha_1'}\cdots\hat{a}^{\dagger}_{\alpha_q'}\hat{a}_{\alpha_q}\cdots\hat{a}_{\alpha_1} \Psi}$.\footnote{Note that to actually calculate the latter case, we have to employ the connection \eqref{eq:photon_wf_representation_connection}.}

In a next step, we see whether these standard ingredients of RDM theories are sufficient to express the energy expectation value of the Hamiltonian of Eq.~\eqref{eq:qed_est:general:Hamiltonian}.
For the purely electronic part, the different contributions can be expressed either explicitly by the electronic 1RDM or by the electronic 2RDM. As the general prescription of Def.~\eqref{eq:est:rdms:pRDM_definition} tells us, all expectation values of the single-particle operators of $\hat{H}$ are given in terms of the 1RDM. These are the standard electronic operators $\hat{T}$ and $\hat{V}$ but also the single-particle part of the dipole self-energy $\hat{H}_{self,1}$. We can thus write
\begin{subequations}
	\begin{align}
	\label{eq:rdms_1body_exp_vals}
	T[\gamma_{e}]&=\int\td^{3}r\left[ -\tfrac{1}{2} \nabla_{\br}^2\right]\gamma_e(\br;\br')|_{\br'=\br},\\
	V[\gamma_{e}]&=\int\td^{3}r\left[\vphantom{\sum} v(\br) \right]\gamma_e(\br;\br),\\
	H_{self,1}[\gamma_{e}]&=\int\td^{3}r\left[\sum_{\alpha=1}^{M}\tfrac{1}{2}(\blambda_{\alpha}\cdot \br)^2\right]\gamma_e(\br;\br)
	\end{align}
\end{subequations}
as functionals of $\gamma_e$. The latter two energies are actually functionals of merely the electronic density $\rho(\br)=\gamma_e(\br;\br)$. As before, the subscript $|_{\br'=\br}$ indicates that $\br'$ is set to $\br$ after the application of the semi-local single-particle operator $-\tfrac{1}{2} \nabla_{\br}^2$.
The expectation value of the electronic interaction energy $\hat{W}$ and the two-body part of the dipole self-energy $\hat{H}_d^{(2)}$ are given in terms of the (diagonal) of the 2RDM by
\begin{subequations}
	\begin{align}
	W[\Gamma^{(2)}_{e}]=&\tfrac{1}{2}\!\int\td^{3}r\td^3r' w(\br,\br')  \Gamma_e^{(2)}(\br,\br';\br,\br'),\\
	H_d^{(2)}[\Gamma^{(2)}_{e}]=&\tfrac{1}{2}\!\int\td^{3}r\td^3r'\left[\sum_{\alpha=1}^M \left(\blambda_{\alpha} \cdot \br\right)\left(\blambda_{\alpha} \cdot \br'\right)\right] \Gamma_e^{(2)}(\br,\br';\br,\br').
	\end{align}
\end{subequations}
Hence, for the electronic operator expectation values little changes in comparison to a purely fermionic problem, except that we have a coupled electron-boson wave function and the extra contributions of the dipole self-energy. For the purely bosonic part of the coupled Hamiltonian we can do the same and find (see App.~\ref{app:symmetry_photon} for further details about the Hamiltonian for different representations)
\begin{align}
\label{eq:qed_est:rdms:PhotonHamiltonian}
H_{ph}[\gamma_{b}] &=\braket{\Psi | \left\{ \sum_{\alpha=1}^M\left[- \tfrac{1}{2} \tfrac{\partial^2}{\partial p_{\alpha}^2}+ \tfrac{\omega^2_{\alpha}}{2}p_{\alpha}^2\right] \right\} \Psi}
= \sum_{\alpha=1}^{M} \left(\omega_{\alpha} + \frac{1}{2} \right) \gamma_{b}(\alpha, \alpha).
\end{align} 
We see that the photon-energy functional resembles structurally exactly the electronic one-body expressions \eqref{eq:rdms_1body_exp_vals}. This is of course due to the construction of RDMs, but nevertheless it is remarkable, since we deal with two different particle classes. The underlying exchange symmetry of an RDM is instead encoded in what we introduced in Sec.\ref{sec:est:rdms:general} as $N$-representability conditions. For the 1RDM of (fermionic or bosonic) ensembles, the conditions are simple and they are most easily expressed in the eigenbases $\gamma_{e/b}=\sum n_i^{e/b} \left(\phi^{i}_{e/b}\right)^{*}\phi^i_{e/b}$, where the $\phi^i_{e/b}$ are called the natural orbitals and the $n^{i}_{e/b}$ the natural occupation numbers. Then, the conditions are
\begin{subequations}
	\begin{align}
	\label{eq:Nrep_cond}
	0\leq \,&n_i^e \leq 1,\\
	0\leq \,&n_i^b,
	\end{align}
\end{subequations}
for fermions and bosons, respectively. Let us recall our example of the $n=n_1+...+n_M$-boson state that describes, e.g., photons occupying mode 1 with $n_1$, mode 2 with $n_2$ photons etc. From the definition of the bosonic 1RDM it is obvious that $n_i\equiv n^b_i$ in the above expression and thus, we see how the bosonic character of photons is directly transferred to the N-representability condition. In practice, this means that \emph{any} positive-semidefinite matrix can be a bosonic 1RDM, which makes the conditions much less stringent than in the fermionic case, where the Pauli-principle translates to the upper bound for the $n_i^e$. 
Additionally, if the particle number $N_{e/b}$ of one species of the system is conserved, the respective sum-rule 
\begin{align}
\label{eq:Nrep_cond2}
\sum_{i=1}^{\infty}n_i^{e/b}=N_{e/b}
\end{align}
becomes a second part of the $N$-representability conditions. Thus, in our case, we have yet another condition for the electronic part of the system, but no further bounds for the photons. We mention this here to stress that the 1RDM $N$-representability conditions for electrons reduce the configuration space of valid 1RDMs quite strongly. If we want to build some kind of RDMFT, this is on the one hand numerically challenging, because we have to test these conditions, but on the other hand it is really helpful, because then any approximation to the interaction functional needs to carry less ``information.'' We discussed in Sec.~\ref{sec:est:rdms} a similar example of the more stringent $N$-representability conditions that refer to pure-states and not only ensembles. \citet{Theophilou2015} showed that enforcing the pure-state conditions in RDMFT yield more accurate results than enforcing only the ensemble conditions for the same functional. This indicates that the functional construction in a theory that is based on the photonic 1RDM might be more difficult.

\subsubsection*{The problem of the electron-photon coupling: we need a new type of RDM}
However, the coupled electron-photon theory provides us even more intricate problems in terms of RDMs and representability conditions. The bilinear coupling term $\hat{H}_{int}$, which is the key quantity of the coupled theory cannot be treated by the  qRDMs, that we have defined in Def.~\eqref{eq:est:rdms:pRDM_definition} and Eq.~\eqref{eq:qed_est:rdms:qRDM_boson_fock}. Formally, we could write
\begin{align}
\label{eq:qed_est:rdms:rdms_3/2body_exp_val}
H_I[\Gamma^{(3/2)}_{e,b}]=&\braket{\Psi |\left[\sum_{\alpha=1}^M-\omega_{\alpha} p_{\alpha}\blambda_{\alpha} \cdot \hat{\bf D}\right]\Psi} \nonumber\\
=&\sum_{\alpha=1}^{M} - \omega_{\alpha} \braket{\Psi| \left[\sqrt{\frac{2}{\omega_{\alpha}}} \left( \hat{a}^{\dagger}_{\alpha} + \hat{a}_{\alpha}\right) \blambda_{\alpha} \cdot \hat{\bf D} \right]\Psi},
\end{align}
with a new reduced quantity that mixes light and matter degrees of freedom and can be interpreted as a ``$3/2$-body'' operator $\Gamma^{(3/2)}_{e,b}(\alpha; \br, \br')$. This can be best understood, if we also lift the continuous fermionic problem into its own Fock space and introduce genuine field operators $\hat{\psi}_e^{\dagger}(\br \sigma)$ and $\hat{\psi}_e(\br \sigma)$ with the usual anti-commutation relations. Similarly to the discussed bosonic case, the electronic RDMs can then be written in terms of strings of creation and annihilation field operators~\cite{Leeuwen2013}. We re-express 
\begin{align*}
	\braket{\Psi| \left[\left( \hat{a}^{\dagger}_{\alpha} \dagger \hat{a}_{\alpha}\right) \blambda_{\alpha} \cdot \hat{\bf D} \right]\Psi} = \sum_{\sigma}\int \td^3 r \braket{\Psi|\left[\left( \hat{a}^{+}_{\alpha} + \hat{a}_{\alpha}\right) \hat{\psi}_e^{\dagger}(\br \sigma) \hat{\psi}_e(\br \sigma) \left(\blambda_{\alpha} \cdot \br\right) \right] \Psi}
\end{align*}
and define 
\begin{align}
\label{eq:qed_est:rdms:3/2RDM}
	\Gamma^{(3/2)}_{e,b}(\alpha; \br, \br') = \sum_{\sigma} \braket{\Psi|\left[\left( \hat{a}^{\dagger}_{\alpha} + \hat{a}_{\alpha}\right) \hat{\psi}_e^{\dagger}(\br \sigma) \hat{\psi}_e(\br' \sigma) \right] \Psi}.
\end{align}
We can now re-write Eq.~\ref{eq:qed_est:rdms:rdms_3/2body_exp_val}
\begin{align}
	H_I[\Gamma^{(3/2)}_{e,b}]= \sum_{\alpha=1}^{M} - \sqrt{2\omega_{\alpha}} \int \td^3 r  \left(\blambda_{\alpha} \cdot \br\right) \Gamma^{(3/2)}_{e,b}(\alpha; \br, \br).
\end{align}
The ``property'' of the bilinear interaction term to create/annihilate bosons by interacting with the electronic subsystem is thus directly connected to the half-integer index of the corresponding RDM. This is clearly the straight-forward generalization of the $q$RDMs. However, this definition is only useful if we understand its properties.  First of all, we note that the $3/2$-body RDM has in general no simple connection to any $q$RDM, similarly to the connection formula \eqref{eq:est:rdms:rdm_connection_formula}, even if we extend the definitions to include combined matter-boson $q$RDMs. If we consider for example the integration $\int\td\br \Gamma^{(3/2)}_{e,b}(\alpha; \br, \br)$, we will have a function that depends only on $\alpha$ and thus will be related to some kind of a bosonic $1/2$-body RDM. We refrain at this point from a precise definition of such half-body objects since we will not further discuss them. However, one obvious reason for this is that $q$RDMs conserve particle numbers, while half-body RDMs do not. 

We want to finish this discussion with a simple example, that shows that the information of the photonic 1RDM $\gamma_{b}(\alpha,\beta) = \braket{\Phi|\hat{a}^{\dagger}_{\beta} \hat{a}_{\alpha} \Phi}$ is different from the photonic $1/2$RDM $\Gamma^{(1/2)}_{b}(\alpha)=\braket{\Phi| \hat{a}^{\dagger}_{\beta} \Phi}$. In the special case that $\ket{\Phi}$ consists only of coherent states for each mode (which essentially means that we have treated the photons in mean field) and since the coherent states are eigenfunctions to the annihilation operators, we find $\gamma_{b}(\alpha,\beta) = d_{\beta}^{*}d_{\alpha} $, where $d_{\alpha}$ is the total displacement of the coherent state of mode $\alpha$. In this case we also know $\braket{\Phi| \hat{a}^{\dagger}_{\beta} \Phi} = d^{*}_{\beta}$. If we now assume all but one mode, say mode 1, having zero displacement, then we only know $\gamma_{b}(1,1) = |d_1|^2$ from the bosonic 1RDM. We do, however, in general not know what $d_1^{*}$ is. For other states, such a connection is even less explicit.

\subsubsection*{Variational RDM theory for NR-QED: The representability problem}
Putting the interrelations among the different RDMs aside for the moment, the minimization for the coupled matter-boson problem can be reformulated by
\begin{align}
\label{eq:var_parinciple_RDM}
E_0 =& \inf_{\Psi} \braket{\Psi|\hat{H}\Psi} \nonumber\\
=&\inf_{\substack{\{\gamma_e,\Gamma_e^{(2)}, \gamma_b, \Gamma_{e,b}^{(3/2)}}\}\rightarrow \Psi} \left\{\left(T+V\right)[\gamma_e]+\left(W+H_d\right)[\Gamma_e^{(2)}]+H_{ph}[\gamma_{b}] +H_I[\Gamma_{e,b}^{(3/2)}] \right\}.
\end{align}
So in principle, we could replace the variation over all wave functions $\Psi$ by their respective set of RDMs needed to define the energy expectation values. Instead of varying over the full configuration space $(\bx_1,...,\bx_N,$ $p_1,...,p_M)$, the above reformulation seems to indicate that we can replace this by varying over $(\br,\br')$ for the diagonal of $\Gamma_{e}^{(2)}$ and also for the 1RDM $\gamma_{e}$ together with a variation over $(\alpha,\beta)$ for $\gamma_{b}$ and over $(\alpha,\br)$ for $\Gamma^{(3/2)}_{e,b}$. Such a reformulation is the basis of any RDM theory, and for electronic systems the properties of RDMs have been studied for more than 50 years~\cite{Coleman2000}. As we have seen in Sec.\ref{sec:est:rdms} for the electronic case, this seeming reduction of complexity is deceptive. The many-body problem has merely been shifted to the representability conditions of the RDMs. Although the conditions are simple for 1RDMs of ensembles, for any higher order RDM, they are extremely complicated. 

Let us transfer the representability question to the present case.
In order to find physically sensible results, we cannot vary arbitrarily over the above RDMs but need to ensure that they are consistent among each other and that they are all connected to \emph{the same} physical wave function. This is indicated in Eq.~\eqref{eq:var_parinciple_RDM}, where $\{\gamma_e,\Gamma_e^{(2)}, \gamma_b, \Gamma_{e,b}^{(3/2)}\}\rightarrow \Psi$ highlights that the RDMs are contractions of a common wave function. Thus, we are confronted with two new difficulties in the coupled case: consistency between the RDMs and representability with respect to one wave function. In the purely electronic problem, all the occurring RDMs are connected and thus, we can calculate the energy expectation value only by one single object, the 2RDM. Consequently, we just need to bother about the representability problem $\Gamma_e^{(2)}\rightarrow \Psi$. For the coupled case, we still have the connection between $\gamma_e$ and $\Gamma_e^{(2)}$, but we cannot connect them to the other RDMs. There is also no kind of coupled higher order RDM that by different contraction yields all the others as we discussed before. Thus, we need to treat all the three objects $\{\Gamma_e^{(2)}, \gamma_b, \Gamma_{e,b}^{(3/2)}\}$ together and find representability conditions for this set. The insights from electronic theory might be helpful here, but they are not generalizable in a straightforward way. One of the crucial new problems is the underlying Fock-space for the photon-degrees of freedom. As the name suggests, $N$-representability conditions are always defined with respect to a constant particle number. These considerations show yet from a different perspective how intricate a many-body description of the coupled-electron photon problem is, and this even in the simplest dipole-approximated case. 

\subsection{QED-RDMFT}
\label{sec:qed_est:rdms:qed_rdmft}
From the analysis of the last subsection, we gained yet another point of view on the difficulties of the electron-photon interaction. 
Transferring concepts from electronic-structure theory is also from an RDM perspective far from trivial and provides us with many interesting research questions. 
In this last part of the section, let us see how far we can get with a ``conservative'' approach that is to generalize electronic RDMFT to its QED version in the same way as \citet{Ruggenthaler2015} generalized DFT. To the best of our knowledge, the proof of the according Hohenberg-Kohn like theorem has not been published in the literature and thus, we present it in the following

We will show the proof for the full Pauli-Fierz theory with minimal coupling and only afterwards show its form for the cavity setting of Hamiltonian \eqref{eq:qed_est:general:Hamiltonian}. We follow Gilbert's proof \citep{Gilbert1975} of a one-to-one mapping between the von-Neumann density matrix of the system and the ``internal'' pair of $(\gamma(\br,\br'),\bA(\br))$, where $\gamma$ is the electronic 1RDM and $\bA$ the vector potential of the coupled electron-photon ground state. The corresponding ``external'' pair is $(v(\br,\br'),\bj(\br))$, where v is a possibly non-local external potential and $\bj$ an external charge current. As in electronic RDMFT, there is no unique $v$ corresponding to a given $\gamma$. For the sake of generality, we will present the proof in SI units.

\subsubsection*{Basic Definitions}
We consider non-relativistic QED in the setting that we introduced in Sec.~\ref{sec:intro:full_qed}, i.e., we neglect any form of spin-dependence of the Hamiltonian (e.g., due to the Stern-Gerlach term) and quantize the theory in the Coulomb gauge where for the vector potential holds $\nabla\cdot\bA=0$. Thus, the corresponding vector-potential operator is purely transversal.\footnote{For notational convenience, we use in this subsection the symbol $\bA$ to denote the transversal vector potential.} We then expand
\begin{align}
\label{eq:rdmft_qed_vector_potential}
\hat{\bA}(\br,t)=\sqrt{\tfrac{\hbar c^2}{\epsilon_0L^3}}\sum_{\bk,\sigma} \tfrac{\bepsilon_{\bk,\sigma}}{\sqrt{2\omega_{q}}}\left(\hat{a}_{\bk,\sigma}e^{\I\bk\cdot\br}+\hat{a}_{\bk,\sigma}^+e^{-\I\bk\cdot\br}\right),
\end{align}
where we assume a quantization box with side-length $L$, leading to allowed wave vectors  $\bk=2\pi \bn/L$ ($\bn\in \mathbb{Z}^3$) and corresponding frequencies $\omega_{\bk}=c|\bk|$. The index $\sigma=\{1,2\}$ denotes the transversal polarization directions and the corresponding polarization vectors obey $\epsilon_{\bk,\sigma}\cdot\epsilon_{\bk,\sigma}=\delta_{\sigma,\sigma'}$ and $\epsilon_{\bk,\sigma}\cdot\bk=0$. The expansion coefficients become the usual creation (annihilation) operators $\hat{a}_{\bk,\sigma}^{(\dagger)}$ with (transversal) commutation relations $[\hat{a}_{\bk,\sigma},\hat{a}_{\bk,\sigma}^{\dagger}]=\delta_{\sigma,\sigma'}\delta^T(\bk-\bk')$, where $\delta^T$ is the transversal delta-distribution. We denoted the vacuum permittivity with $\epsilon_0$, the speed of light with $c$ and the Planck constant with $\hbar$. We consider a classical external charge current $\bj$ that because of the transversality of $\bA$ enters also only by its transversal component. We expand $\bj$ in the modes of the quantization box, i.e.,
\begin{align}
\bj(\br)=\sqrt{\tfrac{\epsilon_0\hbar}{L^3}}\sum_{\bk,\sigma} \omega_{\bk} \tfrac{\epsilon_{\bk,\sigma}}{\sqrt{2\omega_{q}}}\left(j_{\bk,\sigma} e^{\I\bk\cdot\br}+j^*_{\bk,\sigma}e^{-\I\bk\cdot\br}\right),
\end{align} 
with expansion coefficients $\bj_{\bk,\sigma}=\bj_{-\bk,\sigma}^*=(2\omega_{\bk}^3\epsilon_0\hbar L^3)^{-1/2}\int\td^3r \epsilon_{\bk,\sigma}\cdot\bj(\br)\exp(-\I\bk\cdot\br)$. We couple this field to $N$ non-relativistic electrons. The Hamiltonian of the full coupled system reads then (cf. Def.~\ref{def:Pauli-Fierz_Hamiltonian})
\begin{align}
\label{eq:Hamiltonian_rdmftproof}
\hat{H}=&\sum_{i=1}^{N} \left\{\frac{1}{2m} \left(-\I\hbar\nabla_i + \frac{e}{c} \hat{\bA}(\br_i)\right)^2 - e v(\br_i;\br_i') \right\} + \frac{1}{2} \sum_{i,j=1}^{N} \frac{e^2}{|\br_i-\br_k|} \nonumber \\
&+ \sum_{\bk,\sigma} \hbar\omega_{\bk}(\hat{a}_{\bk,\sigma}^{\dagger}\hat{a}_{\bk,\sigma}+\tfrac{1}{2}) - \sum_{\bk,\sigma} \hbar\omega_{\bk} (\hat{a}_{\bk,\sigma} j^*_{\bk,\sigma}+\hat{a}_{\bk,\sigma}^{\dagger} j_{\bk,\sigma}),
\end{align}
where $m$ is the electron mass, and $e$ the elementary charge. We introduced now a general non-local external potential $v(\br;\br')$ that acts on a function $f(\br)$ in an integral sense as $\hat{v}f (\br)=\int\td^3r' v(\br;\br')f(\br')$. Note that for the time-independent case, introducing an external vector potential is equivalent to an external current \citep{Ruggenthaler2015}. The physical charge current that is preserved in a time-independent setting (because it obeys the continuity equation) reads
\begin{align}
\hat{\bJ}(\br)=\hat{\bj}_{P}(\br)-\frac{2}{mc} \hat{n}(\br) \hat{\bA}(\br),
\end{align}
where 
\begin{align}
\hat{\bj}_P(\br)=\frac{\I e\hbar}{2m}\sum_{i=1}^N (\delta(\br-\br_k)\overset{\rightarrow}{\nabla}_k -\overset{\leftarrow}{\nabla}_k\delta(\br-\br_k))
\end{align}
is the paramagnetic current and $\hat{n}(\br)=-e\sum_{i=1}^N \delta(\br-\br_k)$ is the charge density operator. Note that by introducing the term $\frac{e^2}{|\br_i-\br_k|}$, we assumed already the $L\rightarrow\infty$ limit. 

The Hamiltonian \eqref{eq:Hamiltonian_rdmftproof} is a hermitian operator on the Hilbert space 
\begin{align*}
\mathfrak{H}=&\mathfrak{H}_A^N \otimes \mathfrak{F},
\end{align*}
where $\mathfrak{H}_a^N$ is the antisymmetric Hilbert space of $N$ electrons and $\mathfrak{F}$ is the photon Fock space (see Sec.~\ref{sec:qed_est:general:coupled_many_body}).
We denote the eigenstates of $\hat{H}$ with $\ket{\Psi_i}$, i.e., they satisfy the static Schrödinger equation $\hat{H}\ket{\Psi_i}=E_i\ket{\Psi_i}$.

The $\ket{\Psi_i}$ can be parametrized by $N$ 4-dimensional spin-spatial coordinates $\bx=(\br,\sigma)$ and some parametrization of the photon degrees of freedom that we denote by  $\alpha=(\bk,\sigma)$. We choose this representation in analogy to the real-space representation of the electrons. 
Note however that this description is only well-defined in the mode-space and the ``back-transformation'' to real space is problematic.\footnote{The concept of a real-space photon wave function is the object of a long-standing debate and a good summary of this debate can be found in \citep[Chap. 1.5.4]{Scully1999}. For the concrete example of a one-photon-wave function in real-space and its connection to mode-space, the reader is referred to, e.g., Ref.~\citep{Sipe1995}. The author considers there the spontaneous emission process of an atom and shows that the photon wave function diverges at the position of the atom. However, for all other positions, it is well-defined and makes physically sense. This is a good example of the problem: there are several ways to define a photon wave function in real-space that are reasonable, but they are all limited in a similar way as the example of Ref.~\citep{Sipe1995}.}
We then define 
\begin{align}
\Psi(\br_1,...,\br_N,\alpha_1,\alpha_2,...)= \braket{\br_1,...,\br_N,\alpha_1,\alpha_2,...|\Psi},
\end{align}
where we leave the exact number of photons arbitrary, because it is not determined by the Hamiltonian. More generally, we will not only consider pure states, i.e., the eigenstates $\ket{\Psi_i}$ of $\hat{H}$, but ensembles of such eigenstates that are described by the density matrix
\begin{align}
D(\br_1,...,\br_N,\alpha_1,\alpha_2,...;\br_1',...,\br_N',\alpha_1',\alpha_2',...)=\sum_i w_i \Psi_i(\br_1,...,\br_N,\alpha_1,\alpha_2,...) \Psi^*_i(\br_1',...,\br_N',\alpha_1',\alpha_2',...),
\end{align}
where $0\leq w_i \leq 1$, and $\sum_i w_i =1$, making the set $P^N_{\infty}$ of all such $D$ convex (see Sec.~\ref{sec:est:rdms:general}). In the following, we are interested in the electronic one-body reduced density matrix (1RDM)
\begin{align}
\gamma(\br,\br')=&N\int\td\bx_2\cdots\td\bx_N\td\bq_1\td\bq_2\cdots \, D(\br_1,...,\br_N,\alpha_1,\alpha_2,...;\br_1',\br_2...,\br_N,\alpha_1,\alpha_2,...) \nonumber\\
=&\gamma[D],
\end{align}
and the expectation value of the vector potential operator from Eq. \eqref{eq:rdmft_qed_vector_potential}, i.e.,
\begin{align}
\bA(\br)=\trace[D\bA].
\end{align}

\subsubsection*{Gilbert's theorem for NR-QED}
With these definitions, we are now set to show the following correspondence 
\begin{empheq}[box=\fbox]{align}
	(v,\bj) \longrightarrow D_0 \overset{1:1}{\longleftrightarrow} (\gamma_0,\bA_0).
\end{empheq}
We will do this in two steps. First, we show that the pair $(v,\bj)$ determines the (non-degenerate) ground state density-matrix $D_0=\Psi_0\Psi_0^*$ (though the opposite is not true in general) and then, we will show the one-to-one correspondence between $D_0$ and the pair $(\gamma_0,\bA_0)$, where $\gamma_0=\gamma[D_0]$ and $\bA_0=\Trace[D_0\bA]$. This proof is a direct generalization of Gilbert's proof \citep{Gilbert1975} and establishes QED-RDMFT.

The first part is to show that we can associate to every pair $(v,j)$ a corresponding ground state density matrix $D_0$. This is however simple, because the pair $(v,\bj)$ consists of the only external parameters in the Hamiltonian and thus we can label the Hamiltonian as
\begin{align*}
H=H(v,\bj).
\end{align*}
Obviously, we can also label the (unique) ground state of $H(v,\bj)$ in the same way, i.e., $D_0=D_0(v,\bj)$.
However, the opposite is not true, i.e., we cannot label $v(D)$, because there are $v\neq v'$ that have the same ground state density matrix $D_0=D_0'$. Take for instance a non-interacting $N$-electron system that consists only of one-body terms. The Hamiltonian $\hat{H}=\sum_{i=1}^N=-1/2\nabla_i^2+v(\br_i;\br_i')$ of such a system per construction commutes with $\gamma=\sum_i n_i \phi_i^*(\br')\phi_i(\br)$ and thus the ground state is simply the Slater determinant of the lowest $N$ natural orbitals $\phi_1,...,\phi_N$. Consider now the nonlocal potential $v+ \delta v$, with $\delta v(\br;\br')=\sum_{ij=N+1}^{\infty}v_{ij}\phi_i^*(\br')\phi_j(\br)$ that only depends on the unoccupied natural orbitals. Obviously, $v+ \delta v$ is different from $v$, but both potentials lead to the same ground state, since they differ only in the unoccupied subspace.\footnote{Note that conversely, such a construction is not possible, if \emph{all} natural orbitals have a non-zero occupation number. In this case, the mapping is unique. It has been conjectured that this is the case for Coulomb systems, but there is still no general proof~\citep{Giesbertz2013}.}
Thus, there are infinitely many non-interacting systems with different nonlocal potential that have the same ground state and thus, one cannot construct a KS-like auxiliary system in (zero-temperature) RDMFT.\footnote{ If we however generalize the description to grand-canonical ensembles, the mapping between $\gamma$ and $v$ is indeed unique for all systems~\citep{Giesbertz2019}.} In DFT, there is instead a unique KS system for every density, since we consider only local potentials that cannot separate between the occupied and unoccupied space~\citep{Hohenberg1964}. Thus there is a one-to-one correspondence between such potentials and the density.
However, to define RDMFT (or QED-RDMFT), we do not need this unique mapping as Gilbert has pointed out~\citep{Gilbert1975}.

Let us now prove the second part of the theorem, i.e., the one-to-one correspondence between ground state density matrix $D_0$ and the internal variables $(\gamma_0,\bA_0)$. We do so by reductio ad absurdum and show that the opposite assumption leads to a contradiction. Thus, we considering two Hamiltonians $H(v^1,\bj^1),H(v^2,\bj^2)$ that have to two \emph{different} ground states $D_0^1,D_0^2$ but the same $(\gamma_0^1,\bA_0^1)=(\gamma_0^2,\bA_0^2)=(\gamma_0,\bA_0)$. We have then
\begin{align*}
E_0^1=\text{Tr}[D_0^1H(v_1,\bj_1)]<& \text{Tr}[D_0^2H(v^1,\bj^1)] \\
=& E_0^2 + \int\td\br\td\br' \gamma^2(\br,\br') \left(v^1(\br,\br')-v^2(\br,\br')\right) -\int\td\br \bA^2(\br) \left(\bj^1(\br)-\bj^2(\br) \right)\\
E_0^2=\text{Tr}[D_0^2H(v^2,\bj^2)]<& E^2=\text{Tr}[D_0^1H(v^2,\bj^2)]\\
=& E_0^1 + \int\td\br\td\br' \gamma^1(\br,\br') \left(v^2(\br,\br')-v^1(\br,\br')\right) -\int\td\br \bA^1(\br) \left(\bj^2(\br)-\bj^1(\br) \right),
\end{align*}
Since $(\gamma^1,\bj^1)=(\gamma^2,\bj^2)$, the sum of both inequalities leads to
\begin{align*}
E_1+E_2 <E_1+E_2,
\end{align*}
which is obviously a contradiction. Thus, we have proven the assumption.


We can trivially transfer this proof to the dipole-Hamiltonian \eqref{eq:qed_est:general:Hamiltonian} including an external current, cf. Eq. \eqref{eq:Hamiltonian_BO_extcurrent},  that we have employed also for the QEDFT mapping. Therefore, we have
\begin{empheq}[box=\fbox]{align}
	(v(\br,\br'),\{p_{\alpha}\}) \longrightarrow D_0\overset{1:1}{\longleftrightarrow} (\gamma(\br,\br'),\{\dot{j}_{\alpha}\})
\end{empheq}

\subsubsection*{QED-RDMFT in practice}
We want to conclude the subsection with a small discussion of the applicability of QED-RDMFT. We employ our basic setting, i.e., Hamiltonian~\eqref{eq:qed_est:general:Hamiltonian} and consider again wave functions, since the ground state of a closed system is pure and does not require the ensemble construction.

The  proof of the last paragraph is the justification to employ QED-RDMFT in practice. The first step for this is the definition of the energy functional to identify the corresponding unknown parts, especially the photon exchange correlation functional. We can directly conclude from our discussion in Sec.~\ref{sec:qed_est:rdms} that we will have to find approximations for
\begin{align}
\label{eq:qed_est:rdmft:photonXC}
	E_{XC,ph}[\gamma,\{\dot{j}_{\alpha}\}]=&\braket{\Psi[\gamma,\{\dot{j}_{\alpha}\}] |\left[\hat{H}_{int}+\hat{H}_{self,2}\right]\Psi[\gamma,\{\dot{j}_{\alpha}\}]},
\end{align}
where
\begin{align}
	\hat{H}_{int}=&-\sum_i^{N} \sum_{\alpha}^{M_{ph}} \omega_{\alpha} p_{\alpha} \blambda_{\alpha}\cdot \br_i\\
	\hat{H}_{self,2}=&\frac{1}{2} \sum_{i\neq j=1}^{N_e} \left( \blambda_{\alpha}\cdot \br_i \right) \left( \blambda_{\alpha}\cdot \br_j \right)
\end{align}
Since both these terms are local in space, the 1RDM perspective does not directly provide an advantage in comparison to the alternative QEDFT description. However, typical RDMFT functionals are more generic than DFT functionals as we have discussed in Sec.~\ref{sec:est:rdms:rdmft}. We can for example generalize straightforwardly the Müller functional to describe the self-interaction term, simply by exchanging $w(\br,\br')\rightarrow \left( \blambda_{\alpha}\cdot \br \right) \left( \blambda_{\alpha}\cdot \br' \right)$. The reason is that the Müller functional (as many other RDMFT functionals) is basically an approximation of the 2RDM in terms of the 1RDM. 

Unfortunately, since we do not yet have a good understanding of the $3/2$-body RDM (Eq.~\eqref{eq:qed_est:rdms:3/2RDM}), that is connected to the interaction term, we cannot use a similar ``trick'' for the first term in Eq.~\eqref{eq:qed_est:rdmft:photonXC}. However, in contrast to QEDFT, we have some concrete tools that we can try to apply to this problem. For instance, this suggests to investigate the still unknown representability conditions of coupled RDMs (see the Outlook in part~\ref{sec:conclusion}).

	\part{Dressed Orbitals - Old Theory in a New Basis}
\label{sec:dressed}

\begin{aquote}{Alan Watts, 1960~\citep{Watts1960}}
	You can look at something with a microscope and see it a certain way, you can look at it with a naked eye and see it in a certain way, you look at it with a telescope and you see it in another way. Now, which level of magnification is the correct one? Well, obviously, they’re all correct, but they’re just different points of view.
\end{aquote}

\newpage
We now have discussed at length the challenges of accurately describing equilibrium many-body states of non-relativistic QED. Even if we consider only one photon mode, a wave function description of the many-body space is already infeasible for very small systems. Since the electronic many-body problem is a very well-studied topic with a plethora of theories and optimized implementations that are capable to treat many systems extremely accurately, it would be very desirable to extend such methods to the coupled electron-photon problem. 
However, we have shown in Sec.~\ref{sec:qed_est:general:coupled_wavefunction} with the example of Hartree-Fock theory that such an extension is not as straightforward as one might hope. A direct generalization of the Hartree-Fock (single-reference) wave function ansatz is not capable to describe any quantum effects of the photon field. When we try to go beyond that, even a seemingly very restrictive ansatz leads to an (over-)exponentially growing problem.

In order to study to role of the quantum nature of the electron-photon interaction, we cannot avoid dealing with the \emph{multi-reference character} of the coupled states. This is an important difference to electronic problems where a single-reference ansatz, i.e., one Slater determinant already included quantum-mechanical exchange because of its anti-symmetry.
Understanding the role of such quantum effects of the photon field is especially important for polaritonic ground states, because static classical electric and magnetic fields are zero for many equilibrium settings. In the time-dependent case instead, the classical description is very accurate, when it is combined with a quantum mechanical description of the electrons. We made this observation, when we were discussing KS-QEDFT (Sec.~\ref{sec:qed_est:qedft}). QEDFT is in principle capable to treat quantum fluctuations of the photon-field by means of corresponding exchange-correlation potentials. However, it is very difficult to construct an accurate functional and the only known functional that goes beyond the MF, the ``photon OEP,'' can besides being numerically very challenging only account for one-photon processes. For instance, we cannot apply the photon OEP to larger systems and one-photon processes are not sufficient to accurately describe polaritonic ground states in the strong-coupling regime~\citep{Flick2018abinitio}. Consequently, we need new methods to appropriately study such settings.

Let us therefore take one step back and analyze again the origin of the multi-reference character of polaritonic systems. We describe electrons and photons as \emph{separate species} in their own antisymmetric and symmetric many-body spaces. These are then coupled by the interaction operator, which introduces most of the challenges of the coupled problem (see Ch.~\ref{sec:qed_est}). There is a clear analogy to correlated electronic systems, which we also describe starting from non-interacting electrons that then are coupled by the interaction. However, this coupling takes place \emph{within} one antisymmtric many-body space and this is an important difference to the electron-photon coupling. 
The central idea of the dressed(-orbital) construction is now to change the basic description of coupled electron-photon systems such that there is only one (effective) many-body space with one (effective) symmetry. This means specifically that we define a single-particle space with already coupled electron-photon orbitals, i.e., \emph{polaritonic} or \emph{dressed} orbitals and construct the many-body space as a product of such orbital spaces. We can then describe non-interacting polaritons that are coupled due to the corresponding dressed interaction. Such a programme naturally defines polaritons as its own particle class with its own statistics that needs to be respected to construct the many-polariton space. 

Clearly, restructuring the many-electron-photon space in such a way is not straightforward. In fact, the dressed construction considers an auxiliary system that is even higher-dimensional than the cavity-QED Hamiltonian. However, this extension provides the necessary flexibility to introduce an exact reformulation of the cavity-QED systems in terms of dressed orbitals. Most importantly, this allows to approach the challenges of describing coupled light-matter systems from a different perspective. Approximations to the wave function, the density description and reduced density matrices can be defined also in the dressed setting. This is yet another way to utilize these state descriptors and they will have again different properties with respect to the already discussed versions of the previous chapter. A practical consequence of this different characterization of light-matter states is that it allows for relatively simple approximation schemes. For instance, the HF approximation to the many-polariton wave function explicitly accounts for correlation between the electronic and photonic subsystems in the original description.

A key role is hereby played by the \emph{new structure} of the auxiliary Hamiltonian with respect to the standard cavity-QED Hamiltonian. Expressed in terms of polaritonic orbitals, the new Hamiltonian consists only of one-body and two-body operators. It is thus structurally equivalent to the electronic-structure Hamiltonian (Sec.~\ref{sec:est:general}). Consequently, we can generalize in principle every electronic-structure method to a ``polaritonic-structure method'' to describe cavity-QED problems. For instance, this allows to apply RDM methods to coupled light-matter systems, without the need to deal with the little-understood $3/2$-body RDM. Also functional construction for a polariton-based QEDFT is simpler, since one can more directly apply the successful strategies of electronic-structure theory. 

Another important advantage of the structural resemblance between polaritonic and electronic-structure methods is the numerical implementation. Since these methods always require to solve nonlinear equations that do not allow for an analytical treatment, we need numerical solvers to apply any method in practice. By employing the polariton description, we can simply use the existing electronic-structure codes as a basis and extend them accordingly. We do not have to develop entirely new codes.

However, to be able to introduce polaritonic orbitals, we have to increase their dimension by one with every photon mode that we take into account. For instance, to describe the standard case of cavity QED that considers one effective photon mode, the polaritonic orbitals depend on three electronic and one photonic coordinate and are thus four-dimensional. Considering such orbitals (and depending on the system size also with more modes) is still numerically feasible for modern high-performance clusters. Thus, polaritonic-structure methods will be especially useful in such cavity-QED settings.
 
A further important feature of the polariton description is the exchange-symmetry and the according statistics of polaritonic orbitals, which have a \emph{fermion-boson hybrid} character. Such a hybrid statistics is not usual in quantum many-body theory, which leads to interesting new research questions at a very basic level. In fact, we also had underestimated its role when we started the investigation of dressed orbitals. Only after we had obtained reproducible numerical results that violated the Pauli principle, we understood the significance of the hybrid statistics (see part~\ref{sec:numerics}). On the one hand, this leads to new challenges especially with regard to enforcing the statistics in practice. On the other hand, the hybrid statistics are a new tool to describe many-particle spaces of coupled species in general, which could be very valuable also for other coupled problems.

	\newpage

\chapter{The dressed-orbital construction}
\label{sec:dressed:construction}
In this chapter, we introduce the dressed-orbital construction that allows to describe coupled electron-photon systems in terms of new orbitals. This is the basis for all the following discussions. The construction requires several steps, which are nontrivial and we try to explain each of them in a way, as simple and slow as possible. Specifically, we start in Sec.~\ref{sec:dressed:construction:polaritons_correlation} with a brief motivation for a polariton-based description of cavity-QED problems. Then, we introduce the construction with an example (Sec.~\ref{sec:dressed:construction:simple}), that we generalize in Sec.~\ref{sec:dressed:construction:general}. We note that the dressed construction has been discussed in three publications, the first of which was written by my colleagues \citet{Nielsen2018} and in the other two I was the main author (\citep{Buchholz2019,Buchholz2020}). This chapter is based on Ref.~\citep{Buchholz2020}, which is the most recent publication and thus presents the most complete picture of the construction.

\section{Polaritons and correlation}
\label{sec:dressed:construction:polaritons_correlation}
We want to start the discussion with some general comments about polaritons and the multi-reference character of coupled electron-photon systems. We have discussed in the introduction (Sec.~\ref{sec:intro:experiment_strong_coupling}) that the hallmark of strong electron-photon coupling is the emergence of light-matter hybrid states or polaritons. The basic physics of such hybrid states is described by a minimal model (see Sec.~\ref{sec:intro:cavity_qed_models}) that considers two relevant electronic states, labelled by $\ket{m_1}$ (``ground'') and $\ket{m_2}$ (``excited'' state), and two photonic states, the vacuum $\ket{0}$ and one-photon state $\ket{1}$. The resulting polaritons are then denoted as
\begin{align}
\label{eq:Polariton_cQED}
P_{\pm}=\alpha \ket{m_1}\otimes\ket{1} \pm \beta\ket{m_2}\otimes\ket{0},
\end{align}
with coefficients $\alpha,\beta$ that depend on details of the model (see Sec.~\ref{sec:intro:cavity_qed_models} for further details about the family of cavity-QED models). From a first-principles perspective this means that polaritons correspond to \emph{multi-reference} or correlated wave functions, even if we described the matter states within a single-reference approach. We have discussed this in detail in Sec.~\ref{sec:qed_est:general:coupled_wavefunction}. For strongly-coupled ground states which are the focus of this thesis, the results of more general cavity-QED models~\citep{DeLiberato2017} predict that an accurate description requires even more terms, i.e., more references, than the two of the ``minimal'' polariton \eqref{eq:Polariton_cQED}. Systems that require multi-reference states are well known in electronic-structure theory and their accurate description is among the most difficult challenges in the field.\footnote{See Ch.~\ref{sec:est}, especially the last paragraph of Sec.~\ref{sec:est:dft} on strong correlation.} A classical example is the dissociation of the Hydrogen molecule (Sec.~\ref{sec:est:comparison}), where the wave function obtains a multi-reference character for larger bond distances. This prototypical system is commonly used as a challenging test case for first-principle methods~\cite{Lee1993, Lathiotakis2014, Vuckovic2015, Mordovina2020}. 

However, the term multi-reference depends crucially on the basic entities, i.e., the ``single references''  that are used to build the ``multi-reference'' state. Transferred to the problem of polaritonic physics, we face the fundamental problem that separate matter and photon wave functions~\footnote{We want to remind the reader on the subtleties regarding the definition of a photon wave function that we have discussed in Sec.~\ref{sec:qed_est:general:coupled_many_body}} (as in the simple example above) can be very inefficient in describing polaritonic states of the coupled system. We saw this for example in Sec.~\ref{sec:qed_est:general:coupled_wavefunction}, when we were discussing a possible generalized HF wave function. Even a very simplified ansatz brings back the exponential wall. We also saw that the (single-reference) KS system in equilibrium QEDFT reduces the description basically to an effective matter-only system. All effects of the photon field (besides some trivial static shift for matter systems with a permanent dipole) have to be carried by the unknown exchange-correlation potential and there is no known approximation that is accurate in the strong-coupling regime~\citep{Flick2018abinitio}. 

One important challenge for the equilibrium description is that the exchange-correlation functional only depends on matter quantities, since the photon displacement coordinate is trivial (cf. Eq.~\eqref{eq:qed_est:qedft:MWEquations} of Ch.~\ref{sec:qed_est}). From a general perspective, this is rooted in the choice of a non-interacting auxiliary system, where the electron and photon spaces decouple. Thus, we are confronted with a dilemma: On the one hand, strongly-coupled light matter systems require fundamentally a multi-reference description, because polaritons, i.e., hybrid matter-photon particles emerge as the principal degree of freedom. On the other hand, a multi-reference description is especially difficult for coupled systems, because the product space of the separate electron and photon spaces is considerably more difficult to describe than the single-species spaces. Especially, typical (and straightforward) approximations such as an effective single-reference description capture considerably less effects than the equivalent approaches in a matter-only theory.

In this chapter, we want to propose a construction that tries to mediate between the opposing sides of this dilemma. The key idea is to build the theory not as usual from separate electron and photon states but somehow introduce electron-photon hybrid states as basic entities. Before we lay out our construction to introduce such a ``many-polariton'' theory, we want to illustrate the challenges that arise, when we try to modify the standard description. Let us therefore shortly recapitulate how many-body spaces are constructed with the example of the electronic problem. We start with the single-particle Hilbert-space $\tilde{\mathfrak{h}}^1$, that is spanned by a corresponding single-particle basis $\{\tilde{\psi}_i(\br)\}$ (where we always assume bases to be orthonormalized, i.e., $\int\td\br \tilde{\psi}_i^*(\br)\tilde{\psi}_j(\br)=\delta_{ij}$), such that every single-particle state can be expressed by some superposition
\begin{align*}
\Psi(\br)=\sum_{i=1}^{\infty} c_i \tilde{\psi}_i(\br),
\end{align*}
where the $c_i$ are expansion coefficients that need to satisfy the sum rule $\sum_i|c_i|^2=1$. In the case of electrons, we need to include also the spin-degree of freedom, which we do by a tensor-product. The spin-space $\mathfrak{S}$ for a one-electron problem is just two-dimensional and we denote its basis-elements by $\{\alpha(\sigma),\beta(\sigma)\}$. The tensor-product $\mathfrak{h}=\tilde{\mathfrak{h}}\otimes \mathcal{S}$ then combines both sets to one larger basis, taking into account all possible combinations. We subsume both together in a spin-spatial coordinate
$\br\rightarrow \bx=(\br,\sigma)$ where $\sigma$ is the spin-coordinate. The expansion then becomes
\begin{align*}
\Psi(\bx)=&\sum_{j} c_{j} \psi_{j}(\bx) \nonumber\\
=&\sum_{s=\alpha,\beta}\sum_{i=1}^{\infty} c_{i,s} \psi_{i,s}(\br,\sigma) \nonumber\\
=&\sum_{i=1}^{\infty} \left[c_{i,\alpha} \tilde{\psi}_{i}(\br)\otimes\alpha(\sigma) + c_{i,\beta} \tilde{\psi}_{i}(\br)\otimes\beta(\sigma)\right].
\end{align*}
When we now want to describe two particles, we actually go exactly the same way as we have done for the spin and consider the tensor product of the single-particle spaces
\begin{align*}
\mathfrak{h}^2=\mathfrak{h}^1\otimes\mathfrak{h}^1.
\end{align*}
This two-particle Hilbert space is however still too general, because it includes \emph{distinguishable} states \linebreak (Sec.~\ref{sec:est:general}). Since electrons are fermions, we have to constrain $\mathfrak{h}^2$ to only antisymmetric combinations, i.e.,
\begin{align*}
\mathfrak{h}_A^2=\mathfrak{h}\wedge\mathfrak{h}.
\end{align*}
We can make this explicit in the state parametrization by employing Slater determinants $\Psi_{i,j}(\bx_1,\bx_2)=$ \linebreak $\frac{1}{\sqrt{2}}\left(\psi_{i}(\bx_1) \psi_{j}(\bx_2) -  \psi_{j}(\bx_1) \psi_{i}(\bx_2)\right)=|\psi_{i}\, \psi_{j}|_-$ as basis of the two-particle space. If we label every combination of i and j by a new index $I=(i,j)$, we can describe a general state $\Psi\in \mathfrak{h}_A^2$ as
\begin{align*}
\Psi(\bx_1,\bx_2)=\sum_I c_I \Psi_I(\bx_1,\bx_2),
\end{align*}
where $\sum_{I}|c_{I}|^2=1$.
The advantage of this construction is obvious: the basis already takes care about the fundamental particle symmetry. 
We have discussed at length that only this additional symmetry information includes nontrivial quantum effects in the description. 

When we want to describe coupled electron-photon problems, we lose this advantage. Electrons and photons are two different species and thus distinguishable. Accordingly, we have to consider the full tensor product between the matter and photon many-body spaces without further symmetry restrictions (Ch.~\ref{sec:qed_est:general:coupled_many_body}).
Let us contrast this to a hypothetical many-polariton space. In analogy to before, we start with the single-polariton Hilbert-space that is build by objects that have an electron coordinate $\bx$ and a photon coordinate $\bp$. Similarly to the extension of the spatial coordinates by a spin-component for electrons, we could define $\mathfrak{h}_p=\mathfrak{h}_e\otimes\mathfrak{h}_{ph}$, where $\mathfrak{h}_e$ is the electronic one-particle space from before and $\mathfrak{h}_{ph}$ is some one-particle photon space. 
Introducing a photon orbital basis $\{\chi_{\alpha}(\bp)\}$, the elements of $\mathfrak{h}_p$ are thus given by
\begin{align}
\label{eq:dressed:general:polariton_orbital}
\phi(\bx,\bp)=\sum_{i,\alpha} c_{i\alpha} \left(\psi_i(\bx)\otimes \chi_{\alpha}(\bp)\right),
\end{align}
where $\sum_{i\alpha}|c_{i\alpha}|^2=1$. Since electrons and photons are distinguishable, there is no further symmetrization required.

We have discussed in Sec.~\ref{sec:est:general:general_case} that many electronic-structure problems can be simplified, if we assume that the orbitals are the eigenfunctions of the one-body part of the Hamiltonian, ($h(\br)=-\frac{1}{2}\nabla_{\br}^2+v(\br)$). To get a feeling for the challenges of these new kind of orbitals~\eqref{eq:dressed:general:polariton_orbital}, let us try to do find a similar ``one-body'' Hamiltonian. The straightforward choice is to consider the cavity-QED Hamiltonian (Eq.~\eqref{eq:Hamiltonian_BO}) for one electron, i.e.,
\begin{align*}
\hat{h}_{p}=\frac{1}{2} \nabla^2 + v(\br)+ \frac{1}{2}\sum_{\alpha}^{M}\left[-\partial_{p_{\alpha}}^2 + \omega_{\alpha}^2 \left(p_{\alpha} + \frac{\blambda_{\alpha}}{\omega_{\alpha}}\cdot \br \right)^2\right].
\end{align*}
The eigenfunctions $\phi(\bx,\bp)$ of $\hat{h}_p$ would define our photon variable as M-dimensional, i.e., $\bp=(p_1,...,p_M)$. This dimension is determined by the system that we want to describe and as we have discussed, we can often assume even $M=1$ for a cavity system.

%
The next step is to construct the many-body space from this. According to the standard prescription, we consider
\begin{align*}
\mathfrak{h}=\mathfrak{h}_p\otimes\mathfrak{h}_p,
\end{align*}
with elements
\begin{align*}
\Phi(\bx_1\bp_1,\bx_2\bp_2)=\sum_{I,J} c_{I,J} \phi_I(\bx_1\bp_1)\phi_J(\bx_2\bp_2).
\end{align*}
Now we need to understand how we utilize $\Phi$ to describe the ground state of the actual Hamiltonian that describes two electrons, coupled to $M$ modes, i.e.,
\begin{align}
\tag{cf. \ref{eq:Hamiltonian_2e1m_example_physical}}
\hat{H}=\sum_{i=1}^{2}\frac{1}{2} \nabla_{\br_i}^2 + v(\br_i)+ \frac{1}{2}\left[-\partial_{p}^2 + \omega^2 \left(p + \frac{\blambda}{\omega}\cdot (\br_1+\br_2) \right)^2\right].
\end{align}
This confronts us with two fundamental questions: 
\begin{enumerate}
	\item How can we enforce the correct symmetry on this level?
	\item How can we deal with the additional $\bp_2$-coordinates that are not present Eq.~\eqref{eq:Hamiltonian_2e1m_example_physical}?
\end{enumerate}
In the following, we show how the dressed-orbital construction can in principle resolve both of the above questions. This allows to construct a many-polariton space (almost) of the form
\begin{align*}
\mathfrak{h}_p^N=\mathcal{S}_p(\underbrace{\mathfrak{h}_p\otimes\cdots\otimes\mathfrak{h}_p}_{N \text{times}}),
\end{align*}
where $\mathcal{S}_p$ enforces the underlying exchange-symmetry that is of a fermion-boson hybrid nature. 

	\section{Polaritons in the dressed auxiliary system}
\label{sec:dressed:construction:simple}
\begin{figure}[ht]
	\centering
	\includegraphics[width=0.7\columnwidth] {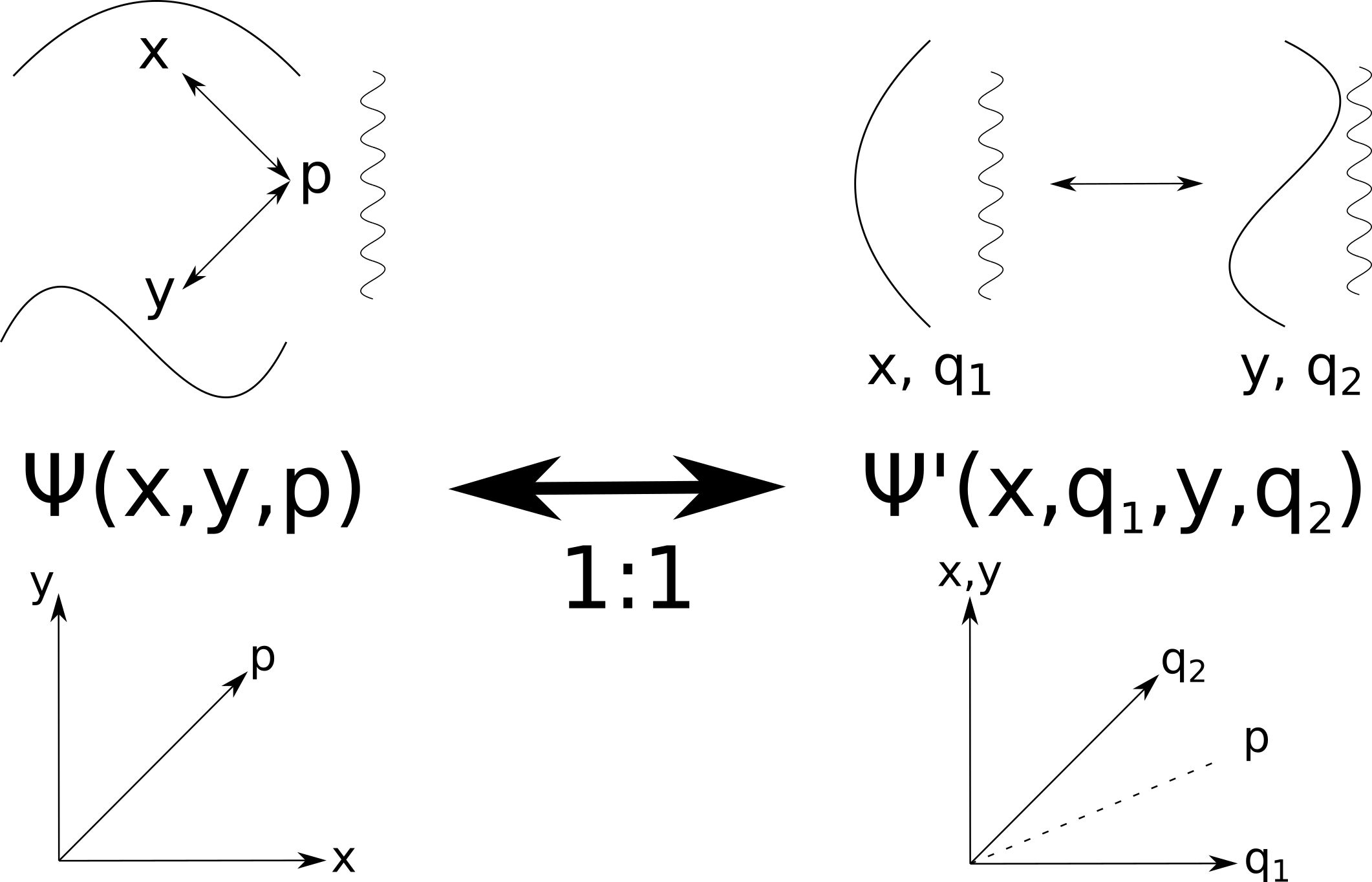}
	\caption{Sketch of the auxiliary construction for an example wave function $\Psi(x,y,p)$ of two one-dimensional electrons $(x,y)$ coupled  to one photon mode with displacement coordinate p. The coupling is indicted by the double arrows $x\leftrightarrow p, y\leftrightarrow p$. The electronic orbital wave functions are symbolized by the ground and first excited state of a box with zero-boundary conditions and the photon mode by a wiggly line. The corresponding dressed wave function $\Psi'(x,q_1,y,q_2)$ has instead two photon-coordinates $q_1,q_2$ that are related to the physical coordinate p by $p=1/\sqrt{2}(q_1+q_2)$. On the wave function level, this connection can be utilized to introduce two polariton orbitals with coordinates $(x,q_1)$ and $(y,q_2)$, respectively, that are interacting. This new interaction is indicated by a double arrow between the two orbitals.}
	\label{fig:dressing_sketch}
\end{figure}

Before we discuss the general case, we want to illustrate the dressed construction with the example of the 2-electron-1-mode system that we have considered already in Sec.~\ref{sec:qed_est:general:coupled_many_body}.\footnote{Note that the contents of this section are part of Ref.~\citep{Buchholz2020}} Since the construction requires many (sometimes quite technical) steps, we reserve this whole section only for the example system. This allows us to go through every step in detail and motivate its purpose. In the next section (Sec.~\ref{sec:dressed:construction:general}), we then recapitulate all these steps for the general case of $N$ electrons coupled to $M$ photon modes. 
The basic idea of the construction is sketched in Fig. \ref{fig:dressing_sketch}. 

\subsection{The auxiliary Hamiltonian: Matching electronic and photonic coordinates}
\label{sec:dressed:construction:simple:auxiliary_hamiltonian}
The Hamiltonian of the 2-electron-1-mode system (introduced in Sec.~\ref{sec:qed_est:general:coupled_many_body}) reads
\begin{align}
\tag{cf. \ref{eq:Hamiltonian_2e1m_example_physical}}
	\hat{H}=\sum_{k=1}^{2}\frac{1}{2} \Delta_{\br_k} + v(\br_k)+ \frac{1}{|\br_1-\br_2|} + \frac{1}{2}\left[-\partial_{p}^2 + \omega^2 \left(p + \frac{\blambda}{\omega}\cdot (\br_1+\br_2) \right)^2\right].
\end{align}
We start by expanding $\hat{H}$ in its single constituents
\begin{align}
\label{eq:dressed:construction_simple:hamiltonian_physical}
\begin{split}
	\hat{H}=&\overbrace{\sum_{k=1}^{2}\left[ -\frac{1}{2} \Delta_{\br_k}+v(\br_k) + \tfrac{1}{2}(\lambda\cdot \br_k)^2 \right]}^{\hat{H}^e=\sum_{k=1}^{2}h^e(\br_k)} \overbrace{-\sum_{k=1}^{2} \omega p \blambda\!\cdot\!\br_k}^{\hat{H}^{ep}=\sum_{k=1}^{2}h^{ep}(\br_k,p)} \\
	&+\underbrace{\tfrac{1}{2}\left(-\frac{\partial^2}{\partial p^2}+ \omega^2p^2\right)}_{\hat{H}^{ph}=h^{ph}(p)} 
	+\underbrace{\frac{1}{|\br_1-\br_2|} + \tfrac{1}{2}\left(\blambda\!\cdot\!\br_1\right)\left(\blambda\!\cdot\!\br_2\right)}_{\hat{H}^{ee}=h^{ee}(\br_1,\br_2)},
\end{split}
\end{align}
where we grouped terms according to the coordinates that are appearing.\footnote{Note that we employ in this section the symbol $\Delta\equiv \nabla^2$ for the Laplacian.} 
We observe that there is an asymmetry between electron- and photon-coordinates: there are two electron coordinates that appear in a completely symmetric way in all the terms. However there is only one mode-displacement coordinate, independently on the number of electrons.

The basics idea of the dressed construction is now to simply introduce as many \emph{artificial} photon coordinates as electron coordinates, which formally ``lifts'' the asymmetry. In this case, this means to introduce one further photon coordinate $p_2$.\footnote{For a system of $N$ electrons, we need to introduce therefore $N-1$ extra photon coordinates (for every mode).} This has to be done such that the original system cannot be influenced. We can imagine another photon cavity that is very far away from our system, but that we want to describe at the same time. If we further assume that this artificial mode has the same frequency as the physical one, the auxiliary system is described by the Hamiltonian (we denote the quantities of the auxiliary system by a prime-symbol)
\begin{align}
\label{eq:dressed:construction_simple:hamiltonian_dressed_p}
	\hat{H}'=&\hat{H} + h^{ph}(p_2) \nonumber\\
	=&\sum_{k=1}^{2}h^e(\br_k) +h^{ee}(\br_1,\br_2) -\sum_{k=1}^{2} \omega p \blambda\!\cdot\!\br_k+ h^{ph}(p) +h^{ph}(p_2).
\end{align}
If we denote by $\chi(p_2)$ an eigenstate of the additional photon Hamiltonian $h^{ph}(p_2)$ and $\Psi(\bx_1,\bx_2,p)$ is an eigenstate of $\hat{H}$, then 
\begin{align}
	\Psi'(\bx_1,\bx_2,p,p_2)=\Psi(\bx_1,\bx_2,p)\chi(p_2),
\end{align}
is an eigenstate of $\hat{H}'$. The ground state of $\hat{H}'$ is then uniquely defined and given by $\Psi'=\Psi\chi_0$ with the harmonic oscillator vacuum state $\chi_0$. 

With $\hat{H}'$, we almost achieved our goal to symmetrize electron and photon coordinates, but there is still the electron-photon interaction term $h^{ep}(\br_k,p)$ that depends only on $p$, but not on $p_2$. The basic idea to remedy this stems from the similarity of the Hamiltonian~\eqref{eq:dressed:construction_simple:hamiltonian_dressed_p} with a molecular center-of-mass Hamiltonian~\citep{Watson1968}.
Transferred to this case, we interpret $p$ as the ``center-of-displacement'' coordinate of two independent displacements $q_1,q_2$. The second key ingredient of the dressed construction is therefore a coordinate transformation $(p,p_2)\rightarrow (q_1,q_2)$ that sets
\begin{align}
\label{eq:dressed_construction_p_condition}
p \propto (q_1+q_2).
\end{align}
Obviously, the transformation should be norm-conserving to not increase the displacement artificially, but more importantly, it needs to keep the photon-energy part form-invariant, i.e.,
\begin{align}
\label{eq:dressed_construction_HOpart}
	h^{ph}(p) +h^{ph}(p_2) \rightarrow h^{ph}(q_1) + h^{ph}(q_2).
\end{align}
Since $h^{ph}(p)=\tfrac{1}{2}\left(-\frac{\td^2}{\td p}+ \omega^2p^2\right)=\left[\tfrac{1}{\sqrt{2}}(\I \frac{\td}{\td p},\omega p)\right] \cdot \left[\tfrac{1}{\sqrt{2}}(\I \frac{\td}{\td p},\omega p)\right]$ has the form of an inner product, we can simply employ \emph{an orthogonal transformation} for that. Specifically, a transformation for a set of variables $(x_1,x_2,...)\rightarrow (x_1',x_2',...)$ is orthogonal, if it leaves any expression of the form $\sum_i x_i^2\rightarrow \sum_i x_i'{}^2$ invariant. In the two-particle case, there is exactly one possible choice for that, which reads (see also Fig. \ref{fig:dressing_sketch})
\begin{align}
\label{eq:dressed:construction_simple:trafo_explicit}
\begin{split}
	p=&1/\sqrt{2}(q_1+q_2)\\
	p_2=&1/\sqrt{2}(q_1-q_2).
\end{split}
\end{align}
As constructed, the transformation changes only $\hat{H}^{ep}=\sum_{k=1}^{2}h^{ep}(\br_k,p)$, which becomes
\begin{align}
\hat{H}^{ep} \rightarrow &-\sum_{k=1}^{2} \omega/\sqrt{2}(q_1+q_2)
 \blambda\!\cdot\!\br_k \nonumber\\
 =&-\sum_{k=1}^{2} \omega/\sqrt{2} q_k \blambda\!\cdot\!\br_k -\omega/\sqrt{2} (q_1 \blambda\!\cdot\!\br_2+ q_2 \blambda\!\cdot\!\br_1).
\end{align}
Again, we grouped the terms according to the particle indices, which we can do now also for the photon coordinates. Since the auxiliary system after the transformation is symmetric under the exchange of photon indices, we can arbitrarily exchange the indices of the $q$-coordinates. Putting all terms together, we get the following new form of the auxiliary Hamiltonian
\begin{align}
\label{eq:dressed:construction_simple:hamiltonian_dressed_q}
	\hat{H}'=& \sum_{k=1}^{2}\left[ -\tfrac{1}{2} \left(\Delta_{\br_k} + \tfrac{\td^2}{\td q_k}\right) + v(\br_k) + \tfrac{\omega^2}{2}q_k^2 - \tfrac{\omega}{\sqrt{2}} q_k \blambda\!\cdot\!\br_k + \tfrac{1}{2}(\lambda\cdot \br_k)^2 \right] \nonumber \\
	&+ \frac{1}{|\br_1-\br_2|} -\tfrac{\omega}{\sqrt{2}} (q_1 \blambda\!\cdot\!\br_2+ q_2 \blambda\!\cdot\!\br_1)
	 +\tfrac{1}{2}\left(\blambda\!\cdot\!\br_1\right)\left(\blambda\!\cdot\!\br_2\right).
\end{align}
Interestingly, this Hamiltonian has the same \emph{structure} as the electronic many-body Hamiltonian~\ref{eq:est:general:hamiltonian}. We can see this more explicitly, when we introduce the new \emph{dressed} coordinate
\begin{align}
	\bz=(\br,q)
\end{align}
and the according dressed Laplacian\footnote{We do not need further differential operators here, but of course we can generalize every differential operator to the new coodinates. The dressed nabla-operator would read for example $\nabla_{\bz}=(\nabla_{\br},\frac{\td}{\td q})$.}
\begin{align}
	\Delta_{\bz}{}'=\Delta_{\br} + \frac{\td^2}{\td q}.
\end{align}
We can write then
\begin{align}
\label{eq:Hamiltonian_dressed_2p1m}
	\hat{H}'=& \sum_{k=1}^{2}\left[- \tfrac{1}{2} \Delta_{\bz_k}' + v'(\bz_k) \right] + \tfrac{1}{2} \sum_{k \neq l=1}^2 w'(\bz_k,\bz_l),
\intertext{where the dressed local potential reads}
	v'(\bz) =& v(\br)+ \tfrac{\omega^2}{2}q_k^2 - \tfrac{\omega}{\sqrt{2}} q_k \blambda\!\cdot\!\br_k + \tfrac{1}{2}(\lambda\cdot \br_k)^2 
\intertext{and the dressed interaction kernel is }
	w'(\bz_k,\bz_l)=& \tfrac{1}{|\br_k-\br_l|} -\tfrac{\omega}{\sqrt{2}} (q_k \blambda\!\cdot\!\br_l+ q_l \blambda\!\cdot\!\br_k)
	+\tfrac{1}{2}\left(\blambda\!\cdot\!\br_k\right)\left(\blambda\!\cdot\!\br_l\right).
\end{align}

\subsection{The auxiliary wave function from a first glance}
\label{sec:dressed:construction:simple:auxiliary_wf}
With this construction, we have found a way to introduce new coordinates and rewrite our basic Hamiltonian in a form that is completely symmetric with respect to the new coordinates. The next step is to define the corresponding wave function. Let us therefore recall the Pauli citation of Sec.~\ref{sec:est:excursion_symmetry}:
\begin{aquote}{\citeauthor{Pauli1933}, \citeyear{Pauli1933} \citep{Pauli1933}}
	\emph{If we deal with many similar (indistinguishable, A/N) particles, special circumstances occur that arise from the Hamilton operator being invariant under any permutations of particles.}
\end{aquote}
According to Pauli, exchange-symmetry is a consequence of the symmetry of the Hamiltonian, which indicates that we are quite close to our goal to formulate a many-polariton theory. We only need to derive the details of the ``special circumstances,'' i.e., the \emph{type} of exchange symmetry that the auxiliary wave function. In the standard description, there are only two choices: symmetric and antisymmetric wave functions.  However, we will see that the situation here is a bit more involved. 

\subsubsection*{Polaritonic symmetry}
Let us briefly recapitulate the  steps to arrive at the dressed Hamiltonian~\eqref{eq:dressed:construction_simple:hamiltonian_dressed_q}.
We start with the physical Hamiltonian (Eq.~\eqref{eq:dressed:construction_simple:hamiltonian_physical}) that depends on the photon coordinate $p$. We then add another harmonic oscillator term, depending on the auxiliary coordinate $p_2$ which is still distinguishable from the original $p$ (because of $\hat{h}^{ep}$) and obtain Eq.~\eqref{eq:dressed:construction_simple:hamiltonian_dressed_p}. Finally, the coordinate transformation~\eqref{eq:dressed:construction_simple:trafo_explicit} introduces new photon coordinates $p_1,p_2$, which enter the transformed Hamiltonian~\eqref{eq:dressed:construction_simple:hamiltonian_dressed_q} in a completely symmetric way.
Since the Hamiltonian is also symmetric in the electronic coordinates, we can deduce the polaritonic symmetry, i.e., the Hamiltonian~\eqref{eq:dressed:construction_simple:hamiltonian_dressed_q} is invariant under the exchange of the two polaritonic coordinates $\bz_1\leftrightarrow\bz_2$. 

With that we have defined all symmetries on the Hamiltonian level. Regarding the wave function \linebreak $\Psi'(\bz_1\sigma_1,\bz_2\sigma_2)$, we still have to understand whether it is symmetric or antisymmetric under $\bz_1\sigma_1\leftrightarrow\bz_2\sigma_2$. For that, we first note that  the photon ground state is inversion-symmetric, i.e.,
\begin{align}
\label{eq:dressed_construction_auxiliary_wf_requirement}
\chi(p_2) = \chi(-p_2).
\end{align}
Further, we want to preserve the antisymmetry of the physical wave function in the electronic coordinates, i.e., $\Psi(\bx_1,\bx_2,p)=-\Psi(\bx_2,\bx_1,p)$. Taken both together, we deduce that
\begin{align}
\Psi'(\bx_1,q_1,\bx_2,p,q_2)=&\hspace{9pt} \Psi(\bx_1,\bx_2,1/\sqrt{2}(q_1+q_2))\chi(1/\sqrt{2}(q_1-q_2)) \nonumber\\
\overset{\bz_1\sigma_1\leftrightarrow\bz_2\sigma_2}{=}&-\Psi(\bx_2,\bx_1,1/\sqrt{2}(q_2+q_1))\chi(1/\sqrt{2}(q_2-q_1)) \nonumber\\
\overset{Eq.~\eqref{eq:dressed_construction_auxiliary_wf_requirement}}{=}&-\Psi(\bx_2,\bx_1,1/\sqrt{2}(q_1+q_2))\chi(1/\sqrt{2}(q_1-q_2)) \nonumber\\
=&-\Psi'(\bx_2,q_2\bx_1,q_1).
\end{align}
The dressed wave function $\Psi'$ is \emph{antisymmetric} with respect to the exchange of dressed coordinates $(\bx,q)=(\br,\sigma, q)=(\bz,\sigma)$. 

Crucially, this allows us to describe the coupled electron-photon Hilbert space in a different many-body basis, i.e., in terms of Slater determinants of dressed orbitals (see also Fig. \ref{fig:dressing_sketch}). 
Therefore, let us go back to our polariton-orbital basis $\{\phi(\bz,\sigma)\}$ of Sec.~\ref{sec:dressed:construction:polaritons_correlation}. We can now explicitly follow the steps of the electronic theory: we consider a basis  $\phi_i(\bz,\sigma)$ of $\mathfrak{h}_p$. For instance, this could be the eigenstates of the single-particles Hamiltonian $h(\bz)=-\tfrac{1}{2} \Delta_{\bz}{}' + v'(\bz)$. The two-particle space is then given as the antisymmetrized tensor-product
\begin{align}
\mathfrak{h}_p^2=\mathcal{A}(\mathfrak{h}_p\otimes\mathfrak{h}_p)
\end{align}
and the elements of $\mathfrak{h}_p^2$ can be parametrized as
\begin{align}
\label{eq:dressed:construction_simple:dressed_wf_fermion_approximation}
\begin{split}
\Psi'(\bz_1,\sigma_1,\bz_2,\sigma_2)=&\sum_{ij} c_{ij} \left[\vphantom{\sum} \phi_i(\bz_1,\sigma_1)\phi_j(\bz_2,\sigma_2)- \phi_j(\bz_1,\sigma_1)\phi_i(\bz_2,\sigma_2)\right]\\
\overset{!}{=}&\sum_{i,j,\alpha,\beta}C'{}_{ij}^{\alpha,\beta}\psi_i(\bx_1)\psi_j(\bx_2) \chi^{\alpha}(q_1)\chi^{\beta}(q_2).
\end{split}
\end{align}
It is clear that since the auxiliary configuration space is much larger than the original configuration space, we made an (beyond simple systems) infeasible numerical problem even more infeasible. What the dressed construction offers us in compensation, is a \emph{new structure} of the configuration space. In terms of the above expansion, we hope that the new coefficients $c_{ij}$ are easier to approximate than the original $C_{ij}^{\alpha}$. The main advantage is that the auxiliary wave function can be expanded in terms of Slater determinants and we have a lot of methods at our disposal that are geared toward such a situation, i.e., first-principles electronic structure theories. Yet the dressed formulation allows for relatively simple approximation schemes, e.g., HF theory in terms of a single polaritonic Slater determinant (see Sec.~\ref{sec:est}). Importantly, such simple wave functions in terms of polaritonic orbitals correspond to correlated (multi-determinant) wave functions in physical space. In Sec.~\ref{sec:dressed:results}, we present calculations based on dressed orbitals which agree remarkably well with exact references even for very strong coupling strengths. We also illustrate there the explicitly correlated character of the electronic subsystem.

\subsubsection*{The explicit form of the auxiliary wave function}
Let us now have a second closer look at the wave function. In fact, we can explicitly construct the dressed wave function in the new coordinates
\begin{align}
	\Psi'(\bx_1,\bx_2,p,p_2)=\Psi(\bx_1,\bx_2,p)\chi(p_2)\rightarrow \Psi(\bx_1,\bx_2,1/\sqrt{2}(q_1+q_2))\chi(1/\sqrt{2}(q_1-q_2))
\end{align}
if we know $\Psi$.
To see this, we consider some electronic orbital-basis $\{\psi_i(\bx)\}$ and the eigenfunctions $\{\chi^{\alpha}(p)\}$ of the photon Hamiltonian $\hat{h}^{ph}$ and expand 
\begin{align*}
	\Psi(\bx_1,\bx_2,p)=\sum_{i,j,\alpha}C_{ij}^{\alpha}\psi_i(\bx_1)\psi_j(\bx_2)\chi^{\alpha}(p).
\end{align*}
The first step of the construction of the auxiliary ground state is simple. We (tensor-)multiply $\Psi$ with the ground state $\chi^0(p_2)$  of the extra Harmonic oscillator that by construction is also an eigenstate of $\hat{h}^{ph}$, i.e.,
\begin{align}
\Psi'(\bx_1,\bx_2,p,p_2)=\left[\sum_{i,j,\alpha}C_{ij}^{\alpha}\psi_i(\bx_1)\psi_j(\bx_2)\chi^{\alpha}(p)\right]\chi^0(p_2).
\end{align}
The nontrivial step is the transformation~\eqref{eq:dressed:construction_simple:trafo_explicit} to the new coordinates. However, since we know the analytical expression of the photon basis,
\begin{align}
	\chi^{\alpha}(p)=\frac{1}{\sqrt{2^{\alpha}\alpha!}}\left(\frac{\omega}{\pi}\right)^{1/4} e^{-\omega p^2/2} H_{\alpha}(\sqrt{\omega}p),
\end{align}
where $H_{\alpha}(z)=(-1)^{\alpha}e^{z^2}\frac{\td^{\alpha}}{\td z^{\alpha}}\left(e^{-z^2}\right)$ are Hermite-polynomials, we can perform the transformation~\eqref{eq:dressed:construction_simple:trafo_explicit} \emph{explicitly}. Specifically, we have to calculate terms of the form 
\begin{align*}
\chi^{\alpha}(1/\sqrt{2}(q_1+q_2))\chi^0(1/\sqrt{2}(q_1-q_2)) \propto& e^{-\omega (q_1+q_2)^2/4} H_{\alpha}(\sqrt{\omega/2}(q_1+q_2)) H_0 e^{-\omega (q_1-q_2)^2/4}\\
=&e^{-\omega (q_1+q_2)^2/4} H_{\alpha}(\sqrt{\omega/2}(q_1+q_2)) e^{-\omega (q_1-q_2)^2/4}
\end{align*}
for all $\alpha$, where we used that $H_0=1$. We start with the Gaussian part of the oscillator states and calculate explicitly
\begin{align*}
	e^{-\omega p^2/2}e^{-\omega p_2^2/2}\rightarrow& e^{-\omega (q_1+q_2)^2/4}e^{-\omega (q_1-q_2)^2/4}\\
	=&e^{-\omega q_1^2/2}e^{-\omega q_2^2/2}.
\end{align*}
The transformation merely exchanges the coordinates, because it is orthogonal (cf. Eq.~\eqref{eq:dressed_construction_HOpart}).\footnote{Thus, also for more than two particles, the Gaussian part of the states remains form-invariant under the transformation.}

For the remaining part involving the Hermite-polynomial, the transformation is more involved, but we can calculate it still analytically. Making use of the identity 
\begin{align}
	H_{\alpha}(z_1+z_2)=2^{-\alpha/2}\sum_{\beta=0}^{\alpha} \binom{\alpha}{\beta}H_{\alpha-\beta}(z_1\sqrt{2})H_{\beta}(z_2\sqrt{2}),
\end{align}
with $z_i=\sqrt{\omega/2}q_i$, we arrive after some algebra at
\begin{align}
\label{eq:dressed:example:Psi_q}
\Psi'(\bx_1,\bx_2,q_1,q_2)=&\sum_{i,j,\alpha}C_{ij}^{\alpha}\psi_i(\bx_1)\psi_j(\bx_2) \sum_{\beta=0}^{\alpha} \sqrt{\frac{\alpha!}{\beta!(\alpha-\beta)!}}\frac{1}{\sqrt{2^{\alpha}}} \chi^{(\alpha-\beta)}(q_1)\chi^{\beta}(q_2) \nonumber\\
=&\sum_{i,j,\alpha,\beta}C'{}_{ij}^{\alpha,\beta}\psi_i(\bx_1)\psi_j(\bx_2) \chi^{\alpha}(q_1)\chi^{\beta}(q_2).
\end{align}
In the last line, we introduced the polaritonic expansion coefficient $C'{}_{ij}^{\alpha,\beta}$, which is uniquely determined by the above equation. 

First of all, Eq.~\eqref{eq:dressed:example:Psi_q} highlights how the value of $p$ is ``distributed'' over the new coordinates $q_1$ and $q_2$. This shows how 2 (or in general $N$) polaritonic orbitals can carry collectively the information of only one mode ($M$ modes).
Beyond that, this explicit construction illustrates how intricate the coordinate transformation acts on the system. It is true that Eq.~\eqref{eq:dressed:example:Psi_q} is antisymmetric under $\bz_1\sigma_1\leftrightarrow\bz_2\sigma_2$ and thus it is covered by our just derived ansatz~$\Psi'(\bz_1\sigma_1,\bz_2\sigma_2)$. However, the opposite is clearly not true, i.e., not all polaritonic wave functions that are antisymmetric will also be of the form~\eqref{eq:dressed:example:Psi_q}. This is a first indication that the ansatz~\eqref{eq:dressed:construction_simple:dressed_wf_fermion_approximation} might be limited.
Nevertheless, there is one generic feature in Eq.~\eqref{eq:dressed:example:Psi_q}: it is symmetric with respect to the new coordinates $q_1,q_2$.\footnote{Note that we can in principle derive a similar (but considerably more cumbersome) expression for $N$ electrons and $M$ modes. To derive Eq.~\eqref{eq:dressed:example:Psi_q}, we only have to make use of the orthogonality and the explicit transformation of $p$. The auxiliary coordinates $p_2,p_3,...$ do not play a role because only for $p$, we have to take Hermite-polynomials $H_{\alpha}$ with $\alpha>0$ into account.} 
We discuss this in the next subsection.

\subsection{The auxiliary wave function on a second glance}
\label{sec:dressed:construction:simple:auxiliary_wf_problem}
In fact, the antisymmetric polariton space $\mathfrak{h}_p^2$ that we constructed in the previous subsection, includes states that do not have a correspondence in the physical system. Let us illustrate this with our two-electron-one-mode example, neglecting no electron-electron and electron-photon interactions. The corresponding \emph{independent particles} (IP) Hamiltonian reads
\begin{align}
\hat{H}_{IP}=\sum_{k=1}^{2}-\tfrac{1}{2}\Delta_{\br_k} + v(\br_k) \tfrac{1}{2}\left(-\frac{\td^2}{\td p}+ \omega^2p^2\right).
\end{align}
The ground state of $\hat{H}_{IP}$ is simply
\begin{align*}
\Psi(\br_1 \sigma_1,\br_2 \sigma_2,p) = \frac{1}{\sqrt{2}}\left[\vphantom{\sum}\psi_1(\br_1 \sigma_1)\psi_2(\br_2 \sigma_2)- \psi_2(\br_1 \sigma_1)\psi_1(\br_2 \sigma_2)\right] \chi_0(p),
\end{align*}
a product of a Slater determinant consisting of the two lowest eigenfunctions $\psi_{1},\psi_2$ of $[-\tfrac{1}{2}\Delta_{\br} + v(\br)]$ and $\chi_0(p)=(\frac{\omega}{\pi})^{1/4} e^{-\omega p^2/4}$, which is the ground state of a harmonic oscillator with frequency $\omega$. As shown in the last subsection, we obtain the dressed version of $\Psi$ by multiplication with another oscillator ground state $\chi_0(p_2)$ with the same frequency. Performing the coordinate transformation~\eqref{eq:dressed:construction_simple:trafo_explicit} in this case is especially simple because also$\chi_0(p)$ is a ground state. Consequently, the transformation does nothing else than replacing the coordinates $(p,p_2)$ with $(q_1,q_2)$.
The complete auxiliary ground state reads then
\begin{align*}
\Psi'(\bz_1 \sigma_1,\bz_2 \sigma_2) &=\frac{1}{\sqrt{2}} \left[\vphantom{\sum}\psi_1(\br_1 \sigma_1)\psi_2(\br_2 \sigma_2)- \psi_2(\br_1 \sigma_1)\psi_1(\br_2 \sigma_2)\right] \chi_0(q_1) \chi_0(q_2) \nonumber \\
&=\frac{1}{\sqrt{2}} \left[\vphantom{\sum}\psi_1(\br_1 \sigma_1)\chi_0(q_1) \psi_2(\br_2 \sigma_2)\chi_0(q_2) - \psi_2(\br_1 \sigma_1)\chi_0(q_1)\psi_1(\br_2 \sigma_2)\chi_0(q_2)\right] \nonumber\\
&=\frac{1}{\sqrt{2}} [\phi_{10}(\bz_1 \sigma_1)\phi_{20}(\bz_2 \sigma_2)- \phi_{20}(\bz_1 \sigma_1)\phi_{10}(\bz_2 \sigma_2)],
\end{align*}
where in the last line we subsumed the electronic and photonic orbitals with the same coordinate index to a dressed orbital, i.e., $\psi_i(\br_j \sigma_j)\chi_n(q_j) \equiv\phi_{in}(\bz_j \sigma_j)$. This wave function is obviously antisymmetric with respect to the exchange of $\bz_1 \sigma_1$ and $\bz_2 \sigma_2$. Now let us consider another wave function, 
\begin{align*}
\tilde{\Psi}'(\bz_1 \sigma_1,\bz_2 \sigma_2)&= \phi_{10}(\bz_1,\sigma_1)\phi_{11}(\bz_2,\sigma_2)- \phi_{11}(\bz_1,\sigma_1)\phi_{10}(\bz_2,\sigma_2) \\
&=\frac{1}{\sqrt{2}} \left[\vphantom{\sum}\psi_1(\br_1 \sigma_1)\chi_0(q_1) \psi_1(\br_2 \sigma_2)\chi_1(q_2) - \psi_1(\br_1 \sigma_1)\chi_1(q_1)\psi_1(\br_2 \sigma_2)\chi_0(q_2)\right] \nonumber\\
&= \psi_1(\br_1 \sigma_1)\psi_1(\br_2 \sigma_2) \left[\vphantom{\sum}\chi_0(q_1) \chi_1(q_2) - \chi_1(q_1)\chi_0(q_2)\right],
\end{align*}
that is a dressed Slater determinant and thus part of $\mathfrak{h}_p^2$. This state is however unphysical, because is \emph{violates the Pauli-principle}: both electrons occupy the same electronic orbital. Depending on their energy eigenvalues $E[\Psi']=\braket{\Psi'|\hat{H}_{IP}^2{}'\Psi'}$, a minimization of the dressed Hamiltonian without ensuring the hybrid statistics could determine either of the two wave functions as the ground state. Only if $E [\Psi']=\epsilon_1+\epsilon_2+\omega< 2 \epsilon_1+ 2\omega =E [\tilde{\Psi}']$, where $\epsilon_1 $ and $\epsilon_2$ are the eigenenergies corresponding to $\psi_1$ and $\psi_2$, a simple minimization would yield the correct solution. If, however, $\epsilon_2-\epsilon_1 > \omega$, then a minimization of the dressed problem would lead to the state $\tilde{\Psi}'$ that violates the Pauli principle. Within the dressed Slater determinant, the antisymmetry can be ``carried'' by the electronic or the photonic part of the orbital, which allows for states like $\tilde{\Psi}'$. For an interacting problem, both cases cannot be separated so easily, but the problem remains in principle the same, as we show numerically in Sec.~\ref{sec:dressed:results}. Note that this is very similar to the violations of the $N$-representability conditions in variational 2RDM theory (see Sec.~\ref{sec:est:rdms:general}). This means that we either have to make sure that $\omega$ is large compared to the electronic excitations, such that the unrestricted minimization with \emph{only} fermionic symmetry in $\bz \sigma$ picks the right wave function~\cite{Buchholz2019} (fermion ansatz), or we have to enforce the hybrid statistics. 

To guarantee the Pauli principle, we actually have to build the dressed many-body space by requiring antisymmetry only with respect to the electronic coordinates together with symmetry in terms of the photonic coordinates. We can summarize this as
\begin{subequations}
\label{eq:dressed:construction_simple:SymmetryPauliDressedFermion2p}
	\begin{alignat}{3}
	\br_1 \sigma_1&\leftrightarrow \br_2 \sigma_2 \quad&\rightarrow\quad \Psi'&\leftrightarrow-\Psi' \label{eq:SymmetryPauliDressedFermion2p_f}\\
	\label{eq:SymmetryPauliDressedFermion2p_b}
	\boldsymbol{q}_1&\leftrightarrow\boldsymbol{q}_2 &\rightarrow\quad \Psi'&\leftrightarrow\Psi'.
	\end{alignat}
\end{subequations}
This does not mean that we have to discard the wave function expansion in terms of dressed Slater determinants~\eqref{eq:dressed:construction_simple:dressed_wf_fermion_approximation}, because from \eqref{eq:dressed:construction_simple:SymmetryPauliDressedFermion2p} follows also
\begin{align}\label{eq:DressedFermionApproximation2p}
\bz_1 \sigma_1\leftrightarrow \bz_2 \sigma_2 \quad&\rightarrow\quad \Psi'\leftrightarrow-\Psi'.
\end{align}
If we require \eqref{eq:DressedFermionApproximation2p}, which builds $\mathfrak{h}_p^2$ and then constrain the space by \emph{either} \eqref{eq:SymmetryPauliDressedFermion2p_f} or \eqref{eq:SymmetryPauliDressedFermion2p_b}, we fulfil \eqref{eq:dressed:construction_simple:SymmetryPauliDressedFermion2p} equivalently. In Sec.~\ref{sec:dressed:construction:general}, we show for the general case of $N$ electrons and $M$ modes that \eqref{eq:dressed:construction_simple:SymmetryPauliDressedFermion2p} is indeed necessary and sufficient to guarantee a \emph{one-to-one correspondence} between the physical and auxiliary ground state. This establishes the polariton description of cavity QED.

Let us now illustrate how the extra condition indeed rules our the unphysical state $\tilde{\Psi}'$. We can enforce the constitutive relations on $\tilde{\Psi}'$ by adding two extra terms that exchange either electronic or photonic coordinates, which leads to
\begin{align*}
\tilde{\Psi}'_{hybrid} (\bz_1 \sigma_1,\bz_2 \sigma_2) {}={}& \overbrace{\phi_{10}(\br_1 q_1 \sigma_1)\phi_{11}(\br_2 q_2 \sigma_2)- \phi_{11}(\br_1 q_1 \sigma_1)\phi_{10}(\br_2 q_2 \sigma_2)}^{	\tilde{\Psi}'} \\
&+ \phi_{10}(\br_1 q_2 \sigma_1)\phi_{11}(\br_2 q_1 \sigma_2)- \phi_{11}(\br_1 q_2 \sigma_1)\phi_{10}(\br_2 q_1 \sigma_2)\\
{}={}& \psi_1(\br_1 \sigma_1)\psi_1(\br_2 \sigma_2) \left[\vphantom{\sum}\chi_0(q_1) \chi_1(q_2) - \chi_1(q_1)\chi_0(q_2)\right] \\
&+ \psi_1(\br_1 \sigma_1)\psi_1(\br_2 \sigma_2) \left[\vphantom{\sum}\chi_0(q_2) \chi_1(q_1) - \chi_1(q_2)\chi_0(q_1)\right] \\
{}={}& 0.
\end{align*}
Thus, we obtain the desired result of excluding the solution that violates the Pauli principle. 
We conclude that we have reached our goal and found an exact representation of the coupled electron-photon ground state in terms of polariton orbitals. 

\subsection{Polariton symmetry and wave functions}
\label{sec:dressed:construction:simple:hybrid_wf_not_possible}
The next step is to construct a theory based on these polaritons. Unfortunately, the constitutive relation ~\eqref{eq:dressed:construction_simple:SymmetryPauliDressedFermion2p} cannot be enforced on the wave function level in a practical way. For that, we would need to generalize the concept of a Slater determinant to polaritonic \emph{and} electronic coordinates, as we have done with $\tilde{\Psi}'$ to construct $\tilde{\Psi}'_{hybrid}$. In the general case, such a ``generalized Slater determinant'' for two particles is given by
\begin{align}
\label{eq:dressed:construction_simple:generalized_slater_determinant}
\Psi_{ab}'(\bz_1,\sigma_1,\bz_2,\sigma_2)=& \phi_{a}(\br_1,q_1,\sigma_1)\phi_{b}(\br_2,q_2,\sigma_2)- \phi_{b}(\br_1,q_1,\sigma_1)\phi_{a}(\br_2,q_2,\sigma_2) \nonumber\\
+& \phi_{a}(\br_1,q_2,\sigma_1)\phi_{b}(\br_2,q_1,\sigma_2)- \phi_{b}(\br_1,q_2,\sigma_1)\phi_{a}(\br_2,q_1,\sigma_2),
\end{align}
where $\phi_{a/b}$ are some (orthonormal) polariton orbitals. To see why such an ansatz is problematic, we calculate the norm-square of $\Psi_{ab}'$, i.e.,
\begin{align*}
||\Psi_{ab}'||^2=& \sum_{\sigma_1,\sigma_2}\int\td\bz_1\bz_2 \Psi_{ab}'{}^*(\bz_1,\sigma_1,\bz_2,\sigma_2) \Psi_{ab}'(\bz_1,\sigma_1,\bz_2,\sigma_2)\\
=& 4 \braket{\phi_a|\phi_a}\braket{\phi_b|\phi_b} -4 \braket{\phi_a|\phi_b}\braket{\phi_b|\phi_a} \\
& +\sum_{\sigma_1,\sigma_2}\int\td\bz_1\bz_2 \phi_{a}^*(\br_1,q_1,\sigma_1)\phi_{b}^*(\br_2,q_2,\sigma_2) \phi_{a}(\br_1,q_2,\sigma_1)\phi_{b}(\br_2,q_1,\sigma_2) + c.c. \\
&- \sum_{\sigma_1,\sigma_2}\int\td\bz_1\bz_2 \phi_{a}^*(\br_1,q_1,\sigma_1)\phi_{b}^*(\br_2,q_2,\sigma_2) \phi_{b}(\br_1,q_2,\sigma_1)\phi_{a}(\br_2,q_1,\sigma_2) - c.c.\\
& + \sum_{\sigma_1,\sigma_2}\int\td\bz_1\bz_2 \phi_{b}^*(\br_1,q_1,\sigma_1)\phi_{a}^*(\br_2,q_2,\sigma_2) \phi_{b}(\br_1,q_2,\sigma_1)\phi_{a}(\br_2,q_1,\sigma_2) + c.c. \\
&- \sum_{\sigma_1,\sigma_2}\int\td\bz_1\bz_2 \phi_{b}^*(\br_1,q_1,\sigma_1)\phi_{a}^*(\br_2,q_2,\sigma_2) \phi_{a}(\br_1,q_2,\sigma_1)\phi_{b}(\br_2,q_1,\sigma_2) - c.c.
\end{align*}
In this expression, only the first line is what would appear in a standard Slater determinant. Since we consider orthonormal orbitals, i.e., $\braket{\phi_a|\phi_b}=\delta_{ab}$, the integrals of the first term are all one and the integrals of the second term are all zero. Thus for an ordinary Slater determinant, the orthonormality condition is sufficient to fix its norm, independently of the orbitals.
However, with the terms that stem from the additional symmetry requirements, new ``mixed-index'' terms arise, whenever one or more of the according orbitals have coordinates with different indices. All these (in this case) 8 terms are actually \emph{two-body} like integrals, because we cannot separate the integrations of one set of coordinates as we have done in the first line. The computation of these integrals is nontrivial and cannot be defined by an orthonormality condition. This means that the norm of $\Psi_{ab}'$ depends on the specific form of $\phi_{a/b}$ and thus has to be calculated explicitly to normalize $\Psi_{ab}'$.

The number of such terms for an $N$-body generalized Slater determinant is given by all permutations of polaritonic and electronic coordinates, which is $N!^2$, minus the $N!$ ``ordinary'' terms. The total number of ``mixed-index'' terms,  $N!^2-N!$, grows therefore factorial with the number of particles. The normalization (and in the same way also the calculation of expectation values) of a wave function that explicitly exhibits the symmetry \eqref{eq:dressed:construction_simple:SymmetryPauliDressedFermion2p} requires the numerical calculation of (over)exponentially many nontrivial terms.
Such an explicit Slater determinant ansatz is thus infeasible in practice. We have found yet another exponential wall.

\subsection{Polariton ensembles}
\label{sec:dressed:construction:simple:hybrid_statistics}
Comparing the challenges of the dressed wave function with the difficulties to construct a ``polariton-like'' wave function within generalized MF theory (Sec.~\ref{sec:qed_est:general:coupled_wavefunction}), it seems that any many-body description based on polaritons leads not only to an exponential but even a ``factorial'' wall.
However, the dressed construction is more flexible than the standard description and we show in the following how we can exploit this flexibility to construct simple but accurate approximation strategies based on polaritonic orbitals. The key-role hereby will be played by the (ensemble) 1RDM that allows the enforce exchange-symmetry without the explicit use of, e.g., Slater determinants.

For that, we generalize the description from pure states (described by single wave functions) to ensembles, characterized by the $N$-body density matrix $\Gamma_E^N = \sum_i w_i \Psi_i^*\Psi_i$, cf. Eq.~\eqref{eq:est:rdms:vonNeumann_density_matrix},
where the $\Psi_i$ are eigenstates of the Hamiltonian and the $w_i$ are weight coefficients with $0\leq w_i \leq 1$ and $\sum_i w_i =1$ (see Sec.~\ref{sec:est:rdms:general}). One important feature of the ensemble description is that it is easier to connect RDMs (Eq.~\eqref{eq:est:rdms:pRDM_definition}) to $\Gamma_E^N$ than to a pure-state density matrix $\Gamma^N=\Psi^*\Psi$. This connection is given by the (ensemble) $N$-representability conditions. 
In other words, if a given matrix is bosonic (fermionic) $N$-representable than a corresponding $N$-body density matrix exists that is composed out of only (anti)symmetric pure state wave functions.
The idea is now to rule out unphysical states such as $\tilde{\Psi}'$ by utilizing the $N$-representability conditions of the 1RDM instead of an explicit ansatz such as the generalized Slater determinant (Eq.~\eqref{eq:dressed:construction_simple:generalized_slater_determinant}).

We start by noting that we can straightforwardly generalize the ensemble description to the dressed setting, if we utilize the eigenstates $\Psi'_i$ of $\hat{H}'$.
However, as we are only interested in the pure ground state $\Psi'$, we will only implicitly make use of the ensemble description. This is comparable to the use of the ensemble $N$-representability conditions in RDMFT (see Sec.~\ref{sec:est:rdms:rdmft}). We therefore define the \emph{dressed 1RDM}
\begin{align}
\gamma[\Psi_i'](\bz \sigma,\bz'\sigma')= \sum_{\sigma_2}\int\td\bz_2 \Psi'{}^*(\bz_1'\sigma_1',\bz_2\sigma_2) \Psi'(\bz_1\sigma_1,\bz_2\sigma_2)
\end{align} 
in terms of the dressed wave function $\Psi'$.\footnote{The generalization to ensembles would simply consist in exchanging $\Gamma^N{}'=\Psi'{}^*\Psi'$ by $\Gamma_E^N{}'=\sum_i w_i \Gamma_i^N{}'$ for a set of pure state density matrices $\Gamma_i^N{}'$.}
Since the pure state $\Psi'$ is an extreme case of an ensemble state, its Fermi statistics \emph{with respect to the polaritonic coordinates} $\bz\sigma$ (cf. Eq. \eqref{eq:DressedFermionApproximation}) is also apparent in $\gamma[\Psi']$ in the form of fermionic $N$-representability conditions.
By using the natural orbitals $\phi_i$ and the natural occupation numbers $n_i$, which are defined by the eigenvalue equation $n_i \phi_i = \hat{\gamma} \psi_{i}$, we represent $\gamma$ in its diagonal form
\begin{align}
\label{eq:Dressed1RDM_diagonal_form_2p}
\gamma[{\Psi'}](\bz \sigma, \bz' \sigma') =\sum_{i=1}^{\infty} n_i \, \phi_i^{*}(\bz' \sigma') \phi_i(\bz \sigma).
\end{align}
The fermionic $N$-representable conditions become especially simple in this representation and are given by (cf. Eq.~\eqref{eq:est:rdms:Nrep_conditions})
\begin{align}
\label{eq:NrepDressed1RDM_2p}
\begin{split}
&0 \leq \,n_i \leq 1, \quad \forall i \\
&\textstyle\sum_i n_i = N.
\end{split}
\end{align}

However, we still need to take care that the antisymmetry remains in the electronic subsystem. For that, we now go one step further and define from the dressed 1RDM the \emph{electronic 1RDM}
\begin{align}
\label{eq:dressed:construction_simple:1rdm_electronic}
\begin{split}
\gamma_e[{\Psi'}](\br \sigma, \br' \sigma') =& \int \td^M \bq \;\gamma[{\Psi'}](\br \bq \sigma,\br' \bq \sigma')\\
=& \sum_{i=1}^{\infty} n_i^{e} \, \psi_i^{e}{}^*(\br' \sigma') \psi_i^{e}(\br \sigma)
\end{split}
\end{align}
by an integration over the photonic coordinates. In the second line, we introduced the diagonal representation of $\gamma_e$ with the according natural orbitals $\psi_i^{e}$ and the natural occupation numbers $n_i^{e}$.
The Fermi statistics with regard to \emph{only the electronic coordinates} $\br\sigma$ thus becomes apparent by considering the electronic natural occupation numbers $n^e_i$\footnote{Note that the normalization of $\gamma_{e}$ to the electron number $N$ is a direct consequence of the dressed construction that considers exactly $N$ polaritons for system with $N$ electrons.}
\begin{subequations}
	\label{eq:Nrepresentability_2p}
	\begin{align}
		\label{eq:Nrepresentability_2p_bounds}
	&0 \leq n_i^{e} \leq \,1, \quad \forall i \\
	&\textstyle\sum_i n_i^{e} = N.
	\end{align}
\end{subequations}
With Eqs.~\eqref{eq:NrepDressed1RDM_2p} and ~\eqref{eq:Nrepresentability_2p} together with the definition of $\gamma_e$ in terms of $\gamma$ (Eq.~\eqref{eq:dressed:construction_simple:1rdm_electronic}), we have derived a set of \emph{necessary conditions} to guarantee the hybrid statistics. Further conditions would be necessary for a sufficient characterization of the set of polaritonic wave functions that exhibit the exchange symmetry~\eqref{eq:dressed:construction_simple:SymmetryPauliDressedFermion2p}. This is comparable to the $N$-representable conditions of the 2RDM~\citep{Mazziotti2012}, since the dressed 1RDM $\gamma(\bz;\bz')=\gamma(\br,q,\br'q')$ written in electronic and photonic variables depends on \emph{four} particle coordinates exactly as the 2RDM. Importantly, the conditions~\eqref{eq:Nrepresentability_2p} are sufficient to guarantee the antisymmetry in the fermionic sector, which is the only symmetry of the physical systems and thus the most important one. We therefore focus in the following on the conditions~\eqref{eq:Nrepresentability_2p}. 

Let us illustrate this new point of view with the example from before. The correct auxiliary ground state $\Psi'$ satisfies the conditions of Eq.~\eqref{eq:Nrepresentability_2p}, since
\begin{align*}
\gamma_e[{\Psi'}]=&\sum_{\sigma_2}\int\td\bz_2\td q_1 {\Psi'}^*(\br' q_1 \sigma',\bz_2 \sigma_2) \Psi'(\br q_1 \sigma,\bz_2 \sigma_2) \\
=& \sum_{i=1}^{2} \psi_i^*(\br'\sigma')\psi_i(\br\sigma).
\end{align*}
The two electronic orbitals are the natural orbitals of $\gamma_e[\Psi']$ with natural occupation numbers $n_1=n_2=1$. If we do the same calculation with $\tilde{\Psi}'$, we get instead
\begin{align*}
\gamma_e[ {\tilde{\Psi}'}] = 2 \psi_1^*(\br' \sigma')\psi_1(\br \sigma),
\end{align*}
which violates the $N$-representability conditions of Eq.~\eqref{eq:Nrepresentability_2p} and thus the Pauli principle. For the wave functions of an interacting system, the diagonalization of $\gamma_e[\Psi']$ will not be as trivial as for this simple example, but nevertheless the conditions \eqref{eq:Nrepresentability_2p} are \emph{sufficient} to ensure the Pauli exclusion principle in the sense that maximally one fermion can occupy a single quantum state (upper bound in Eq.~\eqref{eq:Nrepresentability_2p_bounds}).

Most importantly, we can employ these conditions to obtain a computationally tractable procedure (presented in Sec.~\ref{sec:dressed:est:prescription}) to approximately ensure the polariton statistics implied by Eq.~\eqref{eq:SymmetryPauliDressedFermion}, instead of the \emph{factorially} growing number of ``mixed-index'' orbitals.

\section{Dressed construction: the general case}
\label{sec:dressed:construction:general}
We recapitulate now the dressed construction for the general case of $N$ electrons that are coupled to $M$ photon modes. We follow hereby Ref.~\citep[Sec. 2]{Buchholz2020}. We consider the setting of cavity QED with according Hamiltonian (reordering its terms for the purposes of this section)
\begin{align}
\tag{cf. \ref{eq:Hamiltonian_BO}}
\begin{split}
\hat{H}=&\underbrace{\sum_{k=1}^N \left[ -\tfrac{1}{2} \Delta_{\br_k} + v(\br_{k}) \right] + \tfrac{1}{2}\sum_{k \neq l}^N w(\br_k,\br_l)}_{\hat{H}_m= \hat{T}[t]+ \hat{V}[v]+ \hat{W}[w]}+\underbrace{\sum_{\alpha=1}^M\left(- \tfrac{1}{2} \tfrac{\partial^2}{\partial p_{\alpha}^2}+ \tfrac{\omega_{\alpha}^2}{2}p_{\alpha}^2\right)}_{\hat{H}_{ph}} \\
&{}+ \underbrace{\sum_{\alpha=1}^M-\omega_{\alpha} p_{\alpha}\blambda_{\alpha} \cdot \hat{\mathbf{D}}}_{\hat{H}_I}
+ \underbrace{\sum_{\alpha=1}^M \tfrac{1}{2}\left(\blambda_{\alpha} \cdot \hat{\mathbf{D}}\right)^2}_{\hat{H}_d}.
\end{split}
\end{align}
The first three terms constitute the usual matter Hamiltonian of quantum mechanics, with the kinetic and external one-body parts, $\hat{T}[t]$ and $\hat{V}[v]$, respectively, and the two-body interaction term $\hat{W}[w]$. Here the kinetic term is the usual Laplacian $t(\br)=-\tfrac{1}{2} \Delta_{\br}$, the external potential $v(\br)$ is due to the attractive nuclei/ions and $w(\br,\br')$ is the electron-electron repulsion. Usually this is just taken as the free-space Coulomb interaction $w(\br,\br')=1/|\br-\br'|$, but inside a cavity the interaction can be modified~\cite{Power1982}.
The fourth term $\hat{H}_{ph}$ is the free field-energy of $M$ effective modes of the cavity. The effective modes are characterized by their displacement coordinate $p_{\alpha}$, frequency $\omega_{\alpha}$ and polarization vectors $\blambda_{\alpha}$. The latter include already the effective coupling strength $g_{\alpha}=|\blambda_{\alpha}| \sqrt{\frac{\omega_{\alpha}}{2}}\propto 1/\sqrt{V}$~\cite{Kockum2018, Ruggenthaler2018} that is proportional to the inverse square-root of the cavity mode volume V. In the dipole approximation, the coupling between light and matter is described by the bilinear term $\hat{H}_{int}$ together with the dipole self-energy term $\hat{H}_{self}$. Here the dipole operator is defined by $\hat{\mathbf{D}}=\sum_{k=1}^N \br_k$.

The ground state of $\hat{H}$ is given by a wave function
\begin{align}
\Psi (\bx_1,...,\bx_N,p_1,...,p_{M})
\end{align}
that depends on N spin-spatial electron coordinates and M photon mode-displacement coordinates and that is antisymmetric with respect to the exchange of any two electron coordinates.

\subsection{The dressed auxiliary system}
\label{sec:dressed:construction:general:auxiliary_system}
To turn this coupled electron-photon problem into an equivalent and exact dressed problem we will follow three steps: 
\begin{enumerate}	
	\item For each mode $\alpha=1 ,\dotsc, M$ and all but the first electron $i=2,\dotsc, N$, we introduce extra auxiliary coordinates $p_{\alpha,i}$. This adds $(N-1)M$ extra degrees of freedom to the problem. In this higher-dimensional auxiliary configuration space, we now consider wave functions depending on $4N+NM$ coordinates, i.e., 
	\begin{align*}
	\Psi' (\br_1 \sigma_1,\dotsc,\br_{N} \sigma_{N}, p_1,\dotsc,p_M, p_{1,2},\dotsc,p_{1,N},\dotsc,p_{M,2}, \dotsc , p_{M,N}).
	\end{align*}
	Here and in the following, we will denote all quantities in the auxiliary configuration space with a prime.
	
	\item We next construct an auxiliary Hamiltonian in the extended configuration space of the form
	\begin{align*}
	\hat{H}'=\hat{H}+\sum_{\alpha=1}^{M} \hat{\Pi}_{\alpha}
	\end{align*}
	where
	\begin{align}
	\label{eq:lin_op}
	\hat{\Pi}_{\alpha}
	= \sum_{i=2}^{N}\biggl( -\frac{1}{2} \frac{\partial^2}{\partial p_{\alpha,i}^2} + \frac{\omega_{\alpha}^2}{2}p_{\alpha,i}^2\biggr)
	\end{align}
	depends only on these new auxiliary coordinates.  This construction guarantees that the auxiliary degrees of freedom do not mix with the physical ones, which will ensure a simple and explicit connection between the physical and auxiliary system. 
	\item Finally we perform an orthogonal coordinate transformation of the physical and auxiliary photon coordinates $(p_1,...,p_{M,N}) \rightarrow (q_{1,1},...,q_{M,N})$ such that 
	\begin{align}\label{eq:CenterOfMass}
	p_{\alpha} = \tfrac{1}{\sqrt{N}} &\left( q_{\alpha,1} + \dotsb + q_{\alpha,N}  \right), \nonumber\\
	-\frac{1}{2} \frac{\partial^2}{\partial p_{\alpha}^2} + \frac{\omega_{\alpha}^2}{2} p_{\alpha}^2 + \hat{\Pi}_{\alpha}& = \sum_{i=1}^{N}\biggl(-\frac{1}{2}\frac{\partial^2}{\partial q_{\alpha,i}^2} + \frac{\omega_{\alpha}^2}{2} q_{\alpha,i}^2\biggr).
	\end{align}
	Note that the second line is automatically satisfied for any orthogonal transformation and the first line defines $p_{\alpha}$ as the ``center-of-mass'' of all the $q_{\alpha,i}$ with uniform relative masses $1/\sqrt{N}$. 
\end{enumerate}
In total, we then find the auxiliary Hamiltonian in the higher-dimensional configuration space given as
\begin{align}
\hat{H}' {}={}&\sum_{k=1}^N \left[ -\tfrac{1}{2} \Delta_{\br_k}^2 + v(\br_{k}) \right] + \tfrac{1}{2}\sum_{k \neq l} w(\br_k,\br_l)  -\sum_{\alpha=1}^M\omega_{\alpha} p_{\alpha}\blambda_{\alpha} \cdot \hat{\mathbf{D}}+ \sum_{\alpha=1}^M \tfrac{1}{2}\left(\blambda_{\alpha} \cdot \hat{\mathbf{D}}\right)^2 \nonumber\\
&+\sum_{\alpha=1}^M\left(- \tfrac{1}{2} \tfrac{\partial^2}{\partial p_{\alpha}^2}+ \tfrac{\omega_{\alpha}^2}{2}p_{\alpha}^2\right) + \sum_{\alpha=1}^M \sum_{i=2}^{N}\left( -\tfrac{1}{2} \tfrac{\partial^2}{\partial p_{\alpha,i}^2} + \tfrac{\omega_{\alpha}^2}{2}p_{\alpha,i}^2\right)
\nonumber\\
{}\overset{\eqref{eq:CenterOfMass}}{=}{}&\sum_{k=1}^{N}  \left\{-\tfrac{1}{2} \Delta_{\br_k} + v(\br_{k}) + \sum_{\alpha=1}^{M}\! \left[ -\tfrac{1}{2}\tfrac{\partial^2}{\partial q_{\alpha,k}^2} 
+ \tfrac{1}{2}\omega_{\alpha}^2q_{\alpha,k}^2  
- \tfrac{\omega_{\alpha}}{\sqrt{N}} q_{\alpha,k}(\blambda_{\alpha}\! \cdot \br_{k})  
+\tfrac{1}{2}(\blambda_{\alpha}\! \cdot \br_{k})^2 \right]  \right\} \nonumber\\
&+ \tfrac{1}{2}\sum_{k\neq l} \left[ w(\br_{k}, \br_l) + \sum_{\alpha=1}^{M} \left(  - \tfrac{\omega_{\alpha}}{\sqrt{N}} q_{\alpha,k} \blambda_{\alpha} \cdot \br_{l} - \tfrac{\omega_{\alpha}}{\sqrt{N}} q_{\alpha,l} \blambda_{\alpha} \cdot \br_{k}  +  \blambda_{\alpha}\cdot \br_{k} \blambda_{\alpha} \cdot \br_{l}  \right)\right], \nonumber
\end{align}
where we inserted the definition of the total dipole operator and reordered the expressions, such that the terms with only one index and the terms with two different indices are grouped together. 
Introducing then a $(3+M)$-dimensional polaritonic vector of space and transformed photon coordinates $\bz = \br \bq$ with $\bq \equiv (q_1,\dotsc,q_{M})$, we can rewrite the above Hamiltonian as
\begin{empheq}[box=\fbox]{align}
\label{eq:AuxiliaryHamiltonian2}
\hat{H}' &= \sum_{k=1}^{N}\left[- \tfrac{1}{2} \Delta_k' + v'(\bz_k) \right] + \tfrac{1}{2} \sum_{k \neq l} w'(\bz_k,\bz_l)\\
&= \hat{T}[t'] + \hat{V}[v'] + \hat{W}[w'], \nonumber
\end{empheq}
where we introduced the dressed one-body terms
\begin{align}
\label{eq:dressedkinetic}
t'(\bz)&=- \tfrac{1}{2} \Delta_k'\equiv -\tfrac{1}{2}\sum_{i=1}^{3}\tfrac{\partial^2}{\partial r_i^2}-\tfrac{1}{2}\sum_{\alpha=1}^{M}\tfrac{\partial^2}{\partial q_{\alpha}^2},\\ 
v'(\bz)&=v(\br)+ \sum_{\alpha=1}^{M} \left[ \tfrac{1}{2}\omega_{\alpha}^2q_{\alpha}^2  
- \tfrac{\omega_{\alpha}}{\sqrt{N}} q_{\alpha}\blambda_{\alpha}\!\! \cdot\! \br+\tfrac{1}{2}(\blambda_{\alpha}\!\! \cdot\! \br)^2 \right],\label{eq:dressedpotential}
\intertext{and the dressed two-body interaction term}
\label{eq:dressedinteraction}
w'(\bz,\bz')=& w(\br, \br') + \sum_{\alpha=1}^{M} \left[  - \tfrac{\omega_{\alpha}}{\sqrt{N}} q_{\alpha} \blambda_{\alpha} \!\!\cdot\! \br' - \tfrac{\omega_{\alpha}}{\sqrt{N}} q_{\alpha}' \blambda_{\alpha} \!\!\cdot\! \br  +  \blambda_{\alpha}\!\!\cdot\! \br \blambda_{\alpha} \!\!\cdot\! \br'  \right].
\end{align}
We see here that only the conditions~\eqref{eq:CenterOfMass}, but not the details of the coordinate transformation of step 3 are important for our construction~\cite{Buchholz2019}. The crucial part of this coordinate transformation is the replacement of $p_{\alpha}$ in the interaction terms $p_{\alpha}\blambda_{\alpha} \!\cdot\! \hat{\mathbf{D}}$. Instead of $p_{\alpha}$ only, now all $q_{\alpha,i}$ couple to the dipole of the matter system just with a rescaled coupling-strength by the factor $1/\sqrt{N}$.

\subsubsection*{The hybrid statistics of the dressed wave function}
\label{sec:dressed:construction:general:wf_hybrid_statistics}
Let us next discuss the wave function $\Psi'$ in the auxiliary configuration space. The wave function $\Psi$ in the usual configuration space is a (normalized) solution of the (time-independent) Schrödinger equation $E_0 \Psi = \hat{H} \Psi$. Since $\hat{H}'=\hat{H}+\sum_{\alpha=1
}^M\hat{\Pi}_{\alpha}$ and $\hat{\Pi}_{\alpha}$  acts only  on the auxiliary coordinates, we can simply construct
\begin{align}\label{eq:SimplePolariton}
\Psi'(\br_1 \sigma_1, \dotsc, p_{M, N}) = \Psi(\br_1 \sigma_1, \dotsc, \br_N \sigma_N;p_1,\dotsc,p_M) \chi(p_{1,2},\dotsc, p_{M, N}),
\end{align}
with $\chi$ being the (normalized) ground state of $\sum_{\alpha=1}^M \hat{\Pi}_{\alpha}$, which is a product of individual harmonic-oscillator ground states. Clearly, $\Psi'$ is a normalized solution of the auxiliary Schrödinger equation $E_0' \Psi' = \hat{H}' \Psi'$. In principle any combination of eigenstates of the auxiliary harmonic oscillators would lead to a new eigenfunction for $\hat{H}'$ but since we here focus on the ground state the natural choice is the lowest-energy solution. Rewriting this wave function in the new coordinates and employing the polaritonic coordinates $\bz \sigma \equiv \br \boldsymbol{q} \sigma$, we arrive at
\begin{align}
\label{eq:DressedManyBodyWF}
\Psi'(\br_1 \sigma_1, \dotsc,\br_N \sigma_N, q_{1,1},\dotsc, q_{M, N}) = \Psi'(\bz_1 \sigma_1,\dotsc,\bz_N \sigma_N).
\end{align}
This polaritonic wave function as the ground state of~\eqref{eq:AuxiliaryHamiltonian2} is the reformulation of the original electron-photon problem of~\eqref{eq:Hamiltonian_BO} we were looking for. Since all the new photonic coordinates belong to harmonic oscillator ground states, exchanging $p_{\alpha,i}$ with $p_{\alpha,j}$ does not change the total wave function $\Psi'$ and this property transfers to the exchange of any coordinate $q_{\alpha,i}$ and $q_{\alpha,j}$. Hence we have now a \emph{bosonic} symmetry with respect to the $\boldsymbol{q}$ coordinates. Since the electronic part of the auxiliary system is not affected by the coordinate transformation, the electronic symmetries are the same in the physical and auxiliary system, i.e., we have a \emph{fermionic} symmetry with respect to $\br \sigma$. Together these two fundamental symmetries imply that the polaritonic coordinates $\bz \sigma$ have fermionic character. The symmetries of the polaritonic wave function $\Psi'$ can  be summarized as
\begin{subequations}
	\label{eq:SymmetryPauliDressedFermion}
	\begin{empheq}[box=\fbox]{alignat=3}
	\label{eq:SymmetryPauliDressedFermion_f}
	\br_k \sigma_k&\leftrightarrow \br_l \sigma_l \quad&\rightarrow\quad \Psi'&\leftrightarrow-\Psi' \\
	\label{eq:SymmetryPauliDressedFermion_b}
	\boldsymbol{q}_k&\leftrightarrow\boldsymbol{q}_l &\rightarrow\quad \Psi'&\leftrightarrow\Psi'
	\end{empheq}
\end{subequations}
from which follows
\begin{empheq}[box=\fbox]{align}\label{eq:DressedFermionApproximation}
\bz_k \sigma_k\leftrightarrow \bz_l \sigma_l \quad&\rightarrow\quad \Psi'\leftrightarrow-\Psi'
\end{empheq}
This means that though the dressed wave function has fermionic statistics~\eqref{eq:DressedFermionApproximation} in terms of the polaritonic coordinates $\bz \sigma$, due to the \emph{constitutive relations}~\eqref{eq:SymmetryPauliDressedFermion} it actually consists of two types of particles: one with fermionic character and another with bosonic character. Consequently, the polariton wave function $\Psi'$ has a hybrid Fermi-Bose statistics. As consequences of these symmetries we find the Pauli exclusion principle for the electrons, yet for the auxiliary photon coordinates we find that many photonic auxiliary entities can occupy the same quantum state. 

Indeed, we can prove that the conditions of Eq.~\eqref{eq:SymmetryPauliDressedFermion} are necessary and sufficient to establish a one-to-one mapping between the dressed $\Psi'$ and the physical ground states $\Psi$. The first part  ``$\Psi\rightarrow\Psi'$'' is given by the dressed construction. For the second part ``$\Psi'\rightarrow\Psi$''  we show that the minimal energy state $\Psi'$ has the form $\Psi'=\Psi \chi$, cf. Eq.~\eqref{eq:SimplePolariton}. For that, we consider a trial wave function in the dressed space
\begin{align*}
\Upsilon'(\bz_1 \sigma_1,...,\bz_N \sigma_N) \equiv \Upsilon'(\br_1 \sigma_1,...,p_M; p_{1,2},...,p_{M,N})
\end{align*}
that fulfils \eqref{eq:SymmetryPauliDressedFermion}. Then it holds since $\hat{H}$ only acts on $(\br_1,\dotsc,\br_N,p_1,\dotsc,p_M)$ and $\sum_{\alpha=1}^{M}\hat{\Pi}_{\alpha}$ only acts on $p_{1,2},...,p_{M,N}$ that
\begin{align*}
\inf_{\Upsilon'}\braket{\Upsilon'|\hat{H}' \Upsilon'} &\,\,\,\,\,\geq \inf_{\Upsilon'}\braket{\Upsilon'|\hat{H} \Upsilon'} + \inf_{\Upsilon'}\braket{\Upsilon'|\sum_{\alpha=1}^{M}\hat{\Pi}_{\alpha} \Upsilon'} \\
&\overset{\eqref{eq:SymmetryPauliDressedFermion}}{=} \braket{\Psi|\hat{H} \Psi} + \braket{\chi|\sum_{\alpha=1}^{M}\hat{\Pi}_{\alpha} \chi}\\
&\,\,\,\,\,=\braket{\Psi'|\hat{H}' \Psi'}.
\end{align*}	
We have thus proven the assumption. For excited states the constitutive relations are necessary but not sufficient to single out the eigenfunctions of $\hat{H}'$ that correspond to the original Hamiltonian in terms of simple products. The reason is that we cannot make use of the variational principle as above. Let us illustrate this for the first excited state $\Psi_1$ with corresponding dressed version $\Psi_1'=\Psi_1\chi$. We could try to follow a similar line of proof as above by considering only trail wave functions $\tilde{\Upsilon}'\perp \Psi'$ that are orthogonal to the ground state. If we now search for the lowest energy solution among the $\tilde{\Upsilon}'$, the constitutive relations cannot differentiate the correct state $\Psi_1'=\Psi_1\chi$ from, e.g., the state $\tilde{\Psi}_1'=\Psi\chi_1$, where $\chi_1$ is the first excited state of the auxiliary modes.\footnote{The reader is referred to the work of \citet{Nielsen2018} for further details on excited states and the time-dependent case in general.}
However, for ground states the above proof and thus the one-to-one correspondence between physical and dressed system holds.


\subsection{Enforcing the hybrid statistics in practice: the Pauli principle and $N$-representability}
\label{sec:dressed:construction:general:hybrid_statistic_nrep}
We have discussed in the previous section that enforcing the physical conditions \eqref{eq:SymmetryPauliDressedFermion} on the polaritonic wave function directly, is not practical.
As an alternative, the physical conditions \eqref{eq:SymmetryPauliDressedFermion} are visible in the dressed 1RDM, which is given explicitly by
\begin{align}
\label{eq:Dressed1RDM}
\gamma[{\Psi'}](\bz \sigma,\bz'\sigma')= \sum_{\sigma_2,\dotsc,\sigma_N}\int\td^{3(N-1)}\bz \Psi'{}^*(\bz' \sigma',\dotsc,\bz_N \sigma_N) \Psi'(\bz \sigma,\dotsc,\bz_N \sigma_N).  
\end{align} 
The Fermi statistics of the wave function $\Psi'$ with respect to the polaritonic coordinates $\bz\sigma$~\eqref{eq:DressedFermionApproximation} is also apparent in $\gamma[\Psi']$ in the form of the $N$-representability conditions. By using the \emph{natural orbitals} $\phi_i$ and the \emph{natural occupation numbers} $n_i$, which are defined by the eigenvalue equation $n_i \,\phi_i = \hat{\gamma} \,\psi_{i}$, we represent $\gamma$ in its diagonal form $\gamma[{\Psi'}](\bz \sigma, \bz' \sigma') =\sum_{i=1}^{\infty} n_i \, \phi_i^{*}(\bz' \sigma') \phi_i(\bz \sigma)$.
The fermionic $N$-representability conditions become especially simple in this representation and are given by
\begin{empheq}[box=\fbox]{align}
\label{eq:NrepDressed1RDM}
\begin{split}
&0 \leq \,n_i \leq 1, \quad \forall i \\
&\textstyle\sum_i n_i = N.
\end{split}
\end{empheq}

From the dressed 1RDM, we can define the electronic 1RDM
\begin{align}\label{eq:GammaElectronic}
\begin{split}
\gamma_e[{\Psi'}](\br \sigma, \br' \sigma') = \int \td^M \bq \;\gamma[{\Psi'}](\br \bq \sigma,\br' \bq \sigma')
\end{split}
\end{align}
and the auxiliary photonic 1RDM
\begin{align}
\gamma_p[{\Psi'}](\bq, \bq') = \sum_{\sigma}\int \td^3 \br \, \gamma[{\Psi'}](\br \bq \sigma,\br \bq' \sigma).
\end{align}
Again, we can define the according natural orbitals $\psi_i^{e/p}$ and the natural occupation numbers $n_i^{e/p}$ by the eigenvalue equations $n_i \psi_i^{e/p} = \hat{\gamma}_{e/p} \psi_{i}^{e/p}$ and go into their diagonal representations
\begin{align}
\label{eq:gamma_electronic_eigenrep}
\gamma_e[{\Psi'}](\br \sigma, \br' \sigma') =\sum_{i=1}^{\infty} n_i^{e} \, \psi_i^{e}{}^*(\br' \sigma') \psi_i^{e}(\br \sigma)
\end{align}
and
\begin{align}
\gamma_p[{\Psi'}](\bq, \bq') =\sum_{i=1}^{\infty} n_i^{p} \, \psi_i^{p}{}^*(\bq') \psi_i^{p}(\bq).
\end{align}
The Fermi statistics with regard to only the electronic coordinates $\br\sigma$ thus becomes apparent by considering the electronic natural occupation numbers $n^e_i$
\begin{subequations}
	\label{eq:Nrepresentability}
	\begin{empheq}[box=\fbox]{align}
	\label{eq:Nrepresentability1}
	&n_i^{e} \geq \,0, \quad \forall i \\
	\label{eq:Nrepresentability2}
	&n_i^{e} \leq 1, \quad \forall i \\
	\label{eq:Nrepresentability3}
	&\textstyle\sum_i n_i^{e} = N,
	\end{empheq}
\end{subequations}
where we split the conditions in three parts for later convenience. The equivalent bosonic symmetry of the auxiliary photonic coordinates leads instead to the conditions 
\begin{subequations}
	\label{eq:NrepresentabilityEigenrepresentationPhotons}
	\begin{empheq}[box=\fbox]{align}
	\label{eq:NrepresentabilityEigenrepresentationPhotons1}
	&0 \leq \,n_i^{p} , \quad \forall i \\
	\label{eq:NrepresentabilityEigenrepresentationPhotons2}
	&\textstyle\sum_i n_i^{p} = N.
	\end{empheq}
\end{subequations}
Note that the normalization of $\gamma_{e/b}$ to the electron number $N$ is a direct consequence of the auxiliary construction that considers exactly $N$ polaritons for a system with $N$ electrons. This becomes explicitly visible in the fact that the normalization of $\gamma$ by definition transfers to $\gamma_{e/b}$, since $N=\sum_{\sigma}\int\td\bz\gamma(\bz\sigma,\bz\sigma)=\sum_{\sigma}\int\td\br\gamma_e(\br\sigma,\br\sigma)= \int\td\bq\gamma_p(\bq,\bq)$. Additionally, the lower bounds of $\gamma_{e/b}$, cf. Eqs.~\eqref{eq:Nrepresentability1} and~\eqref{eq:NrepresentabilityEigenrepresentationPhotons1}, transfer from $\gamma$, because the partial trace operation is a completely positive map~\cite{Pillis1967}. We can conclude that if~\eqref{eq:NrepDressed1RDM} is enforced, only the upper bound of the electronic 1RDM, cf. Eq.~\eqref{eq:Nrepresentability2} provides a nontrivial additional constrained.

This now shows explicitly also for an interacting wave function that at most one electron can occupy a specific quantum state, while many auxiliary photon quantities can occupy a single quantum state.  Further, the dressed 1RDM $\gamma[\Psi']$ itself has only natural occupation numbers between zero and one and is therefore fermionic, yet it contains a fermionic and a bosonic subsystem. It is important to note that the original wave function $\Psi$ did not have this simple hybrid statistics but only fermionic symmetry, since the physical $p_\alpha$ did not follow any specific statistics. Further, that we genuinely have formulated the coupled electron-photon problem in terms of hybrid quasi-particles becomes most evident by actually using single-particle (polariton) orbitals $\phi_i(\bz \sigma)$ to expand the dressed 1RDM of $\Psi'$. 

To obtain a computationally tractable procedure, we therefore use the construction presented in Sec.~\ref{sec:dressed:est:prescription} to (approximately) ensure the polariton statistics implied by Eq.~\eqref{eq:SymmetryPauliDressedFermion}, instead of the \emph{factorially} growing number of ``mixed-index'' orbitals. We will consider all fermionic density matrices in the auxiliary configuration space, which we characterize by the conditions of Eq.~\eqref{eq:NrepDressed1RDM} in terms of polaritonic orbitals $\phi_i(\bz \sigma)$. We then constrain this space by enforcing the $N$-representability conditions of Eq.~\eqref{eq:Nrepresentability} for the 1RDM of the electronic subsystem. Since this guarantees that only (ensembles of) fermionic wave functions are allowed, also the minimal energy solution (corresponds to an ensemble that) has fermionic symmetry with respect to $\br \sigma$. We remind the reader that this is the only symmetry of the physical system. Thus, it is especially important to enforce this symmetry also within an accurate approximation scheme in the dressed system. This together with the $\bz\sigma$ antisymmetry guarantees additionally the correct zero-coupling limit.  We call this construction the polariton ansatz for strong light-matter interaction. In the next section, we will based on the polariton ansatz provide a detailed prescription to generalize a given electronic-structure theory to treat ground states of coupled electron-photon systems from first principles.

\subsection{Observables in the dressed system}
\label{sec:dressed:construction:general:observables_auxiliary}
To conclude we want to briefly discuss the role of observables in the auxiliary system. 
Although we constructed the auxiliary space explicitly in a way that the physical wave function $\Psi$ can be reconstructed exactly from its dressed counterpart $\Psi'$ by integration of all auxiliary coordinates, this does not hold for all types of operators. For operators that depend only on electronic coordinates, there is no difference and we have
$\braket{\Psi|\hat{O}\Psi}=\braket{\Psi'|\hat{O}\Psi'}$.
This is not surprising because the coordinate transformation \eqref{eq:CenterOfMass} acts only on the photonic part of the system. For photonic observables instead, the transformation changes the respective operators and thus, the connection between physical and auxiliary space becomes nontrivial in general. However, at least for all observables that depend on photonic $1/2$- or 1-body expressions, there is an analytical connection. For half-body operators, i.e., any operator that depends only on the displacement of 
\begin{align}
p_{\alpha}=\frac{1}{\sqrt{N}}(q_{\alpha,1}+...+q_{\alpha,N})
\end{align}
and its conjugate 
\begin{align}
\frac{\partial}{\partial p_{\alpha}}=\frac{1}{\sqrt{N}}\left(\frac{\partial}{\partial q_{\alpha,1}}+...+\frac{\partial}{\partial q_{\alpha,N}}\right),
\end{align}
the coordinate transformation itself provides us with the connection.
For 1-body operators, this becomes slightly more involved. For example, consider the mode energy operator $\hat{H}_{ph}=\sum_{\alpha=1}^{M}\left[-\frac{1}{2}\frac{\partial^2}{\partial p_{\alpha}^2}+\frac{\omega_{\alpha}^2}{2}p_{\alpha}^2\right]\equiv \sum_{\alpha=1}^{M}\hat{h}_{\alpha}$, that we can straightforwardly generalize in the auxiliary space to $\hat{H}_{ph}'=\sum_{\alpha=1}^{M}\sum_{i=1}^{N} \left[-\frac{1}{2}\frac{\partial^2}{\partial q_{\alpha,i}^2}+ \frac{\omega_{\alpha}^2}{2} q_{\alpha,i}^2\right]$ $\equiv\sum_{\alpha=1}^{M} \hat{h}_{\alpha}'$. The connection between $\hat{H}_{ph}$ and $\hat{H}_{ph}'$ is given by the definition of the coordinate transformation \eqref{eq:CenterOfMass}, 
\begin{align}
	\hat{H}_{ph} = \hat{H}_{ph}'- \sum_{\alpha=1}^{M}\hat{\Pi}_{\alpha}.
\end{align}
Since the expectation value of  $\hat{\Pi}_{\alpha}$ is known analytically, $\braket{\Psi'|\hat{\Pi}_{\alpha}(p_{\alpha,2},...,p_{\alpha,N})\Psi'}=\braket{\chi|\hat{\Pi}_{\alpha}\chi}=(N-1)\sum_{\alpha=1}^{M}\frac{\omega_{\alpha}}{2}$, we have
\begin{align}
	E_{ph}=&\braket{\Psi|\hat{H}_{ph}\Psi}=\braket{\Psi'|\hat{H}_{ph}'\Psi'}-(N-1)\sum_{\alpha=1}^{M}\frac{\omega_{\alpha}}{2},
\intertext{that enters the total energy }
\label{eq:PhotonEnergyConnection}
	E=&\braket{\Psi|\hat{H}\Psi}=\braket{\Psi'|\hat{H}'\Psi'}-(N-1)\sum_{\alpha=1}^{M}\frac{\omega_{\alpha}}{2}.
\end{align}
This can be generalized to any operator that contains terms of the form $p_{\alpha}p_{\beta}, p_{\alpha}\frac{\partial}{\partial p_{\beta}},$ and $\frac{\partial}{\partial p_{\alpha}}\frac{\partial}{\partial p_{\beta}}$, where $\alpha$ and $\beta$ denote any two modes. The reason is that the transformation $\eqref{eq:CenterOfMass}$ preserves the standard inner product of the Euclidean space of the mode plus extra coordinates. This transfers also to their conjugates and combinations of both. From the above, it is straightforward to derive also the expression for the occupation of mode $\alpha$, $\hat{N}^{\alpha}_{ph}=\frac{1}{\omega_{\alpha}}\hat{h}_{\alpha}-\frac{1}{2}$. In the auxiliary system, we have
\begin{align}
	\label{eq:mode_occupation_formula}
	\hat{N}^{\alpha}_{ph}=\frac{1}{\omega_{\alpha}}\hat{h'}_{\alpha}-\frac{N}{2}.
\end{align}
	\newpage
\chapter{Polaritons from first principles}
\label{sec:dressed:est}
In the last chapter, we have defined the theoretical framework to describe coupled matter-photon systems with polaritons as the fundamental entity. Now we need to discuss how to use the framework in practice. The key behind the application of the dressed construction lies in the fact that the polaritons are ``almost electrons'' - with an additional symmetry - and the Hamiltonian in terms of these polaritons is also almost the Hamiltonian of standard electronic-structure theory. In this chapter, we lay out how we can exploit this by providing a general prescription of how to turn \emph{any} given electronic-structure theory into a ``polaritonic-structure theory'' (Sec.~\ref{sec:dressed:est:prescription}). This means that we keep the approximations to the Coulomb-interaction within the electronic theory exactly as they are but use them to approximate the in the previous section derived interaction between the polaritons. Sec.~\ref{sec:dressed:est:prescription} is based on chapter 3 of \citep{Buchholz2020}. Then, we exemplify this prescription for two examples. In Sec.~\ref{sec:dressed:est:HF}, we introduce polaritonic HF, followed by polaritonic RDMFT in Sec.~\ref{sec:dressed:est:RDMFT}. 

\section{Polaritonic structure theory}
\label{sec:dressed:est:prescription}
In this section, we lay out in detail how one can transform a given
electronic-structure theory that meets some minimal requirements into its polaritonic version. The goal of such a ``polaritonic-structure theory'' is to find the ground state of the cavity-QED Hamiltonian of Eq.~\eqref{eq:Hamiltonian_BO} by considering the ground state of the auxiliary Hamiltonian of Eq.~\eqref{eq:AuxiliaryHamiltonian2}. Let us start by defining the according variational principle for the ground-state energy $E_0'$ as
\begin{align}
\label{eq:VariationalPrinciplePauliDressedFermions}
E_0' = \inf_{\Psi' \,\in\, \mathfrak{P}} \braket{\Psi' |\hat{H}'\Psi'},
\end{align}
where $\mathfrak{P}=\{\Psi' : \Psi' \leftrightarrow \text{\eqref{eq:SymmetryPauliDressedFermion}} \}$ is the set of all normalized many-polariton wave functions that obey the constitutive relations of Eq.~\eqref{eq:SymmetryPauliDressedFermion}. For our purposes, as explained in Sec.~\ref{sec:dressed:construction:general}, we will instead consider the larger set $\mathfrak{M}$ of all (mixed-state) density matrices $\Gamma = \sum_{j}w_j \ket{\Psi'_j}\bra{\Psi_j'}$ with $\sum_{j}w_j =1$, that obey the hybrid Fermi-Bose statistics. The minimal energy also in this more general set corresponds to the pure state of Eq.~\eqref{eq:VariationalPrinciplePauliDressedFermions}, i.e., 
\begin{align}
\label{eq:VariationalPrinciplePauliDressedFermionsGamma}
E_0' = \inf_{\Gamma \,\in\, \mathfrak{M}} \Trace\{ \Gamma \hat{H} \}.
\end{align}
The main trick now is in how we approximate this set. We do so by first considering the yet larger set $\tilde{\mathfrak{M}} = \{\tilde{\Gamma} : \tilde{\Psi'_j} = \sum C_{j,K} \Phi_K \}$, i.e., density matrices made of superpositions of Slater determinants $\Phi_K=\text{det}(\phi_{K,1}\cdots\phi_{K,N})/\sqrt{N!}$ of polariton orbitals $\phi_{K,i}$. This guarantees the overall Fermi statistics in terms of the polaritonic coordinates $\bz \sigma$. We then constrain this larger set to
\begin{align}
\label{eq:PolaritonSpace}
\mathcal{M}^\prime=\{\tilde{\Gamma} \in \tilde{\mathfrak{M}} :  n^e_i[{\tilde{\Gamma}}] \leq 1 \},
\end{align}
where $n^e_i[{\tilde{\Gamma}}]$ are the natural occupation numbers, cf. Eq.~\eqref{eq:gamma_electronic_eigenrep}, of the electronic 1RDM $\gamma_e[\tilde{\Gamma}] = \sum_{j} w_j \gamma_e[\tilde{\Psi'_j}]$ that depend on $\tilde{\Gamma}$. This enforces the fermionic statistics with respect to the electronic coordinates $\br \sigma$. The rest of the $N$-representability conditions (Eqs.~\eqref{eq:Nrepresentability1} and~\eqref{eq:Nrepresentability3}) are satisfied automatically by choosing $\Psi'\in \tilde{\mathcal{P}}$ as fermionic with respect to polariton coordinates and thus the corresponding dressed 1RDM satisfies the $N$-representability conditions (Eq. \eqref{eq:NrepDressed1RDM}).  This guarantees that the electronic and photonic 1RDMs of the system are $N$-representable, and thus, e.g., the electronic Pauli-principle is enforced. However, higher order RDMs are not treated exactly, which is an interesting topic for future research. We thus avoid the direct construction of the exponentially growing correlated electron-photon states.

As we have pointed out already, the polariton picture gives any coupled problem that is described by Hamiltonian~\eqref{eq:Hamiltonian_BO} the \emph{same structure} as a purely electronic problem with two-body interactions, i.e., the electronic-structure Hamiltonian (Eq.~\ref{eq:est:general:hamiltonian}). Consequently, we can transfer every type of electronic-structure theory to the coupled electron-photon problem, if the theory provides an expression for the 1RDM (since we need the 1RDM to test the $N$-representability constraints). 
\begin{figure}
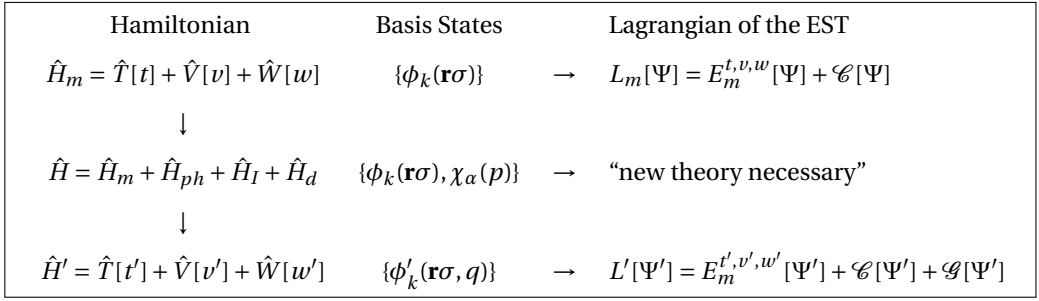

	\centering
	\fbox{
		\begin{tabular}{cccl}
			Hamiltonian & Basis States & & Lagrangian of the EST \\
			\rule{0pt}{15pt}
			$\hat{H}_m = \hat{T}[t] +\hat{V}[v]+\hat{W}[w]$ & 
			$\{\phi_k(\br\sigma)\}$ &$\rightarrow$ 
			& $L_m[\Psi] = E_m^{t,v,w}[\Psi] + \mathcal{C}[\Psi]$ \\
			\rule{0pt}{15pt}
			$\downarrow$ & & & \\
			\rule{0pt}{15pt}
			$\hat{H}=\hat{H}_m + \hat{H}_{ph} + \hat{H}_I + \hat{H}_d$ & 
			$\{\phi_k(\br\sigma),\chi_{\alpha}(p)\}$ &$\rightarrow$ & ``new theory necessary'' \\
			\rule{0pt}{15pt}
			$\downarrow$ & & & \\
			\rule{0pt}{15pt}
			$\hat{H}'= \hat{T}[t'] +\hat{V}[v']+\hat{W}[w']$ &
			$\{\phi_k'(\br\sigma,q)\}$ &$\rightarrow$ & $L'[\Psi'] = E_m^{t',v',w'}[\Psi'] + \mathcal{C}[\Psi'] + \mathcal{G}[\Psi']$
		\end{tabular}
	}
	\caption{Graphical illustration of the polariton construction and its connection to an electronic-structure theory (EST). Here $E_m$ indicates the energy expression of the EST, such as the HF, configuration interaction, or coupled cluster energy functional, and $\mathcal{C}$ indicates the constraints of the EST, such as orthonormality of the orbitals. They are enforced on a (possibly multideterminantal) wave function $\Psi$ constructed from an electronic single-particle basis $\phi_k$. Further, $\mathcal{G}$ indicates the new constraints that arise due to the hybrid statistics of the polaritons, which are now enforced on a (possibly multideterminantal) wave function $\Psi'$ of polaritonic single-particle orbitals $\phi'_k$. The usual coupled electron-photon problem (second line) has a Hamiltonian with a different structure and is built on separate orbitals $\phi_k$ and $\chi_\alpha$. Thus, a new (efficient and accurate) approximate energy expression would be needed.}
	\label{fig:polariton_construction}
\end{figure}
The main steps how to do so are depicted in Fig.~\ref{fig:polariton_construction}. We assume that the theory provides us an energy expression $E_m$ with respect to a set of electronic basis states $\{\phi_k\}$. This requirement is met by basically every electronic-structure theory, as for instance the single-reference methods KS-DFT or HF but also more involved (and numerically expensive) approaches like coupled cluster, valence bond theory or configuration interaction.\footnote{A standard reference for these and other quantum chemical methods is \citet{Helgaker2000}.} Depending on the specific theory, $E_m$ might have quite different forms, but it is always derived from some many-body Hamiltonian $\hat{H}_m=\hat{T}[t]+\hat{V}[v]+\hat{W}[w]$. More specifically, the connection between $\hat{H}_m$ and $E_m$ is given by the particle number $N$ and the integral kernels $(t,v,w)$ of the three energy operators. For the matter Hamiltonian $\hat{H}_m$ of Eq.~\eqref{eq:Hamiltonian_BO} for example, these kernels are given by $t=-\frac{1}{2}\nabla^2_{\br}$, $v=v(\br)$ and $w=w(\br,\br')$. The goal of any electronic-structure theory is then to find the minimum of $E_m^{t,v,w}[\Psi]$, where $\Psi$ is a (possibly multi-determinantal) wave function constructed from the orbital set $\{\phi_k\}$. 
Typically, one needs to impose some constraints on the parametrization of the wave function $\Psi$ to make it physical 
\begin{align}
\label{eq:ConstraintsEquality}
	c_k[\Psi] = 0,
\end{align}
e.g., orthonormality of orbitals or the norm of the CI coefficients.
The \emph{generic electronic-structure minimization problem} is formulated as
\begin{alignat}{2}
\label{eq:MinimizationProblemEST}
&\text{minimize }\, &&E_m^{t,v,w}[\Psi] \nonumber \\
&\text{subject to }\quad &&c_k[\Psi] = 0.
\end{alignat}
We can solve Eq.~\eqref{eq:MinimizationProblemEST} by, e.g., minimizing the Lagrangian
\begin{align}
L^{t,v,w}_m[\Psi,\{\epsilon_k\}] = E_m^{t,v,w}[\Psi] + \mathcal{C}[\Psi,\{\epsilon_k\}],
\end{align}
where $\mathcal{C}[\Psi,\{\epsilon_k\}] = \sum_{k}\epsilon_k c_{k}[\Psi]$ is a Lagrange-multiplier term. Instead of minimizing $E_m$ directly, one minimizes $L_m$ with respect to the orbitals and the Lagrange-multipliers $\epsilon_k$. Today, a plethora of standard electronic-structure codes exist that solve \eqref{eq:MinimizationProblemEST} very efficiently for many different theory levels and thus allow for a highly accurate description of the electronic structure.


If we consider the coupled electron-photon Hamiltonian of Eq.~\eqref{eq:Hamiltonian_BO} instead of the purely electron Hamiltonian $\hat{H}_m$, we find that we need to build new approximation strategies and implementations to deal with the coupled electron-photon Hamiltonian directly. However, by transforming the problem into its dressed counterpart, i.e., we consider~\eqref{eq:AuxiliaryHamiltonian2}, we can utilize the full existing machinery for the electronic case. In particular, this means that we have now polaritonic orbitals $\phi_i'(\bz \sigma)$ as fundamental entities that have as coordinate $\bz \equiv \br \boldsymbol{q}$, where $\boldsymbol{q}$ is an $M$-dimensional (number of photon modes) vector. Additionally, the one- and two-body terms are replaced by their polaritonic counterparts, i.e., $(t,v,w)\rightarrow (t',v',w')$ as given in Eqs.~\eqref{eq:dressedkinetic}, \eqref{eq:dressedpotential} and \eqref{eq:dressedinteraction}. We can then transform straightforwardly the energy expression of a given electronic-structure theory into a polariton energy expression $E_m^{t,v,w}[\Psi]\rightarrow E_m^{t',v',w'}[\Psi']$, because the connection between $E_m$ and $\hat{H}_m$ is defined by the one- and two-body terms and the particle number alone. Also the constraints directly transfer to the polariton system, leading to the Lagrangian term $\mathcal{C}[\Psi,\{\epsilon_k\}] \to \mathcal{C}[\Psi',\{\epsilon_k\}]$. Lastly, since polaritons are particles with a more complicated hybrid statistics than electrons (see Ch.~\ref{sec:dressed:construction}), we need to add to the Lagrangian a further constraint term $\mathcal{G}[\gamma_e]$\footnote{The precise form of this term depends on the method that is used. See Sec.~\ref{sec:numerics:hybrid:pHF_algorithm}} to enforce the constraints
\begin{align}
\label{eq:ConstraintsInequality}
g_i[\gamma_e]= 1 - n_i^e \geq 0 \quad\forall\,i = 1,\dotsc,N.
\end{align}
With this definition, the energy expression $E_m^{t',v',w'}[\Psi']$ and the constraints, we are now able to generalize the minimization problem of Eq.~\eqref{eq:MinimizationProblemEST} to the \emph{generic polaritonic minimization problem}
\begin{alignat}{2}
\label{eq:MinimizationProblem}
&\text{minimize }\, &&E_m^{t',v',w'}[\Psi'] \nonumber\\
&\text{subject to }\quad &&c_k[\Psi'] = 0 \\
& &&g_i\bigl[\gamma_e[\Psi']\bigr] \geq 0.		\nonumber
\end{alignat}
Since there are many possible strategies to solve the minimization problem~\eqref{eq:MinimizationProblem} and a good choice might depend on the specific electronic-structure theory that is considered, we will not further discuss the practical aspects of enforcing the extra conditions. In the next sections, we will exemplify this procedure by generalizing HF and RDMFT to polaritons.

We want to remind the reader that there are scenarios in which the conditions $g_i$ are trivially fulfilled (see Sec.~\ref{sec:dressed:construction:simple:auxiliary_wf_problem}). In this case, a minimization that does not explicitly guarantee $g_i\geq0$ is as accurate as solving the minimization problem~\eqref{eq:MinimizationProblem}. In the following, we will refer to polaritonic minimization problems of the former type as \emph{fermion ansatz} to differentiate them from the generic \emph{polariton ansatz}. We will numerically investigate the settings, where the fermion ansatz and the polariton ansatz lead to the same results in Sec.~\eqref{sec:dressed:results:implementation_lattice:hybrid_influence}.

\section{Polaritonic HF}
\label{sec:dressed:est:HF}
In this section, we will apply the general rules of the previous section to HF theory, which leads to polaritonic HF theory. This means that we approximate the density matrix of the exact dressed wave function of Eq.~\eqref{eq:DressedManyBodyWF} by the density matrix of a single Slater determinant with 
orbitals ${\phi}'_1,\dotsc, {\phi}'_N$, i.e., 
\begin{align*}
\Phi'(\bz_1 \sigma_1,\dotsc,\bz_N \sigma_N)=&\frac{1}{\sqrt{N!}}|{\phi}'_1,\dotsc, {\phi}'_N|_-
\end{align*}
Further, we consider a spin-restricted formalism, i.e., we assume that the number of electrons $N$ is even and define ${\phi}'_{2k-1}(\bz \sigma)=\phi'_k(\bz)\alpha(\sigma), {\phi}'_{2k}(\bz \sigma)=\phi'_k(\bz)\beta(\sigma)$ for $k=1,\dotsc,N/2$, where $\alpha,\beta$ are the usual spin-orbitals (cf. Eqs.~\eqref{eq:est:spin_orbitals_closed_shell}. We again note that we do not necessarily enforce with our constraints that the auxiliary Slater determinant has the right symmetry but rather its 1RDM. In this regard polaritonic HF becomes actually a 1RDM functional theory for polaritonic problems rather than a wave-function based method~\citep{Buchholz2019}. With this ansatz, we calculate the energy expectation value for the Hamiltonian of Eq.~\eqref{eq:AuxiliaryHamiltonian2}, which reads 
\begin{align}
\label{eq:dressed:est:HF:energy}
E_{HF}'=2 \sum_i \braket{\phi_i' | (\hat{T}[t'] + \hat{V}[v']) \phi_i'} + \sum_{i,k} \left[2 \braket{\phi_k' | \hat{J}'_i[w']\phi_k'} - \braket{\phi_k' | \hat{K}'_i[w']\phi_k'}\right],
\end{align}
where we introduced the ``dressed'' Coulomb-operator $\hat{J}_i'$ which acts as
\begin{subequations}
	\begin{align}
	\label{eq:dressed:est:HF:CoulombOperator}
	\hat{J}_i'\phi_k'(\bz) &= \int\td\bz' \phi_i'{}^*(\bz')w'(\bz;\bz') \phi_i'(\bz') \phi'_k(\bz) \\
	\intertext{and the ``dressed'' exchange-operator $\hat{K}_i'$ that acts as}
	\label{eq:dressed:est:HF:ExchangeOperator}
	\hat{K}_i'\phi_k'(\bz) &= \int\td\bz' \phi_i'(\bz)w'(\bz;\bz')\phi_i'{}^*(\bz') \phi_k'(\bz').
	\end{align}
\end{subequations} 
The polaritonic one- and two-body terms are given by~\eqref{eq:dressedkinetic}, \eqref{eq:dressedpotential} and~\eqref{eq:dressedinteraction}, respectively. With this we find that $E_{HF}'=E_{HF}^{t',v',w'}(\{\phi_k'\})$ (see Ch.~\ref{sec:est:hf}). Consequently, we also find structurally the same derivative, which reads
\begin{align}
\label{eq:dressed:est:HF:Fockmatrix}
\nabla_{\phi_k^*} E_{HF}'= \hat{H}^1{}' \phi_k' = 2 (\hat{T}[t']+\hat{V}[v']) \phi_k' + 2 \sum_{i} \left[2  \hat{J}_i'[w']\phi_k' - \hat{K}_i'[w']\phi_k'\right],
\end{align}
where $\hat{H}^1{}'$ is the generalization of the Fock operator (Eq.~\eqref{eq:est:hf:Fock_operator}).
Since we consider only one Slater determinant, the orbitals $\phi_i'$ are also the eigenfunctions of the system's dressed 1RDM $\gamma$. Because of the spin-restriction, it suffices also to consider the spin-summed version $\gamma(\bz,\bz')=2 \sum_{i=1}^{N/2} \phi_i'{}^*(\bz') \phi_i'(\bz)$, which we denote with the same symbol.
We see that $\gamma[\Phi']$ has occupations (eigenvalues) of 2 instead of 1 because of the spin-summation. This transfers to the natural occupation numbers of the electronic 1RDM 
\begin{align*}
\gamma_e(\br,\br')= \int\td\bq \gamma(\br \bq,\br' \bq).
\end{align*}
Now, we have defined all the terms that enter the minimization problem \eqref{eq:MinimizationProblem}. 

\section{Polaritonic RDMFT}
\label{sec:dressed:est:RDMFT}
As mentioned in the last section, we consider polaritonic HF as a test-case. HF theory is conceptually (and numerically almost) the simplest electronic-structure theory and thus it is well-suited to explore the features and possible issues of polaritonic-structure theory. Judging from the first calculations (see Ch.~\ref{sec:dressed:results}), polaritonic HF captures despite its simplicity many phenomena of coupled light matter systems surprisingly well. However, for large coupling strengths, polaritonic HF becomes inaccurate.\footnote{Most of the contents of this section are part of Ref.~\citep{Buchholz2019}}
To improve upon the HF description, a logical next step is DFT based on an auxiliary system of polaritons. For that, we merely have to remove the exchange-part of the HF-implementation and employ a functional of our choice. This has been already tested in a simple setting by \citet{Nielsen2018}. We only want to mention that the employed (simple) functional in terms in the dressed KS system was already more accurate than KS-QEDFT with the photon OEP.

In this section instead, we go from a conceptual point of view yet another step further and derive polaritonic RDMFT, i.e., we construct a theory on the polaritonic 1RDM. This case is especially interesting with regard to the difficulties that we encountered in Sec.~\ref{sec:qed_est:rdms:general} when we tried to express the energy-expectation value of the Hamiltonian in terms of RDMs. By employing polaritonic orbitals, we can let aside the understanding of the complicated 3/2-body RDM that accounts for the electron-photon interaction in the standard picture, including its representability conditions. The polaritonic and electronic 1RDMs are quite similar and most importantly, many concepts that were important for the construction of RDMFT functionals transfer to the polaritonic case. 

\subsubsection*{The RDM perspective on polaritonic-structure theory}
Let us therefore analyze again the structure of $\hat{H}'$, given in Eq.~\eqref{eq:AuxiliaryHamiltonian2}. It consists of only polaritonic one-body terms $\hat{h}^{(1)}(\bz)=- \tfrac{1}{2} \Delta + v'(\bz)$, and two-body terms $ \hat{h}^{(2)}(\bz,\bz')=w'(\bz,\bz')$. It commutes with the polaritonic particle-number operator $\hat{N}'=\int\td^{3+M}z\, \hat{n}(\bz)$, where we used the definition of the polaritonic local density operator $\hat{n}(\bz)=\sum_{i=1}^{N}\delta^{3+M}(\bz-\bz_i)$. This means that the auxiliary system has a constant polaritonic particle number $N$. Additionally, the physical wave function of the dressed system $\Psi'(\bz_1 \sigma_1,\dots,\bz_N \sigma_N)$ is per construction antisymmetric (but it has in addition the $q$-symmetry). These properties \emph{allow} for the definition of the polaritonic (spin-summed) 1RDM in exactly the same way as for an electronic state. We repeat here the definition of Sec.~\ref{sec:dressed:construction:general}, i.e.,
\begin{align}
\tag{cf. \ref{eq:Dressed1RDM}}
\gamma(\bz,\bz') &= N \sum_{\sigma_1,...,\sigma_N}\int \td^{(3+M)(N-1)}  z \\
&\hphantom{+}{\Psi'}^*(\bz' \sigma_1,\bz_2 \sigma_2,...,\bz_{N} \sigma_{N}){\Psi'}(\bz \sigma_1,\bz_2 \sigma_2,...,\bz_{N} \sigma_{N}). \nonumber
\end{align}
Furthermore, we introduce the (spin-summed) dressed 2RDM
\begin{align*}
	\Gamma^{(2)}(\bz_1,\bz_2;\bz_1',\bz_2')=&N(N-1) \sum_{\sigma_1,...,\sigma_N}\int \td^{(3+M)(N-2)} z\\
	&{\Psi'}^*(\bz_1'\sigma_{1},\bz_2'\sigma_2, \bz_3 \sigma_3,..,\bz_{N} \sigma_{N}){\Psi'} (\bz_1\sigma_{1},\bz_2\sigma_2, \bz_3 \sigma_3,..,\bz_{N} \sigma_{N}).
\end{align*}
These dressed RDMs allow for expressing the energy expectation value of the dressed system by
\begin{align*}
E_0'=&\braket{\Psi'|\hat{H}'|\Psi'}=\braket{\Psi'|\sum_{k=1}^{N}\hat{h}^{(1)}(\bz_k) + \tfrac{1}{2} \sum_{k \neq l} \hat{h}^{(2)}(\bz_k,\bz_l)|\Psi'}\\
=& \!\int\!\td^{3+M}z \hat{h}^{(1)}(\bz) \gamma(\bz,\bz')|_{\bz'=\bz}+
\!\tfrac{1}{2}\!\int\!\td^{3+M}z\td^{3+M}z' \hat{h}^{(2)}(\bz,\bz') \Gamma^{(2)}(\bz,\bz',\bz,\bz').
\end{align*}
Thus, we can define the variational principle for the ground state only with respect to \emph{well-defined} reduced quantities,
\begin{align*}
\label{eq:min_principle}
E_0'=\inf_{\{\gamma,\Gamma^{(2)}\}\rightarrow \Psi'} E[\gamma,\Gamma^{(2)}].
\end{align*}
Notice the difference to the according variational principle in the standard picture
\begin{align}
\tag{cf. \ref{eq:var_parinciple_RDM}}
E_0=&\inf_{\substack{\{\gamma_e,\Gamma_e^{(2)}, \gamma_b, \Gamma_{e,b}^{(3/2)}}\}\rightarrow \Psi} \left\{\left(T+V\right)[\gamma_e]+\left(W+H_d\right)[\Gamma_e^{(2)}]+H_{ph}[\gamma_{b}] +H_I[\Gamma_{e,b}^{(3/2)}] \right\},
\end{align}
that we have defined formally in Sec.~\ref{sec:qed_est:rdms:general}. We discussed there that we don't know how to enforce the connection between the RDMs and the wave function, $\{\gamma_e,\Gamma_e^{(2)}, \gamma_b, \Gamma_{e,b}^{(3/2)}\}\rightarrow \Psi$, and thus cannot use it. In contrast to that, we know what we have to do in principle to perform the minimization \eqref{eq:min_principle}: We need to constrain the configuration space to the physical dressed RDMs that connect to an antisymmetric wave function with the extra $q$-exchange symmetry by testing the appropriate $N$-representability conditions of the dressed 2RDM and the dressed 1RDM. Besides the by now well-known conditions for the fermionic 2RDM~\citep{Mazziotti2012} and the fermionic 1RDM~\citep{Coleman1963} we would in principle get further conditions to ensure the extra exchange symmetry. However, already for the usual electronic 2RDM the number of conditions grows exponentially with the number of particles, and it is out of the scope of this work to discuss possible approximations. The interested reader is referred to, e.g., Ref.~\citep{Mazziotti2012}. Instead, we want to stick to the dressed 1RDM $\gamma$ and approximate the 2-body part as a functional of the $\gamma$. We will treat hereby $\gamma$ as approximately fermionic and thus consider only the $N$-representability conditions~\eqref{eq:NrepDressed1RDM}. Additionally, we guarantee the fermionic character $\gamma_e$ as described in Sec.~\ref{sec:dressed:est:prescription}.

\subsubsection*{The polaritonic 1RDM as a basic variable}
The mathematical justification of RDMFT is given by Gilbert's theorem~\citep{Gilbert1975}, which is a generalization of the Hohenberg-Kohn theorem of DFT~\citep{Hohenberg1964}. More specifically, Gilbert proves that the ground state energy of any Hamiltonian with only 1-body and 2-body terms is a unique functional of its 1RDM (see Sec.~\ref{sec:est:rdms:rdmft}. Following this idea, we will express the ground-state energy of the dressed system as a partly unknown energy functional $E'$ of only the system's dressed 1RDM  
\begin{align}
\label{eq:polRDMFT_var_principle_1RDM}
E_0'=&\inf_{\gamma}E'[\gamma],
\end{align}
where 
\begin{align}
\label{eq:dressed:est:RDMFT:energy_functional_general}
	E'[\gamma]=\int\!\td^{3+M}z \hat{h}^{(1)}(\bz) \gamma(\bz,\bz')|_{\bz'=\bz} + \underbrace{\tfrac{1}{2}
		\!\int\!\td^{3+M}z\td^{3+M}z' \hat{h}^{(2)}(\bz,\bz') \Gamma^{(2)}([\gamma];\bz,\bz',\bz,\bz')}_{=W'[\gamma]}
\end{align}
is the dressed RDMFT energy functional.
For this minimization, we need a functional of the diagonal of the dressed 2RDM in terms of the dressed 1RDM as well as adhering to the corresponding $N$-representability conditions when varying over $\gamma$. Analogously to the discussion about the extra symmetry, we need to consider now polariton ensembles to have simple conditions. However, since the ground state is a pure state and pure states are spacial cases of ensembles, the definition \eqref{eq:polRDMFT_var_principle_1RDM} is still exact with the ensemble conditions.

Another advantage of RDMFT in general is the direct access to all one-body observables. This transfers also to polaritonic RDMFT. The calculation of expectation values of purely electronic one-body observables is trivial with the knowledge of the dressed 1RDM,  but also photonic one-body (and half-body) observables can be calculated, using the connection formula shown in the end of Sec.~\ref{sec:dressed:construction:general}. Thus, we are able to calculate very interesting properties of the cavity photons like the mode occupation or quantum fluctuations of the electric and magnetic field.

To see whether our approach is practical and accurate, we employ simple approximations to the unknown part $W'[\gamma]$ that have been developed for the electronic case. To do so, we further, similarly to the electronic case, decompose 
\begin{align*}
W'[\gamma]=E_H[\gamma]+E_{xc}[\gamma]
\end{align*}
into a classical Hartree part
\begin{align*}
E_{H}[\gamma]=\frac{1}{2}\int\int\td^{3+M}z\td^{3+M}z' \gamma(\bz,\bz)\gamma(\bz',\bz')w'(\bz,\bz')
\end{align*}
and an unknown exchange-correlation part $E_{xc}[\gamma]$. Almost all known functionals $E_{xc}[\gamma]$ are expressed in terms of the eigenbasis and eigenvalues of the 1RDM. In our case the dressed natural orbitals $\phi_i(\bz)$ and occupation numbers $n_i$ are found by solving $\int \td^{3+M} z' \gamma(\bz,\bz') \phi_i (\bz')  = n_i \phi_i (\bz)$. One interesting feature of RDMFT is (in contrast to KS-DFT), that it includes HF theory as a special case with the functional
\begin{align}
\label{eq:dressed:est:RDMFT:HF_functional}
E_{xc}[\gamma]=E_{\text{HF}}[\gamma]=-\tfrac{1}{2} \sum_{i,j}n_in_j&\int\!\!\int\td^{3+M}z\td^{3+M}z'\phi^*_i(\bz)\phi^*_j(\bz')\, w'(\bz,\bz') \phi_i(\bz')\phi_j(\bz).
\end{align}
As the HF functional depends linearly on the natural occupation numbers, any kind of minimization will lead to the single-Slater-determinant HF ground state (which corresponds to occupations of 1 and 0)~\cite{Lieb1981}. We have recovered polaritonic HF. We can go beyond the single Slater determinant in polaritonic RDMFT, if we employ a nonlinear occupation-number dependence in the exchange-correlation functional. We have only considered the oldest and best tested functional that was introduced by M\"{u}ller in 1984~\cite{Mueller1984} that reads in the dressed setting (cf. Eq.~\eqref{eq:est:rdms:Mueller_functional})
\begin{align}
\label{eq:dressed:est:RDMFT:Mueller_functional}
E_{xc}[\gamma]=E_{\text{M}}[\gamma]=-\tfrac{1}{2} \sum_{i,j}\sqrt{n_in_j}&\int\!\!\int\td^{3+M}z\td^{3+M}z'\phi^*_i(\bz)\phi^*_j(\bz')\, w'(\bz,\bz') \phi_i(\bz')\phi_j(\bz),
\end{align}
and later re-derived by Bjuise and Baerends~\cite{Buijse2002} from a different perspective. The Müller functional has been studied for many physical systems~\cite{Goedecker1998,Buijse2002} and gives a qualitatively reasonable description of electronic ground states (see also Sec.~\ref{sec:est:comparison}). Additionally, it has many advantageous mathematical properties~\cite{Mueller1984,Frank2007}(see also Sec.~\ref{sec:est:rdms:rdmft}). A thorough discussion of different functionals goes beyond the scope of this work, and we only want to remark that a variety of functionals were proposed after $E_{M}[\gamma]$ and it is likely to have even better agreement with the exact solution by choosing more elaborate functionals.
	
	\clearpage
\chapter{Polaritonic structure methods in practice}
\label{sec:dressed:results}
In this chapter, we finally apply our new polariton tool box to concrete problems and show some first results.\footnote{The chapter is based on Refs.~\citep{Buchholz2019,Buchholz2020}.} 
Specifically, we have two implementations available that are capable to perform calculations with polaritonic orbitals. In Sec.~\ref{sec:dressed:results:implementation_lattice}, we introduce the first one. Here, the polaritonic orbitals  are expanded in a combined basis of a one-dimensional lattice for the electronic subsystem and the Fock-number states for the photon mode (see Sec.~\ref{sec:dressed:results:implementation_lattice:model}). In this setting, we can perform polaritonic HF calculations taking explicitly into account the hybrid statistics of the polaritonic orbitals, i.e., the polariton ansatz (see Sec.~\ref{sec:dressed:est:prescription}). Importantly, this implementation allows to assess our construction to approximatively guarantee the hybrid statistics of polaritonic orbitals (polariton ansatz) and to define settings, in which we can use the fermion ansatz.
In Sec.~\ref{sec:dressed:results:implementation_real_space}, we introduce our second implementation in the real-space electronic structure code \textsc{Octopus}.\footnote{See also the corresponding publication Ref.~[part 4]\citep{Tancogne-Dejean2020}.} It is an extension of the RDMFT routine of \textsc{Octopus} and capable to perform minimizations with the Müller and HF approximation employing the fermion ansatz.\footnote{Note that to ease reading in this section, we only briefly introduce both algorithms, reserving their appropriate introduction for part~\ref{sec:numerics}. The first implementation is discussed in Ch.~\ref{sec:numerics:hybrid}, and for the details on the second implementation, the reader is referred to Ch.~\ref{sec:numerics:dressed}.}

For both implementations, we present accuracy studies, which is a standard step in first-principles theory. Before we can apply a method to a new problem, we have to systematically compare the numerical results to exact references (or analytical limit cases). 
Since this is a standard procedure in electronic structure theory, there are well-defined test sets with accurate reference data. For instance, \citeauthor{Curtiss1997} introduced the frequently used G2/97~\citep{Curtiss1997} or G3/99~\citep{Curtiss2000} theoretical  thermochemistry test sets. For coupled electron-photon systems, we do not have such kind of databases and it will be considerably more difficult to generate them because of the larger configuration space.\footnote{The largest system that has been described exactly considers three particles and one mode~\citep{Sidler2020}.} However, we still can follow the same route and assess our approximations with exact references. With the machinery that we had accessible when we produced our data, we could calculate the (regarding the basis set) exact many-body ground states of one-dimensional two-electron systems that are coupled to one photon mode.
In Sec.~\ref{sec:dressed:results:numerical_perspective}, we conclude the assessment with a brief discussion of the numerical challenges of polaritonic-orbital-based methods, including the scaling of the computational costs. 

The benchmark studies justify then to go beyond problems that we can still treat with exact methods and do some first calculations of nontrivial systems. Our implementations are not yet general enough to treat realistic three-dimensional matter systems, but are constrained to one spatial dimension. Nevertheless, this setting allows already to study nontrivial effects of the electron-photon interaction, which we present in Sec.~\ref{sec:dressed:results:fancy_examples}. 

In Sec.~\ref{sec:dressed:results:summary}, we summarize these results and comment on possible implications with respect to the standard description in terms of cavity-QED models (Sec.~\ref{sec:intro:cavity_qed_models}).

\newpage
\section{Polaritonic orbitals 1: hybrid statistics on the lattice}
\label{sec:dressed:results:implementation_lattice}
In this section, we introduce and asses the setting of our first implementation that numerically minimizes the polaritonic HF energy functional~\eqref{eq:dressed:est:HF:energy} (Sec.~\ref{sec:dressed:est:HF}) within the polariton ansatz. This means that the implementation is capable to explicitly enforce the extra conditions~\eqref{eq:ConstraintsInequality} due to the hybrid statistics of the polaritonic orbitals and allows for (approximately) enforcing the hybrid statistics. For that, we have developed a new algorithm that we briefly outline in Sec.~\ref{sec:dressed:results:implementation_lattice:algorithm}.\footnote{ We present the algorithm with all its details in the last part in Ch.~\ref{sec:numerics:hybrid}. }

In the first subsection (Sec.~\ref{sec:dressed:results:implementation_lattice:model}), we introduce the lattice model that this implementation considers to describe the polaritonic orbitals. 
In Sec.~\ref{sec:dressed:results:implementation_lattice:hybrid_influence}, we assess this implementation by a comparison to exact results and exemplify the influence of the hybrid statistics. 

\subsection{How to enforce the hybrid statistics in practice: The polaritonic HF algorithm}
\label{sec:dressed:results:implementation_lattice:algorithm}
Algorithm-wise, we are confronted with enforcing the additional inequality constraints \eqref{eq:ConstraintsInequality} in extension to the original (HF) minimization problem in~\eqref{eq:MinimizationProblemEST}.
We start by noting that the constraint functions $g_i$ depend on $\gamma\{\phi_k'\}$, which can be directly calculated from the polariton orbitals,
via the eigendecomposition of $\gamma_e$.
Since the diagonalization of $\gamma_e$ is a nontrivial step for large systems (or in real-space) and thus can be a bottleneck of the minimization, it is helpful to consider natural and dressed orbitals as \emph{independent} variables of the minimization and enforce their connection as an additional constraint.\footnote{This is similar to considering $\phi$ and $\phi^*$ as independent.} We thus define $g_i=g_i[\gamma_e\{\phi_k', \psi^{e}_i\}]$ and include the necessary orthonormality of the $\psi^{e}_i$ by a third set of conditions
\begin{align}
\label{eq:ConstraintsNO}
f_{ij}=\braket{\psi^{e}_i|\psi^{e}_j} -\delta_{ij}=0,
\end{align}
that we include in the minimization by a third Lagrange-multiplier term $-\sum_{ij}\bar{\theta}_{ij} f_{ij}$. Note that this construction automatically linearizes the constraints \eqref{eq:ConstraintsInequality} during one minimization step, where the $\psi^e_i$ are fixed.

To enforce now these inequality constraints, we use an \emph{augmented Lagrangian} algorithm, following the book of \citet{Nocedal2006}, part~17.3.
We have chosen this algorithm, since it simply extends a given Lagrangian with penalty terms. Hence, we can make use of any existing implementation that solves the minimization problem of Eq.~\eqref{eq:MinimizationProblemEST} and just add the extra terms with corresponding extra iteration loops. To test this, we employed a standard electronic-structure algorithm~\cite{Payne1992}, which we extended by the augmented-Lagrangian method for the inequality constraints. This extension involves two extra terms. A linear (so-called augmented) term, $-\sum_i \nu_i g_i$ with Lagrange-multipliers $\nu_i$ that are initialized to zero and updated to values $\nu_i>0$ only if the minimization reaches the corresponding boundary of the feasible region where $g_i=0$. And a second nonlinear term, that adds a penalty function $P=\mu / 2 \sum_i ([g_i]^-)^2$, where $[y]^-$ denotes max$(-y, 0)$, which penalizes violations of condition \eqref{eq:ConstraintsInequality} quadratically, but has no effect in the so-called \emph{feasible} region of configuration space, where the conditions \eqref{eq:ConstraintsInequality} are satisfied.

Specifically for our example, the extra Lagrangian term of the translation rules depicted in Fig.~\ref{fig:polariton_construction} is given by
\begin{align}
\label{eq:constraints_inequality_augmented_lagrangian}
\mathcal{G}[\gamma_e\{\phi_k',\psi^{e}_i\}]= - \sum_i \lambda_i g_i[\gamma_e\{\phi_k',\psi^{e}_i\}] + \mu \sum_i ([g_i]^-[\gamma_e\{\phi_k',\psi^{e}_i\}])^2 - \sum_{ij} \bar{\theta}_{ij} f_{ij} [\gamma_e\{\psi^{e}_i\}].
\end{align}
The full Lagrangian for the polaritonic HF minimization problem reads then
\begin{align}
\label{eq:Lagrangian}
L_{HF}'[\gamma\{\phi_k',\psi^{e}_i\}]=&	E_{HF}' -\sum_{ij} \bar{\epsilon}_{ij} h_{ij}[\gamma\{\phi_k'\}] +\mathcal{G}[\gamma_e\{\phi_k',\psi^{e}_i\}]
\end{align}
and the corresponding first order conditions for a minimum (stationary point) of $L_{HF}'$ are
\begin{subequations}
	\label{eq:dressed:HF:stationary_conditions}
	\begin{align}
	\label{eq:GradientPhi}
	0=&\nabla_{\phi_k'{}^*} L_{HF}'= \hat{H}^1\phi_k' - \sum_j\bar{\epsilon}_{kj} \phi_j' + \sum \left[ \lambda_i -\mu [g_i]^-\right]  \hat{G}_i\phi_k'\\
	\label{eq:GradientPsiGamma}
	0=&\nabla_{\psi^{e}_i{}^*}L_{HF}' = (\mu [g_i]^- -\lambda_i )\int\td\br' \gamma_e(\br',\br)\psi^{e}_i(\br') - \sum_j \bar{\theta}_{ij} \psi^{e}_j,
	\end{align}
\end{subequations}
where we considered $\phi_k'$ and $\phi_k'{}^*$ as independent and defined $\hat{G}_i\phi_k'(\br \bq)=n_k^e\int\td\br' \psi^{e}_i{}^*(\br' )\phi_k'(\br' \bq)$ $ \psi^{e}_i(\br \sigma)$.
Additionally, we can diagonalize the Lagrange-multiplier matrices $\bar{\epsilon}_{ij}=\delta_{ij}\epsilon_j$ and $\bar{\theta}_{ij}=\delta_{ij}\theta_j$, since the orbital-dependent Hamiltonian $\hat{H}^1$ and the electronic 1RDM $\gamma_e$ are hermitian. We also want to remark on the second gradient equation, cf. Eq.~\eqref{eq:GradientPsiGamma}, which is much simpler than it looks like on a first glance. In fact, solving Eq.~\eqref{eq:GradientPsiGamma} is equivalent to solving first the eigenvalue equation for $\gamma_e$ (see the paragraph above Eq.~\eqref{eq:Nrepresentability}) and then replacing $\theta_i=n_i^e(\mu [g_i]^- -\lambda_i)$. With these definitions, we are able to perform polaritonic HF calculations by numerically solving the Eqs.~\eqref{eq:GradientPhi} and \eqref{eq:GradientPsiGamma} with the expressions \eqref{eq:dressed:est:HF:energy} and \eqref{eq:dressed:est:HF:Fockmatrix}. 

For a more detailed discussion of this algorithm, the reader is referred to Ch.~\ref{sec:numerics:hybrid}.

\subsection{The Lattice model}
\label{sec:dressed:results:implementation_lattice:model}
\begin{figure}
	\centering
	\includegraphics[width=0.7\columnwidth]{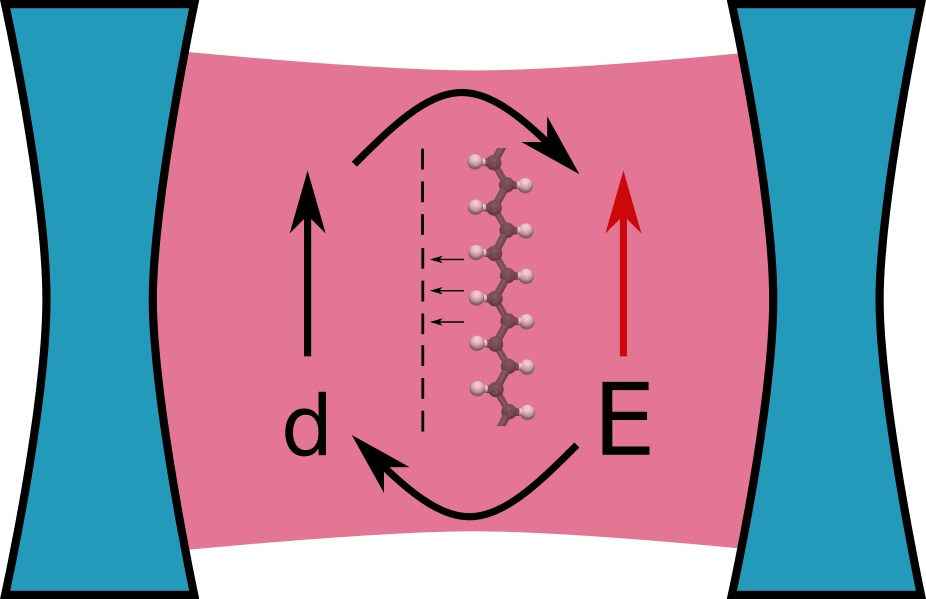}
	\caption{Sketch of our cavity setup. The electrons of our matter subsystem are allowed to move parallel to the electric field $\mathbf{E}$ of the cavity mode yet restricted in the perpendicular directions, i.e., we consider a one-dimensional discretized matter subsystem. Since the extension in the perpendicular directions is small compared to the wave length of the dominant cavity mode, the coupling is mediated via the total dipole $\mathbf{d}$ of the electrons. If the mode volume (distance between the mirrors) is small or the number of particles increased, new hybrid light-matter quasi-particles, i.e., polaritons, emerge.}
	\label{fig:cavity_sketch_polHF}
\end{figure}
For the exemplification of polaritonic HF, we consider a one-dimensional lattice system  that couples to one photon mode in dipole approximation with frequency $\omega$. We depicted a sketch of the setup in Fig. \ref{fig:cavity_sketch_polHF}. We have chosen such a simple lattice model, since this allows us to have still exact numerical reference data for more than 2 electrons to compare to. We stress again that there are no (numerically exact) references solutions currently available for realistic three-dimensional matter plus cavity systems. To the best of our knowledge there are only QEDFT simulations (at several levels of approximations) for such systems~\cite{Flick2018abinitio, Flick2019}. 

The Hamiltonian is of the form of Eq.~\eqref{eq:Hamiltonian_BO} and reads 
\begin{align}
\hat{H}=\overbrace{-t\sum_{i}^{B_m} \sum_{\sigma=\uparrow, \downarrow} (\hat{c}^{\dagger}_{i,\sigma}\hat{c}_{i+1,\sigma} + \hat{c}^{\dagger}_{i+1,\sigma}\hat{c}_{i,\sigma}) +\sum_{i=1}^{B_m}v_i \hat{n}_i}^{\hat{H}_m} + \overbrace{ \frac{\lambda^2}{2} \sum_{i,j=1}^{B_m} \hat{n}_i \hat{n}_j x_i x_j }^{\hat{H}_{self}}\nonumber\\
\underbrace{- \sqrt{\frac{\omega}{2}}(\hat{a}^{\dagger}+\hat{a})\lambda \sum_{i=1}^{B_m} x_i\hat{n}_i }_{\hat{H}_{int}}
+\underbrace{\omega (\hat{a}^+\hat{a} +1/2)}_{\hat{H}_{ph}},
\label{eq:dressed:results:lattice_model:Hamilt_example}
\end{align}
with hopping $t=\frac{1}{2 \Delta x^2}$ corresponding to a second-order finite difference approximation for a grid with spacing $\Delta x$, where we choose $t=0.5$ for all calculations, which corresponds to a spacing between neighbouring sites of $\Delta x=1$ bohr and a local scalar potential with value $v_i$ on site $i$. We have set the Coulomb repulsion to zero in this example to highlight the influence of the matter-photon coupling and how well the polaritonic HF approach can capture it. However, we show results for larger systems including the Coulomb interaction in Sec.~\ref{sec:dressed:results:fancy_examples}. Nevertheless, due to the dipole self-energy term $\hat{H}_d$ we have a mode-induced dipole-dipole interaction among the electrons. This type of interaction is important in many fundamental quantum-optical questions, such as the quest for a super-radiant phase in the strong-coupling case~\cite{Dzsotjan2011,Griesser2016,Plankensteiner2017} (see also Sec.~\ref{sec:intro:experiment_strong_coupling}). 
Further, the electron basis $B_m$ is determined by the number of sites, $x_i=i-x_0$ is the position with respect to the middle of our lattice $x_0$, $\hat{c}^{(\dagger)}_{i,\sigma}$ are the fermionic creation (annihilation) operators that satisfy the anticommutation relation $[\hat{c}_{i,\sigma},\hat{c}^{\dagger}_{j,\sigma'}]_{+}=\delta_{ij}\delta_{\sigma \sigma'}$, and $\hat{n}_i=\hat{c}^{\dagger}_{i,\uparrow} \hat{c}_{i,\uparrow}+\hat{c}^{\dagger}_{i,\downarrow} \hat{c}_{i,\downarrow}$ is the density operator.

For the implementation and to go to the polariton picture, we express the matter $\hat{H}_m$ plus dipole part $\hat{H}_{d}$ of our Hamiltonian in matrix form by using the basis states $\ket{\tilde{\psi}_{i,\sigma}}=\hat{c}^{\dagger}_{i,\sigma}\ket{0}$. As a basis for the photon subsystem, we utilize the eigenstates $\chi_i$ of the photon energy operator, i.e., $\hat{H}_{ph}\chi_{\alpha}=(\alpha + 1/2) \chi_{\alpha}$, which are photon number states. To calculate the coupling term of the energy expression $H_I$, cf. Eq.~\eqref{eq:dressed:est:HF:energy}, we express the displacement operator $p_{\alpha}=1/\sqrt{2\omega_{\alpha}}(\hat{a}^{\dagger}_{\alpha}+\hat{a}_{\alpha})$ in this basis as well. 
To then construct the auxiliary Hamiltonian $\hat{H}'$ to Eq.~\eqref{eq:dressed:results:lattice_model:Hamilt_example} according to the rules from Sec.~\ref{sec:dressed:est:HF}, we would need to define  the auxiliary terms $t',v',w'$, cf. Eqs. \eqref{eq:dressedkinetic}-\eqref{eq:dressedinteraction}. Since in this section we employ a second-quantized picture, it is less convenient to define these kernel-like quantities, but directly the many-body operators $\hat{T}[t'], \hat{V}[v'], \hat{W}[w']$. For the one-body terms $T[t']+V[v']$, this is straightforward and the expression reads 
\begin{align}
\label{eq:dressed:results:lattice_model:TB_Onebody}
T[t']+V[v'] =& \hat{H}_m + \hat{H}_{ph} + \frac{\lambda^2}{2} \sum_{i=1}^{B_m} \hat{n}_i \hat{n}_i x_i^2- \sqrt{\frac{\omega}{N}}(\hat{a}^{\dagger}+\hat{a}) \lambda \sum_{i=1}^{B_m} x_i\hat{n}_i.
\end{align}	
However, the interpretation of the operators is different from before, because we have to apply them to polaritonic basis states. Since we consider the spin-restricted formalism as introduced in Sec.~\ref{sec:dressed:est:HF}, we neglect the spin-dependency of the electronic part of the basis $\tilde{\psi}_{i,\sigma}\rightarrow\psi_{i}$ and define $\ket{\phi'_{i\alpha}}=\ket{\psi_{i}\chi_{\alpha}}$. We can then derive the kernel expression $(t+v)(x,q) \rightarrow(t'+v')_{i\alpha}^{j\beta}=\braket{\psi_{i}\chi_{\alpha}|(\hat{T}' + \hat{V}')\psi_{j}\chi_{\beta}}$ as matrix elements
\begin{align}
(t'+v')_{i\alpha}^{j\beta}=& -t(\delta_{i,j+1}+\delta_{i+1,j}) + v_i \delta_{ij} + \omega (\delta_{\alpha,\beta}+\tfrac{1}{2}) + \tfrac{\lambda^2}{2}(x_i-x_0)^2 \delta_{ij} \delta_{\alpha,\beta} \nonumber\\
&- \lambda\sqrt{\tfrac{\omega}{N}} (x_i-x_0) \delta_{ij} (\sqrt{\beta+1}\delta_{\alpha,\beta+1}+\sqrt{\beta}\delta_{\alpha,\beta-1}).
\end{align}
For the two-body term, a definition analogously to \eqref{eq:dressed:results:lattice_model:TB_Onebody} is more difficult, since we have to differentiate the two polaritonic coordinates. For the sake of the analogy, we formally write
\begin{align}
\label{eq:dressed:results:lattice_model:TB_two_body}
\hat{W}'=&\lambda^2 \sum_{i^1\neq j^2=1}^{B_m} \hat{n}_{i^1} \hat{n}_{j^2} x_{i^1} x_{j^2}- 
\sqrt{\frac{\omega}{N}}(\hat{a}^2{}^{\dagger}+\hat{a}^2)\lambda \sum_{i^1=1}^{B_m} x_{i^1}\hat{n}_{i^1}- 
\sqrt{\frac{\omega}{2}}(\hat{a}^1{}^{\dagger}+\hat{a}^1)\lambda \sum_{^2i=1}^{B_m} x_{i^2}\hat{n}_{i^2},
\end{align}
where the upper indices differentiate the two polaritonic orbitals that both have an electronic and photonic part. This is to be understood in the following sense: To define the corresponding kernel $w(x^1,q^1,x^2,q^2)\rightarrow w^{i^1\alpha^1i^2\alpha^2}_{j^1\beta^1j^2\beta^2}= \braket{\psi_{i^1}\chi_{\alpha^1}\psi_{i^2} \chi_{\alpha^2}|\hat{W}\psi_{j^1}\chi_{\beta^1}\psi_{j^2}\chi_{\beta^2}}$, the operators only act on the basis elements with the same indices. As an example, let us state the kernel $(w_{\text{self}})^{i^1\alpha^1i^2\alpha^2}_{j^1\beta^1j^2\beta^2}$ for the self-interaction part $\hat{W}_{\text{self}}=\lambda^2 \sum_{i^1\neq j^2=1}^{B_m}$ $\hat{n}_{i^1} \hat{n}_{j^2} x_{i^1} x_{j^2}$ that reads
\begin{align}
(w_{\text{self}})^{i^1\alpha^1i^2\alpha^2}_{j^1\beta^1j^2\beta^2}
=&\lambda^2 (x_{i^1}-x_0) (x_{i^2}-x_0) \delta_{i^1j^1} \delta_{i^2j^2}  \delta_{\alpha^1\beta^1} \delta_{\alpha^2\beta^2}.
\end{align}
With these definitions, we can calculate the polaritonic HF energy expression, cf.~\eqref{eq:dressed:est:HF:energy} and the polaritonic HF Fock-matrix, cf.~\eqref{eq:dressed:est:HF:Fockmatrix}. Then, we employ the augmented Lagrangian algorithm as discussed in Sec.~\ref{sec:dressed:results:implementation_lattice:algorithm} to find the polaritonic HF ground state of the model system.

\subsection{Polaritons from first principles: the influence of the hybrid statistics}
\label{sec:dressed:results:implementation_lattice:hybrid_influence}
As a first example, we illustrate the violation of the Pauli principle if we do not enforce the right symmetries (see Sec.~\ref{sec:dressed:est:prescription}).
To this end, we compare ground-state energies, electronic 1RDMs and the photon number of a small 4-electron system obtained with the two different HF ground states, i.e., polaritonic HF using density matrices with the exact symmetry, cf.~\eqref{eq:SymmetryPauliDressedFermion}, and polaritonic HF with only fermionic symmetry (which we call in this section fermionic HF). We can expect deviations between both polaritonic-HF theory levels for systems that contain more than one orbital. In our spin-restricted case this corresponds to more than two electrons and that is why we chose here $N=4$.
Further we set the external potential to zero, i.e., $v_i=0 \,\forall i$. Since we need to calculate the exact coupled electron-photon many-body ground state from a configuration space that grows exponentially fast with the size of the basis sets and the electron number, we choose a small box of length $L=5$ bohr. This corresponds to a matter basis of $B_{m}=6$ spatial sites times two spin states for each electron. For the photon-subsystem, we consider $B_{ph}=5$ photon number states for which all relevant quantities are sufficiently converged.\footnote{For example, deviations in the energy or photon number between $B_{ph}=5$ and $B_{ph}=6$ are maximally of the order of $10^{-4}$.} Despite the small basis sets and electron number employed, the many-body configuration space has the considerable size of $(2B_{m})^{N}*B_{ph}\approx 10^5$, which is already at the edge of standard exact diagonalization solvers: matrices of this size can still be diagonalized without special efforts like parallelization. Since we only aim for a benchmark study here, this limitation is not problematic, but it shows how expensive the exact solutions of coupled electron-photon systems computationally are. The need for numerically manageable approximations is evident here.

\begin{figure}[ht]
	\centering
	\begin{overpic}[width=0.49\columnwidth]{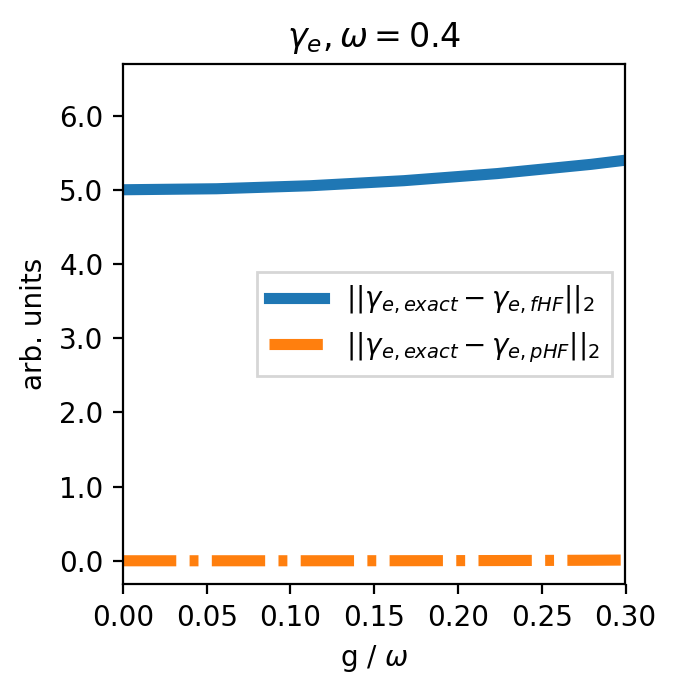}
		\put (22,83) {\textcolor{black}{(a)}}\hfill
	\end{overpic}
	\hfill
	\begin{overpic}[width=0.49\columnwidth]{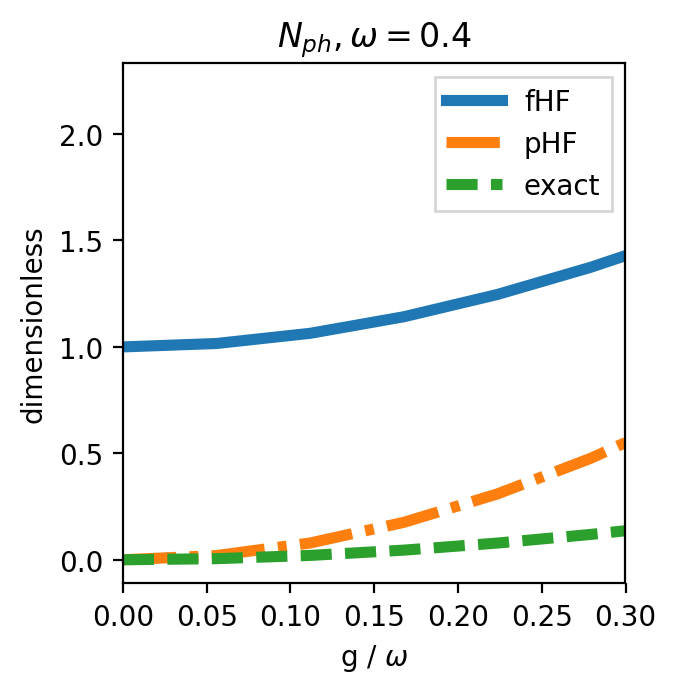}
		\put (22,83) {\textcolor{black}{(b)}}\hfill
	\end{overpic}
	\caption{Comparison of the electronic 1RDM $\gamma_e$ (a) and the photon number $N_{ph}$ (b) for the 4-electron system with $\omega = 0.4 $ hartree for varying coupling strength $g/\omega$. In (a) the norm difference between the exact 1RDM and the polaritonic HF (pHF) 1RDM (dashed-dotted orange line) and between the exact 1RDM and fermionic HF 1RDM (fHF) (solid blue line) are displayed. In (b) the exact photon number (dashed green line) and the polaritonic HF (dashed-dotted orange line) and fermionic HF (solid blue line) photon numbers are shown. In both cases, fermionic HF deviates much stronger from the exact reference than polaritonic HF due to the wrong symmetry.}
	\label{fig:tb_gamma_nphot}
\end{figure}

We first compare the electronic 1RDMs $\gamma_e$, cf. Eq.~\eqref{eq:GammaElectronic}, and the photon numbers $N_{ph}=\braket{\hat{N}_{ph}}$, cf. Eq.~\eqref{eq:PhotonNumber_length} using the connection formula of Eq.~\eqref{eq:PhotonEnergyConnection}, for varying coupling strengths $g/\omega =\lambda/\sqrt{2\omega}$ and $\omega=0.4$. We choose these quantities because they provide us with a consistent measure of how well the electronic part and the photonic part of the system are approximated. The electronic 1RDM determines all electronic one-body quantities and is therefore a very precise measure of the quality of the approximation in the electronic sector. Indeed, the photonic 1RDM in the single mode case corresponds to the photon number $N_{ph}$, which is an equivalently good measure for the quality in the photonic sector. In Fig.~\ref{fig:tb_gamma_nphot} (a), we display the difference of the exact electronic 1RDM from the one of the polaritonic HF and fermionic HF approximations, measured by the Frobenius norm $\lVert A \rVert_2=\sqrt{\sum_{ij} A_{ij}^2}$ for a matrix $A_{ij}$. We see that for all coupling strengths the polaritonic HF 1RDM (dashed-dotted orange line), which enforces the right hybrid statistics, remains very close to the exact solution indicating that the electronic subsystem is captured very well within this approximation. The fermionic HF (solid blue line) approximation, however, deviates strongly due to its wrong purely fermionic character. 

The same behavior is also encountered in the photonic subsector, where in Fig.~\ref{fig:tb_gamma_nphot} (b) the photon number of the exact calculation (dashed green line) is compared to the polaritonic HF (dashed-dotted orange line) and to the fermionic HF photon number (solid blue line). We therefore find, similarly to the simple uncoupled problem in Sec.~\ref{sec:dressed:est:prescription} (for $g/\omega=0$ we recover this case exactly), that for the same energy expression using the wrong statistics leads to sizeable errors.

\begin{figure}[ht]
	\centering
	\begin{overpic}[width=0.49\columnwidth]{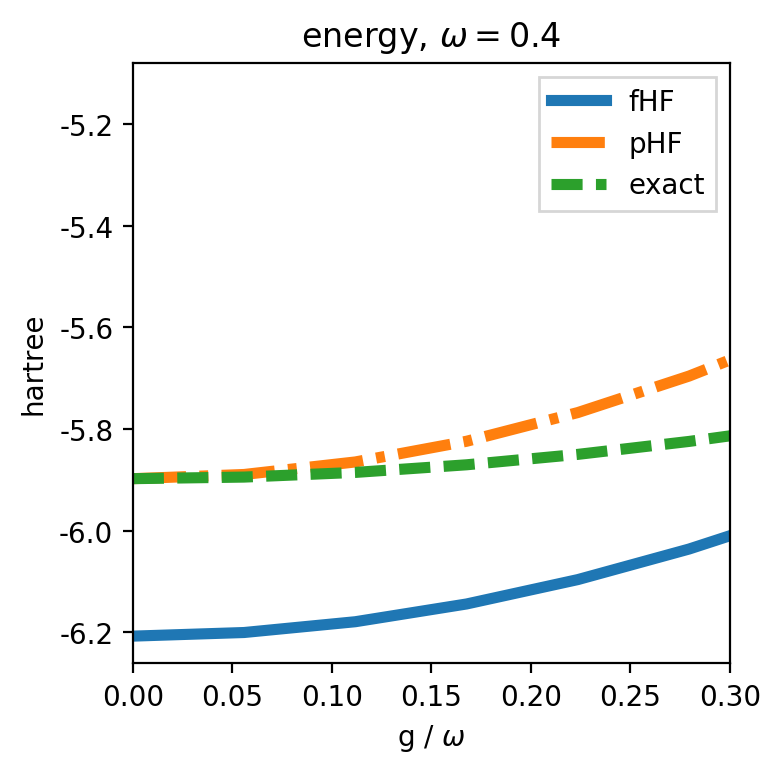}
		\put (22,83) {\textcolor{black}{(a)}}\hfill
	\end{overpic}
	\hfill
	\begin{overpic}[width=0.49\columnwidth]{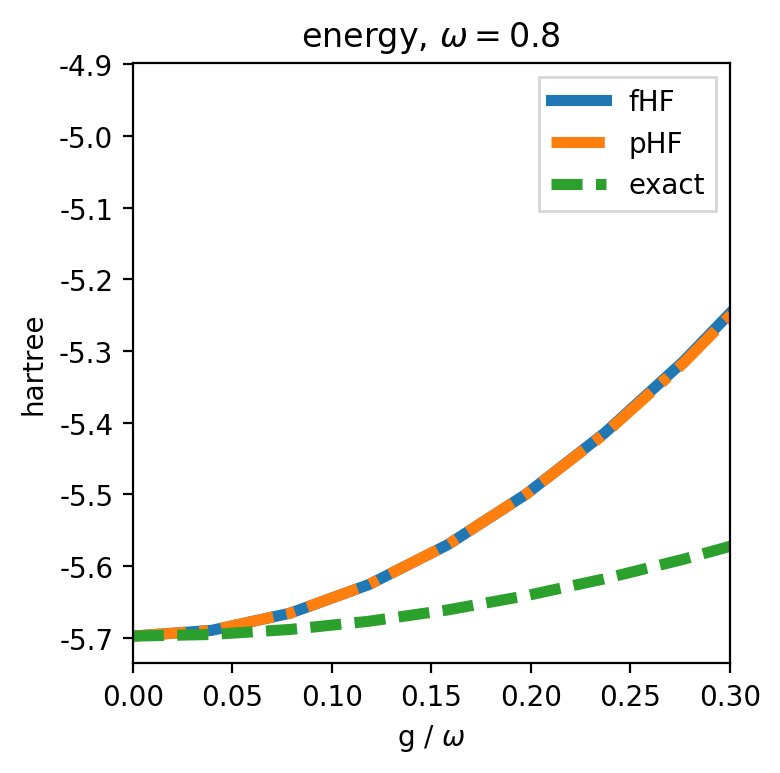}
		\put (22,83) {\textcolor{black}{(b)}}\hfill
	\end{overpic}
	\caption{Total energy for the 4-electron system with $\omega = 0.4 $ hartree (a) and $\omega = 0.8 $ hartree (b) for varying coupling strength $g/\omega$. While in (a) the fermionic HF approximation (solid blue line) can achieve unphysically low energies when compared to the exact solution (dashed green line) due to the wrong statistics, in (b) the minimal-energy condition singles out the right statistics without further constraints. The polaritonic HF (dashed-dotted orange line) by construction always has the right hybrid statistics and thus is variational, i.e., the energy is always above the exact energy.}
	\label{fig:tb_energy}
\end{figure}
The same problem is encountered also in Fig.~\ref{fig:tb_energy} (a), where we display the total energy $E=\braket{\hat{H}}$ of the coupled system as a function of the coupling strength $g/\omega$. While the polaritonic HF (dashed-dotted orange line) is variational, i.e., due to the right statistics we are always equal or above the exact energy (dashed green line), the fermionic HF (blue solid line) breaks the proper symmetry and thus can reach energies below the physically accessible ones. However, again in close analogy to the uncoupled example in Sec.~\ref{sec:dressed:est:prescription}, if we increase the frequency of the photon field such that it is much more costly to excite photons than electrons, the minimal-energy conditions can single out the correct statistics, as displayed in Fig.~\ref{fig:tb_energy}(b). That is, for $\omega$ large enough the constraints $g_i[\gamma\{\phi_k', \psi_i^e \}] \geq 0$ of Eq.~\eqref{eq:MinimizationProblem} are trivially fulfilled. We will exploit this feature later, when we use the other implementation for 4-electron system. We want to stress again that for a 2-electron singlet system, we always trivially satisfy the additional constraints, because there is only one occupied orbital. Thus, in the following benchmark study, we do not have to take care about the photon frequency.

\clearpage
\section{Polaritonic orbitals 2: the fermion ansatz in real space}
\label{sec:dressed:results:implementation_real_space}
In this section, we present our second numerical implementation of polaritonic orbitals that minimizes the RDMFT energy functional~\eqref{eq:dressed:est:RDMFT:energy_functional_general} within the fermion ansatz. We can employ the HF (Eq.~\eqref{eq:dressed:est:RDMFT:HF_functional}) as well as the Müller approximation (Eq.~\eqref{eq:dressed:est:RDMFT:Mueller_functional}) to the energy functional. The implementation is part of the real-space electronic structure code \textsc{Octopus}~\citep{Tancogne-Dejean2020}. This implementation is not yet extended to explicitly guarantee the extra conditions of Eq.~\eqref{eq:ConstraintsInequality} and thus can only be applied in settings, where the conditions are trivially fulfilled (see Sec.~\ref{sec:dressed:results:implementation_lattice:hybrid_influence}). Within this limitation, the implementation is capable to study one-dimensional electronic systems, coupled to one photon mode in full real space. This means the electronic and photonic coordinates of the polaritonic orbitals are approximated on discretized grids and the differential operators are approximated accordingly (see Sec.~\ref{sec:numerics:dressed:implementation}). This allows for a highly accurate description of coupled light-matter systems (see also App.~\ref{sec:numerics:dressed:convergence}).

In Sec.~\ref{sec:dressed:results:implementation_real_space:algorithm}, we  briefly introduce the RDMFT algorithm. The appropriate introduction of the implementation, including a detailed convergence study is presented instead in Ch.~\ref{sec:numerics:dressed} of the numerics part. In Sec.~\ref{sec:dressed:results:implementation_real_space:validation}, we validate the methods for two example systems.

\subsection{Polaritonic orbitals in real-space: the \textsc{Octopus}-implementation of polaritonic RDMFT}
\label{sec:dressed:results:implementation_real_space:algorithm}
An important advantage of the polariton formulation of cavity QED is that we can re-use most of the numerical techniques developed for quantum chemistry and materials science. We demonstrate this explicitly with our working implementation of dressed RDMFT in the electronic-structure code \emph{Octopus}~\citep{Tancogne-Dejean2020} that is publicly available in the actual developer's version.\footnote{\url{https://octopus-code.org/wiki/Developers:Starting_to_develop}.}

Specifically, we rewrite the approximated energy functional in the natural orbital basis as
\begin{align*}
E[\gamma]&=\sum_{i=0}^{\infty} n_i \int \td^{3+M}z \, \phi^*_i (\bz) \left[-\tfrac{1}{2} \Delta + v'(\bz) \right] \phi_i (\bz) + \\
&\tfrac{1}{2} \sum_{i,j}n_in_j\int\!\! \int\td^{3+M}z\td^{3+M}z' \left|\phi_i(\bz)\right|^2 \left|\phi^*_j(\bz')\right|^2 \, w'(\bz,\bz') + E_{xc}[\gamma].
\end{align*}
We use this form to minimize the energy functional by varying the natural orbitals as well as the natural occupation numbers.  To impose fermionic ensemble $N$-representability, we first represent the occupation numbers as the squared sine of auxiliary angles, i.e. $0 \leq n_i=2\sin^2(\alpha_i)\leq 2$, to satisfy Eq.~\eqref{eq:Nrep_cond}.\footnote{Note that the $n_i$ are bounded by 2 because we employed a spin-summed formulation. If we considered natural \emph{spin-orbitals} instead, the upper bound would be 1.}  
The second part of the conditions (Eq.~\eqref{eq:Nrep_cond2}), i.e., $\sum_{i=1}n_i=N$, as well as the orthonormality of the dressed natural orbitals, i.e., $\int\td^{3+M}z\phi^*_i(\bz)\phi_j(\bz)=\delta_{ij}$, are imposed via Lagrange multipliers as, e.g., explained in Ref.~\citep{Andrade2015}. We have available two different orbital-optimization methods, a conjugate-gradient algorithm (see Sec.~\ref{sec:numerics:rdmft:cg}) and an alternative method that was introduced by \citet{Piris2009} (see Sec.~\ref{sec:numerics:rdmft:piris}). The latter expresses the $\phi_i$ in a basis set and can use this representation to considerably speed up calculations in comparison to the conjugate-gradient algorithm. It was used for all results presented in the following. However, it is not trivial to converge such calculations in practice and we developed a protocol to obtain properly converged results. The interested reader is referred to App.~\ref{sec:numerics:dressed:convergence}.

\subsection{Validation of polaritonic RDMFT}
\label{sec:dressed:results:implementation_real_space:validation}
We now validate our real-space implementation by comparing to exact solutions of simple atomic and molecular systems.
The different systems are described by a local potential $v(x)$ and coupled to one photon mode. We transfer the systems in the dressed basis, that leads to a dressed local potential 
\begin{align}
	v'(x,q) = v(x) + \tfrac{1}{2} (\lambda x)^2 + \tfrac{\omega^2}{2} q^2 - \tfrac{\omega}{\sqrt{2}}q (\lambda x).
\end{align} 
Specifically, we consider a one-dimensional model of a helium atom (He), i.e., 
\begin{align}
	v_{He}(x) = -\frac{2}{\sqrt{x^2+\epsilon^2}}	
\end{align} 
and a one-dimensional model of a hydrogen molecule ($H_2$), i.e.,
\begin{align}
\label{eq:results_potential_H2}
	v_{H_2}(x) =-\frac{1}{\sqrt{(x-d)^2+\epsilon^2}} -\frac{1}{\sqrt{(x+d)^2+\epsilon^2}}
\end{align} 
at its equilibrium position $d= d_{eq}=1.628$ a.u.\@  We use the soft Coulomb interaction 
\begin{align}
	w(x,x') = 1/\sqrt{|x-x'|^2+\epsilon_C^2}
\end{align}
for all test systems.\footnote{Note that we introduced $v_{H_2}$ already in Sec.~\ref{sec:est:comparison} (eq.~\eqref{eq:est:comparison:potential_H2}).} Note that the \emph{softening parameters} $\epsilon$/$\epsilon_C$ are a standard tool for one-dimensional models of atoms. In contrast to the 3d case, the divergence of the potential $1/|x|$ for $x=0$ leads to many peculiar and unwanted effects that are discussed since the 1950ies~\citep{Loudon1959}. The softening parameter removes the divergence and can be used as a fitting parameter to give the 1d systems a similar energy structure and wave function behavior to the corresponding 3d-systems.\footnote{See for example \citet{Gebremedhin2014} for an introduction in the topic and further details about the mathematical properties of the soft-Coulomb approximation.} We set $\epsilon=\epsilon_C=1$ as usual for these two systems~\citep{Ruggenthaler2009,Fuks2011}. In Sec.~\ref{sec:dressed:results:fancy_examples} instead, when we study the one-dimensional Beryllium atom we will need to adjust $\epsilon$ to guarantee properly bound electrons (Sec.~\ref{sec:dressed:results:2mores_is_different}). In Sec.~\ref{sec:dressed:results:3confinement}, we will even use $\epsilon$ as an additional parameter to explicitly control the confinement of the electrons. Finally, we choose the photon frequency in resonance\footnote{Note however that such resonance is not an important feature of ground states and we just use these values because we have to choose one.} with the lowest excitations of the respective ``bare'' systems, so outside of the cavity. For that we calculate the ground and first excited state of each system with the exact solver and find the corresponding excitation frequencies $\omega_{He}=0.5535$ a.u. and $\omega_{H2}=0.4194$ a.u.

For both examples, the dressed auxiliary system is 4-dimensional (2 particles with 2 coordinates each), which is still small enough to be solved exactly in a 4-dimensional discretized simulation box, so that we can compare  dressed RDMFT (with the M\"uller functional of Eq.~\eqref{eq:dressed:est:RDMFT:Mueller_functional}), dressed HF (see Eq.~\eqref{eq:dressed:est:RDMFT:HF_functional}) and the exact solutions. We used the box lengths of $L_x=L_q=16$ a.u. and spacings of $dx=dq=0.14$ a.u. to model the electronic x and photonic coordinates q of the two dressed particles in the exact routine. We want to mention that the box length is not entirely converged with these parameters. In a (numerically very expensive) benchmark calculation, we observed a further decrease of energy with larger boxes (the calculations with respect to the spacing are converged), but the changes in energy and density are only of the order of $10^{-5}$ or less. All the following results require a maximal precision of the order of $10^{-2}$ in energy as well as in the density and thus we can safely use the given parameters. Details can be found in Ch.~\ref{sec:numerics:dressed}. For dressed RDMFT and dressed HF instead, we needed to consider 2-dimensional simulation boxes for every dressed orbital and we set $L_x=L_q=20$ a.u. and $dx=dq=0.1$ a.u. We obtained converged results for $\mathcal{M}=41$ ($\mathcal{M}=71$) natural orbitals for He($H_2$). For further details, the reader is referred to App.~\ref{sec:numerics:dressed:convergence} for the details on how to determine these numbers.
\begin{figure}[ht]
	\begin{overpic}[width=0.49\columnwidth]{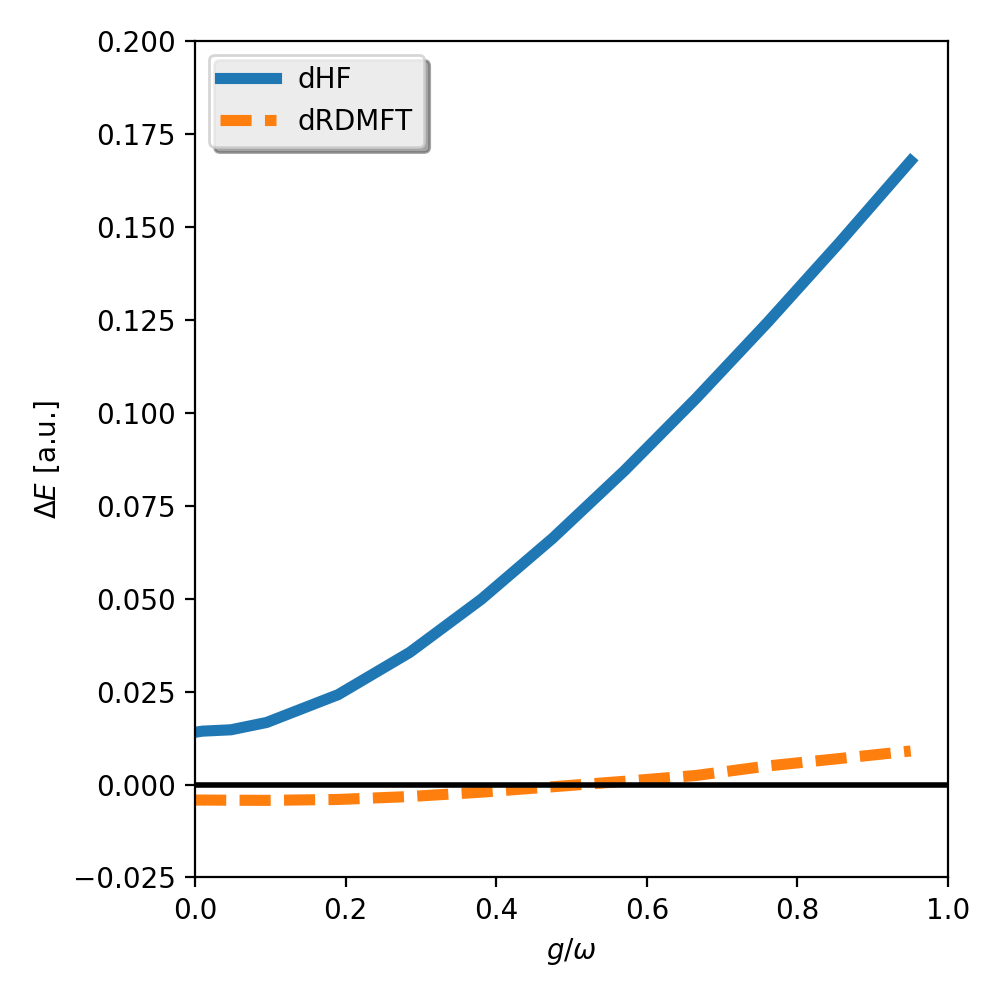}
		\put (50,90) {\textcolor{black}{He}}\hfill
	\end{overpic}
	\begin{overpic}[width=0.49\columnwidth]{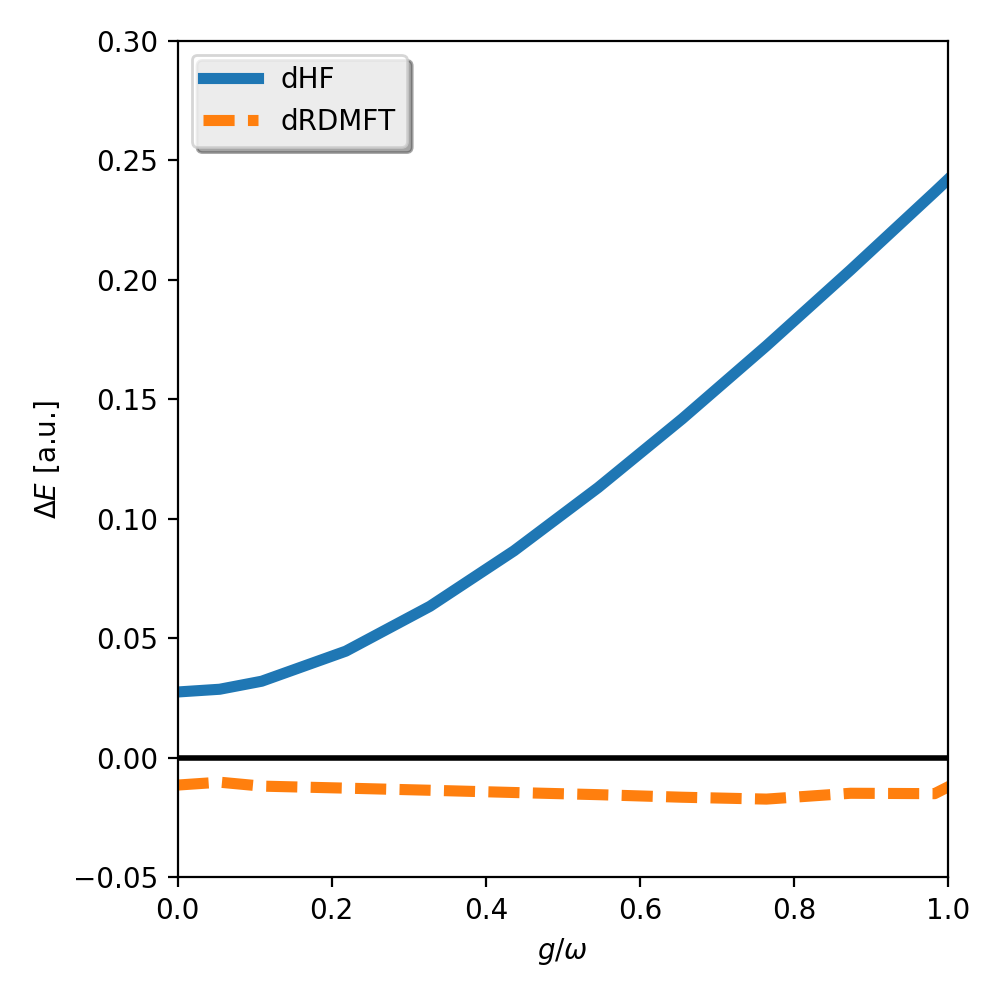}
		\put (50,90) {\textcolor{black}{$H_2$}}
	\end{overpic}
	\caption{Differences of dressed HF (dHF) and dressed RDMFT (dRDMFT) from the exact ground state energies (in Hartree) as a function of the coupling $g/\omega$ for the (one-dimensional) He atom (left) and (one-dimensional) $H_2$ molecule (right) in the dressed orbital description. Dressed RDMFT improves considerably upon dressed HF. For both systems, the energy of dressed RDMFT remains close to the exact one, the error of dressed HF instead increases with the coupling strength.}
	\label{fig:H2_He_lambda}
\end{figure}
We first show (see Fig.~\ref{fig:H2_He_lambda}) the deviations of the ground state energies for dressed RDMFT and dressed HF from the exact dressed calculation as a function of the dimensionless relation between effective coupling strength and photon frequency $g/\omega$ for He and $H_2$, respectively. We thereby go from weak to very strong coupling with $g/ \omega = 1$.\footnote{Such strong coupling strength are sometimes referred to as the \emph{deep-strong} coupling regime~\citep{Kockum2018} that has been observed experimentally in different systems like for instance for Landau polaritons\cite{Bayer2017}. For (organic) molecules the highest reported coupling strengths are in the ultra-strong regime of $g/\omega\approx 0.4$\cite{Kockum2018}.} We see that while dressed HF deviates strongly for large couplings, dressed RDMFT remains very accurate over the whole range of coupling strength. 
Still, a more severe test of the accuracy of our method is if instead of merely energies, we compare spatially resolved quantities like the ground-state density $\rho(x,q)\equiv\gamma(x,q;x,q)$. To simplify this discussion, we separate the electronic and photonic parts of the two-dimensional density by integration, i.e., $\rho(x)=\int\td q\, \rho(x,q)$ and $\rho(q)=\int\td x\, \rho(x,q)$. The exact reference solutions show that with increasing $g/\omega$ the electronic part of the density becomes more localized, while the photonic part becomes broadened.
This behavior is captured qualitatively with dressed HF as well as with  dressed RDMFT. The latter performs for the electronic density considerably better over the whole range of coupling strength, whereas for the photonic densities both levels of theory deviate in a similar way from the exact result. This is shown for $g/\omega=0.1$ in Fig.~\ref{fig:H2_He_dens} for both test systems. Looking at the electronic densities, we can observe a feature that the ground state energy does not reveal. In some cases the effects of the two approximations are contrary to each other as we can see in the He case. Here, the dressed RDMFT electronic density is more localized around the center of charge than the exact reference and the electronic density of dressed HF less. In other cases instead, both theories over-localize $\rho(x)$ (here visible for $H_2$.)

An even more stringent test of the accuracy of the dressed RDMFT approach is to compare the dressed 1RDMs. The essential ingredients of the dressed 1RDMs are their natural orbitals $\phi_{i}(x,q)$. Again, we separate electronic and photonic contributions and show their reduced electronic density $\rho_i(x)=\int\td q\, |\phi_i(x,q)|^2$. Fig.~\ref{fig:H2_NO} depicts the first three dressed natural orbital densities of dressed RDMFT in comparison with the exact ones for both test systems. While it holds that for both systems, the lowest natural orbital density of the dressed RDMFT approximation is almost the same as the exact one, and the second natural orbital densities are only slightly different, the third natural orbital densities of $H_2$ differ even qualitatively. For He, similar strong deviations are visible for the fourth natural orbital. However, as long as such strong deviations only occur for natural orbitals with small natural occupation numbers, like in these cases ($H_2: n_1=1.878, n_2=0.102, n_3=0.015$, He: $n_1=1.978, n_2=0.020, n_3=0.001$), their (inaccurate) contribution to the density and total energy remains small.
\begin{figure}[ht]
	\begin{overpic}[width=0.49\columnwidth]{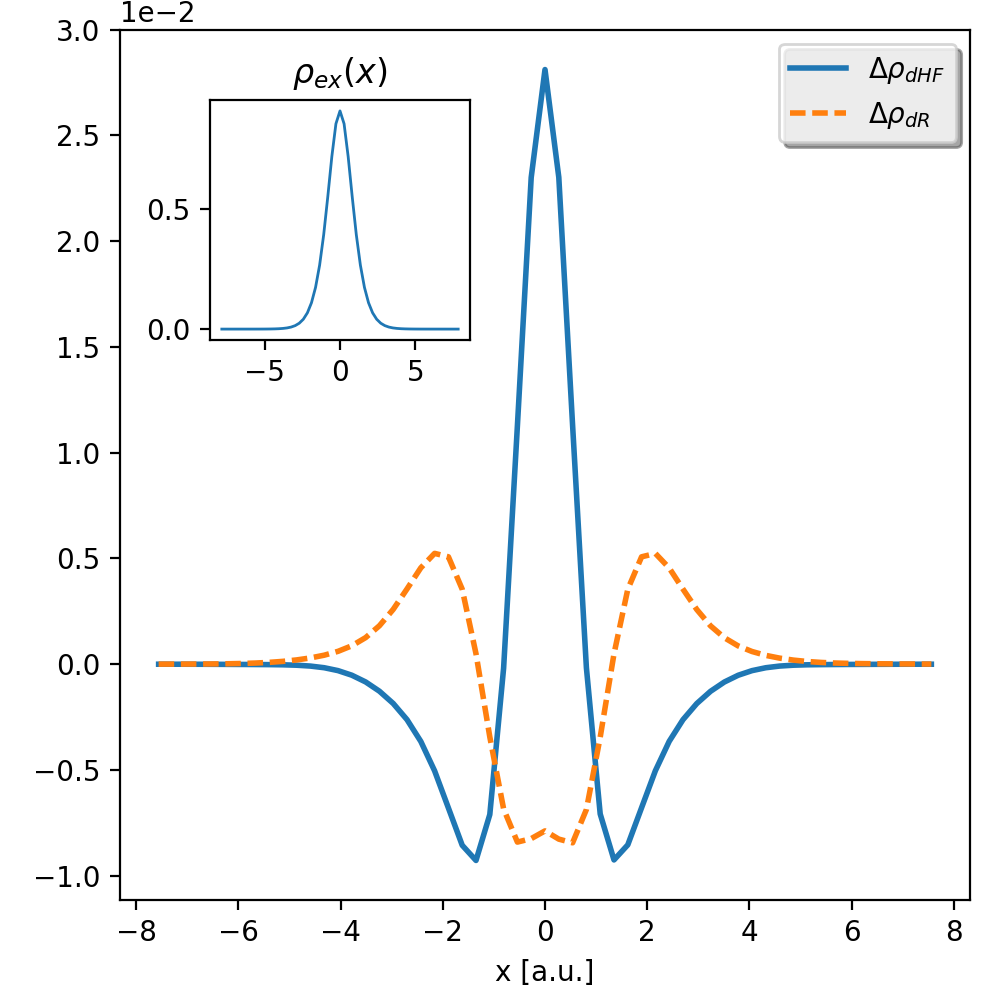}
		\put (85,15) {\textcolor{black}{He}}\hfill
	\end{overpic}
	\begin{overpic}[width=0.49\columnwidth]{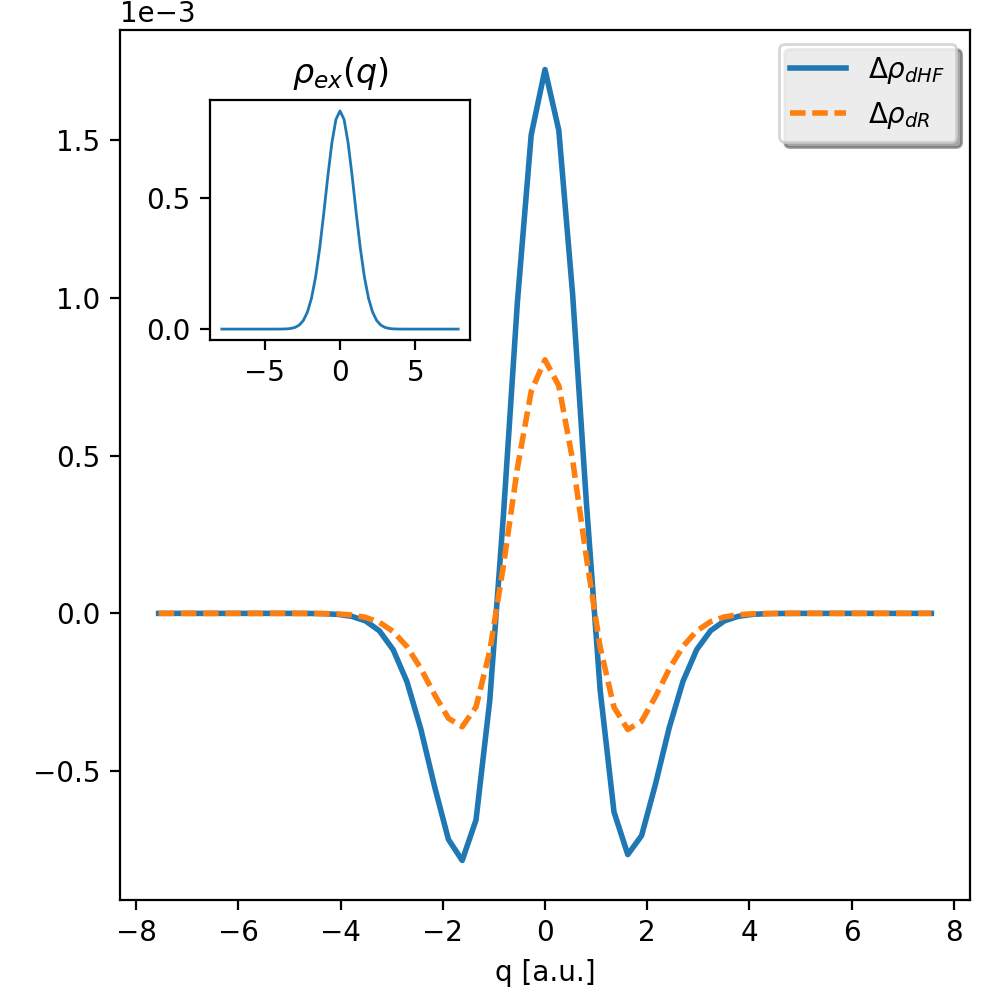}
		\put (85,15) {\textcolor{black}{He}}
	\end{overpic}\\
	\begin{overpic}[width=0.49\columnwidth]{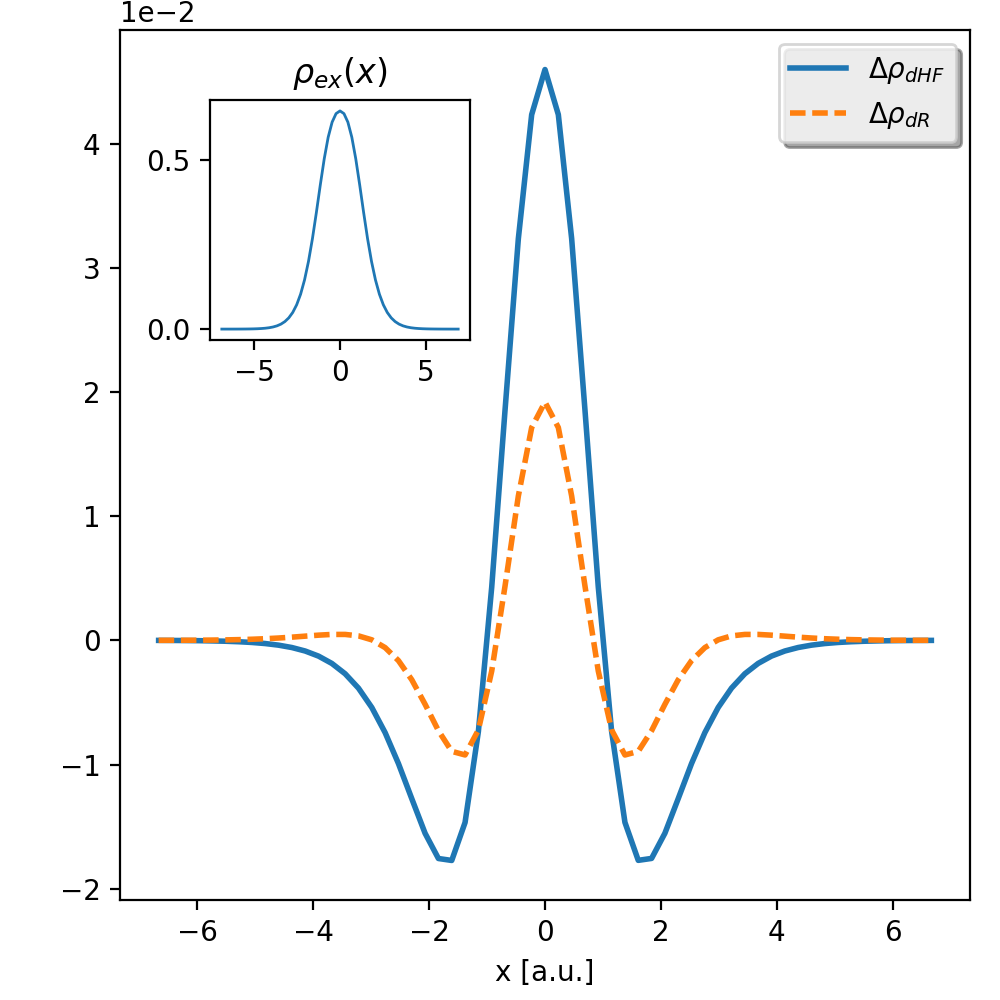}
		\put (85,15) {\textcolor{black}{$H_2$}}\hfill
	\end{overpic}
	\begin{overpic}[width=0.49\columnwidth]{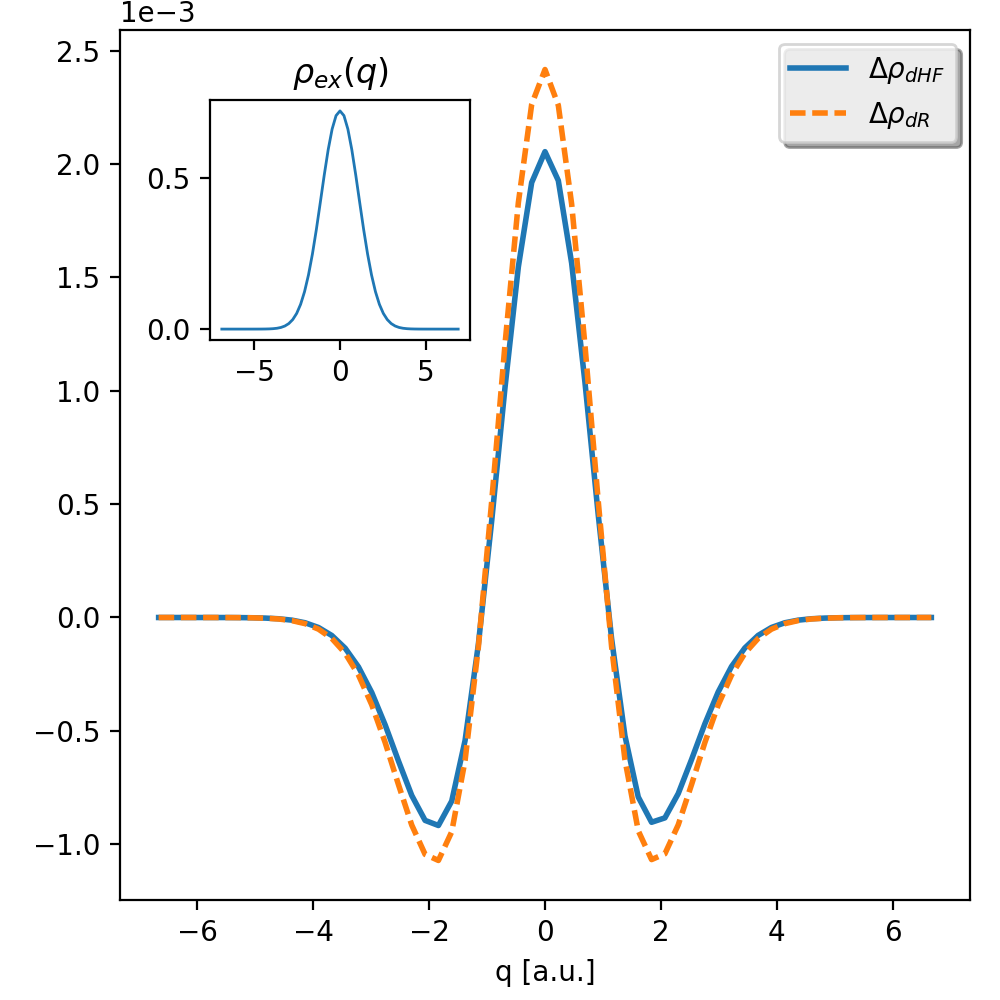}
		\put (85,15) {\textcolor{black}{$H_2$}}
	\end{overpic}
	\caption{Deviations of dressed HF (dHF) and dressed RDMFT (dR) ground state densities from the exact solution (depicted in the insets) for the He atom (top) and the $H_2$ molecule (bottom) with coupling $g/\omega=0.1$. We separate the electronic (x, left) and photonic (q, right) coordinates as explained in the text. For both systems, dressed RDMFT finds a considerably better electronic density than dressed HF, which is consistent with the better result in energy (see Fig.~\ref{fig:H2_He_lambda}.) The photonic densities are reproduced almost exactly for both levels of theory.}
	\label{fig:H2_He_dens}
\end{figure}
\begin{figure}[ht]
	\begin{overpic}[width=0.49\columnwidth]{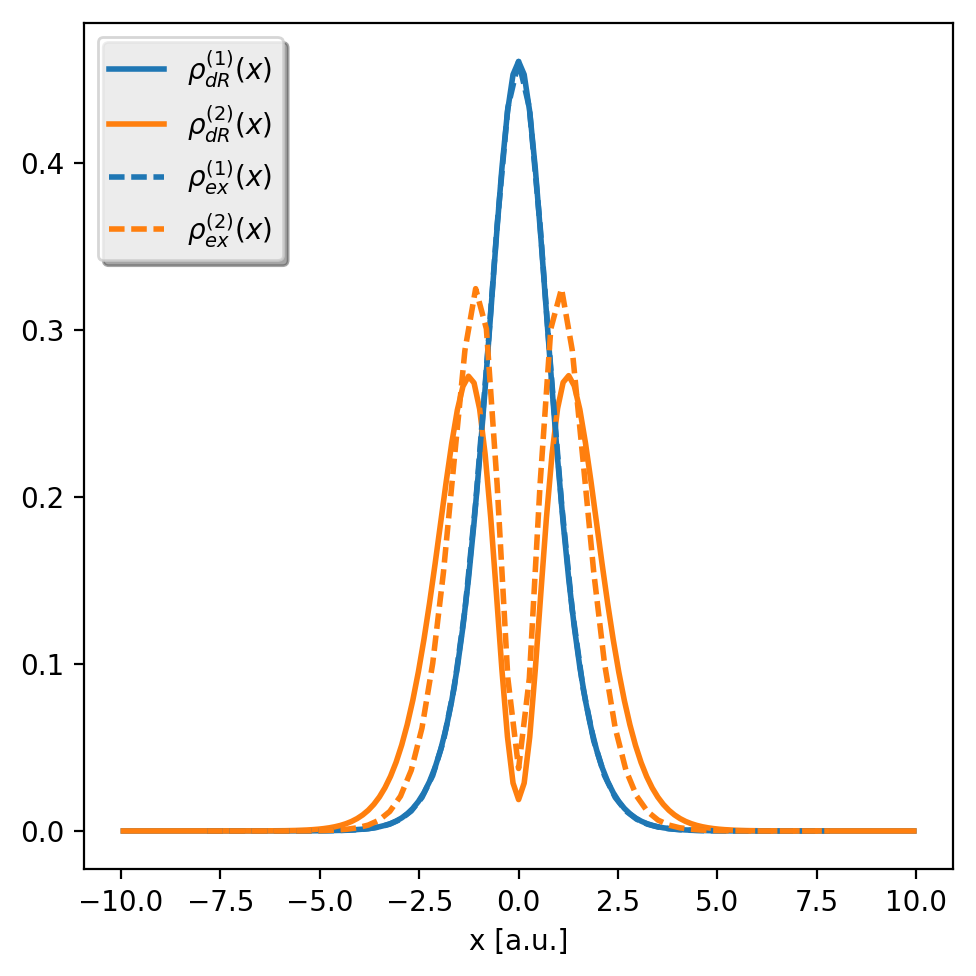}
		\put (85,85) {\textcolor{black}{He}}\hfill
	\end{overpic}
	\begin{overpic}[width=0.49\columnwidth]{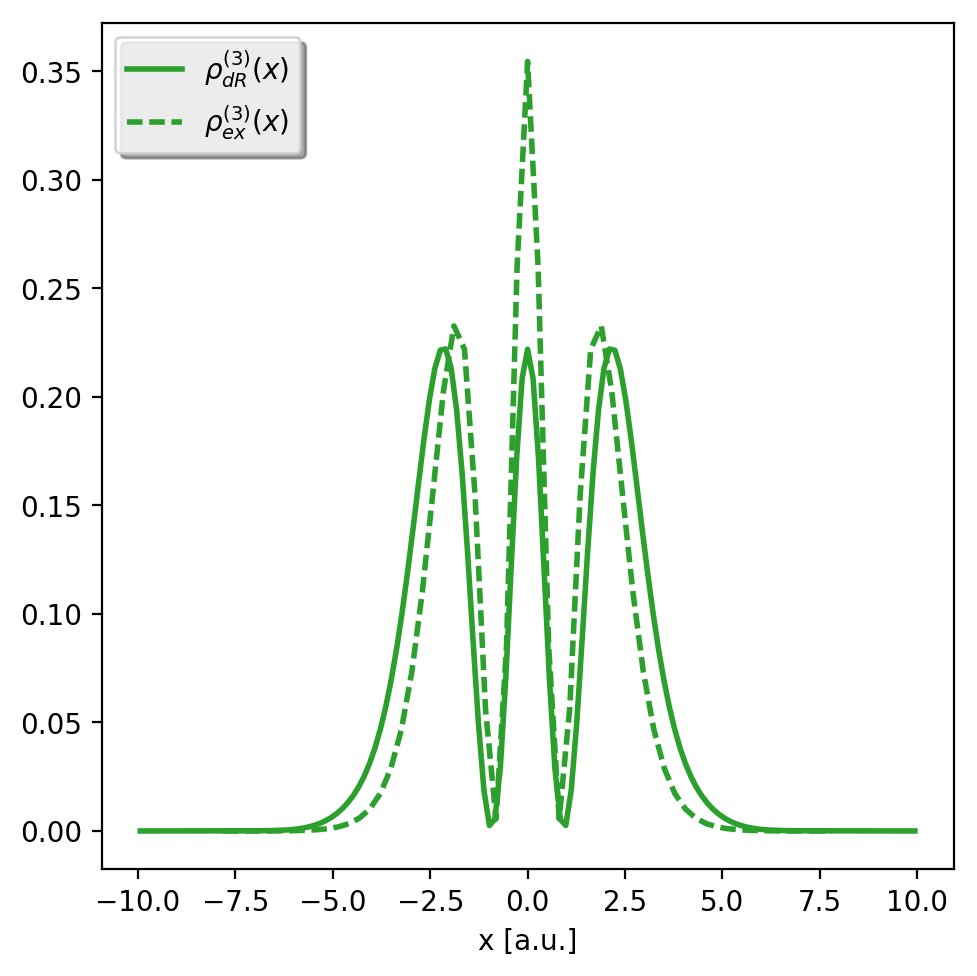}
		\put (85,85) {\textcolor{black}{He}}
	\end{overpic}\\
	\begin{overpic}[width=0.49\columnwidth]{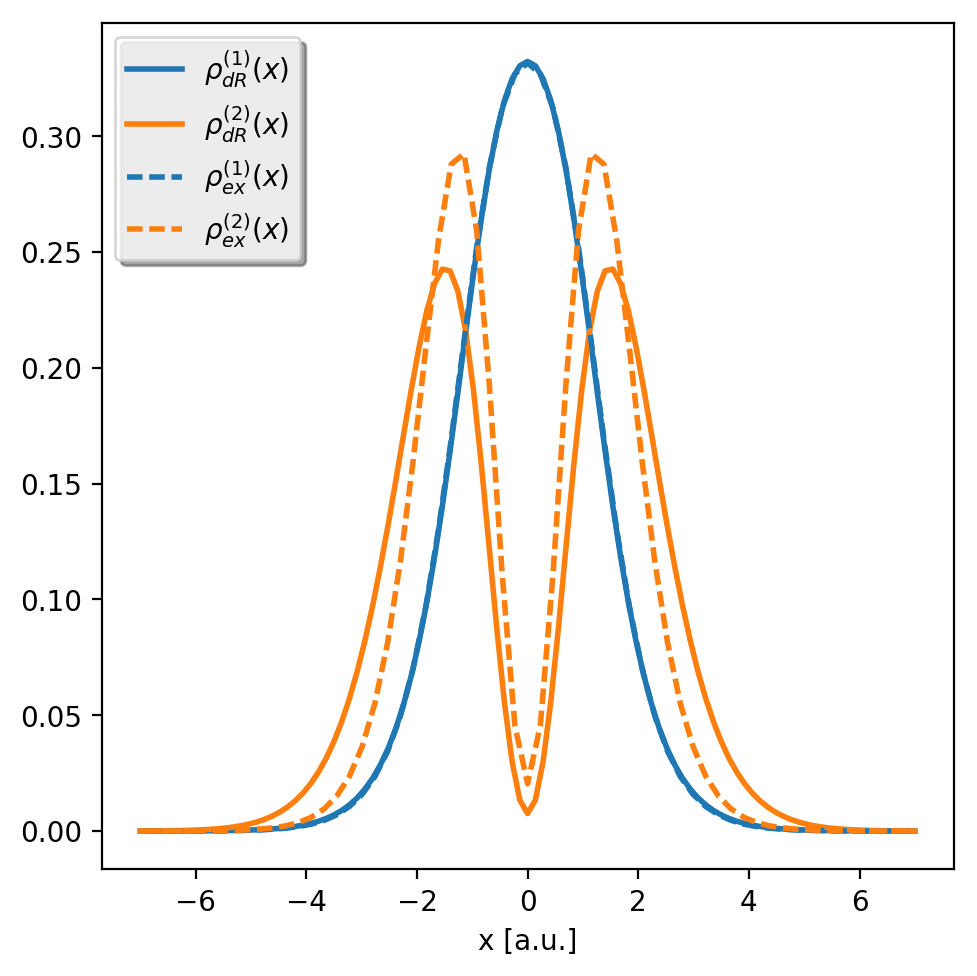}
		\put (85,85) {\textcolor{black}{$H_2$}}\hfill
	\end{overpic}
	\begin{overpic}[width=0.49\columnwidth]{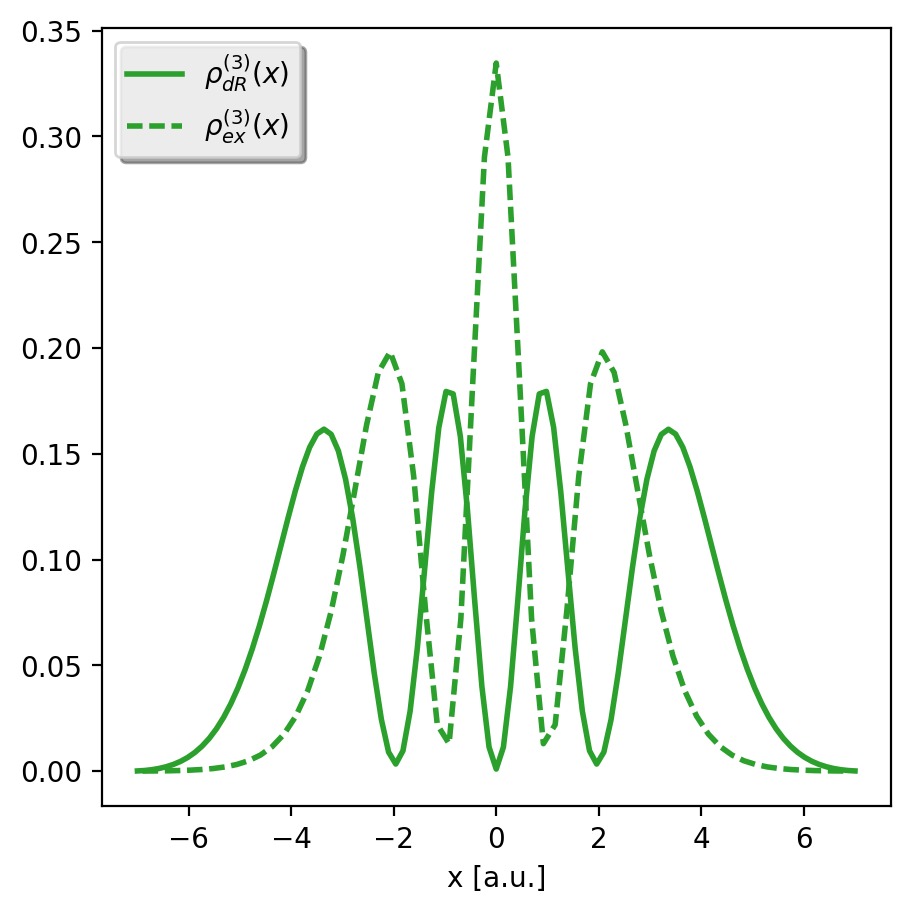}
		\put (85,85) {\textcolor{black}{$H_2$}}
	\end{overpic}
	\caption{The first three natural orbital densities $\rho^{(i)}_{ex/dR}(x)$ of the exact (ex) and dressed RDMFT (dR) calculations are depicted for the He atom (top) and the $H_2$ molecule (bottom) with coupling $g/\omega =0.1$. We see in both cases that $\rho^{(1)}_{ex}(x)$ is almost exactly reproduced by dressed RDMFT, but $\rho^{(2)}_{dR}(x)$ deviates already visibly from $\rho^{(2)}_{ex}(x)$ (left.) However, it is in both cases \emph{qualitatively} correct. This changes for $\rho^{(3)}_{dR}(x)$ of $H_2$, which has one node more than $\rho^{(3)}_{ex}(x)$. Nevertheless, $\rho^{(3)}_{dR}(x)$ of He, is reproduced correctly (right.)}
	\label{fig:H2_NO}
\end{figure}

\begin{figure}[ht]
	\begin{overpic}[width=0.49\columnwidth]{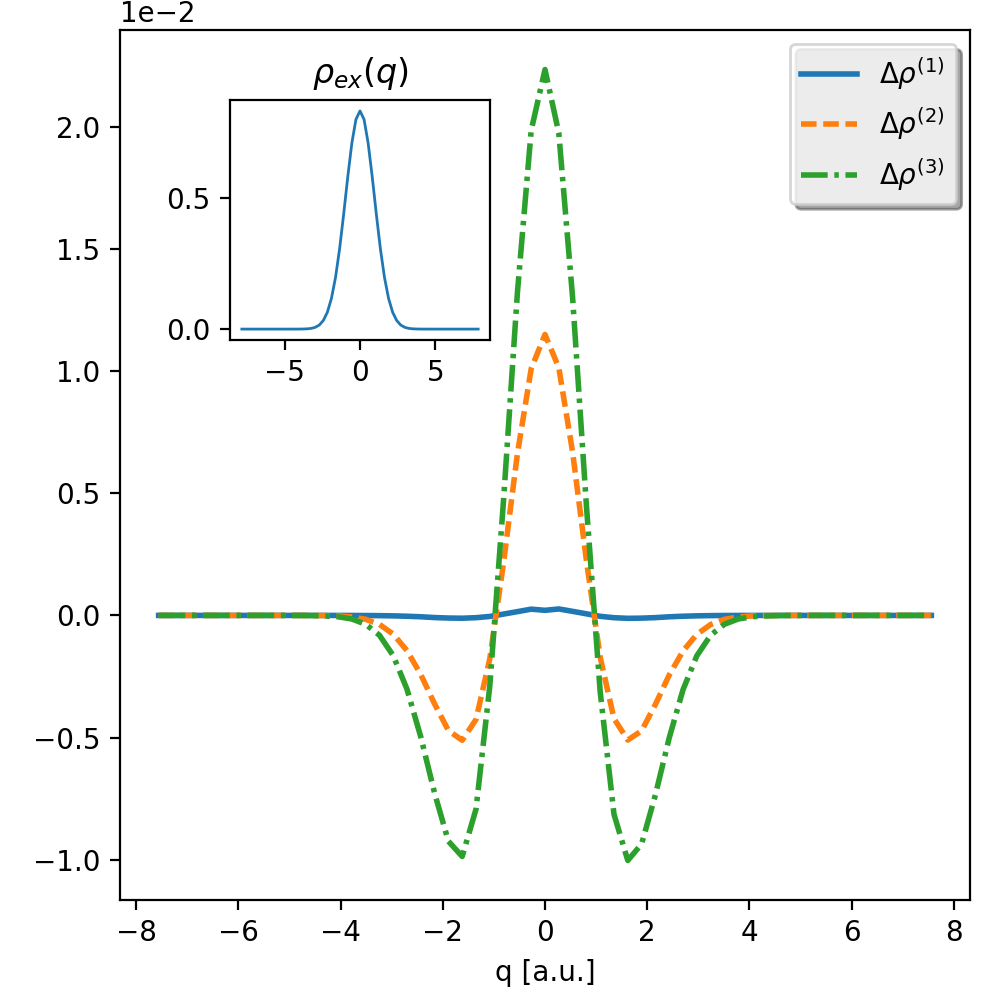}
		\put (80,20) {\textcolor{black}{He}}\hfill
	\end{overpic}
	\begin{overpic}[width=0.49\columnwidth]{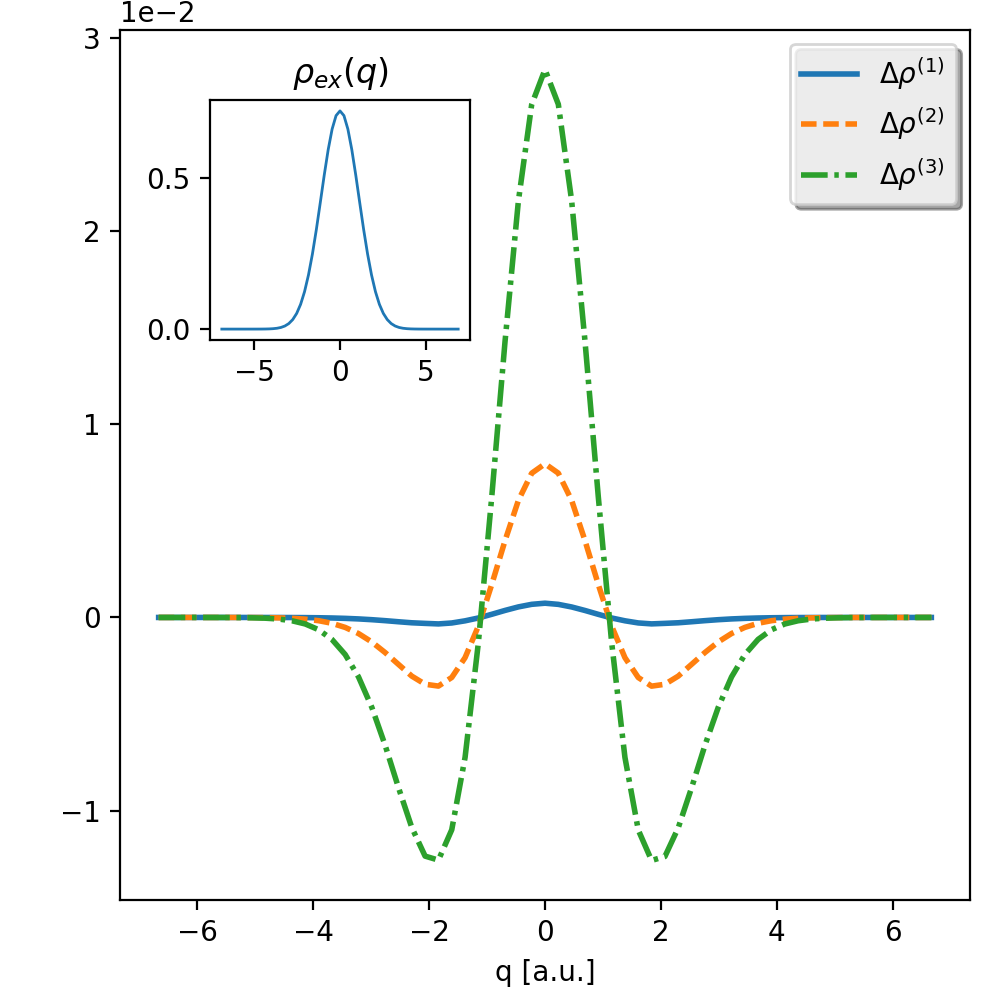}
		\put (80,20) {\textcolor{black}{$H_2$}}
	\end{overpic}
	\caption{We show the differences $\Delta \rho^{(i)}=\rho^{(i)}_{dR}(q)-\rho^{(i)}_{ex}(q)$ between the dressed RDMFT (dR) and the exact (ex) photonic natural orbital densities $\rho^i_{ex/dR}(q)$ for the 3 highest occupied natural orbitals for the He atom (left) and the $H_2$ molecule (right) for coupling strength $g/\omega =0.1$. For both systems, the exact $\rho^{(i)}_{ex}(q)$ have a similar shape as the density (see inset.)  We see in both cases that dressed RDMFT captures the exact solution very well.}
	\label{fig:H2_He_q_orbs}
\end{figure}

To complete the picture, we also look at the photonic natural orbital densities, $\rho_i(q)=\int\td x\, |\phi_i(x,q)|^2$, the first 3 of which are plotted in Fig. \ref{fig:H2_He_q_orbs}, for He and $H_2$. Here, the dressed RDMFT results even agree better with the exact solution than their electronic counterparts. Apparently, dressed RDMFT captures the photonic properties of the tested systems very accurately for the ultra-strong coupling regime. The accuracy drops with increasing $g/\omega $.

\begin{figure}[ht]
	\begin{overpic}[width=0.49\columnwidth]{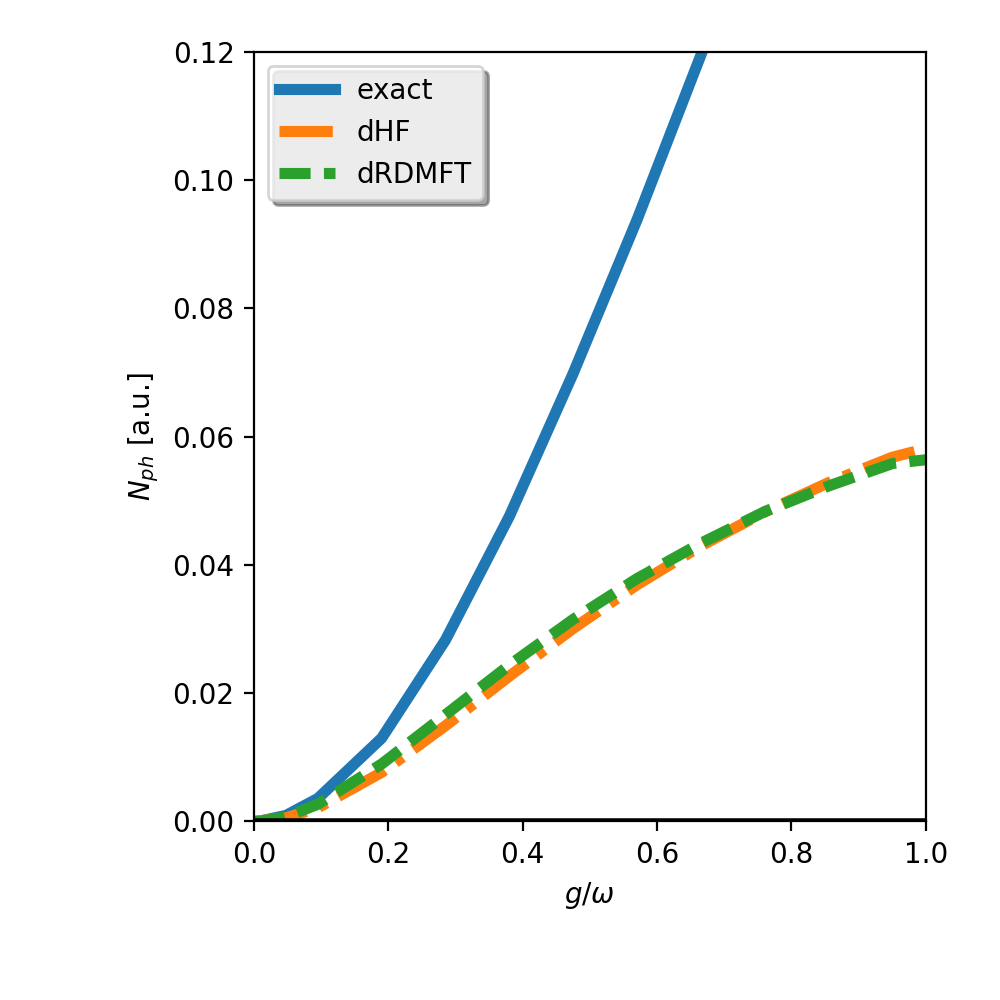}
		\put (80,25) {\textcolor{black}{He}}\hfill
	\end{overpic}
	\begin{overpic}[width=0.49\columnwidth]{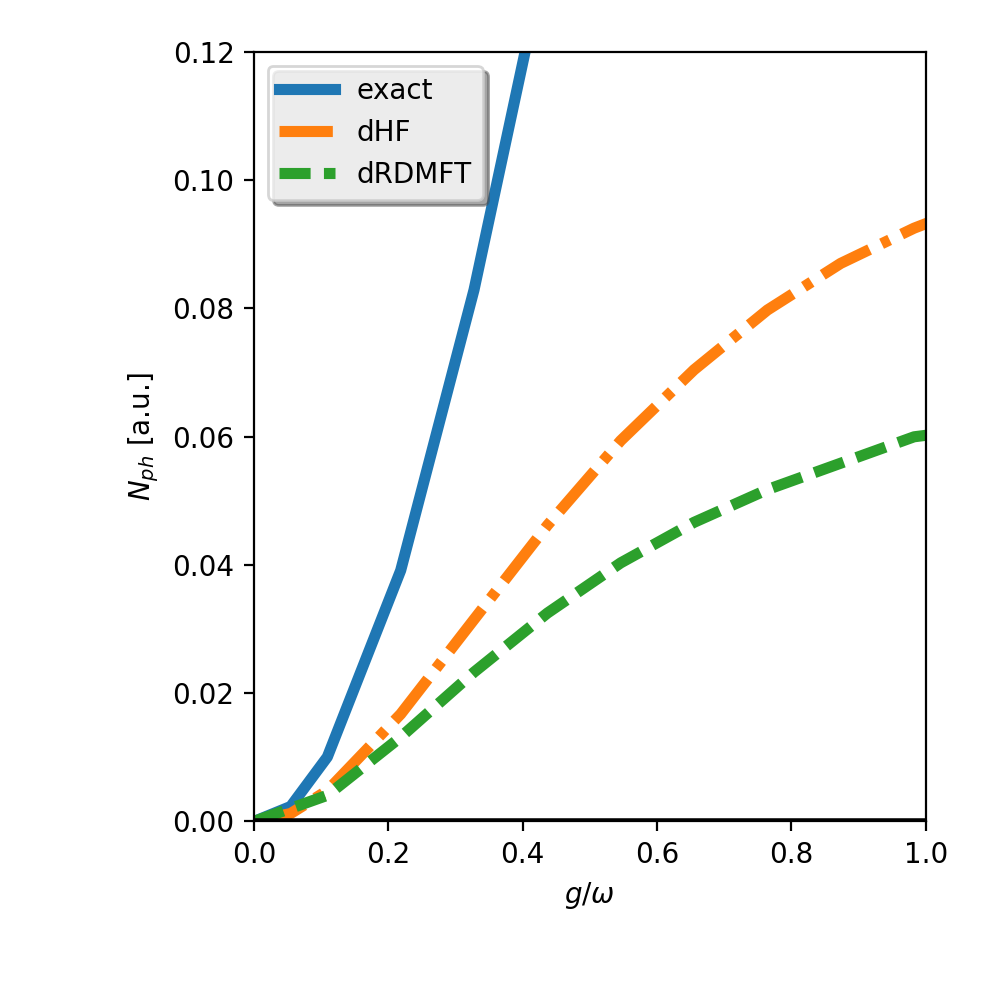}
		\put (80,25) {\textcolor{black}{$H_2$}}
	\end{overpic}
	\caption{The total mode occupation $N_{ph}$, calculated from the exact, dressed HF and dressed RDMFT solutions is shown for He (left) and $H_2$ (right.) We see that both dressed RDMFT and dressed HF underestimate $N_{ph}$. In the ultra-strong coupling regime for $g/\omega >0.3$ both dressed HF and dressed RDMFT (with the Müller functional) deviate strongly from the exact solution.}
	\label{fig:H2_He_nphot}
\end{figure}

As an example for a photonic observable, we show in Fig. \ref{fig:H2_He_nphot} the mode occupation $N_{ph}$\footnote{Note that this is not the usual photon number. See Sec.~\ref{sec:intro:nrqed_longwavelength}.} as a function of the coupling strength $g/\omega $ that we calculated by using Eq. \eqref{eq:mode_occupation_formula}, i.e., $N_{ph}=\frac{E_{ph}}{\omega}-\frac{N}{2}$, with the photon mode energy $E_{ph}=\sum_{i=1}^{\mathcal{M}} n_i \int \td x \td q \, \phi^*_i (x,q)  \left(-\frac{1}{2}\frac{\td^2}{\td q^2} + \frac{w^2}{2}q^2\right) \phi_i (x,q)$. From weak to the beginning of the ultra-strong coupling regime ($g/\omega\approx 0.1$), both dressed HF and dressed RDMFT capture $N_{ph}$ well. For very large coupling strengths, the deviations to the exact mode occupation becomes sizeable. This might sound counter-intuitive, as the photonic density is described comparatively well. The reason is that the photon occupation, in contrast to the density, is mainly determined by the second and third natural orbital, because the first natural orbital resembles a photonic ground state with occupation number zero in the studied cases. Dressed HF does not consider a second orbital (the first instead is doubly occupied) and thus cannot capture the effect. And for dressed RDMFT, the error in the second and third natural orbital is much larger than in the first (see Fig. \ref{fig:H2_He_q_orbs}.) However, it is probable that this can be improved by better functionals.

Although we only can assess our theories on small model systems, we can draw some interesting conclusions. First of all, for coupling strength of $g/\omega \lessapprox 0.2$, both polaritonic HF and polaritonic RDMFT capture the electron photon interaction quite well. For these coupling strengths, the main differences between both levels of theory arise from the different quality of approximation to the Coulomb interaction as the $g/\omega=0$ limit shows. This indicates that a gross part of the interaction is already carried by the polaritons themselves. However, for very large couplings, polaritonic HF considerably overestimates the influence of the interaction (especially the dipole self-energy is too large then), but polaritonic RDMFT remains impressively close to the exact solution. If this is due to some compensation of errors, one has to see. In any case, these results are quite promising and we can definitely use the theories to perform some first studies for larger systems. 

We do not show results for more spatial dimensions, because our implementations are not yet optimized for this case, but we go beyond the 2-particle case. As the results presented in  Sec.~\ref{sec:dressed:results:implementation_lattice:hybrid_influence} have shown, to do so, we either need to enforce the constraints due to the hybrid statistics in the minimization or we consider a scenario, where the photon frequency is large enough to ensure the constraints. We show results for both cases in the following.

\clearpage
\section{Dressed orbitals from a numerical perspective}
\label{sec:dressed:results:numerical_perspective}
We have now discussed the dressed construction in all its details, presented with polaritonic HF and polaritonic RDMFT two examples for its application and showed that these provide an accurate description of a range of simple cavity-QED systems. Let us now briefly analyze the prospects of applying such methods to larger systems, which principally depends on the scaling of their numerical costs with the system size.

We start by remarking that any polaritonic-structure theory scales exactly as the corresponding electronic structure method. This means that the numerically most expensive step of the algorithm has the same dependence on the size of the orbital-basis for the electronic and the polaritonic version of the method. Especially, enforcing the hybrid statistics \emph{does not increase} the scaling of the method. For instance, a very common algorithm to solve the HF equations iteratively diagonalizes the matrix-representation of the Fock operator. If the basis set consists of $B$ elements, the algorithm scales with $B^3$, independently on whether the basis elements correspond polaritonic or electronic orbitals. From a practitioner's perspective, this is probably the most important feature of the polariton description. 

However, despite its importance, the scaling is not sufficient to characterize the numerical challenges of the polariton description. The two most important further challenges are the \emph{increase of the dimension} of the basic orbitals of the theory and the \emph{inclusion of the new constraints} that enforce the right statistics. The former challenge is reflected in the fact that although polaritonic HF has the same scaling as electronic HF with the basis size, we have to consider a larger basis. If the photon mode(s) are satisfactorily described by a basis of size $B_{ph}$ and the matter-part of the system with a basis of size $B_m$, than polaritonic HF scales as $(B_m*B_{ph})^3$. Thus, we get a factor of $B_{ph}^3$ in addition to the matter description. Note that this holds for 1D, 2D and 3D systems equivalently. A similar consideration holds also for a real-space description, where the Fock-matrix cannot be explicitly constructed, but is diagonalized iteratively by, e.g., a conjugate-gradient algorithm like the one of our implementation\cite{Payne1992}. The scaling of such a method in the electronic case is of the order $O(B_m \ln B_m N^2)$ and can even be reduced with state-of-the-art algorithms to $O(B_m \ln B_m N)$~\citep{Lin2016}, where $N$ is the number of electrons.\footnote{Such methods thus scale better than a direct diagonalization of the Fock-matrix, but at the same time the underlying basis size $B_m$ describes the number of grid points and thus is typically much larger than $B_m$ in orbital-based codes. However, for very large systems, both descriptions can have comparable $B_m$. A real-space like description of the displacement coordinates of the photon modes would in a similar manner be quite inefficient for a few photons, but might become even advantageous for large photon numbers.}
As usually one single effective mode is considered, $B_{ph}$ should not become too larger and is therefore not overly numerically expensive.\footnote{Note that we do not want to underestimate the influence of an entirely new dimension in the problem. The given statement depends of course on the necessary size of the photon-basis. However, in the cases that we studied so far, we observed the converged results with photon bases that where of the order of the number of particles.}

The second challenge is instead on the level of the algorithm. We have to solve a nonlinear minimization problem under nonlinear inequality constraints. It is not clear a priori which algorithms are best for such delicate problems and thus, different approaches have to be tested carefully. To develop the algorithm for our implementation (se also Ch.~\ref{sec:numerics:hybrid}), we have already discarded certain approaches, but there are still many open questions (see part~\ref{sec:conclusion}).

	\clearpage


\section{Polaritonic-structure methods: some first results}
\label{sec:dressed:results:fancy_examples}
In this chapter, we present first results for systems that are not (easily) accessible by exact methods to demonstrate the possibilities of our newly developed machinery. 
We aim not (yet) to study experimentally observed effects directly, because our implementations are not yet on the level to treat realistic three-dimensional matter systems. Nevertheless our two implementations allow already to identify a range of nontrivial effects that arise from the complex interplay of electronic correlations and light-matter interaction. Importantly, these effects are local, which makes their description with usual model approaches difficult. Thus, the presented results will allow us to draw some first conclusions on the limitations of such approaches.


\subsection{A first glance in the many-electron-photon space}
\label{sec:dressed:results:0first_glance}
As a starting example, we go one small step beyond our assessment study and present results for a many-body systems that cannot easily be solved exactly: the one-dimensional Beryllium atom Be in a cavity that is described by 
\begin{align}
	v_{Be}(x) = -\frac{4}{\sqrt{x^2+\epsilon^2}}.
\end{align}
In this case, we consider a smaller softening parameter of $\epsilon=0.5$ than for Helium and Hydrogen to make sure that all electrons are properly bound.\footnote{This is a technicality for one-dimensional systems: The softening parameter is a fitting parameter to resemble in 1d a similar energy and wave function behavior to the real 3d-system. For corresponding references, see the second section of Ch.~\ref{sec:dressed:results}.}
Since we use here the \textsc{Octopus} implementation which does not explicitly enforce the hybrid statistics, we cannot choose an arbitrary cavity frequency, but need to give it a \emph{sufficiently} high value such that the constraints are trivially fulfilled (see Sec.~\ref{sec:dressed:est:prescription} and Sec.~\ref{sec:dressed:results:implementation_lattice:hybrid_influence}). We tested this and found that $\omega=\SI{3.0}{\hartree}$ is large enough to make sure that no solutions that violate the Pauli principle can occur.

\begin{figure}[ht]
	\centering
	\includegraphics[width=0.49\columnwidth]{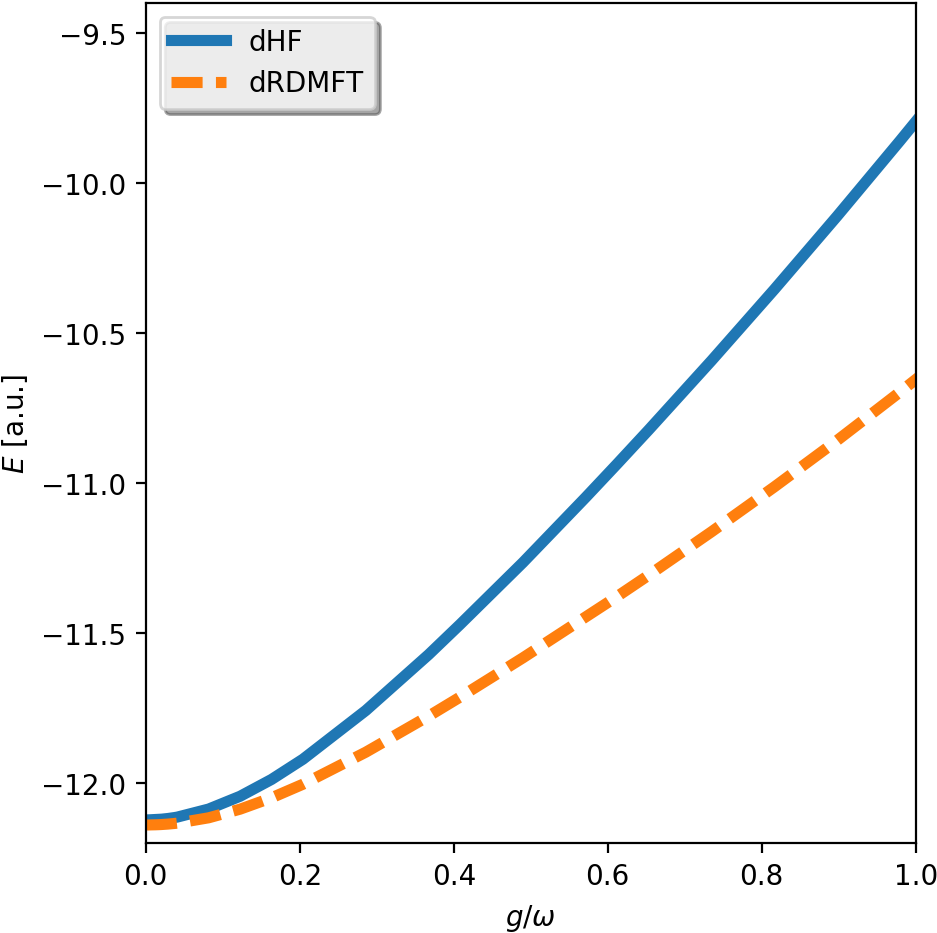}
	\caption{The plot shows the total energy of the dressed HF and dressed RDMFT calculations of Be for increasing $g/\omega $. We observe the same trend as for the two-electron systems: for both levels of theory, the energy grows with increasing $g/\omega $, though for dressed HF faster than for dressed RDMFT.} 
	\label{fig:Be_Etot}
\end{figure}

\begin{figure}[ht]
	\includegraphics[width=0.49\columnwidth]{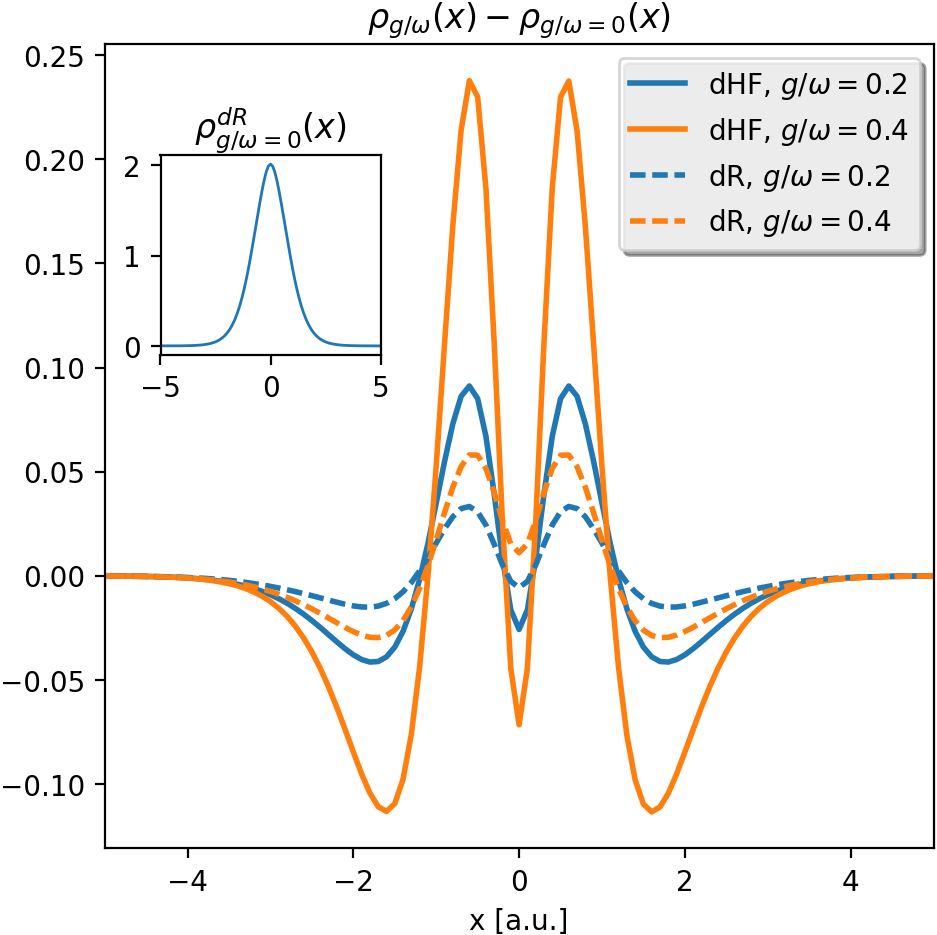}
	\includegraphics[width=0.49\columnwidth]{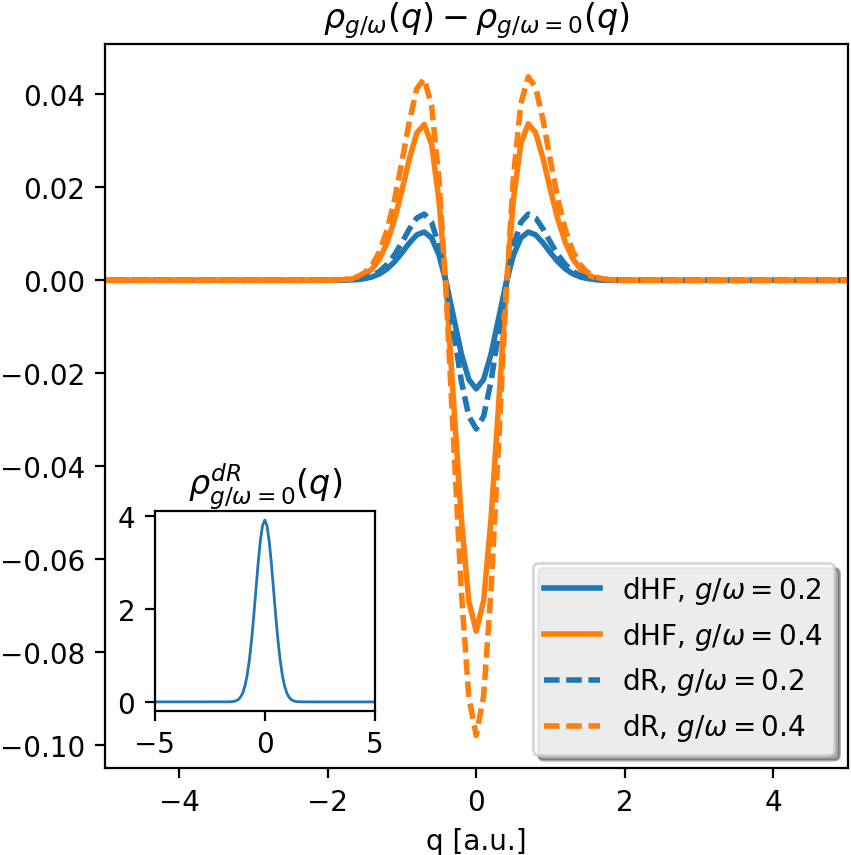}
	\caption{Shown are the electronic ($\rho^{dHF/dR}_{g/\omega}(x)$, left) and photonic ($\rho^{dHF/dR}_{g/\omega}(q)$, right) densities of Be for dressed HF (dHF) and dressed RDMFT (dR) for 2 different coupling strengths subtracted from their counterparts in the no-coupling limit ($\rho^{dHF/dR}_{g/\omega=0}(x/q)$.) We see in the electronic (photonic) case that the dressed RDMFT deviations are less (more) pronounced than for dressed HF.} 
	\label{fig:Be_dens}
\end{figure}

\begin{figure}[ht]
	\centering
	\includegraphics[width=0.49\columnwidth]{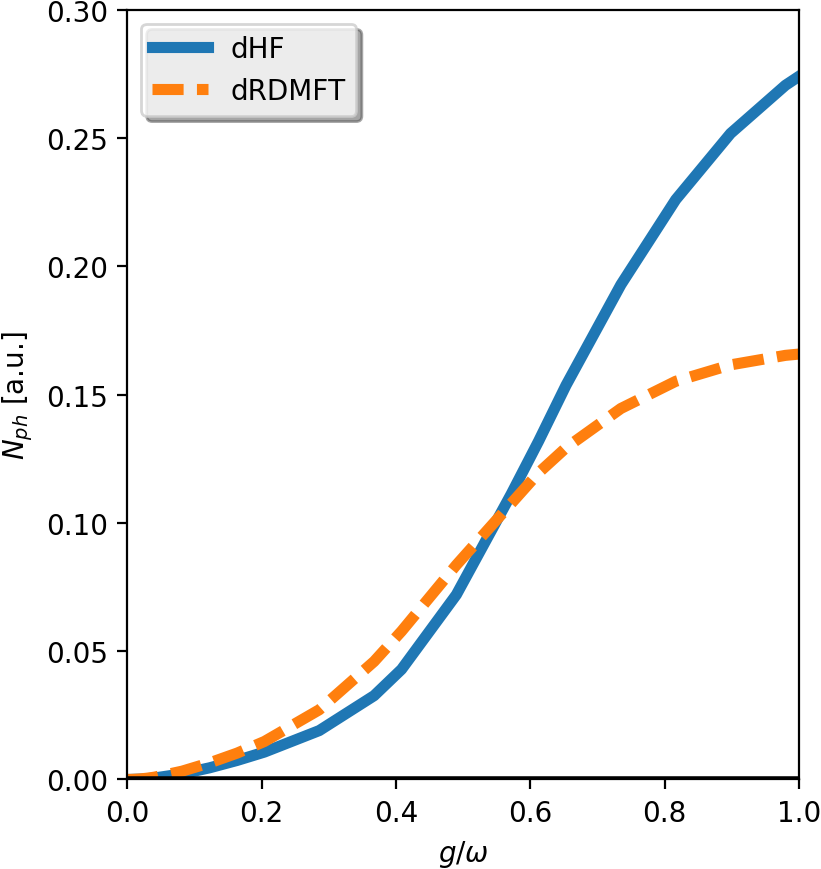}
	\caption{The total mode occupation $N_{ph}$ for dressed HF and dressed RDMFT is shown for Be. We see that dressed RDMFT exhibits larger $N_{ph}$ until a coupling strength of $g/\omega\approx 0.5$. For larger coupling the dressed HF mode occupation becomes higher.} 
	\label{fig:Be_nphot}
\end{figure}

In Fig. \ref{fig:Be_Etot}, we see the total energy as a functional of the coupling strength $g/\omega $ for dressed HF and dressed RDMFT, respectively. Like in the two-electron systems, the deviation between both curves increases for larger $g/\omega $ and as expected the dressed RDMFT energies are lower than the dressed HF results. 
Analyzing the ground-state densities, we see a similar trend as in the 2-particle systems. With increasing $g/\omega$, the electronic (photonic) part of the density becomes more (less) localized, though the details differ as we show in the last part of this chapter (see Fig. \ref{fig:He_Be_comp} and the corresponding part in the main text.)  Comparing dressed RDMFT with dressed HF, we observe that the variation of the electronic (photonic) density with increasing coupling strength is less (more) prounounced for dressed RDMFT, as Fig. \ref{fig:Be_dens} shows.  We conclude the survey of Be with the mode occupation under variation of the coupling strength (see Fig. \ref{fig:Be_nphot}.) We see that the value of $g/\omega\approx 0.5$ separates two regions. For $g/\omega< 0.5$ dressed RDMFT finds a larger mode occupation than dressed HF and for $g/\omega> 0.5$ instead the dressed HF mode is stronger occupied. We found similar behavior also for the 2-particle systems, although the boundary between the 2 regions was considerably different there (He:$ g/\omega\approx0.8$, H$_2:g/\omega\approx0.1$, see Fig. \ref{fig:H2_He_nphot}.)

\clearpage
\subsection{A chemical reaction in real-space}
\label{sec:dressed:results:1chemical_reaction}
\begin{figure}[ht]
	\includegraphics[width=0.49\columnwidth]{{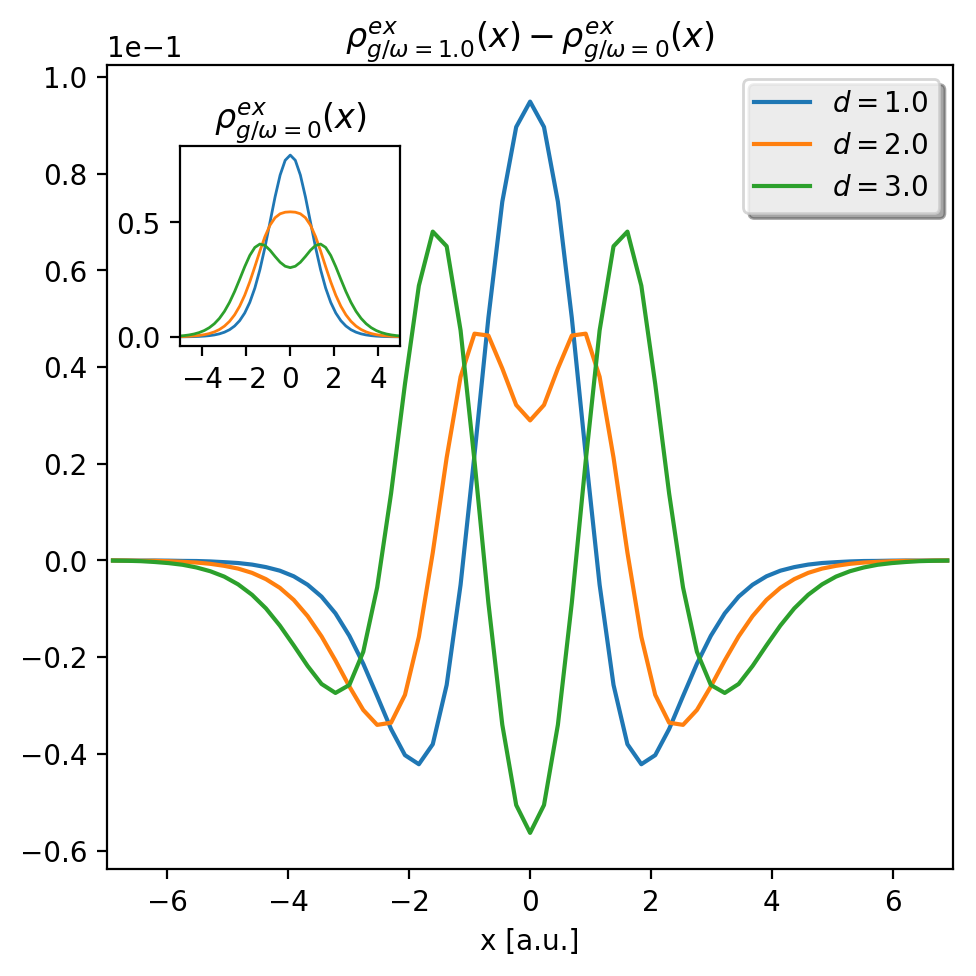}}
	\includegraphics[width=0.49\columnwidth]{{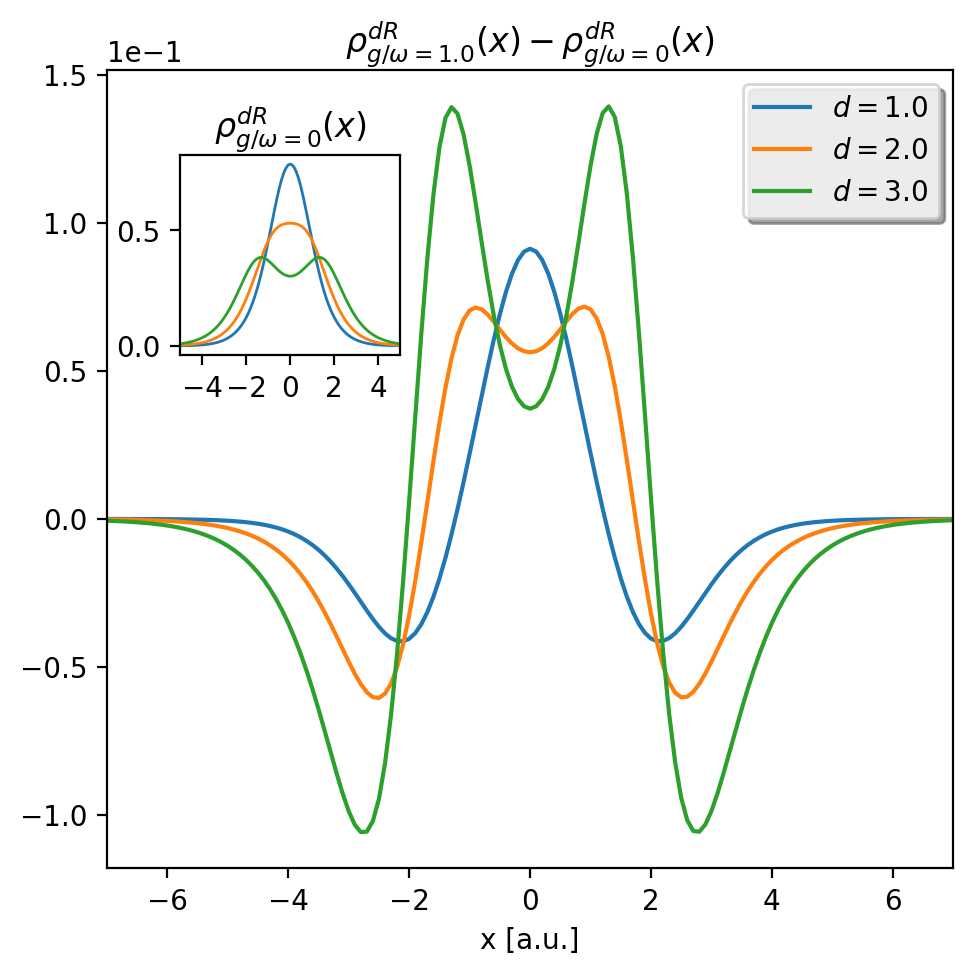}}
	\caption{We show the differences in the electronic density of the H$_2$ molecule for 3 different bond lengths $d$ (as examples of the dissociation) for $g/\omega=1.0$ compared to $g/\omega=0$, calculated exactly ($\rho^{ex}_{g/\omega}(x)$, left) and with dressed RDMFT ($\rho^{dR}_{g/\omega}(x)$, right). We see that for small $d$, the cavity mode reduces the electronic repulsion and localizes the charges at the bond center ($d=1<d_{eq}=1.628$) in comparison to the free molecule (insets.) For larger $d$, the electronic repulsion is locally enhanced such that the charge deviations are separated in two peaks ($d=2$.) For very large $d$, this interplay between local suppresion and enhancement of repulsion becomes more pronounced ($d=3$.) The dressed RDMFT calculations capture the behavior very well.}
	\label{fig:H2_diss}
\end{figure}
We now consider (a simple example of) a chemical reaction, which is a process that mostly can only be understood from a first-principles perspective. Since we are in the fortunate situation to have with polaritonic RDMFT a multi-reference method at hand, we can even study an example that, despite its seeming simplicity, is a challenging test for electronic-structure methods: the dissociation of Hydrogen (see also Sec.~\ref{sec:est:comparison}). When a bond of a molecule is stretched, usually all single reference methods, including HF and basically all known functionals in KS-DFT, become quite inaccurate. Approximate RDMFT functionals are instead often accurate in this limit and we will therefore employ the Müller description.

The dissociation of H$_2$ can be well modelled in a quasi-static picture with the potential 
\begin{align}
\tag{cf. \ref{eq:est:comparison:potential_H2}}
v_{d}(x) =-\frac{1}{\sqrt{(x-d)^2+1}} -\frac{1}{\sqrt{(x+d)^2+1}}
\end{align} 
that depends parametrically on the variable ``bond-length'' $d$. We employed this model already in Sec.~\ref{sec:est:comparison}.

In Fig. \ref{fig:H2_diss}, we see the density of two H-atoms under variation of the distance $d$ with and without the (strong) coupling to the cavity. We see that the influence of the cavity mode strongly depends on the exact electronic structure. The interaction with the cavity mode can locally reduce or enhance the electronic repulsion due to the Coulomb interaction, where the exact interplay between both effects depends on the interatomic distance. Thus, we can observe a number of different effects like pure localization of the density toward the center of charge ($d=1$) or localization combined with a local enhancement of repulsion such that the density deviations exhibit a double peak structure ($d=2$.) The local enhancement of electronic repulsion can grow so strong that the density at the center of charge is reduced but at the same time the density maxima shift closer to each other, which is an effective suppression of electronic repulsion ($d=3$.) This interplay is reflected in the natural orbitals and occupation numbers. The coupling shifts a considerable amount of occupation from the first natural orbital to the second and third one. The contribution to the total density of the former (latter) has the character of enhanced (suppressed) electron repulsion. 
To show the potential of these effects, we present calculations in the deep-strong coupling regime with $g/\omega= 1.0$, where the effects reach the order of 10 \% of the unperturbed density, which is enormous. For smaller coupling strengths of the order of $g/\omega=0.1$, these effects are as diverse, but naturally smaller with density deformations of the order of $10^{-3}$. However, as every observable depends on the density, such deviations are significant. Remarkably, dressed RDMFT reproduces the effects very accurately.
\begin{figure}[ht]
	\begin{overpic}[width=0.49\columnwidth]{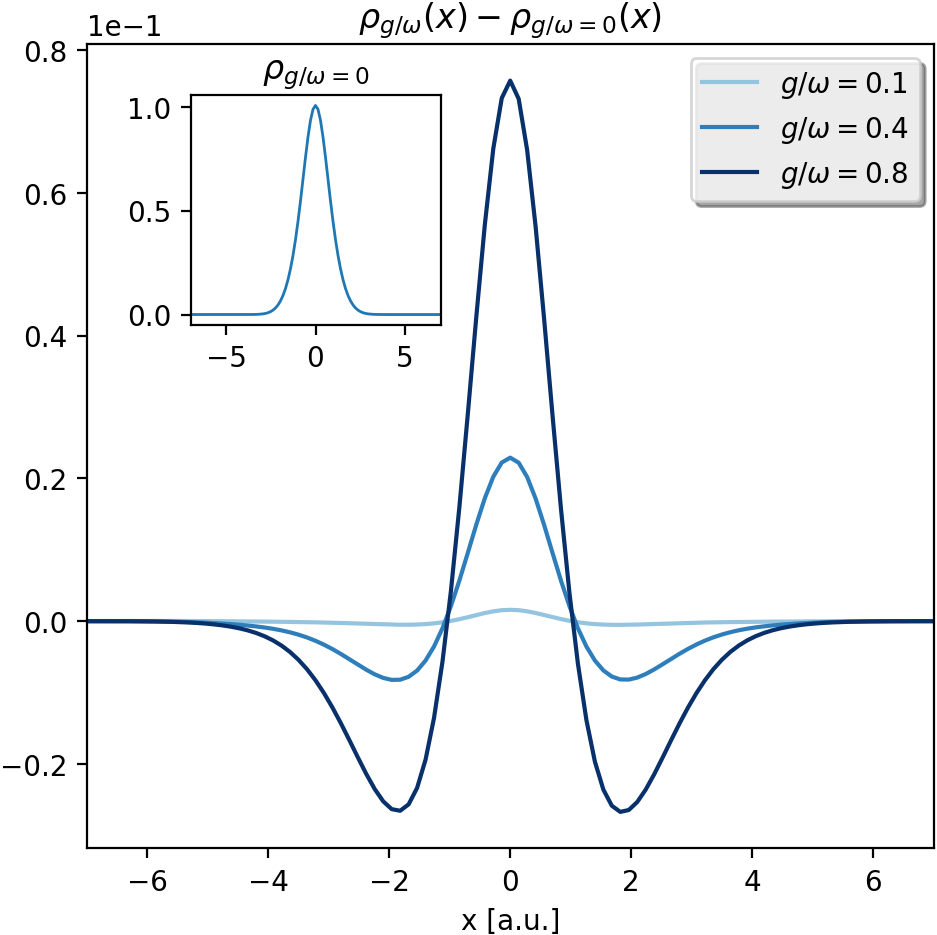}
		\put (85,20) {\textcolor{black}{He}}\hfill
	\end{overpic}
	\begin{overpic}[width=0.49\columnwidth]{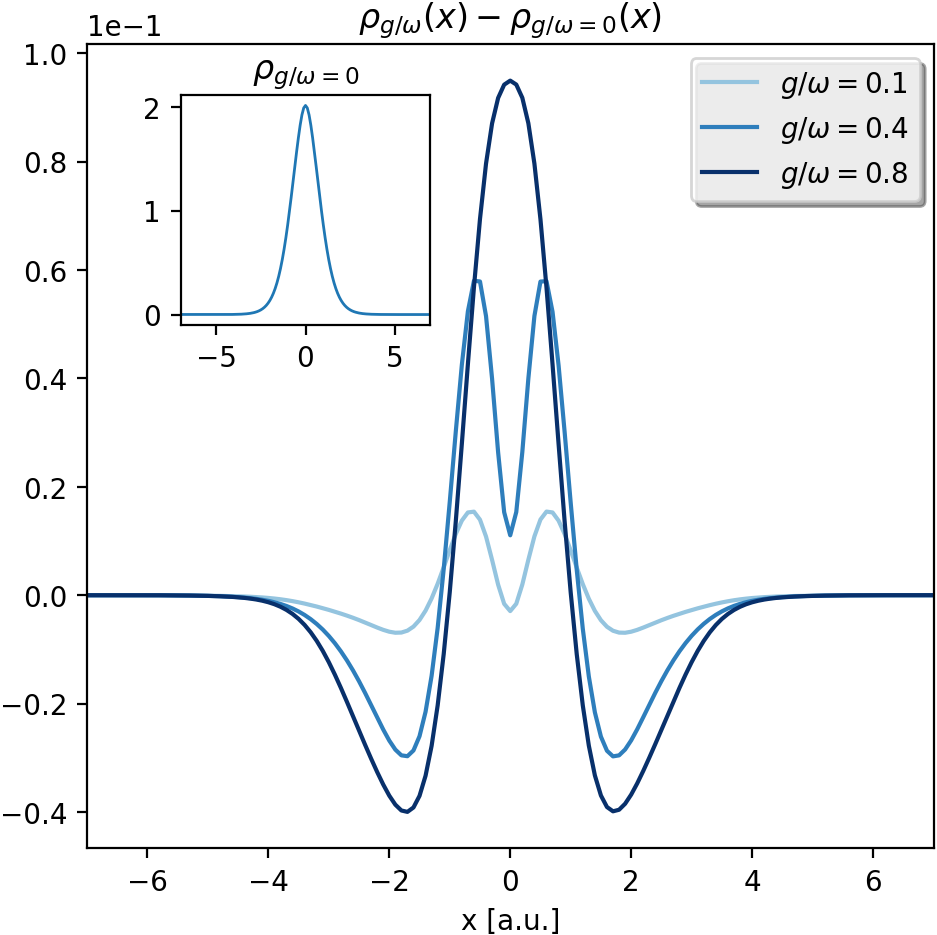}
		\put (85,20) {\textcolor{black}{Be}}\hfill
	\end{overpic}
	\caption{We show the differences in the electronic density ($\rho_{g/\omega}(x)$) of He (left) and Be (right) for 3 different coupling strengths compared to the atoms outside the cavity (insets), calculated with dressed RDMFT. We see that the effect of the cavity is very different for both systems: The strong localization of the electronic density for He indicates the suppression of electronic repulsion for all coupling strengths. For Be instead, we see additionally local enhancement of the repulsion. The interplay of enhancement and suppression changes with increasing coupling strength.}
	\label{fig:He_Be_comp}
\end{figure}
\clearpage
\subsection{More is different: comparing Helium and Beryllium}
\label{sec:dressed:results:2mores_is_different}
In the next example, we compare the behavior of the He and Be atoms under the influence of the cavity. The shapes of the electronic density of the two bare systems are very similar (see insets in Fig. \ref{fig:He_Be_comp}). Since we know that the density determines a system uniquely, one could naively expect that both atoms behave qualitatively similar. However, it is well-known that He is a noble gas which has a very low chemical reactivity and is found in nature almost exclusively as an atomic gas. Beryllium instead occurs naturally only in combination with other elements. It is for example contained in emeralds~\citep{Perera2017}.\footnote{Note that although the one-dimensional setting is very different, there are still many traces of this different behavior as our results will show.} In electronic-structure theory, Beryllium can be described extremely accurately~\citep{Bunge1976} because of its few electrons and it has been studied in many different scenarios. However, from a (cavity-QED) model perspective, atoms are just energy levels that interact with the photon field and all the electronic-structure that contributes to the specific energy is not represented. Accordingly, these (many-body) energy levels are assumed to remain constant even if they interact strongly with the cavity. 

Having this in mind, let us compare the response of our one-dimensional Be and He atoms to the interaction with the cavity-mode. We see in Fig. \ref{fig:He_Be_comp} that they behave actually very differently under the influence of the cavity. The electronic density of He is pushed toward its center of charge with increasing coupling strength, which can be understood as a suppression of the electronic repulsion related to the Coulomb interaction. As He can be understood very well with only one orbital this is to be expected.\footnote{For $g/\omega=0.8$ we still observe $n_1=1.85$. However, it should be noted that the good agreement of dressed RDMFT with the exact calculation in comparison to dressed HF is exactly because of the contribution of the second natural orbital, that is (still considerably) occupied with $n_2=0.14$.}
Things change for Be, where we have several dominant orbitals. With increasing coupling strength, we see like in the dissociation example a subtle interplay between suppression and local enhancement of the electronic repulsion, that depends on the coupling strength. Thus for the same coupling strength, we can observe opposite ($g/\omega=0.1$ and $0.4$ in the plot) but also similar effects ($g/\omega=0.8$ in the plot) in two systems that have almost the same ``bare'' density shape. Like in the dissociation example, this intricate behavior can be understood by the interplay of the different natural orbitals contributing to the electronic density. In this particular case, the main physics happens in the second and third natural orbital, where the former (with a double-peak structure) loses a considerable amount of occupation to the latter (with a triple peak structure) with increasing coupling strength.

These (seemingly simple) examples show how subtle details of the electronic-structure influence the changes induced by the coupling to photons. We see a nontrivial interplay between local suppression and enhancement of the Coulomb induced repulsion between the particles. This is reflected in the natural orbitals and occupation numbers of the light-matter system and thus influences all possible observables.

\clearpage
\subsection{How the microscopic geometry influences light-matter interaction: confinement}
\label{sec:dressed:results:3confinement}
In our last example, we will leave the real-space and go back to our site-based model implementation, which however allows for the study of four-electron systems with arbitrary photon frequencies $\omega$. As we have discussed before, $\omega$ is not very important for polaritonic ground states. Of course different $\omega$ will lead to different results, but qualitatively we can observe the same physics. Since we do not try to study specific experimental setups, we can just use $\omega$ as a kind of ``convergence'' parameter like for the Be atom in the last sections. However, we want to show with this last example that our method is capable to produce results for systems that are not accessible by exact methods. Specifically, we consider a box length of $\SI{30}{\bohr}$ which corresponds to an electronic basis of $B_m=30$ or to a real-space grid from $x=\SI{-14.5}{\bohr}$ bohr to $x=\SI{14.5}{\bohr}$. We consider the case of a 2-electron and a 4-electron system, respectively, and set $\omega=\SI{0.1}{\hartree}$, which is far away from the regime where the fermion ansatz is valid. Thus the right hybrid statistics of the polaritons are crucial. We consider again $B_{ph}=5$, for which all the results are converged. This system might seem quite small but still, it is practically \emph{inaccessible} by exact diagonalization. The corresponding many-body space for 4 particles has a dimension of $(2B_{s})^{N}*B_{ph}= 64.8\cdot10^6$. Only highly optimized methods on a high performance cluster might be able to still explore such a configuration space. In contrast, all the calculations that are presented in the following have been performed on a laptop although the code is by no means optimized.

\begin{figure}[ht]
	\centering
	\begin{overpic}[width=0.49\columnwidth]{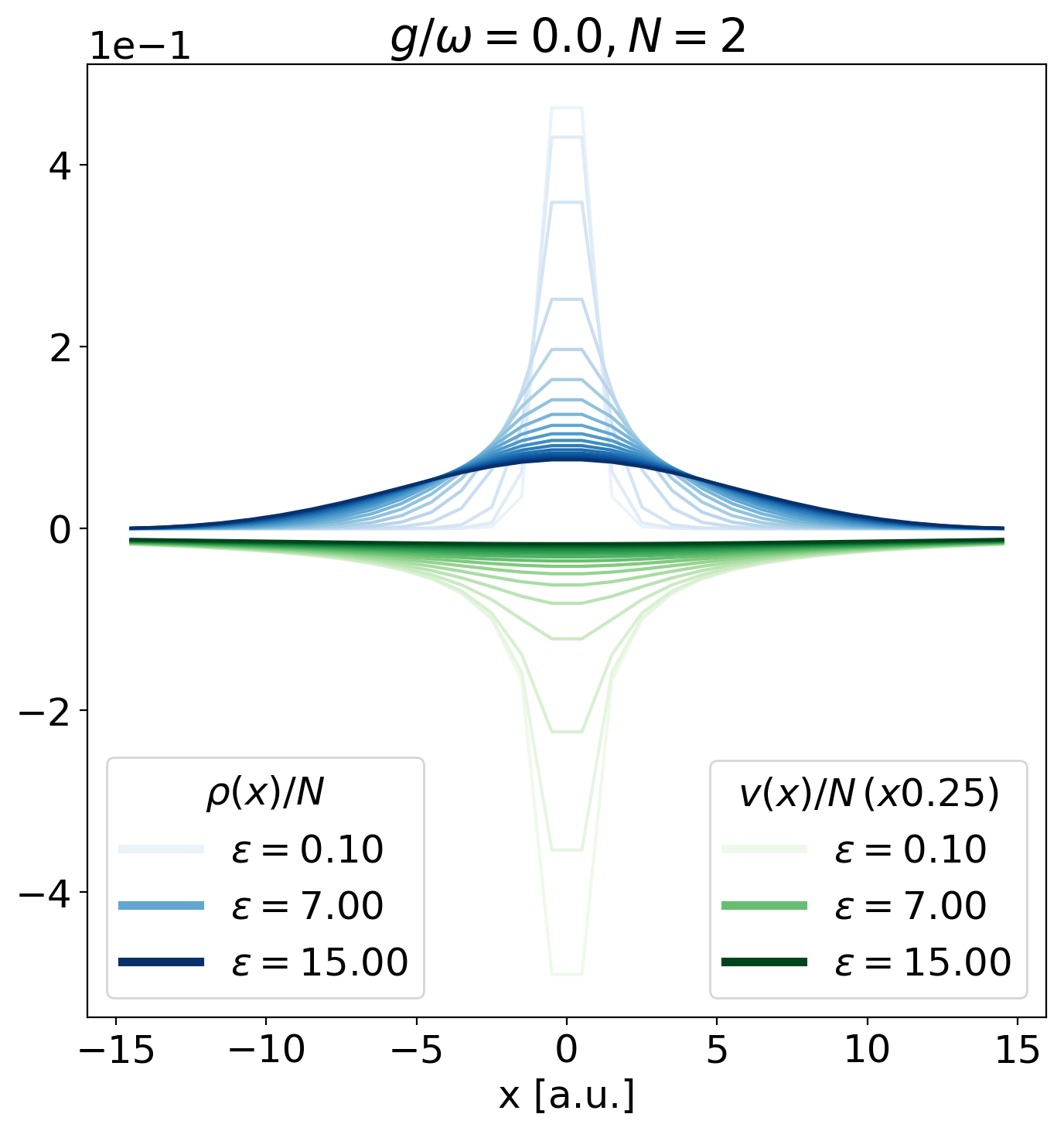}
		\put (13,85) {\textcolor{black}{(a)}}
	\end{overpic}
	\hfill
	\begin{overpic}[width=0.49\columnwidth]{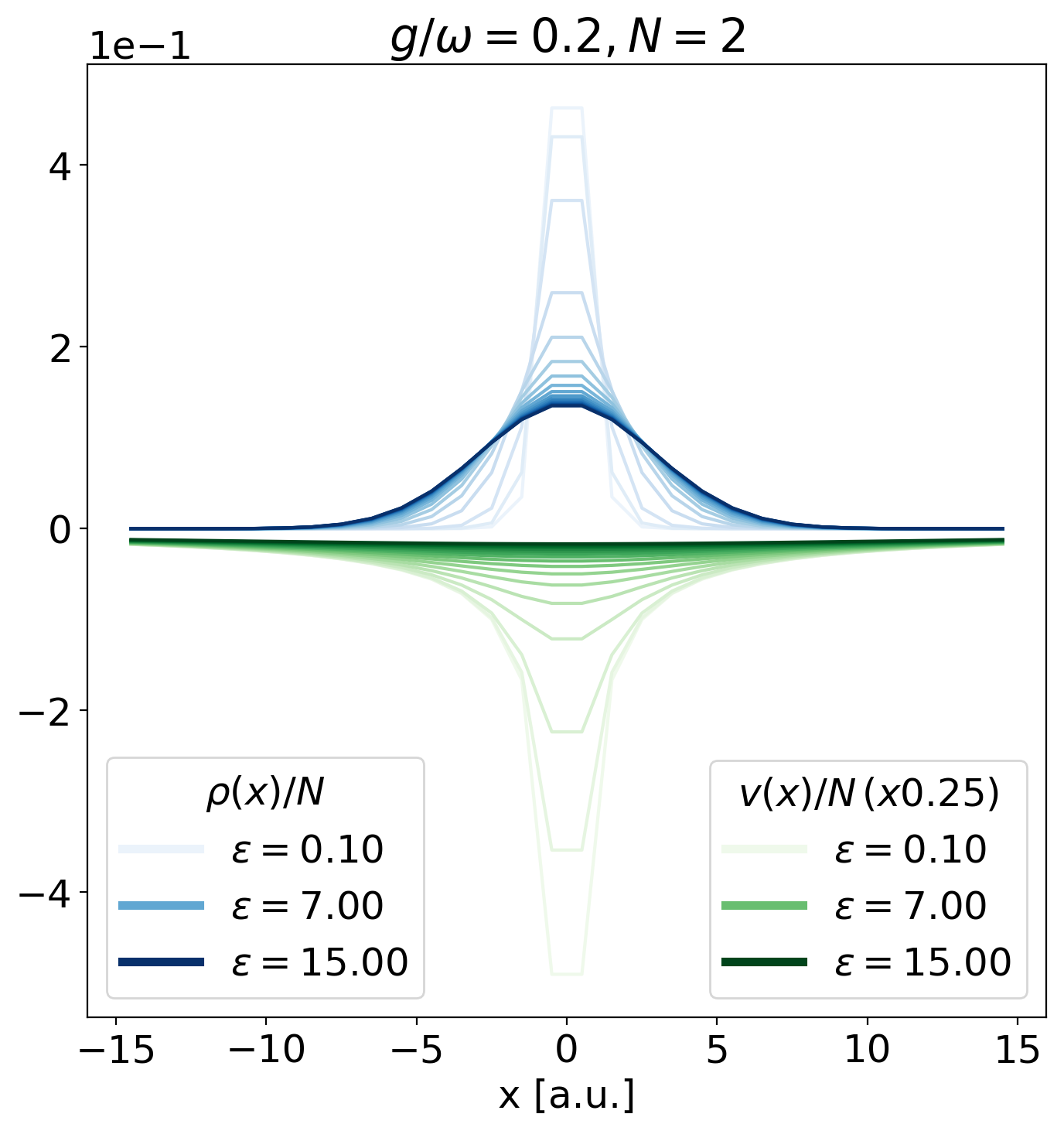}
		\put (13,85) {\textcolor{black}{(b)}}
	\end{overpic}
	\hfill
	\begin{overpic}[width=0.49\columnwidth]{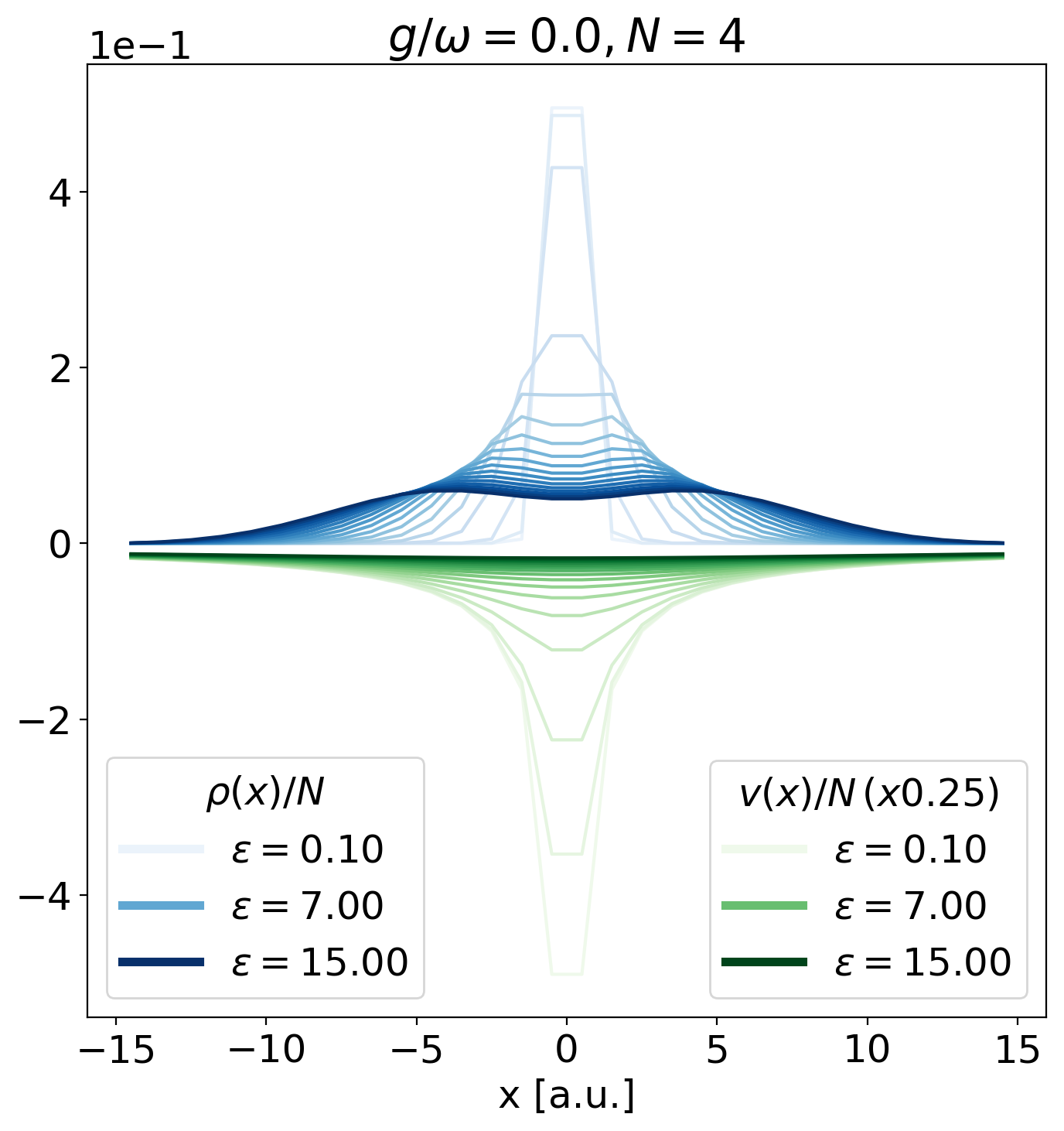}
		\put (13,85) {\textcolor{black}{(c)}}
	\end{overpic}
	\hfill
	\begin{overpic}[width=0.49\columnwidth]{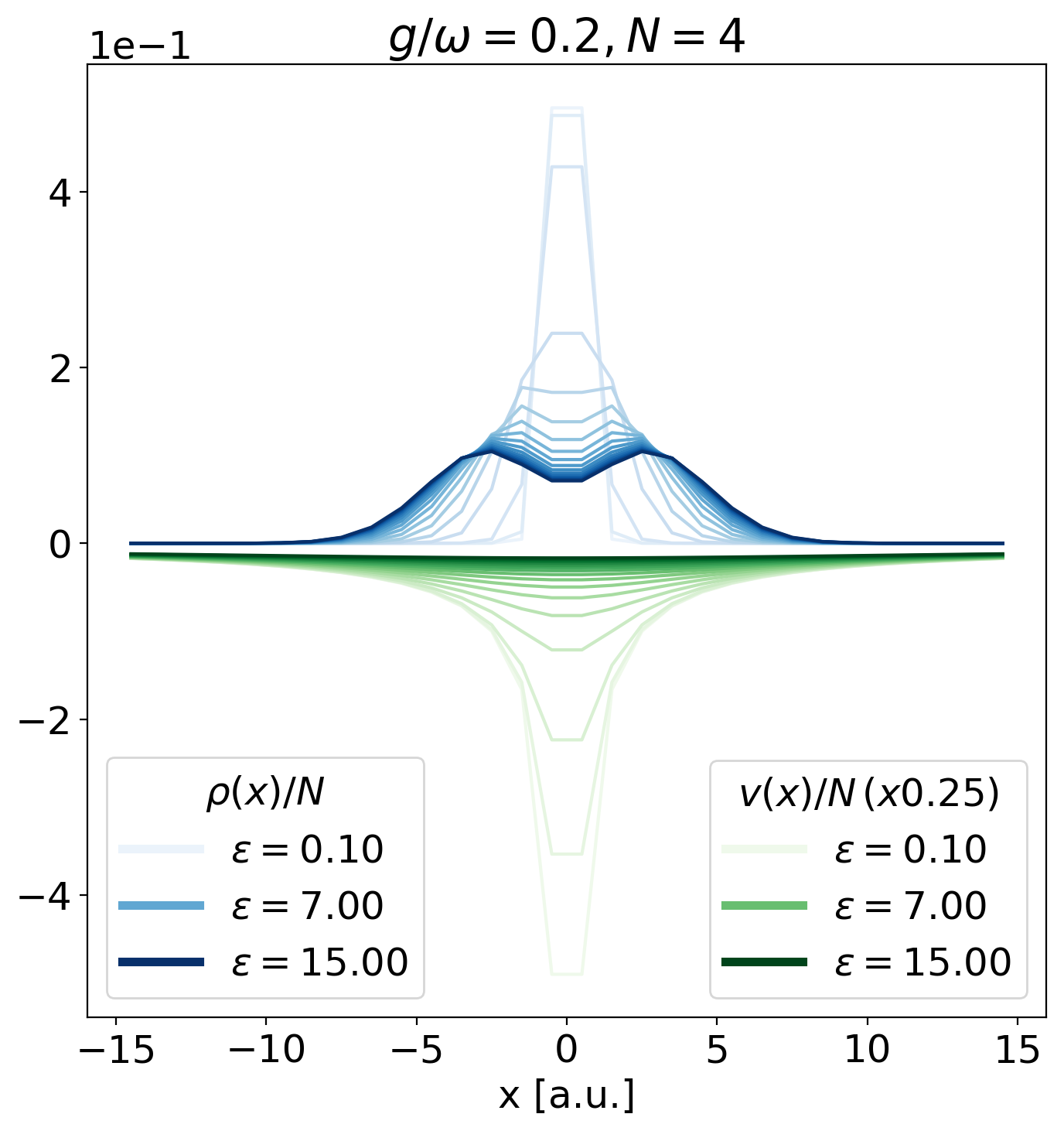}
		\put (13,85) {\textcolor{black}{(d)}}
	\end{overpic}
	\caption{Every plot depicts the normalized electron density $\rho(x)/N$(blue) with corresponding local potentials $v(x)/N=1/\sqrt{x^2 - \epsilon^2}$ (green, rescaled by a factor of $0.25$ for better visibility) for a series of softening parameters $\epsilon$ of the 2- (upper row) and 4-electron (lower row) system and for coupling strength $g/\omega=0$ (left column) and $g/\omega=0.2$ (right column). We see how different both systems respond to the coupling depending on the degree of confinement, measured by $\epsilon$. Note that the legends only depict three example lines (the smallest, largest and middle values of $\epsilon$), respectively.}
	\label{fig:conf_dens_pot}
\end{figure}

In this system, we employ polaritonic HF to study the effect of electron-photon coupling versus electron localization. We consider a matter system with a local potential $v(x)=N/\sqrt{x^2 + \epsilon^2}$, which represents a potential well that is deep (shallow) for small (large) $\epsilon$. The softening-parameter $\epsilon$ thus represents the level of confinement of the potenial $v(x)$ which is depicted in green (for various values of $\epsilon$) in Fig.~\ref{fig:conf_dens_pot}. Note that we need the large simulation box to reduce boundary effects which also represent a form of confinement.
 
Let us first consider how the electronic ground-state density changes when coupling and the localization are varied. To facilitate the comparison between the $N=2$ and $N=4$ case, we plot in Fig.~\ref{fig:conf_dens_pot} the normalized electronic ground-state density $\rho(x)/N$, where $\rho=\gamma_e(x,x)$ is the diagonal of the electronic 1RDM (in blue) and the normalized confinement $v(x)/N$ (in green). In Fig.~\ref{fig:conf_dens_pot} (a) we show the uncoupled 2-particle case and in (b) we use $g/\omega = 0.2$ for the 2-particle case for varying $\epsilon$. In Fig.~\ref{fig:conf_dens_pot} (c) and (d) we show the same plots for the 4-particle case. In both cases we see that for strongly-confined electrons, i.e., for small values of  $\epsilon$, the influence of the strong light-matter coupling on the density is negligible. This is in agreement with the usual assumption underlying, e.g., the Jaynes--Cummings model, that the ground state for atomic systems is only slightly affected by coupling to the photons of a cavity mode. Much higher coupling strengths would need to be employed in order to see a sizeable effect for strong localization. In contrast, once we lift the confinement and the electrons get delocalized, the influence of the light-matter coupling becomes appreciable. The induced changes are not uniform but depend on the details of the electronic-structure, i.e., in the $N=2$ case we have a clear localization effect (Fig.~\ref{fig:conf_dens_pot} (b)) while for $N=4$ we have an enhancement or even emergence of the double-peak structure (Fig.~\ref{fig:conf_dens_pot} (d)). These findings are in good agreement with the He and Be comparison of the previous section and also what \citet{Flick2018abinitio} observed with QEDFT. 

\begin{figure}
	\centering
	\begin{tabular}{ccc}	
		\begin{overpic}[width=0.32\columnwidth]{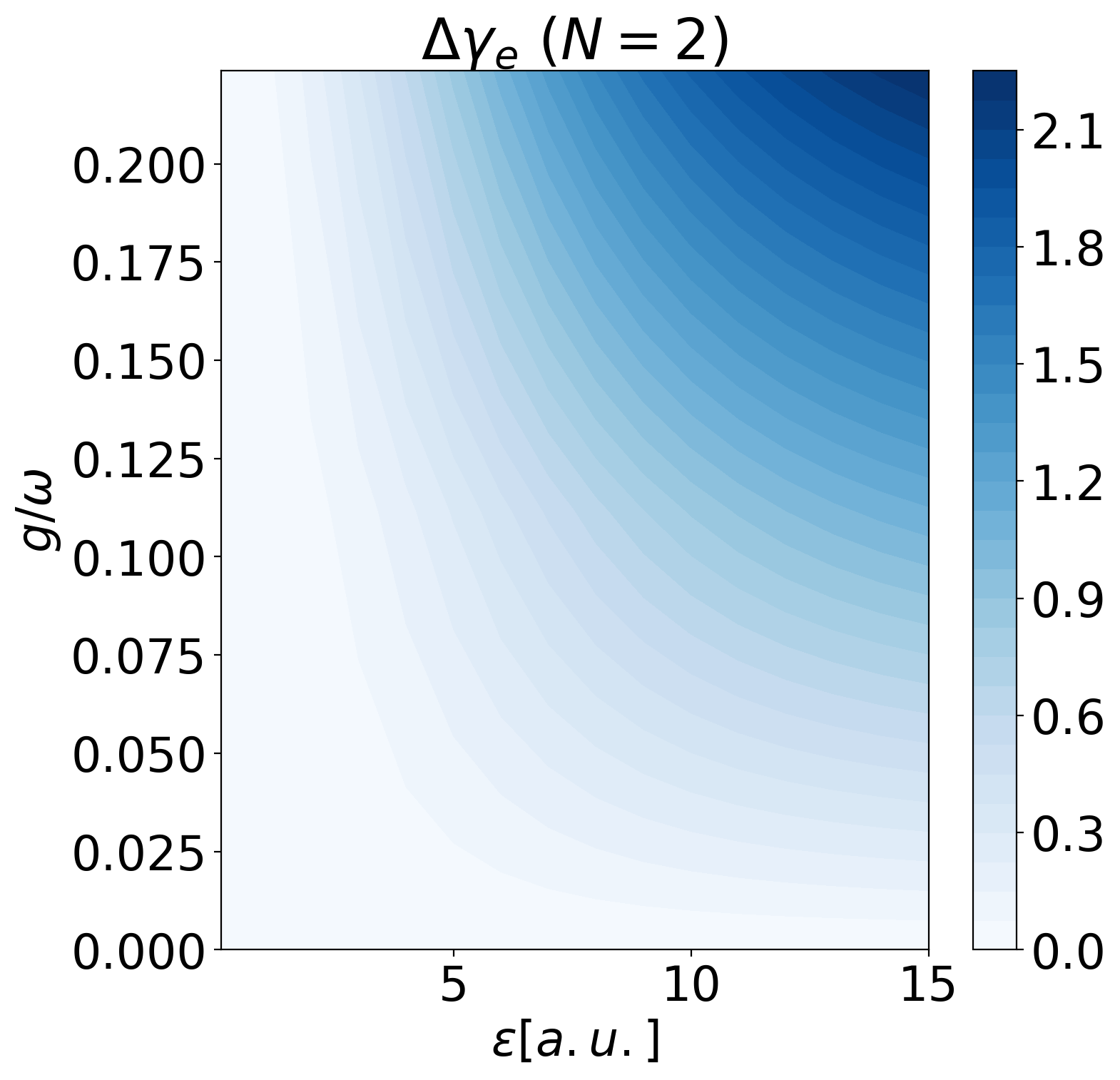}
			\put (21,80) {\textcolor{black}{(a)}}
		\end{overpic}
		&
		\begin{overpic}[width=0.32\columnwidth]{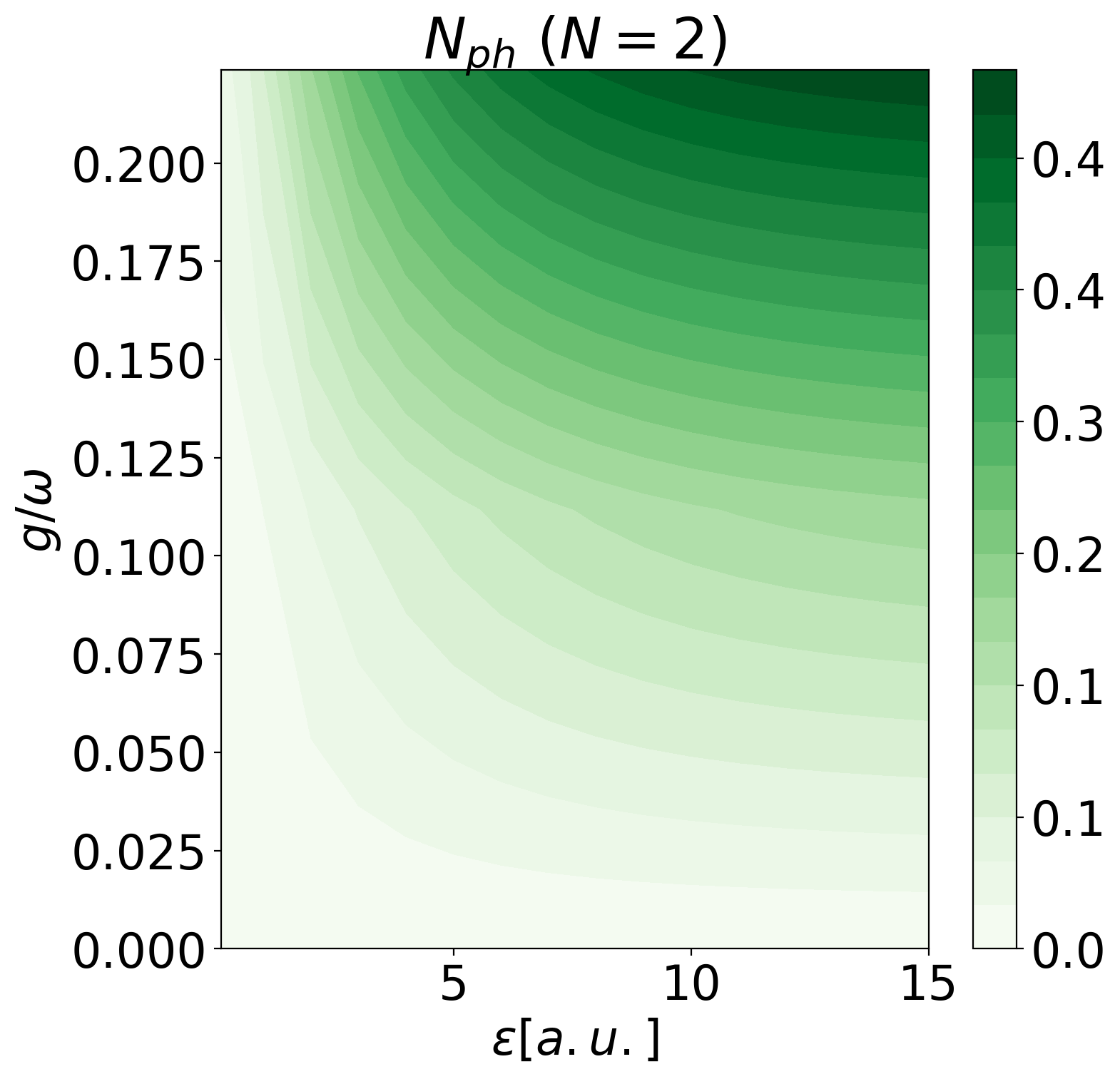}
			\put (21,80) {\textcolor{black}{(b)}}
		\end{overpic}
		&
		\begin{overpic}[width=0.32\columnwidth]{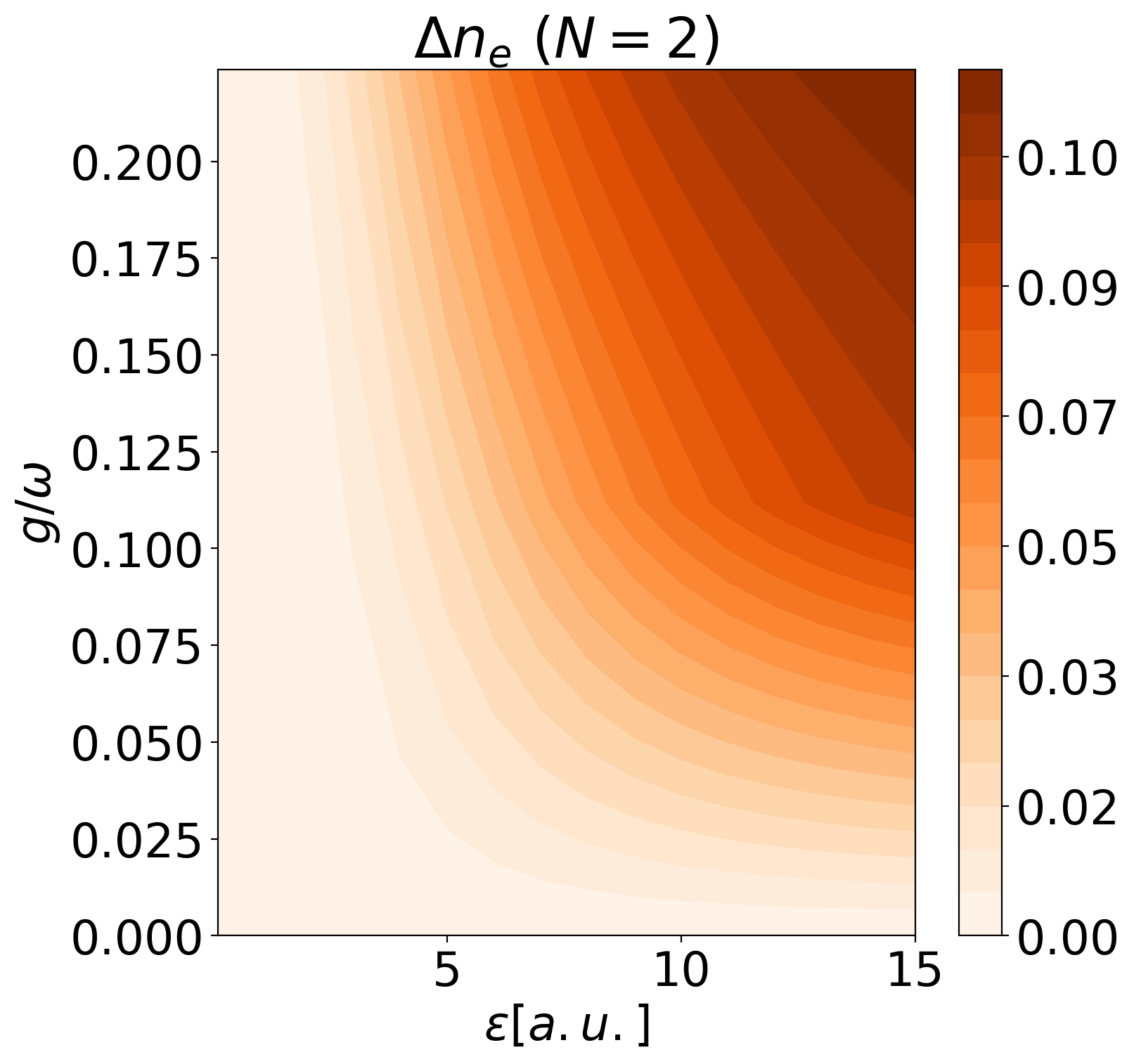}
			\put (21,78) {\textcolor{black}{(c)}}
		\end{overpic}
		\\
		\begin{overpic}[width=0.32\columnwidth]{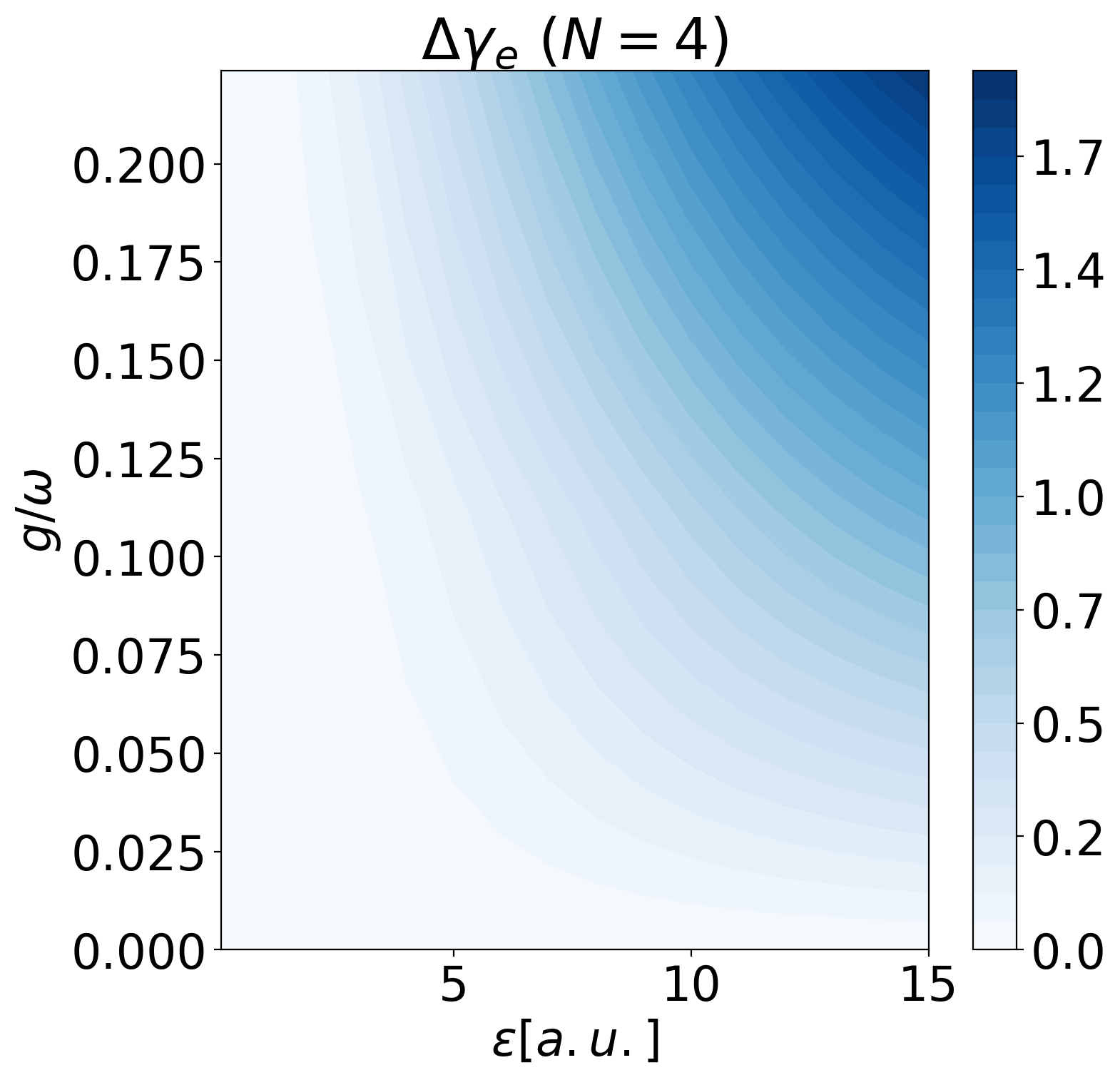}
			\put (21,80) {\textcolor{black}{(d)}}
		\end{overpic}
		&
		\begin{overpic}[width=0.32\columnwidth]{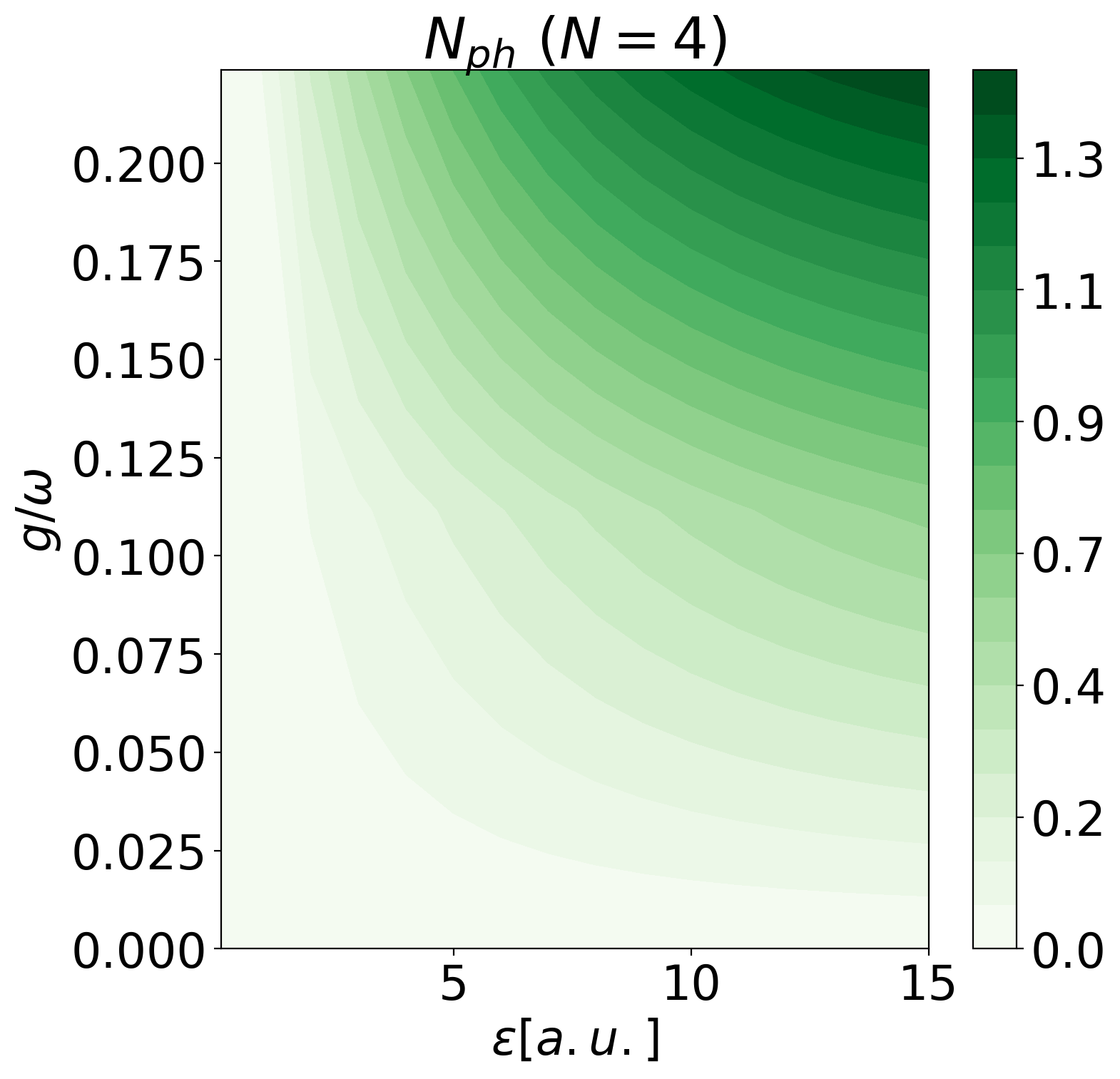}
			\put (21,80) {\textcolor{black}{(e)}}
		\end{overpic}
		&
		\begin{overpic}[width=0.32\columnwidth]{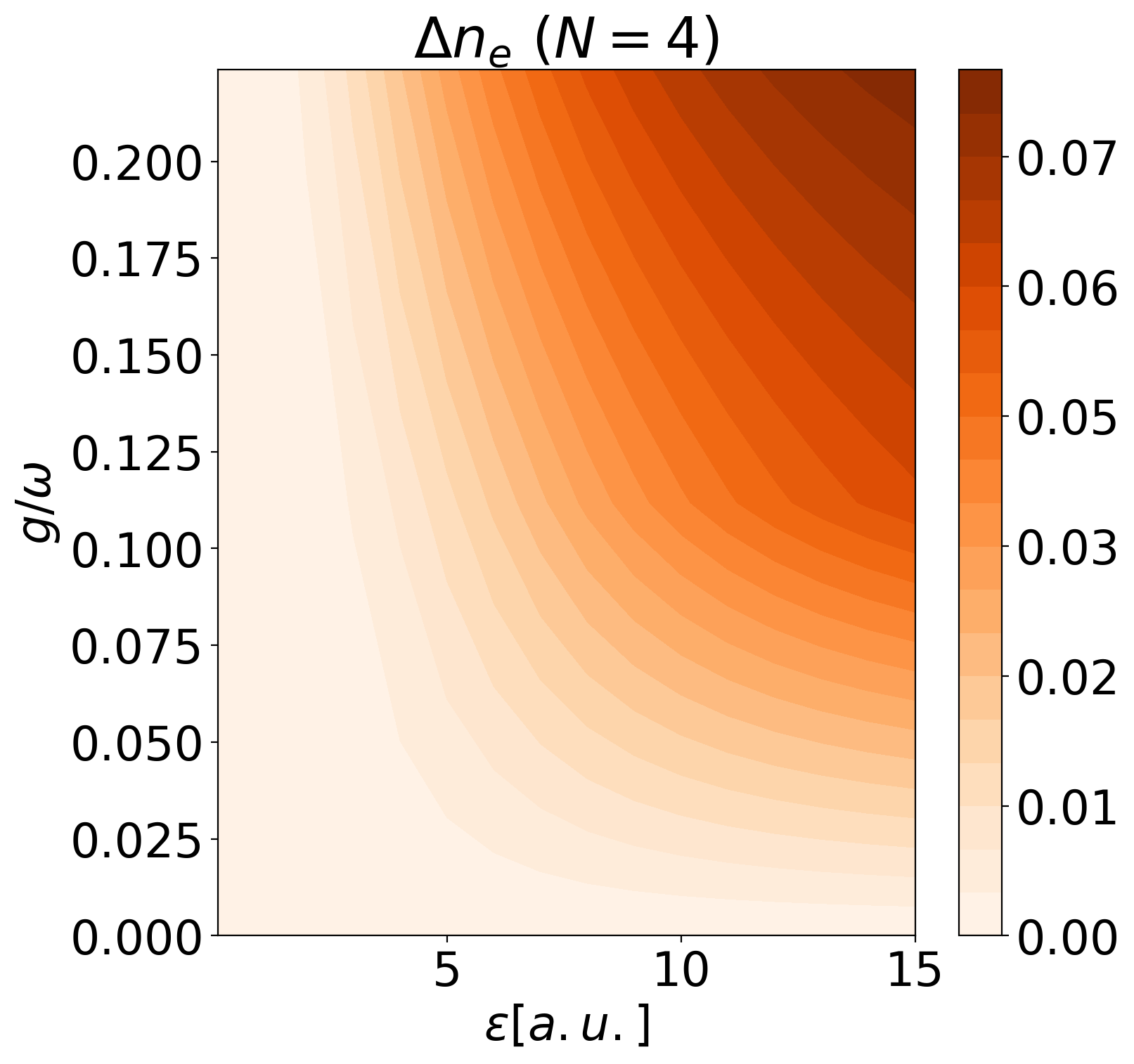}
			\put (21,78) {\textcolor{black}{(f)}}
		\end{overpic}
	\end{tabular}
	\caption{The plots show key quantities of the model system as a function of the coupling strength $g/\omega$ and the localization parameter $\epsilon$ for the 2-electron (upper row) and 4-electron (lower row) case. In the first column, we depict the normalized deviation $\Delta \gamma_e=||\gamma^e_{g/\omega,\epsilon} - \gamma^e_{g/\omega=0.0,\epsilon}||_2/N$ of the electronic 1RDM  to a reference for the same $\epsilon$ and $g/\omega=0$ (blue). In the second column, we show the total photon number $N_{ph}$ (green) and in the third column, the total deviation of the electronic natural occupation numbers $\Delta n_e =\sum_i ||n^e_{i, g/\omega,\epsilon} - n^e_{i, g/\omega=0.0,\epsilon}||_2$ is displayed. This is a measure of the induced electron-photon correlation. We observe in all the cases that when the bare matter wave function becomes more delocalized ($\epsilon$ larger), the modifications due to the matter-photon coupling become stronger.}
	\label{fig:conf_contour}
\end{figure}

\begin{figure}[ht]
	\centering
	\includegraphics[width=0.49\columnwidth]{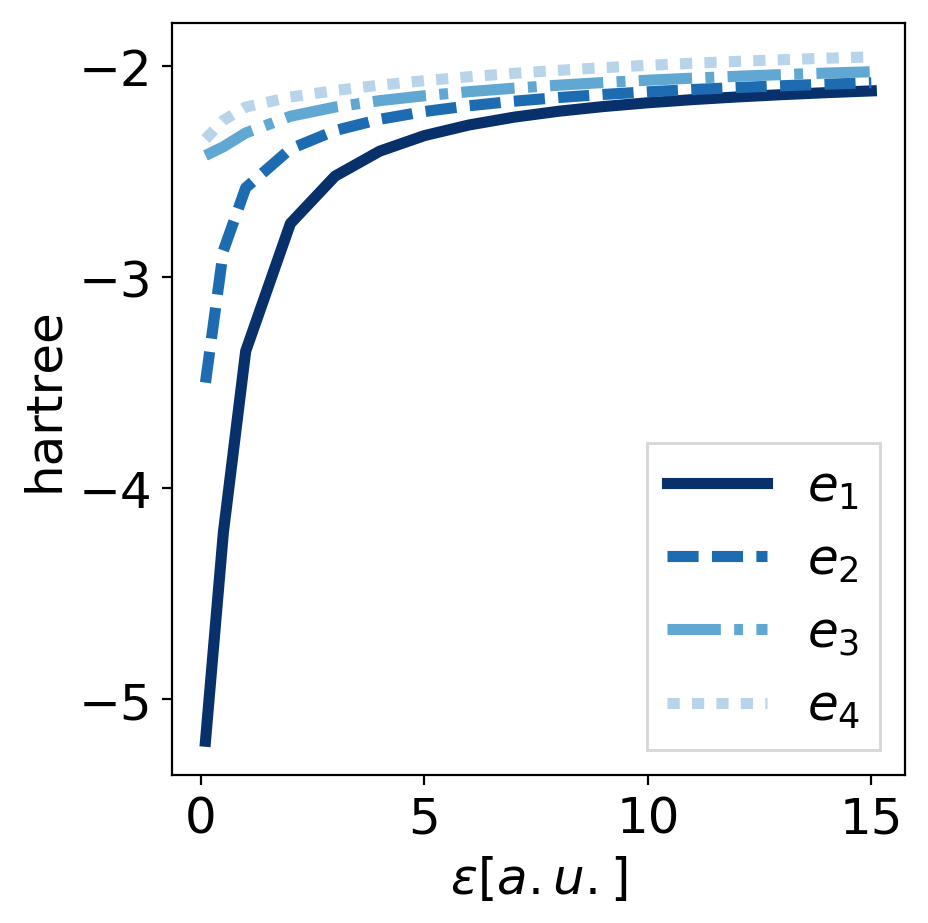}
	
	\caption{The first four eigenenergies $e_1, \dots ,e_4$ of the electronic one-body equation $[-\frac{1}{2}\partial_x^2- v(r)]\psi_i=e_i\psi_i$ for $N=2$ are shown as a function of the confinement parameter $\epsilon$. We see that the energies approach each other with increasing $\epsilon$ and thus decreasing confinement. This means that for a fixed coupling strength, the less confined the system is the more states are available in a small energy range for photon-induced modifications of the ground state.}
	\label{fig:electronic_energy}
\end{figure}

\clearpage
To make these observations more quantitative we display in Fig.~\ref{fig:conf_contour} the normalized changes in the electronic 1RDMs depending on the confinement and the coupling strength, i.e., $\Delta \gamma_e = \lVert \gamma^e_{g/\omega, \epsilon} - \gamma^e_{g/\omega =0, \epsilon}\rVert_2 /N$, in panel (a) and (d) for the 2-particle case and the 4-particle case, respectively. Also, we show the photon number $N_{ph}$ in the ground state in dependence of confinement and coupling strength in (b) and (e) for the 2-particle case and the 4-particle case, respectively. As a third quantity we consider $\Delta n_e =\sum_i ||n^e_{i, g/\omega,\epsilon} - n^e_{i, g/\omega=0.0,\epsilon}||_2$, where $n_i$ are the natural occupation numbers. For the zero-coupling case they are all either zero or one, which corresponds to a single Slater determinant in the electronic subspace. If they are between zero and one they indicate a correlated (multi-determinantal) electronic state. Therefore $\Delta n_e$ measures the photon-induced correlations and also highlights that although polaritonic HF is a single-determinant method in the polaritonic space, for the electronic system it is a correlated (multi-determinantal) method. For both, the 2- and the 4-particle case we find consistently that the more delocalized the uncoupled matter system is, the stronger the coupling modifies the ground state. Although this effect depends on the details of the electronic-structure as we saw in Fig.~\ref{fig:conf_dens_pot}, the plots of Fig.~\ref{fig:conf_contour} indicate that this behavior is quite generic. The reason is that within a small energy range many states with different electronic configurations are available as opposed to a strongly bound (and hence energetically separated) ground-state wave function (see Fig.~\ref{fig:electronic_energy}). A glance on the correlation measure $\Delta n_e$ in panel (c) and (f) of Fig.~\ref{fig:conf_contour} strengthens this explanation: For large $\epsilon$ and $g/\omega$ the electronic correlation is strongest and thus many electronic configurations contribute to the states of this parameter regime. This indicates also that the effective one-body description of the ground state of many cavity-QED models might be inaccurate in this regime.
Additionally, we observe that the maximal values of $\Delta \gamma_e$ in the 2-particle case are slightly larger than in the 4-particle case, which again is related to the different electronic structures of the two systems. 

For the photon numbers, however, which are depicted in panel (b) and (d) of Fig.~\ref{fig:conf_contour}, we see that in general the number of photons is larger in the 4-particle case. This is due to the simple reason that the more charge we have the more photons are created. Nevertheless, the amount of photons does not just double (as expected from a simple linear relation) but is almost three times higher. This highlights the nonlinear regime of electron-photon coupling that we consider here. Again the number of photons increase also with the delocalization and hence the parameter $\epsilon$ is a very decisive quantity.
All these results point toward an interesting parameter in the context of strong light-matter coupling: the localization of the matter wave function. In agreement with a recent case study for simple 2-particle problems~\citep{Schaefer2019}, systems that are less confined react much stronger to a cavity mode.

\clearpage
\section{Summary of the results}
\label{sec:dressed:results:summary}
Let us conclude that although we only have regarded matter systems in one spatial dimension and maximally four electrons, we have made some interesting observations. 

Specifically, we observed in the first sections (Sec.~\ref{sec:dressed:results:1chemical_reaction} and Sec.~\ref{sec:dressed:results:2mores_is_different}) an intricate interplay between Coulomb-induced and photon-induced correlations: in some cases the Coulomb repulsion was suppressed, but in others it was effectively enhanced due to the photon-interaction. Although the changes were seemingly small, one should keep in mind that the correlation contribution to the many-body energy is typically small but nevertheless plays a major role for many important phenomena such as high-temperature superconduction~\citep{Fujita2001}. That this observation transfers to the coupled electron-photon case is highly probable. For instance, \citet{Schaefer2018} have shown some possible implications explicitly for a one-dimensional system coupled to a cavity mode. Whether these modifications of the underlying electronic-structure are indeed a major player in the changes of chemical and physical properties still needs to be seen. However, to capture such modifications in the first place (and study their influence) clearly needs a first-principles theory that is able to treat both types of (strong) correlations accurately and is predictive inside as well as outside of a cavity. We have shown here that polaritonic HF and polaritonic RDMFT are viable options to predict and analyze these intricate structural changes. We want to stress again that polaritonic HF explicitly accounts for electronic correlation and goes thus beyond standard electronic HF.

Besides the interplay between the two types of interaction in coupled light-matter systems, we identified an interesting parameter that especially influences the electron-photon interaction: the localization of the matter wave function. We saw that delocalized wave functions respond considerably stronger to the perturbation by the photon mode than localized ones which is a parameter that is very difficult to be captured by cavity-QED models. This suggests that if we want to observe genuine modifications of the ground state due to strong light-matter coupling, we should consider matter systems that have a spatially extended wave function. One way would be large molecular or solid-state systems, the other way would be an ensemble of emitters. In the latter case the strong influence would lead to local changes, in contrast to the Dicke-like description of collective strong coupling, where we have $N$ independent replicas of the same, perfectly localized system. In both cases, it seems plausible that there are strong modifications of the electronic-structure even if there is only coupling to the vacuum of a cavity. Modifications of chemistry by merely the vacuum seem therefore to be feasible. At the same time an interesting perspective with respect to collective strong-coupling arises: Maybe it is not the simple Dicke-type collectivity~\cite{Kirton2019}, where an excitation is delocalized over many replica, that drives the changes in chemistry, but rather a genuine cavity-photon mediated spatial delocalization of the ensemble wave function. To answer this question, further investigations especially for realistic three-dimensional systems and ensembles including the Coulomb interaction have to be conducted. And polaritonic first-principle methods as introduced in this work seem specifically well-suited to answer these interesting and fundamental questions.

	\part{Numerics}
\label{sec:numerics}

\begin{aquote}{R. Fletcher, 1987 \citep{Fletcher1987}}
	The subject of optimization is a fascinating blend of heuristics and rigour, of theory and experiment. It can be studied as a branch of pure mathematics, yet has applications in almost every branch of science and technology.
\end{aquote}

\newpage
The last part of this thesis highlights the price we pay to make first-principles methods directly applicable to strong matter-photon problems: we need to develop new algorithms as well as to validate the accuracy of well-established algorithms applied to this new setting. Furthermore, this part clarifies how numerical necessity dictates the choice of the particular algorithms that we have presented in part~\ref{sec:dressed}.

Before we start the discussion, we want to remark on the role of numerics as an nowadays indispensable part of research in physics. This is true in all fields, but in particular in first-principles theory, which is impossible to do without the development of efficient algorithms and modern advances in computer technology.
Nowadays, everybody with a minimal knowledge of scripting can use standard implementations to determine, e.g., the equilibrium structure of a molecule. 
Such a service seeks its equals and it is only possible because of the concerted efforts of mathematicians, physicists and software engineers that have painstakingly validated the numerical implementations. Also for our implementations, we have performed exhaustive testing procedures, which are presented in detail in the appendix~\ref{sec:numerics:dressed:convergence}. We are therefore confident that the available codes are as reliable as other standard implementations.

	\chapter{RDMFT in real space}
\label{sec:numerics:rdmft}
Before we come to the coupled problems, we need to explain the matter-only case, which is the basis for polaritonic-structure methods. We therefore discuss the two RDMFT-algorithms of the electronic-structure code \textsc{Octopus}\footnote{Details on the code can be found on the webpage: \url{https://octopus-code.org/}.}. The first one is a standard (and the default) method~\citep{Andrade2015} and the second one was built specifically to investigate the dressed orbitals (but is also useful for matter-only calculations)~\citep[Ch. 14]{Tancogne-Dejean2020}.

Let us briefly explain the difference between the two methods. The RDMFT equations (Sec.~\eqref{sec:est:rdms:rdmft}) consist of two coupled sets of equations for the natural orbitals and the natural occupation numbers that have to be solved self-consistently. Thus, usual algorithms consist of two major parts that are the occupation number optimization and the orbital optimization. 
The default solver for the latter optimization problem cannot use the full flexibility of the real-space grid but requires a two-step procedure. In the first step, one has to perform a preliminary calculation on the real-space grid that generates a basis set for which in the second step Hamiltonian matrix elements are computed. The code performs then the actual optimization by orbital rotations within this basis set (see Ref.~\citep{Andrade2015}). Thus, the flexibility of the underlying real-space grid is only present in the preliminarily generated basis. In principle, this flexibility could be exploited to adopt the basis to specific scenarios and by that, to overcome the limitations of usual quantum-chemistry basis sets (see Sec.~\ref{sec:est:general:example_case}). However, it is very difficult to construct a generic algorithm for this purpose. Currently, there is only one method to generate a basis, that is by a preliminary calculation with a less expensive theory level such as KS-DFT or even independent particles which solves just the non-interacting Schrödinger equation (see Sec.~\ref{sec:est:general:general_case}). Such basis sets are usually considerably less efficient than optimized quantum-chemistry basis sets~\citep[Ch. 14]{Tancogne-Dejean2020} and they have further shortcomings, when we generalize the description to dressed orbitals (see Sec.~\ref{sec:numerics:dressed}). 

For this reason, we have implemented and validated a new orbital optimization method that is a generalization of the conjugate-gradients algorithm by \citet{Payne1992}. This algorithm uses the full flexibility of the grid and needs considerably less natural orbitals for a converged result than the default solver~\citep{Tancogne-Dejean2020}. However, RDMFT calculations with the conjugate gradients algorithm are numerically considerably more expensive than with the orbital-based algorithm. Thus, it is useful to have the choice between both algorithms, which our implementation provides.


\section{The generic RDMFT algorithm}
\label{sec:numerics:rdmft:algorithm}
In this section, we discuss a general strategy to solve an RDMFT minimization problem that is independent of the details of the orbital handling. The algorithm assumes spin-restriction, i.e., the electron-number $N$ is even and all $B>N$ electronic orbitals can be doubly occupied (see the discussion around Eq.~\eqref{eq:est:spin_orbitals_closed_shell}).

\subsection{The RDMFT energy functional}
\label{sec:numerics:rdmft:algorithm:rdmft_functional}
As discussed in Sec.~\ref{sec:est:rdms:rdmft}, all known RDMFT energy functionals (besides the trivial HF functional) are not explicitly given in terms of the 1RDM, but can only be expressed with respect to the natural orbitals $\{\phi_i(\br)\}$ and occupation numbers $\{n_i\}$. The generic RDMFT functional reads in this representation (cf. Eq.~\eqref{eq:est:rdms:rdmft:energy_functional})
\begin{align}
\label{eq:numerics:rdmft:func_general}
\begin{split}
E[\{\phi_i\},\{n_i\}]=& \sum_{i=0}^{B} n_i \int \td\br \, \phi^*_i (\br) h(\br) \phi_i (\br) 
+\frac{1}{2}\sum_{i,j=0}^{B} n_i n_j
\int\td\br\td\br' \left|\phi_i(\br)\right|^2 \left|\phi_j(\br')\right|^2 w(\br,\br')\\
&+E_{xc}[\{\phi_i\},\{n_i\}],
\end{split}
\end{align} 
where $h(\br)=-\frac{1}{2m}\nabla^2 +v(\br)$ is the one-body kernel, $w(\br,\br')$ is the two-body interaction kernel
and $E_{xc}[\{\phi_i\},\{n_i\}]$ is the unknown exchange-correlation part of the functional. In the current state of the \textsc{Octopus} implementation, there are two exchange-correlation functionals available: the HF exchange-only and the M\"{u}ller functional~\citep{Mueller1984} (cf. Eq.~\eqref{eq:est:rdms:Mueller_functional})
\begin{align}
\label{eq:numerics:Mueller}
E_{xc}[\{\phi_i\},\{n_i\}]=E_M[\{\phi_i\},\{n_i\}]\equiv-\frac{1}{2}\sum_{i,j=0}^{B} \sqrt{n_i n_j}\int\td\br\td\br'\phi^*_i(\br)\phi^*_j(\br')\, w(\br,\br') \phi_i(\br')\phi_j(\br).
\end{align}
In the following, we will present all equations  explicitly for the Müller functional. To obtain the equivalent expressions in terms of the HF functional, one simply has to replace $\sqrt{n_in_j}\rightarrow n_in_j$ in Eq.~\eqref{eq:numerics:Mueller}.\footnote{Note that in practice this leads to integer occupation numbers~\citep{Lieb1981} and therefore, the code simply fixes $n_1=...=n_N=1$ and $n_{N+1}=n_{N+2}=...=0$ for all other occupations.} 
We also will assume spin-restriction, which is the standard case in RDMFT.\footnote{Note that treating systems that are not spin-saturated is very challenging for many electronic-structure methods, but in particular for RDMFT.}

\subsection{The constraints of the RDMFT minimization and how to take them into account in practice}
The goal of the RDMFT algorithm is to minimize the energy with respect to the natural orbitals and natural occupation numbers, respecting the following three additional constraints:
\begin{subequations}
\label{eq:numerics:rdmft:constraints}
\begin{align}
	&0\leq n_i\leq 2\quad \forall i\\
	&\mathcal{S}[\{n_i\}]=\sum_i^B n_i - N=0\\
	&\mathcal{C}[\{\phi_i\},\{\phi_j\}]=\int\td^dr\phi_i^*(\br)\phi_j(\br) -\delta_{ij}=0
\end{align}
\end{subequations}
The first two lines are the (ensemble) $N$-representability constraints of the (spin-summed) 1RDM, and the last line guarantees the orthonormality of the natural orbitals. Since the first condition is simple in the natural orbital basis, it is enforced with the help of an auxiliary construction: We set 
\begin{align}
\label{eq:numerics:n_theta}
n_i=2\sin^2(2\pi\theta_i) \quad\forall\,i,
\end{align}
and minimize over the set of $\theta_i$ instead of the $n_i$ directly. The second and the third conditions are imposed by the \emph{Lagrange multiplier} technique. For that, we introduce the Lagrangian
\begin{align}
\label{eq:numerics:rdmft:lagrangian_general}
L[\{\phi_i\},\{\theta_i\};\{\epsilon_{ij}\},\mu]&=E[\{\phi_i\},\{n_i\}]-\mu \mathcal{S}[\{\theta_i\}]-\sum_{i,j}\epsilon_{ij} \mathcal{C}[\{\phi_i\},\{\phi_j\}],
\end{align}
with the Langrange multipliers $\mu$ and $\epsilon_{ij}$. The extrema of $E$ under the constraints $\mathcal{S}$ and $\mathcal{C}$ are then found at stationary points of L,\footnote{Note that since the Müller functional is convex~\citep{Frank2007}, the stationarity condition is sufficient to determine the global \emph{minimum} of $E$.} i.e., for
\begin{align}
\label{eq:numerics:rdmft_stationarity}
\delta L=0.
\end{align}
The condition~\eqref{eq:numerics:rdmft_stationarity} leads to three sets of independent equations (cf. Eq.~\eqref{eq:est:rdms:rdmft:equations}), i.e..
\begin{subequations}
\begin{align}
	0=&\frac{\partial E}{\partial n_i}-\mu \label{eq:numerics:ELE_n}\\
	0=&\frac{\delta E}{\delta \phi_i^*(\br)}-\sum_{k=1}^{B}\epsilon_{ki}\phi_k(\br) \label{eq:numerics:ELE_phi_star}\\
	0=&\frac{\delta E}{\delta \phi_i(\br)}-\sum_{k=1}^{B}\epsilon_{ik}\phi_k^*(\br) \label{eq:numerics:ELE_phi}
\end{align}
\end{subequations}
for all $i$. The first set of equations, cf.~\eqref{eq:numerics:ELE_n} defines the gradient with respect to $n_i$
\begin{align}
\label{eq:numerics:dE_dn}
\frac{\partial E}{\partial n_i}=&\int\td^3r \phi^*_i (\br) h(\br) \phi_i (\br) \nonumber\\
&+\sum_{j=1}^{B} n_j \int\td^3r\td^3r'\, \phi^*_i (\br) \phi^*_j (\br') w(\br,\br') \phi_j (\br') \phi_i (\br) \nonumber\\
&-\sum_{j=1}^{B} \sqrt{\tfrac{n_j}{n_i}} \int\td^3r\td^3r'\, \phi^*_i (\br) \phi^*_j (\br') w(\br,\br') \phi_j (\br) \phi_i (\br') \nonumber	\\
\equiv& \int\td^3r \phi^*_i (\br) h(\br) \phi_i (\br)+\int\td^3r\, \rho_i (\br) v_{H}(\br) - \frac{1}{\sqrt{n_i}} \int\td^3r\, \phi_i^* (\br) v^i_{XC}(\br) ,
\end{align}
where we defined the orbital density
\begin{align}
\rho_i(\br)&= \phi^*_i (\br) \phi_i (\br)= |\phi_i|^2 (\br),
\end{align}
the Hartree-potential
\begin{align}
\label{eq:numerics:v_hartree}
v_{H}(\br)&=\sum_{j=1}^{B} n_j \td^3r'\, \rho_j(\br')  w(\br,\br'),
\end{align}
and the exchange-correlation potential of orbital i
\begin{align}
\label{eq:numerics:v_xc}
	v^i_{XC}(\br)&=\sum_{j=1}^{B} \sqrt{n_j} \int\td^3r'\, \phi_j^*(\br')   w(\br,\br') \phi_j(\br) \phi_i(\br').
\end{align}
The remaining two sets of equations, \eqref{eq:numerics:ELE_phi_star} and \eqref{eq:numerics:ELE_phi}, respectively define the functional derivatives with respect to the orbitals. These ``orbital gradients'' read using the above definitions
\begin{subequations}
\label{eq:numerics:gradients_RDMFT}
\begin{align}
	\frac{\delta E}{\delta \phi_i^*(\br)}=\,&n_i\, h(\br) \phi_i (\br) + n_i\, v_H(\br) \phi_i(\br) - \sqrt{n_i}\, v^i_{XC}(\br) \label{eq:numerics:gradients_RDMFT_phi_star}\\
	\frac{\delta E}{\delta \phi_i(\br)}=\,& n_i\, \phi_i^* (\br) h(\br)  + n_i\, \phi_i^*(\br) v_H(\br)  - \sqrt{n_i}\, (v^i_{XC})^*(\br).
\end{align}
\end{subequations}

\subsection{The basic scheme of the RDMFT algorithm}
\label{sec:numerics:rdmft:algorithm:general}
We have now derived and discussed all the necessary ingredients to define a generic RDMFT minimization algorithm that solves Eq.~\eqref{eq:numerics:ELE_n} and Eq.~\eqref{eq:numerics:ELE_phi_star} (or alternatively Eq.~\eqref{eq:numerics:ELE_phi}) self-consistently. The structure of this minimization problem suggests the separation into the
\begin{enumerate}[label=\alph*]
	\item (natural) \textbf{occupation number optimization}, \\
	where we solve Eq.~\eqref{eq:numerics:ELE_n} for fixed $\{\phi_i\},\{\epsilon_{ij}\}$ and the
	\item (natural) \textbf{orbital optimization}, \\
	where we solve Eq.~\eqref{eq:numerics:ELE_phi_star} for fixed $\{n_i\},\mu$.
\end{enumerate}
Both optimizations have to be performed alternately within an outer loop until self-consistence.

For our convenience in the following, we rewrite the energy functional with the Müller approximation \eqref{eq:numerics:Mueller} as
\begin{align}
\label{eq:numerics:rdmft:Mueller_energy_func}
E[\{\phi_i\},\{n_i\}]=& \sum_{i=0}^{B} n_i \braket{\phi_i | h\phi_i}
+\frac{1}{2}\sum_{i,j=0}^{B} n_i n_j
\braket{\phi_i\phi_i|w|\phi_j\phi_j} -\frac{1}{2}\sum_{i,j=0}^{B} \sqrt{n_i n_j} \braket{\phi_i\phi_j|w|\phi_j\phi_i},
\end{align} 
where we introduced the abbreviations for one-body integrals $\braket{\phi_i | h\phi_i}=\int\td^3r \phi^*_i (\br) h(\br) \phi_i (\br)$ and two-body integrals $\braket{\phi_i\phi_j|w|\phi_k\phi_l}=\int\td\br\td\br' w(\br,\br')  \phi_i^*(\br')\phi_j(\br') \phi_k^*(\br) \phi_l(\br)$. Additionally we subsume  $\bm{\phi}=\{\phi_1,...,\phi_B\}$ and $\bn^0=(n_1,...,n_B)$. We now summarize the algorithm in \emph{pseudocode} form~\citep{Zobel2014}.
\begin{center}
	\begin{fminipage}{0.75\textwidth}
		\begin{algorithm}\label{algorithm:RDMFT} (RDMFT)  \newline
			Set $B$.\newline
			Initialize $\bm{\phi}^0$ (depends on the orbital algorithm).\\
			Initialize $\bn^0,\mu^0$. (see Sec.~\ref{sec:numerics:rdmft:algorithm_non}) \newline
			Calculate the initial matrix elements $h_{ii}^0= \braket{\phi_i^0|h\phi_i^0}, w_{ijkl}^0=\braket{\phi_i^0\phi_j^0|w|\phi_k^0\phi_l^0}$.\\
			\\
			\textbf{for} l=1,2,...
			\begin{enumerate}[label=(\alph*)]
				\item \textbf{Occupation number optimization} (algorithm~\ref{algorithm:RDMFT_non})\\
				Solve Eq.~\eqref{eq:numerics:ELE_n} for fixed $\{\phi_i\},\{\epsilon_{ij}\}$ self-consistently with $h_{ii}^{l-1},w_{ijkl}^{l-1}$ for $\bn^l,\mu^l$
				\item \textbf{Orbital optimization} (algorithm~\ref{algorithm:RDMFT_no} or \ref{algorithm:RDMFT_no_cg})\\
				Solve Eq.~\eqref{eq:numerics:ELE_phi_star} for fixed $\{n_i\},\mu$ self-consistently for $\bm{\phi}^l$\\
				Calculate $h_{ii}^{l},w_{ijkl}^{l}$
				\item[] \textbf{break} if convergence criterion is fulfilled (depends on the orbital algorithm)
			\end{enumerate}
			\textbf{end (for)}	
		\end{algorithm}
	\end{fminipage}
\end{center}

\subsection{The occupation number optimization}
\label{sec:numerics:rdmft:algorithm_non}
We now explain part (a) of algorithm~\ref{algorithm:RDMFT}, i.e., the occupation number optimization. This is independent of the type of orbital optimization routine that we employ for (b) and thus part of the general algorithm. 

We start by combining Eq.~\eqref{eq:numerics:ELE_n} and Eq.~\eqref{eq:numerics:dE_dn} in one set of equations that we aim to solve, i.e.,
\begin{align}
\mu=&\frac{\partial E}{\partial n_i}=\int\td^3r \phi^*_i (\br) h(\br) \phi_i (\br)+\int\td^3r\, \rho_i (\br) v_{H}(\br) - \frac{1}{\sqrt{n_i}} \int\td^3r\, \phi_i^* (\br) v^i_{XC}(\br) \nonumber\\
=& \braket{\phi_i|h\phi_i} + \sum_{j}^B n_j \braket{\phi_i\phi_i|w|\phi_j\phi_j} - \sum_{j}^B \sqrt{n_j/n_i} \braket{\phi_i\phi_j|w|\phi_j\phi_i}. \label{eq:numerics:ELE_n_complete}
\end{align}
Instead of directly solving Eq.~\eqref{eq:numerics:ELE_n_complete} for the set $\bn,\mu$, the algorithm performs a two-step procedure. For that, we define the auxiliary function
\begin{align}
\label{eq:numerics:non_opt_auxiliary_function}
\mathfrak{S}(\mu)=&\min\limits_{\bn}L(\bn;\mu)
\end{align}
with the ``occupation Lagrangian''
\begin{align}
\label{eq:numerics:rdmft:non_optimization:lagrangian}
L(\bn;\mu) =E[\bn]- \mu\left(\sum_{i=1}^{B}n_i-N\right)
\end{align}
where $E[\bn]$ is the energy functional \eqref{eq:numerics:rdmft:Mueller_energy_func} with fixed orbitals. The advantage of this construction is that we can solve Eq.~\eqref{eq:numerics:non_opt_auxiliary_function} with a standard (unconstrained) minimization algorithm. Since we find a set of optimal $\bn^*$ for every given $\mu$, we can label $\bn^*=\bn^*(\mu)$. To find the $\mu=\mu^*$ that extremalizes $\tilde{S}$, we use then the side condition
\begin{align}
\mathcal{S}(\mu)=\left(\sum_{i=1}^{B}n_i^*(\mu)-N\right).
\end{align}
Specifically, we solve
\begin{align}
\mathcal{S}(\mu)=0
\end{align}
with a bisection algorithm. Altogether, we employ the following algorithm to optimize the occupation numbers (part (a) in algorithm \ref{algorithm:RDMFT})
\begin{center}
	\begin{fminipage}{0.7\textwidth}
		\begin{algorithm}\label{algorithm:RDMFT_non} (Occupation number optimization)  \newline
			Set the convergence criterion $\epsilon_{\mu}>0$.\\
			Find an initial interval $[\mu_1^0,\mu_2^0]$ that satisfies $S(\mu_1^0)<0$ and $S(\mu_20)>0$.\\
			\\
			\textbf{for} k=1,2,...
			\begin{enumerate}[label=\roman*]
				\item Calculate the interval center $\mu_m^{k-1}=\tfrac{\mu_1^{k-1}+\mu_2^{k-1}}{2}$,\\
				calculate $\tilde{\mathfrak{S}}(\mu_m^{k-1})$ and \\
				check the side condition $\mathcal{S}(\mu_m^{k-1})$.
				\item Set the interval for the next iteration by the prescription \\
				$S(\mu_m^{k-1})S(\mu_1^{k-1})<0 \rightarrow \mu_1^{k}=\mu_1^{k-1}, \mu_2^{k}\rightarrow\mu_m^{k-1}$\\
				$S(\mu_m^{k-1})S(\mu_1^{k-1})\geq 0 \rightarrow \mu_1^{k}\rightarrow\mu_m^{k-1}, \mu_2^{k}=\mu_2^{k-1}$
				\item[] \textbf{break} if  $\min (S(\mu),|\mu_1-\mu_2|/2)<\epsilon_{\mu}$.
			\end{enumerate}
			\textbf{end (for)}	
		\end{algorithm}
	\end{fminipage}
\end{center}
We want to remark that algorithm \ref{algorithm:RDMFT_non} is still not ready to be implemented as it is written here. For instance, we need yet another small algorithm to determine a good guess for the initial interval and take care that too small occupation numbers are discarded in the optimization since they occur in the denominator in Eq.~\eqref{eq:numerics:ELE_n_complete}. We refer the reader for these minor details to the \textsc{Octopus}-code directly.

We see here impressively how solving even a comparatively simple equation as \eqref{eq:numerics:ELE_n_complete} may require an involved algorithm. The reason for this is its \emph{nonlinearity}. There are basically no ``safe'' algorithms for nonlinear equations, for which, e.g., convergence con be guaranteed and debugging nonlinear solvers is thus always a very delicate matter. Since we had to debug the occupation number optimization during the implementation of the dressed orbitals in \textsc{Octopus}, we can provide a concrete example for the latter statement which also serves well to visualize the algorithm. We present this in Sec.~\ref{sec:numerics:dressed:non_assessment}.

\section{Orbital-based RDMFT in real space}
\label{sec:numerics:rdmft:piris}
In this section, we present the default RDMFT algorithm of \textsc{Octopus} that is publicly available from version 10.0 onwards~\citep{Andrade2015,Tancogne-Dejean2020}. The algorithm has been introduced by \citep{Piris2009} for orbital-based codes and is adopted to fit also the real-space setting. 


\subsection{The challenge of the RDMFT orbital minimization}
\label{sec:numerics:rdmft:challenge_orbital_minimization}
Before we lay out the algorithm (Sec.~\ref{sec:numerics:rdmft:algorithm_no_piris}, we start this section with a brief discussion about the special challenges of the orbital minimization in RDMFT.

\subsubsection*{The difference of the RDMFT in comparison to single-reference methods: the single-particle Hamiltonian is not hermitian}
In single-reference methods, the orbital gradients of Eq.~\eqref{eq:numerics:gradients_RDMFT} would define effective one-particle Hamiltonians, which are hermitian. We have discussed in Sec.~\ref{sec:est:rdms:rdmft} that this is not the case in RDMFT, which has strong implications for the numerical minimization. 
To see this, let us briefly recapitulate the HF gradient equation that defines the Fock operator (cf. Eq.~\eqref{eq:est:hf:Fock_operator}). In the here employed notation, the HF gradient reads
\begin{align*}
\frac{\delta E_{HF}}{\delta \phi_i^*(\br)}=&\hat{H}^1_{HF}\phi_i(\br)\\
=& \,n_i\, h(\br) \phi_i (\br) + n_i\, v_H(\br) \phi_i(\br) - n_i v^i_{X}(\br),
\end{align*}
where $v^i_{X}(\br)=\sum_{j=1}^{B} n_j \int\td\br'\, \phi_j^*(\br')   w(\br,\br') \phi_j(\br) \phi_i(\br')$ is the ``exchange-only'' version of Eq.~\eqref{eq:numerics:v_xc}.\footnote{Note that in standard text books, e.g., Ref.~\citep{Helgaker2000,Gross1991} the occupations are usually neglected, since they are fixed. We explicitly include the $n_i$ here to highlight the difference to RDMFT.} The stationarity condition leads then to the $N$ coupled equations (cf. Eq.~\eqref{eq:est:hf:HF_equations})
\begin{align}
\label{eq:numerics:HF_equations}
\hat{H}^1_{HF}\phi_i(\br)=\sum_{k=1}^{N}\epsilon_{ki}\phi_k(\br) \quad i=1,...,N,
\end{align}
which we can transform to a nonlinear eigenvalue equation by a division by $n_i$ and a projection on the orbital $\phi_l^*$, i.e.,
\begin{align*}
\epsilon_{li}/n_i=& \int\td\br \phi_l^*\hat{H}^1_{HF}\phi_i(\br)\\
=&\int\td\br \phi_l^* \, h(\br) \phi_i (\br) + \, \int\td\br \phi_l^*v_H(\br) \phi_i(\br) - \int\td\br \phi_l^*v^i_{X}(\br) \nonumber\\
\equiv &(\hat{H}_{HF}^1)_{il}
\end{align*}
Crucially, the matrix-form of the single-particle Hamiltonian on the right-hand side is hermitian
\begin{align}
	(\hat{H}_{HF}^1)_{il}=(\hat{H}^1)_{li}^*.
\end{align}
This is obvious for the one-body and the Hartree part. However, also for the exchange term $(\hat{H}_{HF,X})_{il}=\int\td\br \phi_k^*v^i_{X}(\br)$ we find
\begin{align}
	(\hat{H}_{HF,X})_{li}^*=& \left(\int\td\br \phi_i^*(\br)v^l_{X}(\br)\right)^*\nonumber\\
	=&  \left(\sum_{j=1}^{B} n_j \int\td\br\td\br' \phi_i^*(\br) \phi_j^*(\br')   w(\br,\br') \phi_j(\br) \phi_l(\br')\right)^* \nonumber\\
	=& \sum_{j=1}^{B} n_j \int\td\br\td\br' \phi_l^*(\br') \phi_j^*(\br)   w(\br,\br') \phi_j(\br') \phi_i(\br)\nonumber\\
	=& \int\td\br' \phi_l^*(\br') \sum_{j=1}^{B} n_j \int\td\br \phi_j^*(\br)   w(\br,\br') \phi_j(\br') \phi_i(\br) \nonumber\\
	=& \int\td\br' \phi_l^*(\br') v^i_{X}(\br')\nonumber\\
	=& (\hat{H}_{HF,X})_{ik}.
\end{align}
Solving the HF equations, cf. Eq. \eqref{eq:numerics:HF_equations}, is thus equivalent to diagonalizing the Fock-matrix $(\hat{H}_{HF}^1)_{li}$ self-consistently: we start with a guess for the orbitals $\{\phi_i^0\}$, calculate $(\hat{H}_{HF}^1)_{li}[\{\phi_i^0\}]$ and diagonalize it to obtain new orbitals $\{\phi_i^1\}$ for the next iteration. This procedure is repeated until some convergence criterion is  satisfied (see also Sec.~\ref{sec:est:hf}). In a similar way, we can solve the equations of other single-reference methods, e.g., the KS equations in KS-DFT. 

The crucial step in the above derivation to turn the HF equations into a nonlinear eigenvalue problem was the division by the occupation numbers $n_i$, which appear linear in the gradient equation. For an RDMFT description beyond HF, this is not possible, because the exchange-correlation functional \emph{must} depend nonlinearly on the occupation numbers~\citep{Lieb1981}. We can see this explicitly in  Eq.~\eqref{eq:numerics:gradients_RDMFT} for the Müller functional. We therefore cannot reformulate the RDMFT equations as an eigenvalue problem and consequently, we have to employ also different algorithms for the RDMFT orbital minimization. 

\subsubsection*{An effective eigenvalue equation for the RDMFT minimization}
Nevertheless, we can define the one-body Hamiltonian from the gradient equation
\begin{align}
\label{eq:numerics:RDMFT_HamiltonianSingleParticle}
\hat{H}^1\phi_i(\br)\equiv&\frac{\delta E}{\delta \phi_i^*(\br)} \nonumber\\
=\,& n_i\, h(\br) \phi_i (\br) + n_i\, v_H(\br) \phi_i(\br) - \sqrt{n_i}\, v^i_{XC}(\br),
\end{align}
but it is \emph{not a hermitian operator} (see also Sec.~\ref{sec:est:rdms:rdmft}).\footnote{One might think now that this is a technicality and that there should be a way to ``cure'' this unusual behavior by, e.g., a more general definition of the gradient. Indeed, \citet{Pernal2005} investigated this and connected issues quite thoroughly considering a generic RDMFT functional. She showed there that the problem is directly connected to a fundamental property that most common RDMFT functionals have: they depend only \emph{implicitly} on the 1RDM, i.e., we can only express their functional form explicitly in terms of natural orbitals and natural occupation numbers. One way to explain this is that the eigenset of the 1RDM is unique besides possible rotations in the space of orbitals that have an identical occupation number. Thus, we cannot apply a unitary transformation to find the eigenbasis of the one-body Hamiltonian. In a single-reference theory instead, the 1RDM is \emph{idempotent}, i.e., all eigenvalues are one (or two in the spin-restricted case) or zero, respectively and hence such a transformation can be applied within the occupied space, which is the only relevant for the one-body Hamiltonian.}
However, it is possible to manipulate the orbital equations, cf. Eq.~\eqref{eq:numerics:gradients_RDMFT} such that we can solve an equivalent auxiliary eigenvalue problem~\citep{Pernal2005,Piris2009}. The RDMFT routine of \textsc{Octopus} employs such an auxiliary construction, which specifically was introduced by \citet{Piris2009} and which we want to briefly present in the following.

We start by noting that in order to solve Eq.~\eqref{eq:numerics:gradients_RDMFT}, we have to consider the full Langrange-multiplier matrix $\epsilon_{ij}$, since we cannot diagonalize $\hat{H}^1$. Although $(\hat{H}^1)_{ij}$ is not hermitian, the matrix $\epsilon_{ij}$ that fulfils the stationarity condition is hermitian. To show this, we first note that the two energy gradients with the same index
\begin{align}
	\left(\frac{\delta E}{\delta \phi_i(\br)}\right)^*=\frac{\delta E}{\delta \phi_i^*(\br)}
\end{align}
are connected by complex conjugation.\footnote{Note that at the stationary point also $(\psi^*)^*=\phi$ must hold, although we treat $\phi$ and $\phi^*$ in principle as independent.} Comparing now the two gradient equations \eqref{eq:numerics:ELE_phi_star} and \eqref{eq:numerics:ELE_phi}, we have for all $i$
\begin{align*} 
	\sum_{k=1}^{B}\epsilon_{ik}\phi_k^*(\br)
\overset{\eqref{eq:numerics:ELE_phi}}{=}&
	\frac{\delta E}{\delta \phi_i(\br)}\\
\overset{\hphantom{\eqref{eq:numerics:ELE_phi}}}{=}&
	\left(\frac{\delta E}{\delta \phi_i^*(\br)}\right)^*\\
\overset{\eqref{eq:numerics:ELE_phi_star}}{=}&
	\sum_{k=1}^{B}\epsilon_{ki}^*\phi_k^*(\br).
\end{align*}
Since the $\phi_k$ are all orthogonal, we have proven that
\begin{align}
	\epsilon_{ik}=\epsilon_{ki}^*
\end{align}
is symmetric at the stationary point, where $\delta L=0$. 
This suggests to define the matrix
\begin{align*}
\tilde{F}_{ki}=\epsilon_{ik} - \epsilon_{ki}^*,
\end{align*}
that must be zero at the stationary point. We therefore have
\begin{empheq}[box=\fbox]{align}
\label{eq:numerics:hermiticity_LagrangeMultiplier}
	\tilde{F}_{ki}= 0 \quad \Longleftrightarrow\quad \delta L =0
\end{empheq}
If we express $\epsilon_{ki}$ by projecting one of the gradient equations, say \eqref{eq:numerics:ELE_phi_star} on orbital $\phi_k ^*$,
\begin{align}
\label{eq:numerics:Lambda}
\epsilon_{ki}=&\int\td^3r\, \phi_k^*(\br) \frac{\delta E}{\delta \phi_i^*(\br)} \nonumber\\
=& n_i\, \int\td^3r\, \phi_k^*(\br)  h(\br) \phi_i (\br) + n_i\, \int\td^3r\, \phi_k^*(\br) v_H(\br) \phi_i(\br) -  \sqrt{n_i}\, \int\td^3r\, \phi_k^*(\br), v^i_{XC}(\br) 
\end{align}
then Eq.~\eqref{eq:numerics:hermiticity_LagrangeMultiplier} is equivalent to the orbital-gradient conditions, cf. Eqs.~\eqref{eq:numerics:ELE_phi_star}-\eqref{eq:numerics:ELE_phi}.

The method of \citeauthor{Piris2009} exploits this by considering an auxiliary matrix
\begin{align}
\label{eq:numerics:Piris_F_matrix}
	F_{ki}=\begin{cases}
		\epsilon_{ik} - \epsilon_{ki}^* \quad &i\neq k\\
		F_i \quad &i=k
	\end{cases}
\end{align}
with the same entries as $\tilde{F}_{ki}$ on the off-diagonals but additionally with some in principle arbitrary diagonal terms. The stationarity condition $\tilde{F}_{ki}=0$ is thus equivalent to $F_{ki}=F_i\delta_{ki}$ being in diagonal form and we can diagonalize $F$ iteratively to minimize the RDMFT functional.

\subsubsection*{A remark on hermiticity, the one-body Hamiltonian and the Lagrange-multiplier matrix}
Before we lay out the concrete algorithm for this minimization, we want to address the difference between the hermiticity of $(\epsilon_{ik})$ and $\hat{H}^1$, which is subtle but crucial. When we define $(\epsilon_{ik})$ by the gradient equation
\begin{align*}
	\epsilon_{ki}=&\int\td^3r\, \phi_k^*(\br) \frac{\delta E}{\delta \phi_i^*(\br)}\\
	=& \int\td^3r\, \phi_k^*(\br)\hat{H}^1\phi_i(\br),
\end{align*}
it seems as if $\epsilon_{ki}\overset{?}{=}(\hat{H}^1)_{ki}$ would just be the matrix representation of $\hat{H}^1$ and thus from $\epsilon_{ki}=\epsilon_{ik}^*$ we can follow that $\hat{H}^1$ is indeed hermitian, contrary to what we claimed. 
The resolution of this seeming contradiction is that $(\hat{H}^1)_{ki}=(\hat{H}^1)_{ik}^*$ holds only for exactly one point of the definition space of $\hat{H}^1$, which is the stationary point of $L$. Only if the indices $i,k$ correspond exactly to the natural orbitals that extremalize $L$, the matrix $(\hat{H}^1)_{ki}$ is symmetric. For all other points, this is not the case. To show this, let us calculate $\epsilon_{ki}$ again, but this time from the other gradient equation, cf. Eq.~\eqref{eq:numerics:ELE_phi}. We have
\begin{align}
\label{eq:numerics:Lambda_alternative}
	\tilde{\epsilon}_{ik}=&\int\td^3r\, \frac{\delta E}{\delta \phi_i(\br)}  \phi_k(\br)\\
	=& n_i\, \int\td^3r\, \phi_i^*(\br)  h(\br) \phi_k (\br) + n_i\, \int\td^3r\, \phi_i^*(\br) v_H(\br) \phi_k(\br) -  \sqrt{n_i}\, \int\td^3r\, v^i_{XC}(\br) \phi_k(\br)
\end{align}
and thus
\begin{align*}
	\tilde{\epsilon}_{ki}=& n_k\, \int\td^3r\, \phi_k^*(\br)  h(\br) \phi_i (\br) + n_k\, \int\td^3r\, \phi_k^*(\br) v_H(\br) \phi_i(\br) -  \sqrt{n_k}\, \int\td^3r\, v^k_{XC}(\br) \phi_i(\br)\\
	\neq& n_i\, \int\td^3r\, \phi_k^*(\br)  h(\br) \phi_i (\br) + n_i\, \int\td^3r\, \phi_k^*(\br) v_H(\br) \phi_i(\br) -  \sqrt{n_i}\, \int\td^3r\, \phi_k^*(\br) v^i_{XC}(\br)\\
	=&\epsilon_{ki}.
\end{align*}
Obviously, both definitions cannot agree in general. In fact, we could rephrase the task of the RDMFT orbital optimization as  ``to search for the one set of $n_i$ and $\phi_i$ (assuming no degeneracy) for that $\tilde{\epsilon}_{ki}=\epsilon_{ki}$.''

\subsection{The Piris and Ugalde orbital-optimization in \textsc{Octopus}}
\label{sec:numerics:rdmft:algorithm_no_piris}
We have now derived and discussed all the necessary ingredients to complete algorithm \ref{algorithm:RDMFT} and introduce the orbital optimization. Therefore, we employ the algorithm due to \citet{Piris2009}.

The goal of the routine is to diagonalize the matrix
\begin{align}
\tag{cf. \ref{eq:numerics:Piris_F_matrix}}
F_{ki}=\begin{cases}
\epsilon_{ik} - \epsilon_{ki}^* \quad &i\neq k\\
F_i \quad &i=k
\end{cases},
\end{align}
for which we still need to specify the $F_i$, which in principle may have arbitrary values. However, a good choice will crucially influence the stability of the algorithm. In their original paper, \citet{Piris2009} describe this issue and provide a prescription for initializing and updating them. A good choice are the eigenvalues of $(\epsilon_{ik})$ in ascending order and we refer for more details to the publication. The complete algorithm is summarized in pseudocode as
\begin{center}
	\begin{fminipage}{0.7\textwidth}
		\begin{algorithm}\label{algorithm:RDMFT_no} (Orbital optimization 1)  \newline
			Set the convergence criteria $\epsilon_{F}>0, \epsilon_{E}>0$.\\
			Construct $\epsilon_{ij}^0$ by Eq.~\eqref{eq:numerics:Lambda} from the initial orbitals $\bm{\phi}^0$.\\
			Diagonalize $(\epsilon_{ij}^0)$ to obtain the eigenvalues $F_i$\\
			\\
			\textbf{for} k=1,2,...
			\begin{enumerate}[label=\roman*]
				\item Construct $\epsilon_{ij}^{k}$ by Eq.~\eqref{eq:numerics:Lambda} from $\bm{\phi}^{k-1}$.
				\item Set the off-diagonals $F_{ij}^{k}=\epsilon_{ij}^{k}-\epsilon_{ji}^{k}{}^*$ and\\
				determine $F_{max}^{k}=\max F_{ij}^{k}$
				\item Diagonalize $F^{k}$ to obtain a new set of orbitals $\bm{\phi}^k$
				\item Calculate the total energy $E^{k+1}$ by Eq.~\eqref{eq:numerics:rdmft:Mueller_energy_func}
				\item[] \textbf{break} if  $F_{max}^{k}<\epsilon_{F}$ and $k>1$ $|E^{k+1}-E^{k}|/E^k<\epsilon_{E}$.
			\end{enumerate}
			\textbf{end (for)}	
		\end{algorithm}
	\end{fminipage}
\end{center}
Again, we skipped some additional steps that assure a better convergence of the code and that are discussed in \citep{Piris2009}.

\subsubsection*{The explicit RDMFT algorithm of \textsc{Octopus}}
Now we have discussed the major steps of the orbital-based RDMFT algorithm that is implemented in \textsc{Octopus}. We want to conclude the section by mentioning the last missing pieces, which are the initialization of $\bm{\phi}$ and the convergence criteria in algorithm~\ref{algorithm:RDMFT}. 

For the former, we first do a preliminary calculation with the independent particles routine of \textsc{Octopus}, that is we solve the orbital equation
\begin{align}
\label{eq:numerics:rdmft:IP_equation}
h(\br)\phi_i(\br)=e_i\phi_i(\br) \quad i=1,...,B
\end{align}
to generate a set of orbitals $\bm{\phi}^{-1}$. We denote them with the index ``-1,'' because we perform then a further step before we start with the iteration. Additionally, we need a starting value $\bn^{-1}$ for the occupations, which we obtain by the rule
\begin{alignat}{3}
\label{eq:numerics:rdmft:initial_n}
	n_i^{-1}=&2-n_{thresh} \quad &&\forall\, i=1,...,N/2 \nonumber\\
	n_i^{-1}=&n_{thresh} \quad\quad &&\forall\, i=N/2+1,...,B
\end{alignat}
where $n_{thresh}$ is a threshold that is fixed to $n_{thresh}=10^-5$ and that stabilizes the algorithm. We perform one occupation number optimization with the $\bm{\phi}^{-1}$ and obtain $n_i^{0}$. Then, we calculate $\epsilon_{ij}$ and diagonalize $(\epsilon_{ij}^0+\epsilon_{ji}^0{}*)/2$ to obtain $\bm{\phi}^{0}$. As an overall convergence criterion we calculate the total energy after every occupation number optimization $E_{occ}^k$ and then slightly generalize the criterion of the orbital optimization as
\begin{align*}
	F_{max}^{k-1}<\epsilon_{F} \quad\text{and}\quad |E^k-E_{occ}^{k}|/E^k<\epsilon_{E}
\end{align*}
with the \emph{same} thresholds $\epsilon_{F},\epsilon_{E}$.

The complete algorithm is then summarized as
\begin{center}
	\begin{fminipage}{0.7\textwidth}
		\begin{algorithm}\label{algorithm:orbital_RDMFT} (RDMFT)  \newline
			Set convergence criteria $\epsilon_E,\epsilon_F$.\\
			Set $B$.\newline
			Generate $\bm{\phi}^{-1}$ by solving Eq.~\eqref{eq:numerics:rdmft:IP_equation} and calculate $h_{ii}^{-1}, w_{ijkl}^{-1}$.\newline
			Initialize $\bn^{-1}$ according to Eq.~\eqref{eq:numerics:rdmft:initial_n}.\newline
			Solve Eq.~\eqref{eq:numerics:ELE_n_complete} self-consistently with $h_{ii}^{l-1},w_{ijkl}^{l-1}$ for $\bn^0,\mu^0$.\newline
			Calculate $\epsilon_{ij}$ according to Eq.~\eqref{eq:numerics:Lambda}.\newline
			Diagonalize $(\epsilon_{ij}^0+\epsilon_{ji}^0{}*)/2$ to obtain $\bm{\phi}^{0}$ and calculate $h_{ii}^{0}, w_{ijkl}^{0}$.\newline
			\\
			\textbf{for} l=1,2,...
			\begin{enumerate}[label=\alph*]
				\item \textbf{Occupation number optimization}\\
				Solve Eq.~\eqref{eq:numerics:ELE_n_complete} self-consistently with $h_{ii}^{l-1},w_{ijkl}^{l-1}$ for $\bn^l,\mu^l$, employing algorithm~\ref{algorithm:RDMFT_non}\\
				Save total energy as $E_{occ}$
				\item \textbf{Orbital optimization}\\
				Solve Eq.~\eqref{eq:numerics:hermiticity_LagrangeMultiplier} with $\bn^l$ for $\bm{\phi}^l$, employing algorithm~\ref{algorithm:RDMFT_no}\\
				Calculate $h_{ii}^{l},w_{ijkl}^{l}$\\
				Save total energy $E$ and the $F_{max}=\max F_{ij}$ (Eq.~\eqref{eq:numerics:Piris_F_matrix}) 
				\item[] \textbf{break} if $F_{max}^{k-1}<\epsilon_{F}$ and $|E^k-E_{occ}^{k}|/E^k<\epsilon_{E}$
			\end{enumerate}
			\textbf{end (for)}	
		\end{algorithm}
	\end{fminipage}
\end{center}

This algorithm has been assessed for a 3d $H_2$-molecule comparing to another RDMFT code based on Gaussian orbitals and for further details we refer to Ref.~\citep{Andrade2015}. Since we employed this implementation to study dressed orbitals, we did additional high-accuracy benchmark studies for one-dimensional model systems. The results are presented in App.~\ref{sec:numerics:dressed:convergence}.
	\newpage
\section{Conjugate-gradients algorithm for RDMFT}
\label{sec:numerics:rdmft:cg}
As we have discussed in the introduction, the implementation of the orbital-based algorithm can not take advantage of the full flexibility of the real space grid. The straightforward way to remedy this, is to employ another orbital optimization algorithm that does not depend on an orbital basis.

Before we discuss this, we want to briefly remark on the \emph{fundamental difference} between orbital and real-space-based electronic-structure codes. In principle, one could see the grid-points as an effective orbital basis, since all (approximate) bases of a Hilbert space are theoretically equivalent for properly converged calculations. However, the number of grid points that are necessary to accurately describe even the smallest molecular systems is orders of magnitude larger than any standard orbital basis. Imagine for instance that we approximate one spatial dimension with only 10 grid points, which is not sufficient for an accurate description of any realistic scenario. Despite this low quality of the description, we have in 3d already $10^3=1000$ grid points and thus a basis with 1000 elements. Consequently, the matrix-representation of, e.g., a one-body Hamiltonian in this ``grid-basis'' consists of $1000^2=1$ million matrix elements. Diagonalizing such a matrix is already at the limits of the capabilities of modern computers. Instead, if we wanted to describe a systems like Helium by RDMFT with, e.g., Gaussian orbitals and wanted to employ a very large basis set for such systems, we could choose \emph{aug-cc-pVQZ}~\citep{Dunning1989} that consists of merely 55 elements. However, one would consider such a basis set only for numerical tests or comparisons. In practice, considerably smaller basis sets suffice to obtain very accurate RDMFT results for Helium. From this example, it is obvious that both approaches require entirely different numerical techniques.

The standard algorithm for the orbital optimization for most theory levels, including standard KS-DFT or HF in \textsc{Octopus} is the \emph{conjugate gradients} algorithm by \citet{Payne1992}. Originally proposed and optimized for DFT calculations of solids, which require a parametrization of the reciprocal or $k$-space, it is ideally suited for the real space, which is very similar from a numerical point of view~\citep{Beck2000}. The main complication of a real-space description is the accurate description of the core-region of atoms, close to the divergence of the Coulomb potential. The strongly increasing slope would require a very fine grid, which would spoil the numerical efficiency. An accurate and efficient solution of this problem is given by so-called \emph{pseudopotentials}, that describe the effective (classical) electrostatic potential of the nuclei together with the core-electrons. Pseudopotentials do not diverge at the position of the nuclei and have in general much smaller slopes in the core-region than the exact Coulomb potential. The quasi-classical treatment of the interaction between these core-electrons and the remaining so-called valence electrons is in most cases very accurate.\footnote{For a good introduction in the theory of pseudopotentials, the read is referred to the review of \citet{Schwerdtfeger2011}.} An in-depth discussion about the advantages and disadvantages of real-space codes is beyond the scope of this thesis and the reader is for further details referred to the review of \citet{Beck2000} and the first \textsc{Octopus} paper by \citet{Marques2003} and reference therein. 

In conclusion, we have to describe in a real-space code $N$ active or valence electrons under the influence of pseudopotentials, instead of all the $N_{all}>N$ electrons of the matter system in the local Coulomb potential of the nuclei (which is called an \emph{all-electron} calculation). In single-reference methods, this corresponds to the calculation of $N$ orbital wave functions $\bm{\phi}=(\phi_1,...,\phi_N)$ on the (real-space) grid. The conjugate-gradients algorithm by \citet{Payne1992} accomplishes this by a direct (iterative) minimization of the corresponding energy functional $E[\bm{\phi}]$. This means that the code minimizes $E[\bm{\phi}]$ iteratively along the direction of the gradient or the steepest-descent $\bm{\zeta}=\delta E[\bm{\phi}]/\delta \bm{\phi}$, taking into account prior steps by what is called \emph{conjugation}~\citep[part 5]{Nocedal2006}. Since the calculation of the gradient with respect to orbital $i$ is still for the whole grid, it is usually the bottle neck of the algorithm. Taking into account, e.g., the (approximate) Hessian is usually computationally too expensive. Nevertheless, modern conjugate-gradients methods have become a standard tool in nonlinear programming~\citep{Nocedal2006} and the algorithm by \citeauthor{Payne1992} seems to be very reliable for electronic structure calculations.\footnote{It is difficult to make this statement general, because there is no database that compares the performance of certain algorithms in standard codes. For other features there are indeed such databases, see, e.g., \url{https://www.nomad-coe.eu/externals/codes}. However, at least for \textsc{Octopus} we can confirm that the algorithm is usually reliable.}

The idea behind the real-space optimization algorithm for RDMFT is to use the already existing well-tested conjugate-gradients implementation for single-reference methods to a maximal degree. The hope is that the good convergence properties of the algorithm persist also in the more involved theory. Importantly, the algorithm minimizes directly the energy functional and not, e.g., the KS or HF eigenvalue equation. Thus, we just had to replace the respective functional with the RDMFT energy and redo the subsequent derivations. We present this derivation in the following subsection, which in shortened form is published in Ref.~\citep{Tancogne-Dejean2020}.

\subsection{Generalizing the conjugate-gradients method to RDMFT}
We now derive and comment the conjugate-gradients method for the orbital optimization in real-space RDMFT. For that, we recapitulate the most important steps of the algorithm by \citet{Payne1992}, explicitly highlighting the necessary modifications for the RDMFT minimization. As the general definition of the RDMFT minimization problem and the according equations have been presented in Sec.~\ref{sec:numerics:rdmft:piris}, we recapitulate here only the most important equations.

The goal is to find the set of natural occupation numbers $n_i$ and natural orbitals $\phi_i$ that extremalize the RDMFT energy functional \eqref{eq:numerics:rdmft:func_general} under the constraints \eqref{eq:numerics:rdmft:constraints}. Within the Müller approximation~\eqref{eq:numerics:Mueller}, the functional reads for $M$ natural orbitals
\begin{align}
\label{eq:numerics:rdmft_Mueller_energy_func_cg}
\begin{split}
E[\{\phi_i\},\{n_i\}]=& \sum_{i=0}^{B} n_i \int \td\br \, \phi^*_i (\br) h(\br) \phi_i (\br) 
+\frac{1}{2}\sum_{i,j=0}^{M} n_i n_j
\int\td\br\td\br' \left|\phi_i(\br)\right|^2 \left|\phi_j(\br')\right|^2 w(\br,\br')\\
&-\frac{1}{2}\sum_{i,j=0}^{B} \sqrt{n_i n_j}\int\td\br\td\br'\phi^*_i(\br)\phi^*_j(\br')\, w(\br,\br') \phi_i(\br')\phi_j(\br).
\end{split}
\end{align}
We will only discuss the orbital optimization and thus assume that $n_i$ and $\mu$ are constant. Consequently, we neglect the occupation-number dependence of $E$ and consider the reduced version of the Lagrangian~\eqref{eq:numerics:rdmft:lagrangian_general}, i.e.,
\begin{equation}
L[\{\phi_i\};\{\epsilon_{ij}\}]= E[\{\phi_i\}]-\sum_{j,k=1}^B \epsilon_{jk}\left(\int d\br \phi_j^*(\br)\phi_k(\br)-\delta_{jk}\right)\,,
\end{equation}
To present the algorithm, we assume that we are at a certain step $m$ of algorithm~\ref{algorithm:RDMFT} with the current orbitals $\bm{\phi}^m=(\phi_1^m,...,\phi_B^m)$. To minimize $L$, we want to vary $\bm{\phi}^m$ along the direction of the local steepest-descent of $L$, which is given by the negative gradient with respect to, e.g., $\phi_i^*$, cf. Eq.~\eqref{eq:numerics:gradients_RDMFT_phi_star}. The steepest-descent vector therefore reads
\begin{align}
\label{eq:numerics:StDesVector}
\zeta_i^m=-\frac{\delta L}{\delta \phi_i^m{}^*(\br)}&=-\left(\hat{H}^1\phi_i^m(\br) -\sum_{k=1}^{B}\epsilon_{ki}^m\phi_k^m(\br)\right),
\end{align}
where we used the definition of the one-body Hamiltonian $\hat{H}^1\phi_i^m(\br)=\frac{\delta E}{\delta \phi_i^m{}^*(\br)}$, cf. Eq.~\eqref{eq:numerics:RDMFT_HamiltonianSingleParticle}. 
Here, we have to take the whole sum $\sum_{k=1}^{B}\epsilon_{ki}^m\phi_k^m(\br)$ into account to find the steepest descent, which is in contrast to the corresponding definition for single-reference, where we can assume $\epsilon_{ki}=\delta_{ki}\epsilon_i$ (cf.\ Eq.\ (5.10) of Ref.~\citep{Payne1992}). This is the \emph{first of three modifications} that are necessary to extend the algorithm the RDMFT. 

Next, we need to find an estimate for $\epsilon_{ki}^m$, for which we proceed in an analogous way to \citep{Payne1992} by exploiting the stationarity conditions $\delta L=0$ that determine the solution. The corresponding equations read
\begin{align}
	0=&\frac{\delta E}{\delta \phi_i^*(\br)}-\sum_{k=1}^{B}\epsilon_{ki}\phi_k(\br) \tag{cf. \ref{eq:numerics:ELE_phi_star}}\\
	0=&\frac{\delta E}{\delta \phi_i(\br)}-\sum_{k=1}^{B}\epsilon_{ik}\phi_k^*(\br) \tag{cf. \ref{eq:numerics:ELE_phi}}.
\end{align}
We can now derive an expression for $\epsilon_{ki}$ from both equations, cf. Eqs.~\eqref{eq:numerics:Lambda} and~\eqref{eq:numerics:Lambda_alternative}. As we have discussed in Sec.~\ref{sec:numerics:rdmft:challenge_orbital_minimization}, these expressions are not equal, because the single-particle ``Hamiltonian'' in RDMFT is not hermitian. In the single-reference case that is considered by \citet{Payne1992}, the single-particle Hamiltonian is instead hermitian and thus the corresponding two sets of orbital equations are equivalent. This is equivalent to the fact that we can diagonalize $\epsilon_{ik}=\delta_{ki}\epsilon_i$ for single-reference methods.
Thus, in the RDMFT case, we have to make sure that the ``information''  of both equations, \eqref{eq:numerics:ELE_phi_star} and \eqref{eq:numerics:ELE_phi} enters the algorithm. This is the \emph{second modification} of the RDMFT-algorithm. 

Whereas in the single-reference case, we merely have to evaluate the diagonal element $\epsilon_i^m$ to calculate the steepest descent with respect to orbital $\psi^m_i$ (Eq.\ (5.11) of Ref.~\citep{Payne1992}), we now have to calculate the $B$ elements $\epsilon_{ki}$ (with $k=1,...,B$). For that, we have to choose the ``correct'' equation from our two choices, Eq.~\eqref{eq:numerics:ELE_phi_star} and Eq.~\eqref{eq:numerics:ELE_phi}. Since we have used the gradient with respect to $\phi_i^*$ (Eq.~\eqref{eq:numerics:ELE_phi_star}) for the definition of $\zeta_i$, we have to employ the equation from the \emph{other gradient} with respect to $\phi_i$, i.e., Eq.~\eqref{eq:numerics:ELE_phi}, to define\footnote{Note that this definition is analogous to Eq.~\eqref{eq:numerics:Lambda_alternative}, where we used the tilde $\tilde{\epsilon}_{ik}$ to stress the difference to $\epsilon_{ki}$ from the other equation. We drop the tilde in this section.}
\begin{align}
	\epsilon_{ki}^m=\int\td^3r\, \frac{\delta E}{\delta \phi_k^m(\br)}  \phi_i^m(\br).
\end{align}
We stress this point, because it is very unusual to work with non-hermitian operators in electronic structure theory. 
As a matter of fact, the RDMFT algorithm cannot converge to the minimum, if not both equations are taken into account. 

This can be understood from another perspective, if we derive the equations assuming real orbitals $\phi^*=\phi$. In this case, there is only one gradient equation and thus no ambiguity. The derivation can be found in the appendix~\ref{sec:app:cg_rdmft_real_orbitals}.

Having determined the steepest-descent vector, we need to perform certain orthogonalization steps, the \emph{preconditioning} step for a faster convergence, and the conjugation. All these steps are independent of the explicit theory and thus, we do not present them in this subsection. We summarize these steps in the next subsection, when we present the full algorithm in pseudo-code. For the details, reader is referred to Ref.~\citep{Payne1992}. 

We conclude this subsection with the last missing modification to extend the algorithm to RDMFT that concerns the \emph{line-minimization}. We therefore assume that we have calculated the normalized conjugate-gradients vector $\xi_i^m$, which determines the ``direction'' in which we want to minimize the energy functional. However, we do not know ``how long'' we need to ``go'' in this direction and thus we need to perform a line-minimization, i.e., we need to find the minimal value of the energy $E$ along the line that is defined by $\xi_i^m$.
 
In principle, the vector $\xi_i^m$ defines an infinite line that we can parametrize by, e.g., a multiplication with a scalar. However, we can here explicitly take the normalization constraint into account. All orbitals are normalized to one (and orthogonal to each other) and thus they can be seen as unit or basis vectors of an $B$-dimensional subspace of the Hilbert space. The algorithm takes care that this property is preserved at a iteration step, i.e., $\braket{\phi_i^m|\phi_j^m}=\delta_{ij}$. Additionally, when we optimize orbital $\phi_i^m$ in the direction $\xi_i^m$, it is taken care that also $\braket{\phi_i^m |\xi_i^m}=0$,\footnote{Note that there is a further subtle difference between the standard algorithm and the RDMFT version. In the former, we can additionally orthogonalize $\xi_i^m$ to all other states $\phi_j^m$ for $j\neq i$, because arbitrary rotations are allowed within this subspace. This is not possible in RDMFT, because the index of every $\phi_j^m$ is uniquely defined by its occupation number. This is another consequence of the impossibility to diagonalize the Lagrange-multiplier matrix in RDMFT.} and thus, we can parametrize all points of the possible descent by the \emph{angle} $\Theta$ between $\phi_j^m$ and $\xi_i^m$. This reduces the line-minimization problem to merely a closed interval of $[0,\pi/2]$, which can be exploited by the algorithm.
The optimized state has the form 
\begin{align}
\label{eq:numerics:cg_phi_parametrization}
\tilde{\phi}_i^{m}(\Theta)=\phi_i^m\cos\Theta  + \xi_i^m\sin\Theta.
\end{align}
We can now evaluate the energy as a function of $\Theta$ by inserting 
\begin{align}
	E[\Theta]=E[\{\phi_k^{m}, k\neq i\}, \tilde{\phi}_i^{m}(\Theta)]
\end{align}
and find the optimized state by a minimization over $\Theta$, i.e.,
\begin{align}
	\phi_i^{m+1}=\phi_i^m\cos\Theta^*  + \xi_i^m\sin\Theta^*,
\end{align}
where $\Theta^*$ is determined by
\begin{align}
\label{eq:numerics:line_min_def_theta}
\min\limits_{\Theta} E[\Theta].
\end{align}
Since the evaluation of $E[\Theta]$ requires the application of operators to orbitals, it is as expensive as calculating the gradient, which is the bottleneck of the whole procedure. Consequently, we need to find an approximate solution to $\eqref{eq:numerics:line_min_def_theta}$. The form of the parametrization $\eqref{eq:numerics:cg_phi_parametrization}$ suggests an expansion of $E$ in a Fourier series and for single-reference theories, a truncation after the first order is already quite accurate~\citep{Payne1992}. We assume that this also holds for RDMFT and thus make the ansatz
\begin{align}
	E[\Theta]= E_0 + A_1 \cos(2\Theta)  + B_1\sin(2\Theta).
\end{align}
First test-calculations have confirmed that this assumption is reasonable~\citep{Tancogne-Dejean2020}. We can now analytically determine the stationary point $\Theta^*$ of $E[\Theta]$ by the usual derivative condition $\td E/\td \Theta=0$ (Eq./ (5.30) of Ref.~\citep{Payne1992}), i.e., 
\begin{align}
\label{eq:numerics:LineMinThetaOpt}
	\Theta^*=\tfrac{1}{2}\tan^{-1}\left(\frac{B_1}{A_1}\right).
\end{align}
There are several possibilities to determine the coefficients $A_1$ and $B_1$ (see Ref.~\citep{Payne1992}) and we employ the one that is implemented in \textsc{Octopus},  which reads
\begin{align}
\label{eq:numerics:LineMinDef}
	\frac{B_1}{A_1}=-\frac{2\left.\frac{\td E}{\td \Theta}\right|_{\Theta=0}}{\left.\frac{\td^2 E}{\td \Theta^2}\right|_{\Theta=0}}.
\end{align}
To calculate the derivatives, occurring in this expression, we have to consider  the RDMFT energy functional, which leads to the third and last modification with respect to the single-reference algorithm. We have
\begin{subequations}
\label{eq:numerics:LineMinCoeff}
\begin{align}
	\left.\frac{\td E}{\td \Theta}\right|_{\Theta=0}=&\int\td\br \left(\frac{\delta E}{\delta\phi_i^*}
	\frac{\td\phi_i^*}{\td \Theta}+\frac{\delta E}{\delta\phi_i}\frac{\td\phi_i}{\td \Theta}\right)\nonumber\\
	=&\braket{\xi_i|\hat{H}^1\phi_i}+\braket{\phi_i|\hat{H}^1\xi_i}- 
	\sum_{k}\left(\epsilon_{ki} \braket{\xi_i|\phi_k}+\epsilon_{ik}\braket{\phi_i|\xi_k}\right)\\
\intertext{for the first derivative and }
	\left.\frac{\td^2 E}{\td \Theta^2}\right|_{\Theta=0} =& \int \frac{\delta E^2}{\delta\phi_i^*{}^2} 
	\left(\frac{\td\phi_i^*}{\td \Theta}\right)^2 +\int \frac{\delta E^2}{\delta\phi_i^2} \left(\frac{\td\phi_i}{\td \Theta}\right)^2   + 2\int \frac{\delta E^2}{\delta\phi_i^*\delta\phi_i} \frac{\td\phi_i^*}{\td \Theta}\frac{\td\phi_i}{\td \Theta} + \int \frac{\delta E}{\delta\phi_i^*}\frac{\td^2\phi_i^*}{\td \Theta^2} +\int \frac{\delta E}{\delta\phi_i}\frac{\td^2\phi_i}{\td \Theta^2}\nonumber\\
	=& 2 \left(\braket{\xi_i|\hat{H}^1\xi_i} - \braket{\phi_i|\hat{H}^1\phi_i}\right) + \alpha_i^{H} + \alpha_i^{XC}
\end{align}
\end{subequations}
for the second derivative.\footnote{Note that the former is denote by $\beta$ and the latter by $\alpha$ in the source code of \textsc{Octopus}.}
Here, we defined
\begin{subequations}
\begin{align}
	\alpha_i^{H}=&n_i \int\td\br 2\Re[\phi_i^*(\br)\xi_i(\br)] \left(n_i \int\td\br'2\Re[\phi_i^*(\br')\xi_i(\br')] w(\br,\br') \right)\\
	\alpha_i^{XC}=& -n_i\int\td^3r\td^3r'\, 2\Re\left[\phi_i^* (\br) \phi_i^* (\br') w(\br,\br') \xi_i (\br) \xi_i (\br')\right]-n_i\int\td^3r\td^3r'\, \phi_i^* (\br) \xi_i^* (\br') w(\br,\br') \xi_i (\br) \phi_i (\br') \nonumber\\
	&-\sqrt{n_i}\sum_{j=1}^{B} \sqrt{n_j} \int\td^3r\td^3r'\, \xi_i^* (\br) \phi_j^* (\br') w(\br,\br') \phi_j (\br) \xi_i (\br'),
\end{align}
\end{subequations}
and used 
$\frac{\td\phi_i^{(*)}(\Theta)}{\td\Theta}|_{\Theta=0} =\xi_i^{(*)}$ and $\frac{\td^2\phi_i^{(*)}(\Theta)}{\td\Theta^2}|_{\Theta=0} =\phi_i^{(*)}$.
We see that $\epsilon_{ik}$ appears only in the first derivative term, which hence is the only one that needs to be modified for the RDMFT algorithm, which we summarize in the next subsection.

\subsection{The conjugate-gradients algorithm for RDMFT in \textsc{Octopus}}
\label{sec:numerics:rdmft:cg:concrte_algorithm}
We now want to formulate the complete algorithm for the orbital optimization by conjugate-gradients of the RDMFT routine of \textsc{Octopus}. The routine can be chosen in the developer's version of \textsc{Octopus} as alternative to algorithm~\ref{algorithm:RDMFT_no}. The general part (algorithm~\ref{algorithm:RDMFT}) including the occupation number optimization (algorithm~\ref{algorithm:RDMFT_non}) are not modified.

Before we formulate the complete algorithm, some remarks are appropriate. The crucial advantage of the whole routine is that we do not need to explicitly calculate the one-body Hamiltonian $\hat{H}^1$ in some basis, but only need to \emph{apply} it to a state $\phi_i$ according to
\begin{align}
\tag{cf. \ref{eq:numerics:RDMFT_HamiltonianSingleParticle}}
\hat{H}^1\phi_i(\br)
=\,& n_i\, h(\br) \phi_i (\br) + n_i\, v_H(\br) \phi_i(\br) - \sqrt{n_i}\, v^i_{XC}(\br).
\end{align}
This operation can be done very efficiently on the grid for a couple of reasons.
One important of these reasons is a crucial property of all the occurring operators in Eq.~\eqref{eq:numerics:RDMFT_HamiltonianSingleParticle} with the exception of $v^i_{XC}$. The operators $h(\br)$ and $v_H(\br)$ are (semi)local, i.e., they effectively depend only on one coordinate and thus their application to a state has the cost of an inner product, instead of a full matrix-vector multiplication. For the one-body part
\begin{align}
	h(\br)=-\tfrac{1}{2}\nabla^2+v(\br),
\end{align}
the locality is obvious for the local potential $v(\br)$. The Laplacian $\nabla^2$ is instead what is called \emph{semi-local}. This means for vectors on the grid that we can approximate $\nabla^2$ with, e.g., finite-differences of some order and then apply it to a state by a so-called \emph{stencil} that is the same for every point and thus ``almost'' local~\citep[Sec. IV.A]{Beck2000}. The Hartree-potential
\begin{align}
\tag{cf. \ref{eq:numerics:v_hartree}}
v_{H}(\br)&=\sum_{j=1}^{B} n_j \int\td^3r'\, \rho_j(\br')  w(\br,\br') \nonumber\\
=& \int\td^3r'\, \rho(\br')  w(\br,\br'),
\end{align}
with the total density $\rho(\br)=\sum_{j=1}^{B} n_j\rho_j(\br')$ is also local \emph{during every iteration step}, which is the basic approximation of our self-consistent field procedure. It is crucial here that we can exchange the sum with the integration and thus have to calculate the integral only once, which in practice is done in $k$-space, i.e., after a Fourier transformation. This is considerably faster for a large number of grid-points and is a main advantage of the conjugate-gradients algorithm.\footnote{For this reason, also orbital-based electronic structure codes employ conjugate gradients algorithms for systems that require a very large basis sets.}
However, the exchange-correlation potential that we employ, e.g., in RDMFT, HF or for hybrid functionals in KS-DFT (see Sec.~\ref{sec:est:dft}) is truly nonlocal. We recall the definition 
\begin{align}
\tag{cf. \ref{eq:numerics:v_xc}}
v^i_{XC}(\br)&=\sum_{j=1}^{B} \sqrt{n_j} \int\td^3r'\, \phi_j^*(\br')   w(\br,\br') \phi_j(\br) \phi_i(\br').
\end{align}
Importantly, we \emph{cannot} exchange summation and integration here, because both operations involve \emph{different} orbitals. This is a severe bottleneck of the algorithm and prevented real-space and $k$-space codes for a long time to treat larger systems with these methods~\citep{Lin2016}. In RDMFT this limitation is especially pronounced, because we need to calculate the exchange-correlation potential not only for the one orbital that we optimize but also for all the other orbitals to determine $(\epsilon_{ik})$. Consequently, it is still very expensive to do RDMFT calculations with the conjugate-gradients algorithm and we have so far mostly performed tests with one-dimensional systems. However, it has been shown recently that this problem can be solved with a newly developed approximation for the exchange-correlation operator~\citep{Lin2016}. Although the author explicitly considers HF theory, the method should be generalizable to RDMFT in a straightforward way.

Next, we want to remark briefly on \emph{preconditioning}, which we mentioned already in the first part of this section. This is a very common part of large-scale minimization algorithms and a thorough discussion is beyond the scope of this text. The basic problem in the realm of electronic structure calculations is that the kinetic energy operator that we apply to a state to calculate the gradient has a larger error for energetically higher than lower lying states, because the former have more nodes and thus, in general bigger slopes. The standard preconditioner in \textsc{Octopus} thus applies a kind of low-pass filter on the gradient. We denote the corresponding operator with $\hat{P}$. For further details, we refer to \citep{Payne1992} and references therein.

As a final remark, before we present the algorithm, we want to refer the reader to the excellent explanation of the conjugate-gradients method in Sec. 5.A of Ref.~\citep{Payne1992}. The authors explain there how a conjugation of the steepest-descent vector leads to a maximally fast convergence of the algorithm, because it cannot get ``trapped'' in a zig-zag trajectory. This is illustrated in Fig. 14 of the reference. Note that there are different possibilities to do the conjugation and in the following we only present the \emph{Fletcher-Reeves}~\citep{Fletcher1964} method, but there is also the \emph{Polak-Ribiere}\footnote{The original paper \citep{Polak1969} is unfortunately only available in French.} scheme available in \textsc{Octopus}. Both methods are elucidated in, e.g., part 5 of \citet{Nocedal2006}.

We now summarize the complete algorithm in the pseudocode form that we used already before.

\begin{center}
	\begin{fminipage}{0.7\textwidth}
		\begin{algorithm}\label{algorithm:RDMFT_no_cg} (Orbital optimization 2)  \newline
			Set the convergence criterion $\epsilon_{\phi}>0$.\\
			\\
			\textbf{for} k=1,...,B
			\begin{enumerate}[label=\alph*]
				\item Minimization of orbital with index k. Initialize $\phi_k^{l,0}{}'=\phi_k^l$ from the current set of orbitals $\bm{\phi}^l$
				\item Orthogonalize to previously optimized states:\\ $\phi_k^{l,0}=\phi_k^{l,0}{}'- \sum_{i=1}^{k-1}\braket{\phi_i^{l}|\phi_k^{l,0}{}'} \phi_i^{l}$
				\item Calculate relevant entries of $(\epsilon_{ik})$:\\
					$\epsilon_{ik}^{l,0}=\braket{\hat{H}^1\phi_i^{l}|\phi_k^{l,0}} \,\forall\,i=1,...,B$ and\\
					$\epsilon_{ki}^{l,0}=\braket{\phi_k^{l,0}|\hat{H}^1\phi_i^{l}} \,\forall\,i=1,...,B$
			\end{enumerate}
			\begin{adjustwidth}{0.5cm}{}
					\textbf{for} m=0,1,2,...
					\begin{enumerate}[label=\roman*]
						\item \textbf{if} $k>1$, update $\epsilon_{in}^{l,m},\epsilon_{ni}^{l,m}$ for all $n<k$
						\item Calculate the steepest-descent vector (Eq.~\eqref{eq:numerics:StDesVector}):\\
						$\zeta_k^m=-\left(\hat{H}^1\phi_k^{l,m} -\sum_{i=1}^{B}\epsilon_{ik}^{l,m}\phi_i^l\right)$
						\item Apply the preconditioner: $\eta_k^m{}'=\hat{P}\zeta_k^m$
						\item Orthogonalize to the current and previously optimized states:\\ 
						$\eta_k^m=\eta_k^m{}'- \braket{\phi_k^{l,m}|\eta_k^m{}'} \phi_k^{l,m}- \sum_{i=1}^{k-1}\braket{\phi_i^{l}|\eta_k^m{}'} \phi_i^{l}$
						\item Calculate the conjugation factor:\\ $\gamma_k^m=\braket{\eta_k^m|\zeta_k^m}/\braket{\eta_k^{m-1}|\zeta_k^{m-1}} \,\forall\,m>0$\\
						$\gamma_k^0=0$
						\item Calculate conjugate-gradients vector:\\
						$\xi_k^m=\eta_k^m + \gamma_k^m \xi_k^{m-1}$
						\item Calculate $A_1/B_1$ according to Eq.~\eqref{eq:numerics:LineMinDef} with the expressions of Eq.~\eqref{eq:numerics:LineMinCoeff} determine the optimal angle $\Theta^*$ 
						\item Calculate the optimized orbital
						$\phi_k^{m+1}=\phi_k^m\cos\Theta^*  + \xi_k^m\sin\Theta^*$
						\item Calculate the residue $R^{m+1}=|\hat{H}^1\phi_k^{m+1}-\epsilon_{kk}\phi_k^{m+1}|^2$
						\item[] \textbf{break} if $m>0$ and $R^{m+1}<\epsilon_{\phi}$ and $R^{m}<\epsilon_{\phi}$
					\end{enumerate}
					\textbf{end (for)}	
			\end{adjustwidth}
			\textbf{end (for)}		
		\end{algorithm}
	\end{fminipage}
\end{center}
We want to remark that we left out some details that are not crucial for the algorithm to work. We refer here again to \citep{Payne1992} or directly to the source-code of \textsc{Octopus}.

\newpage
\section{Comparison of both orbital optimization methods}
\label{sec:numerics:rdmft:comparison:cg_piris}
Having two different orbital-optimization techniques at hand, we are able to study the implications of the real-space description for practical RDMFT calculations. We want to stress here again that there is no other real-space implementation of RDMFT, which naturally leads to new types of problems that have to be solved. 

The most pronounced limitation of the conjugate-gradients algorithm is the high computational cost of evaluating the exchange-correlation term that we have mentioned in Sec.~\ref{sec:numerics:rdmft:cg:concrte_algorithm}. This has prevented us so far from performing all-electron real-space calculations in three spatial dimensions. This would be necessary for a comparison with an orbital-based code. However, there are new developments that might overcome this limitation (see Sec.~\ref{sec:numerics:rdmft:cg:concrte_algorithm}). Besides that, all-electron calculations without pseudo-potentials are very inefficient in a real-space code, because they require a very high resolution around the divergence of the Coulomb potential of the nuclei (see the introduction of this section). In RDMFT, this problem becomes even more pronounced than in single-reference methods, because we need to determine considerably more orbitals. Since all orbitals are orthogonal to each other, the number of nodes increases with the number of orbitals, which requires again a finer grid for a good representation. This is not only problematic for all-electron calculations, but could turn out a general limitation of real space RDMFT. 

Nevertheless, the RDMFT implementation in \textsc{Octopus} works and we have tested it extensively for one-dimensional systems, where we basically can afford arbitrarily fine grids. We want to finish this chapter about real-space RDMFT thus with a comparison of the different methods for a 1d model-system. The following discussion is based on Ref.~\citep[Ch. 14]{Tancogne-Dejean2020}. To do such a calculation, we need to converge the two basic numerical parameters for every real space calculation, i.e., the box-length $L_x$ and the spacing $\delta_x$. In RDMFT, we additionally have to converge the number of natural orbitals $B$, which are system-dependent but not equal to the particle number $N$ in contrast to, e.g., the number of KS- or HF-orbitals. Since we have two methods for the orbital optimization at hand, it is especially interesting to compare these in terms of the convergence with respect to $B$. 

The default method due to Piris and Ugalde (Piris method in the following) uses the orthonormality constraint of the natural orbitals, which implies that the ``F-matrix'' constructed from the Lagrange multipliers $\epsilon_{jk}$ is diagonal at the solution point (see Sec.~\ref{sec:numerics:rdmft:piris}). As an immediate consequence of the Piris method, the natural orbitals at the solution point are linear combinations of the orbitals used as the starting point for the minimization. In other words, the initial orbitals serve as a basis and the convergence of the method will depend on this basis. The conjugate gradient algorithm also requires a set of initial orbitals to start the self-consistent calculation, but at convergence the results are \emph{independent} of that starting point. Therefore, while a calculation using the Piris method requires a set of initial states which serve as the basis, the conjugate gradient algorithm can be used starting from a initial set of random states. In our tests of the conjugate gradient implementation, the quality of the initial
states only had an influence on the number of iterations necessary for the
convergence, but not on the final result. We suggest to use the orbitals
obtained from an independent particle calculation as initial states since they can be obtained for a low numerical cost and simultaneously can serve as a basis set in the Piris implementation.

\begin{figure}
	\centering
	\includegraphics[width=0.45\textwidth]{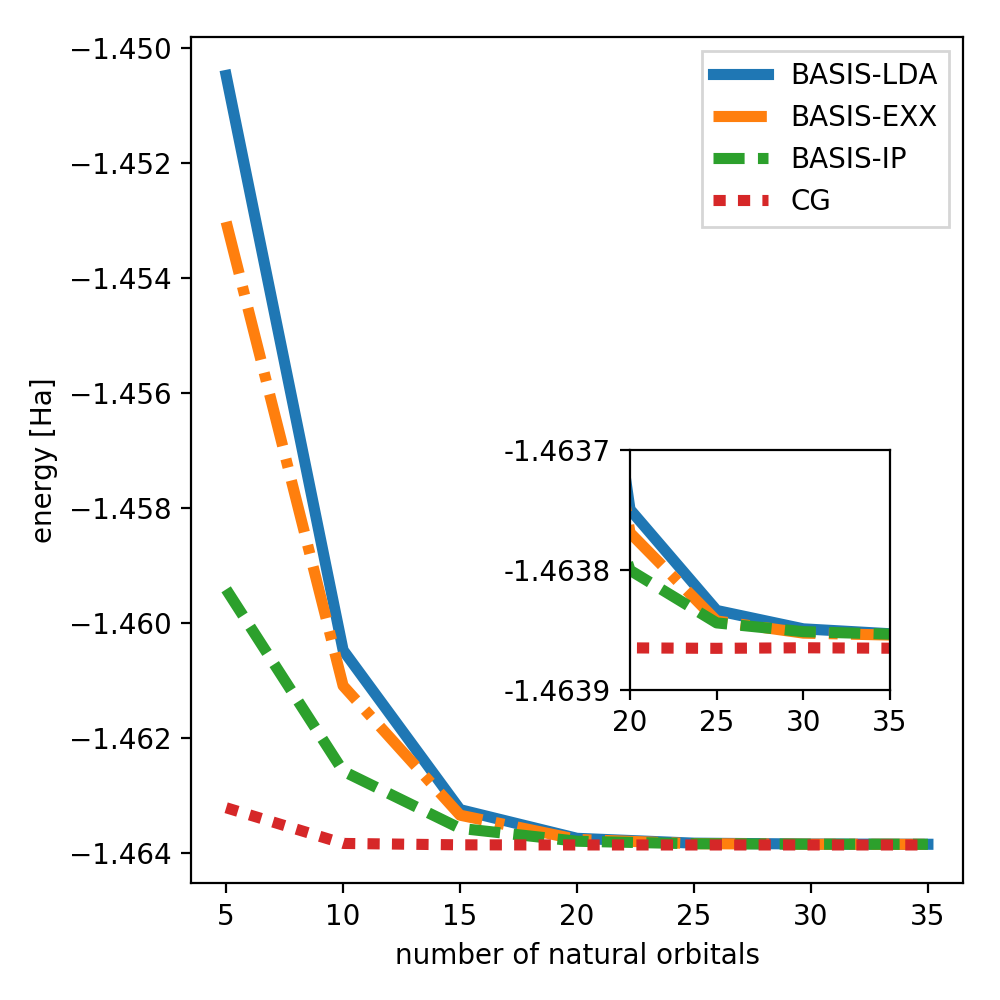}
	\caption{Total energy for the RDMFT calculation for one-dimensional
		H$_2$ using the Piris method with different basis sets and using the conjugate gradient implementation. The inset shows a zoom into the area where convergence is reached.}
	\label{fig:numerics_1dH2}
\end{figure}

We thus will compare different starting points for the Piris method with the conjugate-gradients implementation. For the former, the initial orbitals are taken to be the solutions obtained with a different level of theory, like independent particles (IP) or KS-DFT. In order to better understand the effect of the choice of basis, we tested the following choices: (i) independent particles, KS-DFT within (ii) the local density approximation (LDA) or (iii) the exact exchange (EXX) approximation, as well as (iv) HF theory. In all cases we have to ensure that the number of unoccupied states in the calculation is sufficient to cover all the natural orbitals which will obtain a significant occupation in the following RDMFT calculation. The results for the convergence of the total energy of a one-dimensional (1D) hydrogen molecule (see Sec.~\ref{sec:est:comparison}) using the M\"uller functional~\cite{Mueller1984} are given in Fig.~\ref{fig:numerics_1dH2}. The calculations were performed on a 1D grid extending from $-12.0$ to $12.0$ bohr, with a grid spacing of $0.03$ bohr. The nuclear potential for the 1D molecule reads
\begin{align}
\tag{cf. \ref{eq:est:comparison:potential_H2}}
v(x) = -\frac{1}{\sqrt{(x-d)^2+1}} -\frac{1}{\sqrt{(x+d)^2+1}}\,,
\end{align}
with $d=1.628$ bohr, which corresponds to the equilibrium geometry. The
electron-electron interaction in one dimension is described by the soft-Coulomb interaction
\begin{align}
\tag{cf. \ref{eq:est:comparison:interaction_soft_coulomb}}
w(x,x') = \frac{1}{\sqrt{(x-x')^2+1}}\,.
\end{align}

Since Octopus performs all calculations on a finite grid, we typically obtain a finite number of bound states in the calculations for the basis set and any additional orbitals extend over the whole grid, i.e., they are unbound and therefore delocalized. However, all natural orbitals with non-zero occupation are localized close to the center of mass, because they decay with the ionization potential of the system~\citep{Morrell1975}. Hence, the extended basis states will only contribute with very small coefficients, if at all, and their inclusion in the basis set does not lead to a significant improvement of the results. Consequently, the fastest convergence and lowest total energies are obtained by using the results from an independent particles calculation, as those yield the largest number of localized orbitals from all the different basis sets that were tested, as shown in Fig.~\ref{fig:numerics_1dH2}.

The convergence study with respect to the number of natural orbitals for the conjugate-gradients algorithm is also included in Fig.~\ref{fig:numerics_1dH2}. As one can see, a smaller number of natural orbitals needs to be included in the calculation than for the basis set implementations with the Piris method. As the number of natural orbitals equals the number of basis functions in these calculations, this is mostly due to the fact that the available basis sets are of rather poor quality. In addition, the converged total energy for the
conjugate gradient algorithm is slightly lower than for all the basis sets using the Piris method (see inset of Fig.~\ref{fig:numerics_1dH2}), which shows that the conjugate gradient algorithm exploits the full flexibility of the grid implementation, allowing for contributions to the natural orbitals which are not covered by any of the basis sets. We have also verified that the converged result of the conjugate gradient method is indeed independent of the choice of initial state for the M\"uller functional~\cite{Mueller1984} employed in our calculations. However, one should note that the M\"uller functional is known to be convex when all infinitely many natural orbitals are included~\cite{Frank2007}. This property is most likely not shared by all available RDMFT functionals. In addition, the number of natural orbitals in any practical calculation is always finite. Consequently, in practice one needs to test the convergence for the different starting points, as the appearance of local minima cannot be excluded.
	
	\clearpage
\chapter{Polaritonic structure theory:  a numerical perspective}
\label{sec:numerics:dressed}
This second chapter of the numerics part deals with the extension of electronic-structure methods to coupled light-matter problems. The principal strategy to do this has been laid out in Sec.~\ref{sec:dressed:est:prescription}. In this chapter, we show how to do this in practice in two steps. We first present how to extend an electronic structure code to treat dressed orbitals with the modified interaction and potential, which is already sufficient to describe specific situations such as the two-polariton case (fermion ansatz, Sec.~\ref{sec:numerics:dressed:implementation}). We exemplify this simple way to implement polaritonic-structure methods with our implementation in \textsc{Octopus} of polaritonic HF and polaritonic RDMFT~\citep[Ch. 4]{Tancogne-Dejean2020}. Then, we show how to do this also in the general case with the example of a purpose-built implementation (Sec.~\ref{sec:numerics:hybrid}).

At this point, we want to remark briefly on the particular numerical challenge of modifying well-known electronic-structure algorithms. We present in App.~\ref{sec:numerics:dressed:non_assessment} an example that highlights that even simple modifications (such as adding the dressed term to the interaction kernel) of a sufficiently complex code require extensive testing. 
Besides revealing such issues, the validation procedure of a numerical implementation often influences also the \emph{development} of theory itself. For instance, we have only realized the importance of the hybrid statistics during the exhaustive convergence studies that are presented in App.~\ref{sec:numerics:dressed:convergence}.

\section{The implementation of dressed orbitals in \textsc{Octopus}}
\label{sec:numerics:dressed:implementation}
In this section, we present how to apply the general prescription to turn an electronic-structure into a polaritonic-structure method of Sec.~\ref{sec:dressed:est:prescription} with the example of our implementation of dressed orbitals in \textsc{Octopus}.\footnote{This section is based on Ref.~\citep[part 4]{Tancogne-Dejean2020}.} 
The actual state of the code supports the inclusion of one photon mode and matter systems of one spatial dimension.\footnote{Note that the implementation is in principle general enough to treat 3d-matter systems, but this has not yet been tested.} We therefore do the presentation explicitly for the 1+1 dimensional case. The method is part of \textsc{Octopus} version 10.0 or higher, which is publicly available (\url{https://octopus-code.org/wiki/Octopus_10}).

As we have mentioned already, our implementation of the dressed orbitals is based on the RDMFT implementation in \textsc{Octopus} and can thus only be used for RDMFT and HF calculations. 
However, a generalization to other methods is possible with little effort, because the only crucial change is in the \emph{Poisson solver} that calculates integrals of the form
\begin{align}
\label{eq:numerics:poisson_integral}
	\tilde{v}(\br)=\int\td^zr' \tilde{\rho}(\br')\tilde{w}(\br,\br').
\end{align}
Here $\br\in V\subset \mathrm{R}^z$  (in the present case $\br=(x,q)$ and thus $z=2$) and $\tilde{w}(\br,\br')$ is the two-body integral kernel. In the standard 3d-case, this is the Coulomb kernel $\tilde{w}(\br,\br')=w(\br,\br')=1/|\br-\br'|$, which makes Eq.~\eqref{eq:numerics:poisson_integral} the solution of the Poisson's equation $\nabla^2\tilde{v}=\tilde{\rho}$ and explains the name of the routine. For 1d, it is by default the soft-Coulomb kernel $w(x,x')=1/\sqrt{(x-x')^2+\epsilon^2}$ with $\epsilon=1$ (see also the discussion in Sec.~\ref{sec:est:comparison}). Additionally, we have defined $\tilde{\rho}(\br)$, which can be some orbital density $\tilde{\rho}(\br)=\rho_{ij}(\br)=\phi_i^*(\br)\phi_j(\br)$ that occurs in the exchange-correlation potential (cf. Eq.\eqref{eq:numerics:v_xc}) or the total density $\tilde{\rho}=\rho(\br)=\sum_{i=1}^{M}n_i\rho_{ii}(\br)$ that occurs in the Hartree-potential  (cf. Eq.\eqref{eq:numerics:v_hartree}). The Poisson-solver is thus utilized by all methods in \textsc{Octopus} to include the contribution of the (approximate) Coulomb interaction. 

To generalize the code to dressed orbitals, we have to define one additional coordinate $q$ that parametrizes the photon-degrees of freedom and replace the Coulomb-kernel
\begin{align}
\label{eq:numerics:coulomb_kernel_dressed}
	w(x,x')\rightarrow w'(xq,x'q')=w(x,x')+ \underbrace{\left[-\tfrac{\omega}{\sqrt{N}}\lambda qx' -\tfrac{\omega}{\sqrt{N}}\lambda q'x +\tfrac{1}{2}(\lambda x)^2\right]}_{w_d(xq,x'q')},
\end{align}
where $\omega$ is the frequency of the photon-mode, $\lambda=\blambda_x$ the component of the polarization vector in the direction of the spatial dimension of the matter system $x$. Obviously, the extra term $w_d(xq,x'q')$ is not related to Poisson's equation and thus, it does not really ``belong'' in the Poisson solver. To avoid changing the structure of the code too much, the application of $w_d(xq,x'q')$ is nevertheless implemented in this part of the code. As a consequence, it is not possible to choose the specific method to evaluate Eq.\eqref{eq:numerics:poisson_integral}, when dressed orbitals are used.  One is constraint to the (inefficient) \emph{direct sum} method, which just calculates the integral as a summation on the grid. 
We want to stress here that the evaluation of Eq.\eqref{eq:numerics:poisson_integral} for the extra part $w_d(xq,x'q')$ is considerably less involved than solving Poisson's equation, because its 2-coordinate dependence is only due to simple \emph{products}. To calculate for example
\begin{align*}
	\int\td x'\td q' \rho_{ij}(x',q')[-\tfrac{\omega}{\sqrt{N}}\lambda q'x]
	=& -\tfrac{\omega}{\sqrt{N}} x \int\td x'\td q' \rho_{ij}(x',q') q',
\end{align*}
we can pull out $x$ and thus need to evaluate the integral only once for every value of $x$, which has merely the (very low) computational cost of in inner product.\footnote{The calculation of an inner product scales linear with the grid-size $B$} 
For the Coulomb-kernel instead, we have to evaluate the integral for \emph{every} value of $x$, when we perform a naive direct-sum calculation. This operation scales \emph{quadratically} on the grid and is the bottleneck of basically every method. More sophisticated methods can reduce this scaling considerably. For example, the standard Poisson solver of \textsc{Octopus} performs a Fourier transformation of Eq.\eqref{eq:numerics:poisson_integral} to turn the integral into a multiplication and then transforms back to real space. Using the \emph{fast Fourier transformation}~\citep{Beck2000}, this procedure reduces the scaling to $B\ln B$~\citep{Beck2000}.

Besides the two-body integral kernel, also the local potential has to be modified by adding a dressed extra term, as discussed in Sec.~\ref{sec:dressed:est:prescription}. This is not yet implemented explicitly in \textsc{Octopus}. Instead, the user is forced the make use of the option of a \emph{user-defined potential}, following, e.g., the example of the test-suite of \textsc{Octopus}.\footnote{See \url{https://octopus-code.org/wiki/Developers:Starting_to_develop\#Testsuite}.} 
Finally, we have also implemented the output of some photonic expectation values at the end of the calculation. They are written in the \texttt{static\_information}.

The just describes modifications allow to perform polaritonic RDMFT and HF calculations with the fermion ansatz that does not explicitly guarantee the Pauli-principle (see Sec.~\ref{sec:dressed:est:prescription}).  This means that in order to guarantee results that do not violate the Pauli principle, we either have to restrict the system to one active orbital (thus the two-particle singlet case) or make sure that the photon frequency $\omega$ is large enough to guarantee the Pauli-principle trivially (see Secs.\ref{sec:dressed:construction:simple:auxiliary_wf_problem} and Fig.~\ref{fig:tb_energy}). 

To go beyond that and consider the polariton ansatz, we need to enforce the $N$-representability of the electronic 1RDM with an extra routine, which would require a considerable change of the code. We have so far only tested this in a small purpose-built code that we present in the next section.

	\newpage
\section[Hybrid Statistics]{The hybrid statistics of polaritons in practice}
\label{sec:numerics:hybrid}
In the following, we will show how to go beyond the fermion ansatz of dressed orbitals in practice, i.e., how to (approximately) enforce the hybrid statistics of polaritonic orbitals in a numerical minimization (polariton ansatz, see Sec.~\ref{sec:dressed:est:prescription}). The additional challenge in comparison to the fermion ansatz is here to enforce a set of nonlinear (only implicitly given) inequality constraints during the minimization of the (also nonlinear) energy functional of a given method. Including such complicated constraints requires considerable modifications of standard first-principles algorithms and it is a priori not clear, which algorithm is suited. We therefore analyze the problem from a general perspective and present specifically the augmented-Lagrangian algorithm that is accurate, yet simple to implement and validate (Sec.\ref{sec:numerics:hybrid:inequality_challenge}). 

We then show with the example of polaritonic HF how to obtain a general polaritonic-structure method that accounts for the hybrid statistics (polariton ansatz) by combining the augmented-Lagrangian with a standard electronic-structure algorithm. To test the algorithm, we have implemented it in a simple purpose-built code,\footnote{The code is written in the programming language \textsc{Python}(\url{https://www.python.org/}) employing the numerical routines of the scientific programming library \textsc{NumPy} (\url{https://numpy.org/}).} which we have employed to show that the polariton ansatz indeed overcomes the limitations of the fermion ansatz (see Sec.~\ref{sec:dressed:results:implementation_lattice}) and to highlight the influence of localization on the light-matter interaction (Sec.~\ref{sec:dressed:results:3confinement}).
To get accustomed to the new challenges, it is important to reduce the complexity of the minimization problem as much as possible. Therefore, we have developed and tested the algorithm in a simplified setting with 2 electron orbitals (corresponding to $N=4$ electrons) in one spatial dimension and one photon mode (see also Sec.~\ref{sec:dressed:results:implementation_lattice}). We also employ one of the simplest strategies to enforce inequality constraints, that is the augmented Lagrangian approach. It is likely that in order to extend a standard electronic-structure code to treat dressed orbitals with the polariton ansatz, a more sophisticated algorithm will be necessary. The here proposed algorithm is only a first step toward this goal. We therefore not only present the final form of the algorithm, but also explain the rationale behind its different parts.

\subsection{The new challenge: enforcing inequality constraints}
\label{sec:numerics:hybrid:inequality_challenge}
To start the discussion, let us briefly recapitulate the generic minimization problem~\eqref{eq:MinimizationProblem} that we have to solve to guarantee the Pauli principle within polaritonic-structure theory (polariton ansatz). 
Irrespective of the particular method that is considered, the principal numerical challenge of the polariton ansatz is to enforce a set of $M$ nonlinear inequality constraints $g_i \geq 0 (i=1,...,M)$ during the minimization. 
Without loss of generality, we assume that the state of the system is described $N$ orbitals $\bm{\phi}=\{\phi_1,...,\phi_N\}$.\footnote{For instance, this applies to HF and KS-DFT. However, we can straightforwardly generalize the argument to, e.g., RDMFT if we consider additionally the natural occupation numbers. As we have shown in Sec.~\ref{sec:numerics:rdmft}, this only influences the equality constraints, but not the inequality part. For other methods, a similar argument holds.}

The goal is to minimize the energy 
\begin{align}
	E=E[\bm{\phi}].
\end{align}
that is a (nonlinear) functional of $N$ polaritonic orbitals $\bm{\phi}=(\phi_1,...,\phi_{N})$ under a set of equality constraints (cf. Eq.~\eqref{eq:ConstraintsEquality})
\begin{align}
\label{eq:numerics:hybrid:ConstraintsEqualityGeneric}
	c_{ik}[\bm{\phi}]=0 \quad\forall i,k
\end{align}
and the $B_m$ inequality constraints\footnote{Note that there are in principle as many $g_i$ as matter basis states that we have denoted in the previous section by $B_m$. In practice, it is however not necessary to consider all these constraints as we discuss below.} (cf. Eq.~\eqref{eq:ConstraintsInequality})
\begin{align}
\label{eq:numerics:hybrid:ConstraintsInequalityGeneric}
	g_i[\bm{\phi}]=1-n_i[\bm{\phi}]\geq 0 \quad\forall i,
\end{align}
where the $n_i$ are the (electronic) natural occupation numbers, i.e., the eigenvalues of the electronic 1RDM
\begin{align}
\label{eq:numerics:gamma_electronic_definition}
\gamma_e[\bm{\phi}](\br,\br')=& \int\td\bq \gamma[\bm{\phi}](\br \bq,\br' \bq),
\end{align}
where $\gamma[\bm{\phi}](\bz,\bz') =\sum_{k=1}^{N} \phi_k^*(\bz') \phi_k(\bz)$ is the polaritonic 1RDM.
With these definitions, the minimization problem reads
\begin{alignat}{2}
\label{eq:numerics:hybrid:MinimizationProblem}
&\text{minimize }\, &&E[\bm{\phi}] \nonumber\\
&\text{subject to }\quad &&c_{ik}[\bm{\phi}] = 0 \\
& &&g_i[\bm{\phi}] \geq 0.		\nonumber
\end{alignat}

Because of the involved diagonalization, the functional dependence $g_i=g_i[\bm{\phi}]$ is only \emph{implicitly} known. Thus, the nonlinearity of the constraints $g_i$ is comparable to, e.g., the nonlinearity of the HF equations, which suggests a similar solution strategy, i.e., another SCF procedure. Additionally, we have to account for the inequality character of the new constraints, which, as we will see below, requires different methods than equality constraints. The polariton ansatz thus demands to nest both types of algorithms, which is nontrivial and requires considerable modifications of the standard algorithms. 

\subsubsection{Minimization under inequality constraints}
\label{sec:numerics:hybrid:inequality_challenge:general}
In order to solve problem~\eqref{eq:numerics:hybrid:MinimizationProblem}, it is possible to generalize the Lagrange-multiplier technique and derive a set of necessary conditions for a stationary point, which are known as the \emph{Karush-Kuhn-Tucker} (KKT) conditions.\footnote{For small summary of the original publications and later generalizations, the reader is referred to the ``Notes and References'' of \citep[part 12, p349f.]{Nocedal2006}} Additionally, we will assume that our functional has only one stationary point, which is a minimum. In this case, the KKT conditions are not only necessary but also sufficient to characterize this minimum. As usual in first-principles theory, it is very difficult to proof such an assumption and we can only hope that it is justified in most cases. We remark that in the tested scenarios, the code always converged to the same solution, independently of the starting point.

\subsubsection*{The KKT conditions}
To present the KKT conditions, we introduce the Lagrange multipliers $\bar{\bm{\epsilon}}=\bar{\epsilon}_{11},\bar{\epsilon}_{12},...,\bar{\epsilon}_{N,N}$ and assuming  the KKT multipliers $\bar{\bm{\nu}}=(\bar{\nu}_1,...,\bar{\nu}_{B_m})$ and define the Lagrangian
\begin{align}
	L[\bm{\phi};\bar{\bm{\epsilon}},\bar{\bm{\nu}}]=E[\bm{\phi}] - \sum_{ij=1}^{N} \bar{\epsilon}_{ij} c_{ij}[\bm{\phi}] - \sum_{i=1}^{B_m} \bar{\nu}_i g_i[\bm{\phi}].
\end{align}
The KKT theorem\footnote{See \citep[part 12.3]{Nocedal2006} for the mathematical details.} states then  that if $\bm{\phi}^s$ is a stationary point of the minimization problem~\eqref{eq:numerics:hybrid:MinimizationProblem}, there exist $(\bar{\bm{\epsilon}}^s, \bar{\bm{\nu}}^s)$ such that the following four sets of conditions hold (under some regularity conditions on $E$)
\begin{subequations}
\label{eq:numerics:KKT}
\begin{align}	
	\intertext{\hspace{1cm} 1. \emph{Stationarity}\footnote{Here, we regard $\phi_k$ and $\phi_k^*$ as independent as usual and thus have in principle a second set of equations for the derivative with respect to the $\phi_k$.}}
		\label{eq:numerics:KKT:stationarity}
		\frac{\partial}{\partial \phi_k^*} E[\bm{\phi}^s] &= \sum_{ij} \bar{\epsilon}^s_{ij} \frac{\partial}{\partial \phi_k^*}c_{ij}[\bm{\phi}^s] +\sum_i \bar{\nu}_i^s \frac{\partial}{\partial \phi_k^*}g_i[\bm{\phi}^s],
	\intertext{\hspace{1cm} 2. \emph{Primal feasibility}}
		g_i[\bm{\phi}^s]&\geq 0 \quad\forall i,\\
		h_{ij}[\bm{\phi}^s]&=0 \quad\forall i,j,
	\intertext{\hspace{1cm} 3. \emph{Dual feasibility}}
		\nu_i^s&\geq 0 \quad\forall i,
	\intertext{\hspace{1cm} 4. \emph{Complementary slackness}}
		\nu_i^s g_i [\bm{\phi}^s]&= 0 \quad\forall i.
\end{align}
\end{subequations}
Note that the KKT conditions reduce to the Lagrange conditions for equality constraints if all the $g_i=0$. 

Before we discuss solution strategies to determine the set $(\bm{\phi}^s, \bar{\bm{\epsilon}}^s, \bar{\bm{\nu}}^s)$ for a given problem in practice, let us have a closer look to the KKT conditions. From a practical point of view, stationarity is the most important KKT condition, because it provides us the basic equations that have to be solved in order to determine the minimum. Primal feasibility is the trivial KKT condition that is defined by the minimization problem itself and primal feasibility together with stationarity are the equivalent to the Lagrange conditions for equality constraints. The KKT conditions that are not present in Lagrange theory are dual feasibility and complementary slackness. The former is necessary to fix the sign of the term containing the inequality constraints (if $	\nu_i\leq 0$ were allowed, we could not differentiate $g_i\leq 0$ from $g_i\geq 0$). But the most important KKT condition that differentiates inequality from equality constraints is the latter. Complementary slackness states that \emph{either} $\nu_i^s=0$ \emph{or} $g_i (\bm{\phi}^s)=0$ at the solution point. This defines the role of the new KKT multipliers: if $\bm{\phi}^s$ is in the \emph{strictly feasible} region where $g_i>0$, there is no need for ``correction'' and thus the corresponding KKT multiplier $\nu_i^s=0$. Only if the solution is on the boundary of the feasible region, i.e., $g_i (\bm{\phi}^s)=0$ then we have a non-zero $\nu_i^s>0$. This is exactly the case, when the unconstrained minimization would have a solution $\bm{\phi}^s{}'$ that is \emph{infeasible}, i.e., when $g_i(\bm{\phi}^s{}')<0$. In this case, the KKT multiplier plays \emph{the same role} as the Lagrange multiplier and in this sense KKT theory is a straightforward generalization of Lagrange theory.

The major difference between dealing with inequality constraints in comparison to the equality constraints is thus that we do not need to determine as many KKT multipliers as there are constraints. For our specific minimization problem, most of the multipliers will be indeed zero. The reason is that the particle number is conserved, i.e., $\sum_{i=1}^{B_m}n_i=N$, and consequently we can maximally have $g_i=2-n_i=0$ for $N/2$ conditions (which corresponds to the non-interacting case, where the first $n_1=...=n_{N}=1$ and all other $n_{N+1}=...=n_{B_m}=0$). This is a very important fact, when we think about describing larger systems, where $B_m$ is big. We will never need to determine all $g_i$, which would require to determine all the $B_{m}$ eigenvalues of the electronic 1RDM, but it will be sufficient to calculate say the lowest $M$ eigenvalues, where $M \lessapprox N$. 

\subsubsection*{The basic approaches to solve nonlinear (inequality) constrained minimization problems}
However, we still need a method to determine the necessary KKT multipliers and the fact that we do not know a-priori the \emph{active set}, i.e., the $\bar{\nu_i}$ that are non-zero, crucially influences the corresponding algorithm. Nowadays on can find a plethora of such algorithms, that have been implemented and applied successfully. For an overview of the topic, we refer the reader to the book of \citet{Nocedal2006} that we also use as a main reference for this chapter. The minimization problem~\eqref{eq:numerics:hybrid:MinimizationProblem} considers a nonlinear energy functional and nonlinear constraints and thus belongs to the \emph{hardest class} of minimization problems. To solve such problems, there are basically three different classes of algorithms.
\begin{itemize}
	\item The \emph{sequential quadratic programming} approach:\\
	The idea of this approach is to constrain the minimization to a subspace such that a quadratic approximation of the energy functional becomes very accurate. The challenge is then to choose this subspace in a smart way taking the constraints into account.
	\item The \emph{barrier} or \emph{interior-point} methods:\\
	Here, one enforces the inequality constraints during the minimization by using a barrier function that depends on the \emph{barrier parameter} $\mu$. During the minimization, $\mu$ is reduced successively until the KKT conditions are fulfilled.
	\item The \emph{penalty} methods:\\
	Instead of employing a barrier that constrains the minimization to the feasible region and needs to be reduced for convergence, here violations of the constraint are \emph{penalized}. For convergence, the penalty is successively increased.
\end{itemize}
All of these have certain advantages and disadvantages, which are in principle well-studied. However, it is not easy to apply this knowledge to our specific problem. We therefore make the following pragmatic choices:
Since our aim is to extend the framework of a working electronic-structure code such as \textsc{Octopus}, we try to choose a method that is as less invasive as possible. Therefore, we exclude sequential quadratic programming approaches, which are \emph{conceptually very different} from the standard electronic-structure algorithms. For instance, it is not straightforward to combine active sets together with the line-minimization of the conjugate-gradients method by \citet{Payne1992}. Unfortunately, this means to exclude the class of methods that are considered to ``show their strength when solving  problems with significant nonlinearities in the constraints,'' \citep[part 18, p. 529]{Nocedal2006}.

The other two classes of methods can in principle be employed to extend standard electronic-structure methods. Conceptually, barrier and penalty methods are very similar, but in practice, the latter are easier to handle. Usual penalty functions are simpler to implement and to debug than typical barrier functions. Therefore, the final version of our here proposed algorithm employs the \emph{augmented Lagrangian method} that is based on a barrier function. Before we discuss the full algorithm in Sec.~\ref{sec:numerics:hybrid:pHF_algorithm}, we briefly outline this method in the following.


\subsubsection{The augmented Lagrangian method}
\label{sec:numerics:hybrid:inequality_challenge:penalty_augmented}
\begin{figure}
	\centering
	\includegraphics[width=0.5\columnwidth] {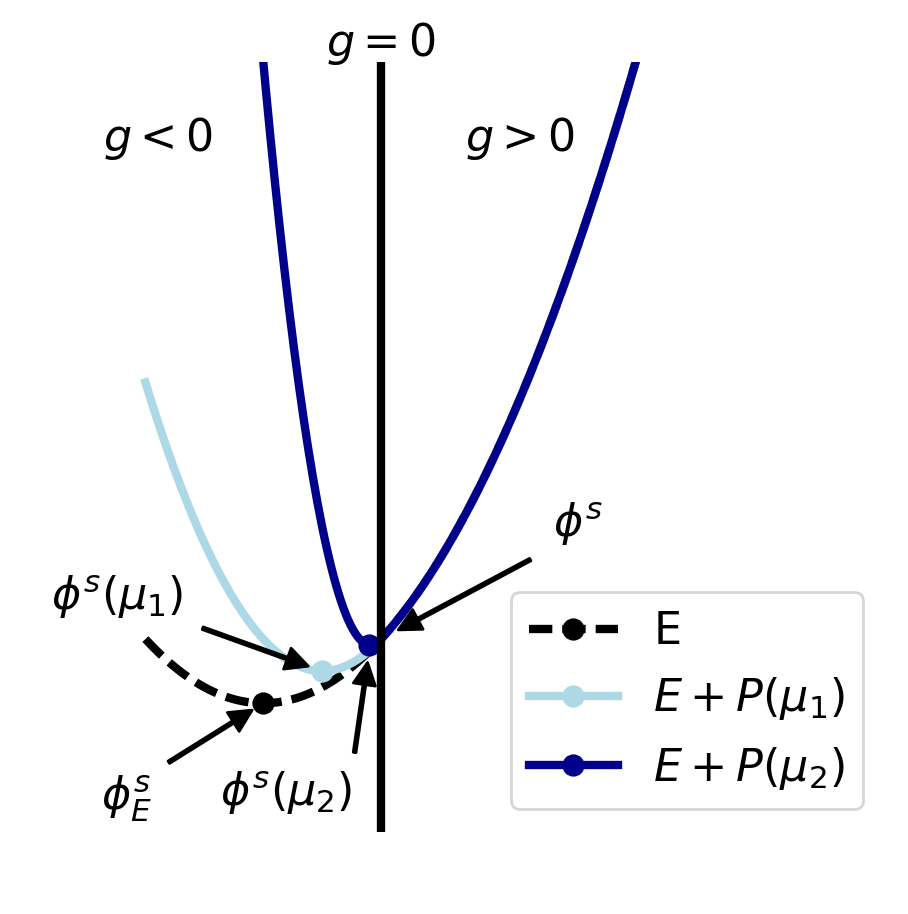}
	\caption{Illustration of the penalty method. The (unconstrained) function E (black dashed line) has its minimum $\phi^s_E$ in the \emph{infeasible region}, where $g<0$ (left of the black vertical line that marks $g=0$). The true minimum $\phi^s$ of the constrained problem is thus on the boundary, where $g=0$. Penalty methods construct an auxiliary minimization problem by adding a penalty function $P(\mu)$ to the energy (blue solid lines). For every value of the penalty parameter $\mu$, we can find the minimum $\phi^s(\mu)$ of the auxiliary problem, which approaches the true minimum $\phi^s$ for increasing $\mu$. This is illustrated for two example values $\mu_1$ (light-blue) and $\mu_2$ (dark-blue), where $\mu_2>\mu_1$.
	}
	\label{fig:numerics:penalty_illustration}
\end{figure}

The basic quantity of every penalty method is a function that \emph{penalizes} violations of the constraints. For simplicity, we discuss in the following only the case of inequality constraints,\footnote{Note that penalty functions can also be applied to equality constrained minimization problems.} 
i.e., we consider the minimization problem
\begin{alignat}{2}
\label{eq:numerics:hybrid:MinimizationProblemInequality}
&\text{minimize }\, &&E(\bm{\phi}) \nonumber \\
&\text{subject to }\quad &&g_i(\bm{\phi}) \geq 0.		
\end{alignat}
For that, we define the \emph{quadratic penalty function}
\begin{align}
\label{eq:numerics:hybrid:Penalty_Function}
P(\bm{\phi};\mu)=\frac{\mu}{2} \sum_{i}  ([g_i(\bm{\phi})]^-)^2,
\end{align}
where $\mu>0$ is the so-called \emph{penalty parameter} and $[\cdot]^-$ is defined as  $[x]^-=\max(-x,0)$ for any real number $x$. Therefore, $P$ penalizes constraint violations quadratically ($g_i<0$) and proportional to $\mu$, but has no effect, if $g_i\geq 0$. 
It can be shown that a minimization with this penalty function converges to the exact solution of the minimization problem~\eqref{eq:numerics:hybrid:MinimizationProblemInequality} for $\mu\rightarrow\infty$~\citep[Theorem 17.3]{Nocedal2006}.	The nonsmoothness of $P$ is a necessary price to pay for this property, but in practice, this is not a big issue. Most importantly, one can generalize the stationarity KKT conditions (Eq.~\eqref{eq:numerics:KKT:stationarity}) to this case by making use of directional derivatives.\footnote{For details about this, see \citep[part 17.2]{Nocedal2006}.} In a concrete algorithm, we consider the Lagrangian
\begin{align}
\label{eq:numerics:hybrid:lagrangian_penalty_simple}
L[\bm{\phi};\mu] = E[\bm{\phi}] + P[\bm{\phi};\mu],
\end{align}
and sequentially minimize $L[\bm{\phi};\mu^m]$ to obtain approximate minima $\bm{\phi}^m=\bm{\phi}^s(\mu^m)$ for increasing $\mu^{m}>\mu^{m-1}$ until convergence. We have illustrated this procedure in Fig.~\ref{fig:numerics:penalty_illustration}.


To calculate the derivative of $P$, we have to deal with its non-differentiability. For the contribution of condition $i$, we can calculate
\begin{align}
	\frac{\partial}{\partial \phi_k^*} \left[\tfrac{\mu}{2}([g_i[\bm{\phi}]]^-)^2\right]=\begin{cases}
		-\mu g_i[\bm{\phi}] \frac{\partial}{\partial \phi_k^*}g_i[\bm{\phi}] &g_i < 0\\
		0 & g_i > 0,
	\end{cases}
\end{align}
but for $g_i=0$, there is no well-defined derivative. As we have mentioned before, the formally correct way to define the stationarity condition~\eqref{eq:numerics:KKT:stationarity} for nonsmooth penalty functions is by means of the directional derivative. Sophisticated algorithms take care of this by ``smoothing procedures''~\citep[part 17.2]{Nocedal2006} that effectively remove the non-differentiable point. However, a simpler way to deal with this issue is to explicitly keep the step in the gradient and to \emph{define} 
\begin{align}
\label{eq:numerics:hybrid:penalty_function_nondifferential_point}
	\frac{\partial}{\partial \phi_k^*} \left[\tfrac{\mu}{2}([g_i(\bm{\phi})]^-)^2\right]_{g_i=0}=-\mu g_i[\bm{\phi}] \frac{\partial}{\partial \phi_k^*}g_i[\bm{\phi}],
\end{align}
where $-\mu g_i[\bm{\phi}]$ is assumed to remain finite (see below). We therefore have the following stationarity conditions
\begin{align}
\label{eq:numerics:hybrid:Penalty_stationarity}
\frac{\partial}{\partial \phi_k^*}E[\bm{\phi}]  =  \sum_i 
\Theta(-g_i[\bm{\phi}])\, \mu g_i[\bm{\phi}]\, \frac{\partial}{\partial \phi_k^*}g_i[\bm{\phi}],
\end{align}
where we employed the Heaviside step-function $\Theta(x)=x$ for $x\geq 0$ and $\Theta(x)=0$ for $x< 0$.

Algorithms based on Lagrangian~\eqref{eq:numerics:hybrid:lagrangian_penalty_simple} usually require very large barrier parameters $\mu$ to guarantee the equality constraints. 
To see this, we compare Eq.~\eqref{eq:numerics:hybrid:Penalty_stationarity} with the KKT conditions~\eqref{eq:numerics:KKT:stationarity}. The KKT multiplier corresponding to the inequality constraint $g_i$ reads for the penalty method (assuming
\begin{align}
\label{eq:numerics:hybrid:Penalty_KKTmultiplier}
	\bar{\nu}_i^p= -\mu g_i(\bm{\phi}).
\end{align}
When $g_i\rightarrow 0$, we need $\mu\rightarrow\infty$ to have a finite KKT multiplier $\bar{\nu}_i$. This can be problematic in practice, because $\mu$ is part of the gradient $\frac{\partial L}{\partial \phi_k^*}=\frac{\partial}{\partial \phi_k^*}E[\bm{\phi}]-\sum_i 
\Theta(-g_i[\bm{\phi}])\, \mu g_i[\bm{\phi}]\, \frac{\partial}{\partial \phi_k^*}g_i[\bm{\phi}] $ (and of the higher derivatives such as the Hessian). If $\mu$ is large, even small errors in $\frac{\partial L}{\partial \phi_k^*}$ are amplified strongly, which might prevent an algorithm that employs $\frac{\partial L}{\partial \phi_k^*}$  (or other derivatives) to determine the minimization steps from convergence. This well-known problem of penalty (and many other) methods is called \emph{ill conditioning}.\footnote{See p. 505f of \citep{Nocedal2006} for a more general discussion of ill conditioning.}


The standard way to reduce the ill conditioning of the quadratic penalty method is to  ``augment'' the Lagrangian by a further linear term. This \emph{augmented Lagrangian method} has considerably better convergence properties than the penalty method. The corresponding Lagrangian reads
\begin{align}
\label{eq:numerics:hybrid:Augmented_Lagrangian}
L[\bm{\phi};\bm{\nu},\mu]=E[\bm{\phi}] - \sum_{i} \nu_{i} g_{i} + \frac{\mu}{2} \sum_{i}  ([g_i]^-)^2,
\end{align}
where we introduced a new set of Lagrange-multipliers $\bm{\nu}=(\nu_1,...,\nu_{B_m})$. The corresponding stationarity conditions read
\begin{align}
\label{eq:numerics:hybrid:Augmented_Stationarity}
\frac{\partial}{\partial \phi_k^*} E[\bm{\phi}] &= \sum_i \left[\nu_i -\mu g_i[\bm{\phi}]\right] \frac{\partial}{\partial \phi_k^*}g_i[\bm{\phi}],
\end{align}
which we can again compare to Eq.~\eqref{eq:numerics:KKT:stationarity} to obtain an expression for the KKT multiplier
\begin{align}
\label{eq:numerics:hybrid:Augmented_KKTmultiplier}
	\bar{\nu}_i^a= \nu_i -\mu g_i[\bm{\phi}].
\end{align}
Assuming that the algorithm is in the $m$-th iteration step close to the solution point, we have
\begin{align*}
	g_i[\bm{\phi}^m]\approx - \frac{1}{\mu} (\bar{\nu}^a_i - \nu_i^m ).
\end{align*}
Thus, if $\nu_i^m \approx \bar{\nu}^a_i$, the constraint $g_i\approx0$ is (approximately) satisfied also for small $\mu$.
Instead, in the simple penalty method, we have close to the solution
\begin{align*}
g_i[\bm{\phi}^m]\approx - \frac{1}{\mu} \bar{\nu}^p_i,
\end{align*}
and accordingly need much larger $\mu$ to satisfy $g_i\approx0$. We therefore see how the problem of ill conditioning is considerably reduced by the linear term of the augmented Lagrangian method.
 
To guarantee $\nu_i^m \rightarrow \bar{\nu}^s_i$ in practice, Eq.~\eqref{eq:numerics:hybrid:Augmented_KKTmultiplier} provides us even with an update formula, i.e.,
\begin{align}
\label{eq:numerics:hybrid:Augmented_UpdateFormula}
	\nu_i^{k+1}= \nu_i^k -\mu^k g_i[\bm{\phi}^k].
\end{align}
The additional Lagrange-multipliers $\bm{\nu}$ are initialized to zero and updated to values $\nu_i>0$ according to Eq.~\eqref{eq:numerics:hybrid:Augmented_UpdateFormula} only if the minimization reaches the corresponding boundary of the feasible region where $g_i=0$. The $\nu_i$ hence play the same role as the usual Lagrange multipliers for equality constraints.
	\subsection{The polariton ansatz in practice}
\label{sec:numerics:hybrid:pHF_algorithm}
Having introduced the basic ingredients of the augmented Lagrangian method to guarantee inequality constraints in minimization algorithm, we present in this section a concrete algorithm to solve the minimization problem~\eqref{eq:numerics:hybrid:MinimizationProblem}. We have implemented this algorithm in a purpose-built code to produce the results of Ref.~\citep{Buchholz2020} that we have presented and discussed in Sec.~\ref{sec:dressed:results}. 
Specifically, we combine the augmented Lagrangian method with a simplified version of an algorithm of the \textsc{Lancelot} software package\footnote{See \url{http://www.numerical.rl.ac.uk/lancelot/blurb.html} or the corresponding book by the head developers \citet{Conn1992}.} following Ref.~\citep[algorithm 17.4]{Nocedal2006}.


\subsubsection{The basic algorithm}
\label{sec:numerics:hybrid:pHF_algorithm:basic}
We want to present now our newly developed algorithm to solve the minimization problem~\eqref{eq:numerics:hybrid:MinimizationProblem} to solve the polaritonic HF equations with the polariton ansatz (polariton-HF algorithm). We start by recapitulating the most important definitions and reformulating the inequality constraints~\eqref{eq:numerics:hybrid:ConstraintsInequalityGeneric} in a more suitable form. We then apply the augmented Lagrangian method introduced in the previous subsection to the specific problem. We derive all the necessary equations and present the basic idea of an algorithm to solve these.

\subsubsection*{Reformulating the minimization problem}
We describe the polaritonic orbitals in a basis-set with finite dimension $B=B_mB_{ph}$, where $B_m (B_{ph})$ is the size of the matter (photon) basis with corresponding index $i (\alpha)$. We assume spin-restriction (see Sec.~\ref{sec:est:general:general_case}) and thus consider for a system of $N$ electrons $N/2$ polaritonic orbitals $\bm{\phi}=(\phi_1,...,\phi_{N/2})$ that are represented as vectors in the given basis, i.e., $\phi_k=(\phi_k^{i\alpha})\in\mathrm{R}^{B_mB_{ph}}$. Consequently, $\bm{\phi}$ can be regarded as a super-vector (a vector of vectors) or equivalently as a matrix. The goal is to minimize the HF energy functional
\begin{align}
\tag{cf. \ref{eq:dressed:est:HF:energy}}
E[\bm{\phi}]=2 \sum_i \braket{\phi_i | (\hat{T}[t'] + \hat{V}[v']) \phi_i} + \sum_{i,k} \left[2 \braket{\phi_k | \hat{J}_i[w']\phi_k} - \braket{\phi_k | \hat{K}_i[w']\phi_k}\right]
\end{align}
where we used the definitions of the  ``dressed'' Coulomb-operator $\hat{J}_i$ which acts as $\hat{J}_i\phi_k(\br \bq) = \int\td\bz' \phi_i'{}^*(\bz')$ $w'(\bz;\bz')$ $\phi_i(\bz') \phi_k(\bz)$ (cf. Eq.~\eqref{eq:dressed:est:HF:CoulombOperator}), and the ``dressed'' exchange-operator $\hat{K}_i$ which acts as $\hat{K}_i\phi_k(\bz) = \int\td\bz' \phi_i(\bz)$ $w'(\bz;\bz')\phi_i{}^*(\bz') \phi_k(\bz')$ (cf. Eq.~\eqref{eq:dressed:est:HF:ExchangeOperator}). Here $t',v'w'$ are the dressed integral kernels, cf. Eqs.~\eqref{eq:dressedkinetic},\eqref{eq:dressedpotential} and \eqref{eq:dressedinteraction}. Using the definition of the polaritonic 1RDM 
\begin{align}
\gamma^{i\alpha,i'\alpha'}=\sum_{k=1}^{N/2} \phi_k^*{}^{i'\alpha'} \phi_k^{i\alpha},
\end{align}
we define
\begin{subequations}
	\label{eq:numerics:pHF_JK_gamma}
	\begin{align}
	\hat{J}[\gamma] \phi_k(\bz) \equiv\sum_{i=1}^{N/2} \hat{J}_i \phi_k(\bz) &= \sum_{i=1}^{N/2} \int\td\bz' \phi_i^*(\bz')w'(\bz;\bz') \phi_i(\bz') \phi_k(\bz) = \int\td\bz' w'(\bz;\bz') \gamma[\bm{\phi}](\bz',\bz') \phi_k(\bz)\\
	\hat{K}[\gamma] \phi_k(\bz)\equiv\sum_{i=1}^{N/2}\hat{K}_i \phi_k(\bz) &= \sum_{i=1}^{N/2}\int\td\bz' \phi_i(\bz)w'(\bz;\bz')\phi_i^*(\bz') \phi_k(\bz') = \int\td\bz' w'(\bz;\bz') \gamma[\bm{\phi}](\bz,\bz') \phi_k(\bz')
	\end{align}
\end{subequations}
Thus, the energy can be expressed as
\begin{align}
\label{eq:numerics:hybrid:pHF_energy}
E[\bm{\phi}]=2 \sum_i \braket{\phi_i | (\hat{T} + \hat{V}) \phi_i} + \sum_{k} \left[2 \braket{\phi_k | \hat{J}[\gamma]\phi_k} - \braket{\phi_k | \hat{K}[\gamma]\phi_k}\right],
\end{align}
where we shortened $\hat{T}\equiv\hat{T}[t']$ and $\hat{V}\equiv\hat{V}[v']$.
The equality constraints~\eqref{eq:numerics:hybrid:ConstraintsEqualityGeneric} guarantee the orthonormality of the polaritonic orbitals and read explicitly
\begin{align}
\label{eq:numerics:hybrid:constraint_equality}
c_{ik}[\bm{\phi}]=\braket{\phi_i|\phi_k} - \delta_{ik}=0 \quad k=1,...,N/2.
\end{align}
Because of the spin-restriction, the $M$ inequality constraints~\eqref{eq:numerics:hybrid:ConstraintsInequalityGeneric} read
\begin{align}
\label{eq:numerics:hybrid:constraint_inequality}
g_i(\bm{\phi})=2-n_i[\bm{\phi}]\geq 0 \quad i=1,...,M,
\end{align}
where the $n_i$ are the (electronic) natural occupation numbers. The number $M\leq B_m$ is smaller or equal to the size of the electronic basis $B_m$ and depends on the specific problem (see Sec.\ref{sec:numerics:hybrid:inequality_challenge}).
To determine the $n_i$ and therefore the $g_i$ from a given a set of states $\bm{\phi}$, we first calculate the polaritonic 1RDM 
\begin{align}
\label{eq:numerics:hybrid:gamma_definition}
\gamma^{i\alpha,i'\alpha'}=\sum_{k=1}^{N/2} \phi_k^*{}^{i'\alpha'} \phi_k^{i\alpha}.
\end{align}
From $\gamma$, we obtain the electronic 1RDM (cf. Eq.~\eqref{eq:numerics:gamma_electronic_definition})
\begin{align}
\label{eq:numerics:hybrid:gamma_electronic_definition}
\gamma_e^{i,i'}=\sum_{\alpha=1}^{B_{ph}} \gamma^{i\alpha,i'\alpha}= \sum_{\alpha=1}^{B_{ph}} \phi_k^*{}^{i'\alpha} \phi_k^{i\alpha}
\end{align}
by a contraction over the photon index, which means that we set $\alpha=\alpha'$ and perform the sum $\sum_{\alpha=1}^{B_{ph}}$. Then, we have to diagonalize $\gamma_e$ by solving the system of equations
\begin{align*}
\sum_{i'=1}^{B_{ph}}\gamma_e^{i,i'}\psi_j^{e,i'} = n_j \psi_j^{e,i}
\end{align*}
for the natural occupation numbers $n_j$ and the natural orbitals $\psi^e_j$. Diagonalizing a matrix can be performed numerically very efficiently, but there is no explicit formula that relates $\gamma_e$ to its eigenvectors, i.e., the $n_i$ are \emph{implicit} functions of $\bm{\phi}$.\footnote{Note that this is very similar to nonlinearity of the HF equations themselves, which we have to solve by a self-consistent diagonalization (see the discussion in Sec.~\ref{sec:numerics:rdmft:piris} and below).}

To disentangle the diagonalization step of $\gamma_e$ from the rest of the rest of the algorithm, we consider natural orbitals and dressed orbitals as \emph{independent} variables of the minimization and enforce their connection as an additional constraint.\footnote{This is similar to considering $\phi$ and $\phi^*$ as independent.} 
We collect the natural orbitals in the vector $\bm{\psi}^e=(\psi_1^e,...,\psi_{B_m}^e)$ and define (cf. Eq.~\eqref{eq:gamma_electronic_eigenrep})
\begin{align}
\label{eq:numerics:gamma_electronic_evp}
n_i[\bm{\phi}, \bm{\psi}^e] \psi_i^{e} = \hat{\gamma}_{e}[\bm{\phi}] \psi_{i}^{e},
\end{align}
where
\begin{align}
\label{eq:numerics:pHF_def_ni}
n_i[\bm{\phi}, \bm{\psi}^e ] = \braket{\psi^e_i|\hat{\gamma_e}\psi^e_i}
\end{align}
The inequality constraints~\eqref{eq:numerics:hybrid:constraint_inequality} therefore become explicit functionals of $\bm{\phi}$ and $\bm{\psi}^e$, i.e.,
\begin{align}
\label{eq:numerics:constraint_inequality_byNO}
	g_i=g_i[\bm{\phi}, \bm{\psi}^e ]=2 - n_i[\bm{\phi}, \bm{\psi}^e ].
\end{align}
Since the $\bm{\psi}^e$ are independent variables now, we also have to enforce their orthonormality by a third set of conditions
\begin{align}
\label{eq:numerics:constraints_NO}
\bar{f}_{ij}=\braket{\psi^{e}_i|\psi^{e}_j} -\delta_{ij}=0.
\end{align} 
Importantly, this construction allows us to consider either $\bm{\phi}$ or $\bm{\psi}^e$ as constant, while optimizing the other and connect both in a self-consistent field procedure.

We thus have transformed the original minimization problem~\eqref{eq:MinimizationProblem} with the implicit inequality constraints \eqref{eq:numerics:hybrid:constraint_inequality} into the new minimization problem 
\begin{alignat}{2}
\label{eq:numerics:pHF:MinimizationProblemAugmented}
&\text{minimize }\, &&E[\bm{\phi}] \nonumber \\
&\text{subject to }\quad &&c_{ij}[\bm{\phi}] = 0 \nonumber \\
&									&&g_i[\bm{\phi},\bm{\psi}^e] \geq 0 \nonumber \\
&									&&f_{ij}[\bm{\psi}^e] = 0,
\end{alignat}
with the explicit inequality constraints~\eqref{eq:numerics:constraint_inequality_byNO}.

\subsubsection*{Enforcing the inequality constraints: the augmented Lagrangian method} 
To enforce the inequality constraints, we consider an \emph{augmented Lagrangian} algorithm, which we have introduced and motivated in the previous section (especially Sec.~\ref{sec:numerics:hybrid:inequality_challenge:penalty_augmented}).
The augmented Lagrangian method extends a given Lagrangian with extra penalty terms, which allows us to do a similar ``trick'' as we have shown before for the RDMFT generalization (see Sec.~\ref{sec:numerics:rdmft:cg}): we use the conjugate-gradients method by \citet{Payne1992}, but exchange the Lagrangian. However, we will see in the following that the inequality constraints require a considerably stronger modification of the algorithm (and several further steps) than the generalization to RDMFT, where we only had to consistently exchange the diagonal elements with the full Lagrange-multiplier matrix.

Consequently, instead of minimizing $E$ directly, we consider the Lagrangian 
\begin{align}
\label{eq:numerics:pHF_lagrangian}
	L[\bm{\phi}, \bm{\psi}^e]=&	E[\bm{\phi}] + \mathcal{C}[\bm{\phi}] +\mathcal{G}[\bm{\phi}, \bm{\psi}^e ],
\end{align}
where 
\begin{align}
\label{eq:numerics:Lagrange_C}
\mathcal{C}[\bm{\phi}]=-\sum_{ij} \bar{\epsilon}_{ij} c_{ij}[\bm{\phi}]
\end{align}
is a standard Lagrange term to enforce the equality constraints \eqref{eq:numerics:hybrid:constraint_equality} that introduces the Lagrange multipliers $\bar{\epsilon}_{ij}$.
We used here the notation according to the translation rules depicted in Fig.~\ref{fig:polariton_construction}.
The other term $\mathcal{G}$ includes all extra terms that are related to the inequality constraints. For our specific construction, these are on the one hand a standard Lagrange term, $-\sum_{ij}\bar{\theta}_{ij} f_{ij}$, to guarantee the constraints \eqref{eq:numerics:constraints_NO} and on the other hand, the two terms, $-\sum_i \nu_i g_i + \mu / 2 \sum_i ([g_i]^-)^2$, that are introduced by the augmented Lagrangian method (see Sec.~\ref{sec:numerics:hybrid:inequality_challenge:penalty_augmented}). The $\bar{\theta}_{ij}$ and $\nu_i$ are Lagrange multipliers, $\mu$ is the penalty parameter and $[y]^-=\max(-y, 0)$.
Collecting all the three terms, we have 
\begin{align}
\label{eq:numerics:constraints_inequality_augmented_lagrangian}
\mathcal{G}[\bm{\phi}, \bm{\psi}^e ]= - \sum_i \nu_i g_i[\bm{\phi}, \bm{\psi}^e ] + \frac{\mu}{2} \sum_i ([g_i]^-[\bm{\phi}, \bm{\psi}^e ])^2 - \sum_{ij} \bar{\theta}_{ij} f_{ij} [\bm{\psi}^e],
\end{align}
which completes the definition of the Lagrangian~\eqref{eq:numerics:pHF_lagrangian}.

To derive the according stationarity conditions, we need to evaluate functional derivatives of the form
\begin{align*}
\frac{\partial}{\partial \phi_k^*}g_i(\bm{\phi})=\frac{\partial}{\partial \phi_k^*} (2-n_i) =-\frac{\partial}{\partial \phi_k^*}n_i,
\end{align*}
which is nontrivial to compute. Using operator perturbation theory, one can derive
\begin{align}
\label{eq:numerics:hybrid:derivative_inequality}
\frac{\partial}{\partial \phi_k^*}n_i= 2 \sum_{j'=1}^{B_m} \psi_i^{e,j'}{}^* \phi_k^{j',\alpha} \psi_i^{e,j},
\end{align}
which can be seen as a projection of only the electronic part of $\phi_k$ on the to $n_i$ corresponding natural orbital $\psi_i^{e}$. The derivation of Eq.~\eqref{eq:numerics:hybrid:derivative_inequality} is shown in appendix~\ref{sec:app:gradient_non} for the general case of real-space orbitals. For convenience, we introduce the operator $\hat{G}_i$ that acts as
\begin{align}
\label{eq:numerics:hybrid:definition_Goperator}
\hat{G}_i \phi_k = 2 \sum_{j'=1}^{B_m} \psi_i^{e,j'}{}^* \phi_k^{j',\alpha} \psi_i^{e,j}
\end{align}
on a given polaritonic orbital $\phi_k$ and thus $\frac{\partial}{\partial \phi_k^*}g_i=-\hat{G}_i \phi_k$.

The complete stationarity conditions (cf. Eq.~\eqref{eq:dressed:HF:stationary_conditions},  Eq.~\eqref{eq:numerics:KKT:stationarity}) read
\begin{subequations}
	\label{eq:numerics:pHF_gradients}
	\begin{align}
	\label{eq:numerics:pHF_gradientPhi}
	0=&\frac{\partial}{\partial \phi_k^*} L= \hat{H}^1\phi_k - \sum_j\bar{\epsilon}_{kj} \phi_j + \sum_i \left[ \nu_i -\mu [g_i]^-\right]  \hat{G}_i\phi_k\\
	\label{eq:numerics:pHF_gradientPsiGamma}
	0=&\frac{\partial}{\partial \psi^{e}_i{}^*}L = (\mu [g_i]^- -\nu_i ) \hat{\gamma}_e\psi^{e}_i - \sum_j \bar{\theta}_{ij} \psi^{e}_j,
	\end{align}
\end{subequations}
where we considered $\phi_k,\phi_k{}^*,\psi^e_i,\psi^e_i{}^*$ as independent variables and employed the definition of the Fock operator (cf. Eq.~\eqref{eq:dressed:est:HF:Fockmatrix})\footnote{The several factors of 2 that appear in this expression are the occupation numbers of the polaritonic orbitals, which arise from the spin-summation in restricted setting. Note that these numbers also occur in electronic HF. However, they are usually neglected, because they enter every term of the HF equations linearly. Here, this is not possible anymore because of the nonlinear penalty function.}
\begin{align}
\label{eq:numerics:hybrid:pHF_FockMatrix}
\hat{H}^1[\gamma]= 2 (\hat{T}+\hat{V}) + 2 \hat{J}[\gamma] - \hat{K}[\gamma].
\end{align}
Looking at Eq.~\eqref{eq:numerics:pHF_gradientPhi}, we observe a structural similarity to the stationarity conditions for electronic single-reference methods. We have a Lagrange-multiplier matrix $\bar{\epsilon}_{kj}$ and nonlinear operators that depend on the orbitals $\bm{\Phi}$ of a Slater determinant. Since the one-body Hamiltonian $\hat{H}^1$ and the new operators $\hat{G}_i$ are all hermitian, we can also here diagonalize the Lagrange-multiplier matrix $\bar{\epsilon}_{ij}=\delta_{ij}\epsilon_j$ and bring Eq.~\eqref{eq:numerics:pHF_gradientPhi} into the form of an eigenvalue equation
\begin{subequations}
\begin{align}
\label{eq:numerics:pHF_gradientPhi_ev}
	\epsilon_k \phi_k=\hat{H}^1\phi_k + \sum_i \left[ \lambda_i -\mu [g_i]^-\right]  \hat{G}_i\phi_k.
\intertext{The same is possible for the second gradient equation, cf. Eq.~\eqref{eq:numerics:pHF_gradientPsiGamma}, because also the electronic 1RDM $\gamma_e$ is hermitian. We choose $\bar{\theta}_{ij}=\delta_{ij}\theta_j$ and rewrite}
\label{eq:numerics:pHF_gradientPsiGamma_ev}
	\theta_i\psi^{e}_i = (\mu [g_i]^- -\lambda_i ) \hat{\gamma}_e\psi^{e}_i.
\end{align}
\end{subequations}
This equation is in principle nontrivial to solve. However, as the $\bm{\phi}$ and $\bm{\psi}^e$ are treated as independent variables, we can \emph{equivalently} solve the eigenvalue problem for $\gamma_e$ (Eq.~\eqref{eq:numerics:gamma_electronic_evp}) an then simply replace $\theta_i=n_i^e(\mu [g_i]^- -\lambda_i)$. With these definitions, we are able to perform HF calculations with the polariton ansatz by numerically solving the Eqs.~\eqref{eq:numerics:pHF_gradientPhi_ev} and \eqref{eq:numerics:pHF_gradientPsiGamma_ev} with the expressions \eqref{eq:numerics:hybrid:pHF_energy} and \eqref{eq:numerics:hybrid:pHF_FockMatrix}. 

\subsubsection*{The basic polariton-HF algorithm}
We split the full algorithm in the following two principal parts that we discuss separately in the following.
\begin{enumerate}
	\item The inner part that (approximately) minimizes the subproblem:
		\begin{align*}
			L^{(\mu_l, \bm{\nu}_l)}[\bm{\phi}, \bm{\psi}^e; \bm{\epsilon},\bm{\theta}]=L[\bm{\phi}, \bm{\psi}^e; \bm{\epsilon},\bm{\theta}; \mu=\mu_l, \bm{\nu}=\bm{\nu}_l]
		\end{align*}
		for fixed penalty parameters $(\mu_l, \bm{\nu}_l)$ that are updated by the outer part. This part can be solved in principle by any minimization method and we will employ a modified version of the conjugate-gradients algorithm by \citet{Payne1992}. We call this \emph{penalty-corrected conjugate-gradients} (PCG) method.
	\item The outer part that constitutes the actual augmented Lagrangian method. Here we iteratively update $(\mu_l, \bm{\nu}_l)\rightarrow(\mu_{l+1}, \bm{\nu}_{l+1})$ until the approximate solution $(\bm{\phi}^{(\mu_l, \bm{\nu}_l)},\bm{\psi}^e{}^{(\mu_l, \bm{\nu}_l)})$ of the subproblem is sufficiently feasible, i.e., all constraints are
	\begin{align*}
		g_{i}^{(\mu_l, \bm{\nu}_l)}=g_{i}[\bm{\phi}^{(\mu_l, \bm{\nu}_l)},\bm{\psi}^e{}^{(\mu_l, \bm{\nu}_l)}]\gtrapprox 0.
	\end{align*}
	We employ here a simplified version of an algorithm of the \textsc{Lancelot} software package.\footnote{See \url{http://www.numerical.rl.ac.uk/lancelot/blurb.html} or the corresponding book by \citet{Conn1992}.} following Ref.~\citep[algorithm 17.4]{Nocedal2006}
\end{enumerate}

\subsubsection{The outer part: the augmented Lagrangian method}
\label{sec:numerics:hybrid:pHF_algorithm:outer}
We start with the second part of the algorithm, i.e., the augmented Lagrangian method that constitutes the ``penalty loop'' with iteration index $l$. This outer algorithm has two tasks. First, it needs to provide a set a of rules to update
\begin{align*}
(\mu_l,\bm{\nu}_l)\rightarrow(\mu_{l+1},\bm{\nu}_{l+1}),
\end{align*}
and second, it needs to control the convergence threshold $\epsilon_{PCG}$ of the inner part of the algorithm. As long as we are far away from the overall solution, there is no point in finding the minimum of $ L^{(\mu_l, \bm{\nu}_l)}[\bm{\phi}, \bm{\psi}^e; \bm{\epsilon},\bm{\theta}]$ with a high precision. On the contrary, a too small $\epsilon_{PCG}$ far away from the solution can even prevent the code from convergence, especially due to the strong nonlinear character of $L$. Thus, the penalty loop should increase or decrease $\epsilon_{PCG}$, depending on the constraint functions $g_i^{(\mu_l, \bm{\nu}_l)}$ and the gradient of $ L^{(\mu_l, \bm{\nu}_l)}$.

Let us first discuss the updates of the penalty parameters. These are determined by the ``measure of feasibility,'' i.e., the value of the constraint functions
\begin{align*}
g_{i}^l=g_i[\bm{\phi}^{(\mu_l, \bm{\nu}_l)},\bm{\psi}^e{}^{(\mu_l, \bm{\nu}_l)}]
\end{align*}
for the current approximate set of orbitals. If we have not found the overall solution, where $\delta L =0$, there must be at least one function $g_{i}^l<0$ that violates the constraint. Thus, $g_i^l$ indicates how far we are away from the feasible region. If the violation is weak, i.e., if $-g_{i}^l<\epsilon_{g}^l$ is smaller than a threshold $\epsilon_{g}^l>0$ that also is updated during the minimization process, we are ``sufficiently close'' to the feasible region. The current value of $\mu_l$ is producing an acceptable level of constraint violation and we update $\mu_l\rightarrow\alpha_{\mu}\mu_l$ by a factor $\alpha_{\mu}\gtrsim 1$ equal or slightly bigger than one. In our implementation, we set $\alpha_{\mu}=1.5$, which reduces the number of iterations until convergence in comparison to the most conservative choice $\alpha_{\mu}=1$. At this point, the ``augmented'' Lagrange multiplier $\nu_i$ comes into play, which we update according to the rule (cf. \eqref{eq:numerics:hybrid:Augmented_UpdateFormula} and \citep[Eq. 17.39]{Nocedal2006})
\begin{align}
\label{eq:numerics:pHF_penalty_lm_update_formula}
\nu_{i,l+1}=\nu_{i,l}-\mu_lg_i^{l}.
\end{align}
As discussed in Sec.~\ref{sec:numerics:hybrid:inequality_challenge:penalty_augmented}, this update formula is not chosen arbitrarily but is prescribed by augmented Lagrangian formalism. It is a crucial ingredient for the convergence of the method (see also \citep[part 17.3]{Nocedal2006}).
If instead $-g_{i}^l\geq\epsilon_{g}^l$, the minimization has not been ``enough'' penalized and thus, $\mu_l\rightarrow\beta_{\mu}\mu_l$ is raised by a factor $\beta_{\mu}>1$. We choose a conservative value of $\beta_{\mu}=20$, but very stable codes can even afford values until $\beta_{\mu}=100$ to speed up the convergence~\citep[part 17.3]{Nocedal2006}. If $\mu$ is updated, we should not update the Lagrange-multipliers $\bm{\nu}_l\rightarrow\bm{\nu}_l$, because $L$ is nonlinear and the update formula~\eqref{eq:numerics:pHF_penalty_lm_update_formula} is only valid close to the solution. We would risk to increase $\bm{\nu}_l$ too much. 

Now, we still need to define the threshold $\epsilon_{g}^l$ that determines whether $\mu$ or $\bm{\nu}$ is updated. 
Clearly, the value of $\epsilon_{g}^l$ should depend on how close we are to the solution and one can show that a good measure for that is the current value of $g_i^l$ and $\mu_l$~\citep[Theorem 17.5 and 17.6]{Nocedal2006}. 
We employ the update formulas of \citep[Algorithm 17.4]{Nocedal2006}, which have a form that can be derived analytically\footnote{This derivation is presented in  \citep[part 14.4]{Conn2000}.} but involve certain free parameters, which have been determined empirically. Additionally, \citep[Algorithm 17.4]{Nocedal2006} provides us with an update rule for the convergence criterion $\epsilon_{PCG}$ of the inner part of the algorithm. In total, we have to following update formulas
\begin{align}
\label{eq:numerics:pHF_update_penalty_paramters}
\begin{split}
&\text{if } g_i^l \leq \epsilon_{g}^l \\
&\quad \begin{cases}
\nu_{i,l+1}=\nu_{i,l}-\mu_lg_i^l\\
\mu_{l+1}=\alpha_{\mu}\mu_l\\
\epsilon_{g}^{l+1}= (1/\mu^{l+1})^{0.9}\\
\epsilon_{PCG}^{l+1}= \epsilon_{PCG}^{l}/\mu^{l+1}
\end{cases}\\
&\text{else}\\
&\quad \begin{cases}
\nu_i^{l+1}=\nu_i^{l}\\
\mu^{l+1}=\beta_{\mu}\mu^{l}\\
\epsilon_{g}^{l+1}= (1/\mu^{l+1})^{0.1}\\
\epsilon_{PCG}^{l+1}= \epsilon_{PCG}^{l}/\mu^{l+1}
\end{cases}
\end{split}
\end{align}
The above prescription is completed by the initial values for penalty parameter, for which we chose $\mu_0=10$. The Lagrange multipliers have to be initialized $\bm{\nu}$ to zero, because they are strictly positive and can only grow due to the update formula~\eqref{eq:numerics:pHF_penalty_lm_update_formula}. Thus, if we choose them too big in the beginning, the algorithm cannot converge. 

The last missing piece of the polaritonic HF algorithm is the overall convergence criterion $\epsilon_{total}$, which must be set by the user. 
Since we aim to find the stationary point of $L$, the eigenvalue equations~\eqref{eq:numerics:pHF_gradientPhi_ev},\eqref{eq:numerics:pHF_gradientPsiGamma_ev} provide the natural convergence criterion. However, Eq.~\eqref{eq:numerics:pHF_gradientPsiGamma_ev} is much simpler than Eq.~\eqref{eq:numerics:pHF_gradientPhi_ev} and therefore it is sufficient to test only Eq.~\eqref{eq:numerics:pHF_gradientPhi_ev} (see below). For that, we define the residue sum\footnote{The quantity  $\sqrt{}\braket{\phi_k | \frac{\partial}{\partial \phi_k^*}L}$ is called the residue of the $k$-th gradient $\nabla_{\phi_k^*}L$.}
\begin{align}
\label{eq:numerics:hybrid:residue_sum}
R^l= \sum_{k=1}^{N/2} \sqrt{\braketm*{\phi_k^{(\mu_l,\bm{\nu}_l)} }
	{\frac{\partial}{\partial \phi_k^*}L |_{\phi_k=\phi_k^{(\mu_l,\bm{\nu}_l)}}}}
\end{align}
which should go to zero, when we approach the overall solution. However, it may happen that $R^l$ is very small although the code has not yet converged.\footnote{This situation occurs for instance, when we are close to the unconstrained minimum of $E$ in the \emph{infeasible} region, i.e., the solution of the polaritonic-HF minimization under the fermion approximation.} Thus, we consider a second convergence test based on the value of the Lagrangian function $L^l= L^{(\mu_l,\bm{\nu}_l)}$ that we (approximately) calculate in the PCG algorithm (see below). We simply compare the differences
\begin{align}
\label{eq:numerics:hybrid:Lagrangian_difference}
\Delta L^l= |L^{(\mu_{l-1},\bm{\nu}_{l-1})}-L^{(\mu_l,\bm{\nu}_l)}|
\end{align}
between subsequent iterations.
The overall convergence test reads
\begin{align}
R^l  <& \epsilon_{total,R} \\
\Delta L^l <& \epsilon_{total_L},
\end{align}
where we usually simply set $\epsilon_{total}=\epsilon_{total,R}+\epsilon_{total,L}.$

\begin{center}
	\begin{fminipage}{0.99\textwidth}
		\begin{algorithm}\label{algorithm:pHF} (augmented Lagrangian)  \newline
			Set the overall convergence criterion $\epsilon_{total}>0$.\\
			Initialize $\bm{\phi}^{(\mu_0,\bm{\nu}_0)}$.\\
			Initialize the penalty parameters $\mu^0=10,\bm{\nu}^0=\bm{0}$.\\
			Initialize  $\epsilon_{g}^0,\epsilon^0_{PCG}$ according to \eqref{eq:numerics:pHF_update_penalty_paramters}.\\
			(Approximately) calculate and save $L^{l}$ for the overall convergence test.\\
			\textbf{for} $l=0,1,2,...$ (penalty loop)
			\begin{adjustwidth}{0.5cm}{}
%
					\textbf{Inner part:} (PCG algorithm)\\
					Solve $\frac{\partial}{\partial \phi_k^*} L < \epsilon_{PCG}^l$ and $\frac{\partial}{\partial \psi^e_i{}^*} L < \epsilon_{PCG}^l$ to obtain $\bm{\phi}^{(\mu_{l+1},\bm{\nu}_{l+1})}, \bm{\psi}^{e,(\mu_{l+1},\bm{\nu}_{l+1})}$
			\end{adjustwidth}
			Then,
				\begin{enumerate}[label=(\Alph*)]
					\item Calculate $g_i=2-n_i$.
					\item Update  $\mu_{l+1},\bm{\nu}_{l+1},\epsilon_{g}^{l+1},\epsilon_{PCG}^{l+1}$ according to \eqref{eq:numerics:pHF_update_penalty_paramters}.
					\item Calculate $L^{l+1}$ (see PCG algorithm) and the residue sum $R^{l+1}$ according to Eq.~\eqref{eq:numerics:hybrid:residue_sum}
					\item[] \textbf{break} if $\max(\Delta L^{l+1},R^{l+1})<\epsilon_{total}$
				\end{enumerate}
			\textbf{end for} (penalty loop)
		\end{algorithm}
	\end{fminipage}
\end{center}

\subsubsection{The inner part: orbital optimization with the penalty-corrected conjugate-gradients algorithm}
\label{sec:numerics:hybrid:pHF_algorithm:inner}
Let us now discuss the inner part of the pHF-algorithm, i.e. the PCG algorithm. The goal is to approximately solve the two intermediate coupled eigenvalue problems
\begin{subequations}
\begin{align}
\label{eq:numerics:pHF_gradientPhi_sub}
	\epsilon_k \phi_k\approx&\hat{H}^1[\gamma[\bm{\phi}]]\phi_k + \sum_i \left[ \nu_{l,i} -\mu_l [g_i[\bm{\phi},\psi^e_i]]^-\right]  \hat{G}_i[\psi^e_i]\phi_k,\\
\label{eq:numerics:pHF_gradientPsiGamma_sub}
	\theta_i\psi^{e}_i \approx& (\mu_l [g_i[\bm{\phi},\psi^e_i]]^- -\nu_{l,i} ) \hat{\gamma}_e\psi^{e}_i,
\end{align}
\end{subequations}
up to the threshold $\epsilon_{PCG}$ (we show below how $\epsilon_{PCG}$ enters in the algorithm).
We have explicitly highlighted the functional dependence of the occurring operators:
\begin{itemize}
	\item $\hat{H}^1=\hat{H}^1[\gamma]$ depends only on the polaritonic 1RDM $\gamma$ (Eq.~\eqref{eq:numerics:hybrid:pHF_FockMatrix},
	\item $\hat{G}_i=\hat{G}_i[\psi^e_i]$ depends on the natural orbitals $\psi^e_i$ (Eq.~\eqref{eq:numerics:hybrid:definition_Goperator}) and
	\item $g_i=g_i[\bm{\phi},\psi^e_i]=2-n_i[\bm{\phi},\psi^e_i]$ depends on both $\bm{\phi}$ and $\psi^e_i$ (Eq.~\eqref{eq:numerics:pHF_def_ni}) and thus couples the two eigenvalue problems~\eqref{eq:numerics:pHF_gradientPhi_sub} and~\eqref{eq:numerics:pHF_gradientPsiGamma_sub}.
\end{itemize}
These intricate dependencies suggest to split the minimization in three parts that define the following three nested loops with according indices $m_1,m_2,m_3$ and convergence thresholds $\epsilon_{\phi},\epsilon_{\psi^e},\epsilon_{PCG}$.
\begin{enumerate}
	\item DO loop, $m_1,\epsilon_{\phi}$:\\
	In the innermost loop  Eq.~\ref{eq:numerics:pHF_gradientPhi_sub} is (approximately) solved for fixed $\bm{\psi^e}$ \emph{and} fixed $\hat{H}^1[\gamma]$ to update the dressed orbitals $\bm{\phi}^{m_1}\rightarrow\bm{\phi}^{m_1+1}$.
	\item NO loop, $m_2$:\\
	The next loop (approximately) solves Eqs.~\eqref{eq:numerics:pHF_gradientPsiGamma_sub} for fixed $\hat{H}^1[\gamma]$ and $\bm{\phi}$ to update the natural orbitals $\bm{\psi^{e,m_2}} \rightarrow \bm{\psi^{e,m_2+1}}$.
	\item PCG loop, $m_3$:\\
	In the utmost loop of the PCG algorithm, $\hat{H}^1[\gamma^{m_3}]\rightarrow \hat{H}^1[\gamma^{m_3+1}]$ is updated.
\end{enumerate}

In principle, every quantity that is updated depends on 3 indices, e.g., $\bm{\phi}=\bm{\phi}^{m_1,m_2,m_3}$. To simplify the notation, we keep within one loop only explicitly track of the one corresponding index.
For example, we denote the polaritonic orbitals for the orbital optimization by $\phi_k^{m_1}$. Once, we have completed the orbital optimization, we can update the set $\phi^{m_2}\rightarrow\phi^{m_2+1}$ in the next higher loop, which is the NO loop in this case. Once the NO loop has converged, we can update the set $\phi^{m_3}\rightarrow\phi^{m_3+1}$ of the utmost loop of the inner minimization, the SCF loop. Once the SCF loop is converged, we can update the penalty parameters $(\mu_l, \bm{\nu}_l)\rightarrow (\mu_{l+1}, \bm{\nu}_{l+1})$ in the second part of the algorithm. Correspondingly, we have to update the polaritonic orbitals in this loop, i.e., $\bm{\phi}^{(\mu_l, \bm{\nu}_l)}\rightarrow \bm{\phi}^{(\mu_{l+1}, \bm{\nu}_{l+1})}$.

\subsubsection*{The dressed-orbital optimization (DO loop)}
Instead of solving Eq.~\ref{eq:numerics:pHF_gradientPhi_ev} directly, we employ a modified version of the conjugate-gradients algorithm of \citet{Payne1992} to minimize the Lagrangian $L[\bm{\phi}]$ with fixed $\bm{\psi}^e$. To ease reading, we denote the approximate orbitals during the orbital optimization only by $\bm{\phi}^{m_1}=(\phi_1^{m_1},...,\phi_{N/2}^{m_1})$, where $m_1$ denotes the iteration step. We reserve the upper index $(\mu_l, \bm{\nu}_l)$ only for the solutions of the complete inner minimization. The steepest-descend vector for orbital $\phi_k^{m_1}$ is given by the negative gradient (cf. Eq.~\eqref{eq:numerics:StDesVector})
\begin{align}
\label{eq:numerics:pHF_steepest_descend_vector}
\zeta_k^{m_1} =-\frac{\partial}{\partial \phi_k^*} L^{(\mu_l, \bm{\nu}_l)}= -\left(\hat{H}^1 + \sum_i \left[ \nu_{l,i} -\mu_l [g_i]^-\right]  \hat{G}_i  - \epsilon_{k}^{m_1} \right) \phi_k^{m_1},
\end{align}
where the Lagrange multiplier $\epsilon_k^{m_1}$ is estimated in every step by projecting the eigenvalue equation \eqref{eq:numerics:pHF_gradientPhi_sub} on $\phi_k^{m_1}$, i.e.,
\begin{align}
\label{eq:numerics:hybrid:formula_epsilon}
\epsilon_{k}^{m_1} = \braket{\phi_k^{m_1}|\hat{H}^1\phi_k^{m_1}} + \sum_i \left[ \nu_{l,i} -\mu_l [g_i]^-\right]  \braket{\phi_k^{m_1}|\hat{G}_i\phi_k^{m_1}}.
\end{align}
Until this point, we have generalized the conjugate gradients algorithm in a very similar way as we have shown for RDMFT in Sec.~\ref{sec:numerics:rdmft:cg}. We changed the basic Lagrangian in comparison to the originally considered one in Ref.~\citep{Payne1992} and accordingly, calculated a modified steepest-descend vector. We then follow the procedure of Ref.~\citep{Payne1992} to obtain the corresponding conjugate-gradients vector $\xi_k^{m_1}$, which is analogous to how we have proceeded for RDMFT. The most important steps for that are summarized in algorithm~\ref{algorithm:RDMFT_no_cg} of Sec.~\ref{sec:numerics:rdmft:cg} and we do not repeat them here.\footnote{We remark that we skip in our test-implementation the preconditioning step, which is not crucial for small systems and the tight-binding-like approximation of the kinetic energy.}

However, we have to modify the line-minimization to properly account for the penalty. This is nontrivial and we have developed a new algorithm for this task that we present below (see algorithm~\ref{algorithm:line_minimization}).

To test the convergence of state $\phi_k$ in the DO loop, we define
\begin{align}
	|\phi_k^{m_1+1}-\phi_k^{m_1}|<\epsilon_{PCG}^l.
\end{align}

\subsubsection*{The natural orbital optimization (NO loop)}
The next higher iteration loop is the ``NO loop'' that updates the natural orbitals $\bm{\psi}^{e,m_2}$, after the $\bm{\phi}$-optimization (DO loop) has converged for fixed $\hat{H}^1$ and $\bm{\psi}^{e,m_2}$. Instead of solving Eq.~\eqref{eq:numerics:pHF_gradientPsiGamma_sub} directly, we determine the first $M$ eigenvalues of $\gamma_e^{m_2}=\gamma_{e}[\bm{\phi}^{m_2}]$, i.e., we solve
\begin{align}
\tag{cf. \ref{eq:numerics:gamma_electronic_evp}}
n_i^{m_2+1} \psi_i^{e,m_2+1} = \hat{\gamma}_{e}[\bm{\phi}^{m_2}] \psi_{i}^{e,m_2+1},
\end{align}
for a set of natural occupation numbers $\bm{n}^{m_2+1}=(n_1^{m_2+1},...,n_M^{m_2+1})$ and the new set of natural orbitals $\psi_i^{e,m_2+1}$. This is a linear eigenvalue problem and thus, we can employ any standard eigensolver to solve it. We use the function ``eigh'' of the \textsc{NumPy} library. 

If we then set $\theta_i^{m_2+1}=n_i^{m_2+1} (\mu [g_i]^- -\nu_i )$, we have solved Eq.~\eqref{eq:numerics:pHF_gradientPsiGamma_ev}. In practice, we do not need to perform the latter step explicitly, because we only utilize the updated natural orbitals in the rest of the algorithm.

The convergence criterion of the NO loop is
\begin{align}
	|\gamma_e^{m_2+1}-\gamma_e^{m_2}|< \epsilon_{NO}^l.
\end{align}

\subsubsection*{The update of the Fock-matrix (PCG loop)}
After the NO loop has converged, we have obtained a set of orbitals $(\bm{\phi}^{m_3},\bm{\psi}^{m_3})$ that (approximately) solve both, Eq.~\eqref{eq:numerics:pHF_gradientPhi_ev} and Eq.~\eqref{eq:numerics:pHF_gradientPsiGamma_ev}, for a fixed$ \hat{H}^{1,m_3}$. The PCG loop updates then 
\begin{align*}
\hat{H}^{1,m_3+1}=\hat{H}^{1}[\gamma^{m_3}]=\hat{H}^{1}[\gamma[\bm{\phi}^{m_3}]].
\end{align*}
After the PCG loop has converged, we have found an approximate solution $(\bm{\phi}^{(\mu_l, \bm{\nu}_l)}, \bm{\psi}^e{}^{(\mu_l, \bm{\nu}_l)})$ for the current values of the penalty parameters $(\mu_l,\bm{\nu}_l)$.

To test the convergence of the PCG loop and thus of the inner part of the algorithm, we employ the criterion
\begin{align}
	|\gamma^{m_3+1}-\gamma^{m_3}|<\epsilon_{SCF}^l.
\end{align}

\subsubsection*{Convergence thresholds}
Since the outer algorithm only determines the $\epsilon_{PCG}^l$, we need to adopt the convergence thresholds of the other loops accordingly.
Since the loops are nested, it is crucial that criteria of inner loops are stricter than the thresholds of outer loops. In our implementation, we employ
\begin{align}
\label{eq:numerics:pHF_update_tolerances}
\begin{split}
	\epsilon_{NO}^l=&10^{-1}\epsilon_{PCG}^{l}\\
	\epsilon_{DO}^l=&10^{-1}\epsilon_{NO}^l=10^{-2}\epsilon_{PCG}^{l}\\
\end{split}
\end{align}

\newpage
\subsubsection*{The full PCG algorithm}
We now summarize the full PCG-algorithm in pseudo-code.
\begin{center}
	\begin{fminipage}{0.99\textwidth}
		\begin{algorithm}\label{algorithm:pHF_inner_part} (PCG)  \newline
			Initialize $\bm{\phi}^{0}$ and calculate
			\begin{adjustwidth}{0.5cm}{}
				$\gamma^0=\gamma[\bm{\phi}^0]$ from Eq.~\eqref{eq:numerics:hybrid:gamma_definition},\\
				$\hat{H}^{1,0}=\hat{H}^1[\gamma^0]$ from Eq.~\eqref{eq:numerics:hybrid:pHF_FockMatrix}.\\
				$\gamma_e^0=\gamma_e[\gamma^0]$ from Eq.~\eqref{eq:numerics:gamma_electronic_definition}.\\
				$\bm{\psi}^{e,0}$ by solving Eq.~\eqref{eq:numerics:gamma_electronic_evp}.
			\end{adjustwidth}
			Initialize $\epsilon_{DO}^0,\epsilon_{NO}^0$ according to \eqref{eq:numerics:pHF_update_tolerances}.\\
			\\
				\textbf{for} $m_3=0,1,2,...$ (\textbf{PCG loop})
					\begin{adjustwidth}{0.5cm}{}
					\textbf{for} $m_2=0,1,2,...$ (\textbf{NO loop})
						\begin{adjustwidth}{0.5cm}{}
						Initialize $\bm{\phi}^{m_2=0}=\bm{\phi}^{m_3}$ $\bm{\psi}^{e,m_2=0}=\bm{\psi}^{e,m_3}$.\\
						\textbf{for} $k=1,...,N/2$ (state iteration)
							\begin{adjustwidth}{0.5cm}{}
							Initialize $\phi_k^{m_1=0}=\phi_k^{m_2}$.\\
							\textbf{for} $m_1=0,1,2,...$  (\textbf{DO loop})
							\begin{adjustwidth}{0.5cm}{}
							\begin{enumerate}[label=(\alph*)]
								\item Calculate $\zeta_k^{m_1}$ from Eq.~\eqref{eq:numerics:pHF_steepest_descend_vector} with the current $H^{m_3}$, $\hat{G}_i^{m_2}$ and $\bm{\psi}^{e,m_2}$.
								\item Calculate $\xi_k^{m_1}$ according to the algorithm by \citet{Payne1992} (see algorithm~\ref{algorithm:RDMFT_no_cg}).
								\item perform a line-minimization along the direction of $\xi_k^{m_1}$ to obtain $\phi_k^{m_4+1}$ (algorithm~\ref{algorithm:line_minimization}).
								\item[] \textbf{break} if $|\phi_k^{m_1+1}-\phi_k^{m_1}|<\epsilon_{PCG}^l$
							\end{enumerate}
							\end{adjustwidth} 
							\textbf{end for} (DO loop)
							\end{adjustwidth}	
						\textbf{end for} (state iteration)
						\begin{enumerate}[label=(\roman*)]
							\item Collect all final sates of the PCG loop into the new $\bm{\phi}^{m_2+1}$.
							\item Calculate $\gamma_e^{m_2+1}=\gamma_e[\bm{\phi}^{m_2+1}]$ from Eq.~\eqref{eq:numerics:gamma_electronic_definition}.
							\item Diagonalize $\gamma_e^{m_2+1}$, i.e., solve Eqs.~\eqref{eq:numerics:gamma_electronic_evp}, to obtain the new set $\bm{\psi}^{e,m_2+1}$.
							\item[] \textbf{break} if $|\gamma_e^{m_2+1}-\gamma_e^{m_2}|< \epsilon_{NO}^l$
						\end{enumerate}
						\end{adjustwidth} 
						\textbf{end for} (NO loop)
					\begin{enumerate}[label=(\arabic*)]
						\item Update $\bm{\phi}^{m_3+1}, \bm{\psi}^{e,m_3+1}$ with the final set of $\bm{\phi}^{m_2}\bm{\psi}^{e,m_2}$ for which the NO loop has converged.
						\item Update $\gamma^{m_3+1}=\gamma[\bm{\phi}^{m_3+1}]$.
						\item Calculate $\hat{H}^{1,m_3+1}=\hat{H}^1[\gamma^{m_3+1}]$ from Eq.~\eqref{eq:numerics:hybrid:pHF_FockMatrix}.
						\item[] \textbf{break} if $|\gamma^{m_3+1}-\gamma^{m_3}|<\epsilon_{SCF}^l$
					\end{enumerate}
					\end{adjustwidth} 
				\textbf{end for} (PCG loop)
		\end{algorithm}
	\end{fminipage}
\end{center}

\subsubsection*{The line-minimization}
In order to employ the conjugate gradient algorithm by \citet{Payne1992} together with the augmented Lagrangian method, we need to modify the line-minimization (LM). The first-order Fourier approximation of the energy functional that \citet{Payne1992} employ (see Sec.~\ref{sec:numerics:rdmft:cg}) is not justified here because we have to take the nonlinear penalty-term into account. Thus, instead of minimizing $E$, we consider the ``penalized energy'' functional
\begin{align}
\label{eq:numerics:pHF_penalized_energy}
\tilde{E}^{(\mu_l, \bm{\nu}_l)}=& E + \mathcal{P}^{(\mu_l, \bm{\nu}_l)},
\end{align}
where $\mathcal{P}^{(\mu_l, \bm{\nu}_l)}=- \sum_i^{M} \nu_{l,i} g_i[\bm{\phi}, \bm{\psi}^e ] + \frac{\mu_l}{2} \sum_i^M ([g_i]^-[\bm{\phi}, \bm{\psi}^e ])^2$. The goal of the LM is then to minimize $\tilde{E}^{(\mu_l, \bm{\nu}_l)}$ along the direction that is defined by the conjugate-gradients vector, where we need to take especially the nonlinearity of the penalty part into account. 

Specifically, we parametrize the conjugate-gradients vector $\xi_k^{m_1}$ by the angle $\Theta_k\in [0,\pi]$ (cf. Eq.~\eqref{eq:numerics:cg_phi_parametrization})
\begin{align*}
\tilde{\phi}_k^{m_1}(\Theta_k)=\cos\Theta_k 	\phi_k^{m_1} + \sin\Theta_k \xi_k^{m_1},
\end{align*}
and accordingly $\tilde{E}^{(\mu_l, \bm{\nu}_l)}=\tilde{E}^{(\mu_l, \bm{\nu}_l)}(\Theta_k)$, which we approximate by
\begin{align}
\label{eq:numerics:pHF_line_min_functional}
\tilde{E}^{(\mu_l, \bm{\nu}_l)}(\Theta_k) \approx& \tilde{E}^{(\mu_l, \bm{\nu}_l)}(0)+ \braketm*{\tilde{\phi}_k^{m_1}(\Theta_k)} {\nabla_{\phi_i^*}\tilde{E}^{(\mu_l, \bm{\nu}_l)}|_{\phi_k=\tilde{\phi}_k^{m_1}(\Theta_k)}} \nonumber\\
=& \tilde{E}^{(\mu_l, \bm{\nu}_l)}(0)+ \braketm*{\tilde{\phi}_k^{m_1}(\Theta_k)} {\hat{H}^1 \tilde{\phi}_k^{m_1}(\Theta_k)} + \sum_i \left[ \nu_{l,i} -\mu_l [g_i[\tilde{\phi}_k^{m_1}(\Theta_k)]]^-\right] \braketm*{\tilde{\phi}_k^{m_1}(\Theta_k)}  {\hat{G}_i \tilde{\phi}_k^{m_1}(\Theta_k)}.
\end{align}
Note that this is not a first-order Taylor expansion, as one might think at a first glance. Expression~\eqref{eq:numerics:pHF_line_min_functional} has been  developed specifically for this algorithm and it cannot be understood out of its context. We explain this in detail in the following.

We start by remarking that  the operators $\hat{H}^1$ and $\hat{G}_i$ are constant during the DO optimization. Nevertheless Eq.~\eqref{eq:numerics:pHF_line_min_functional} is nonlinear, because of the  constraint functions $g_i=2-n_i$ with (cf. Eq.~\eqref{eq:numerics:pHF_def_ni})
\begin{align*}
n_i[\bm{\phi}]= \braket{\psi^e_i| \hat{\gamma}_e[\bm{\phi}]\psi^e_i}.
\end{align*}
This formulation allows us to keep the natural orbitals $\psi_i^e$ fixed during the optimization of the $\bm{\phi}^{m_1}$, which is consistent with the rest of the algorithm, but to consider
\begin{align*}
\gamma_e=\gamma_e[\tilde{\phi}_k^{m_1}(\Theta_k)]=\gamma_e[\Theta_k]
\end{align*}
as a function of $\Theta_k$. To find the optimal angle $\Theta_k^*$, the natural next step would be to derive some approximate expression similar to Eq.~\eqref{eq:numerics:LineMinThetaOpt} in Sec.~\ref{sec:numerics:rdmft:cg}, which would be crucial for larger systems. However, for our test-implementation, we could afford to employ an algorithm that directly calculates $\tilde{E}^{(\mu_l, \bm{\nu}_l)}(\Theta_k)$ with the approximate expression~\eqref{eq:numerics:pHF_line_min_functional} for a not too large set of test-values of $\Theta_k$. It is important to realize that to do so,  we only need to construct but \emph{not} to diagonalize $\gamma_e[\Theta_k]$, which is numerically cheep for not too large bases. 

In practice, we perform the LM by a newly developed method that is inspired by so-called \emph{divide-and-conquer} algorithms. The idea is to sample the interval of possible values of the angle $\Theta$, such that we can locate the (global) minimum of $\tilde{E}^{(\mu_l, \bm{\nu}_l)}(\Theta)$ roughly. We define then a smaller interval around the approximate position of the minimum, which we sample in the next iteration with the same number of test-values to localize the minimum with a higher precision. We repeat this procedure until the interval is smaller than a prescribed threshold $\epsilon_{LM}$ that we adopt with respect to the threshold of the DO-loop, i.e.,
\begin{align}
\label{eq:numerics:hybrid:threshold_LM}
	\epsilon_{LM}^l=&10^{-1}\epsilon_{PCG}^l=10^{-3}\epsilon_{SCF}^{l}.
\end{align}
Specifically, we define the interval
\begin{align*}
\Theta\in I^j=[\Theta_{min}^j,\Theta_{max}^j],
\end{align*}
and initialize the search for the maximal possible interval, i.e., $\Theta_{min}^j=0$ and $\Theta_{max}^0=\pi/2$. 
We sample then the energy $\tilde{E}^{(\mu_l, \bm{\nu}_l)}(\Theta)$ over a set of $N_{LM}$ linearly distributed test-values in $I^j$. Therefore we define $\Delta^j=(\Theta_{max}^j-\Theta_{min}^j)/N_{LM}$ and the test set
\begin{align}
\label{eq:numerics:pHF_lm_Tinterval}
T^j=\{\Theta_{min}^j,\Theta_{min}^j+\Delta^j,\Theta_{min}^j+2\Delta^j,...,\Theta_{max}^j\},
\end{align}
calculate
\begin{align}
\label{eq:numerics:pHF_lm_Einterval}
E^j=\{ \tilde{E}^{(\mu_l, \bm{\nu}_l)}(\Theta) | \Theta\in T^j\}.
\end{align}
and determine the approximate minimum $E_{min}^j=\min E^j$. The corresponding approximate minimum $\Theta^j$ for which holds that
\begin{align}
\label{eq:numerics:pHF_lm_Thetamin}
E_{min}^j=\tilde{E}^{(\mu_l, \bm{\nu}_l)}(\Theta^j)
\end{align}
defines then the \emph{center} of the interval for the next iteration. We define the new (reduced) interval half-length $D_{\Theta}^j=\alpha_{LM}|\Theta_{min}^j-\Theta_{max}^j|/2$, where $0<\alpha_{LM}<1$ and then choose
\begin{align}
\label{eq:numerics:pHF_lm_updates}
\begin{split}
\Theta_{min}^{j+1}=&\max( \Theta^j-D_{\Theta}^j, 0 )\\
\Theta_{max}^{j+1}=&\min( \Theta^j+D_{\Theta}^j, \pi/2 ).
\end{split}
\end{align}
Here the min and max functions are necessary to constrain $\Theta$ within the maximally allowed interval $[0,\pi/2]$. Depending on the values for $N_{LM}=10$ and $\alpha_{LM}$, the algorithm may perform quite differently and these values should be tested with care. We found that $N_{LM}$ and $\alpha_{LM}=0.5$ lead to very precise minima, but remained numerically feasible for our small test systems. The advantage of the algorithm is that we in principle can locate the minimum of $\tilde{E}^{(\mu_l, \bm{\nu}_l)}(\Theta)$ with a very high precision, which was very important in the debugging process. The algorithm is summarized in pseudo-code as follows.
\begin{center}
	\begin{fminipage}{0.7\textwidth}
		\begin{algorithm}\label{algorithm:line_minimization} (PCG line-minimization)  \newline
			Set the convergence threshold $\epsilon_{LM}$ according to Eq.~\eqref{eq:numerics:hybrid:threshold_LM}, the sampling number $N_{LM}$ and $\alpha_{LM}$ for the interval reduction.\newline
			Initialize $\Theta_{min}^0=0$ and $\Theta_{max}^0=\pi/2$ and set $I^0=[\Theta_{min}^0,\Theta_{max}^0]$.\newline
			\\
			\textbf{for} j=0,1,2,...
			\begin{enumerate}[label=\alph*]
				\item Sample $I^j$ according to Eq.~\eqref{eq:numerics:pHF_lm_Tinterval}
				\item Calculate $E^j$ according to Eq.~\eqref{eq:numerics:pHF_lm_Einterval}
				\item Determine the approximate minimum $\Theta^j$ according to Eq.~\eqref{eq:numerics:pHF_lm_Thetamin}
				\item Update $\Theta_{min}^{j+1}$ and $\Theta_{max}^{j+1}$ according to Eq.~\eqref{eq:numerics:pHF_lm_updates}
				\item[] \textbf{break} if $(\Theta_{max}^{j+1}-\Theta_{min}^{j+1})/2<\epsilon_{LM}$
			\end{enumerate}
			\textbf{end (for)}	
		\end{algorithm}
	\end{fminipage}
\end{center}

\subsubsection*{Remarks}
We want to conclude the section with some final remarks regarding the algorithm and our implementation.
\begin{itemize}
	\item The convergence threshold $\epsilon_{LM}$ of the line-minimization should be smaller than the threshold
	\item Although we formally defined all different sets of states for all iteration loops, we do not have to save them separately. Thus, when we write, e.g., $\bm{\phi}^{m_3=0}=\bm{\phi}^{(\mu_l,\bm{\nu}_l)}$ we do just pass a pointer between the routines.
	\item We want to stress that we could converge the code for basically every tested setting up to $\epsilon_{total}=10^{-4}$. Higher accuracies are possible for small systems, but become difficult with increasing system size, such as a large number of lattice sites. A possible reason could be the simple treatment of the non-differential point of the penalty function (see Eq.~\eqref{eq:numerics:hybrid:penalty_function_nondifferential_point}).
	\item The current version of the code works only for 4 particles and with the additional approximation do not apply the penalty to the lowest occupied orbital, but only the second one. Without this approximation, we only converge very small test systems (see also the Outlook in Sec.~\ref{sec:conclusion}).
\end{itemize}

	\part{Concluding Remarks}
\label{sec:conclusion}
\newpage
\chapter{Conclusions}
In the course of this thesis, we have analyzed the new challenges that arise in a first-principles description of (equilibrium) coupled light-matter systems in comparison to the well-studied problem of describing matter from first principles. 
Motivated from this, we have proposed the dressed-orbital construction that allows to overcome many of these challenges by using established (matter-only) methods to also investigate strongly-coupled electron-photon systems. We have illustrated by several examples that this approach stays accurate even for very strong coupling between light and matter, while being numerically feasible.

This still very young research area confronted us with many unexpected challenges. Already the starting point of our analysis was not clear. For instance, the usual cavity-QED model Hamiltonian (see Sec.~\ref{sec:intro:cavity_qed_models}) that has been used extensively to study coupled light-matter systems, is not useful for a first-principles description because it is not bounded from below. On the other hand, the full theory of QED is also not useful, because we do not know how to renormalize the theory non-perturbatively. 
To account for the specific needs of a first-principles approach, we have then identified Pauli-Fierz theory as a suitable theoretical framework that includes cavity QED as a well-defined limit case. We have presented most of our discussion for this simpler setting that constitutes a good starting point to describe many phenomena of strong-coupling experiments. 

The next challenge expected us when we tried to generalize standard electronic-structure methods to the cavity-QED setting: we first had to understand how such a generalization can be done. Therefore, we analyzed three conceptually different approaches to describe the many-electron ground state, i.e., HF theory (as an example for wave-function methods), KS-DFT and RDM theory. We then showed that all these approaches have a clear counterpart in the coupled setting. However, it turned out that especially for equilibrium scenarios, many of these generalizations are considerably less efficient. For instance, the MF description, i.e., the generalization of HF theory cannot account for the quantum nature of the electron-photon interaction. This is a direct consequence of the fact that there is no exchange symmetry between the electronic and photonic Hilbert space, and thus no ``Fock-term''. The remaining classical electron-photon interaction consists of merely electrostatic fields induced by the total dipole of the matter. This means that in order to understand nontrivial effects, such as the formation of polaritons and their influence on equilibrium properties of the system, we fundamentally have to describe photon-induced correlation.

This is in principle possible in terms of the MF wave function, if we consider (KS-)QEDFT. The challenges of the coupled space manifest here in the unknown photon-exchange-correlation potential, for which we still do not have useful approximations. To describe the electron-photon interaction energy instead within the RDM approach, we had to introduce the $3/2$-body RDM, which is a new type of RDM of mixed-electron photon nature. The $3/2$-body RDM has very different properties than the usual RDMs, e.g., it is not connected to the usual RDMs by a sum rule, because of its photonic half-body part. Without investigating this and other issues such as the unknown representability conditions, it is very difficult to make practical use of this $3/2$-body RDM. Nevertheless, it is possible to define QED-RDMFT that describes the system's state with the electronic 1RDM and the displacement coordinates of the photon modes (or the vector potential for full Pauli Fierz-theory). We have explicitly shown that this description is in principle exact by generalizing Gilbert's theorem to Pauli-Fierz theory. However, to capture the quantum nature of the electron-photon interaction with QED-RDMFT, one faces similar challenges as in KS-QEDFT, since the according exchange-correlation functional is also not known.

Our thorough analysis allowed us to identify a similarity in these challenges of the many-electron-photon problem: many of the difficulties are directly connected to the \emph{structure} of the electron-photon interaction operator, which couples both particle species and leads to a non-conserved photon number. Established tools to approximate the Coulomb interaction between electrons often seem to be less powerful if there is a second particle species involved. And most approaches are directly geared toward the description of systems with a fixed particle number. 
We have thus proposed an alternative strategy that approaches the problem on a more fundamental level. Instead of further investigating the interaction between the two species directly, we have presented the dressed-orbital construction, which allows to describe cavity-QED systems in terms of only one polaritonic particle species. We have shown that we can describe the ground state of an $N$-electron-$M$-mode system in terms of an $N$-polariton wave function. These polaritons are hybrid particles that depend on electronic and photonic coordinates and adhere to a Fermi-Bose hybrid statistics. Only the possibility to introduce a new particle class is a striking feature of the coupled space, which to the best of our knowledge has never been explored before, and opens up many interesting research directions. 

One important consequence of describing the coupled problem with only one particle species is the structural equivalence between the dressed version of the cavity-QED Hamiltonian and the electronic-\linebreak{}structure Hamiltonian: both consist of only one-body and two-body terms that conserve the particle number. Thus, the restructuring of the many-body space allows to circumvent basically all fundamental problems of the light-matter first-principles approaches that we have discussed. The polaritonic-HF ansatz leads to an exchange contribution and explicitly accounts for correlation between the electronic and photonic subspaces. In polaritonic DFT, we can in principle employ any known exchange-correlation functional to describe ``polaritonic correlation,'' i.e., all effects beyond polaritonic HF. And the necessary RDMs to calculate the polaritonic energy conserve the particle number. Thus, they are connected to each other in an analogous way as in the single-species case. Although the polaritonic 1RDM is not as simple as the electronic or photonic one, we can straightforwardly employ it in polaritonic RDMFT as basic variable together with known exchange-correlation functionals.
But the dressed construction is not limited to these three particular approaches. We have explicitly shown how to turn basically \emph{any} electronic-structure theory into a polaritonic-structure theory. Importantly, this also means that we can generalize existing implementations of electronic-structure methods in a straightforward way to describe coupled systems.

However, the advantages of dressed orbitals come also with a price. First, we have to raise the dimension of the electronic orbitals by one for every photon mode that is considered. This clearly limits the application of polaritonic-structure methods to settings, where the explicit description of only one or a few modes is sufficient. Then, we have to add additional terms to the kernels of the potential and interaction operators. This should not be a big issue in most cases, as we have shown explicitly for our implementation in \textsc{Octopus} (see Sec.~\ref{sec:numerics:dressed}). Finally, we have to enforce the hybrid statistics of the polaritons, which is an entirely unexplored issue in the realm of first-principles theory. The standard approach that is to expand the many-polariton state in a properly symmetrized basis (such as Slater determinants for many-fermion systems) is not applicable in practice because of the occurrence of ``mixed-index'' orbitals (see Sec.~\ref{sec:dressed:construction:simple:auxiliary_wf_problem}). Thus, new strategies have to be explored. As a first step, we have proposed two practical approximations. 
The simplest one is the fermion ansatz, which treats polaritons as fermions and thus completely neglects the bosonic part of the statistics. This allows for the most straightforward implementation of dressed orbitals and to study the practical consequences of the hybrid statistics. With the help of the comprehensive convergence studies with our first implementation of the fermion ansatz (see App.~\ref{sec:numerics:dressed:convergence}), we could understand that the ansatz allows for violations of the Pauli principle in the electronic subsystem. Conversely, it seems that polaritonic-structure methods with the fermion ansatz describe coupled light matter systems very accurately in all cases, where the Pauli principle is satisfied trivially (such as two-polariton singlet states that are described by one doubly-occupied orbital). 

This observation motivates the second proposed approximation to the hybrid statistics. The idea of the polariton ansatz is to remain as close as possible to the fermion ansatz, but remedy its most severe shortcoming by explicitly enforcing the Pauli principle. To do this in practice, the wave function description is not helpful, because enforcing the Pauli principle is here basically equivalent to enforcing antisymmetry. To define the polariton ansatz, we therefore consider the electronic (ensemble) 1RDM to guarantee the Pauli principle explicitly (but not strict antisymmetry) by enforcing its ensemble $N$-representability conditions. 
To test this idea, we have developed and implemented an algorithm to perform polaritonic-HF calculations with the polariton ansatz. Although this algorithm is considerably more involved than any standard HF algorithm, our first results have confirmed that the polariton ansatz is not only numerically feasible but also accurate. In scenarios where the fermion ansatz trivially fulfils the Pauli-principle and provides a very accurate description, the polariton ansatz reproduces these results. However, the polariton ansatz remains accurate also in the scenarios, where violations of the Pauli principle may occur with the fermion ansatz. We also have explicitly shown how to ``switch'' between both cases by varying the photon mode frequency $\omega$.
This illustrates yet from another perspective the differences between a single-species and a coupled two-species problem: a concept such as the electronic 1RDM might be useful for both problems, but in considerably different ways.


Besides the interesting new perspective on many-species problems that the dressed construction offers, we have shown that polaritonic-structure methods have a big potential to shed light on the complicated mechanisms of polaritonic physics. For instance, we have demonstrated a nontrivial interplay between local suppression and enhancement of the Coulomb induced repulsion between the electrons (Sec.~\ref{sec:dressed:results:1chemical_reaction} and~\ref{sec:dressed:results:2mores_is_different}). This is reflected in the natural orbitals and occupation numbers of the light-matter system and thus influences all possible observables. Such small changes have been shown to theoretically strongly affect chemical properties and reactions, which are determined by an intricate interplay between Coulomb and photon induced correlations~\cite{Schaefer2018}. Whether these modifications of the underlying electronic-structure are indeed a major player in the changes of chemical and physical properties still needs to be seen. However, to capture such modifications in the first place (and study their influence) clearly needs a first-principles theory that is able to treat both types of (strong) correlations accurately and is predictive inside as well as outside of a cavity.

As a last example, we have applied polaritonic HF to investigate the interplay of electron localization and electron-photon coupling. We found for nontrivial problems that the more delocalized the uncoupled matter wave function is, the stronger it reacts to the modes of a cavity, which comes along with an increase of electronic correlation. This is a first result that directly contributes to the debate on whether the ground state can be measurably influenced by only coupling to the vacuum of a cavity (see Sec.~\ref{sec:intro:experiment_strong_coupling}). The findings indicate that the spatial extension of the electronic wave function might be another important parameter that influences the light-matter coupling. The ground state of spatially extended systems could indeed be strongly modified by the vacuum field of an optical cavity. Importantly, the influence of localization cannot be studied easily with standard approaches such as cavity-QED models, which is one probable explanation why it has not yet played a role in the debate.
Additionally, this raises the question whether an ensemble of emitters with a spatially extended wave function might show collective strong-coupling effects that also modify the local electronic-structure, in contrast to the Dicke-type collective coupling picture that does not consider electronic correlation.

These first results illustrate the prospects of the new first-principles perspective on coupled light-matter systems that the polariton picture provides (see Ch.~\ref{sec:outlook}). However, this comes with the price of an increased numerical complexity, which demands the development of efficient and robust algorithms, a careful implementation and comprehensive validation procedures. Nevertheless, with the example of our polaritonic-RDMFT implementation in \textsc{Octopus}, we have shown that even complex electronic-structure code environments can be extended to describe polaritons. Although the nonlinearities of the involved equations may lead to hardly predictable side effects (App.~\ref{sec:numerics:dressed:non_assessment}), our convergence studies have revealed that polaritonic-structure methods are as accurate as their electronic counter parts (App.~\ref{sec:numerics:dressed:convergence}). Importantly, many synergetic effects arise from such an extension of standard routines. For instance, we understand now the accuracy and limitations of the default RDMFT routine of \textsc{Octopus} (Sec.~\ref{sec:numerics:rdmft:algorithm_no_piris}) much better because of the convergence studies with polaritonic RDMFT. This understanding lead to the development and implementation of the new conjugate-gradients methods (Sec.~\ref{sec:numerics:rdmft:cg}, which allows to perform both electronic-RDMFT and polaritonic-RDMFT calculations with the full flexibility of the grid. On the one hand, this opens new interesting research directions in the entirely unexplored field of real-space RDMFT (see Ch.~\ref{sec:outlook}), while on the other hand it crucially has contributed to our understanding of the hybrid statistics and eventually to the development of the polariton ansatz.

Although generalizing an electronic-structure algorithm to treat general cavity-QED systems with the polariton ansatz is numerically nontrivial, we have shown with the example of polaritonic HF that it is indeed possible and also has affordable computational costs (Sec.~\ref{sec:numerics:hybrid}). This algorithm and its successful implementation represents an important first step toward the goal of describing realistic strongly-coupled light-matter systems from first principles. However, to reach this goal, further optimization of the algorithm is still necessary (see Ch.~\ref{sec:outlook}). We have therefore carefully analyzed the numerical challenge of enforcing nonlinear inequality constraints in addition to the nonlinear minimization problems of typical electronic-structure methods.

\newpage
\chapter{Outlook: open questions and prospects}
\label{sec:outlook}
During the years of my PhD, I have faced many interesting research questions. Some of them have been answered in this manuscript or by my colleagues, but many others are still open. The present work will be an important step forward to understand and resolve some of these open questions. We want to address these in the following.

\subsubsection*{Extension of existing dressed-orbital implementation to 3d}
The next obvious step is to extend the actual implementation of dressed orbitals in \textsc{Octopus}~\cite[Ch. 4]{Tancogne-Dejean2020} to treat matter in three spatial dimensions. 
Most prerequisites are already fulfilled for that and the last missing piece is to optimize the calculation of the dressed interaction integrals, which should be done separately from the Coulomb integrals (see Sec.~\ref{sec:numerics:dressed:implementation}).\footnote{Note that the calculation of the dressed-interaction integrals is much simpler than the solution of the Poisson equation. The separation of both computations thus allows to do the latter more efficiently with (already implemented) standard solvers.} 
Once this straightforward modification is accomplished, one can directly perform polaritonic-HF and polaritonic-RDMFT calculations with the fermion ansatz. For instance, systems with only two active electrons, such as many diatomic molecules can be described without further limitations, but as we have shown, also systems with more electrons can be accurately described if the precise value of the photon-mode frequency plays a subordinate role and can be adjusted (see Sec.~\ref{sec:dressed:results:implementation_lattice:hybrid_influence}). Additionally, such an extension allows to investigate the accuracy of standard exchange-correlation functionals for polaritonic QEDFT without further modifications of the code. In \textsc{Octopus}, we have a direct access to the full \textsc{LibXC}-library to do so.\footnote{\url{https://www.tddft.org/programs/libxc/}}
It is crucial to realize here that besides the necessarily larger orbital bases, the scaling of these methods is not affected by the inclusion of the photon field (see Sec.~\ref{sec:dressed:results:numerical_perspective}).

\subsubsection*{Large-scale calculations with the polariton ansatz}
The next evident research questions is how to extend a standard electronic-structure code to treat dressed orbitals with the polariton ansatz. Our in Sec.~\ref{sec:numerics:hybrid} presented polariton-HF algorithm is in principle applicable, if not too large electronic bases are considered  and a direct diagonalization of the electronic 1RDM is feasible. This applies to, e.g., medium-sized systems described by standard quantum-chemistry basis sets. Note that in contrast to the matter description, the extra photon orbitals do not necessarily have to be constructed in the code. The analytically known structure of the photon-number states allows to calculate all necessary matrix elements analytically. We have exploited this also in our implementation (see Sec.~\ref{sec:dressed:results:implementation_lattice:model}). 

Nevertheless, to apply the polariton-HF algorithm to systems involving many particles, it clearly needs to be optimized. 
The current algorithm of the line-minimization should be replaced by a similar approximation as laid out in Ref.~\citep{Payne1992}. Additionally, we have to investigate new practical ways to determine the Lagrange multipliers $\epsilon_i$ and $\nu_i$ corresponding to the equality and inequality constraints, respectively. In the current state of the algorithm, the $\nu_i$ are calculated according to the prescription of the augmented Lagrangian method, which works well. However, the $\epsilon_i$ are determined according to a direct generalization of the conjugate-gradients algorithm by \citet{Payne1992} (see Sec.~\ref{sec:numerics:hybrid:pHF_algorithm:inner}). As a consequence the $\epsilon_i$ depend nontrivially on the $\nu_i$, which may lead to numerical instabilities. This issue requires a careful analysis, but should be resolvable by employing a different method to determine the $\epsilon_i$.

Once, we have optimized the algorithm, a reasonable next step is to extend the existing implementation of the dressed construction in \textsc{Octopus}~\cite[Ch. 4]{Tancogne-Dejean2020} to the polariton ansatz. For that, one would not explicitly diagonalize the electronic 1RDM, which is numerically prohibitively expensive but resort to \emph{semidefinite programming} methods~\cite{Vandenberghe1996} that exploit the reformulation of the constraints in terms of positivity-conditions (see also Sec.~\ref{sec:dressed:results:numerical_perspective}).

It is clear that developing, implementing and validating such an optimized polariton-HF algorithm is nontrivial. We believe however that such extensions, though they require a certain amount of work, are indeed worthwhile.
At the moment, polaritonic-structure methods are one of the most promising candidates to describe molecule-cavity systems even for very large coupling strengths. The principal reason for that is that we do not have to develop entirely new functionals, but can ``recycle'' the ones of the corresponding electronic-structure methods. This will be extensively tested with the planned extension of the \text{Octopus} implementation to 3d systems with the fermion ansatz, before we develop the generalized algorithm.

\subsubsection*{Describing vibrational strong coupling with dressed orbitals}
The fact that many strong-coupling phenomena have been experimentally observed by coupling the vibrational degrees of freedom of molecular systems to a cavity~\citep{George2016,Lather2019,Ojambati2019} poses the question whether we can generalize the dressed-orbital approach to this setting. This means that we have to include the motion of the nuclei and their coupling to the cavity mode in our description.

To answer this question, it is important to realize that in the cavity-QED setting the coupling of the photon modes to electrons structurally identical and to the coupling to the nuclear degrees of freedom (see Sec.~\ref{sec:intro:nrqed_longwavelength}). Thus, in principle, we can employ the same strategy and dress the nuclear orbitals with the photon modes. The straightforward way to do so is to employ the Born-Oppenheimer approximation to decouple the electronic from the nuclear-photonic degrees of freedom (cavity Born-Oppenheimer approximation~\citep{Flick2017cavity}). However, one could also investigate different approaches such as discussed in Ref.~\citep{Schaefer2018}. Note that also in such settings, the dressed construction should allow to transfer accurate first-principles methods for matter-only problems to describe coupled systems.

\subsubsection*{Polaritonic structure methods for extended systems}
Another interesting research question is whether we can apply the polaritonic-structure methods also to extended systems. This would allow to study strong-coupling phenomena of solid-state systems. For instance, \citet{Rokaj2019} recently have presented an approach based on the cavity-QED Hamiltonian that allows to describe interesting phenomena such as Landau physics from first principles. Polaritonic-structure methods based on this approach would allow to study, e.g., the modification due to quantum fluctuations of phenomena such as the Landau levels~\citep{Landau2013}, the integer~\citep{Klitzing1980,Kohmoto1985} and the fractional quantum Hall effect~\citep{Laughlin1983}.    
Another very promising research direction is to investigate how properties of solids can be controlled by the interaction with the cavity. Examples are the stabilization of topological states~\citep{Hubener2017} or the control of the electron-phonon coupling strength and thus phonon-mediated superconductivity~\citep{Sentef2018}. Since such phenomena usually depend on a complex interplay of different effects, for their investigation there are few alternatives  to an unbiased description from first principles. Especially, if strongly-correlated materials are considered, the dressed-orbital approach might be advantageous in comparison to other methods, simply because of its flexibility. Following our prescription of Sec.~\ref{sec:dressed:construction:general}, one could generalize an established accurate method for strongly-correlated electrons to the cavity setting.

To extend the dressed-orbital approach to such cases, one has to consider periodic local potentials. Usually, this setting is simplified according to the Bloch theorem by a kind of mode expansion that takes the periodicity of the potential into account. Since the (dipole) coupling to the photon modes breaks the translational symmetry, the Bloch theorem cannot be applied straightforwardly. Here, a similar strategy to the recently proposed generalized Bloch theorem~\citep{Rokaj2019} should be applicable within the dressed-orbital construction. If this is indeed the case, one could straightforwardly make use of the $k$-space routines of \textsc{Octopus} to perform, e.g., polaritonic QEDFT calculations with extended systems (strongly) coupled to a cavity mode.


\subsubsection*{Development of a KS-QEDFT photon-exchange-correlation functional}
Although methods based on dressed orbitals are a promising candidate to describe the strong-coupling regime of the light-matter interaction, their practical applicability is limited to a few modes. The most promising first-principles method to describe systems with many relevant modes is KS-QEDFT (Sec.~\ref{sec:qed_est:qedft}). One important reason is that calculations with KS-QEDFT have the same numerical costs as the semiclassical Maxwell-KS approach, which in equilibrium settings often even reduces further to the cost of a standard KS-DFT calculation (see Sec.~\ref{sec:qed_est:general:coupled_wavefunction}). However, to account for the quantum nature of the electron-photon interaction, new approximation strategies to the unknown photon-exchange-correlation (pXC) functional have to be investigated. Known strategies of KS-DFT either cannot be generalized in a straightforward way, such as the LDA, or the generalizations are less useful, such as the KLI approximation~\citep{Flick2018abinitio} for the photon-OEP (see Sec.~\ref{sec:qed_est:qedft} for details). 

For that, the insights from the dressed construction can contribute in two practical ways. First, we can try to derive useful pXC functionals by investigating the connection between polaritonic QEDFT and KS-QEDFT. For instance, such a connection can be established via the equations of motion of both approaches, assuming the same density~\citep{Tchenkoue2019}. Another option is to study the polaritonic and KS auxiliary systems of an analytically solvable reference systems, such as the homogeneous electron gas coupled to a few cavity modes~\citep{Rokaj2020}. This would allow to study the connection between both auxiliary constructions by comparing how they describe the same system. In the best case, we will identify some general rules that allow to derive pXC functionals from polaritonic exchange-correlation functionals.

Another very interesting approach to construct an approximate pXC functional is to employ the Breit approximation to the light-matter description. As the dressed construction, the Breit approximation removes the difficult electron-photon interaction term from the theory. However, this is done very differently in the latter case by an approximate expression of the vector potential in terms of the electric current, which is derived from the Maxwell's equations. This in turn allows to remove the photon-degrees of freedom entirely from the description and leads to a (purely electronic) Hamiltonian that conserves the particle number. The electron-photon interaction manifests here in an additional 2-body term that consequently can be approximated with standard methods for the Coulomb interaction. This idea was motivated by the encouraging results of polaritonic-structure methods that also employ approximations for the Coulomb-interaction operator to describe the electron-photon interaction. The research in this direction is already ongoing and the first results are promising.

\subsubsection*{Dressed orbitals for full minimal coupling}
Another research question is to understand whether we can generalize the dressed-orbital approach in a useful way beyond the cavity-QED setting and consider, e.g., full minimal coupling (Pauli-Fierz theory). It is clear that such a generalization could not be directly applicable in the sense of a polaritonic-structure theory, because the corresponding dressed orbitals would depend on too many photon coordinates. However, if we generalize the approach and construct the corresponding many-polariton space, this can already be useful to analyze this still very little understood setting from a different perspective. 

To apply the dressed construction to, e.g., the Pauli-Fierz Hamiltonian that considers full minimal coupling, we have to consider the full spatially-dependent (transversal) vector potential $\bA_{\perp}(\br)$. The polaritonic coordinates read then $(\br,\bA_{\perp}(\br))$ and the polaritonic density is $n(\br,\bA_{\perp}(\br))$ including the information of both the electronic one-body density and $\bA_{\perp}(\br)$. We would then only need to generalize the coordinate transformation that symmetrizes the auxiliary $\bA_{\perp}$-fields in a reasonable way (see Sec.~\ref{sec:dressed:est:prescription}). This should in principle be possible.

\subsubsection*{The dressed-orbital construction in different contexts}
Another very interesting application of the dressed-orbital construction is with regard to other multi-species situations. It seems possible, provided we can define species with the same number of particles, that one can instead of working with complicated multi-species wave functions, work with the combined density matrices and enforce the ensemble representability conditions on the subsystems. This does not necessarily need any dressed construction. For instance, think about the Schrödinger equation for electrons and nuclei/ions. Assuming that we have one kind of nuclei/ions we could express the combined density matrix in terms of electron-nuclei/ion pairs. It seems interesting to investigate the above procedure also in the context of such cases.  

\subsubsection*{The connection between hybrid statistics and RDM representability conditions}
A research question, that popped up when we developed the polariton ansatz, concerns the representability conditions of the polaritonic 1RDM $\gamma$ (and other polaritonic RDMs). With respect to the combined coordinates, $\gamma$ is a 1RDM, but if we separate the electronic and photonic parts, $\gamma$ becomes a $(1,1)$-body RDM (one electronic and one photonic coordinate) that is very similar to the electronic 2RDM. Importantly, only for the latter point of view, we know how the hybrid statistics of the polaritonic orbitals manifest in terms of the coordinates. This suggests that $\gamma$ has more complicated representability conditions than the simple ones that we enforce in polaritonic RDMFT (Sec.~\ref{sec:dressed:est:RDMFT}). Understanding these conditions would be valuable to improve polaritonic-structure methods. For instance, we have observed that all polaritonic HF calculations were variational, although the corresponding wave-function ansatz may violate the hybrid statistics (even if the Pauli-principle is fulfilled).\footnote{Specifically, one can construct polaritonic Slater determinants with the polariton ansatz(, i.e., every electronic orbital has an occupation number between zero and one), that do not correspond to an ensemble of wave functions that adhere to the full hybrid statistics.} 
If this is not a coincidence (which further calculations with polaritonic HF will show), it suggests that the single-reference ansatz in polaritonic coordinates with the polariton ansatz reduces the configuration space of the minimization with respect to the full many-polariton space. This is different from electronic variational 2RDM theory, which usually leads to energies that are a lower bound to the exact one. The reason is that not all $N$-representability conditions can be tested in practice and thus, the configuration space of the minimization is too large and includes elements that do not correspond to an  $N$-electron wave function. 

To answer such questions, we need to understand the representability conditions of the $(1,1)$RDM. Although it depends on electronic and photonic coordinates, the $(1,1)$RDM is much simpler than the $3/2$-body RDM, because it can be connected to all the other relevant RDMs of the system by sum rules. The $(1,1)$RDM is also a positive-semidefinite matrix,\footnote{Note that this follows directly from its definition.} such as the usual single-species RDMs. Thus, we believe that we can derive the representability conditions of the $(1,1)$RDM and more generally of any $(p,q)$RDM with a similar strategy as \citet{Mazziotti2012} employed to derive the $N$-representability conditions of the 2RDM.

\subsubsection*{Representability conditions of the $3/2$-body RDM}
A research question that accompanied me from the beginning of my PhD and that is directly connected to the previous question is about the representability conditions of the $3/2$-body RDM. In Sec.~\ref{sec:qed_est:rdms:general}, we have briefly discussed the RDM approach to study the coupled electron-photon space that to the best of our knowledge has never been investigated. We believe that understanding the representability conditions for the $3/2$-body RDM could be very valuable for further progress in the field. For instance, one could perform variational minimizations of the energy without the wave function analogously to variational 2RDM theory (see Sec.~\ref{sec:qed_est:rdms:general}). Importantly, such a method would not be limited to cavity QED, but could be generalized straightforwardly to the Pauli-Fierz level. 
A very interesting opportunity that opens up on this level is to investigate the full gauge freedom of the theory. There are gauges such as the temporal gauge~\citep{Barut1984}, for which the Coulomb interaction does not explicitly occur in the Hamiltonian, but instead is carried by the longitudinal part of the vector potential. Thus, the energy could be completely described only in terms of the electronic and photonic 1RDMs and the $3/2$-body RDM. If the representability conditions of the $3/2$-body RDM were easier to handle than the $N$-representability conditions of the 2RDM, this reformulation could be useful not only for coupled light-matter systems but also for difficult electronic-structure problems. For instance, most methods that accurately describe the strong-correlation regime can efficiently only take short-range interactions into account, but there are many systems, where this approximation breaks down. To include also the long-range part of the Coulomb interaction, there have been proposals to introduce additional artificial degrees of freedom (see, e.g., Ref.~\citep{Lanata2020}). The ``natural'' candidates for this are longitudinal photons that could be described by the $3/2$-body RDM. 

Although the $3/2$-body RDM has very different properties than the 2RDM, one could derive its representability conditions with a similar approach as proposed by Ref.~\citep{Mazziotti2012}. The first step of this programme would be to generalize the bipolar theorem~\citep{Kummer1967} that is the basis of the construction in Ref.~\citep{Mazziotti2012} to the more general case of coupled electron-photon systems. Based on this theorem, one can derive representability conditions from the positivity of linear combinations of RDMs. There is already work in progress on this direction and the first results are promising.

\subsubsection*{RDMFT in real space}
Finally, we want to address another question that came up during my PhD and that concerns our real-space implementation of RDMFT in \textsc{Octopus}. With the new conjugate-gradients algorithm, we are in the position to explore natural orbitals with the full flexibility of a real-space grid. This would allow to study, e.g., the limitations of standard quantum-chemical basis sets in RDMFT. We suppose that there are many scenarios in which the flexibility of the grid would allow to achieve converged results with less natural orbitals than such basis sets. The results presented in Sec.~\ref{sec:numerics:rdmft:comparison:cg_piris} provide a first indication for this although the there employed basis sets where considerably less optimized than usual quantum chemistry ones. To really compare real-space RDMFT to orbital-based RDMFT, we have to perform 3d calculations.  
However, in its current state the algorithm is numerically too expensive for that, because of the inefficient calculation of the exchange-integrals (see Sec.~\ref{sec:numerics:rdmft}). This is a known issue that also severely limits the applicability of HF and hybrid functionals of KS-DFT in real space (or for solids that are described in $k$-space). 
Another limitation of the current RDMFT implementation is the lack of pseudo potentials. These are crucial for real-space first-principles calculations, because they allow to converge the results with considerably larger grid spacings than an all-electron calculation (see Sec.~\ref{sec:numerics:rdmft}).

Recently, a new method has been proposed to calculate exchange integrals considerably more efficiently \citep{Lin2016}. Although this method has been tested only for HF theory and hybrid functionals of KS-DFT, it is highly probable that we can also apply it to RDMFT calculations, since the corresponding exchange integrals are equivalent to the Fock-exchange ones (see Sec.~\ref{sec:est:rdms:rdmft}). There is already work in progress in this direction (by a colleague) and the first tests seem to confirm this assumption. Naturally, this method would then also be applicable to polaritonic RDMFT.

Once, the implementation of this new method is finished, we can straightforwardly generalize the usual strategies to generate pseudo-potentials~\citep{Schwerdtfeger2011} to RDMFT functionals.

	\appendix
	\renewcommand\chaptername{Appendix}
	\chapter{Appendix}
\section{The bosonic symmetry of the photon wave function}
\label{app:symmetry_photon}
In this appendix we go into a little more detail and show how the mode-representation, which makes the bosonic symmetry explicit and which we discussed in Sec.~\ref{sec:qed_est:general:coupled_many_body}. We introduce in this setting the usual bosonic density matrices~\cite{Giesbertz2019}. Instead of starting with the displacement representation we start with the definition of the single-particle Hilbert space and its Hamiltonian. We choose the single-particle Hilbert space $\mathcal{H}_1$ to consist of $M$ orthogonal states $\ket{\alpha}$. These states are defined by the eigenstates of the Laplacian with fixed boundary conditions and geometry and correspond to the Fourier modes of the electromagentic field~\cite{Greiner2013, Ruggenthaler2014}. This real-space perspective is a natural choice if one either wants to connect to quantum mechanics and deduce the Maxwell field from gauge independence of the electronic wave function~\cite{Greiner2013}, or when deducing the theory in analogy to the Dirac equation~\cite{Gersten1999}. It is this analogy of Maxwell's equations as a single-photon wave function with spin 1 that makes the appearance of a bosonic symmetry most explicit when quantizing the theory~\cite{Keller2012}. We note, however, that in general the concept of a photon wave function can become highly  nontrivial~\cite{Scully1999}. Since we work directly in the dipole approximation we do not go through all the steps of the usual quantization procedure of QED but from the start assume that we have chosen a few of these modes $\ket{\alpha}$ (with a certain frequency and polarization) in Coulomb gauge~\cite{Ruggenthaler2014}. The single-particle Hamiltonian in this representation is then given by
\begin{align*}
\hat{h}_{ph}' = \sum_{\alpha=1}^{M} \omega_{\alpha} \ket{\alpha} \bra{\alpha}.
\end{align*} 
Since a total shift of energy does not change the physics and for later reference, we can equivalently use $\hat{h}_{ph} = \sum_{\alpha=1}^{M} \left( \omega_{\alpha} + \tfrac{1}{2} \right)\ \ket{\alpha} \bra{\alpha}$. Therefore, the energy of a single-photon wave function $\ket{\phi} = \sum_{\alpha=1}^{M} \phi(\alpha) \ket{\alpha}$ (corresponding to the classical Maxwell field in Coulomb gauge~\cite{Keller2012}) is given by
\begin{align*}
E[\phi] = \sum_{\alpha,\beta =1}^{M} \phi^{*}(\beta) \underbrace{\braket{\beta|\hat{h}_{ph}|\alpha}}_{= \hat{h}_{ph}(\beta,\alpha)} \phi(\alpha) = \sum_{\alpha, \beta}  \hat{h}_{ph}(\beta,\alpha) \gamma_{b}(\alpha, \beta) = \sum_{\alpha} \left(\omega_{\alpha}+ 
\frac{1}{2}\right)\underbrace{\gamma_b(\alpha,\alpha)}_{|\phi(\alpha)|^2}.
\end{align*}
Here we have introduced the single-particle photonic 1RDM $\gamma_{b}(\alpha, \beta) = \phi^{*}(\beta) \phi(\alpha)$. We can then extend the single-particle space and introduce photonic many-body spaces $\mathcal{H}_{N_{b}}$ which are the span of all \textit{symmetric} tensor products of single-particle states of the form~\cite{Giesbertz2019,Leeuwen2013}
\begin{align*}
\ket{\alpha_1,...,\alpha_{N_b}} = \frac{1}{\sqrt{N_b!}} \sum_{\wp} \ket{\wp(\alpha_1)}...\ket{\wp(\alpha_{N_b})},
\end{align*}
where $\wp$ goes over all permutations of $\alpha_1,..., \alpha_{N_{b}}$. This construction is completely analogous to the typical construction of the fermionic many-body space with the only difference having minus sings in front of odd permutations. Such a many-body basis is not normalized for bosons, as states can be occupied with more than one particle. Thus, the normalization factor occurs in the corresponding resolution of identity, i.e. $\mathbb{1}=\frac{1}{N_b!}\sum_{\alpha_1,...,\alpha_{N_b}=1}^{M}\ket{\alpha_1,...,\alpha_{N_b}}$ $\bra{\alpha_1,...,\alpha_{N_b}}$. This approach is explained in great detail in Ref. \citep{Stefanucci2013}. An $N_b$-particle Hamiltonian is then given by a sum of individual single-particle Hamiltonians (interactions among the photons will only come about due to the coupling with the electrons.) Introducing for a general $N_b$ photon state $\ket{\tilde{\phi}} =  \tfrac{1}{\sqrt{N_b!}}\sum_{\alpha_1,...,\alpha_{N_b}=1}^{M} \tilde{\phi}(\alpha_1,...,\alpha_{N_b}) \ket{\alpha_1,...,\alpha_{N_b}}$ with $\tilde{\phi}(\alpha_1,...,\alpha_{N_b})=\frac{1}{\sqrt{N_b!}}\braket{\alpha_1,...,\alpha_{N_b}|\tilde{\phi}}$ the corresponding 1RDM according to Eq.~\eqref{eq:qed_est:rdms:1rdm_photonic} as $\gamma_{b}(\alpha,\beta) = N_{b} \sum_{\alpha_2,...,\alpha_{N_{b}}} $ $\tilde{\phi}^*(\beta, \alpha_2,...,\alpha_{N_b})$ $\tilde{ \phi}(\alpha, \alpha_2,...,\alpha_{N_b})$, the energy of that state is given by
\begin{align*}
E[\tilde{\phi}] = \sum_{\alpha, \beta=1}^{M} \hat{h}_{ph}(\beta,\alpha) \gamma_{b}(\alpha,\beta) = \sum_{\alpha=1}^{M} \left(\omega_{\alpha} + \frac{1}{2} \right) \gamma_{b}(\alpha,\alpha).
\end{align*}
Such a state can be constructed, for instance, as a permanent of $N_b$ single-photon states $\phi(\alpha)$. Note further that the 1RDM of an $N_b$ photon state obeys $N_b = \sum_{\alpha} \gamma_{b}(\alpha,\alpha)$.

Finally, since we want to have a simplified form of a field theory without fixed number of photons, we make a last step and represent the problem on a Hilbert space with indetermined number of particles, i.e., a Fock space. By defining the vacuum  state $\ket{0}$, which spans the one-dimensional zero-photon space, the Fock space is defined by a direct sum of $N_b$-photon spaces $\mathcal{F}=\bigoplus_{N_b=0}^{\infty}\mathcal{H}_{N_b}$. Introducing the ladder operators between the different photon-number sectors of $\mathcal{F}$ by~\cite{Giesbertz2019}
\begin{align*}
&\hat{a}_{\alpha}^{+} \ket{\alpha_1,...,\alpha_{N_{b}}} = \ket{\alpha_1,...,\alpha_{N_{b}}, \alpha}
\\
&\hat{a}_{\alpha} \ket{\alpha_1,...,\alpha_{N_{b}}} = \sum_{k=1}^{N_b} \delta_{\alpha_k,\alpha} \ket{\alpha_1,..., \alpha_{k-1}, \alpha_{k+1},...,\alpha_{N_{b}}}
\end{align*}
with the usual commutation relations, we can lift the single-particle Hamiltonian to the full Fock space and arrive at Eq.~\eqref{eq:qed_est:rdms:PhotonHamiltonian}. The Fock space 1RDM for a general Fock space wave function $\ket{\Phi}$ can then be expressed as
\begin{align*}
\gamma_{b}(\alpha,\beta) = \braket{\Phi| \hat{a}^{+}_{\beta} \hat{a}_{\alpha} \Phi},
\end{align*}
and $\sum_{\alpha=1}^{M} \gamma_{b}(\alpha,\alpha) = N_b$ now corresponds to the \emph{average} number of photons. And finally, since we know that Eq.~\eqref{eq:qed_est:rdms:PhotonHamiltonian} is equivalent to $\hat{H}_{ph}= \sum_{\alpha=1}^M\left(- \tfrac{1}{2} \tfrac{\partial^2}{\partial p_{\alpha}^2}+ \tfrac{\omega_{\alpha}^2}{2}p_{\alpha}^2\right)$, we also see that the Fock space $\mathcal{F}$ is isomorphic to $L^{2}(\mathbb{R}^{M})$, which closes our small detour.

\section{Conjugate gradients algorithm for real orbitals}
\label{sec:app:cg_rdmft_real_orbitals}
In this appendix, we present an alternative derivation for the conjugate-gradients algorithm for RDMFT that is presented in Sec.~\ref{sec:numerics:rdmft:cg}.
Since we are only interested in ground states, we can also restrict the configuration space of the RDMFT Lagrangian to only real orbitals, so we have $\phi_i=\phi_i^*$ and 
\begin{align}
\tilde{L}=\tilde{L}[\phi_i,n_i]=&\tilde{E}[n_i,\phi_i]-\mu \left(\sum_{i=1}^{M}n_i-N\right)-\sum_{i,j=1}^{M}\lambda_{ji}\left(\int\td^3r\,\phi_i(\br)\phi_j(\br)-\delta_{ij}\right),
\end{align}
where we have to define the introduced total energy functional and the respective Hartree and XC-Potentials,
\begin{align}
\tilde{E}[n_i,\phi_i]=&-\frac{1}{2m}\sum_{i=1}^{M} n_i \int\td^3r \phi_i (\br) h(\br) \phi_i (\br)
+\frac{1}{2}\sum_{i=1}^{M} n_i \int\td^3r\, \phi_i (\br) \tilde{v}_{H}(\br)\phi_i (\br) \nonumber\\
&-\frac{1}{2}\sum_{i=1}^{M} \sqrt{n_i} \int\td^3r\, \phi_i (\br) \tilde{v}^i_{XC}(\br). \nonumber\\
\tilde{v}_{H}(\br)&=\sum_{j=1}^{M} n_j \td^3r'\, \phi_j(\br')  w(\br,\br') \phi_j(\br')\\
\tilde{v}^i_{XC}(\br)&=\sum_{j=1}^{M} \sqrt{n_j} \int\td^3r'\, \phi_j(\br) \phi_i(\br')  w(\br,\br') \phi_j(\br')
\end{align}
So the differential of L reads:
\begin{align}
\delta \tilde{L}=&4\sum_{i=1}^{M}\sin(\theta_i)\left[\frac{\partial \tilde{E}}{\partial n_i}-\mu\right] \td\theta_i +\sum_{i=1}^{M}\int\td^3r\, \left[\frac{\delta \tilde{E}}{\delta \phi_i(\br)}-\sum_{k=1}^{M}\lambda_{ik}\phi_k(\br)\right] \delta\phi_i(\br)
\end{align}
When we now require stationarity, $\delta \tilde{L}=0$, we arrive at two different Euler-Lagrange equations for every orbital i:
\begin{subequations}
\begin{align}
	\frac{\partial \tilde{E}}{\partial n_i}-\mu&=0 \\
	\frac{\delta \tilde{E}}{\delta \phi_i(\br)}- \sum_{k=1}^{M}(\lambda_{ik}+\lambda_{ki})\phi_k(\br)&=0 \label{eq:app:ELE_phi_real}
\end{align}
\end{subequations}
So in fact, only the (real-)symmetric part of $\Lambda$ enters the equations, which is the \emph{main} differences between the two formulations. For the sake of completeness, we also provide the modified orbital equations (the equations for $n_i$ remain structurally unchanged we just remove the stars from the orbitals):
\begin{align}
\frac{\delta \tilde{E}}{\delta \phi_i(\br)}=&n_i\left(\phi_i(\br)h(\br) +h(\br)\phi_i(\br)\right) + 2n_i\tilde{v}_H(\br)\phi_i(\br) + 2 \sqrt{n_i} \tilde{v}^i_{XC}(\br) \nonumber\\
\overset{h=h^+}{=}& 2 \left[n_i h(\br)\phi_i(\br)+ n_i\tilde{v}_H(\br)\phi_i(\br) + \sqrt{n_i} \tilde{v}^i_{XC}(\br)\right]\\
=&2\frac{\delta E}{\delta \phi_i(\br)}
\end{align}

The steepest descent-vector is now given as
\begin{align}
\label{eq:app:StDesVectorReal}
\tilde{\zeta}_i=-\frac{\delta \tilde{F}}{\delta \phi_i(\br)}&=-\left(2\frac{\delta \tilde{E}}{\delta \phi_i(\br)}-\sum_{k=1}^{M}(\lambda_{ki}+\lambda_{ik})\phi_k(\br)\right).
\end{align}
We can calculate $\lambda_{ki}+\lambda_{ik}$ then just by Eq.~\eqref{eq:app:ELE_phi_real}.
We have tested also this version in \textsc{Octopus} and it seems to work equivalently well than the other definition of Sec.~\ref{sec:numerics:rdmft:cg}.

\section{Gradient of the electronic natural occupation numbers}
\label{sec:app:gradient_non}
In this appendix, we show how to derive expression~\eqref{eq:numerics:hybrid:derivative_inequality} for the gradient of the inequality constraints~\eqref{eq:numerics:hybrid:constraint_inequality}.

We consider a symmetrized perturbation of the electronic 1RDM $\gamma_e$ by the $k$-th polaritonic orbital
\begin{align}
\delta \gamma_k(\br,\br') = 2\int\td q \phi_k^*(\br',q)\delta\phi_k(\br, q) + 2\int\td q \delta\phi_k^*(\br',q)\phi_k(\br, q),
\end{align}
such that $\delta \gamma_k(\br,\br') = \delta \gamma_k^*(\br,\br')$. We perform first-order perturbation theory on the equation of the i-th eigenvalue of $\gamma_e$,
\begin{align}
2 n_i \phi_i^e(\br) = \int\td\bx [\gamma(\br,\br') + \delta \gamma_k(\br,\br')] \phi_i^e(\br),
\end{align} 
which results in a correction of $n_i$, i.e.,
\begin{align*}
\delta n_i =& \int\td\br\td\br' \phi^e_i{}^*(\br) \delta \gamma_k(\br,\br')\phi_i^e(\br')\\
=& 2\int\td\br\td\br'\td q \phi^e_i{}^*(\br) \phi_k^*(\br',q)\delta\phi_k(\br, q) \phi_i^e(\br') + 2\int\td\br\td\br'\td q \phi^e_i{}^*(\br) \delta\phi_k^*(\br',q)\phi_k(\br, q) \phi_i^e(\br')\\
=& 2\int\td\br\td q \phi^e_i{}^*(\br) [\int\td\br'\phi_k^*(\br',q) \phi_i^e(\br')] \,\delta\phi_k(\br, q) + 2\int\td\br\td q \phi_i^e(\br) [\int\td\br'\phi^e_i{}^*(\br') \phi_k(\br',q) ] \, \delta\phi_k^e{}^*(\br, q)
\end{align*}
From this expression, we can derive
\begin{align}
\frac{\delta n_i}{\delta\phi_k(\br, q)} &= 2\phi^e_i{}^*(\br) [\int\td\br'\phi_k^*(\br',q) \phi_i^e(\br')] \\
\frac{\delta n_i}{\delta\phi_k^*(\br, q)} &= 2\phi_i^e(\br) [\int\td\br'\phi^e_i{}^*(\br') \phi_k(\br',q) ].
\end{align}

\newpage
\chapter{Validation of the occupation number optimization in RDMFT}
\label{sec:numerics:dressed:non_assessment}
In this appendix, we discuss the validation of the occupation number optimization part (algorithm~\ref{algorithm:RDMFT_non}) of the RDMFT routine. Although the generalization from electronic to dressed orbitals directly influences only the orbital optimization, there might still be an indirect influence on other parts of the routine. This is a typical challenge of nonlinear programming. 

In fact, during the implementation of the polaritonic RDMFT routine, an issue of this type has occurred: some approximations of the occupation number optimization algorithm can lead to pronounced inaccuracies in polaritonic RDMFT, but a similar problem has not been observed in electronic RDMFT. Specifically, the algorithm converges to a set of occupation numbers $\bn=(n_1,...,n_M)$ that do not sum up to the total particle number $N$. Depending on the specific scenarios, this sum might deviate very strongly from $N$ including $\sum_i n_i=0$. This issue is so severe that the polaritonic RDMFT routine cannot converge. If we apply instead the electronic RDMFT routine to the same matter system but outside the cavity, we find a converged result.

We want to utilize this issue to exemplify the challenges of the numerical implementation of first-principles methods and how we can confront them. A very important role is hereby played by the proper definition of a test case (Sec.~\ref{sec:numerics:dressed:non_assessment:test_setting}) that allows for a simple visualization (Sec.~\ref{sec:numerics:dressed:non_assessment:visualization}) of the involved quantities. This considerably facilitates the identification of the origin of the issue, i.e., the ``bug'' (Sec.~\ref{sec:numerics:dressed:non_assessment:identification_error}). In this and many other cases, it is not difficult to fix the bug, once it is identified (Sec.~\ref{sec:numerics:dressed:non_assessment:resolution}).


\section{Definition of the test setting}
\label{sec:numerics:dressed:non_assessment:test_setting}
The issue occurs in all versions of \textsc{Octopus} before 10.0 that support RDMFT and can be reproduced in the following test setting. We consider for the matter part a 1d-Helium-atom model, i.e.,  2 electrons confined in a soft-coulomb potential $v(x)=2/\sqrt{x^2+1}$ and interacting via $w(x,x'))=1/\sqrt{(x-x')^2+1}$ (see also Sec.~\ref{sec:dressed:results:implementation_real_space:validation}).  This serves as a test system for the purely electronic routine. Additionally, we couple this system to a photon mode with frequency $\omega=0.55$ and $\lambda=0.1$ as a test system for the dressed routine. 

We constrain the natural orbital basis in both cases by $M=3$ natural orbitals. We remind the reader that the orbitals $\{\phi_1,\phi_2,\phi_3\}$ are fixed during the occupation number optimization. Thus, the energy functional reduces \eqref{eq:numerics:rdmft:Mueller_energy_func} to
\begin{align}
E[\bn]=\sum_{i=1}^{3}n_i \epsilon^h_i+\tfrac{1}{2}\sum_{i,j=1}^{3}n_in_j\epsilon^{H}_{ij}-\tfrac{1}{2}\sum_{i,j=1}^{3}\sqrt{n_in_j}\epsilon^X_{ij},
\end{align}
where we introduced the vector of the occupation numbers $\bn=(n_1,n_2,n_3)$, and $\epsilon^{one}_i=\braket{\phi_i | h\phi_i}$, $\epsilon^H_{ij}=\braket{\phi_i\phi_i|w|\phi_j\phi_j}$, $\epsilon^X_{ij}=\braket{\phi_i\phi_j|w|\phi_j\phi_i}$ (see definitions below Eq.~\eqref{eq:numerics:rdmft:Mueller_energy_func}). In the dressed setting, the one-body kernel $h(x,x')$ is replaced by its dressed version $h'(xq,x'q')=t'(xq,x'q')+v'(xq,x'q')$ (Eq.~\eqref{eq:dressedkinetic} and~\eqref{eq:dressedpotential}) and the Coulomb kernel $w(x,x')$ that enters the latter two integrals is replaced by $w'(xq,x'q')$ as discussed in the previous section. 
The Lagrangian \eqref{eq:numerics:rdmft:lagrangian_general} that we denote in this section by $F$
becomes\footnote{Note that this is equivalent to  Eq.\eqref{eq:numerics:rdmft:non_optimization:lagrangian}, where we denoted $F$ with $L$.}
\begin{align}
F[\bn;\mu]=& E[\bn]-\mu S[\bn]\\
=& E[\bn]-\mu \left(\sum_{i=1}^{3}n_i-2\right)
\end{align}
and we have stationary conditions
\begin{align*}
\delta F =0 \Longleftrightarrow \quad \nabla_{\bn}E[\bn]&=\mu\\
S[\bn]&=0.
\end{align*}

\section{Visualization of the algorithm}
\label{sec:numerics:dressed:non_assessment:visualization}
Let us begin the discussion with a small illustration of algorithm~\ref{algorithm:RDMFT_non} with the electronic test system. As discussed in Sec.~\ref{sec:numerics:rdmft:algorithm_non}, the minimization of the $F[\bn;\mu]$ is performed with the help of an auxiliary function $\tilde{F}$\footnote{Note that this expression is equivalent to Eq.~\eqref{eq:numerics:non_opt_auxiliary_function} for $M=3$, where the auxiliary function is denoted by $\mathfrak{S}$.} that reads for this test-case
\begin{align}
\label{eq:numerics:dressed:non_problem:auxiliary_lagrangian}
\tilde{F}(\mu)=&\min_{n_1,n_2,n_3}F[\bn;\mu]\\
\equiv&F[\bn^*;\mu],
\end{align}
where in the second line we indicated with $\bn^*$ the set of occupation numbers that minimize $\tilde{F}(\mu)$ for a prescribed $\mu$. The problem of finding $\nabla_{\bn,\mu} L=0$ is thus transformed in a two-step procedure that involves solving a minimization problem to determine $\tilde{F}(\mu)$ and then finding the stationary point of $F[\bn^*;\mu]$ that fulfils $\tfrac{\td F}{\td \mu}=0$. The algorithm in \textsc{Octopus} exploits this reformulation by using a bisection algorithm to find the set of occupation numbers ${n_1^*(\mu),n_2^*(\mu),n_3^*(\mu)}$ that satisfies the side condition $S(\mu)=\sum_{i=1}^{3}n_i^*(\mu)-2=0$ (see algorithm~\ref{algorithm:RDMFT_non}).

We now want to visualize this procedure for our test system. For that we employ the computer-algebra software \textsc{Mathematica}\footnote{\url{https://www.wolfram.com/mathematica/}} to solve $\min_{n_1,n_2,n_3}F[\bn;\mu]$ with its highly accurate internal routine. The matrix elements $\epsilon^{one}_i,\epsilon^H_{ij},\epsilon^X_{ij}$ that enter F are extracted from the RDMFT routine of \textsc{Octopus}.  

\begin{figure}[ht]
	\begin{overpic}[width=0.49\columnwidth] {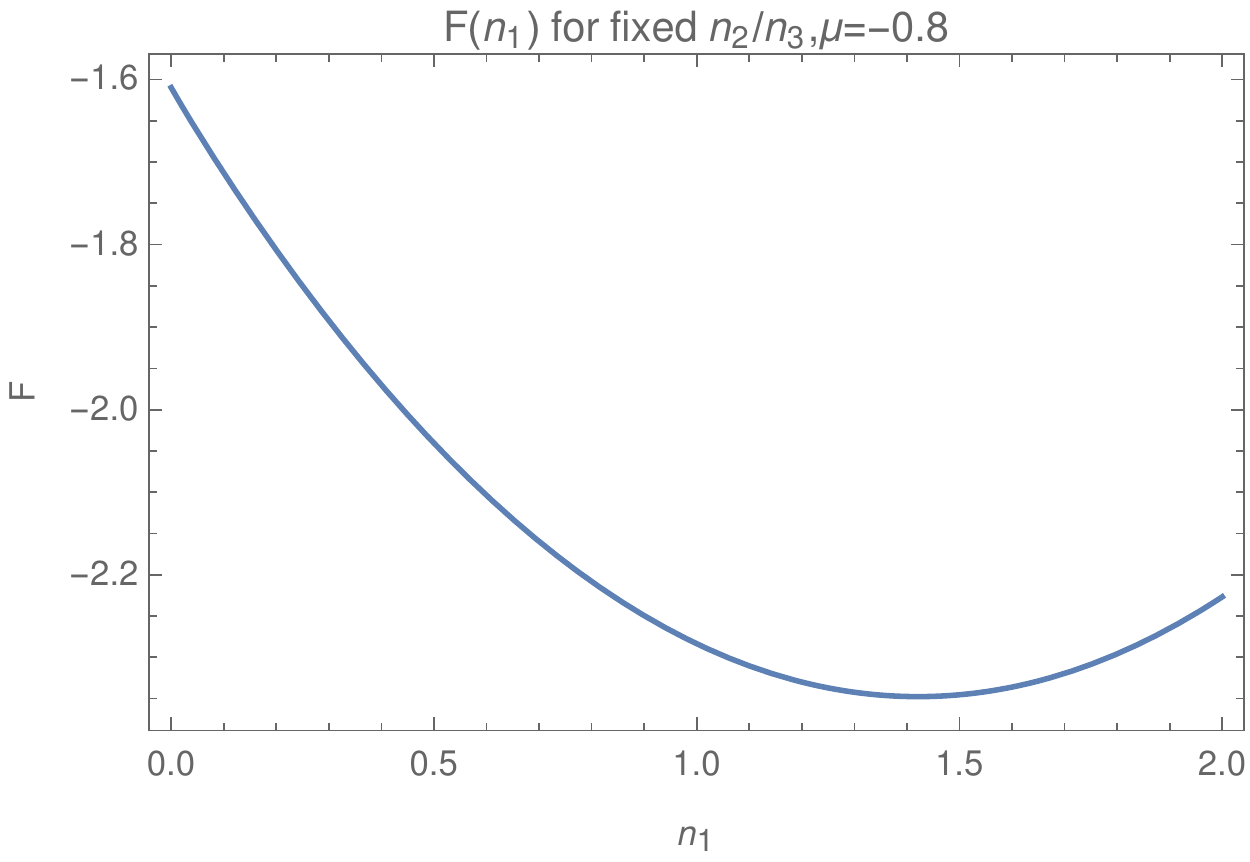}
		\put (85,55) {\textcolor{black}{(a)}}
	\end{overpic}
	\hfill
	\begin{overpic}[width=0.49\columnwidth] {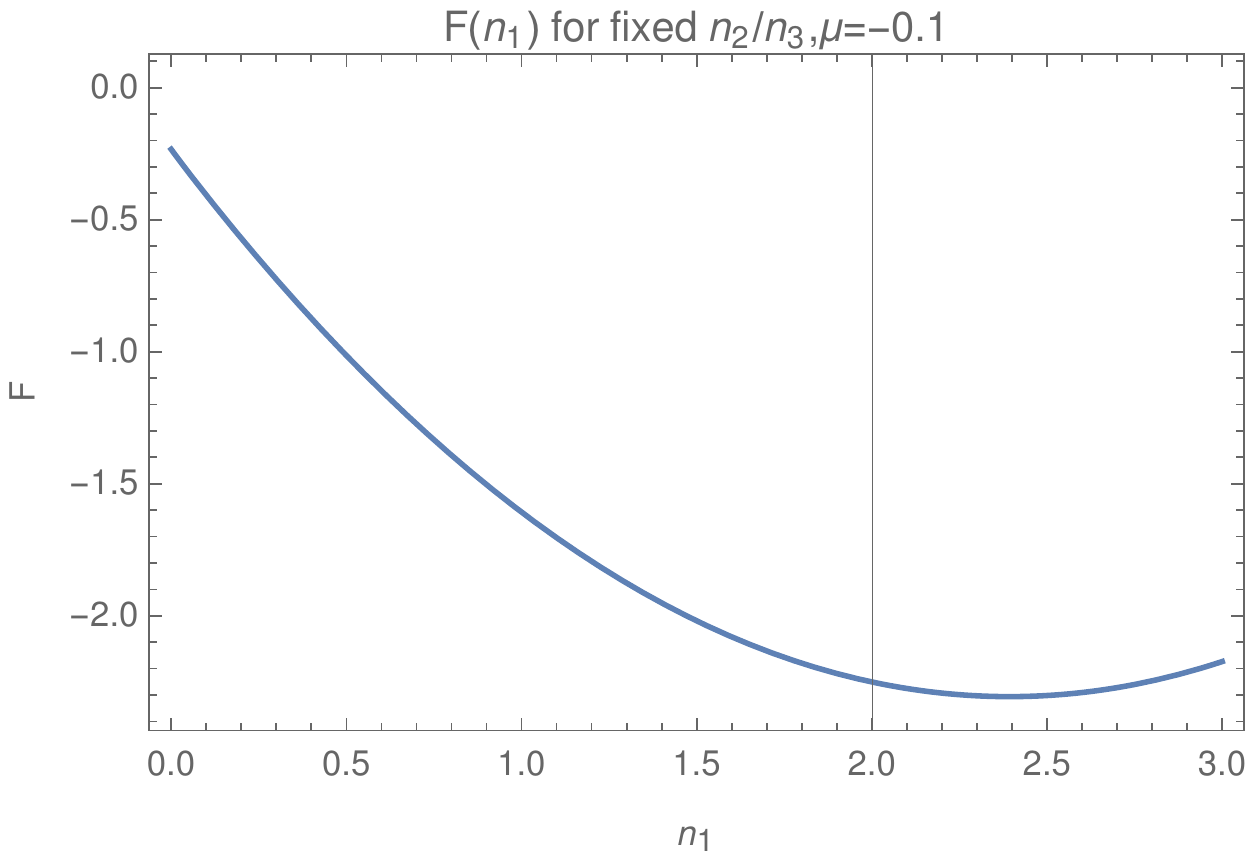}
		\put (85,55) {\textcolor{black}{(b)}}
	\end{overpic}
	\caption{Shown is the Lagrangian $F$ as a function of the first occupation number $n_1$ keeping the other occupation numbers $n_2,n_3$ and the Lagrange multiplier $\mu$ fixed. We see the two different types of minima that occur depending on the value of $\mu$. In figure (a), we see a normal minimum for $\mu=-0.8$, and in figure (b) the border minimum  for $\mu=-0.1$ that occurs because the occupation numbers are constraint to values between 0 and 2 (the vertical line denote the latter border). In this case, we say that the occupation numbers are pinned. Note that in the chosen setting $n_2/n_3$ do not vary much with $\mu$ and can therefore be ignored.}
	\label{fig:border_vs_normal_minimum}
\end{figure}

We start by noting that $F[\bn;\mu]$ for constant $\mu$ \emph{must} always have at least one (and possibly more than one) minimum because of the constraint $0\leq n_i \leq 2$. Put differently, $F$ can have ``normal'' minima, where $\nabla F=0$ or border minima and both cases are visualized in Fig.~\ref{fig:border_vs_normal_minimum}. In the latter case, we call the occupation number pinned.

\begin{figure}[ht]
	\centering
	\includegraphics[width=0.48\columnwidth]{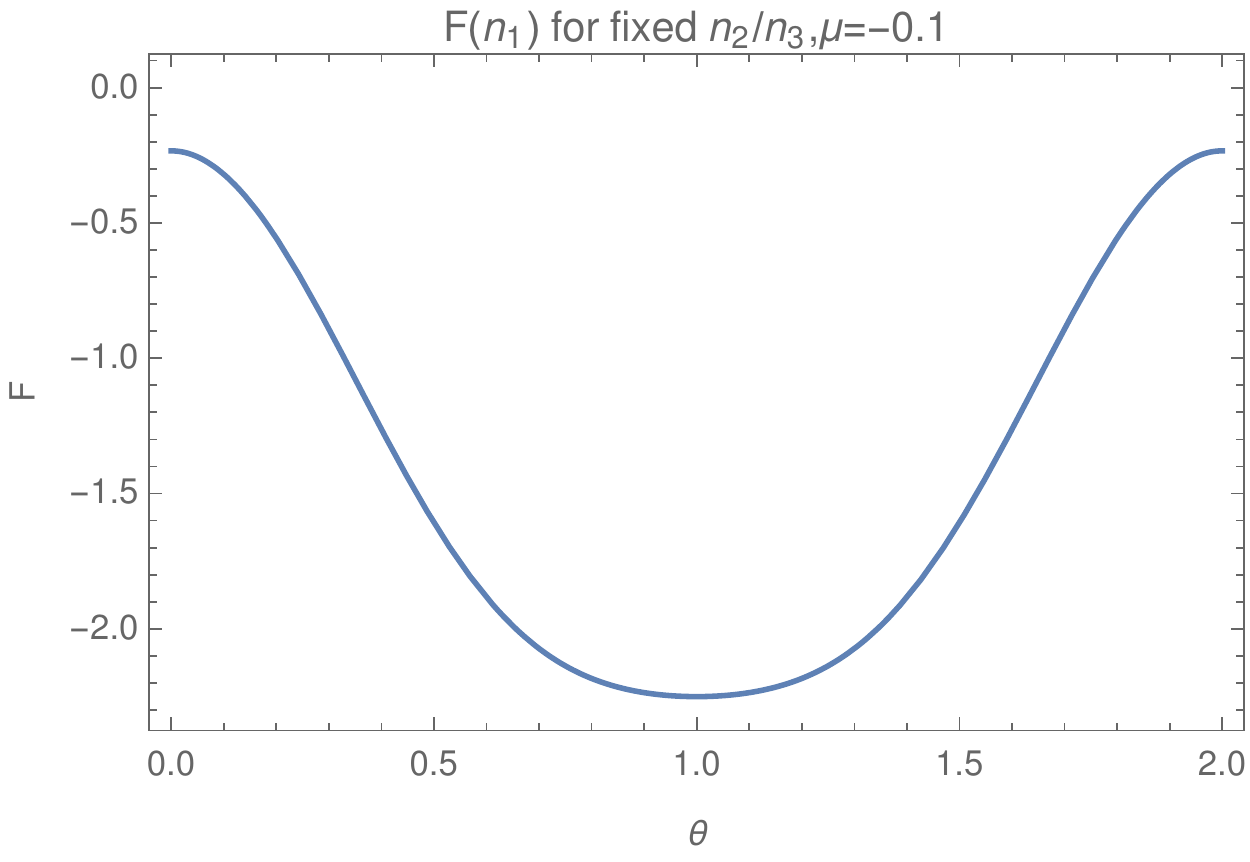} 
	\caption{We show the Lagrangian F for the same parameters as in Fig.\ref{fig:border_vs_normal_minimum} (b), but re-expressing the $n_i$ by Eq.~\eqref{eq:numerics:n_theta}. We see that the border minimum of F($\bn$) turns to a regular minimum of $F(\bm{\theta})$.}
	\label{fig:border_minimum_theta}
\end{figure}

To enforce these constraints, we either have to choose an algorithm that can perform a \emph{constrained minimization} or we replace the occupations 
\begin{align}
\tag{cf. \ref{eq:numerics:n_theta}}
n_i=2\sin^2(2\pi\theta_i) \quad\forall\,i,
\end{align}
as variables by the angles $\theta_i$ and perform an unconstrained minimization. The latter is a very common approach, because there are many standard algorithms for unconstrained optimization. In this case, $F$ as a function of $\bm{\theta}=(\theta_1,\theta_2,\theta_3)$ has a true minimum at the boundary, as shown in Fig.~\ref{fig:border_minimum_theta}.

\begin{figure}[ht]
	\begin{overpic}[width=0.49\columnwidth] {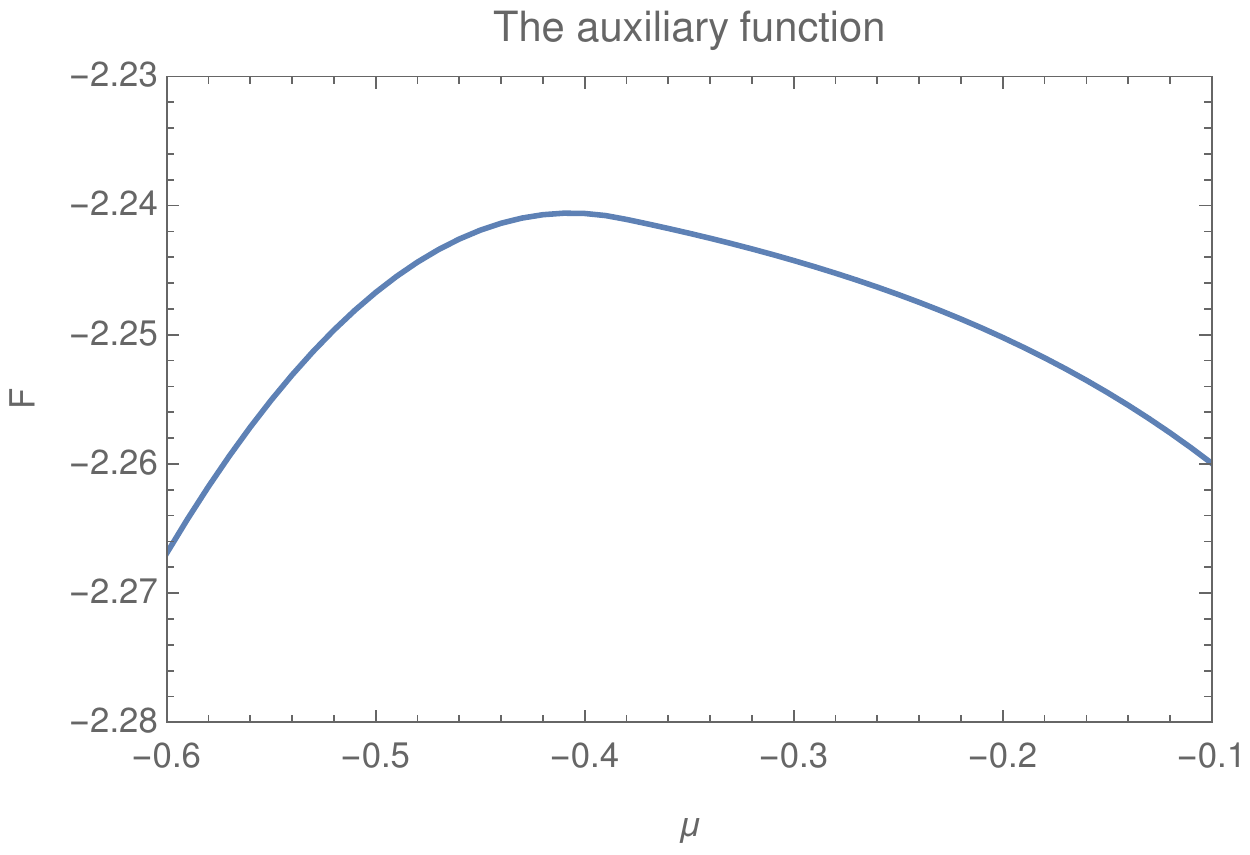}
		\put (20,50) {\textcolor{black}{(a)}}
	\end{overpic}
	\hfill
	\begin{overpic}[width=0.49\columnwidth] {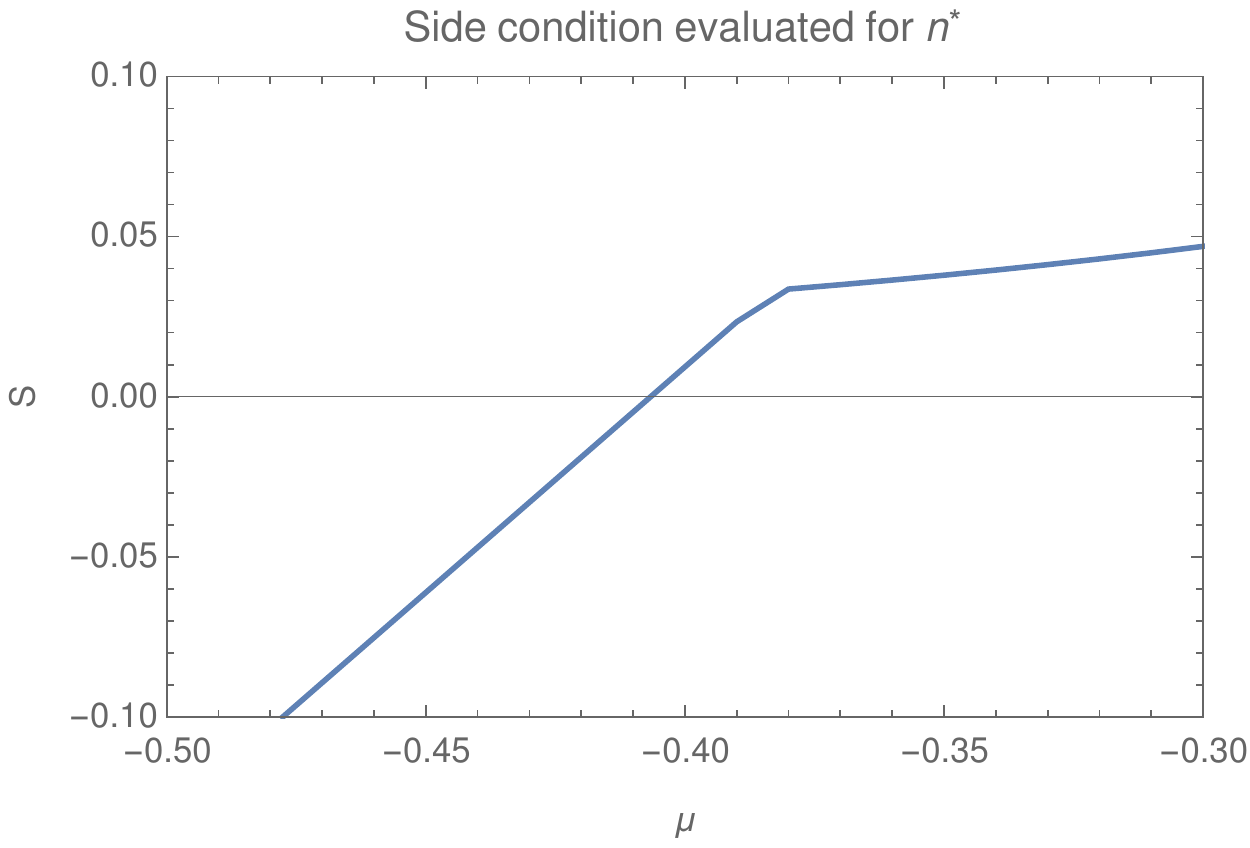}
		\put (20,50) {\textcolor{black}{(b)}}
	\end{overpic}
	\caption{The auxiliary function $\tilde{F}_{\bn^*}(\mu)$ (a) and the side condition $S_{\bn^*}(\mu)$ (b) are shown. $\tilde{F}_{\bn}(\mu)$ has its maximum at the point $\mu^*$, where the side condition is satisfied, i.e., $S_{\bn}(\mu^*)=0$ has a root. We also see that $S_{\bn}(\mu)$ is monotone over the whole region of $\mu$ and has a kink at $\mu\approx-0.38$. For this point of the normal minimum $\bn^*(\mu)$  becomes a border minimum, i.e., the first occupation number reaches $n_1^*=2$  and keeps this value also for larger value of $\mu$ because of the constraint (``pinning'').}
	\label{fig:F_S_mu}
\end{figure}

Since our test system is very small, we can afford enough calculations to actually plot the auxiliary function $\tilde{F}(\mu)$ (shown in Fig.~\ref{fig:F_S_mu} (a)), which allows us to visualize its properties. The bisection algorithm of \textsc{Octopus} calculates $\tilde{F}(\mu)$ of course only for certain values of $\mu$.
We see in Fig.~\ref{fig:F_S_mu} (a) that $\tilde{F}(\mu)$ has indeed a stationary point at $\mu^*\approx-0.41$, but it is actually a maximum  and not a minimum as one could have expected naively. In fact, that stationary point might correspond even to a saddle point in certain settings. The simple reason is that $\tilde{F}(\mu)$ is not related to the energy, in fact  $\tilde{F}(\mu)$ corresponds per definition to a minimal energy solution for \emph{all} $\mu$. The advantage of the algorithm is that it does not matter, which type of stationary point $\tilde{F}(\mu)$ has, because we do not directly consider $\tilde{F}(\mu)$. Instead, the problem is transferred to finding the root of the side condition $S(\mu)$. If $S(\mu)$ is \emph{monotone}, the bisection algorithm will always converge. This cannot be proven rigorously, but judging from the experience with another code\footnote{The algorithm was copied from the well-tested \textsc{Hippo}-code. It is not publicly available and the interested reader is referred to \url{lathiot@eie.gr}.} it seems to be a reasonable assumption. For our test-case, $S(\mu)$ is indeed monotone as we can see clearly in Fig.~\ref{fig:F_S_mu} (b).
Another interesting observation is that, while $\tilde{F}(\mu)$ is everywhere smooth, $S(\mu)$ has a kink, which occurs exactly, when the minimum of $F(\bn,\mu)$ touches the border. 

\section{The identification of a resolved bug of polaritonic RDMFT}
\label{sec:numerics:dressed:non_assessment:identification_error}
We now apply the just introduced visualization tools to study the failure of the occupation number optimization algorithm of polaritonic RDMFT. This means that the algorithm converges to $\bn^*$ with $S[\bn^*]\neq 0$. For a better comparison with the results of \textsc{Mathematica}, we have manipulated the occupation number routine slightly to evaluate the function $S(\mu)$ for a larger range of $\mu$ than the algorithm would require. 

\begin{figure}[ht]
	\begin{overpic}[width=0.49\columnwidth] {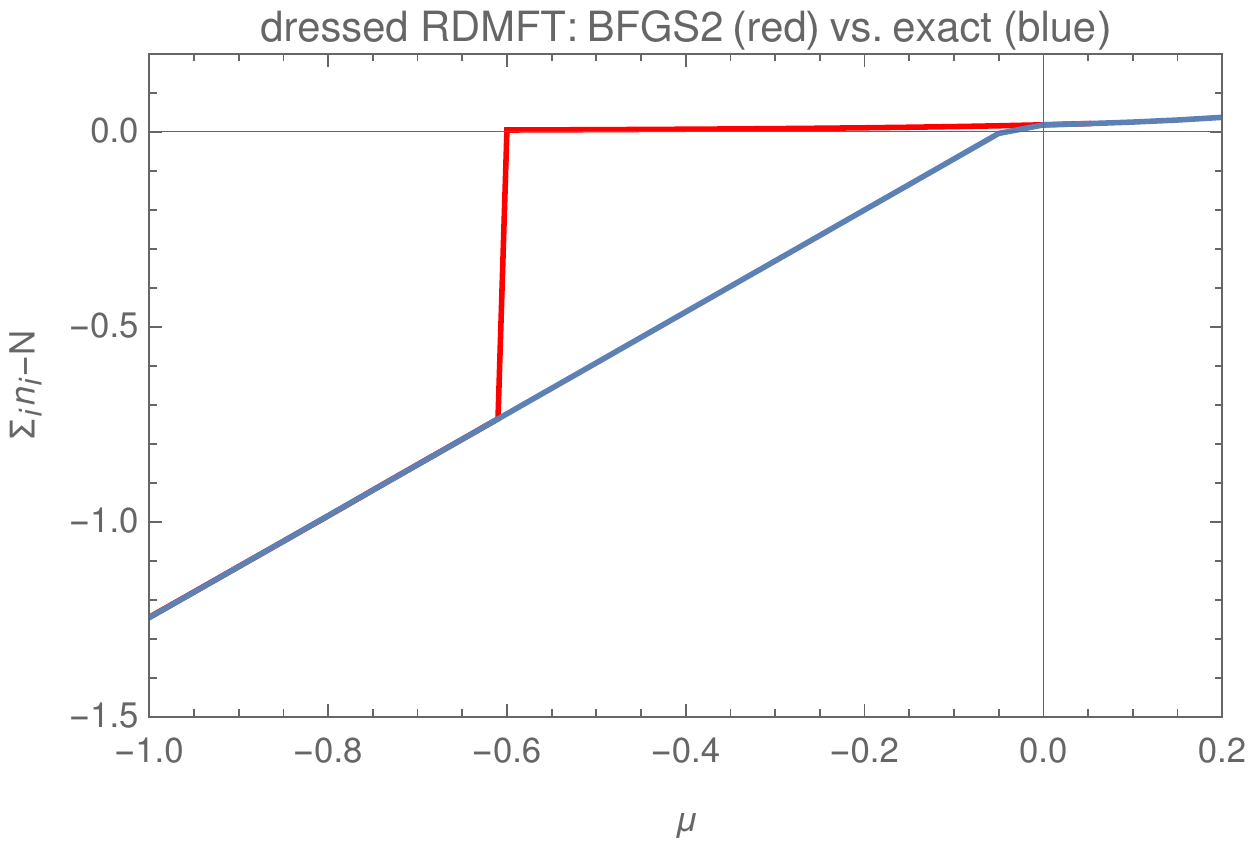}
		\put (20,50) {\textcolor{black}{(a)}}
	\end{overpic}
	\hfill
	\begin{overpic}[width=0.49\columnwidth] {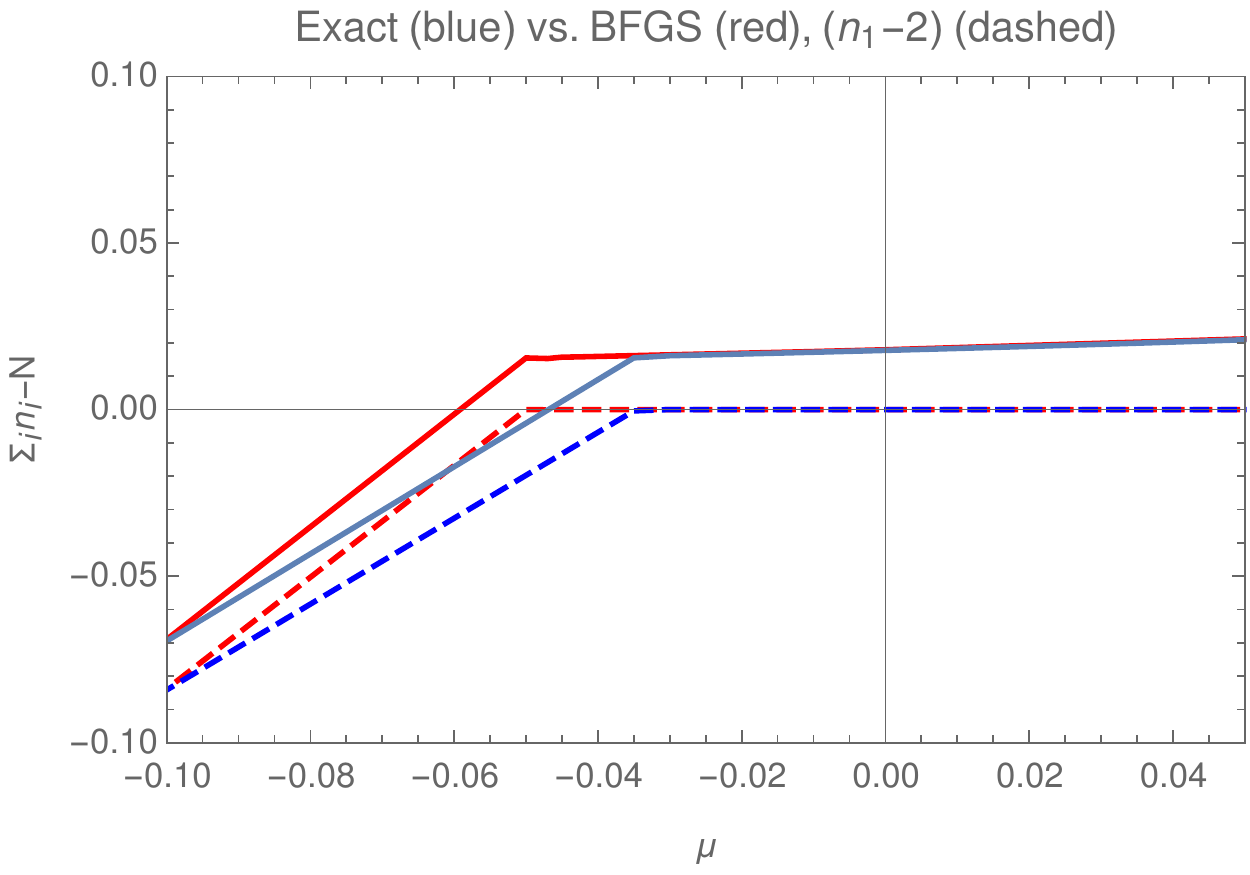}
		\put (20,50) {\textcolor{black}{(b)}}
	\end{overpic}
	\caption{(a) Comparison of the function $S(\mu)$ , calculated by the very accurate routine of \textsc{Mathmatica} (exact, blue) and the BFGS2 method (red) from the GSL-library. The latter fails to find the correct minimum of $\tilde{F}[\bn;\mu]$ for $\mu>-0.6$, which leads to a non-smooth behavior of $S(\mu)$.\newline
		(b) Same as (a) but comparing \textsc{Mathematica} (exact, blue solid line) with the BFGS method (red solid line). Additionally the function $n_1-2$ is plotted in both cases (\textsc{Mathematica}: dashed (dark) blue line, BFGS: dashed red line). We see that BFGS leads to a continuous $S(\mu)$, but it still predicts a slightly wrong root. Additionally, it predicts a boundary minimum of $n_1$ ($n_1 -2=0$) for too small values of $\mu$.
	}
	\label{fig:S_exact_BFGS_BFGS2}
\end{figure}

The RDMFT routine of \textsc{Octopus} 9.2 and below uses the fast version of the Broyden-Fletcher-Goldfarb-Shanno (BFGS) method\footnote{See for example \citep[part 6.1]{Nocedal2006}.} of the GSL\footnote{\url{https://www.gnu.org/software/gsl/doc/html/multimin.html}} library (called \emph{BFGS2}). Fig.~\ref{fig:S_exact_BFGS_BFGS2} (a) shows the comparison of BFGS2 with the results from the accurate routine of \textsc{Mathematica}: We see that while the function $S(\mu)$, obtained from \textsc{Mathematica} smoothly increases with $\mu$ (blue line), $S(\mu)$ obtained from the BFGS2 algorithm jumps for $\mu>-0.6$ to $S\approx 0$. This coincides with a jump in the solution occupation number to $n_1\approx 2$ (not shown). This comparison shows that the BFGS2 algorithm fails in certain cases drastically to solve the minimization problem~\eqref{eq:numerics:dressed:non_problem:auxiliary_lagrangian}. 

If we employ instead the standard BFGS algorithm of the same library, the accuracy increases, as Fig.~\ref{fig:S_exact_BFGS_BFGS2} (b) shows. We see that BFGS reproduces the curve of \textsc{Mathematica} considerably better than BFGS2, but still not exactly. Also here the occupation number $n_1$ gets pinned for too small values of $\mu$. 

Nevertheless, this comparison shows that the minimization algorithm, employed in the RDMFT routine is the origin of the inaccuracy.

\section{A simple resolution of the issue}
\label{sec:numerics:dressed:non_assessment:resolution}
Let us therefore briefly discuss the BFGS-method.
The basic idea behind the algorithm is the Newton method for minimization: we approximate the minimum of a function (stationary point) with a second order Taylor expansion. In the 1d-case, this requires merely the second order-derivative of the function (since the first-order derivative is zero by definition), i.e., the calculation of one term. 
However, in a multi-dimensional setting, the derivative has to be generalized to the Hessian $H$. In thise case, the second-order Taylor expansion around the extremum $\bn^*$ reads
\begin{align*}
F(\bn^*+\bn)-F(\bn)=F(\bn^*)+\tfrac{1}{2}\bn^T \,H_F\,\bn.
\end{align*}
For the very high-dimensional minimization problems in electron structure theory, only the calculation of the gradient is the bottleneck, i.e., the most expensive step, of the whole method. Calculating the Hessian is usually numerically too expensive and hence, one either needs to employ directly gradient-based methods or approximate the Hessian. The conjugate-gradients algorithm that is employed for the orbital optimization in \textsc{Octopus} is an example for the former option and the BFGS and BFGS2 algorithms are examples of the latter category. In particular, the BFGS algorithms are so-called \emph{quasi-Newton methods}, which approximate the Hessian $H_F$ by the gradient $\nabla F$, such that the so-called \emph{secant equation},
\begin{align*}
\nabla F(\bn^*+\bn)=\nabla F(\bn_0)+H_F\,\bn,
\end{align*}
is satisfied.
As this equation does not fully determine $H_F$, there are many different quasi-Newton methods that are adopted to certain minimization problems.\footnote{For further details, the reader is referred to any standard textbook about optimization, e.g., the book by \citet{Nocedal2006}.} Note that typical quasi-Newton methods (including BFGS and BFGS2) can only perform \emph{unconstrained} minimizations.

We can conclude that certain approximations of the quasi-Newton algorithms BFGS and BFGS2 must lead to the observed inaccuracies. The best way to avoid this issue is to employ a method that does not approximate the Hessian. 
However, the inaccuracies are much more pronounced for BFGS2 than for BFGS and thus, in a first attempt to resolve the issue, one could simply employ BFGS. We have tested this extensively and indeed, we have not observed convergence issues of the RDMFT routine. We have concluded that the identified inaccuracies of BFGS (Fig.~\ref{fig:S_exact_BFGS_BFGS2} (b)) are small enough to still ensure the overall convergence. From \textsc{Octopus} version 10.0 onwards, BFGS is therefore employed as default in the RDMFT occupation number optimization routine.

\chapter{Convergence of dressed orbitals in \textsc{Octopus}}
\label{sec:numerics:dressed:convergence}
In this appendix, we present in detail how we have validated the dressed orbital implementation of \textsc{Octopus}. We therefore have developed a strategy to systematically converge the numerical calculations with dressed orbitals for HF theory and RDMFT, respectively. All numerical results of Ch.~\ref{sec:dressed:results} in the in real-space setting have been converged according to this strategy. We present the validation and the convergence study with an example system (Sec.~\ref{sec:numerics:dressed:convergence:exact}-\ref{sec:numerics:dressed:convergence:RDMFT}) and then provide a summary in the form of a ``HowTo'' for such calculations in \ref{sec:numerics:dressed:convergence:protocol}. This appendix is based on the supporting information of Ref.~\citep{Buchholz2019}.

\subsubsection*{Validation with the help of a second implementation}
To find the accuracy threshold of our implementation we compared the \textsc{Octopus} results to calculations of a private code \textsc{Dynamics} that was used and validated for Ref. \citep{Nielsen2018} by S.E.B. Nielsen.\footnote{Contact: \href{mailto:soerenersbak@hotmail.com}{soerenersbak@hotmail.com}}
In \textsc{Dynamics}, the polaritonic orbitals are approximated on discretized real-space boxes in the $x$ and $q$ coordinate exactly as in \textsc{Octopus}. However, both codes employ different boundary conditions, a different level of finite differences for the approximation of the differential operators (fourth order in \textsc{Octopus} vs sixth order in \textsc{Dynamics}), a different method of the grid point evaluation (on-point in \textsc{Octopus} vs. mid-point in \textsc{Dynamics}), and also a different orbital optimization technique. Both codes are only able to perform calculations under the fermion ansatz, i.e., solutions may violate the Pauli-principle for electrons. We have in \textsc{Dynamics} the option to calculate the exact many-body ground state (for very small systems) as well as the KS-DFT ground state under the exact exchange (EXX) approximation. As for a two-electron spin-singlet system EXX \emph{coincides} exactly with HF theory, we can assess the electronic HF and the dressed HF routine (that we abbreviate in the following by dHF) with \textsc{Dynamics}.

\subsubsection*{The test system} 
We present the convergence studies at the example of the one-dimensional Helium atom inside a cavity (He) that we employed already before (Sec.~\ref{sec:dressed:results:implementation_real_space:validation} and Sec.~\ref{sec:numerics:dressed:non_assessment:test_setting}). The He atom is described by the nuclear potential $v_{He}(x)=-\frac{2}{\sqrt{x^2+1}}$ and is coupled to one cavity mode with frequency $\omega$ and coupling parameter $\lambda$. The modified local potential due to the dressed auxiliary construction reads $v_{He}'(x,q)=v_{He}(x) + v_d(x,q)$ with $v_d(x,q)=\tfrac{1}{2} (\lambda x)^2 + \tfrac{\omega^2}{2} q^2 - \tfrac{\omega}{\sqrt{2}}q (\lambda x)$ and the modified interaction kernel is $w'(xq,x'q')= w(x,x')+w_d(xq,x'q')$ with $w(x,x')= \frac{1}{\sqrt{(x-x')^2+1}}$ and
$w_d(xq,x'q')= -\frac{\omega\lambda}{\sqrt{2}}(xq'+x'q) + \lambda^2 xx'$. 

\subsubsection*{The principal steps of the convergence study}
The convergence study and validation of our implementation in \textsc{Octopus} requires several steps, which correspond to the following sections. We start with converging the exact dressed many-body ground-state and compare it to the results of \textsc{Dynamics} (Sec.~\ref{sec:numerics:dressed:convergence:exact}). Although we are limited to a very small Hilbert space for such exact calculations, we only have to solve a linear eigenvalue problem, so it is easy to converge the results to a very high precision within the accessible parameter range. Thus, we obtain an \emph{upper bound} for the accuracy from these exact calculations. 
In the next section, we present the convergence of dressed HF (dHF)\footnote{Note that we use the term dressed HF/RDMFT in stead of polaritonic HF/RDMFT in this appendix to stress the fact that we employ dressed orbitals with the fermion approximation.} in all its details and compare the converged ground-state to \textsc{Dynamics} (Sec.~\ref{sec:numerics:dressed:convergence:HF}). This allows us to determine the accuracy for dHF calculations, that is (because of its nonlinear nature) harder to converge than the exact solver. Additionally, the comparison of the two codes is a good validation of the correct implementation of the dressed modifications. Since these modifications are exactly the same for dressed RDMFT (that we abbreviate by dRDMFT in the following) and dHF, validating the dHF implementation validates also the corresponding changes for dRDMFT. In Sec.~\ref{sec:numerics:dressed:convergence:RDMFT}, we discuss the convergence of dRDMFT, which requires two different minimization procedures that are interdependent, and thus again is harder to converge than dHF.

All these sections are organized similarly. We start with the separated problem of the electronic system outside the cavity and converge first the electronic counterparts of the routines, i.e., the electronic exact solver, HF or RDMFT. 
The purely photonic problem of the separated problem remains the same at all the discussed levels of theory. We therefore discuss it only once in the first section. Then, we analyze the convergence of the dressed theories in the no-coupling ($\lambda=0$) limit, which theoretically means that the electronic and photonic problems are perfectly separated, but solved in only one large calculation. By comparison of only the electronic (photonic) part of the dressed solutions to the results of the purely electronic (photonic) theories, we can measure if there is a decrease in accuracy due to the simultaneous description of electronic and photonic coordinates. We finish the sections with a convergence study for $\lambda>0$.

\section{Validation of the exact dressed many-body ground-state}
\label{sec:numerics:dressed:convergence:exact}
We start with the validation of the exact many-body ground-state. In both codes, this is calculated by minimizing directly the energy expression of the full many-body wave function (denoted by $\Psi'$ in the main text), discretized on the grid. For the He test system, $\Psi'=\Psi'(x_1,q_1,x_2,q_2)$ is four-dimensional. The actual minimization is performed in \textsc{Octopus} by conjugate-gradients (see Sec.~\ref{sec:numerics:rdmft:cg}), whereas \textsc{Dynamics} makes use of a Lanczos algorithm.\footnote{This algorithm can be seen as a generalization of the conjugate-gradients methods. It was originally proposed in 1950 by \citet{Lanczos1950} and is widely used today. It is explained in most standard textbooks about optimization, e.g., in \citet{Nocedal2006}.}

\begin{figure}[ht]
	\centering
	\begin{overpic}[width=0.49\columnwidth]
		{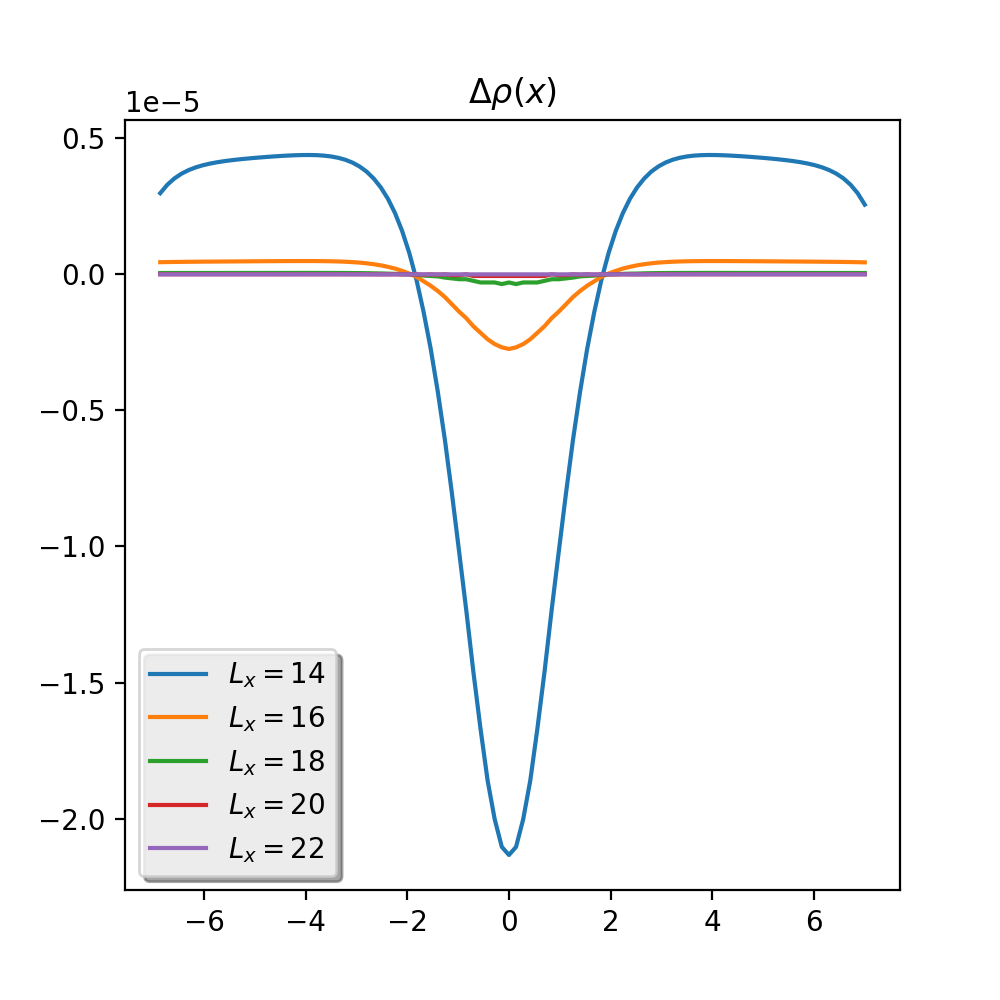}
		\put (85,13) {\textcolor{black}{(a)}}
	\end{overpic}
	\hfill
	\begin{overpic}[width=0.49\columnwidth]
		{img/numerics/convergence_octopus/example_lx_series.png}
		\put (85,13) {\textcolor{black}{(b)}}
	\end{overpic}
	\caption{As an example for the convergence tests, we show here the two spatially-resolved convergence parameters $\Delta\rho_{L_x}$ and $\Delta\rho_{\Delta_x}$ in (a) and (b), respectively. We see for in figures that the deviations can change the sign and are typically most pronounced in the center region.  We also observe how the deviations $\Delta\rho_{L_x}$ ($\Delta\rho_{\Delta_x}$) decrease at all positions with increasing (decreasing) $L_x$ ($\Delta_x$). These are generic features of \emph{all} performed series.
	However, this decrease continues until $\Delta\rho_{L_x}<10^{-8}$, reaching the limit set by the convergence parameter $\epsilon_{\rho}$ for the $L_x$-series (not shown). The $\Delta_x$-series instead saturates at $\Delta\rho_{\Delta_x}\leq 10^{-7}$. The reason is the finite difference approximation that effectively leads to different Hamiltonians for every value $\Delta_x$ of the spacing. To push the accuracy beyond $\epsilon_{\rho_{\Delta_x}}=10^{-7}$, one would need to adjust the finite-differences stencil accordingly. However, this accuracy is clearly beyond our needs, since the goal is to converge the numerically more involved theory levels HF and RDMFT, which are in general less accurate.}
	\label{fig:example_dx_dl_series}
\end{figure}

At the beginning of every calculation, we need to find the proper grid, which is defined by the box sizes $L_x,L_q$ and the spacings $\Delta_x,\Delta_q$ in both dimensions.  For the minimization in \textsc{Octopus}, we use two different convergence criteria, $\epsilon_E=10^{-9}$ and $\epsilon_{\rho}=10^{-8}$. The former tests the energy deviations and the latter the integrated absolute value of the density deviations between subsequent iteration steps.\footnote{Details can be found on \url{http://octopus-code.org/doc/develop/html/vars.php?page=alpha} with the keywords \emph{EigensolverTolerance} and \emph{ConvRelDens}.} These are the criteria already available in \textsc{Octopus}, so we here show that these are sufficient to produce reliable results. Note that we choose $\epsilon_E=\frac{1}{10}\epsilon_{\rho}$, because $\epsilon_{\rho}$ is a much stricter criterion.\footnote{This is also recommended by the \textsc{Octopus} Variable Reference.}  For the box size and spacing convergence, we perform series $\mathcal{C}=\{ \mathcal{C}^1,\mathcal{C}^2,\dots \}$, where $\mathcal{C}$ can be $L_x,L_q,\Delta_x,$ or $\Delta_q$ in the following. We perform two types of convergence tests for every parameter $\mathcal{C}$. In the first one, we investigate the deviations in the energy between subsequent elements $\Delta E_{\mathcal{C}^i}=E_{\mathcal{C}^i}-E_{\mathcal{C}^{i-1}}$. We denote the corresponding thresholds with $\epsilon_{E_{\mathcal{C}}}$. The second type of convergence series considers the maximal deviations in the absolute-value of the electronic/photonic part of the polaritonic density $\Delta \rho_{\mathcal{C}^i}=\max_{x/q}|\rho_{\mathcal{C}^i}(x/q)-\rho_{\mathcal{C}^{i-1}}(x/q)|$, with $\rho_{\mathcal{C}^i}(x)=\int\td q\rho_{\mathcal{C}^i}(x,q)$ for the series $L_x,\Delta_x$ and $\rho_{\mathcal{C}^i}(q)=\int\td x\rho_{\mathcal{C}^i}(x,q)$ for the series $L_q,\Delta_q$ (see Fig.~\ref{fig:example_dx_dl_series} for an illustration of these quantities).  We denote the corresponding thresholds with $\epsilon_{\rho_{\mathcal{C}}}$. 

We start with the electronic part of the example (corresponding to the He atom outside the cavity), choose $\Delta_x=0.14$ and vary $L_x=\{8,10,12,\dots\}$. The ground-state energy $E_{L_x}$ drops with increasing $L_x$ (because boundary effects become less important) and we find $\Delta E_{L_x}<10^{-8}$ and $\Delta\rho_{L_x}<10^{-7}$ for $L_x\geq 20$. For illustration, we have depicted $\Delta\rho_{L_x}$ for a choice of $L_x$ in Fig.~\ref{fig:example_dx_dl_series} (a). In the following, we choose $L_x=20$ because the thresholds of $\epsilon_{\rho_{L_x}}=10^{-7}$ and $\epsilon_{E_{L_x}}=10^{-8}$ are already stricter than the maximal accuracy of the later nonlinear (dHF, dRDMFT) calculations. Next, we perform  $\Delta_x=\{0.2,0.19,0.18,\dots,0.06\}$. We find that $\Delta E_{\Delta_x}$ also drops with decreasing $\Delta_x$ until $\Delta E_{\Delta_x}<10^{-8}$ for $\Delta_x = 0.15$ and does not decrease any more (as the convergence criteria of the minimization are no more precise). We show $\Delta\rho_{\Delta_x}$ for a choice of $\Delta_x$ in Fig.~\ref{fig:example_dx_dl_series} (b).
$\Delta\rho_{\Delta_x}$ instead decreases (slowly) until the lowest tested value of $\Delta_x=0.06$. We find $\Delta \rho_{\Delta_x}<10^{-7}$ already for $\Delta_x<0.14$, but to reach $\Delta \rho_{\Delta_x}<10^{-8}$, we need to decrease the spacing until $\Delta_x=0.07$. Such a small spacing is numerically unfeasible for larger boxes and thus we will not try to go beyond $\epsilon_{\rho_{\Delta_x}}=10^{-7}$.

We repeat the same series for the harmonic oscillator system of the q-coordinate with frequency $\omega=\omega_{res}\approx0.5535$ (resonance with the transition between ground and first excited state of the He atom outside the cavity) and find that the box size is converged with $\Delta E_{L_q}< 10^{-8}$ and $\Delta \rho_{L_q}<10^{-8}$ for $L_q \geq 14$. 
The spacing is converged in the energy $\Delta E_{\Delta_q}<10^{-8}$ and in the density $\Delta \rho_{\Delta_q}< 10^{-7}$ for $\Delta_q\leq0.20$.

From these preliminary calculations, we can infer the parameters for the dressed calculations: $L_x=20,L_q=14,\Delta_x=0.14, \Delta_q=0.20$. To be sure that these parameters are still sufficient for non-zero coupling ($\lambda>0$), we perform another box length series in the four-dimensional space of the exact ground-state, e.g., for $\lambda=0.1$ and $\omega=\omega_{res}$. To have a uniform grid distribution, we also set $\Delta_x=\Delta_q$, although our preliminary calculations suggest that we could choose a larger $\Delta_q$. Unfortunately, we cannot explore this space completely, but we are limited with $L_x,L_q\leq 18$.\footnote{The precise reasons is that the memory of one node of the cluster we are using is too small. We would consequently need to distribute the wave function over several nodes, which is possible, but would demand a considerable programming effort.} 
However, we can confirm that the energy is converged in the q-direction for $L_q=14$ and in the x-direction, we have $\Delta E_{L_x}\approx 10^{-7}$ for $L_x=18$.

Finally, we compare the \textsc{Octopus} results with the results from the \textsc{Dynamics} code. For that, we consider the differences in the total energy $E_{OD}=|E_{Dynamics}-E_{Octopus}|$ and the maximum deviations in the polaritonic density $\rho_{OD}\equiv\max_{x,q}|\rho_{Octopus}(x,q)-\rho_{Dynamics}(x,q)|$ between the two codes. Due to the optimization of \textsc{Dynamics}, we are limited to a box length of $L_x=L_q=14$ to have a spacing of $\Delta_x=\Delta_q=0.14$. 
With these parameters, all the energy and density errors in \textsc{Octopus} increase to $\Delta E_{L_x,L_q,\Delta_x,\Delta_q}=\Delta\rho_{L_x,L_q,\Delta_x,\Delta_q}\approx 10^{-5}$. When we compare the ground-state energies of both codes for this maximum possible mesh, we find $E_{OD}\approx 10^{-5}$ and $\rho_{OD}\approx 10^{-5}$.  
We thus were able to confirm that both codes agree on the level of accuracy that we estimated from the \textsc{Octopus} calculations before and this although they are quite different from a numerical perspective. We conclude from these calculations that the implementation of the exact many-body routine for dressed two-electron systems\footnote{Larger systems are in principal also possible, but numerically infeasible in real space.} in \textsc{Octopus} is reliable and we can use it as benchmark for the dHF and dRDMFT approximations within \textsc{Octopus}.


\section{Validation of the dHF routine}
\label{sec:numerics:dressed:convergence:HF}
In this section, we present the validation of dHF, which requires several steps. We start with the electronic HF routine and converge the He atom model (outside the cavity) in box size and spacing, where we proceed as in the previous section. Then, we converge the second HF routine (that we call HF$_{basis}$ in this section) that is implemented in \textsc{Octopus}, which is based on the RDMFT implementation and thus makes use of a basis set (Sec.~\ref{sec:numerics:rdmft:algorithm_no_piris}). We have mentioned there already that the difference of such a basis-set implementation is that the routine calculates all the integral kernels of the total energy as matrix elements of a chosen basis and then searches for the energy minimum by only varying the corresponding coefficients. This routine can be considerably faster than the standard one of \textsc{Octopus}, i.e., the conjugate-gradients algorithm, because the calculation of the problematic exchange-term needs only to be performed once in the beginning for the matrix elements. Such a basis-set implementation requires a convergence study with respect to the basis size, which we explain in detail in the second part of this section together with a comparison between HF and HF$_{basis}$.
We conclude the section with the discussion of dHF, which also uses a basis set. However, the basis-set convergence for dHF is more involved than for the purely electronic HF$_{basis}$ and we explain this in detail. Afterwards, we compare dHF in the no-coupling limit to HF/HF$_{basis}$ and we discuss the comparison of \textsc{Octopus} and \textsc{Dynamics} on the level of dHF, where we follow the same strategy as in the previous section.

\subsection{HF minimization on the grid} 
We start with the He atom outside the cavity and calculate the HF ground-state with the standard routine of \textsc{Octopus}, which uses a conjugate gradients algorithm for the orbital optimization. The convergence criteria for the conjugate gradient algorithm are defined exactly as before for the exact many-body ground-state. We set them to $\epsilon_E=10^{-9}$ and $\epsilon_{\rho}=10^{-8}$ and determine $L_x$ and $\Delta_x$. We start with a spacing of $\Delta_x=0.1$ and find that $\Delta E_{L_x}\approx 10^{-8}$ and $\Delta \rho_{L_x}<10^{-7}$ for all $L_x\geq 20$. 
We choose $L_x=20$ and perform the spacing series, which is converged in energy with $\Delta E_{\Delta_x}<10^{-8}$ for $\Delta_x\leq 0.15$ and in the density with $\Delta \rho_{\Delta_x}< 10^{-7}$ for $\Delta_x\leq0.14$. These calculations suggest that all $\Delta_x\leq0.14$ are sufficient and we choose $\Delta_x=0.1$, which is numerically feasible for the test system, we consider. We see that although we solve a nonlinear equation in HF, we achieve the same accuracy than for the exact calculations of the previous section. We want to remark that the for the above accuracy necessary parameters correspond to considerable computational costs and therefore, it will be very difficult to achieve similar results for larger systems. 

\subsection{HF minimization with a basis set}
\label{sec:numerics:dressed:convergence:HF:HFvsHFbasis}
In this section, we present the convergence of the He test system in the HF$_{basis}$ routine with respect to the basis set. It is based on the orbital-based RDMFT implementation that we introduced and discussed in Sec.~\ref{sec:numerics:rdmft:piris}, employing the HF functional (see paragrpah below Eq.~\ref{sec:numerics:rdmft:algorithm:rdmft_functional}). 
To generate a basis, we perform a preliminary calculation with the \emph{independent particle} (IP) routine, which solves a simple 1-body Schrödinger equation to generate the basis ($(1/2 \nabla^2 +v_{He}')\phi_i=e_i\phi_i$, see Sec.~\ref{sec:numerics:rdmft:comparison:cg_piris}).
To allow for enough variational freedom, we calculate besides the occupied orbitals\footnote{Note that in our case, we have always $GS=\frac{N}{2}$, where $N$ is the number of electrons, because we only look at closed-shell systems, which distribute two electrons to every spatial orbital.} (that form the HF ground state and are denoted by $GS$) also excited orbitals, which are called \emph{extra states} ($ES$) in \textsc{Octopus}. The so generated basis set has the size $M=GS+ES$.

As discussed in Sec.~\ref{sec:numerics:rdmft:piris}, the routine performs the minimization by representing the coefficients of the basis-set expansion in a matrix and diagonalizing it repeatedly until self-consistence. The corresponding energy convergence criterion $\epsilon_E$ remains the same in HF$_{basis}$ like in HF.
But the second convergence criterion $\epsilon_{F}$ of the orbital minimization that tests the hermiticity of the Lagrange multiplier matrix needs to be adapted. As $\epsilon_{F}$ is a considerably stricter criterion than $\epsilon_E$ (and $\epsilon_{\rho}$), it is set as default to $\epsilon_{\Lambda}=10^3\cdot\epsilon_E$.

For the convergence of the size of the basis set $M$ or equivalently the parameter $ES$, we perform a series $ES=\{10,20,\dots,100\}$ and test deviations in energy $\Delta E_{ES, ES_{ref}}=E_{ES}-E_{ES_{ref}}$ and density $\Delta \rho_{ES,ES_{ref}}=\max_x|\rho_{ES}(x)-\rho_{ES_{ref}}(x)|$ from the reference value $ES_{ref}$ that yields the lowest total energy and that typically occurs for the largest basis. We use $L_x$ and $\Delta_x$  that we determined before in HF and find that $\Delta E_{ES,100}$ and $\Delta \rho_{ES,100}$ decrease with increasing $ES$ and for $ES\geq40$, $\Delta E_{ES,100}<10^{-8}$ and $\Delta \rho_{ES,100}\approx 10^{-5}$. The latter value is considerably smaller than the one we obtained with the standard solver and the reason is the bad quality of the basis set. We see that for $ES\geq40$ the extra basis states cannot improve the accuracy anymore but instead add only noise. We will see later for RDMFT, where we also have to optimized the occupation numbers that this effect leads to a generic \emph{non-variational} behavior, i.e., the energy does not always decrease or stay constant with increasing $ES$, but will first drop and than rise again for very large $ES$. Thus, we define there what we have called an \emph{optimal region} for the basis size, in which the energy remains constant within a certain threshold. For the sake of completeness, we want to mention that after we had published the paper, we found such kind of non-variationality even for HF, however only for very large $ES\gtrapprox 120$. Thus, in principle there is also such an optimal region for HF, but it is so large that in practice it does not matter.

In summary, HF$_{basis}$ cannot be as strictly converged as HF, which is especially visible for the electron density deviations that saturate of the order of $\Delta \rho_{ES,100}\approx 10^{-5}$ even for large basis sets. Nevertheless, this accuracy is sufficient to interpret and compare numerical data in the following. Comparing HF with HF$_{basis}$ with these parameters, we find that $|E_{HF}-E_{HF_{basis}}|\approx 10^{-8}$ and $\max_x|\rho_{HF}(x)-\rho_{HF_{basis}}(x)|\approx 10^{-5}$. So both methods are consistent.

\subsection{Validation of dHF}
After having properly understood the convergence of the electronic HF methods, we can now turn to dHF. All the calculations shown in Ch.~\ref{sec:dressed:results} are done with the basis-set type of implementation,  which is more involved than in the purely electronic HF$_{basis}$ case. Thus, we start in Sec. \ref{sec:numerics:dressed:convergence:HF:dHF_basis_conv} 
with discussing the new issues that enter when one needs to converge a system in the dressed space with respect to the basis set. In Sec. \ref{sec:numerics:dressed:convergence:HF:dHF_no_coupling}, we present the convergence of dHF with zero-coupling ($\lambda=0$), which means that the electronic and photonic part of the system completely decouple such that we can compare the results of the electronic part to electronic HF. 
In the last section, \ref{sec:numerics:dressed:convergence:HF:dHF_comp_dynamics}, we conclude the discussion by comparing the dHF results from \textsc{Octopus} with \textsc{Dynamics}.

\subsection{Basis-set convergence in the dressed auxiliary space}
\label{sec:numerics:dressed:convergence:HF:dHF_basis_conv}
In the dressed auxiliary space that we explore with dHF and dRDMFT, the basis-set convergence is even more difficult than in the HF$_{basis}$ case. We illustrate this additional difficulty for dHF in the no-coupling ($\lambda=0$) case. $\lambda=0$ makes the discussion much simpler, because we have two perfectly separated problems, the atomic system and harmonic oscillators that are just calculated at the same time. However, we use a combined basis that consequently needs to include the appropriate degrees of freedom for both systems. But the composition of the basis set is strongly influenced by the parameters of the system, especially $\omega$, because it is generated by a preliminary calculation. This can be illustrated as follows:

For $\lambda=0$, the electronic and photonic part of the system separate. Thus, the coupled (dHF) Hamiltonian is a direct sum of the electronic (photonic) Hamiltonian $\hat{H}^e$ $(\hat{H}^p)$ and the two-dimensional dressed orbitals $\psi_{i\alpha}(x,q)$ can be exactly decomposed in their one-dimensional electronic  $\phi_i(x)$ and photonic $\chi_{\alpha}(q)$ constituents, $\psi_{i\alpha}(x,q)=\phi_i(x)\otimes \chi_{\alpha}(q)$. Here, $\phi_i$ $( \chi_{\alpha})$ is an eigenfunction with eigenvalue $e^e_i$ $(e^p_{\alpha})$ of the electronic (photonic) Hamiltonian $\hat{H}^e$ $(\hat{H}^p)$. Consequently, we know that we can calculate the eigenvalues of the dressed orbitals as sum of the uncoupled ones, i.e., $e^{ep}_{i\alpha}=e^e_i +e^p_{\alpha}$. The basis set for a dHF calculation is then constructed using the ground and the first $ES$ excited orbitals of a preliminary IP calculation. These orbitals are ordered by their eigenvalue $e^{ep}_{i\alpha}$ and for that, the relation between the individual energies of the electron and photon space is crucial. 

Table \ref{tab:basis_contributions} shows the decomposition of such a basis with $ES=6$ (and thus $M=7$). We see that slightly more electronic states contribute to the basis, but already 3 out of 7 states describe excited photon contributions. So if we wanted to use that basis for an dHF calculation with for example $\lambda=0$, that trivially needs only ground-state contributions for the photonic coordinate, these 3 states would be entirely unnecessary and waste computational resources. Before we discuss this further, we want to explain how $\omega$ influences this distribution between electronic and photonic contributions: If we chose for example a larger $\omega\approx1.3$, the first 4 states of the combined basis would only vary in the electronic contribution. If we instead lowered $\omega$, the opposite would happen and the first states would vary in the photonic contribution. A similar kind of argument could be done for all other ingredients of the Hamiltonian of course, but for $\omega$ this influence is most directly visible and comparatively strong. Note that this effect is \emph{independent} from the issue of the fermion ansatz that we discussed in Sec.~\ref{sec:dressed:construction:simple:auxiliary_wf_problem}. Also there, we can increase $\omega$ to trivial satisfy the extra conditions due to the hybrid statistics of the polaritonic orbitals. 
However, it is very difficult to disentangle both effects (employing the Piris orbital-optimization routine). 

At this point, one might ask the question why at all we use the basis-set implementation and the answer remains the same as for the purely electronic case: Although we need large basis sets for the properly converged dHF calculations which makes them numerically very expensive, these computations are still relatively inexpensive compared to calculations with a conjugate gradients algorithm that calculates all integrals on the grid. In the case of the He atom, the dHF calculation with conjugate gradients and $ES=15$ takes still 4 times longer than the same calculation with a basis set and $ES=50$.\footnote{Note that both algorithms are not yet fully optimized and thus, the efficiency might still significantly change. See also the Outlook in part~\ref{sec:conclusion}.}

Finally, we want to mention that the statements of this subsection carry over straightforwardly to the coupled ($\lambda>0$) case, we just do not have the direct connection to the decoupled spaces and thus cannot visualize this case as well. It is clear that we need more variational freedom for the photonic subspace than in the $\lambda=0$ case but the basis will still depend on the system parameters. One can expect that also for $\lambda>0$ many basis states will be unnecessary.
\begin{table}
	\centering
	\begin{tabular}{@{}ccc|ccc@{}}
		&&&&\multicolumn{2}{c}{contribution} \\
		index & $e^{e}_i$ & $e^{p}_{\alpha}$ & $e^{ep}_{i\alpha}$ & $i$ & $\alpha$\\
		\hline
		1 & -1.483 & 0.277 & -1.207 & 1 & 1 \\
		2 & -0.772 & 0.830 & -0.653 & 1 & 2 \\
		3 & -0.461 & 1.384 & -0.495 & 2 & 1 \\
		4 & -0.263 &       & -0.184 & 3 & 1 \\
		5 &  &  					 & -0.100 & 1 & 3 \\
		6 &  &  					 & 0.014  & 4 & 1 \\
		7 &  &  					 & 0.058  & 2 & 2 \\
	\end{tabular}
	\caption{Basis set structure (IP calculations) with $ES=6$ (thus $M=N/2+ES=7$ basis states) for the dressed He atom with mode frequency $\omega\approx 0.5535$, but no coupling ($\lambda=0$). On the left side of the table are the first eigenenergies $e^e_i$ ($e^p_{\alpha}$) of the pure electronic (photonic) Hamiltonian $\hat{H}^e$ $(\hat{H}^p)$ shown. On the right side, we see the orbital energies $e^{ep}_{i\alpha}=e^e_i+e^p_{\alpha}$ of the combined Hamiltonian $\hat{H}^{ep}$ and the decomposition of the combined index: The first eigenenergy $e^{ep}_1=e^{ep}_{11}=e^e_1+e^p_1$ is the sum of the two first uncoupled energies. The second energy $e^{ep}_2=e^{ep}_{12}$ instead is formed from the electronic ground but photonic first excited state, etc. In this example, we see that both contributions are similar, although there are slightly more electronic orbitals included. This tendency continues also for increasing $ES$.}
	\label{tab:basis_contributions}
\end{table}

\subsection{Validation of dHF for $\lambda=0$}
\label{sec:numerics:dressed:convergence:HF:dHF_no_coupling}
Now we can address the $ES$-convergence of dHF, which we present for $\lambda=0$, such that we can compare the electronic part of the converged result to HF afterwards. As we set $\lambda=0$, we know that for the photon component, the IP ground-state orbital $\chi_1(q)$ is already the exact solution, because the photons do not interact directly with each other and HF (or any other level of approximation) will not change this. The dHF ground-state orbital thus is $\phi(x,q)=\psi_{HF}(x)\otimes \chi_1(q)$ with the electronic HF ground-state orbital $\psi_{HF}(x)$. As we saw in the last section, when we converged the He test system using the HF$_{basis}$ routine, $\psi_{HF}(x)$ can be approximated by an IP basis with $ES=40$. So we know that in the no-coupling limit of dHF, we do not need a larger basis set, because Kronecker-multiplying the basis-states of the HF$_{basis}$-calculation with $\chi_1(q)$ would be exactly sufficient.
However, here and in the following, we do not want to choose by hand such a well-adapted state space (which for very strong coupling strength is nontrivial), but instead rely on the polaritonic IP-calculations. From our considerations before, we expect to need more basis states to reach the same level of accuracy than using HF$_{basis}$ and the exact number will depend on $\omega$. Indeed, for $\omega=\omega_{res}$, we need $ES=80$ to have $\Delta E_{ES,100}<10^{-7}$ and $\Delta \rho_{ES,100}\approx10^{-5}$, where we calculate the electron density of the dHF-calculation by $\rho(x)=\int\td q\, \rho(x,q)$.
When we instead choose a very high value of $\omega=5.0$, we reach $\Delta E_{ES,100}<10^{-7}$  and $\Delta \rho\approx10^{-5}$ already for $ES\approx 40$. For both calculations, we do not reach $\Delta E_{ES,100}<10^{-8}$ even for bigger $ES$. This is due to the many "unnecessary" states that are taken into account in the minimization. These then only introduce numerical noise without providing useful variational information. We find this confirmed by analyzing the ``photonic density'' $\rho(q)=\int\td x\, \rho(x,q)$, which deviates from the correct density $\rho_{ex}(q)=2|\chi_1|^2(q)$ although $\chi_1(q)$ is explicitly part of the basis. Adding basis states consequently cannot improve the photonic orbital. For instance, for $\omega=\omega_{res}$, the density deviations remain at approximately $\Delta \rho_{ES, ES_{ref}}\approx10^{-4}$ for \emph{all} $ES$ \emph{and} $ES_{ref}$. When we instead choose $\omega=5.0$, we increase the accuracy to $\Delta \rho_{ES,100}\approx10^{-5}$ for $ES\geq 40$. Despite being less accurate, we used $\omega=\omega_{res}$ for the results in Ch.~\ref{sec:dressed:results}, because $\Delta \rho_{ES}\approx10^{-4}$ is still two orders of magnitude smaller than the deviations from the exact or the dRDMFT solution. Again, we can safely use these ``less'' accurate results, because we know ``where it comes from,'' i.e., we understand how the accuracy is controlled.

This is confirmed by the comparison of the electronic part of dHF to HF that we are now able to perform. For the energy comparison, we need to subtract the photon part $E_{p}=2\frac{\omega}{2}$ from the total dHF energy $E_{dHF}$, $E_{dHF,e}=E_{dHF}-E_p$. However, this analytical expression is also not exact, because of the just mentioned error in the photonic part of the orbitals. Consequently, we have deviations for $\omega=\omega_{res}$ of $|E_{HF}-E_{dHF,e}^{\omega=\omega_{res}}|\approx 3\cdot10^{-6}$ and $\max_x|\rho_{HF}(x)-\rho_{dHF,e}^{\omega=\omega_{res}}(x)|\approx 10^{-4}$. However, for the ``better'' value of $\omega=5.0$, we find $ |E_{HF}-E_{dHF,e}^{\omega=5.0}|\approx4\cdot10^{-7}$ and $\max_x|\rho_{HF}(x)-\rho_{dHF,e}^{\omega=5.0}(x)|\approx 10^{-5}$.

\subsection{Validation of dHF with \textsc{Dynamics}}
\label{sec:numerics:dressed:convergence:HF:dHF_comp_dynamics}
We  conclude this section by comparing the results of our implementation of dHF in \textsc{Octopus} with the \textsc{Dynamics} code, which uses an imaginary time propagation~\citep{Feit1982} algorithm to calculate the HF ground-state. This comparison allows us to validate also the case of $\lambda>0$. We choose $\omega=\omega_{res}$ and $\lambda=0.1$ for the comparison and perform another $ES$-convergence, confirming that we need $ES=90$ for the same level of accuracy like before in the no-coupling case. For the sake of completeness, we want to mention that for $\omega=5.0$ we need again a significantly smaller basis set for even better converged results, exactly as we found before in the $\lambda=0$ case. However, already for $\omega=\omega_{res}$ the level of convergence is an order of magnitude better than the expected deviations between the codes that we estimated before.  

So we compare the ground-state for these parameters to the result of \textsc{Dynamics} and find for the energy the expected deviations of $E_{OD}\approx 10^{-5}$. For the density, we find instead deviations of $\rho_{OD}\approx 10^{-3}$. This discrepancy is probably due to the different convergence criteria of the two codes. \textsc{Dynamics} tests the eigenvalue equation of the one-body Hamiltonian for a certain subset of all the grid points, which is a much stronger criterion than the one of \textsc{Octopus}, explained before. The influence of these different criteria on a self-consistent calculation are naturally stronger than on the calculation of a linear eigenvalue-problem like the many-body calculation. Still, density errors of the order of $10^{-3}$ are small enough for our purposes.
We find similar errors also for other values of $\lambda$ and conclude that both codes are sufficiently consistent.


\section{Validation of dRDMFT}
\label{sec:numerics:dressed:convergence:RDMFT}
In this section, we can finally turn to dRDMFT. This is the first implementation at all of this theory, so we cannot validate it with a reference code any more. However, the difference between HF and RDMFT (using the Müller functional) on the implementation-level essentially is in the treatment of the occupation numbers, which are fixed to 2 and 0 in the former case but are allowed to be non-integer for the latter. The 1-body and 2-body terms, which implementation-wise are the only modifications due to the dressed auxiliary construction ($v(\br)\rightarrow v'(\br,q)$, $w(\br,\br')\rightarrow w'(\br q,\br'q')$) are the same for HF and RDMFT and thus also for dHF and dRDMFT. This means that the validation of dHF that we presented in the previous section at the same time largely validates the implementation of dRDMFT.

Still, we need to analyze and understand the convergence with respect to the basis set in dRDMFT and check for the consistency between the results of RDMFT and dRDMFT in the no-coupling limit.

\subsection{Basis-set convergence of RDMFT}
In RDMFT, we have to perform two minimizations that are interdependent: for the natural orbitals $\phi_i$ \emph{and} the natural occupation numbers $n_i$ (where always $i=1,\dots,M$). This is done by fixing alternately $\phi_i$ or $n_i$, while optimizing the other until overall convergence is achieved (see Sec.~\ref{sec:numerics:rdmft:algorithm:general}).

We have the possibility to define different convergence criteria for each minimization, $\epsilon_{E}$ (which is connected to $\epsilon_{\Lambda}$, see Sec.~\ref{sec:numerics:dressed:convergence:HF:HFvsHFbasis}) and $\epsilon_{\mu}$.
The latter tests the convergence of the Lagrange multiplier $\mu$ that appears in the RDMFT functional to fix the total number of particles in the system (see algorithm~\ref{algorithm:RDMFT_non}). The occupation number optimization routine at iteration step m sets $\mu=\mu^m$, minimizes the total energy with respect to the $n_i=n_i^m$ and calculates the particle number $N^m=\sum_{i=1}^{M}n_i$. Based on the deviation to the correct system's particle number, $\mu^{m+1}$ is increased or decreased. The routine exits if $|\mu^{m}-\mu^{m-1}|<\epsilon_{\mu}$. For the remainder of this section, we set $\epsilon_E=\epsilon_{\mu}=10^{-8}$.

However, the $ES$-convergence of RDMFT is again more difficult than in the HF$_{basis}$-case. Contrary to the (largely) monotonic dependence between $ES$ and the energy that we found for HF$_{basis}$, the current RDMFT implementation in \textsc{Octopus} shows a clear non-variational behavior with respect to the number of basis states. For all tested systems, the energy went down with increasing $ES$ until a certain value $ES_{min}<100$ and then up again. Therefore, it seems that the interplay of the two minimization processes and the relatively soft types of convergence criteria introduce for big $ES$ such large errors that they exceed the gain of accuracy due to more variational freedom.
In Tab. \ref{tab:HF_RDMFT},  we show the variation of the ground-state energy of He (outside the cavity) for a series $ES=\{10,20,\dots,80\}$. The energy first decreases until its lowest value for $ES_{min}\approx 40$ and then increases again. 
\begin{table}[H]
	\centering
	\begin{tabular}{c|c|c}
		$ES$ & Energy & $\Delta E_{ES}=E_{ES}-E_{ES-10}$\\ 
		\hline
		10 & -2.2421837 & -\\ 
		20 & -2.2426908 & $-5.1\cdot10^{-4}$\\ 
		30 & -2.2427080 & $-1.7\cdot10^{-5}$\\ 
		40 & -2.2427085 & $-4.9\cdot10^{-7}$\\ 
		50 & -2.2427049 & $+3.6\cdot10^{-6}$\\ 
		60 & -2.2427035 & $+1.3\cdot10^{-6}$\\ 
		70 & -2.2426928 & $+1.1\cdot10^{-5}$\\ 
		80 & -2.2426937 & $-9.9\cdot10^{-8}$	
	\end{tabular} 
	\caption{Ground state energies of the He atom (outside the cavity) calculated by RDMFT with parameters mentioned in the main text. The energy goes first down with increasing number of ES until it reaches its lowest value at $ES_{min}\approx 40$ and then up again.}
	\label{tab:HF_RDMFT}
\end{table}

This non-strictly variational behavior makes a clear definition of the convergence difficult. However, we find that for every considered system there is an \emph{optimal region} $\mathcal{ES}_{opt}=\{ES\,|\,ES_{min}\leq ES\leq ES_{max}\}$. By optimal region, we mean an interval of $ES$ in which the solutions vary minimally among each other, i.e., their energy and density deviations are minimal. For the energies, we define $\Delta E_{ES,ES'}=|E_{ES}-E_{ES'}|$, the corresponding threshold $\epsilon_{E_{opt}}$, and require $\Delta E_{ES,ES'}<\epsilon_{E_{opt}}$ for \emph{all} $ES,ES'\in\mathcal{ES}_{opt}$. For the densities, we define the point-wise density deviations $\rho_{ES,ES'}(x)=\rho_{ES}(x)-\rho_{ES'}(x)$, their maximal deviations $\Delta \rho_{ES,ES'}=\max_x |\rho_{ES,ES'}(x)|$, and the corresponding threshold $\epsilon_{\rho_{opt}}$. As second condition on  $\mathcal{ES}_{opt}$ we require $\Delta \rho_{ES,ES'}<\epsilon_{\rho_{opt}}$ for \emph{all} $ES,ES'\in\mathcal{ES}_{opt}$.

\begin{figure}[ht]
	\centering
	\includegraphics[width=0.48\columnwidth]{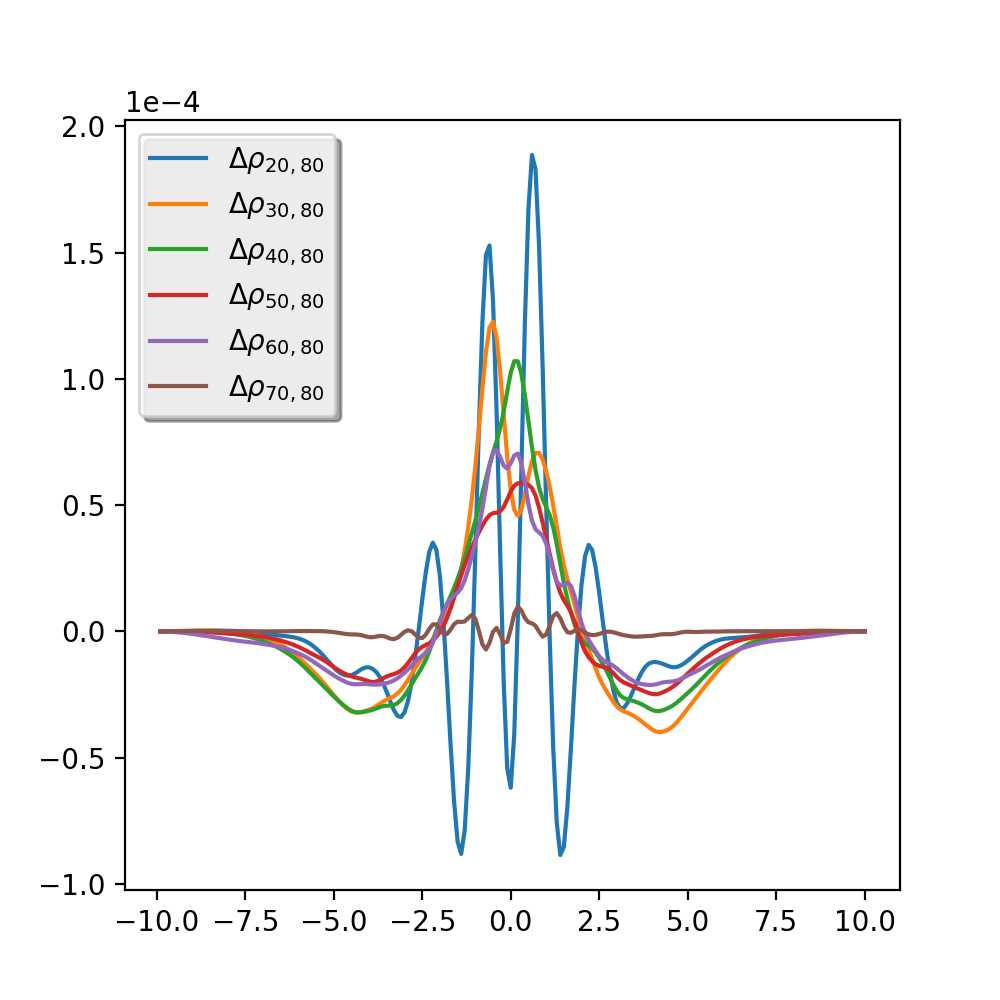} \hfill
	\includegraphics[width=0.48\columnwidth]{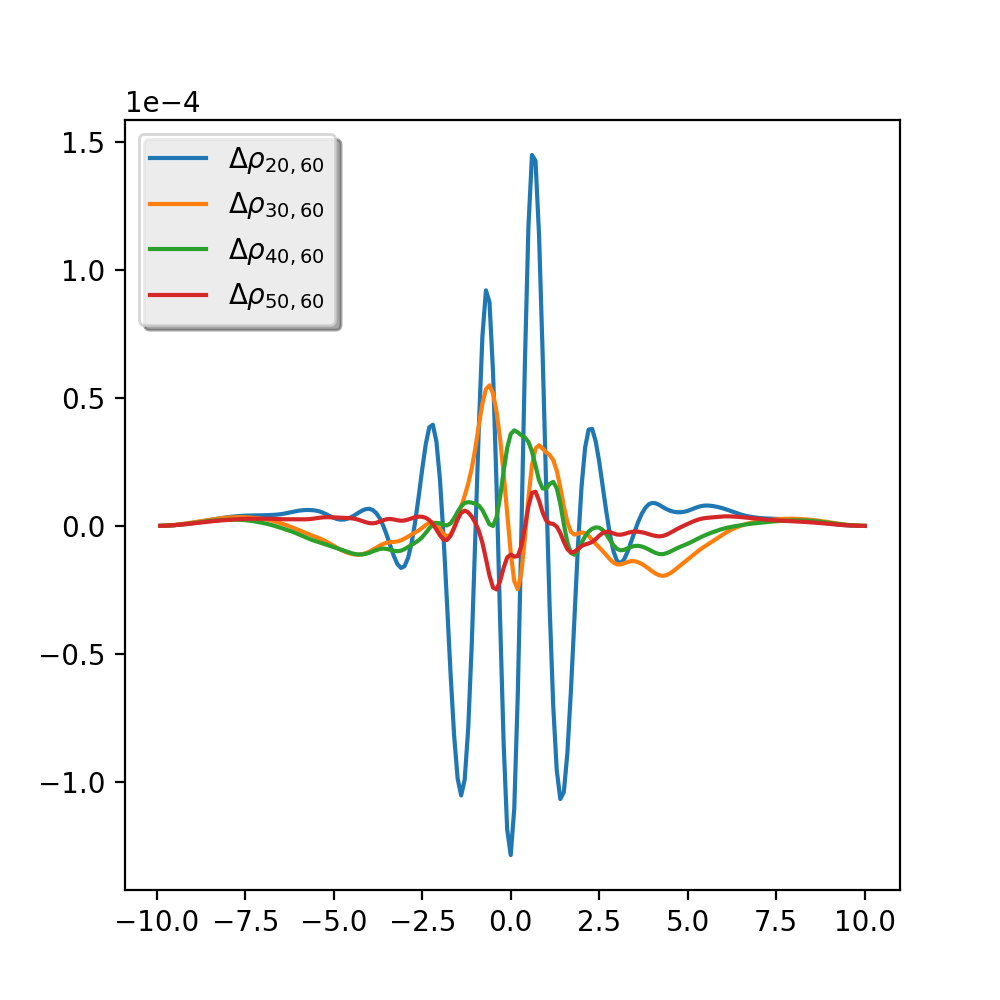}
	\caption{Differences in the ground-state density $\Delta\rho_{ES,ES'}(x)$ for RDMFT calculations of the He atom with $ES'=80$ (left) and $ES'=60$ (right). $\rho_{ES=20}(x)$ deviates similarly for both reference points, but for $ES_{ref}=80$, we see a systematic (instead of random) $\Delta\rho_{ES}(x)$ for $30\leq ES\leq 60$. The deviations merely drop under $10^{-4}$, except for $ES=70$, which is very close to the reference. When we instead use $ES'=60$, the calculations for $ES=30$ to $ES=60$ only deviate of the order of $10^{-5}$ and the deviations have a random character.}
	\label{fig:HE_RDMFT_DeltaN}
\end{figure}

We start with the investigation of the energies and find that $\Delta E_{ES,ES'}<5 \cdot 10^{-6}$ for all combinations $30 \leq ES,ES'\leq 60$, but $\Delta E_{ES,ES'}>10^{-5}$ when we choose $30 \leq ES\leq 60$ and $30 \leq ES'\leq 60$ or $10 \leq ES'\leq 20$. We conclude that the first condition for $\mathcal{ES}_{opt}$ is met by the interval $30 \leq ES \leq 60$ with threshold $\epsilon_{E_{opt}}= 5 \cdot 10^{-6}$. 
For the investigation of the second condition, we depict in Fig. \ref{fig:HE_RDMFT_DeltaN} the density deviations $\rho_{ES,ES'}(x)$ with $ES'=60$, the upper boundary of the just found interval and $ES'=80$, that corresponds to the largest basis set of this example. For a better visibility, the curve for $ES=10$ is not shown, but it deviates stronger than all the other curves from both $ES'$. We conclude that $\Delta\rho_{ES,ES'}$ goes down until $ES=20$, independently of the reference. However, for $ES\geq 30$ this is not the case any more. We find  $\Delta\rho_{ES,80}\approx10^{-4}$ but $\Delta\rho_{ES,60}< 5\cdot 10^{-5}$. Additionally, we observe two different \emph{types of deviations}: The curves $\rho_{ES,80}(x)$ have a similar form for all $30\leq ES\leq 60$, but when we change the reference point to $ES'=60$, we cannot find pronounced similarities which is what we would expect from fluctuations. When we test also the other possible values for $ES'$, we find $\Delta\rho_{ES,ES'}< 5\cdot 10^{-5}$ for all $30\leq ES,ES'\leq 60$. Thus, the second condition for the optimal region is met by the same interval like the first condition with threshold $\epsilon_{\rho_{opt}}=5\cdot 10^{-5}$. Therefore, we have $\mathcal{ES}_{opt}=\{ES\, |\, 30 \leq ES \leq 60\}$ and conclude that the maximum possible accuracy of RDMFT calculations is already reached for $ES=30$ and it is generally lower than for HF$_{basis}$. 
%

The strong decrease in accuracy of RDMFT in comparison to HF$_{basis}$ suggests that the occupation number minimization adds a significant error to the calculation. As we have discussed in Sec.~\ref{sec:numerics:dressed:non_assessment}, also the more accurate BFGS algorithm that is now the default solver exhibits certain inaccuracies. It is very difficult to estimate the exact error that is introduced by the method, because of the strong nonlinear character of the minimization (especially the interdependence between the $\phi_i$ and $n_i$ optimizations).
For a thorough understanding of this issue, one needs to implement and test different numerical solvers. However, for the purposes of this text, we consider the current accuracy as sufficient.

\subsection{Basis-set convergence of dRDMFT and comparison to RDMFT}
For the $ES$-series of dRDMFT, we need to deal with the combination of the inaccuracies introduced by the $n_i$-minimization, explained in the previous subsection and the additional errors due to extra large basis sets that contain many redundant degrees of freedom, that we found for dHF before (Sec.~\ref{sec:numerics:dressed:convergence:HF}). 
We know already that a large photon frequency is advantageous in terms of the latter. So we choose $\omega=5.0$ to make this error as small as possible. 

As expected, it is much harder to define a convergence region in the way we did before for electronic RDMFT. We find the lowest energy at $ES=50$, which deviates from $ES=40$ and $ES=60$ by $|E_{ES=40/60}-E_{ES=50}| > 10^{-5}$. So we cannot find a region, where the energy is converged \emph{better} than $10^{-5}$. Nevertheless, if we accept an accuracy of $\Delta E_{ES}<5\cdot 10^{-5}$ as sufficient, we find a region as large as $20\leq ES\leq 100$ that satisfies the criterion. A look at the density deviations reveals that we can slightly tighten this region and exclude $ES=20$, such that we have deviations $\Delta \rho_{ES,50}\approx10^{-4}$ for all $30\leq ES\leq 100$. We found similar values also for other examples and we can conclude that we lose about one order of magnitude of accuracy for dRDMFT in comparison to RDMFT. Careful fine-tuning of the numerical parameters could improve this, but the current accuracy suffices for our purposes. For the calculations, presented in Ch.~\ref{sec:dressed:results}, we employed much smaller frequencies $\omega < 1.0$ and consequently, we had to employ larger bases. Nevertheless, we obtained the same level of accuracy.

Finally, we present the consistency check between RDMFT and dRDMFT in the $\lambda=0$ limit. We find $|E_{RDMFT}-E_{dRDMFT}|\approx 10^{-5}$ and $\max_x|\rho_{RDMFT}(x)-\rho_{dRDMFT}(x)|\approx 10^{-4}$, which means that the deviations between the levels of theory are of the same order as the maximal accuracy that dRDMFT provides. We conclude that both theories are consistent.


\section{Protocol for the convergence of a dHF/dRDMFT calculation}
\label{sec:numerics:dressed:convergence:protocol}
We conclude the convergence study with a step-by-step guide for the proper usage of the dressed orbital implementation in Octopus. All the real-space calculations presented in Ch.~\ref{sec:dressed:results}, presented in the main part of this paper were performed according to this protocol
\begin{enumerate}
	\item Box length $L_x$ and spacing $\Delta_x$ convergence for the purely electronic part of the system on the level of IP and electronic HF and for the uncoupled photonic system on the level of IP. 
	\begin{itemize}
		\item Test the deviations in energy $\Delta E_{L_x}<\epsilon_{E_{L_x}}$ and density $\Delta \rho_{L_x}<\epsilon_{\rho_{L_x}}$, as explained in Sec. \ref{sec:numerics:dressed:convergence:exact}. We chose $\epsilon_{E_{L_x}}=10^{-8}$ and $\epsilon_{\rho_{L_x}}=10^{-5}$ to exclude any numerical artefacts. However, as the dHF and dRDMFT calculations do typically not reach such precisions, one can relax these criteria in general.
		\item For the $\Delta_x$-series, test only the deviations in energy $\Delta E_{\Delta_x}<\epsilon_{E_{\Delta_x}}$ due to the larger density errors. We chose $\epsilon_{E_{\Delta_x}}=10^{-8}$, but like for the box length, this criterion can be relaxed.
	\end{itemize}
	\item Basis size convergence for the HF$_{basis}$ routine with the purely electronic part of the system.
	\begin{itemize}
		\item  Perform an $ES$-series and test the deviations in energy $\Delta E_{ES,ES_{ref}}<\epsilon_{E_{ES}}$ and in density $\Delta \rho_{ES,ES_{ref}}<\epsilon_{\rho_{ES}}$ as explained in Sec. \ref{sec:numerics:dressed:convergence:HF:HFvsHFbasis}. Here, we were typically able to reach $\epsilon_{E_{ES}}=10^{-8}$ and $\epsilon_{\rho_{ES}}=10^{-4}$.\footnote{Note that in Sec. \ref{sec:numerics:dressed:convergence:HF:HFvsHFbasis}, we wrote instead $\Delta \rho_{ES,ES_{ref}}\approx 10^{-5}$ because some values were slightly larger than $10^{-5}$. Thus, $\epsilon_{E_{ES}}=10^{-4}$ for sure is satisfied.}
		Again, these criteria can be relaxed.
		\item Compare the converged HF$_{basis}$ and the electronic HF results in energy and density and make sure that both are consistent on their level of accuracy.
	\end{itemize}
	\item Basis size convergence of the dressed theory that is wanted (dHF or dRDMFT) in the no-coupling ($\lambda=0$) limit 
	\begin{itemize}
		\item Perform an $ES$-series like for HF$_{basis}$. Note that one needs to expect considerably larger basis sets for the same level of convergence (see Sec. \ref{sec:numerics:dressed:convergence:HF:dHF_no_coupling} for details).
		\item Check consistency of the electronic sub-part of the system with HF$_{basis}$ as mentioned in Sec. \ref{sec:numerics:dressed:convergence:HF:dHF_no_coupling}. If this check fails drastically, this is very probably due to the violation of the extra exchange symmetry in the photonic coordinates (see Sec.~\ref{sec:dressed:est:prescription}). 
	\end{itemize}
	\item The convergence study is finished with another basis-set convergence for $\lambda>0$. Typically, we also performed another small box length series with the converged basis set to make sure that the coupling does not increase the size of the system crucially such that boundary effects could influence the results.
\end{enumerate}
	
	\bibliographystyle{phd_thesis_style} 
	\bibliography{thesis_literature}
	
\end{document}